\documentclass[12pt,german,english,a4paper
  ,openright,twoside 
]{report}

\usepackage[ngerman, english]{babel}	
\usepackage[latin1]{inputenc}		
\usepackage[T1]{fontenc}				
\usepackage{graphicx}						
\usepackage{natbib}
\usepackage{url} 
\usepackage{mathptmx}
\usepackage[scaled=.90]{helvet}
\usepackage{courier}
\usepackage{amsmath}
\usepackage{amsfonts} 
\usepackage[margin=9pt,font={small},labelfont=bf]{caption}[2003/12/20]
\usepackage{listings}
\usepackage{hyphenat}
\usepackage{verbatim}
\usepackage[totalwidth=16cm,totalheight=23.7cm, dvips,pdftex, vcentering]{geometry}
\usepackage{fancyhdr}

\newcommand{\dcauthorsurname}{Breitling} 
\newcommand{\dcauthorname}{Frank}

\newcommand{\dctitle}{Detection of VHE Gamma Radiation from the Pulsar Wind
Nebula MSH\,15-5{\it2} with H.E.S.S.}

\newcommand{\dcapprovala}{} 
\newcommand{\dcapprovalb}{} 
\newcommand{\dcapprovalc}{} 

\newcommand{\dcfaculty}{Mathematisch-Naturwissenschaftlichen Fakult\"at I}

\newcommand{\dcdean}{Prof. Dr. Christian Limberg}
\newcommand{\dcpresident}{Prof. Dr. Dr. h.c. Christoph Markschies}

\newcommand{\dcdatesubmitted}{2008-04-02} 
\newcommand{\dcdateexam}{} 

\newcommand{\dckeydea}{MSH 15-52, PSR B1509-58, G320.4, RCW 89, HESS J1514-591}
\newcommand{\dckeydeb}{Gammaastronomie, abbildende atmosph\"arische Cherenkov-Teleskope, H.E.S.S.} 
\newcommand{\dckeydec}{Pulsarwind-Nebel, Plerion}
\newcommand{\dckeyded}{Bildentfaltung, Richardson-Lucy Algorithmus}

\newcommand{\dckeywordsde}{\vfill \raggedright {\textbf{Schlagw\"orter:}}\\ \dckeydea, \dckeydeb, \dckeydec, \dckeyded \\}

\newcommand{\dckeyena}{MSH 15-52, PSR B1509-58, G320.4, RCW 89, HESS J1514-591}
\newcommand{\dckeyenb}{Gamma-ray astronomy, imaging atmospheric Cherenkov
  telescopes, H.E.S.S.}
\newcommand{\dckeyenc}{pulsar wind nebula, plerion}
\newcommand{\dckeyend}{image deconvolution, Richardson-Lucy algorithm}

\newcommand{\dckeywordsen}{\vfill \raggedright {\textbf{Keywords:}}\\ \dckeyena, \dckeyenb, \dckeyenc, \dckeyend \\}

\newcommand{\dcpdfsubject}{}

\usepackage{ifpdf}

\ifpdf

\usepackage[%
	pdftitle={\dctitle},
	pdfauthor={\dcauthorsurname\ \dcauthorname},
	pdfsubject={\dcpdfsubject}, 
	pdfkeywords={\dckeydea, \dckeydeb, \dckeydec, \dckeyded},
	pdfpagemode=UseOutlines,
  colorlinks=true,					
	linkcolor=black,					
	filecolor=black,					
	urlcolor=black,						
	citecolor=black,					
	pdftex=true,              
	plainpages=false,         
	hypertexnames=false,      
	pdfpagelabels=true,       
	hyperindex=true]{hyperref}

\usepackage{epstopdf}           
\DeclareGraphicsExtensions{.pdf,.eps,.jpg,.png} 

\else
	\newcommand{\texorpdfstring}[2]{#1}
\fi

\newcommand{\X}{X\texorpdfstring{\protect\nobreakdash}{}}
\newcommand{\g}{\texorpdfstring{\ensuremath{\gamma}\protect\nobreakdash}{Gamma}}
\newcommand{\MSH}{MSH 15\ensuremath{-}5{\it2}}
\newcommand{\PSR}{PSR B1509\ensuremath{-}58}
\newcommand{\SNR}{G320.4\ensuremath{-}1.2}
\newcommand{\HESS}{HESS J1514\ensuremath{-}591}
\newcommand{\RCW}{RCW 89}

\newcommand\fs{\mbox{$.\!\!^{\mathrm s}$}} 
\newcommand\farcs{\mbox{$.\!\!^{\prime\prime}$}} 
\newcommand\farcm{\mbox{$.\mkern-4mu^\prime$}} 

\begin{document}
\renewcommand{\thefootnote}{\arabic{footnote})}
\pagenumbering{roman}


%
%
%
%

\author{Frank Breitling \\\small{fbreitling (at) aip.de}}
\date{2007-09-03
\vspace{2.5cm}
\begin{figure}[h!]
  \centering
    \includegraphics[width=.8\textwidth]{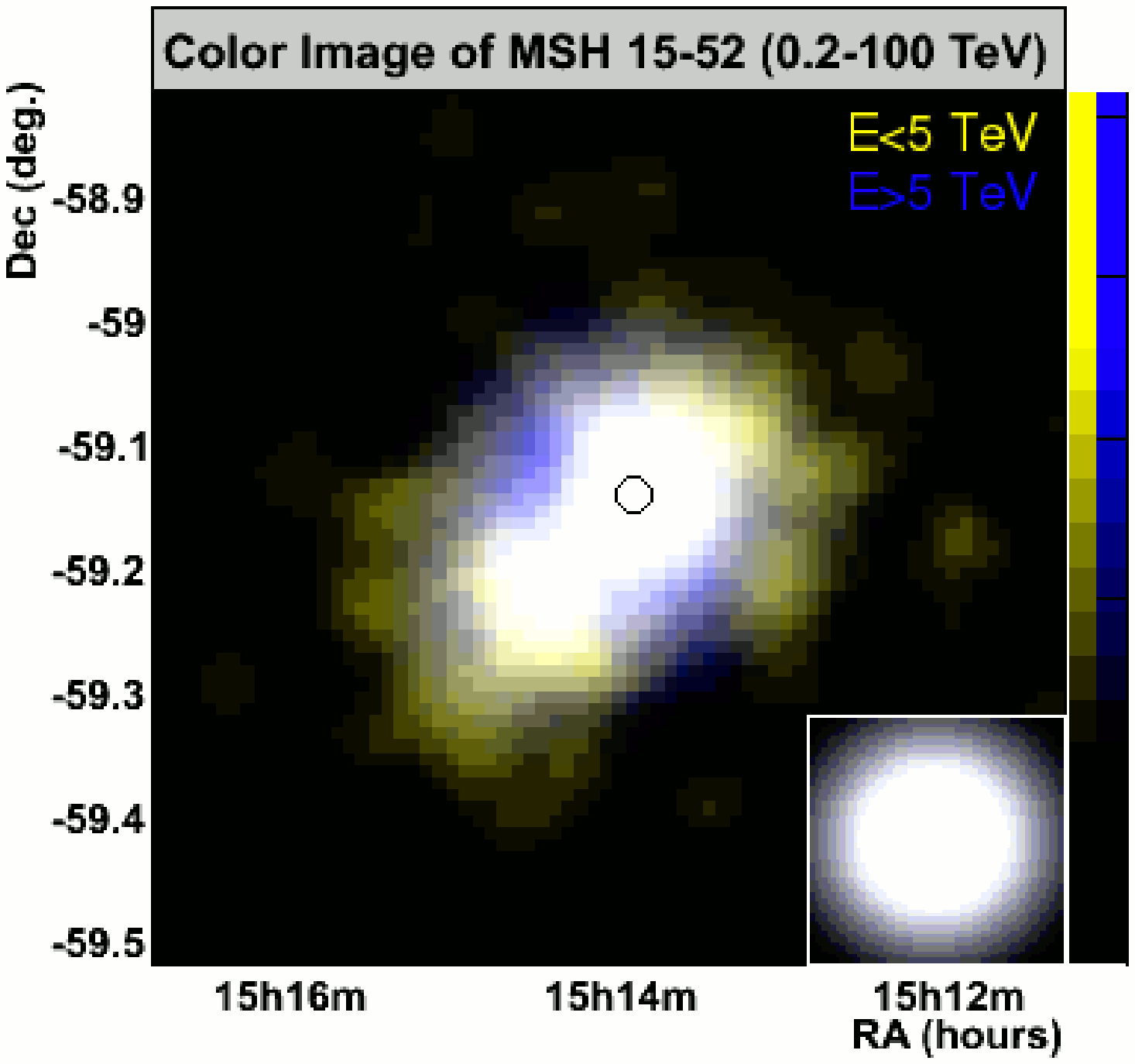}
\end{figure}
}
\title{ \vspace{-2cm} \dctitle \vspace{1cm} }

\maketitle


\selectlanguage{english}

\begin{abstract} \setcounter{page}{2}
This work reports on the discovery of \HESS, a VHE \g-ray source found at the
pulsar wind nebula (PWN) \MSH\ and its associated pulsar \PSR. The discovery
was made with the High Energy Stereoscopic System (H.E.S.S.), which currently
provides the most sensitive measurement in the energy range of about
0.2--100\,TeV. This analysis is the first to include all H.E.S.S. data from
observations dedicated to \MSH. The data was taken in 2004 from March 26 to
July 20, with a total live-time of 26.14\,h. The \g-ray signal was detected
with a statistical significance of 32 standard deviations. The intensity
distribution shows an elliptical extension with the major axis oriented in a
southeast direction. The standard deviations of a Gaussian fit function are
$\rm 6.5'\pm0.5'_{stat}\pm0.1'_{syst}$ and $\rm
2.3'\pm0.4'_{stat}\pm0.1'_{syst}$ for the major- and minor axis, respectively.
The \g-ray emission extends in direction of the pulsar jet, previously resolved
in \X-rays. This becomes more apparent after image deconvolution. The emission
region along the jet axis decreases with increasing energy. The corresponding
flux above 1\,TeV is $\rm (4.4\pm0.2_{stat}\pm1.0_{syst}) \times
10^{-12}cm^{-2}s^{-1}$. The energy spectrum obeys a power law with a
differential flux at 1\,TeV of $\rm (5.8\pm0.2_{stat}\pm1.3_{syst}) \times
10^{-12}cm^{-2}s^{-1}TeV^{-1}$ and a photon index of $\rm
2.32\pm0.04_{stat}\pm0.10_{syst}$. The \g-ray light curve with periodicity
according to \PSR\ yields a uniform distribution. An upper limit of 11.0$\rm
\times 10^{-12} cm^{-2}s^{-1}$ for the pulsed \g-ray flux from \PSR\ was
calculated with a confidence level of 99\%.

In addition to these results the following subjects are discussed: previous
observations of \MSH\ and \PSR, the theory of pulsars, PWNs and their \g-ray
production, the imaging air Cherenkov technique for the detection of
\g\ radiation in the earth's atmosphere, the H.E.S.S. experiment and its data
analysis, the first (Richardson-Lucy) deconvolution of VHE gamma-ray maps, the
analysis of H.E.S.S. data for pulsed emission from pulsars using radio
ephemeris.

The results are discussed within the framework of PWNs and are explained by
inverse Compton scattering of leptons. A hadronic component in \MSH\ is not
excluded, but its \g-ray emission would not be significant. Moreover, it is
concluded that advection is the dominant transport mechanism over diffusion in
the magnetized flow of the pulsar wind from \PSR. A correlation analysis with
the Chandra \X-ray data suggests that the \g\ radiation is emitted from the
region of \PSR, but not from the neighboring optical nebula \RCW.

\dckeywordsen				
\end{abstract}

\selectlanguage{english}				


\chapter*{Dedication}
To whom it may concern


\tableofcontents

\listoffigures

\listoftables

\clearpage 
\pagenumbering{arabic}

\pagestyle{fancy}
\setlength{\headheight}{14.5pt}
\renewcommand{\chaptermark}[1]{
  \markboth{\MakeUppercase{\chaptername}~\thechapter\ \ #1}{}}
\renewcommand{\sectionmark}[1]{\markright{\thesection\ \ #1}}


\hbadness=10000
\setlength{\abovecaptionskip}{.2cm}
\chapter{Introduction}

\begin{quote} \it
  ``We owe our existence to stars, because they make the atoms of which we are
  formed. So if you are romantic you can say we are literally starstuff. If
  you're less romantic you can say we're the nuclear waste from the fuel that
  makes stars shine.

  We've made so many advances in our understanding. A few centuries ago, the
  pioneer navigators learnt the size and shape of our Earth, and the layout of
  the continents. We are now just learning the dimensions and ingredients of
  our entire cosmos, and can at last make some sense of our cosmic habitat.''
  \begin{flushright} \rm
    --- Sir Martin Rees, British astrophysicist and president of the Royal
      Society
  \end{flushright}
\end{quote}
\bigskip
Astronomy is and always has been a central discipline of natural science,
driven by fundamental questions and exiting answers. The first recorded
astronomical achievements date back to early cultures such as the Babylonians,
Egyptians and Chinese. Further progress was made in the Renaissance, when the
heliocentric model of the solar system was proposed by Nicolaus Copernicus,
Galileo Galilei and Johannes Kepler. The use of the telescope for astronomic
observations by Galilei marks the beginning of experimental astronomy. Since
then, astronomy has evolved rapidly. For example, the introduction of
spectroscopy and photography by Joseph Fraunhofer in 1814 laid the foundations
for a ``New Astronomy'' and astrophysics by providing the means for determining
the chemical composition of astronomical objects. Moreover, it paved the way
for the determination of red shifts by Vesto Slipher in 1912, which allowed for
such far-reaching conclusions as the expansion of the universe by Hubble in
1929. An astrophysical revolution began in the second half of the 20th century,
when new types of telescopes became available, owed to technological advances,
with which the full range of the electromagnetic spectrum could be explored.
Radio telescopes permitted the discovery of the cosmic microwave background
radiation in 1965 and pulsars in 1967, both of which were honored with the
Nobel prize; infrared telescopes revealed the view through vast dust clouds to
previously hidden objects; \X- and \g-ray satellites provided pictures of the
non-thermal universe and its most violent processes, such as \g-ray bursts
first observed in 1967, active galactic nuclei or super nova remnants; also new
experiments in the rising field of astroparticle physics provided fresh
insights into the non-thermal universe, e.g. Kamiocande, the
Irvine-Michigan-Brookhaven detector and the scintillator experiment at Baksan
by the detection of neutrinos from the supernova SN~1987A. Although these
examples only name some of astrophysical milestones, they demonstrate that the
exploration of new fields of astronomy can lead to outstanding discoveries with
great impact on the understanding of the universe.

In this respect, imaging atmospheric Cherenkov telescopes (IACT) have been
developed for the exploration of the very high energy (VHE) \g-ray sky which
extends from 10\,GeV to 100\,TeV. Therefore, IACTs currently provide a window
to the highest available \g-ray energies. Since this \g\ radiation is produced
where highly accelerated particles interact with their environment, TeV \g\
radiation provides important information about the acceleration mechanisms for
the primary particles. In comparison, \g-ray astronomy using IACTs is a
relatively young field which achieved its breakthrough in 1989 with the
discovery of TeV \g\ radiation from the Crab Nebula by the Whipple
collaboration. Since then, IACT based \g\ astronomy has progressed
significantly. While it took several weeks for the first detection of the Crab
Nebula which is known as the strongest TeV \g-ray source, nowadays IACTs can
detect the Crab Nebula within less than a minute. Moreover, while previously
only a handful of TeV sources could be detected, today about 50 TeV \g-ray
sources have been established and almost monthly the detection of a new source
is reported. Many of these new detections are owed to the High Energy
Stereoscopic System (H.E.S.S.), which is currently one of the most sensitive
IACTs worldwide. During its first few years of operation, starting in 2002,
H.E.S.S. has detected or confirmed more than 40 sources of TeV \g\ radiation
(cf. \citet{hofmann_ICRC2005} and \citet{PlaneScanI}).

One of these sources is the pulsar wind nebula (PWN) \MSH, at a distance of
about 5.2\,kpc from earth. PWNs, also called Plerions, are very unique but also
very rare objects of which only about 50 have been identified, all within the
Galaxy. The central object in a PWN is a pulsar which exposes extreme condition
to its environment and generates \g\ radiation in its vicinity, in particular
by the emission of a wind of VHE particles. The pulsar associated with \MSH\ is
\PSR, which is one of the most energetic pulsars known. Many observations of
\MSH\ have been conducted from radio to \g-ray energies since its discovery in
1961. They have shed light on the violent emission processes. However, the
observations at the highest energies in the TeV range, which are crucial for
the understanding of the processes in \MSH\ and in PWNs in general, were rather
incomplete, since no experiment had been able to provide them. Therefore, when
H.E.S.S. came into operation with its unprecedented sensitivity, \MSH\ was
scheduled as one of the first targets for thorough observation over a period of
several month. This work presents this data from the first H.E.S.S. studies. It
discusses details of the analysis, the detection, and further results as well
as possible implications for the astrophysical processes involved. As it turns
out, \MSH\ is one of the strongest TeV \g-ray sources that have ever been
detected. For a comprehensive discussion, also an introduction to the system of
\MSH\ and \PSR, the imaging atmospheric Cherenkov technique and the H.E.S.S.
experiment is given in advance in the following chapters.

\chapter{\MSH\ and \PSR} \label{chp:MSHandPSR}

\section{Review of Previous Observations}
The supernova remnant (SNR) \MSH, also known as \SNR, was first discovered by
\cite{Mills:1961} as an extended radio source. Later, radio observations by
\cite{Caswell:1981} did resolve individual, non-thermal radio features of the
SNR, one of them coinciding with the optical nebula \RCW\ in the northwest.
X-ray observations in the energy range from 0.2-4\,keV with the Einstein
satellite by \cite{Seward:1982} revealed a possibly associated pulsar with an
increasing period of about 150\,ms, a characteristic age of about 1.6\,kyr and
a surrounding nebula. Calculations showed that the nebula could easily be
powered by the pulsar. Subsequent radio observations by \cite{Manchester:1982}
confirmed the existence of the pulsar \PSR. Nevertheless, it was still unclear
whether \MSH\ and \PSR\ are associated and whether \MSH\ was a pulsar wind
nebula (PWN) similar to the Crab Nebula. A main argument against an association
was the unexplained difference between the apparent age of the pulsar and the
supernova remnant. However, these questions have been resolved in later years
in favor of an association. In 1983 \cite{Seward1983} reported on infrared
spectral lines in the region of \RCW\ and provided an optical image in which
\RCW\ and a new northerly filament became apparent. The detection of TeV
\g\ radiation from the region of \MSH\ was reported by \cite{Sako:2000}.
\cite{Caraveo:1994} also suggested an optical counterpart of \PSR. However,
\cite{Kaplan:2006} found a more likely infrared counterpart which was hidden by
the object proposed by \cite{Caraveo:1994}.

A historical association of \MSH\ with the {\it guest star} \footnote{The
ancient Chinese term for a star that newly appears and is visible for a short
time.} of AD 185 December 7, which was witnessed by Chinese astronomers
according to the Houhanshu, \footnote{The Houhanshu are Chinese records of the
Later Han dynasty.} has been pointed out by \cite{Strom:1994} and
\cite{Schaefer:1995}. The age and position of the {\it guest star} would allow
for an interpretation of it as the supernova of \MSH.

\begin{figure}[tbh]
  \centering
  \includegraphics[width=.67\textwidth]{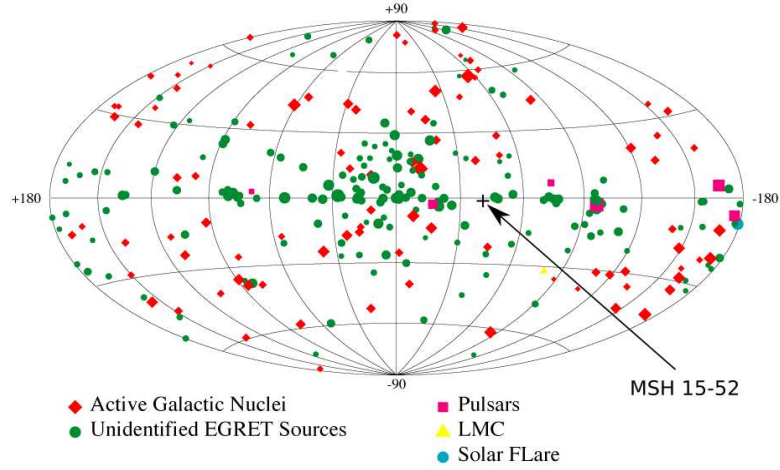}
  \caption[Galactic Map Showing \MSH]{Galactic map of the Third EGRET Catalog
    overlaid with the location of \MSH. (Figure taken from
    \cite{EGRET:3rdCatalog}.)}
  \label{fig:EgretCatalog}
\end{figure}

\MSH\ is located in the galactic plane with an offset of $40^\circ$ from the
galactic center. Fig.~\ref{fig:EgretCatalog} shows this location overlaid to
the galactic map of the Third EGRET Catalog, which was obtained by the
Energetic Gamma Ray Experiment Telescope (EGRET) and which covers the energy
range from 100\,MeV to 30\,GeV. \MSH\ does not coincide with any of the sources
shown. The estimated distance of \MSH\ from earth is 5.2\,kpc.

Tbl.~\ref{tbl:MSHPrameters} and~\ref{tbl:PSRPrameters} summarize basic
parameters of \MSH\ and \PSR\ which have been determined from observations at
different wavelengths discussed below.

\begin{table}[ht]
  \centering
  \caption[Parameters of \MSH]{Parameters of the supernova remnant \MSH\ as
    found by \cite{Gaensler:1999}.}
  \bigskip
  \begin{tabular}{lcc}
    \hline \hline
    Parameter                    & Variable    & Value \\
    \hline
    Celestial coordinates        & RA (J2000)  & $\rm 15^h14^m27^s$ \\
                                 & Dec (J2000) & $-59^{\circ}16\farcm3$ \\
    Galactic coordinates         & $l$         & 320.31$^{\circ}$ \\
                                 & $b$         & $-1.31^{\circ}$ \\
    Diameter                     & $D$         & 5' \\
    Age                          & $T_N$       & (2-20)$\times10^{3}$\,yr \\
    Distance                     & $d$         & ($5.2\pm1.4$)\,kpc \\
    Magnetic field of the PWN    & $B_{PWN}$   & $\sim$5-8\,$\mu$G \\
    \hline \hline
  \end{tabular}
  \label{tbl:MSHPrameters}
\end{table}

\begin{table}[ht]
  \centering
  \caption[Parameters of \PSR]{Parameters of \PSR\ for epoch 48355.0000\,MJD
    determined by \cite{Kaspi:1994} from radio observations at 843\,MHz with
    the Molonglo Observatory Synthesis Telescope (MOST, \cite{Mills:1981}). A
    $^*$ marks parameters found by \cite{Gaensler:2002}.}
  \bigskip
  \begin{tabular}{lcc}
    \hline \hline
    Parameter                    & Variable   & Value \\
    \hline
    Celestial coordinates        & RA (J2000) & $\rm15^h13^m55\fs62\pm0\fs09$\\
                                 & Dec (J2000)& $-59^{\circ}08'09\farcs0\pm1\farcs0$ \\    
    Period                       & $P$        & 151\,ms \\
    Frequency                    & $f$        & 6.6375697328(8)\,s$^{-1}$ \\
    First Frequency derivative   & $\dot{f}$  & $-6.7695374(4)\times10^{-11}$s$^{-2}$ \\
    Second Frequency derivative  & $\ddot{f}$ & $1.9587(9)\times10^{-21}$s$^{-3}$ \\
    Third Frequency derivative   & $\dddot{f}$& $-1.02(25)\times10^{-31}$s$^{-4}$ \\
    Breaking index               & $n$        & $2.837\pm0.001$ \\
    Second deceleration parameter& $m$        & $14.5\pm3.6$ \\
    Dispersion measure           &            & $(253.2\pm1.9)$\,pc cm$^{-3}$\\
    Characteristic age           & $T_c$      & 1700\,yr \\
    Spin-down luminosity$^*$     & $\dot{E}$  & $1.8\times10^{37}$\,erg\,s$^{-1}$ \\
    Surface magnetic field$^*$   & $B_p$      & 1.5$\times10^{13}$\,G \\
    \hline \hline
  \end{tabular}
  \label{tbl:PSRPrameters}
\end{table}
\clearpage

\subsection{\MSH}
\subsubsection{Radio Observations}
\begin{figure}[hb!]
  \begin{minipage}[c]{0.47\linewidth}
    \includegraphics[width=\textwidth]{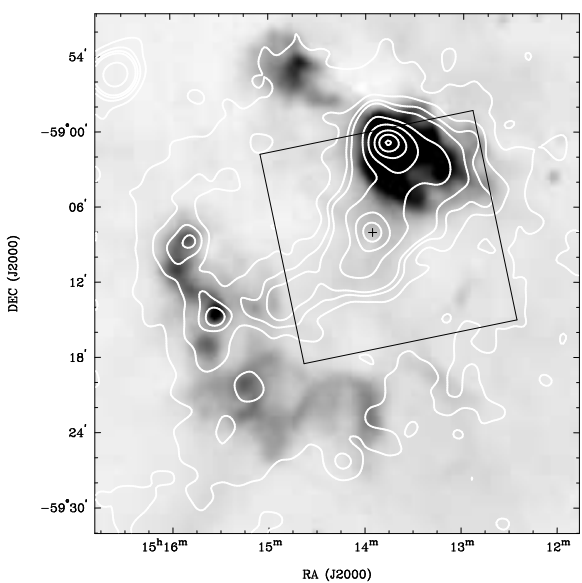}
  \end{minipage}\hfill
  \begin{minipage}[c]{0.47\linewidth}
    \caption[Radio and \X-ray Image of \MSH]{Radio and \X-ray image of \MSH.
    The gray scale corresponds to the MOST radio observation at 843\,MHz by
    \cite{Whiteoak:1996}. The white contour lines at the levels of 0.5, 1, 1.5,
    2, 5, 10, 20, 30 and 40 in arbitrary units represent the ROSAT observations
    from \cite{Trussoni:1996}. The position of \PSR\ is marked by the cross.
    The black box marks the Chandra ACIS-I field of view. (Figure taken from
    \cite{Gaensler:2002}.)}
    \label{fig:Gaensler2002Radio}
  \end{minipage}
\end{figure}

The radio observations of \MSH\ show an unusual appearance of two separated
regions with non-thermal emission. The situation is illustrated in
Fig.~\ref{fig:Gaensler2002Radio}, which shows the radio map from
\cite{Whiteoak:1996} overlaid with \X-ray contours from \cite{Trussoni:1996}.
While the southern region approximates a partial shell, the northern region is
rather compact and coincides with the optical nebula \RCW, which contains an
unusual ring of radio clumps. From the region of \PSR\ no signal is seen. With
standard parameters for the supernova and the interstellar medium, the size of
the SNR suggests an age of $\sim$6-20\,kyr, which is an order of magnitude
larger than the spin-down age of the pulsar of 1.7\,kyr and which is in
contradiction to the association of \MSH\ and \PSR. However,
\cite{Gaensler:1999} have confirmed by H\,{\sc i}
\footnote{H\,{\sc i} and H\,{\sc ii} denote neutral and ionized hydrogen,
respectively. H\,{\sc i} measurements look for the 21\,cm line of the forbidden
hyper fine transition. It is used to determine the density and velocity of
hydrogen.} absorption measurements that the observed components are part of a
single SNR.

\subsubsection{Infrared Observations}
\cite{Seward1983} have detected infrared spectral lines of many elements from
the region of \RCW. Among them Fe\,{\sc ii} at 1.76\,$\mu$m, which was detected
for the first time from an SNR. The spectrum is clearly non-thermal and typical
for a reddened, high density and collisionally excited nebula at a distance of
5\,kpc.

\subsubsection{Optical Observations}
\begin{figure}[ht!]
  \begin{minipage}[c]{0.43\linewidth}
    \includegraphics[width=\textwidth]{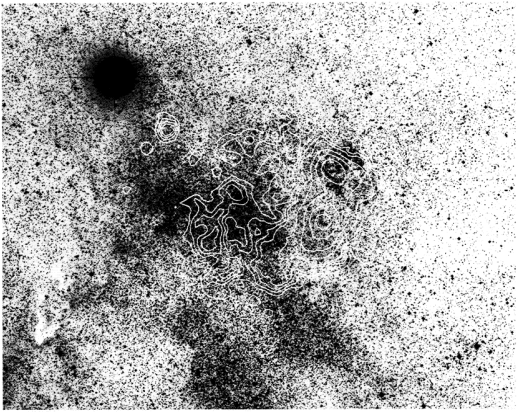}
  \end{minipage}\hfill
  \begin{minipage}[c]{0.5\linewidth}
    \caption[Optical Image of \MSH]{Optical image of the region of \MSH\ from a
    90\,min R-band ($\lambda=$610-700\,nm) plate overlaid with \X-ray contours
    from the Einstein satellite. The region of \RCW\ and a filament $9'$ to the
    northeast are apparent. (Image taken from \cite{Seward1983}.)}
    \label{fig:Seward1983}
  \end{minipage}
\end{figure}

Fig.~\ref{fig:Seward1983} from \cite{Seward1983} shows the area of \MSH\ from a
90\,min R-band ($\lambda=$610-700\,nm) plate overlaid with the \X-ray contours
from the Einstein satellite. The region of \RCW\ and a filament $9'$ to the
northeast are apparent. The filament coincides with the radio arc of \MSH,
which extends eastwards from \RCW. Due to the good agreement with the radio
observations, \cite{Seward1983} concluded that the whole northwest arc is a
single entity. They also pointed out that the association of the SNR and \PSR\
can still be explained if a local low density of the interstellar medium is
assumed, in which the SNR could have expanded unusually rapidly. The low
density could be explained by an earlier supernova in the same region.

\subsubsection{X-ray Observations} \label{sec:X-rays}
The \X-ray emission from \MSH\ has several components. The component in the
region of \PSR\ shows highest intensity and pulsed emission. Another component
is elongated and extending southeast of the pulsar. They have been identified
as a PWN and as a jet by several authors including \cite{Gaensler:2002} and
\cite{Goret:2006}. A third component is in the region of \RCW, which also
reaches a considerable peak intensity, but in contrast to the former has a
thermal spectrum. There is also a faint diffuse \X-ray emission throughout the
whole region of the SNR, which can be considered as a fourth component.
According to \cite{Trussoni:1996}, the spectrum is compatible with thermal and
non-thermal emission and could therefore correspond to thermal emission from
the SNR blast wave (\cite{Gaensler:2002}).

\begin{figure}[tb!]
  \begin{minipage}[c]{0.52\linewidth}
  \setlength{\captionmargin}{.0cm}
    \includegraphics[width=\textwidth]{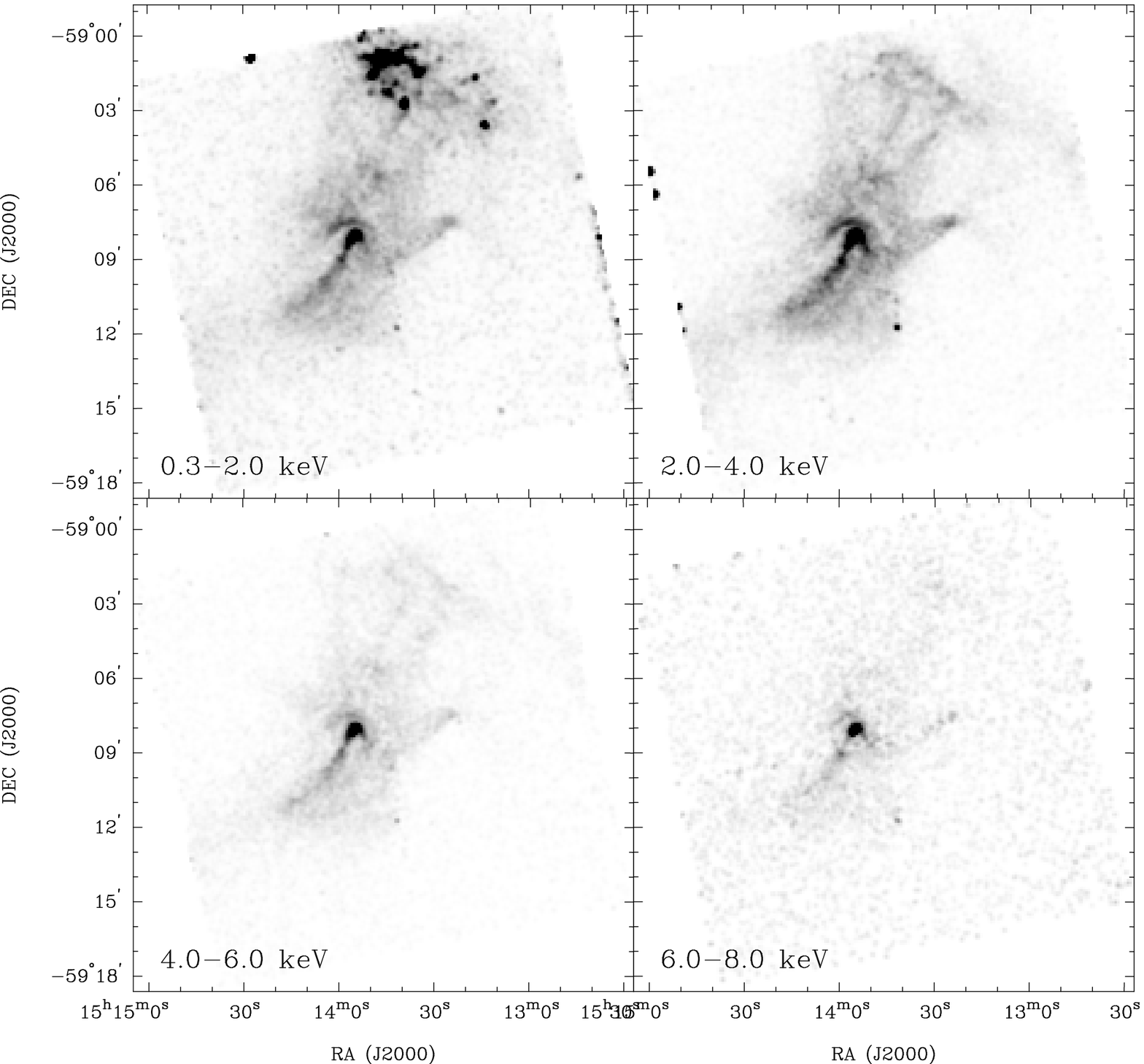}
    \caption[Chandra \X-ray Images of \MSH\ for different Energy Bands]{Chandra
      \X-ray images of \MSH\ showing different energy bands. The images are
      exposure corrected, convolved with a Gaussian of FWHM of $10''$ and
      displayed using a linear transfer function. (Figure taken from
      \cite{Gaensler:2002}.)}
    \label{fig:Gaensler2002}
  \end{minipage}\hfill
  \begin{minipage}[c]{0.43\linewidth}
    \setlength{\captionmargin}{.0cm}
    \setlength{\abovecaptionskip}{.4cm}
    \includegraphics[width=\textwidth]{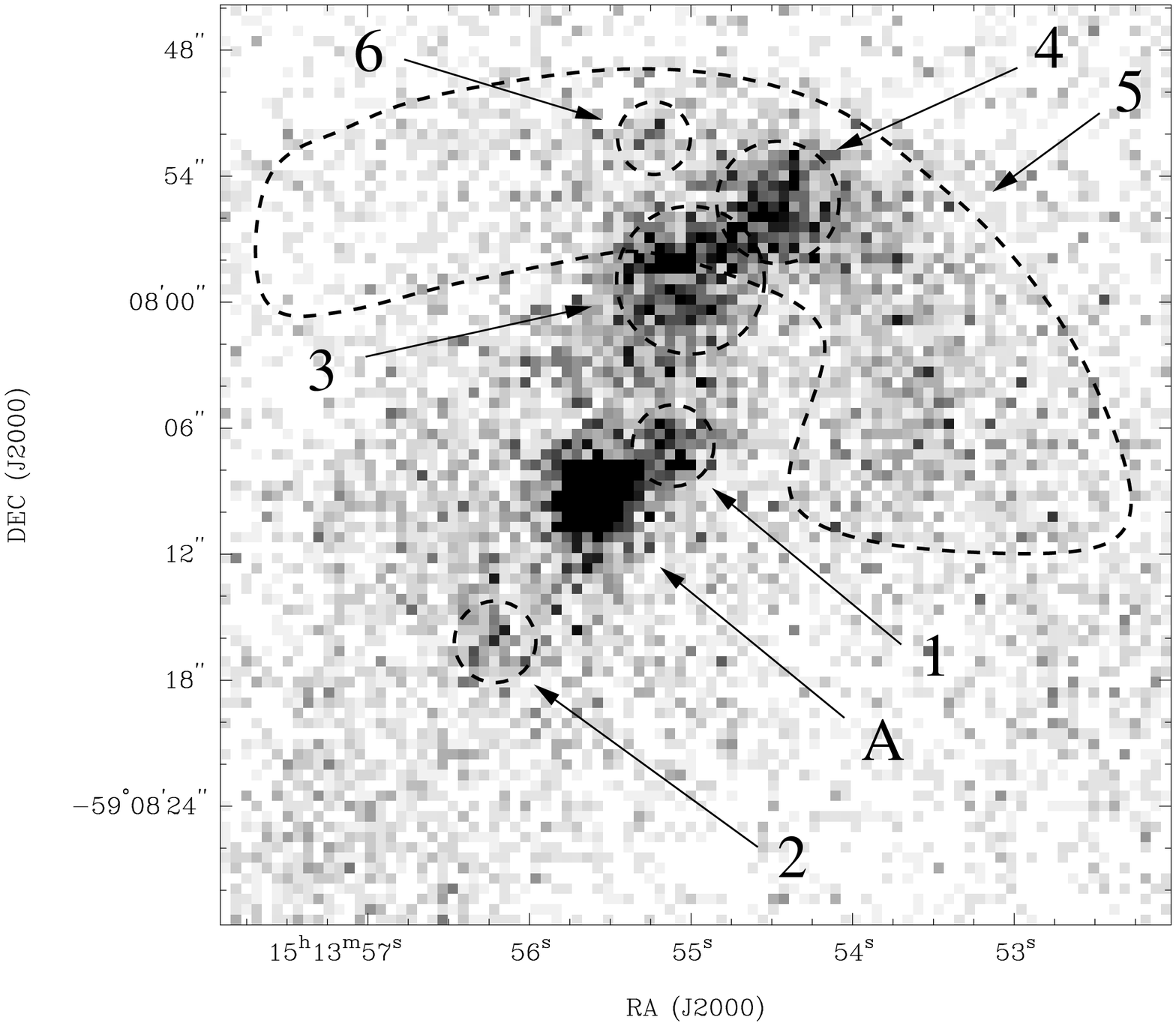}\hfill
    \caption[High Resolution Chandra \X-ray Image of \MSH]{High resolution
      Chandra \X-ray image of \MSH\ showing the immediate surrounding of \PSR\
      (corresponding to feature A) in the energy range 0.3-8.0\,keV. The image
      is exposure corrected but no smoothing was applied. The labeled features
      are discussed in the text. (Figure taken from \cite{Gaensler:2002}.)}
    \label{fig:Gaensler2002f5}
  \end{minipage}
\end{figure}

The four Chandra satellite images of different energy bands in
Fig.~\ref{fig:Gaensler2002} from \cite{Gaensler:2002} provide detailed insights
into the structure of the PWN and the energy spectrum of the individual
components. Clearly, the pulsar has the hardest spectrum, followed by the jet
and the PWN. The region of \RCW\ and the contained \X-ray and radio clumps
disappear with increasing energy. A closer analysis by \cite{Gaensler:2002}
showed that the spectrum in this region is dominated by emission lines, in
contrast to the spectrum in the pulsar region. The energy spectrum of the
diffuse PWN was determined with a photon index of $2.05\pm0.04$.
Fig.~\ref{fig:Gaensler2002f5} shows the same data in a smaller region
surrounding \PSR\ in the energy range from 0.3-8.0\,keV. It reveals several
distinct features which are discussed in \cite{Gaensler:2002}. Feature A
corresponds to the pulsar. Noteworthy is feature 5 which refers to a faint
circular arc of emission. It is approximately centered at the pulsar, with a
radius of $17''$ and an angle subtended at the pulsar of $\sim110^\circ$. 

Moreover \cite{Gaensler:2002} concluded that the PWN has a magnetic field of
$B_{PWN}\sim8\,\mu$G and that the outflow of the jet has a velocity of 0.2$c$,
carrying away at least 0.5\% of the pulsar's spin-down luminosity.

Further studies of the Chandra and ROSAT data led \cite{DeLaney:2006} to the
discovery that bright knots within $20''$ of the pulsar show an even higher
outflow velocity, up to $0.6c$. Also, significant time variability of the
brightness has been found. For example, the brightness of the jet has increased
by 30\% within 9 years.

Recent observations with the INTEGRAL X-ray satellite in 2005 in the energy
range from 20-200\,keV reported by \cite{Goret:2006} predict a similar high
outflow velocity of 0.3-0.5$c$ corresponding to parent electron energies of
400-730\,TeV and a mean magnetic field strength of 22-33\,$\mu$G with a
systematic uncertainty of the later of 27\%. Comparisons with other X-ray
measurements show a jet length $(L)$ scaling with the X-ray energy $(E)$ as
$L\propto E^{-\frac12}$. The analysis also shows the source extension with a
standard deviation of the major axis of $5\farcm53\pm0\farcm07$. The energy
spectrum of the PWN was found to obey a power law with a photon index of
$2.12\pm0.05$ in agreement with the Chandra observations. However, in contrast
to previous measurements by the BeppoSAX satellite the INTEGRAL data suggests a
spectral break near 160\,keV. Also pulsed emission from \PSR\ was clearly
resolved.


\subsubsection{TeV \g-Ray Observations}
\begin{figure}[t!]
  \medskip
  \begin{minipage}[c]{0.52\linewidth}
    \includegraphics[width=\textwidth]{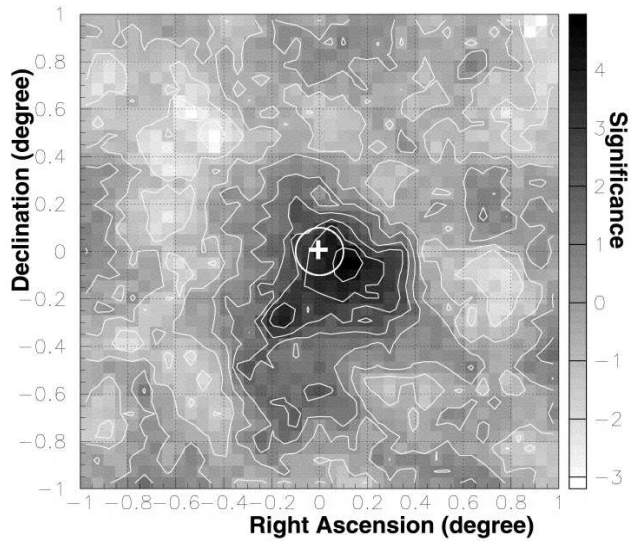}
  \end{minipage}\hfill
  \begin{minipage}[c]{0.44\linewidth}
    \caption[\g-Ray Map of \MSH]{\g-Ray map of \MSH\ showing the first evidence
    for TeV \g\ radiation from this region. (Figure taken from
    \cite{Sako:2000}.)}
    \label{fig:Sako2000}
  \end{minipage}
\end{figure}

\g\ radiation from \MSH\ was first predicted by \cite{DuPlessis:1995} based on
similarities to the Crab Nebula. These were the first prediction of VHE \g\
radiation from a PWN other than the Crab nebula. \cite{Sako:2000} reported the
first evidence for \g\ radiation from \MSH\ based on observations with the IACT
CANGAROO. An excess with a significance of 4.1 standard deviations was found in
the region of \PSR\ (Fig.~\ref{fig:Sako2000}). The corresponding flux above
1.9\,TeV was determined to $(2.9\pm0.7)\times 10^{-12}\rm cm^{-2}s^{-1}$. A
magnetic field strength of $5\mu$G was estimated. No pulsed \g-ray emission was
found.

\subsection{\PSR} \label{sec:PSR1509}


\PSR\ is one of the most energetic pulsar known, having very high spin-down
luminosity and magnetic fields. It is also one of the youngest pulsars, and for
a young pulsar its spin is extraordinarily stable. No glitches have occurred
within 24 years since its discovery in 1982. Owing to this stability,
\cite{Kaspi:1994} have determined the spin parameters with very high precision
from radio observations over six years (Tbl.\ref{tbl:PSRPrameters}). The second
deceleration parameter $m$ is consistent with a constant breaking index and
magnetic moment on timescales of kyr.

\begin{figure}[tb!]
  \vspace{-.3cm}
  \setlength{\captionmargin}{.1cm}
  \begin{minipage}[c]{0.33\linewidth}
    \includegraphics*[scale=.3, viewport=35 30 530 760]{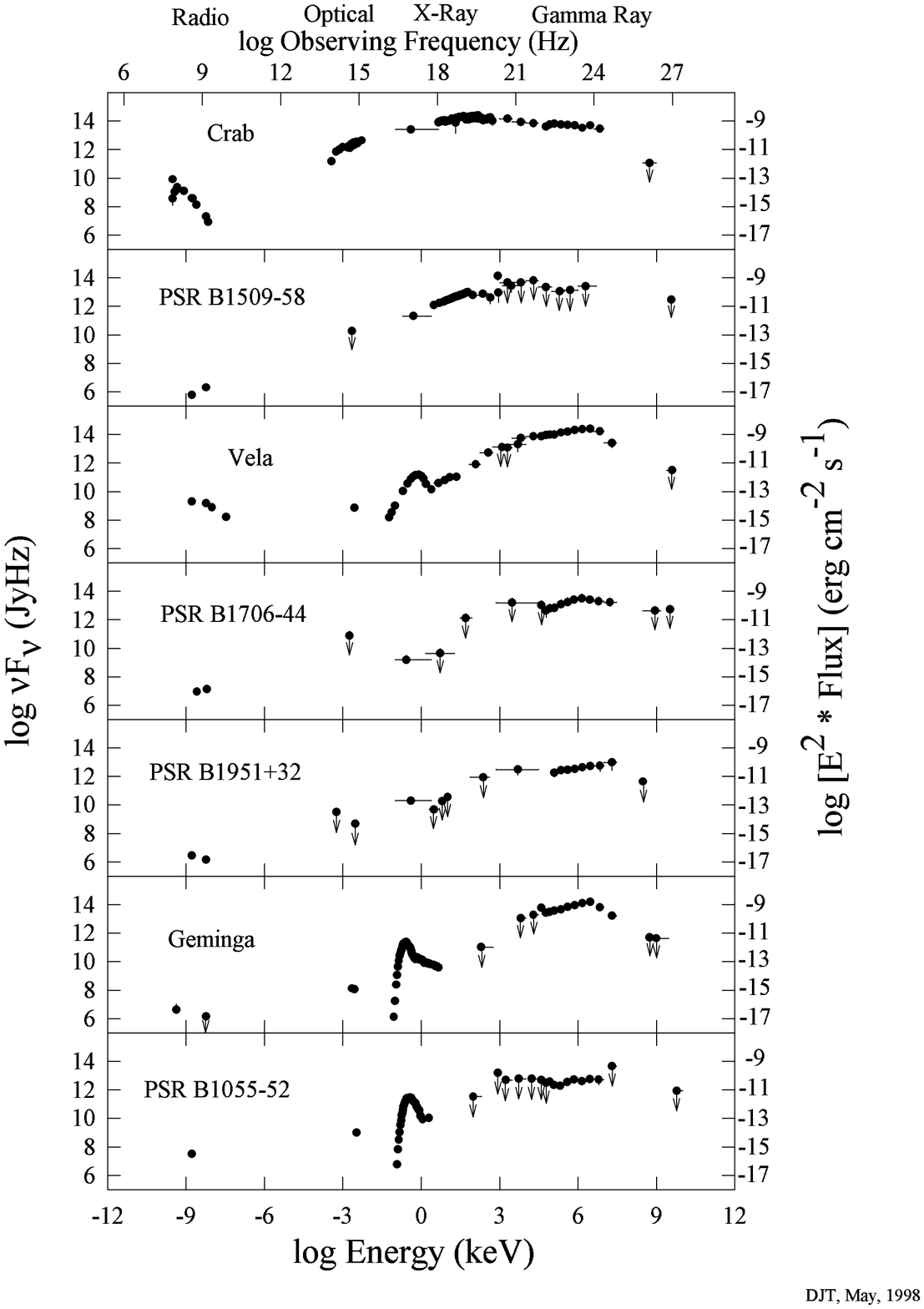}
  \end{minipage}\hfill
  \begin{minipage}[c]{0.32\linewidth}
    \includegraphics[width=\textwidth]{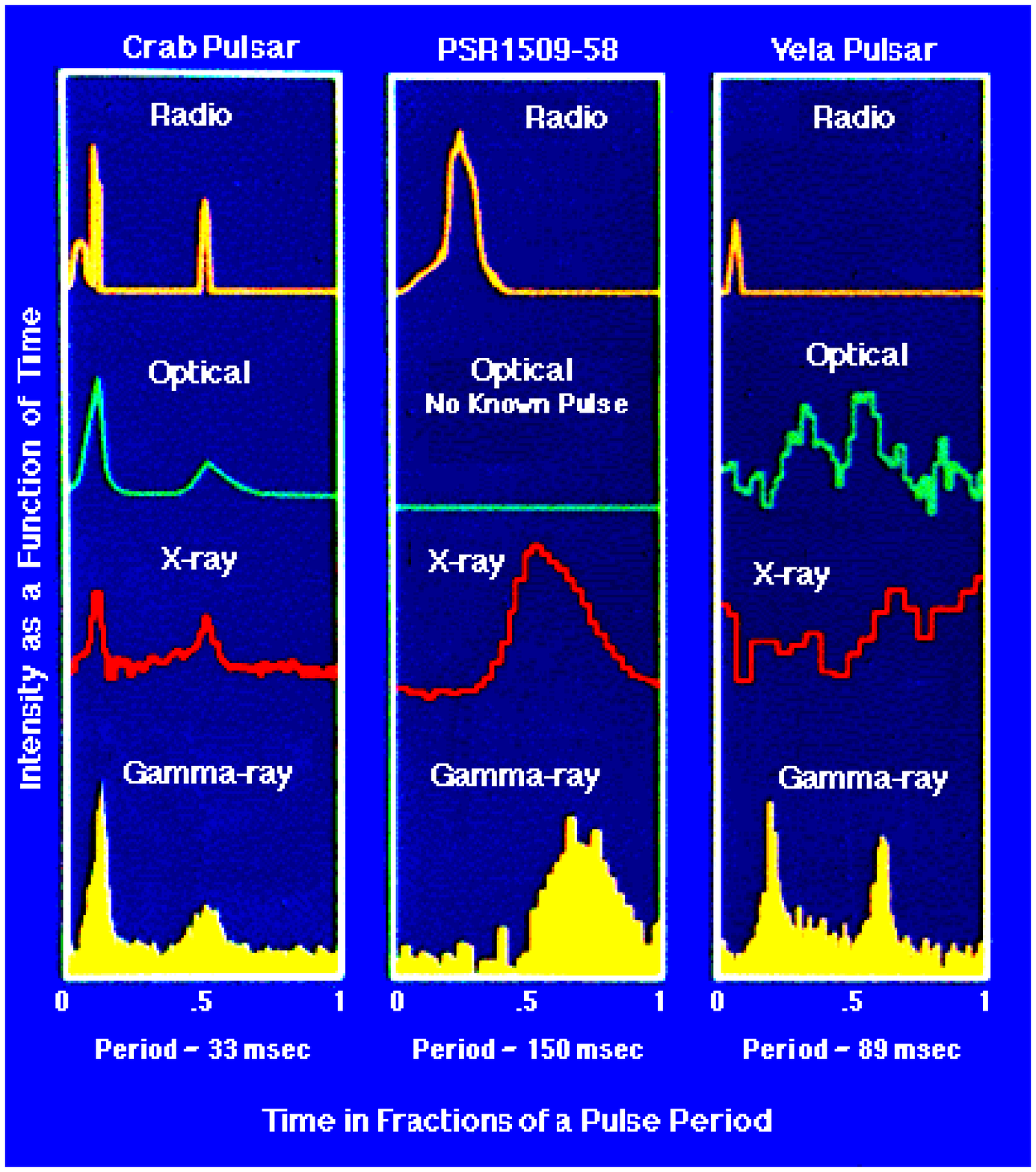}
    \smallskip
  \end{minipage}\hfill
  \begin{minipage}[c]{0.33\linewidth}
    \hspace{.3cm} \vspace{-.7cm}
    \includegraphics[width=.8\textwidth]{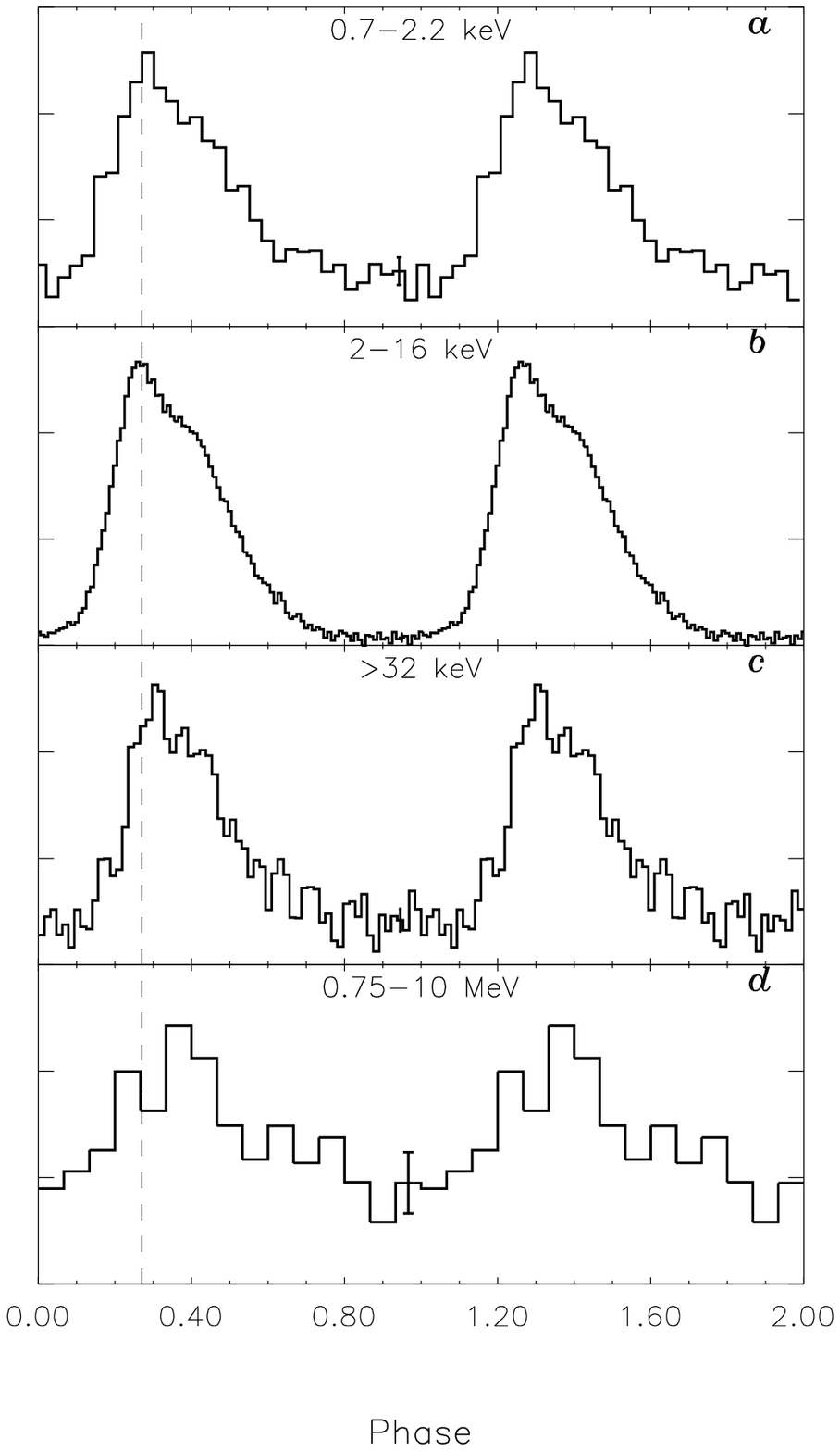}
  \end{minipage}\hfill
  \begin{minipage}[t]{0.32\linewidth}
    \caption[Spectral Energy Distribution of \PSR]{Spectral energy distribution
      of \PSR\ from multiwavelength observation in comparison to other pulsars.
      Note that for \PSR\ the optical data point and the data points at
      energies $>$1\,MeV only indicate upper limits. (Figure taken from
      \cite{Thompson:1999}.)}
    \label{fig:Thompson1999}
  \end{minipage}\hfill
  \begin{minipage}[t]{0.32\linewidth}
    \caption[Light Curves of \PSR, the Crab and the Vela Pulsar]{Light curve of
      \PSR\ in comparison to the light curves of the Crab and the Vela Pulsar.
      Characteristic for \PSR\ is a phase lag increasing with energy and a low
      optical emission. (Figure taken from \cite{NASA:PulsarLightCurves}.)}
    \label{fig:PulsarLightCurves}
  \end{minipage}\hfill
  \begin{minipage}[t]{0.32\linewidth}
    \caption[Light Curves of \PSR\ at High Energies]{Light curves of \PSR\ at
      high energies: 0.7-2.2\,keV (ASCA, \cite{Saito:1997}), 2-16\,keV (RXTE,
      \cite{Rots:1998}), $>$32\,keV (BATSE, \cite{Rots:1998}) and 0.75-10\,MeV
      (COMTEL, \cite{Kuiper:1999}). (Figure taken from \cite{Kuiper:1999}.)}
    \label{fig:Kuiper1999}
  \end{minipage}\hfill
\end{figure}

\PSR\ has been detected from radio to \g-ray energies i.e. in the range from
$10^{-12}$ to $10^7$\,eV. Fig.~\ref{fig:Thompson1999} from \cite{Thompson:1999}
shows the spectral energy distribution in comparison to other known \g-ray
pulsars. Characteristic for \PSR\ is the low optical and \g-ray flux indicated
by the upper limits in Fig.~\ref{fig:Thompson1999}. The light curve of \PSR\ is
shown in Fig.~\ref{fig:PulsarLightCurves} from radio to low \g-ray energies in
comparison to light curves of the Crab and Vela Pulsar. A phase lag is apparent
which increases with energy. A detailed investigation of the light curve of
\PSR\ near its cutoff energy was done by \cite{Kuiper:1999}. With an analysis
of CGRO \footnote{CGRO is the acronym for the Compton Gamma Ray Observatory,
which was a satellite mission from 1991 to 2000. Two of its instruments were
the Imaging Compton Telescope (COMPTEL) and the Energetic Gamma Ray Experiment
Telescope (EGRET).} data from COMPTEL and EGRET, they detected a signal with a
modulation significance of 5.6 standard deviations in the energy range from
0.75-30\,MeV and found the cutoff energy at $\sim$10\,MeV. They also found
indications for a double-peaked profile at \X-ray energies.
Fig.~\ref{fig:Kuiper1999} shows the light curve of this analysis in the energy
range from 0.75-10\,MeV along with the \X-ray light curves. The phase lag of
about 0.3 to the radio phase is also indicated.

According to \cite{Harding:1997}, an explanation for the low \g\ radiation at
high energies could be the extraordinary strong magnetic field, which causes a
spectral cutoff at a few MeV due to photon splitting.

\begin{figure}[b!]
  \begin{minipage}[c]{0.5\linewidth}
    \includegraphics[width=\textwidth]{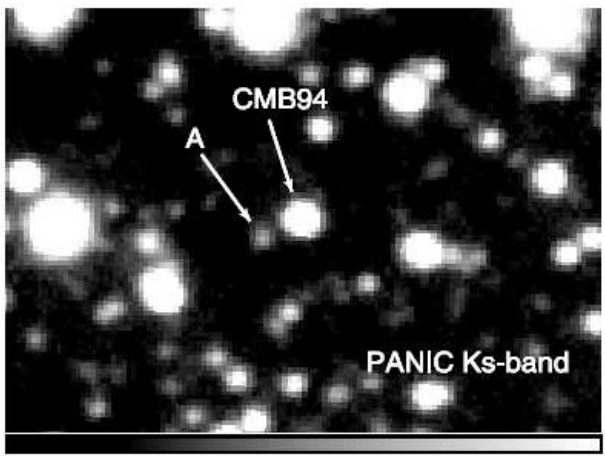}
  \end{minipage}\hfill
  \begin{minipage}[c]{0.47\linewidth}
    \caption[Potential Infrared Counterpart of \PSR]{Potential infrared
      counterpart of \PSR, visible with a magnitude of $\simeq19.4$ on a K$\rm
      _s$-band (1.99-2.30\,$\mu$m) image from the PANIC infrared camera
      reported by \cite{Kaplan:2006}. Source ``A'' is the proposed counterpart.
      The object labeled ``CMB94'' is probably the optical object previously
      discovered by \cite{Caraveo:1994}.}
    \label{fig:Kaplan2006}
  \end{minipage}
\end{figure}

The infrared counterpart of \PSR\ is faint but was probably discovered recently
by \cite{Kaplan:2006}. Fig.~\ref{fig:Kaplan2006} shows the K$\rm _s$-band
(1.99-2.30\,$\mu$m) of these observations with the Persson's Auxiliary Nasmyth
Infrared Camera (PANIC, \cite{Martini:2004}). The source labeled with ''A'' is
the proposed counterpart with a magnitude of $\simeq19.4$, which coincides with
the pulsar's \X-ray position determined by \cite{Gaensler:2002}.
\cite{Kaplan:2006} have also pointed out that the previously proposed optical
counterpart by \cite{Caraveo:1994} is likely to be the object labeled ''CMB94''
and not \PSR. The optical measurement should therefore rather be considered as
an upper limit. So the detection of an optical counterpart remains a task for
future observations.

Further pulsar parameters derived from different observations are presented in
Tbl.~\ref{tbl:PSRPrameters}.

\clearpage
\section{Model of the Pulsar}
To understand how the pulsar parameters of Tbl.~\ref{tbl:PSRPrameters} can be
derived from observations, it is necessary to be familiar with pulsar models.
The basic concepts are discussed in this section. Further details can be found
in a book by \cite{Lyne:PulsarAstronomy}.

The models assume that pulsars are rotating neutron stars with high magnetic
fields. The rotation of the magnetic fields and the supply of particles from
the neutron stars' surface can eventually lead to the periodic emission of
strong electromagnetic radiation analogous to a lighthouse.

\subsection{Formation and Inner Structure}
Although only about a dozen out of more than 700 known pulsars appear to be
convincingly associated directly with supernova remnants, neutron stars are
commonly believed to originate in a supernova from the collapse of a star's
core. Thereby the mass of the core determines its density and whether the core
forms a white dwarf, a neutron star or a black hole. An important criterion is
the Chandrasekhar limit $(M_{Ch})$, which is given by
\begin{equation}
  M_{Ch} \approx \left ( \frac{\hbar c}{G}\right)^{\frac32}\frac{1}{m_{p}^{2}}
  =1.44 M_\odot,
\end{equation}
where $\hbar$ is the reduced Planck constant, $c$ is the speed of light, $G$ is
the gravitational constant, $m_{p}$ is the mass of a proton and
$M_\odot=2\times 10^{30}$\,kg is the mass of the sun. For masses exceeding
$M_{Ch}$, the gravitational pressure exceeds the electron degeneracy
pressure\footnote{The electron degeneracy pressure is caused by the Pauli
exclusion principle, which states that two electrons cannot occupy the same
quantum state at the same time.} inside the core, such that the atoms are
compressed. Electron capture by the nuclei is the consequence that eventually
leads to the formation of neutrons by inverse $\beta$ decay $(p^++e^-
\rightarrow n+\overline\nu_e)$, resulting in an extremely compact state of
matter --- a neutron superfluid. With an increasing mass of the core, the
ratio of neutrons to atoms also increases, while the radius decreases. However,
an upper mass limit for a neutron star is reached at $\sim2.5 M_\odot$, where
gravitation compresses the neutron star below its Schwarzschild radius,
converting it in a black hole. So the radius of a neutron star ranges from
between 10 to 20\,km only. The moment of inertia $(I)$ is therefore of the
order of $3\times10^{44}$\,g/cm$^2$, which approximately equals the moment of
inertia of the earth.

\begin{figure}[tb!]
  \begin{minipage}[c]{0.5\linewidth}
    \includegraphics[width=\textwidth]{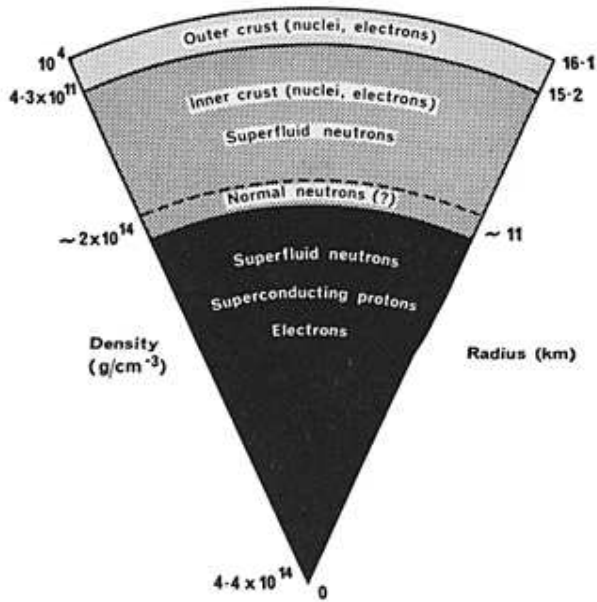}
  \end{minipage}\hfill
  \begin{minipage}[c]{0.5\linewidth}
    \caption[Inner Structure of a Neutron Star]{Inner structure of a neutron
    star. (Figure taken from \cite{NeutronStar}.) }
    \label{fig:NeutronStar}
  \end{minipage}
\end{figure}

The assumed structure of a neutron star is shown in Fig.~\ref{fig:NeutronStar}.
It consists of different layers. The outer layer, i.e. the crust, presumably
contains a rigid lattice of heavy atoms, such as iron with a density of
$\sim10^{11}$\,g/cm$^3$. The middle layer contains a neutron superfluid with
a density of $\sim10^{14}$\,g/cm$^3$. The inner region might contain a solid
core with a density up to $10^{15}$\,g/cm$^3$.

\subsection{Spin-down Luminosity}
The conservation of the angular momentum and magnetic flux of the progenitor
star can lead to high angular velocities $(\Omega=2\pi f)$, i.e. short
rotational periods $(P=1/f)$ and extremely high magnetic fields $(B)$ of the
pulsar. Typical values range from a few milliseconds to a few seconds for $P$
and from $10^8$ to $10^{13}$\,Gauss for $B$. Such magnetic fields are extremely
high. Their energy density is equivalent to a mass density of 1\,kg/cm$^3$
(\cite{Ruderman:1974}).

By the rotation of their strong magnetic fields, pulsars lose a significant
amount of their rotational energy $(E_{rot}=\frac12I\Omega^2)$ by
electromagnetic dipole radiation. The total loss of rotational energy is called
spin-down luminosity $(\dot E)$ and is given by
\begin{equation}
  \dot E=-\frac{d}{dt}E_{rot}=-I\Omega\dot\Omega=4\pi^2I\frac{\dot P}{P^3}.
\end{equation}
Thus, it can be determined by a measurement of $\Omega$ and $\dot\Omega$. For a
magnetic field with the moment $(M_\perp)$ perpendicular to the axis of
rotation, the electromagnetic radiation is $\frac23 M_\perp^2 c^{-3}
\Omega_\perp^4$ and the relation between $\dot \Omega$ and $\Omega$ reads
\begin{eqnarray}
  I\Omega\dot\Omega&=&-\frac23\frac{M_\perp^2}{c^3}\Omega_\perp^4 \\
  \Leftrightarrow \dot\Omega&=&-\frac23\frac{M_\perp^2}{Ic^3}\Omega_\perp^3.
\end{eqnarray}

\subsection{Characteristic Age and Breaking Index}
Observations have shown that pulsars dissipate their energy not exclusively by
electromagnetic dipole radiation. In a more general spin-down model, the
relation between $\dot \Omega$ and $\Omega$ is determined by a power law as
\begin{equation}
  \dot\Omega=-k\Omega^n, \label{eqn:BreakingIndex}
\end{equation}
where $k$ is a constant and $n$ is the breaking index. The integration of
spin-down law of Eqn.~\ref{eqn:BreakingIndex} yields
\begin{equation}
  t=-\frac{\Omega}{(n-1)\dot\Omega}
  \left(1-\frac{\Omega^{n-1}}{\Omega_0^{n-1}}\right),
\end{equation}
where $t$ is the time that passes until the angular velocity has changed from
the initial value $\Omega_0$ to the current value $\Omega$. This equation can
be used to estimate a pulsar's age. With the reasonable assumption $\Omega_0
\gg \Omega$, the last term can be neglected and one obtains the characteristic
age $(\tau)$ as
\begin{equation}
  \tau=-\frac{1}{(n-1)}\frac{\Omega}{\dot\Omega}
  =-\frac{1}{(n-1)}\frac{f}{\dot f}
  =\frac{1}{(n-1)}\frac{P}{\dot P},
\end{equation}
where $f$ and $P$ are the frequency and period of rotation.

Since for many pulsar $n\approx3$ as expected for magnetic dipole breaking,
$n=3$ is assumed as a common definition for the characteristic ages, i.e.
\begin{equation}
  \tau=\frac{f}{2\dot f}=\frac{P}{2\dot P}.
\end{equation}
Nevertheless, the breaking index can be correctly determined if $\ddot\Omega$
can be measured. It can be obtained by the differentiation of the spin-down law
of Eqn.~\ref{eqn:BreakingIndex} and replacing $k$ by the same equation as
\begin{equation}
  n=\frac{\Omega \ddot\Omega}{\dot\Omega^2}=2-\frac{P\ddot P}{\dot P^2}.
\end{equation}
Similarly, one can determine the second deceleration parameter $m$ through a
second differentiation of Eqn.~\ref{eqn:BreakingIndex} as
\begin{equation}
  \dddot f=\frac{n(2n-1) \dot f^3}{f^2}=\frac{m \dot f^3}{f^2} \Leftrightarrow
  m=n(2n-1)=\frac{f^2 \dddot f}{\dot f^3},
\end{equation}
according to which $m$ is given through the third derivative of the pulsar
frequency. Both deceleration parameters are therefore of high theoretical
interest, since they reflect a pulsar's spin-down mechanism and deviations
could indicate variations of the magnetic moment.

\subsection{Magnetosphere and Emitting Regions} \label{sec:EmittingRegions}
The magnetosphere is determined by the strength and the geometry of the
magnetic field. A general estimate for the magnetic field ($B_0$) is given by
\begin{equation}
  B_0=3.3\times10^{19}(P\dot P)^\frac12 \rm G.
\end{equation}
The magnetosphere's geometry is determined by the light cylinder --- sometimes
also called velocity-of-light cylinder. The light cylinder is a cylinder
oriented parallel to the axis of rotation of the pulsar and its radius is given
by the distance $r_L=c/\Omega$ at which a co-rotating particle would exceed the
velocity of light $(c)$. The magnetic field lines, which extend beyond this
cylinder, are called $open$ field lines, and the others $closed$. Inside the
light cylinder, the pulsar contains a high energy plasma, as well as a
relativistic stream of electrons and positrons --- most of which co-rotate with
the pulsar as they are confined to the field lines. Outside this cylinder the
particle density is much lower, and the particles cannot maintain co-rotation
due to the limit imposed by the velocity of light. Therefore, co-rotation is
limited to particles on closed field lines inside the light cylinder, while
particles and field lines at greater distances are wrapped in spirals around
the pulsar. This geometry and the strong magnetic fields lead to two distinct
regions in the magnetosphere where the radiation of the pulsar is generated:
the polar cap and the outer gap. Both regions produce very high energy
particles by different acceleration mechanisms and thus exhibit different
radiation characteristics. Evidence for the polar cap model is seen in
beamwidth and polarization, while evidence for the outer gap model is seen in
the high energy radiation from young pulsars.

\begin{figure}[tb!]
  \centering
    \includegraphics[width=.9\textwidth]{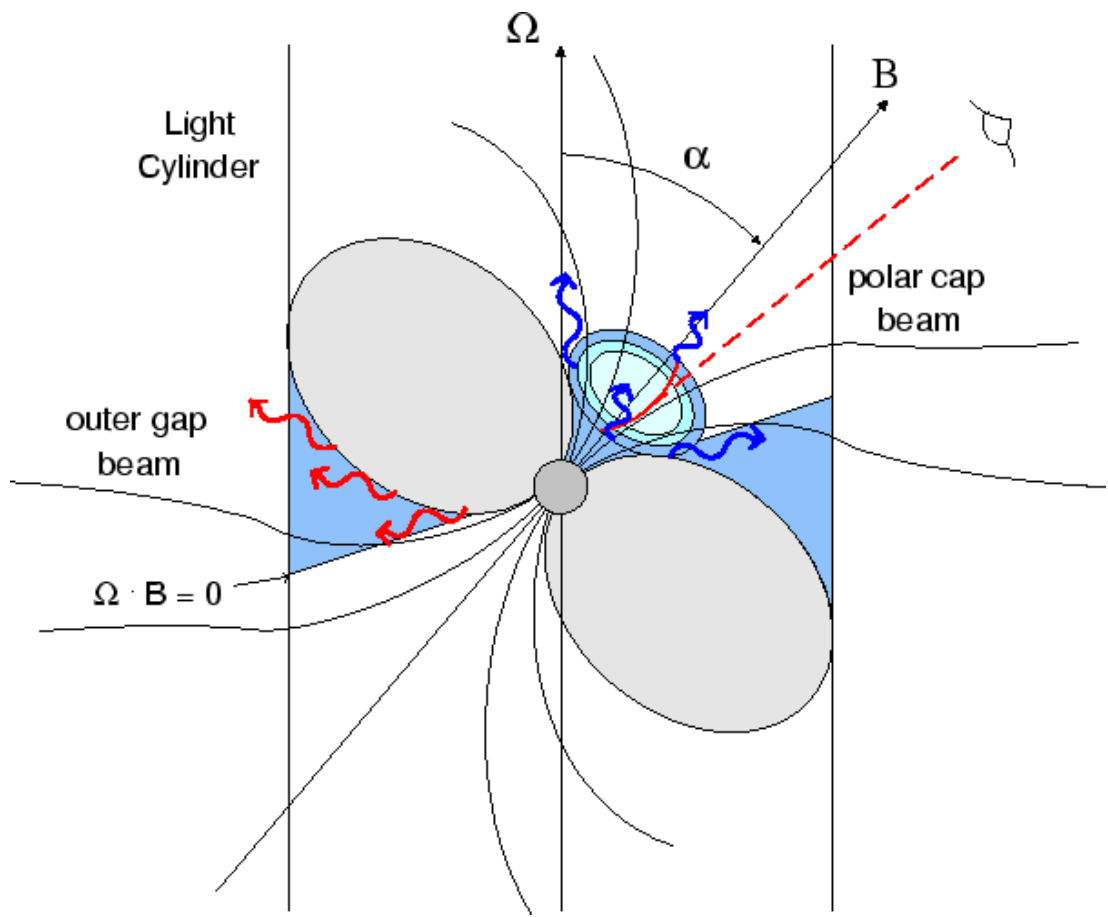}
    \caption[Magnetosphere of a Pulsar]{Magnetosphere of a pulsar showing the
      regions of the polar cap and outer gap which are presumably the regions
      of high particle acceleration and emission of radiation. (Figure taken
      from \cite{Magnetosphere}.) }
    \label{fig:Magnetosphere}
\end{figure}

The polar cap region is indicated for one pole in Fig.~\ref{fig:Magnetosphere}.
It is a small vacuum region near the magnetic poles in which charged particles
can be accelerated. This region is believed to be the origin of highly
polarized narrow radio beams. The radiation is coherent, and also circular
polarization is often observed as the dominant component of polarization.

The outer gap is located far out in the magnetosphere, close to the velocity of
light cylinder as shown in Fig.~\ref{fig:Magnetosphere}. It is assumed to be
responsible for a very similar beam pattern which can extend over many decades
of the electromagnetic spectrum as e.g. observed for the Crab Pulsar. The beam
direction and the timing of the pulses is determined by the geometry of the
magnetosphere near the light cylinder. Electromagnetic radiation is generated
by charged particles which have been accelerated by the high electric fields
in the charge-depleted region of the outer gap. Moreover, this radiation can
also be amplified by cascades of pair creation (\cite{Sturrock:1971}).

\section{Model of the Pulsar Wind Nebula}
If one assumes that \MSH\ and \PSR\ are associated, this system has to be
understood in the framework of PWNs. Therefore it is important to understand
the concept of PWNs for a correct interpretation of its observations. This
section gives an introduction to PWNs. Further details about recent
experimental results can be found in the review of \cite{Gaensler:2006}.
Further theoretical discussions are presented by \cite{Swaluw:2001}.

The term pulsar wind denotes the flow of particles, which is generated by the
pulsar. It mainly consists of relativistic and ultrarelativistic electrons
which are ejected from the pulsar surface, or which are produced in cascades of
pair creation in the emitting regions of the pulsar's magnetosphere
(Sec.~\ref{sec:EmittingRegions}). Therefore, the term electrons equally refers
to electrons and positrons, i.e. to the leptonic component, in the context of
pulsar winds. Also, the importance of the contribution from VHE hadronic
particles to pulsar winds is debated. The hadrons are mainly nuclei which are
emitted from the pulsar's surface or provided by the SNR.

\subsection{Evolution}
Since a PWN is typically embedded in an SNR, its evolution is determined by the
evolution of the SNR. At early times ($t<\sim$1\,kyr) the SNR expands freely at
velocities $>5-10\times 10^3$\,km/s. The expansion of the relativistic pulsar
wind occurs rapidly and symmetrically within the unshocked ejecta. This early
stage is described by the magneto-hydrodynamic (MHD) model of
\cite{Kennel:1984}, which was initially developed for the Crab Nebula. This
model discussed in the next section is likely to apply to \MSH\ as well, which
is a similar young PWN. The ages of the Crab Nebula and \MSH\ are 1.0\,kyr and
1.7\,kyr respectively.

\begin{figure}[b!]
  \begin{minipage}[c]{0.4\linewidth}
    \includegraphics[width=\textwidth]{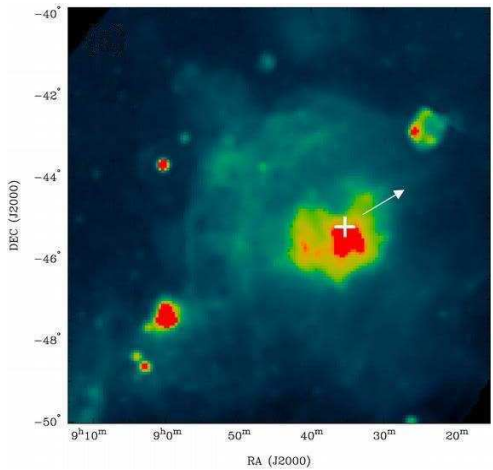}
  \end{minipage}\hfill
  \begin{minipage}[c]{0.58\linewidth}
    \caption[Parkes 2.4\,GHz Map of the Vela PWN]{Parkes 2.4\,GHz map of the
      Vela SNR (\cite{Duncan:1996}) showing the embedded Vela PWN. The cross
      indicates the location of the associated pulsar
      B0833\nobreakdash\ensuremath{-}45, while the white arrow indicates its
      direction of motion (\cite{Dodson:2003}). The fact that the pulsar is
      neither located at nor moving away from the PWN's center suggests that a
      reverse shock interaction has taken place. (Figure taken from
      \cite{Gaensler:2006}.)}
    \label{fig:Gaensler2006}
  \end{minipage}
\end{figure}

\begin{figure}[tb!]
  \centering
  \includegraphics[width=.8\textwidth]{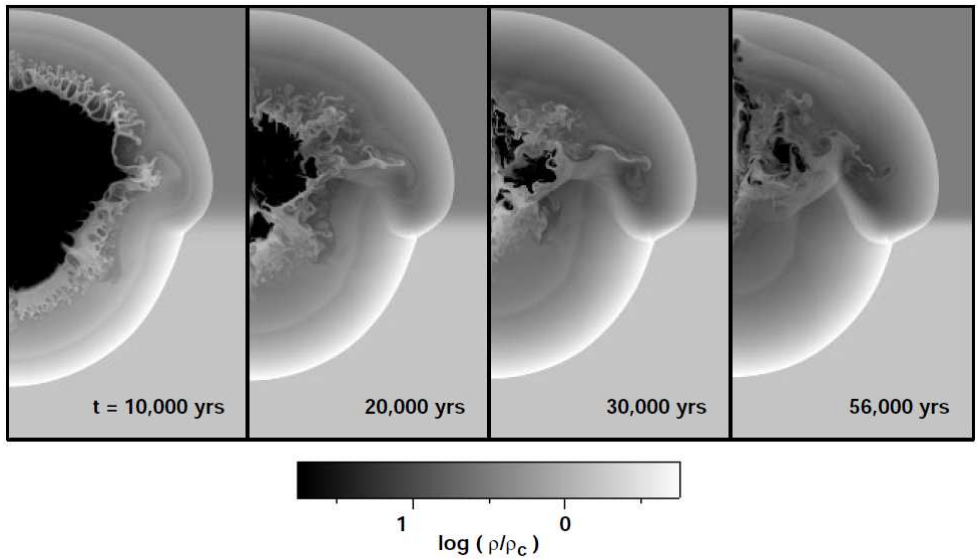}
  \caption[Simulated Evolution of a PWN]{Simulated evolution of a PWN in the
    reverse shock of an SNR. The reverse shock compresses the PWN to a fraction
    of its earlier size. Since the expansion of the SNR is simulated in a
    nonuniform medium with a density gradient (in the vertical direction), this
    simulation is able to reproduce the displacement of the PWN from the center
    as observed for the Vela SNR in Fig.~\ref{fig:Gaensler2006}. (Figure taken
    from \cite{Blondin:2001}.)}
  \label{fig:Blondin2001}
\end{figure}

The evolution of a PWN takes a turn when the SNR evolves into the Sedov-Taylor
phase ($t>\sim$1\,kyr). At this stage, the SNR also develops a reverse shock in
addition to its forward shock. The reverse shock first moves outward behind the
forward shock and eventually moves inward, leading to a compression of the PWN
typically at time spans of several kyr. It is assumed that the Vela PWN is in
this stage (Fig.~\ref{fig:Gaensler2006}). The compression of PWNs by reverse
shock has been modelled by \cite{Blondin:2001}. Fig.~\ref{fig:Blondin2001}
shows that these simulations can reproduce this evolution. However, \MSH\ does
not yet appear affected by a reverse shock. For comparison, the Vela pulsar has
an estimated age of $\sim$10\,kyr, which is about an order of magnitude above
the age of the Crab Nebula and \MSH.

\subsection{The Model by Kennel and Coroniti} \label{sec:KennelCoroniti}
\begin{figure}[tb!]
  \begin{minipage}[c]{0.5\linewidth}
    \includegraphics[width=\textwidth]{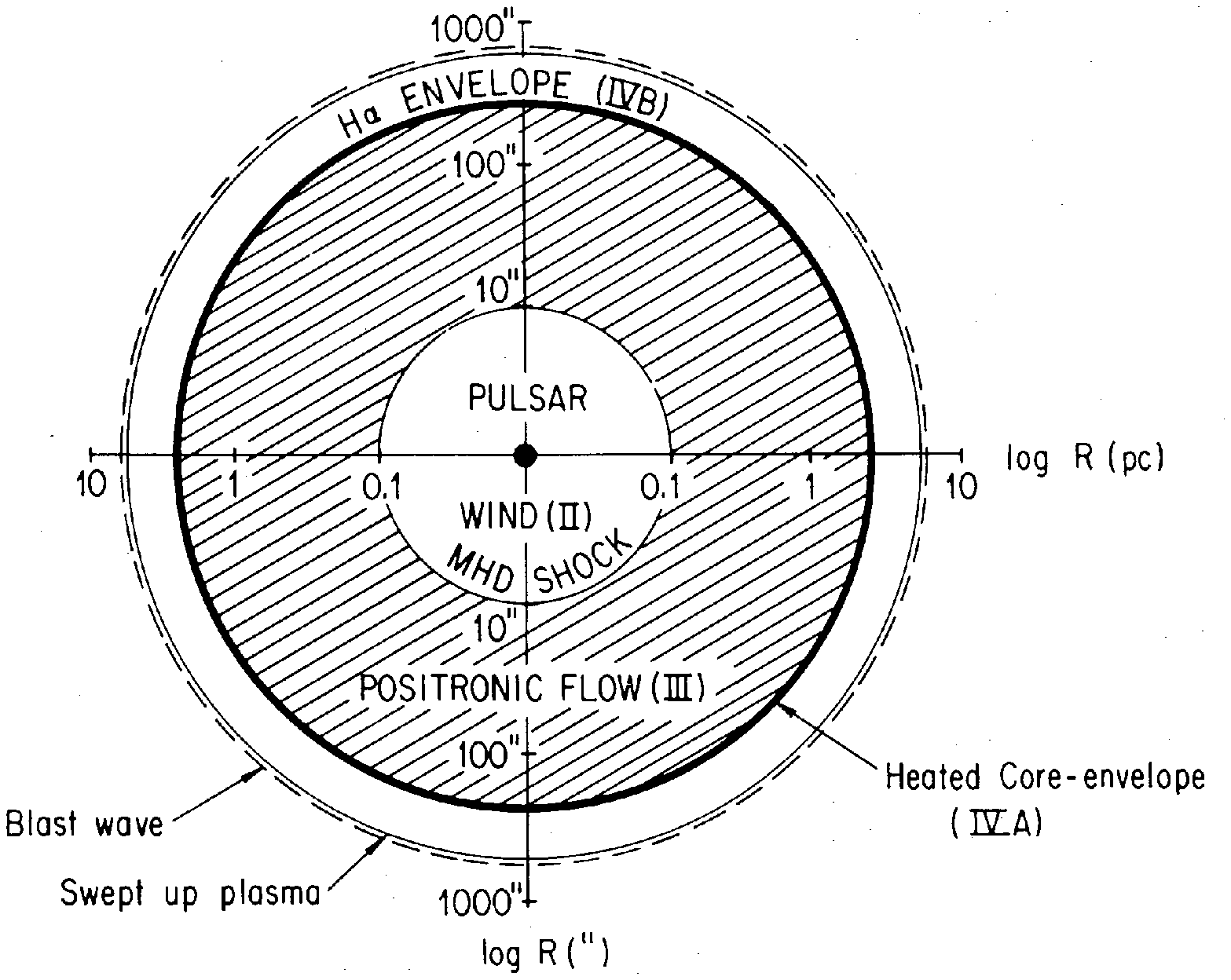}
  \end{minipage}\hfill
  \begin{minipage}[c]{0.5\linewidth}
    \centering
    \includegraphics[width=\textwidth]{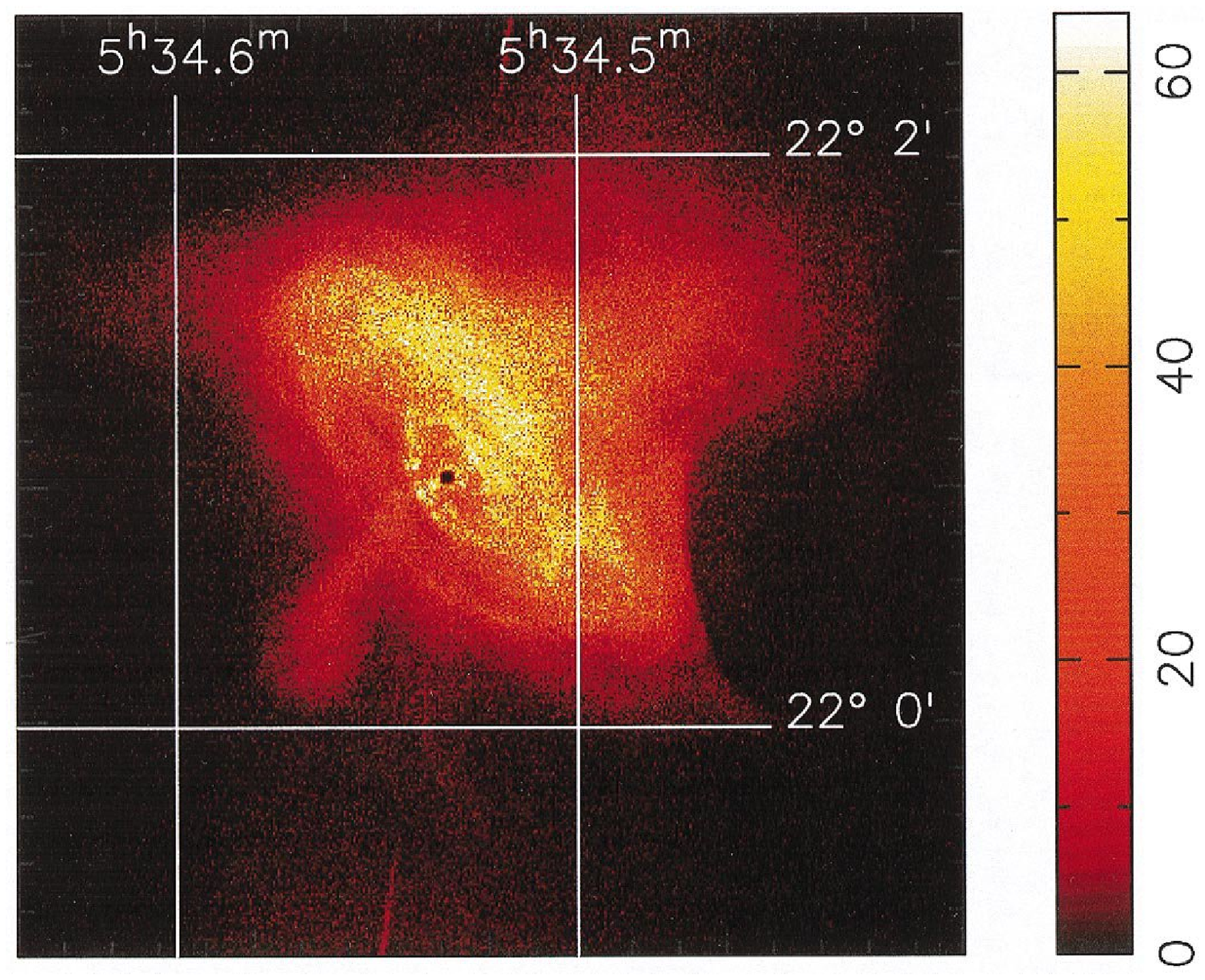}
  \end{minipage} \\
  \begin{minipage}[t]{0.5\linewidth}
    \caption[The Model of a PWN by \cite{Kennel:1984}]{Sketch of the
      theoretical MHD model by \cite{Kennel:1984} for the Crab Nebula. The
      model predicts different regions of the nebula. (Figure taken from
      \cite{Kennel:1984}.)}
    \label{fig:Kennel1984}
  \end{minipage}\hfill
  \begin{minipage}[t]{0.5\linewidth}
    \caption[Chandra \X-ray Image of the Crab Nebula]{Chandra \X-ray data of
      the Crab Nebula showing features predicted by the model of
      \cite{Kennel:1984}. For example the inner ring corresponds to the shock
      front. (Figure taken from \cite{Weisskopf:2000}.)}
    \label{fig:Weisskopf2000}
  \end{minipage}
\end{figure}

The evolution of a PWN at early times ($t<\sim$1\,kyr), when the SNR is
expanding freely at velocities grater than $5-10\times 10^3$\,km/s into the
ambient medium, is described in the model by \cite{Kennel:1984}. It is a
self-consistent, spherically symmetric MHD model which was initially developed
for the Crab Nebula. It describes the flow of relativistic plasma and the
magnetic field from the pulsar to the boundary of the nebula within an SNR.
Hadronic components are neglected. The model distinguishes six different
concentric regions of different astrophysical properties. They are illustrated
in Fig.~\ref{fig:Kennel1984} and characterized in the order of increasing
radius ($r$) as follows:

\begin{itemize}
\item Region I is contained in a small region within the light cylinder
  ($r<r_L$), where the pulsar wind is created. It is indicated by the small
  spot at the center of Fig.~\ref{fig:Kennel1984} and \ref{fig:Weisskopf2000}.

\item Region II extends from the light cylinder to the standing shock front
  ($r_L<r<r_s$) which forms where the highly relativistic pulsar wind expands
  supersonically ($v>c/\sqrt3$) into the SNR ejecta. This region is
  under-luminous. The corresponding region in Fig.~\ref{fig:Weisskopf2000}
  extends from the light cylinder to the bright ring with a radius of
  $\sim10''$. According to \cite{Weisskopf:2000} and \cite{Hester:2002}, the
  ring, which is located between the pulsar and the torus, corresponds to the
  shock front where the cold relativistic wind converts into a more slowly
  moving synchrotron-emitting plasma.

\item Region III contains the synchrotron-emitting plasma, which extends from
  the standing shock front to the boundary of the PWN ($r_s<r<r_N$, $r_N
  \approx 2$\,pc for the Crab Nebula.) It represents the main portion of the
  PWN and emits most of the radio, optical and \X-ray synchrotron radiation. In
  Fig.~\ref{fig:Weisskopf2000} this region corresponds to the torus of the PWN
  beyond the inner ring.

\item Region IV lies outside the PWN but still inside the SNR. For the Crab
  Nebular this region is at a distance of $\sim5$\,pc from the pulsar. This
  region is outside the visible area of Fig.~\ref{fig:Weisskopf2000}.

\item Region V is the region of material swept up by the blast wave of the
  supernova.

\item Region VI lies outside the SNR ($r>5$\,pc for the Crab Nebula). It only
  contains the interstellar medium.
\end{itemize}

The experimental evidence from recent Chandra \X-ray observations for the
existence of these theoretically predicted regions is discussed in greater
detail by \cite{Weisskopf:2000} and \cite{Hester:2002}. The agreement between
Fig.~\ref{fig:Kennel1984} and Fig.~\ref{fig:Weisskopf2000} from
\cite{Weisskopf:2000} is remarkable. In particular the bright ring between
pulsar and its torus is conspicuous.

\subsection{\g-Ray Production in PWNs}
\begin{figure}[bh!]
  \begin{minipage}[c]{0.5\linewidth}
    \includegraphics*[scale=.55,viewport=0 70 430 570]{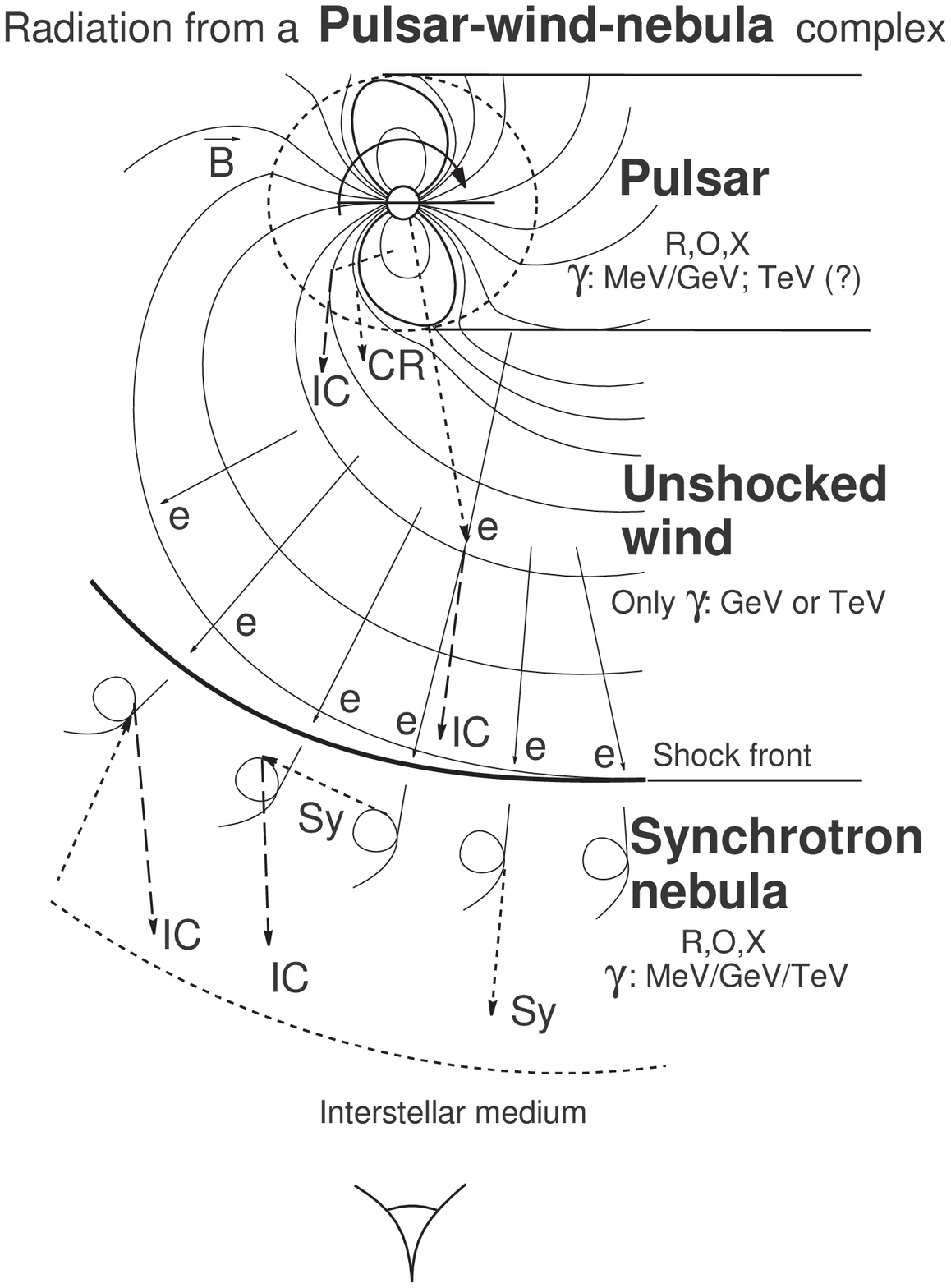}
  \end{minipage}\hfill
  \begin{minipage}[c]{0.47\linewidth}
    \caption[\g-Ray Production in PWNs]{\g-ray production by electrons and
    positrons (e) from a pulsar wind. The schematic representation shows the
    different production mechanisms in the inner three regions of a PWN as
    described by the model by \cite{Kennel:1984}. R, O, X and \g\ stand for
    radio, optical, X-ray and \g-ray emission, respectively. CR, IC and Sy
    stand for curvature, inverse Compton and synchrotron radiation,
    respectively. The orientation of the magnetic field lines (B) is also
    indicated. (Figure taken from \cite{Bogovalov:2003}.)}
    \label{fig:Bogovalov2003}
  \end{minipage}
\end{figure}

The model by \cite{Kennel:1984} can explain the observed radiation from PWNs of
leptonic origin. As the regions differ in their astrophysical properties, they
also differ in the production mechanism of electromagnetic radiation. Most
important is the contribution of the inner three regions (I-III), which are
illustrated in Fig.~\ref{fig:Bogovalov2003}. Although the radiation of PWNs
extends from radio to TeV \g-ray energies, this section will mainly discuss the
production of VHE \g\ radiation.

\begin{itemize}
\item In region I, i.e. inside or close to the light cylinder, the \g\ emission
  is dominated by pulsed curvature, synchrotron and IC radiation. According to
  the polar cap model, electrons with an energy of $\sim10$\,TeV produce
  \g-rays of $\sim10$\,GeV. At 10\,GeV a cutoff in the \g-ray spectrum is
  expected, since the optical depth increases drastically with the \g\ energy,
  and the radiation from the inner magnetosphere is heavily absorbed due to the
  strong magnetic fields.

\item Region II is the regime of an ultrarelativistic, cold and under-luminous
  pulsar wind. The reduced luminosity is due to a reduction of curvature and
  synchrotron radiation. While it is obvious that curvature radiation decreases
  with decreasing curvature of the magnetic field lines, the decrease of
  synchrotron radiation can be explained by the fact that the wind and the
  magnetic field move together, since the field is frozen into the wind. On the
  other hand, an increase of IC radiation is unavoidable due to scattering of
  photons e.g. from the cosmic microwave background radiation (CMB), the
  interstellar radiation field, the emission from the magnetosphere and the
  thermal emission from the pulsar surface ($\sim10^6$\,K)
  (\cite{Bogovalov:2000}, \cite{Bogovalov:2003}). The IC spectrum is primarily
  determined by the wind's Lorentz factor. Typical values of the latter range
  from $10^4$ to $10^7$ and result in IC \g\ radiation from 10\,GeV to 10\,TeV.

\item In region III the shocked pulsar wind produces synchrotron and IC
  radiation. Since this region is much larger than the previous regions, it
  contributes the largest fraction of the observable emission from PWNs. Also,
  if a VHE hadronic component contributes to the \g\ production through
  $\pi^0$-decay, its main contribution would be expected from this region.
\end{itemize}

Although the confirmation of \g\ radiation of hadronic origin from $\pi^0$
decay appears more difficult, a contribution from hadrons is also discussed.
Such considerations are interesting since they, in comparison to leptonic
models, explain a larger fraction of the unexplained energy loss rate of
pulsars which is derived from their spin-down luminosity (\cite{Horns:2006}).
While the leptonic wind observable by its IC radiation only transports a
smaller fraction of the pulsar's spin-down luminosity, the acceleration of a
nucleonic wind could absorb a larger fraction. The nucleonic interaction
regions are inside the nebula and at the boundary to the surrounding
interstellar medium.

\section{Radiation Mechanisms}
Various very high energy radiation mechanisms can explain the energy spectrum
from PWNs observed. The important leptonic mechanisms are synchrotron,
curvature and inverse Compton radiation. Leptons in this case again refer to
electrons and positrons. The hadronic radiation mechanism is mainly the decay
of $\pi^0$, which are produced in interactions of protons and nuclei. Here the
individual mechanisms are introduced and compared.

\subsection{Synchrotron Radiation}
Synchrotron radiation does not only dominate in many PWNs but also in many high
energy astrophysical processes. Although in principle any charged high energy
particle can emit synchrotron radiation in a magnetic field, the main
contribution is observed from electrons and positrons. Synchrotron radiation is
emitted when a charged particle with relativistic energy is forced to a spiral
path by a magnetic field. A detailed discussion can be found in
\cite{Ginzburg:1965}, \cite{Blumenthal:1970} and \cite[Chp. 18]{Longair2}. Here
only a few important aspects are presented.

The direction of the emission depends on the pitch angle $\theta$, i.e.
the angle between the velocity of the particle with respect to the field lines.
The mean energy loss of an electron (or positron) is
\begin{equation}
  -\left<\frac{dE_e}{dt}\right>_{sy}=2\sigma_TcU_{mag}\left(\frac
  vc\right)^2\gamma^2\sin^2\theta, \label{eqn:SynchrotronLoss}
\end{equation}
where $c$ and $v$ are the velocity of light and the electron respectively and
$\gamma$ is the Lorentz factor
\begin{equation}
  \sigma_T=\frac{8\pi}{3}r_e^2=0.665\times10^{-28}\,\rm m^2
\end{equation}
is the Thomson cross section, $r_e$ is the classical electron radius,
\begin{equation}
  U_{mag}=\frac{B^2}{2\mu_0}
\end{equation}
is the energy density of the magnetic field $B$ and $\mu_0$ is the permeability
of free space. For an isotropic distribution of the pitch angles, the mean
radiation loss of an electron is
\begin{equation}
  -\left<\frac{dE_e}{dt}\right>_{sy}=\frac43\sigma_TcU_{mag}\left(\frac
  vc\right)^2\gamma^2= 6.6\times10^4\gamma^2B^2\rm\,eVs^{-1}.
\end{equation}
The intensity spectrum emitted by a single electron can be written as
\begin{equation}
  \frac{d\Phi_\gamma}{dE_e}=\frac{\sqrt{3}e^3}{8\pi^2\epsilon_0cm_e}
  B\sin{\theta}F\left(\frac{\nu}{\nu_c}\right), \label{eqn:SynchrotronSpectrum}
\end{equation}
where $e$ and $m_e$ are the charge and mass of an electron, $\epsilon_0$ is the
permitivity of free space, $\nu_c$ is the critical frequency for synchrotron
radiation and $F(x)$ is given by
\begin{equation}
  F(x)=x\int_x^\infty K_{5/3}(z)dz \approx
  \begin{cases}
    \frac{4\pi}{\sqrt{3}\Gamma(\frac13)} \left(\frac{x}{2}\right)^\frac13,
    & \mbox{if } x \ll 1
    \medskip \\
    \left(\frac{\pi}{2}\right)^\frac12 x^\frac12 \exp(-x), & \mbox{if } x \gg1,
  \end{cases}
\end{equation}
where $K_{5/3}$ and $\Gamma$ denote the modified Bessel and the Gamma function
respectively. The critical frequency is given by
\begin{equation}
  \nu_c=\frac32 \left(\frac cv \right)\gamma^2\nu_g\sin\theta,
\end{equation}
where $\nu_g$ is the gyrofrequency. Therefore,
\begin{equation}
  -\left<\frac{dE_e}{dt}\right>_{sy} \approx 0.29h\nu_c,
\end{equation}
where $h$ is Planck's constant. The corresponding intensity distribution is
shown in Fig.~\ref{fig:Synchrotron}.

The energy spectrum of synchrotron radiation for a sample of electrons,
which has an energy distribution according to a power law, i.e.
\begin{equation}
  \frac{d\Phi_e}{dE_e}=\kappa E^{-\alpha}, \label{eqn:PowerLaw}
\end{equation}
where $E_e$ is the electron energy, $\kappa$ is a normalization constant and
$\alpha$ is the constant of the spectral index, results in a $\gamma$-ray
spectrum $(\Phi_\gamma)$ of the form
\begin{equation}
  \frac{d\Phi_{\gamma}}{dE_\gamma}\propto \kappa
  B^{\frac{(\alpha+1)}{2}}E_\gamma^{-\frac{(\alpha-1)}{2}},
\end{equation}
where $E_e$ is the energy of the photon and $B$ is the magnetic field. So in
this common case the relation between the photon index $(\Gamma)$ of the \g-ray
spectrum and the spectral index of parent electron distribution $(\alpha)$ is
\begin{equation}
  \Gamma=\frac{\alpha-1}{2}.
\end{equation}
The cooling time, i.e. the time for a particle to lose all its energy, is
calculated as
\begin{equation}
  \tau=\frac{E_e}{-\left<\frac{dE_e}{dt}\right>}.
\end{equation}
The formalism of synchrotron radiation from protons with energy $E_p$ is
identical and a comparison can be found in \cite[Sec. 3.3.2]{Aharonian:2004}.
Due to the larger proton mass $(m_p=1836 m_e)$, the radiation loss and
therefore the cooling times from electrons to protons differs by orders of
magnitudes. The ratio is given by
\begin{equation}
  \frac{\tau_p}{\tau_e}
  =\frac{\left<\frac{dE_e}{dt}\right>_e}{\left<\frac{dE_p}{dt}\right>_p}
  =\left(\frac{m_p}{m_e}\right)^\frac52=1.5\times10^8.
\end{equation}
Therefore synchrotron radiation from electrons dominates in most high energy
astrophysical processes.

\subsection{Curvature Radiation}
Curvature radiation is similar to synchrotron radiation. It is also caused by
acceleration of charged particles that pass through a magnetic field. In
contrast to synchrotron radiation however, where the acceleration due to
gyration is perpendicular to the field lines, curvature radiation is associated
with the acceleration parallel to, i.e. along, the field lines. Although
curvature radiation is usually exceeded by synchrotron radiation by orders of
magnitude, curvature radiation becomes relevant for extremely strong and
magnetic fields, such as those in the vicinity of pulsars. Since the radiation
process is in many aspects similar to that of synchrotron emission, the same
equations apply, if the cyclotron radius $r_{cy}=\gamma mv/eB$ is replaced by
the radius of curvature $\rho$ of the magnetic field lines. Then, similar to
Eqn.\ref{eqn:SynchrotronLoss}, one finds the radiation loss given by
\begin{equation}
  -\left<\frac{dE_e}{dt}\right>_{cu}
  =\frac{e^2c}{6\pi\epsilon_0}\frac{\gamma^4}{\rho^2}\left(\frac vc\right)^4.
\end{equation}
The intensity spectrum follows the distribution of synchrotron radiation given
by Eqn.~\ref{eqn:SynchrotronSpectrum} if the critical frequency for curvature
radiation $(\nu_c)$ is used, i.e.
\begin{equation}
  \frac{d\Phi_\gamma}{dE_\gamma}=
  \frac{\sqrt3e^2\gamma}{8\pi\epsilon_0\rho}F\left(\frac {\nu}{\nu_c}\right)^2,
\end{equation}
where 
\begin{equation}
  \nu_c=\frac32 \left(\frac{c}{\rho_c} \right)\gamma^3.
\end{equation}
Therefore, Fig.~\ref{fig:Synchrotron} also represents the intensity spectrum of
curvature radiation.

A sample of electrons with a power law energy distribution given by
Eqn.~\ref{eqn:PowerLaw} results in a \g-ray spectrum with a photon index of
\begin{equation}
  \Gamma=\frac{\alpha-1}{3}
\end{equation}
(cf. \cite[pg. 180]{Lyne:PulsarAstronomy}).

\subsection{Inverse Compton Radiation}
Inverse Compton (IC) radiation is produced in the scattering process of high
energy particles with photons. The energy $(E_\gamma)$ of a photon after
scattering with an electron at rest is given by
\begin{equation}
  E_\gamma(E_{ph},\theta) = E_{ph} P(E_{ph}, \theta),
\end{equation}
where $\theta$ is the scattering angle, $E_{ph}$ is the initial energy of
the photon and
$P(E_{ph},\theta)$ is the ratio of the photon energy after and before the
collision given by
\begin{equation}
  P(E_{ph},\theta) = \frac{1}{1 + \frac{E_{ph}}{m_e c^2}(1-\cos\theta)},
\end{equation}
where $m_e$ is the mass of an electron (\cite[pg. 99]{Longair1}). The cross
section $(\sigma_{KN})$ is according to the Klein-Nishina formula
\begin{equation}
  \frac{d\sigma_{KN}}{d\Omega} = \frac12 r_e^2 \left[P(E_{ph},\theta) -
    P(E_{ph},\theta)^2 \sin^2(\theta) + P(E_{ph},\theta)^3\right],
\end{equation}
where $r_e$ is the classical electron radius, $c$ is the velocity of light.

\begin{figure}[tb!]
  \setlength{\abovecaptionskip}{-0.2cm}
  \begin{minipage}[b]{0.5\linewidth} 
    \includegraphics[width=.8\textwidth]{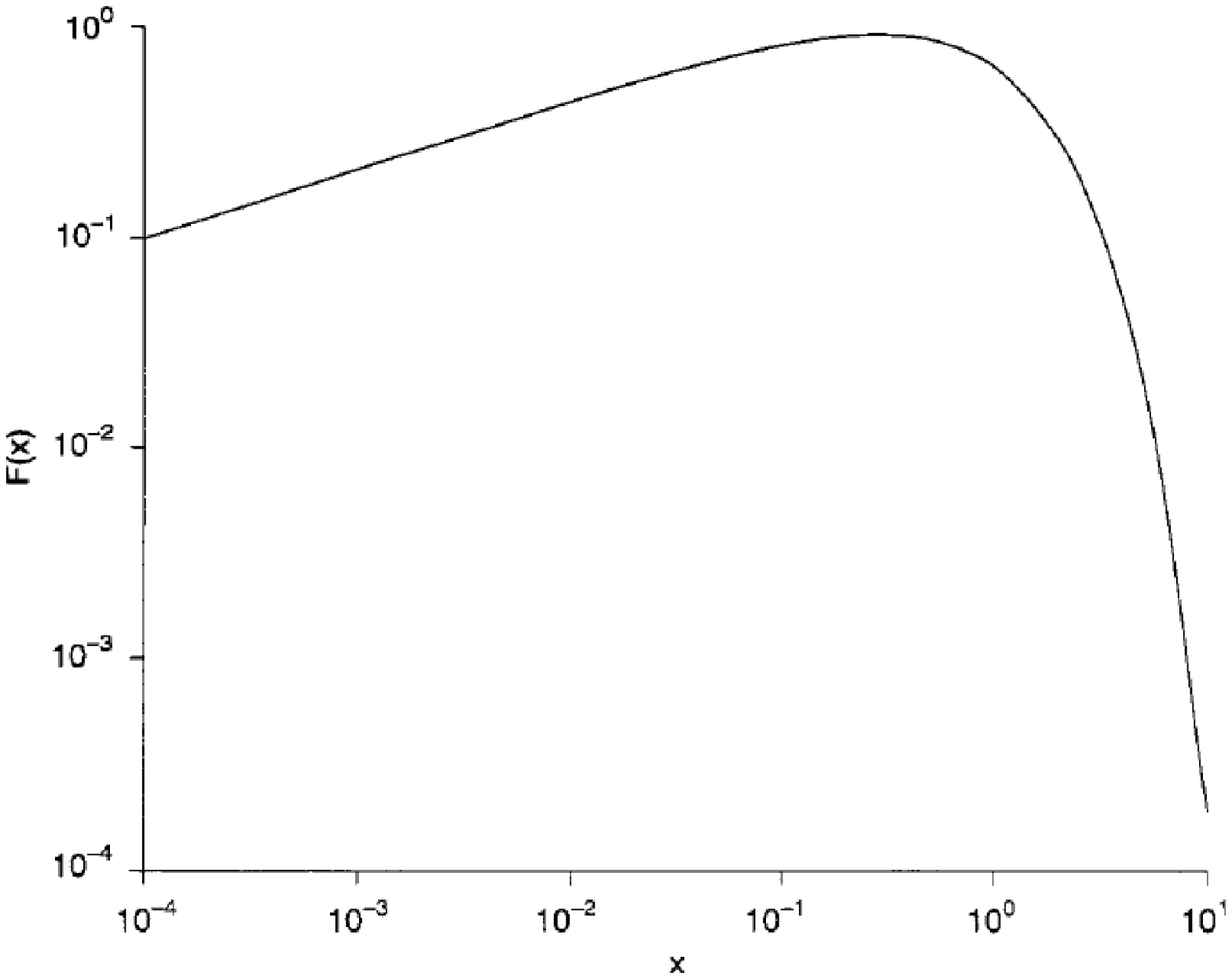} 
    \smallskip \smallskip \smallskip
  \end{minipage}\hfill
  \begin{minipage}[b]{0.5\linewidth}
    \includegraphics[width=\textwidth]{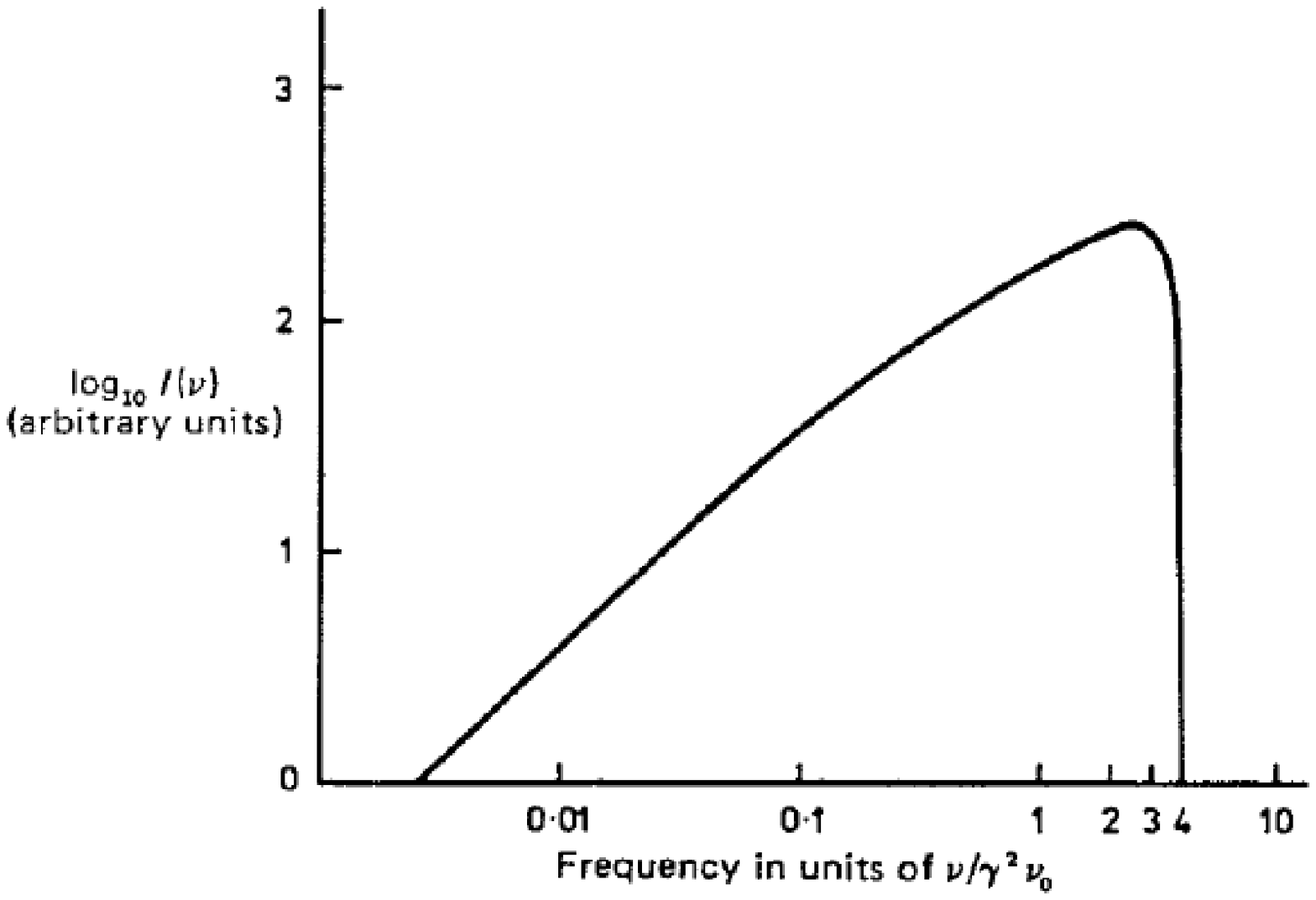}
  \end{minipage}
  \begin{minipage}[t]{0.5\linewidth} 
    \caption[Intensity Spectrum of Synchrotron and Curvature
      Radiation]{Intensity spectrum of synchrotron and curvature radiation
      emitted by a charged particle of a fixed energy in a magnetic field of
      constant strength and curvatures. The spectrum is shown as a function of
      $x=\nu/\nu_c$, where $\nu_c$ is the critical frequency. (Figure taken
      from \cite[pg. 247]{Longair2}.)}
    \label{fig:Synchrotron}
  \end{minipage}\hfill
  \begin{minipage}[t]{0.5\linewidth}
    \caption[Intensity Spectrum of Inverse Compton Radiation]{Intensity
      spectrum of inverse Compton radiation emitted by a charged particle.
      $\nu_0$ is the frequency of the unscattered photon. (Figure taken from
      \cite[pg. 104]{Longair1}.)}
    \label{fig:InverseCompton}
  \end{minipage}
\end{figure}

From these principles \cite{Blumenthal:1970} calculated the IC energy spectrum
which is produced in the interactions of accelerated electrons with photons as
\begin{equation}
  \frac{d\Phi_{\gamma}}{dE_\gamma}=\frac{2\pi
    r_0^2m_ec^3}{\gamma}\frac{n(E_{ph})dE_{ph}}{E_{ph}}\times
  \left[2q\ln q+(1+2q)(1-q)+\frac12\frac{(pq)^2}{1+pq}(1-q) \right],
\end{equation}
where the constants $p$ and $q$ are given as
\begin{equation}
  p=\frac{4E_{ph}\gamma}{m_ec^2} \quad \mathrm{and} \quad
  q=\frac{E_e}{p(1-E_e)},
\end{equation}
$\gamma$ is the Lorentz factor and $E_e$ is the initial electron energy.
Fig.~\ref{fig:InverseCompton} shows the corresponding intensity distribution.

Thomson scattering is obtained if $E_\gamma(E_{ph},\theta) \approx E_{ph}$,
i.e.
\begin{equation}
  P(E_{ph},\theta) \approx 1 \Leftrightarrow \frac{E_{ph}}{m_e c^2}
  (1-\cos\theta) < 1.
\end{equation}
In IC scattering the electron is not at rest but has the Lorentz factor of
$\gamma$. Then the energy of the photon is $\approx 2\gamma E_{ph}$ (for $v
\approx c$) in the rest frame of the electron. Therefore one distinguishes
\begin{equation}
  \mbox{if} \quad
  \begin{cases}
    \quad 4 \frac{\gamma E_{ph}}{m_e c^2} < 1 \Leftrightarrow \gamma <
    \frac{m_e^2c^4}{4E_{ph}} & \mbox{the Thomson limit,} \\ \quad 4
    \frac{\gamma E_{ph}}{m_e c^2} > 1 \Leftrightarrow \gamma >
    \frac{m_e^2c^4}{4E_{ph}} & \mbox{the Klein-Nishina limit.}
    \label{eqn:Regimes}
  \end{cases}
\end{equation}
In the Thomson limit the maximum $(\hat E_\gamma)$ and mean energy $(\overline
E_\gamma)$ of the scattered photons are
\begin{equation}
  \hat E_\gamma \approx 4\gamma^2E_{ph}
\end{equation}
and
\begin{equation}
 \overline E_\gamma \approx \frac13 \hat E_\gamma \approx \gamma^2 E_{ph} =
 \left(\frac{E_e}{m_ec^2}\right)^2 E_{ph}, \label{eqn:ThomsonE}
\end{equation}
where $E_e$ is the energy of the electrons. The Thomson limit is valid for many
astrophysical processes. For example, for the CMB $(E_{ph}=2.35 \times
10^{-4}$\,eV) the Thomson limit is fulfilled for
\begin{equation}
  \gamma < \frac{m_e^2c^4}{4E_{ph}} = \rm \frac{0.5 \times 10^6\,eV}{4\times
  2.35\times 10^{-4}eV}=0.5\times 10^9 \quad \mathrm{i.e.} \quad E_e = \gamma
  m_ec^2 < 250\,TeV.
\end{equation}
In the Thomson limit the radiation loss is
\begin{equation}
  -\left<\frac{dE_e}{dt}\right>_{IC}=\frac43\sigma_TcU_{ph}\left(\frac
vc\right)^2\gamma^2,
\end{equation}
where $U_{ph}$ is the energy density of the photon field. The ratio of IC to
synchrotron radiation is immediately given by the ratio of the photon to the
magnetic energy density as
\begin{equation}
  \frac{\left<\frac{dE_e}{dt}\right>_{IC}}{\left<\frac{dE_e}{dt}\right>_{sy}}=
  \frac{U_{ph}}{U_{mag}}.
\end{equation}
According to \cite{Ginzburg:1965}, IC radiation in the Thomson limit of parent
particles with an energy distribution according to a power law
(Eqn.~\ref{eqn:PowerLaw}) passing through a monochromatic photon field has a
photon index of
\begin{equation}
  \Gamma=\frac{\alpha+1}{2}.
\end{equation}
The radiation loss in the Klein-Nishina regime is
\begin{equation}
  -\left<\frac{dE_e}{dt}\right>_{IC}=\frac38\sigma_Tm_e^2c^5\int
  n{(E_{ph})}{E_{ph}}\left(\ln
  \frac{4E_{ph}\gamma}{m_ec^2}-\frac{11}{6}\right)dE_{ph}.
\end{equation}
According to \cite{Blumenthal:1970}, IC radiation in the Klein-Nishina limit of
parent particles with an energy distribution according to a power law
(Eqn.~\ref{eqn:PowerLaw}) passing through a monochromatic photon field has a
photon index of
\begin{equation}
  \Gamma=\alpha+1.
\end{equation}
Calculations for IC radiation in the intermediate regime at relativistic
energies $(\gamma \approx \frac{m_e^2c^4}{4E_{ph}})$ can be found in
\cite{Aharonian:1981}.

Although high energy protons can also produce IC radiation, IC scattering of
protons with the same energy as electrons is suppressed by many orders of
magnitude as
\begin{equation}
  \left(\frac{m_e}{m_p}\right)^4=9\times10^{-14}
\end{equation}
(cf. \cite[Chp. 3.2]{Aharonian:2004}). Therefore, IC radiation from electrons
dominates in high energy astrophysics.

\subsection{Hadronic \g\ Radiation}
\begin{figure}[t!]
  \setlength{\abovecaptionskip}{-0.3cm}
  \begin{minipage}[b]{0.5\linewidth}
    \centering
    \includegraphics[width=.8\textwidth, height=.65\textwidth]
                    {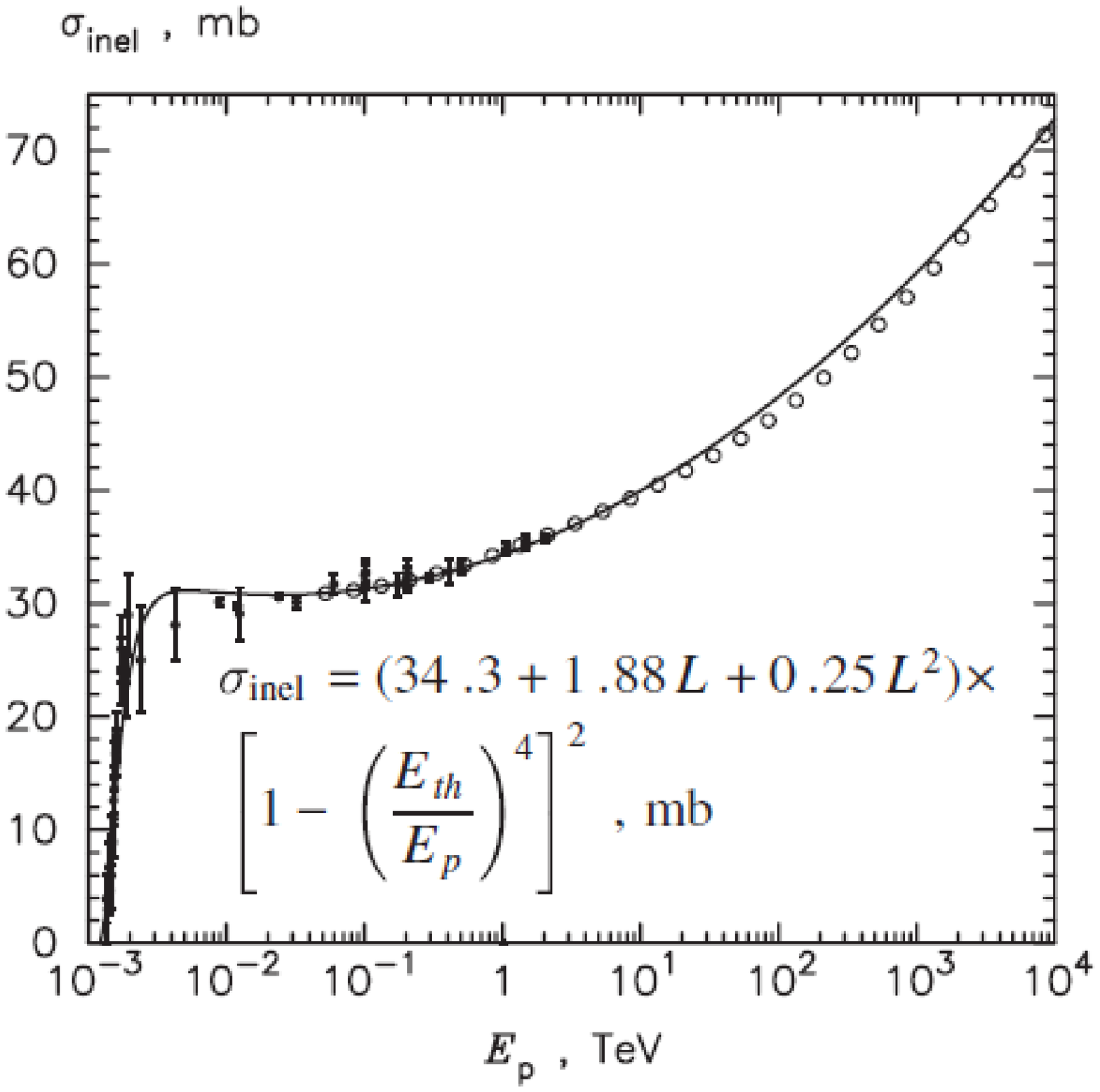} 
  \end{minipage}\hfill
  \begin{minipage}[b]{0.5\linewidth}
    \centering
    \includegraphics[width=.9\textwidth, height=.6\textwidth]
                    {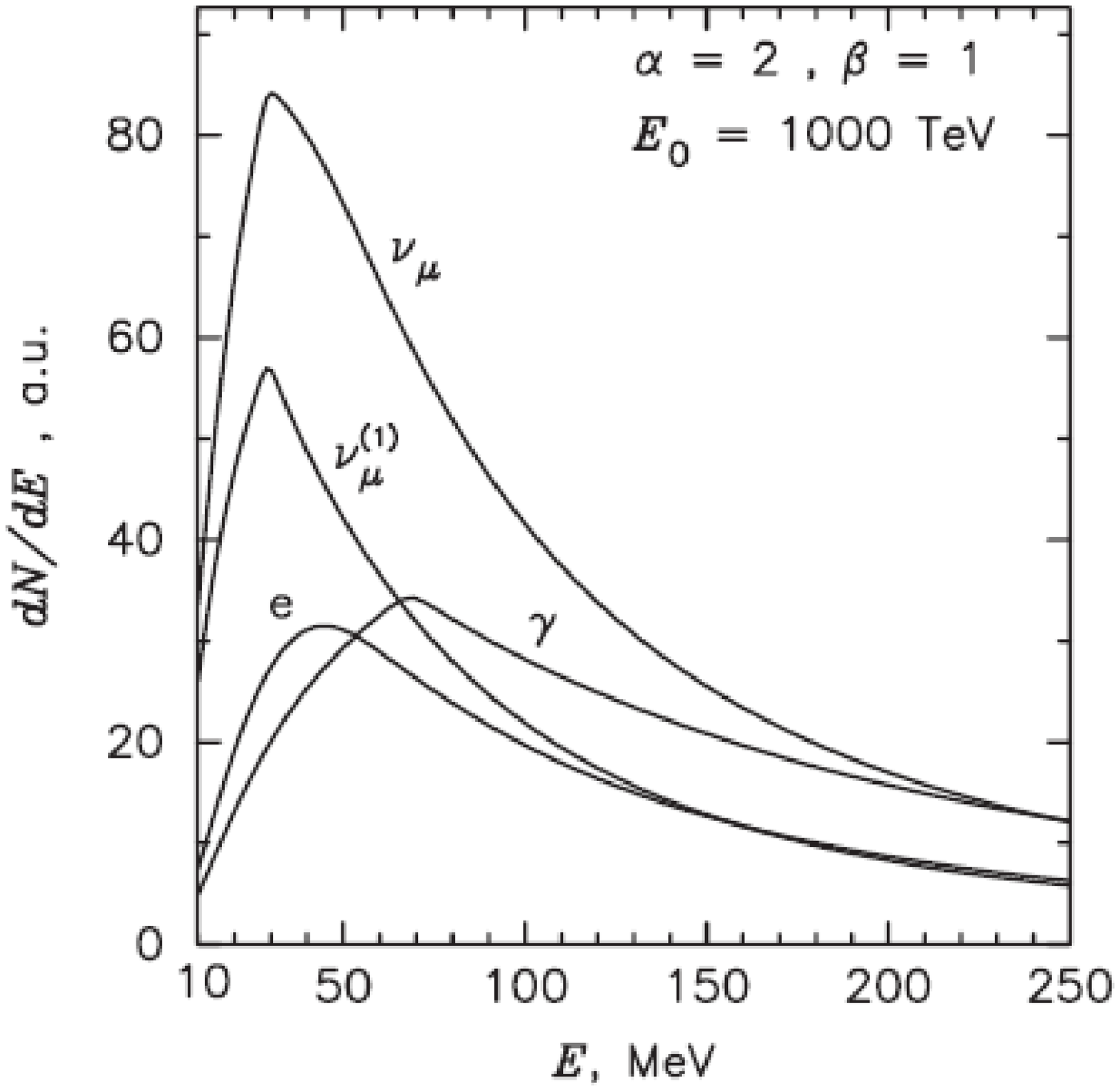}
  \end{minipage}
  \begin{minipage}[t]{0.5\linewidth}
    \caption[Total Cross Section for Inelastic Proton-Proton Collisions]{Total
      cross section for inelastic proton-proton collisions. The threshold
      energy $E_{\rm th}$ is $m_pc^2+2m_\pi c^2(1+m_\pi/4m_p)$, where the
      kinetic energy of the proton exceeds $\sim$280\,MeV, i.e. twice the mass
      of the $\pi^0$. (Figure taken from \cite{Kelner:2006}.)}
    \label{fig:KelnerProtonCrossSection}
  \end{minipage}\hfill
  \begin{minipage}[t]{0.5\linewidth}
    \caption[Energy Spectra for Proton-Proton Collisions]{Energy spectra of
      \g-rays and other secondary products in proton-proton collisions. The
      proton energy spectrum obeys the power law of
      Eqn.~\ref{eqn:ProtonPowerLaw}. The electron and neutrino spectrum
      coincide. (Figure taken from \cite{Kelner:2006}.)}
    \label{fig:KelnerProtonSpectrum}
  \end{minipage}
\end{figure}

\g\ radiation from hadronic interactions is mainly produced through the decay
of secondary particles from inelastic nucleon collisions. While the secondary
particles are mainly pions, i.e. $\pi^+$, $\pi^-$ and $\pi^0$ with equal
probability, only the $\pi^0$ mesons decay into two photons and contribute to
the \g-ray spectrum. The $\pi^\pm$ mesons decay into muons, electrons and
neutrinos. The majority of nucleonic interactions are produced by highly
accelerated protons which collide with ambient hydrogen, i.e. by
proton-proton collision. Therefore, the inelastic part of the total
proton-proton cross section $(\sigma_{\rm inel}(E_p))$ determines the hadronic
\g-ray spectrum. According to \cite{Kelner:2006} $\sigma_{\rm inel}(E_p)$ is
approximated as
\begin{equation}
  \sigma_{\rm inel}(E_p)=(34.3+1.88L+0.25L^2) \times\left[1-\left(\frac{E_{\rm
  th}}{E_p}\right)^4\right]^2\rm\,mb,
\end{equation}
where $E_p$ is the energy of the proton, $L=\ln(E_p/1$\,TeV) and $E_{\rm th}$
is the threshold energy of the proton for the production of $\pi^0$ mesons.
Since the kinetic energy of the proton has to exceed twice the rest mass of
the pion ($m_\pi=135$\,MeV) $E_{\rm th} = m_pc^2+2m_\pi c^2(1+m_\pi/4m_p)=
1.22\times10^{-3}$\,TeV. $\sigma_{\rm inel}(E_p)$ is shown in
Fig.~\ref{fig:KelnerProtonCrossSection}.

\noindent The \g-ray energy spectrum is then given as
\begin{equation}
  \frac{d\Phi_\gamma}{dE_\gamma}=q_\gamma(E_\gamma)=2\int_{E_{min}}^\infty
  \frac{q_{\pi}(E_\pi)}{\sqrt{E_\pi^2-m_\pi^2c^4}}dE_{\pi},
\end{equation}
where $E_{min}=E_\gamma+m_\pi^2c^4/4E_\gamma$, $E_\pi$ and $q_\pi(E_\pi)$ are
the energy and the emissivity of secondary pions with
\begin{equation}
  q_\pi(E_\pi)=\frac{cn_H}{\kappa_\pi}\sigma_{\rm inel}
  \left(m_pc^2+\frac{E_\pi}{\kappa_\pi}\right)
  J_p\left(m_pc^2+\frac{E_\pi}{\kappa_\pi}\right).
\end{equation}
Here $n_H$ is the density of the ambient hydrogen, $c$ is the speed of light,
$\kappa$ is the mean fraction of kinetic energy of the proton transferred to
\g\ photons or the $\pi^0$ mesons per collision and $J_p$ is the energy
spectrum of the protons (\cite{Aharonian:2004}).

\begin{figure}[tb!]
  \centering
  \includegraphics[width=.9\textwidth]{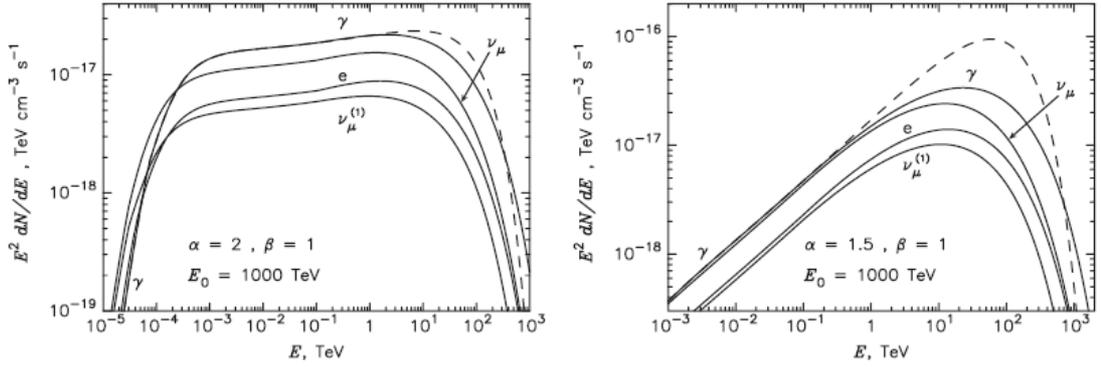}
  \caption[SED for Proton-Proton Collisions]{Spectral energy distribution
  corresponding to Fig.~\ref{fig:KelnerProtonSpectrum} (left) and for a harder
  proton spectrum with a spectral index $\alpha=1.5$ (right). The latter
  spectrum is closer to the nucleonic emission model for \MSH\ by
  \cite{Bednarek:2003}. (Figure taken from \cite{Kelner:2006}.)}
  \label{fig:KelnerProtonSED}
\end{figure}

A distinct feature of the hadronic \g-ray spectrum is a pronounced peak in the
energy spectrum at $E=m_\pi c^2/2\simeq68$\,MeV, which is independent of the
energy distribution of the $\pi^0$ mesons and therefore of the proton.
Fig.~\ref{fig:KelnerProtonSpectrum} shows the energy spectrum of the \g-ray
photons and the other secondary particles for protons with a spectral
distribution
\begin{equation}
  J_p(E_p)=\frac{A}{E_p^\alpha}
  \exp\left[-\left(\frac{E_p}{E_0}\right)^\beta\right], 
  \label{eqn:ProtonPowerLaw}
\end{equation}
where $E_p$ is the energy of the proton, the spectral index $\alpha=2$, the
cut-off energy $E_0=1000$\,TeV and $\beta=1$. The peak in the \g-ray spectrum
at $\simeq68$\,MeV is apparent. Fig.~\ref{fig:KelnerProtonSED} shows the
corresponding SED for the same spectral index and for a different index of
$\alpha=1.5$. Proton spectra with a harder index ($\alpha=1.5$) become
important in the models of hadronic particle acceleration in PWNs by
\cite{Bednarek:2003}.

The average energy loss of a proton of about 50\% at each collision is
described by the coefficient of inelasticity $(f=0.5)$. Taking this energy loss
into account and assuming an approximately constant $\sigma_{\rm inel}$ at high
energies as \cite{Aharonian:2004}, the \g-ray spectrum reproduces the proton
spectrum, and the photon index reads
\begin{equation}
  \Gamma\approx\alpha.
\end{equation}
Moreover, for a hydrogen density of $n_0=n/\rm cm^3$ the cooling time is
\begin{equation}
  \tau_{pp}=\frac{1}{n_0\sigma_{pp}fc}\approx \frac {\rm
  cm^3}{n}5.3\times10^7\rm\,yr
\end{equation}
and
\begin{equation}
  \left<\frac{dE_p}{dt}\right>_{pp}=\frac{E_p}{\tau_{pp}}=n_0\sigma_{pp}fcE.
\end{equation}

\subsection{Energy Spectra from PWNs} \label{sec:PWNSpectra}
The radiation mechanisms discussed above can provided important information for
the understanding of PWNs. Since they lead to different energy spectra from a
common primary electron distribution, the radiation at different wavelength has
to result in a consistent picture of the astrophysical conditions at the source
region. The different photon indices $\Gamma$ for the same power law electron
distribution with index $\alpha$ are summarized in
Tbl.~\ref{tbl:RadiationSpectrum}.

Moreover, for a typical magnetic field strength of $10^{-5}$\,G in a PWN the
electron energy $E_e$ for producing synchrotron radiation is approximately
given by
\begin{equation}
  E_e = (70\,\mathrm{TeV})B_{-5}^{\frac12}E_\mathrm{keV}^{\frac12},
\end{equation}
were $B_{-5}=B/(10^{-5}G)$ is a number for the transverse magnetic field
strength $(B)$ and $E_\mathrm{keV}$ is the number for the mean energy of the
synchrotron radiation in units of keV (\citet{Okkie:2006}). The corresponding
value for IC radiation produced by scattering of CMB photons is approximately
given by
\begin{equation}
  E_e = (18\,\mathrm{TeV})E_\mathrm{TeV}^{\frac12}, \label{eqn:OkkieIC}
\end{equation}
where $E_\mathrm{TeV}$ is the number for the mean energy of the IC radiation in
units of TeV (\citet{Okkie:2006}). From these two equations one immediately
obtains the relation between the magnetic field strength, the synchrotron and
the IC radiation as
\begin{equation}
  E_\mathrm{keV}=0.06B_{-5}^{-2}E_\mathrm{TeV}.
\end{equation}
It allows to infer the magnetic field strength if the synchrotron and IC
component are both known.

Also, one can explain spectral steepening with increasing distance form the
center of extended PWNs if lifetimes are considered. For example (cf.
\cite{HESSJ1825I}, \citet{Okkie:2006}), the lifetime $\tau(E_\gamma)$ of VHE
\g-ray emitting electrons in a magnetic field is given by
\begin{equation}
  \tau(E_\gamma)=(4.8\,\mathrm{kyr})B_{-5}^{-2}E_\mathrm{TeV}^{-\frac12}.
\end{equation}
The corresponding lifetime $\tau(E_X)$ for keV emitting electrons is
\begin{equation}
  \tau(E_\mathrm{X})=
  (1.2\,\mathrm{kyr})B_{-5}^{-\frac32}E_\mathrm{keV}^{-\frac12}.
\end{equation}
In both cases the lifetime of electrons with higher energy is shorter. So with
increasing distance less high energy electrons and therefore a steeper photon
index is expected as observed in \cite{HESSJ1825II}.

Nevertheless, the modelling of astrophysical processes in a PWN often remains a
difficult task, since many parameters are often not well constrained allowing
for different explanations. However, simulations can often provide plausible
solutions.

\begin{figure}[ht!]
  \begin{minipage}[t]{0.52\linewidth}
    \vfill
    \flushright
    \includegraphics[width=.92\textwidth, height=.22\textheight]
                    {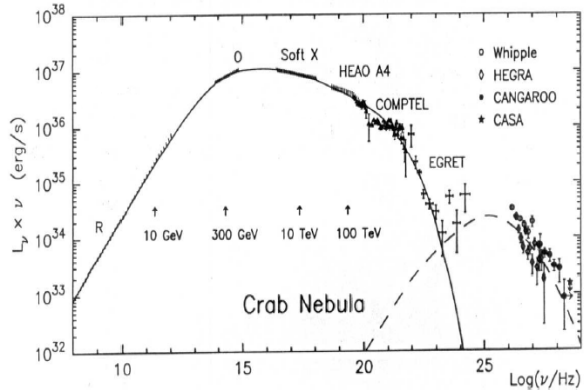}
  \end{minipage}\hfill
  \begin{minipage}[t]{0.45\linewidth}
    \caption[Non-thermal radiation from the Crab Nebula]{Non-thermal radiation
      observed from the Crab nebula. The solid and dashed curves represent the
      synchrotron and IC component radiation calculated for a spherical
      symmetric MHD model. The vertical arrows indicate the contribution to
      synchrotron radiation which is produced by parent electrons with energies
      of the corresponding label. (Figure taken from \cite[pg.
      47]{Aharonian:2004}.)}
    \label{fig:Aharonian2004}
  \end{minipage}
\end{figure}

Fig.~\ref{fig:Aharonian2004} shows the energy spectrum of the Crab Nebula. It
is an example of a PWN with a well determined energy spectrum over more than 20
decades. The synchrotron peak at keV energies and the IC peak at TeV energies
are visible. The corresponding energies of the parent electrons producing the
radiation are indicated by the labeled arrows.

\begin{figure}[ht!]
  \begin{minipage}[t]{0.54\linewidth}
  \vfill 
  \includegraphics[width=.965\textwidth]
                  {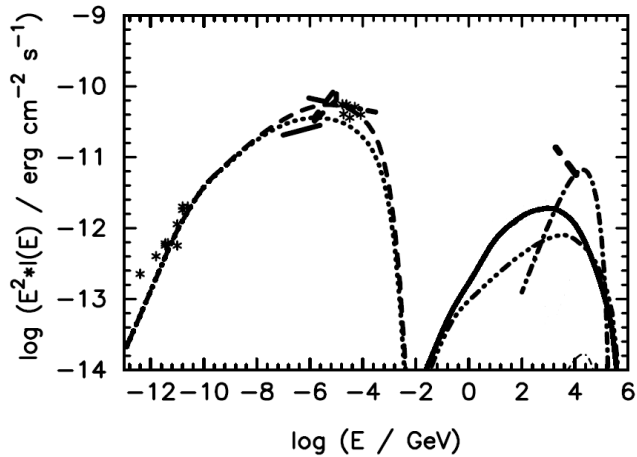}
  \end{minipage}\hfill
  \begin{minipage}[t]{0.45\linewidth}
    \caption[Non-thermal radiation from \MSH]{Non-thermal radiation from \MSH.
      The dashed and dot-dot-dot-dashed curves show the calculated synchrotron
      and IC spectra, respectively. The dotted and full curves show the same
      calculations with additional infrared photons. The nucleonic radiation
      from $\pi^0$ decay for the density of the interstellar medium of
      0.3\,cm$^{-3}$ (thin dot-dashed curve) and for a high density of
      300\,cm$^{-3}$ (thick dot-dashed curve) is also shown. (Figure taken from
      \citet{Bednarek:2003}.)}
    \label{fig:Bednarek2003}
  \end{minipage}
\end{figure}

\begin{figure}[htb!]
  \begin{minipage}[t]{0.52\linewidth} 
    \vfill
    \flushright
    \includegraphics[width=.94\textwidth]{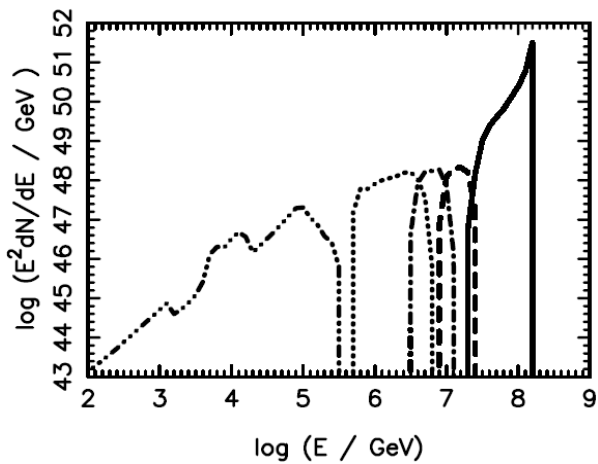}
  \end{minipage}\hfill
  \begin{minipage}[t]{0.45\linewidth}
    \caption[Parent Energy Spectra of Different Nuclei in a PWN]{Parent energy
      spectra of nuclei of different mass numbers $(A)$ in a PWN after 1\,kyr.
      Similar spectra have been used for the calculations in
      Fig.~\ref{fig:Bednarek2003}. The curves represent: $A=1$
      (dot-dot-dot-dashed curve), 2-10 (dotted), 11-20 (dot-dashed), 21-40
      (dashed) and 41-55 (full). The corresponding \g-ray spectra are shown in
      Fig.~\ref{fig:Bednarek2003} for different densities (thin and thick
      dot-dashed curve). (Figure taken from \citet{Bednarek:2003}.) }
    \label{fig:BednarekProtonSED}
  \end{minipage}
\end{figure}

The energy spectrum of \MSH\ is less well constrained by measurements. However,
different spectra have been predicted. A few are shown in
Fig.~\ref{fig:Bednarek2003} as calculated by \cite{Bednarek:2003}. These
calculations show the contributions of the leptonic and also of nucleonic
components for different densities of the ambient medium. The nucleonic energy
spectra are similar to those used in the calculations shown in
Fig.~\ref{fig:BednarekProtonSED}. They represent the equilibrium spectra of
different nuclei after 1\,kyr. The corresponding \g-ray spectra are represented
in Fig.~\ref{fig:Bednarek2003} by the thin and thick dot-dashed curve.

\begin{table}[ht!]
  \centering
  \caption[Relation between Photon Index and Spectral Index of Parent
  Particles]{Relation between the photon index $(\Gamma)$ and spectral index of
  a power law parent particle distribution $(\alpha)$ for different radiation
  mechanisms.}
  \bigskip
  \begin{tabular}{lccccc}
    \hline \hline
    & synchrotron & curvature 
    & IC radiation & IC radiation & $\pi^0$ decay \\
    & radiation   & radiation
    & Thomson limit & Klein-Nishina limit \\
    \hline
    $\Gamma$ & $(\alpha-1)/2$ & $(\alpha-1)/3$ & $(\alpha+1)/2$ 
    & $\alpha+1$ & $\alpha$ \\
    \hline \hline
  \end{tabular}
  \label{tbl:RadiationSpectrum}
\end{table}

\chapter[\g-Ray Astronomy with IACTs]{\g-Ray Astronomy with Imaging Atmospheric
  Cherenkov Telescopes}

Astroparticle physicists and astronomers have long been interested in observing
cosmic VHE \g\ radiation. After many years of research, imaging atmospheric
Cherenkov telescopes (IACTs) were developed and established as useful
instruments in VHE \g\ astronomy. IACTs can detect cosmic \g\ radiation through
its interaction with the earth's atmosphere. Since the IACT technique allows
for building systems with large effective areas, it is possible to detect a
significant amount of cosmic \g\ radiation in the energy range from about
100\,GeV to 100\,TeV. At this energy range even the strongest sources have a
flux of less than $\rm 10^{-10}cm^{-2}s^{-1}$. IACTs are currently the most
sensitive instruments for \g-ray astronomy at these energies.

\section{Air Showers} \label{sec:AirShowers}
To understand the IACT technique it is useful to know about the physics of air
showers, which develop in the atmosphere, and about the Cherenkov light which
is emitted. One can distinguish between two types of showers, electromagnetic
and hadronic showers.

\subsection{Electromagnetic Showers}
Electromagnetic showers are electron-photon cascades which are initiated when a
photon of high energy enters the atmosphere. The interaction with the molecules
of the atmosphere leads to pair production. In turn, the produced
electron-positron pairs emit photons via bremsstrahlung. The sequence of pair
production and bremsstrahlung results in a cascade with an exponential increase
of particles. The maximum of particles is reached when their energy has reduced
to critical energy ($E_c$). At $E_c$ the particles' energy is not sufficient to
sustain the pair production, and the remaining energy is finally dissipated by
ionization. A full shower cascade develops within $\sim$50 micro seconds. The
frequency of the interactions is determined by the radiation length $X_0$.
$X_0$ is defined as the mean distance the particles travel when they lose all
but $\exp(-1)$ of their energy. Similarly, the interaction length $(\lambda_0)$
is defined as the distance a particle traverses until the probability is
$\exp(-1)$ that no interaction will occur. The interaction length for pair
production is about $\frac97$ of $X_0$ for bremsstrahlung. After crossing the
distance $X$ a particle's energy is given by
\begin{equation}
  E(X)=E_0\exp\left(-\frac{X}{X_0}\right),
\end{equation}
where $E_0$ is the initial energy. Bremsstrahlung and pair production mainly
happen in interactions with the nuclei of the atmosphere, since the probability
is proportional to the square of the atomic number.

\begin{figure}[tb!]
  \centering
  \begin{minipage}[c]{0.65\linewidth}
    \includegraphics[width=\textwidth, height=.6\textwidth]{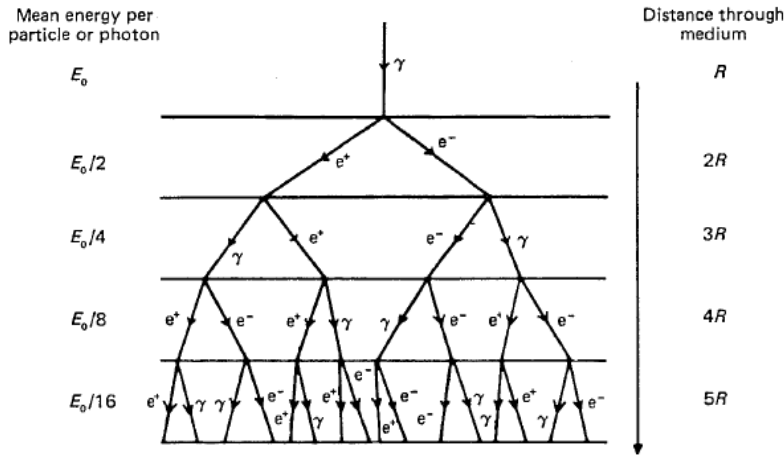}
  \end{minipage}\hfill
  \begin{minipage}[c]{0.35\linewidth}
  \caption[Model of an Electromagnetic Air Shower]{Simple model of an
    electromagnetic air shower according to \cite{Heitler}. The two
    interactions are bremsstrahlung and pair production. (Figure taken from
    \cite[pg. 120]{Longair1}.)}
  \label{fig:ElectromagneticShower}
  \end{minipage}
\end{figure}

In a simple model by \cite{Heitler} (Fig.~\ref{fig:ElectromagneticShower}), the
differences between the radiation and the interaction length are neglected and
it is assumed that bremsstrahlung and pair production occur with the same
frequency. Then, the distance $R$, after which on average half the particles
interact, is defined through
\begin{equation}
  \exp\left(\frac{R}{X_0}\right)=\frac12
\end{equation}
\begin{equation}
  \Leftrightarrow R=X_0\ln2.
\end{equation}
The total number of particles $N$ after $n$ steps of interaction is $N=2^n$.
Assuming that the particles split their energy at each interaction, the
critical energy is reached after the maximal number of $(n_{max})$
interactions. Therefore,
\begin{equation}
  E_c=\frac{E_0}{2^{n_{max}}},
\end{equation}
which yields the relation between $n_{max}$ and $E_c$ as
\begin{equation}
  n_{max}=\frac{\ln(E_0/E_c)}{\ln2}.
\end{equation}
Moreover, one obtains the relation for the number of particles at the shower
maximum $N_{max}$ by
\begin{equation}
  N_{max}=2^{n_{max}}=\frac{E_0}{E_c},
\end{equation}
with a composition of $\frac13N_{max}$ photons and $\frac23N_{max}$ electrons
and positrons. Also, the depth of the shower maximum $t_{max}$ in the atmosphere
is given by
\begin{equation}
  t_{max}=n_{max}\xi_0,
\end{equation}
where $\xi_0$ is the radiation length measured in matter per cm$^2$.
$\xi_0=\rho X_0$, where $\rho$ is the density.

\begin{figure}[tb!]
  \centering
  \begin{minipage}[c]{0.55\linewidth}
    \includegraphics[width=\textwidth]{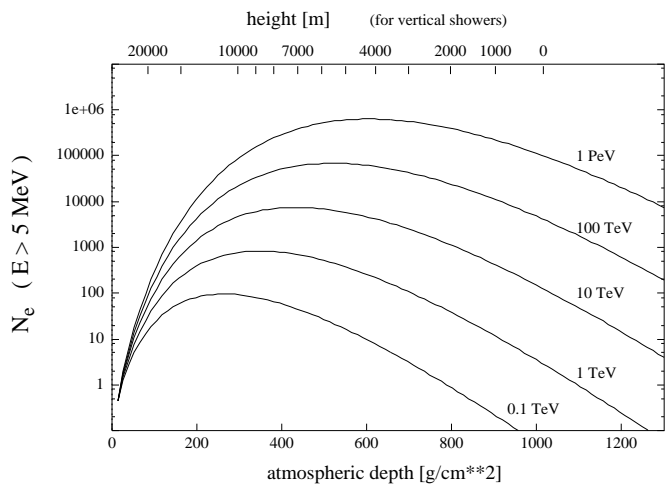}
  \end{minipage}\hfill
  \begin{minipage}[c]{0.4\linewidth}
    \caption[Longitudinal Profile of Electromagnetic Air Showers]{Average
      longitudinal profile of electromagnetic air showers with energies for the
      primary \g-ray interaction between 0.1\,TeV and 1\,PeV. $N_e$ is the
      number of electron-positron pairs with energy above 5\,MeV as determined
      from Monte Carlo simulations. The relation between the shower depth and
      shower height is given by the top and bottom scales. (Figure taken from
      \cite{Bernloehr:1996}.)}
    \label{fig:LongitudinalGammaShowers}
  \end{minipage}
\end{figure}

With the values of $E_c$ \footnote{For air $E_c=87$\,MeV
(\cite{Wigmans:2000}).} and $\xi_0$ \footnote{For air $\xi_0=37$\,gcm$^{-2}$
(\cite{ParticleDataBooklet:2006}).} for air, one can calculate these shower
parameters for the different energies of the primary particle. A few values are
given in Tbl.~\ref{tbl:GammaShowers}. In comparison with the more accurate
Monte Carlo simulation of the longitudinal shower development in
Fig.~\ref{fig:LongitudinalGammaShowers}, these values already provide a good
estimate. A more accurate description is given by the Nishimura-Kamata-Greisen
(NKG) formula (\cite{NKG}).

The height of the shower maximum depends on the atmospheric depth which is
determined through the atmospheric density profile
\begin{equation}
  \rho(h)=1.3\times10^{-3}{\rm gcm}^{-3}\exp \left(\frac{-h}{8\,{\rm
  km}}\right).
\end{equation}
The relation between height and atmospheric depth is represented by the top and
bottom scales of Fig.~\ref{fig:LongitudinalGammaShowers}.

The lateral distribution of a \g\ shower is determined by Moli\`ere scattering
(\cite{Bethe:1953}), which describes the multiple Coulomb scattering of
electrons and positrons in the atmosphere. The characteristic parameter is the
Moli\`ere radius. The spread caused by bremsstrahlung is negligible since
bremsstrahlung is emitted in a cone in forward direction with an angle
$\theta\sim \frac{m_ec^2}{E}=\frac1\gamma$. At sea level, 90\% of the shower
energy is deposited in a cylinder around the shower axis with a radius of
80\,m. Fig.~\ref{fig:ShowerLateral} and Fig.~\ref{fig:ShoweronGround} show the
lateral profile of a simulated shower of a photon of 300\,GeV and the
corresponding Cherenkov light at the ground, respectively.

\begin{table}[tb!]
  \centering
  \caption[Parameters of Electromagnetic Air Showers]{Parameters of
  electromagnetic air showers. $n_{max}$, $N_{max}$ and $t_{max}$ are the
  number of interactions, the number of particles and the depth at the shower
  maximum, respectively. $t_{max}$ is obtained by a simple estimate using the
  Heitler model. The corresponding height ($h$) is determined from
  Fig.~\ref{fig:LongitudinalGammaShowers}.}
  \bigskip
  \begin{tabular}{ccccc}
    \hline \hline
    $E_0$ [TeV] & $n_{max}$ & $N_{max} [10^{5}]$ & $t_{max} [X_0]$ & $h$ [km]\\
    \hline
    0.1 & 10 & 0.01 & 370 & 10 \\
    1   & 13 & 0.1  & 500 & 6  \\
    10  & 17 & 1    & 620 & 4  \\
    \hline \hline
  \end{tabular}
  \label{tbl:GammaShowers}
\end{table}

\subsection{Hadronic Showers}
Hadronic showers are initiated by nuclei of cosmic radiation $(N_{\rm cosmic})$
which penetrate the atmosphere. The cosmic radiation consists mainly of protons
(87\%), $\alpha$ particles (12\%) and a small fraction of heavier atomic
nuclei. Electrons, \g-rays and high energetic neutrinos only constitute a minor
fraction. When entering the atmosphere the nuclei produce spallation fragments
and new particles $(X)$ in inelastic collisions with the nuclei of the
atmosphere ($N_{\rm atm.}$). The new particles are mainly pions in the ratios
$\frac13\pi^0$, $\frac13\pi^+$, $\frac13\pi^-$. The $\pi^0$ have a lifetime of
$8.4\times 10^{-17}$\,s and decay into two photons which can initiate
electromagnetic showers. The $\pi^\pm$ have a longer lifetime of
$2.6\times10^{-8}$\,s and can produce other particles, mainly pions, in
inelastic collisions. The charged pions can also decay into muons and
neutrinos. The muons decay into neutrinos and electrons, which can start
electromagnetic showers. Therefore, a hadronic shower also has an
electromagnetic shower component. Also the primary particles and the spallation
fragments can form hadronic subshowers. So the main reactions are
\begin{eqnarray}
  N_{\rm cosmic} +N_{\rm atm.} & \rightarrow & {\rm hadrons} + X \\
  \pi^0 & \rightarrow & \gamma+\gamma \\
  \pi^- & \rightarrow & \mu+\overline\nu_\mu \\
  \pi^+ & \rightarrow & \mu^+\nu_\mu \\
  \mu^- & \rightarrow & e^-+\overline\nu_e+\nu_\mu \\
  \mu^+ & \rightarrow & e^++\nu_e+\overline\nu_\mu.
\end{eqnarray}
\begin{figure}[b!]
  \begin{minipage}[b]{0.5\linewidth}
    \includegraphics*[width=.85\textwidth]{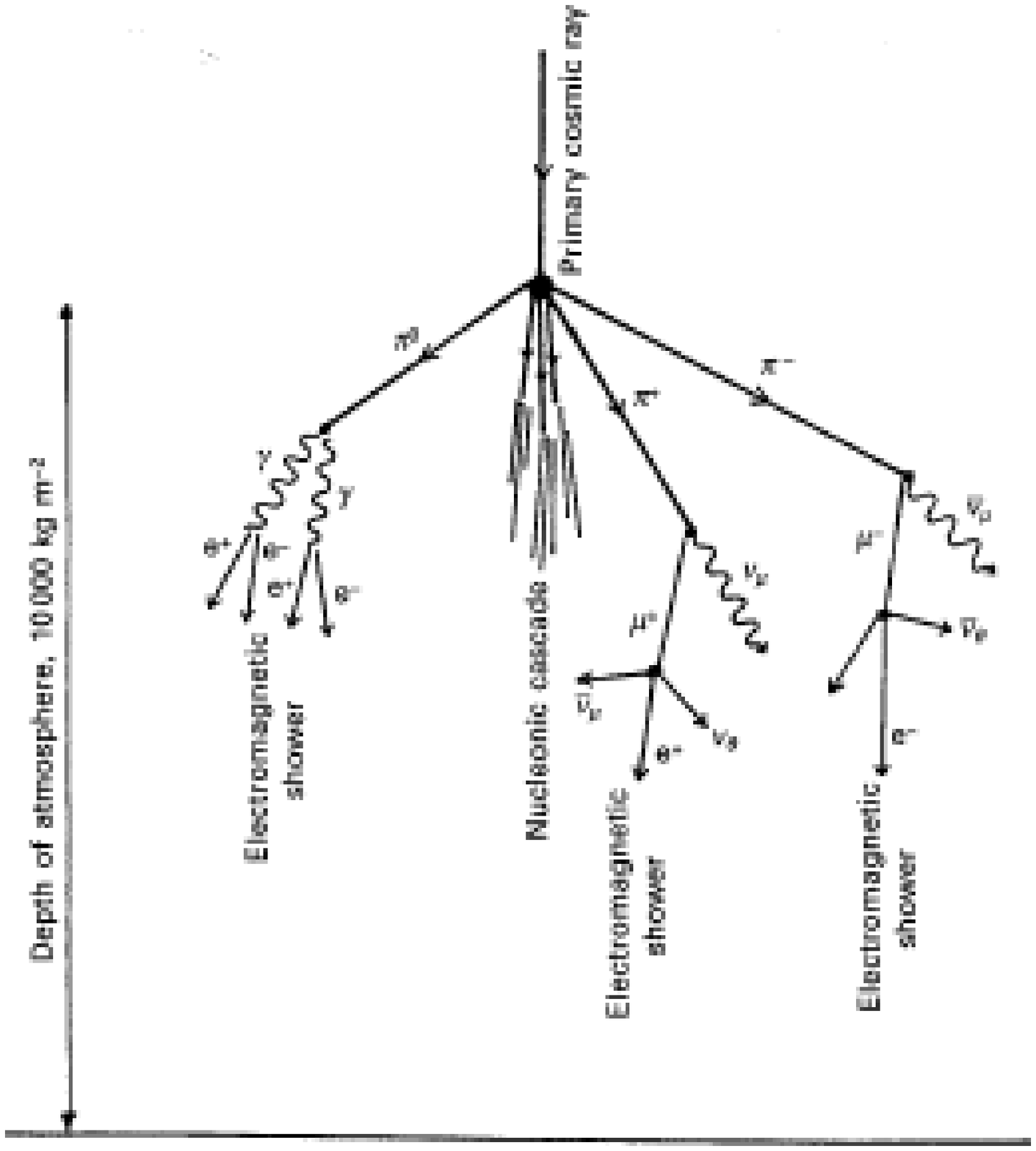}
    \caption[Model of a Hadronic Air Shower]{Model of a hadronic air shower.
      Characteristic is the hadronic component of pions and muons.
      Electromagnetic subshowers are also part of a hadronic shower. (Figure
      taken from \cite[pg. 149]{Longair1}.)}
    \label{fig:HadronicShower}
  \end{minipage}\hfill
  \begin{minipage}[b]{0.5\linewidth}
    \includegraphics[width=\textwidth]{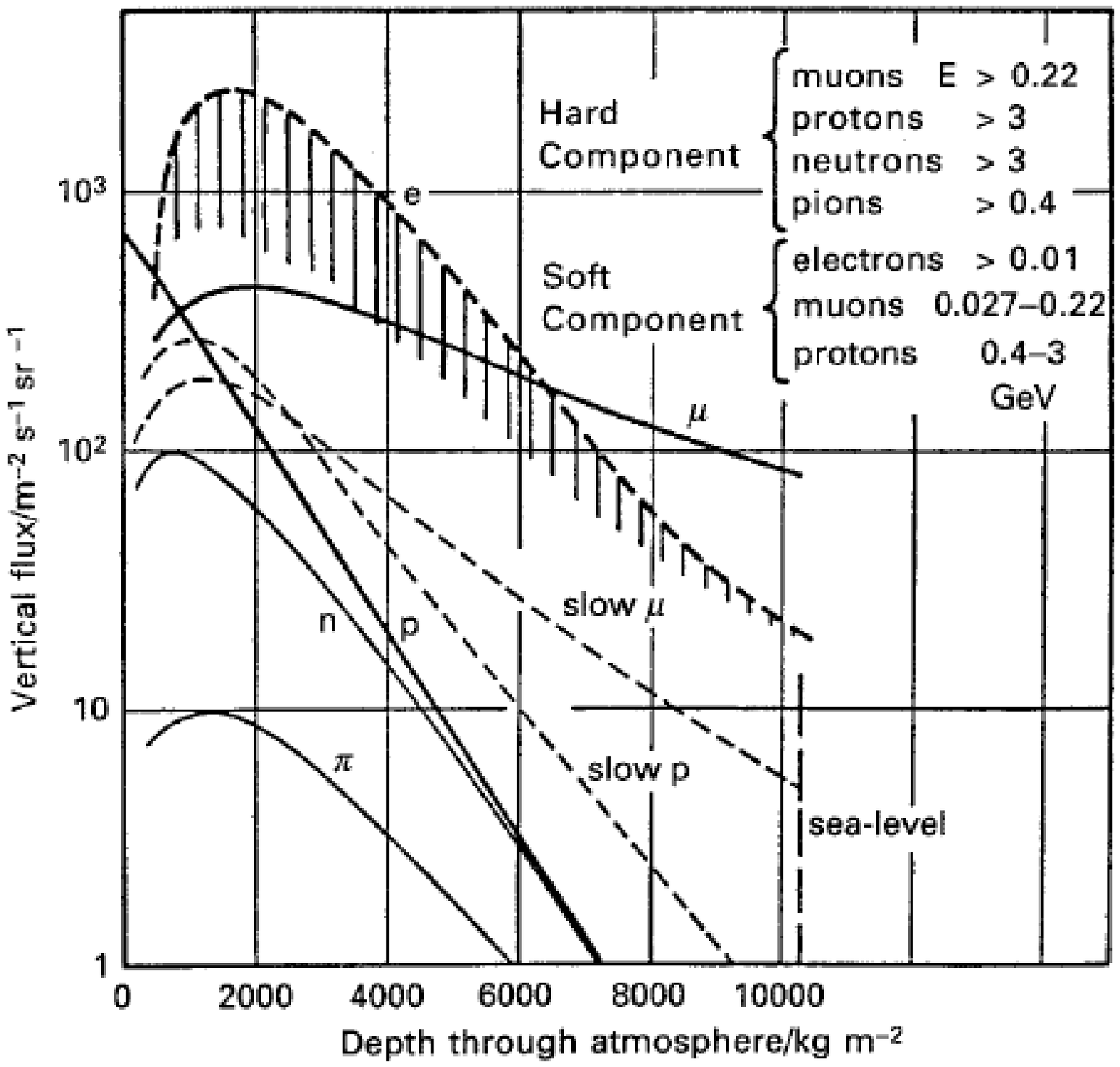}
    \caption[Flux of Different Particles in the Atmosphere]{The vertical fluxes
      of different components of particles in the atmosphere as presented by A.
      M. Hillas (1972, Cosmic rays, Page 50, Oxford: Pergamon Press.) (Figure
      taken from \cite[pg. 150]{Longair1}.)}
    \label{fig:HadronicShowerComposition}
  \end{minipage}
\end{figure}
The development of a hadronic shower is sketched in
Fig.~\ref{fig:HadronicShower}. Fig.~\ref{fig:HadronicShowerComposition} shows
the distribution of high energy particles in the atmosphere. It also represents
the composition of hadronic shower cascades which are the most frequent in the
atmosphere.

The mean free path of the hadrons is given by
\begin{equation}
  \lambda_i=\frac{1}{n\sigma_i},
\end{equation}
where $n$ and $\sigma_i$ are the particle density and the target particles'
cross section in the medium. The atmospheric depth for hadrons corresponds to
about $12\times\lambda_i$. Since $\lambda_i$ is about twice the radiation
length of electromagnetic showers, hadronic showers penetrate deeper into the
atmosphere.

Due to the transversal momentum of the secondary particles, the hadronic
showers have a larger lateral spread than electromagnetic showers, and are more
irregular. Their lateral distribution can be used for the separation from
electromagnetic showers. Fig.~\ref{fig:ShowerLateral} and
Fig.~\ref{fig:ShoweronGround} show the longitudinal profile of a simulated
proton shower of 1\,TeV and the corresponding Cherenkov light at the ground,
respectively.

\section{Cherenkov Light}
Air showers can be detected by their Cherenkov light. Cherenkov light is
emitted from charged particles which travel faster through a medium than the
speed of light in that medium. For a medium with an index of refraction $n$,
the condition for the velocity $v$ to produce Cherenkov light is
\begin{equation}
  v\ge\frac{c}{n}, \quad \Leftrightarrow \quad \beta\ge \frac1n,
  \label{eqn:CherenkovCondition}
\end{equation}
where $c$ denotes the speed of light. This condition implies an energy
threshold $(E_{th})$ which a particle has to exceed before it can emit
Cherenkov light, namely
\begin{equation}
  E_{th}(m_0)=\gamma_{th} m_0 c^2=\frac1{\sqrt{1-\beta_{th}^2}}m_0c^2 = 
  \frac1{\sqrt{1-\frac1{n^2}}}m_0c^2.
\end{equation}
Therefore, particles with a low mass, such as electrons, dominate Cherenkov
emission. The threshold energies for electrons and muons are 21\,MeV and
4.3\,GeV respectively.

Cherenkov radiation is emitted at an angle of
\begin{equation}
  \theta_C={\rm acrcos}\frac{1}{n\beta}
\end{equation}
relative to the direction of the particle's velocity. For air with the index of
refraction $n_0 \sim 1+3\times 10^{-4}$ and the condition of
Eqn.~\ref{eqn:CherenkovCondition}, the maximal opening angle of 1.4$^\circ$ is
obtained by $\beta=1$, i.e. $v = c$.

In air, Cherenkov radiation is emitted at wavelengths $\lambda$ between 400\,nm
and 700\,nm. About 30 photons per meter are produced by a single charged
particle. Although it takes about 50 microseconds for a shower cascade to
develop, the front of Cherenkov light is only visible within $\sim$10\,ns at
the ground, since the cascade develops nearly along the light pass. At each
point within the Cherenkov cone the light is only visible for 5\,ns. Within
100\,m from the shower axis, the light front reaches the ground with about 100
photons per m$^2$. Therefore, IACTs require cameras with high sensitivity and
short exposure times.

\begin{figure}[h!]
  \setlength{\abovecaptionskip}{0.4cm}
  \begin{minipage}[c]{0.5\linewidth}
    \centering \fbox{\centering
      \includegraphics[width=.97\textwidth, height=10.8cm]
                      {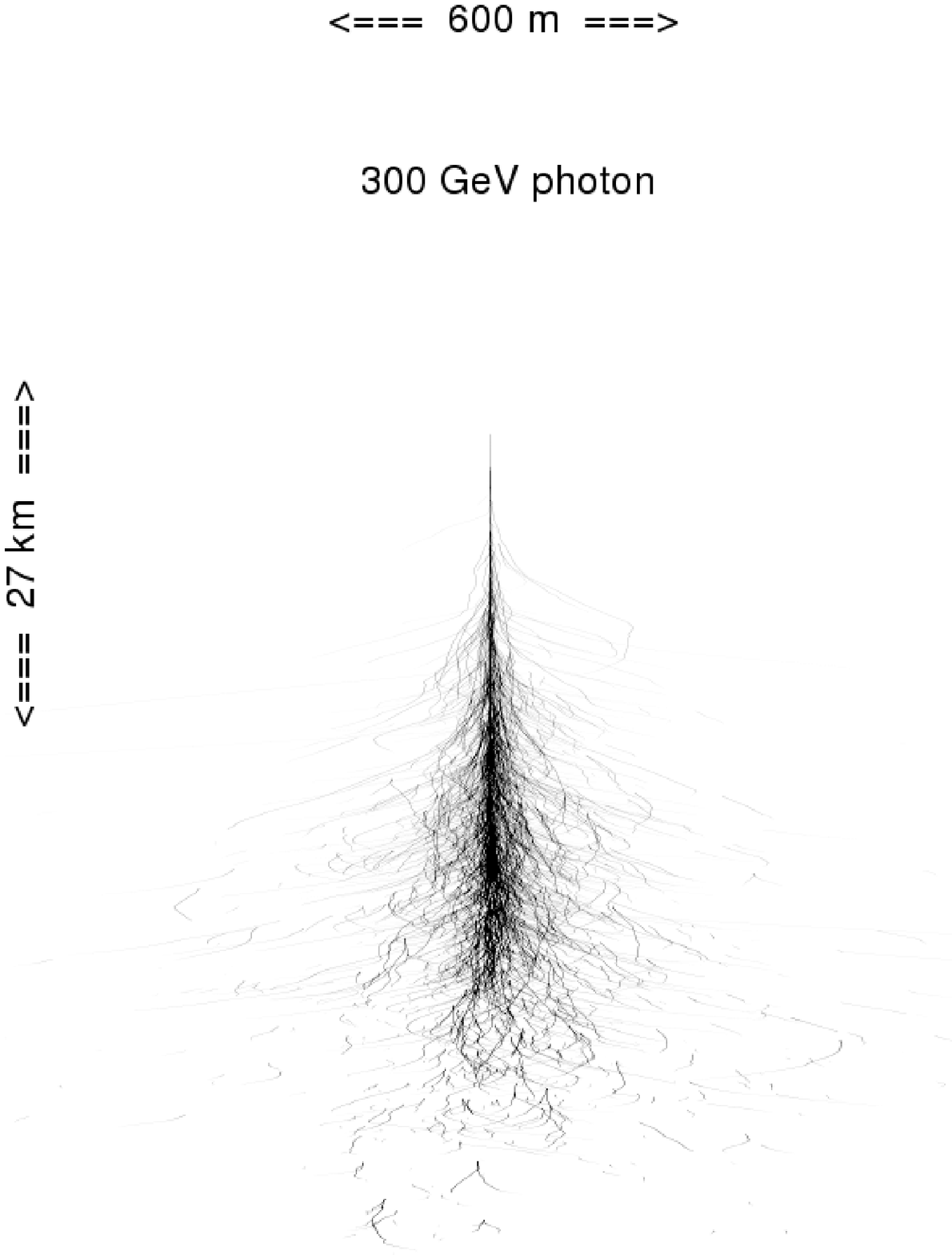}}
  \end{minipage}\hfill
  \begin{minipage}[c]{0.5\linewidth}
    \centering
    \fbox{\centering
      \includegraphics[width=.97\textwidth, height=10.8cm]
                    {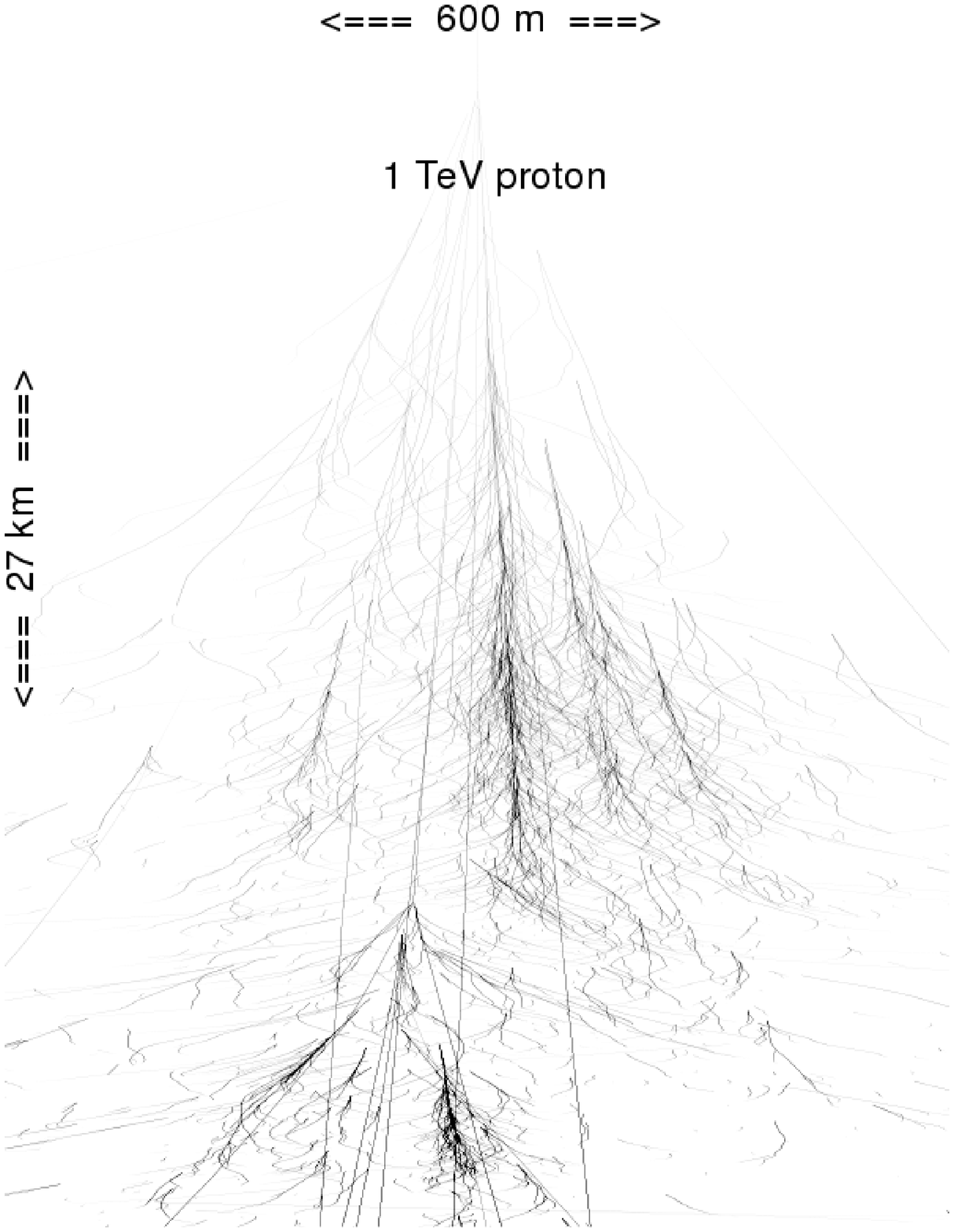}}
  \end{minipage}
  \begin{minipage}[c]{\linewidth}
    \caption[Monte Carlo Simulations of Longitudinal Shower Profiles]{Monte
    Carlo simulations of the longitudinal profile of a \g-ray shower (left) and
    proton shower (right) within 0 to 27\,km a.s.l. (Figure taken from
    \cite{Bernloehr}.)}
    \label{fig:ShowerLateral}
  \end{minipage} \vspace{.7cm} \\
  \begin{minipage}[c]{0.5\linewidth}
    \centering
    \fbox{\includegraphics[width=.97\textwidth]{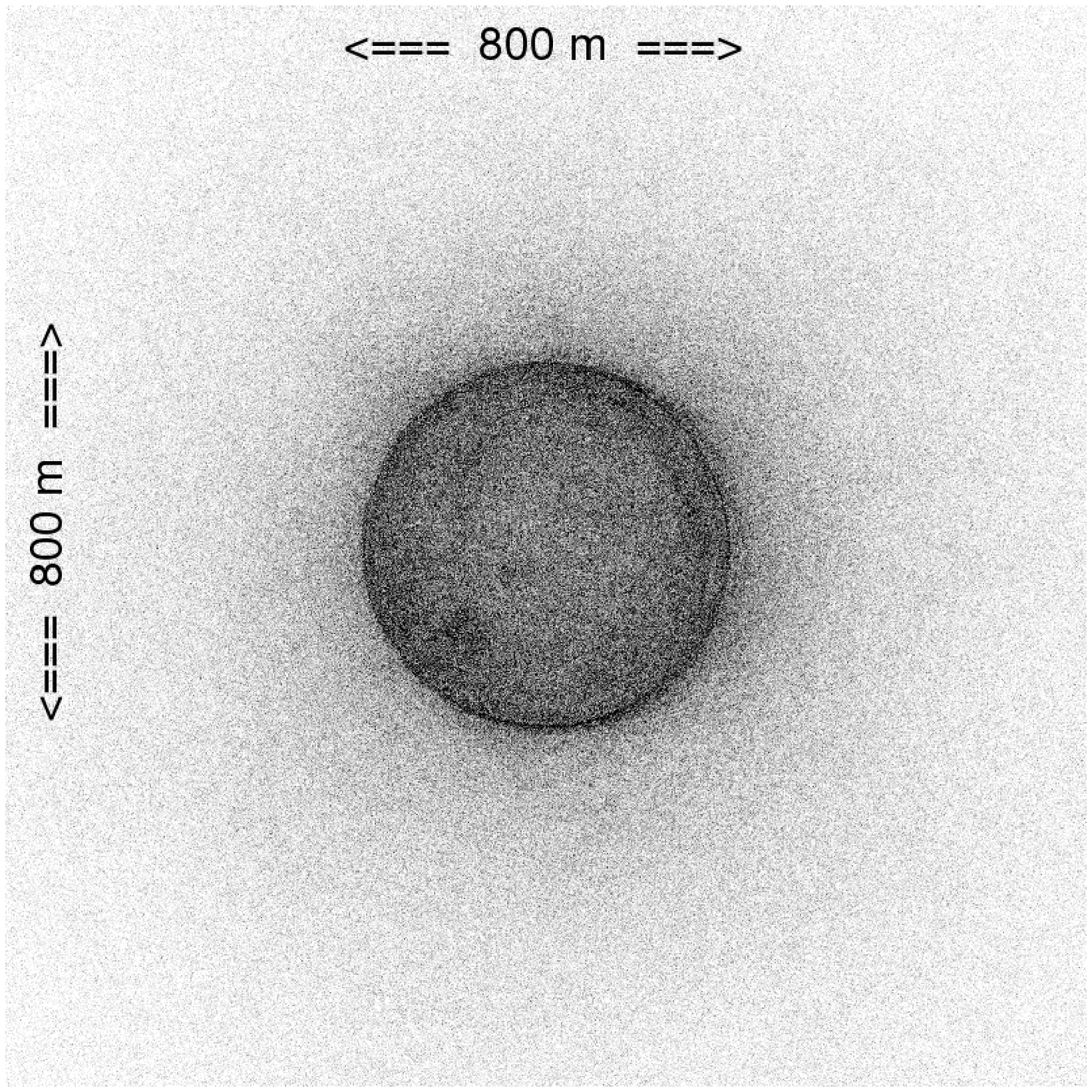}}
  \end{minipage}\hfill
  \begin{minipage}[c]{0.5\linewidth}
    \centering
    \fbox{\includegraphics[width=.97\textwidth]{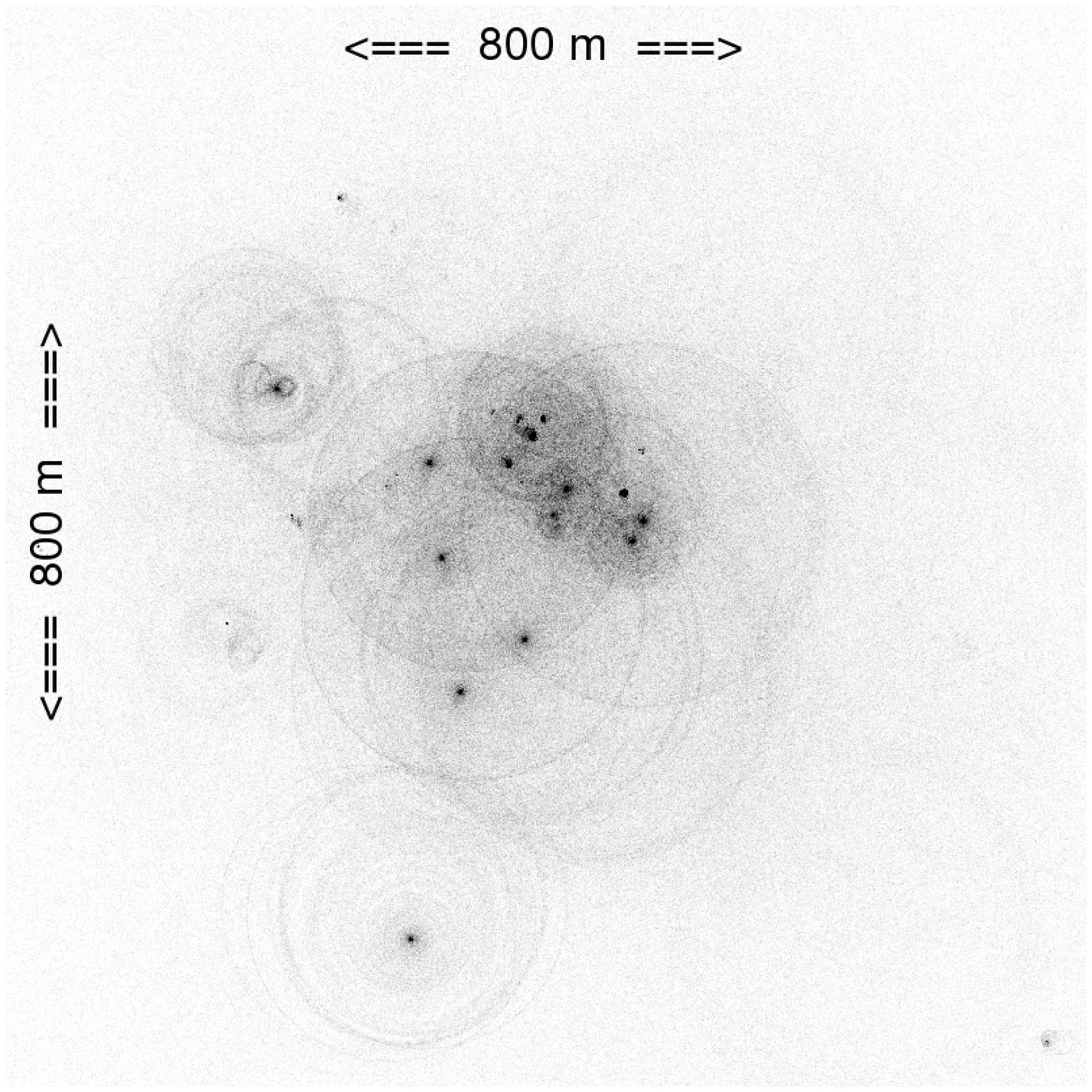}}
  \end{minipage}
  \begin{minipage}[c]{\linewidth}
    \caption[Monte Carlo Simulations of Cherenkov Light Reaching the
    Ground]{Monte Carlo simulations of Cherenkov light at the ground emitted
    from a \g\ shower (left) and proton shower (right). The \g\ shower shows
    the typical Cherenkov light pool with a radius of $\sim$120\,m. The proton
    shower is rather irregular. (Figure taken from \cite{Bernloehr}).}
    \label{fig:ShoweronGround}
  \end{minipage}
\end{figure}
\clearpage

Fig.~\ref{fig:ShoweronGround} (left) shows the distribution of Cherenkov
light at the ground, simulated for a \g\ shower of 300\,GeV. The light is
rather homogeneously concentrated within a radius of $\sim$120\,m from the
shower axis and attenuates outside. This radius is similar for \g\ showers of
different zenith angles and energies. Fig.~\ref{fig:ShoweronGround} (right)
shows the same simulation for a proton shower of 1\,TeV. The light is
inhomogeneously distributed and has several intensity maxima.

\section{Imaging Atmospheric Cherenkov Technique}
\begin{figure}[b!]
    \includegraphics[width=\textwidth]{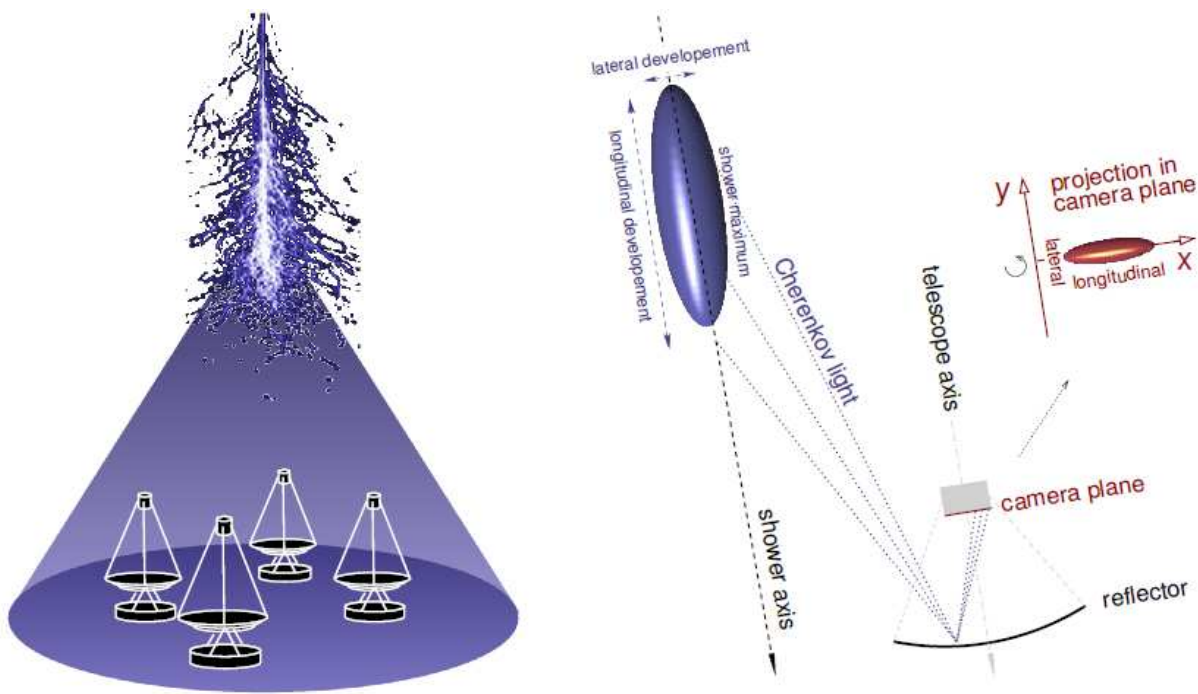}
    \caption[Illustration of the Imaging Atmospheric Cherenkov
    Technique]{Illustration of the imaging atmospheric Cherenkov technique. An
    air shower produces a light cone of Cherenkov light close to the center of
    a telescope array (left). The shower image is recorded by the telescope.
    The geometric optics of the mapping to the camera plane and the camera
    image are illustrated (right). (Illustration taken from \cite[pg.
    39]{Schlenk}.)}
    \label{fig:IACTechnique}
\end{figure}

With IACTs it is possible to detect and identify the Cherenkov light from
\g-ray showers in the atmosphere and thus to determine the energy and arrival
direction of the \g-rays. Detection is possible with an optical system which
consists of a mirror dish and a camera close to the focal plane. If a shower
develops close to the telescope, the shower profile is reflected to the camera
and recorded. "Close" means within the radius of the Cherenkov light pool of
$\sim$120\,m at the ground. Fig.~\ref{fig:IACTechnique} illustrates the mapping
of an air shower by the optical system of a telescope through geometric optics.

Since the atmosphere is an essential part of the detection, IACTs are sensitive
to the atmospheric conditions during observations. On the other hand, the
collection areas do not have theoretical limits and they scale linearly with
the mirror surface and the number of telescopes.

IACTs can be combined into arrays for observation in a stereoscopic mode,
meaning that a shower is recorded by at least two telescopes from different
viewing angles. The advantages are e.g. a higher accuracy in the shower
reconstruction and an improved rejection rate of background events.

The separation of \g\ showers from the majority of hadronic showers, the shower
reconstruction and data analysis are accomplished with Monte Carlo simulations
of the air showers and the system's response.

\section{Overview of Current IACTs}
IACTs have been in use for about two decades now. The Whipple IACT is
considered the first IACT which was widely recognized among \g-ray astronomers.
The Whipple collaboration established the IACT technique and detected TeV \g\
radiation from the Crab Nebula in 1989 (\cite{Weekes:1989}). In subsequent
years new IACTs were developed, among them Cat, HEGRA and Cangaroo. These are
considered as IACTs of the first generation. They detected new TeV \g-ray
sources and improved their technique. In 2002 H.E.S.S. came into operation
which is considered as the first IACT of the second generation, for it is based
on the same technique but has a greatly increased sensitivity and precision.
H.E.S.S. is currently one of the most sensitive IACTs and has more than doubled
the number of known TeV \g-ray sources by 2006. Other similar sensitive IACTs
of the second generation are Magic, Veritas and Cangaroo~III.

\begin{figure}[ht!]
    \includegraphics[width=\textwidth]{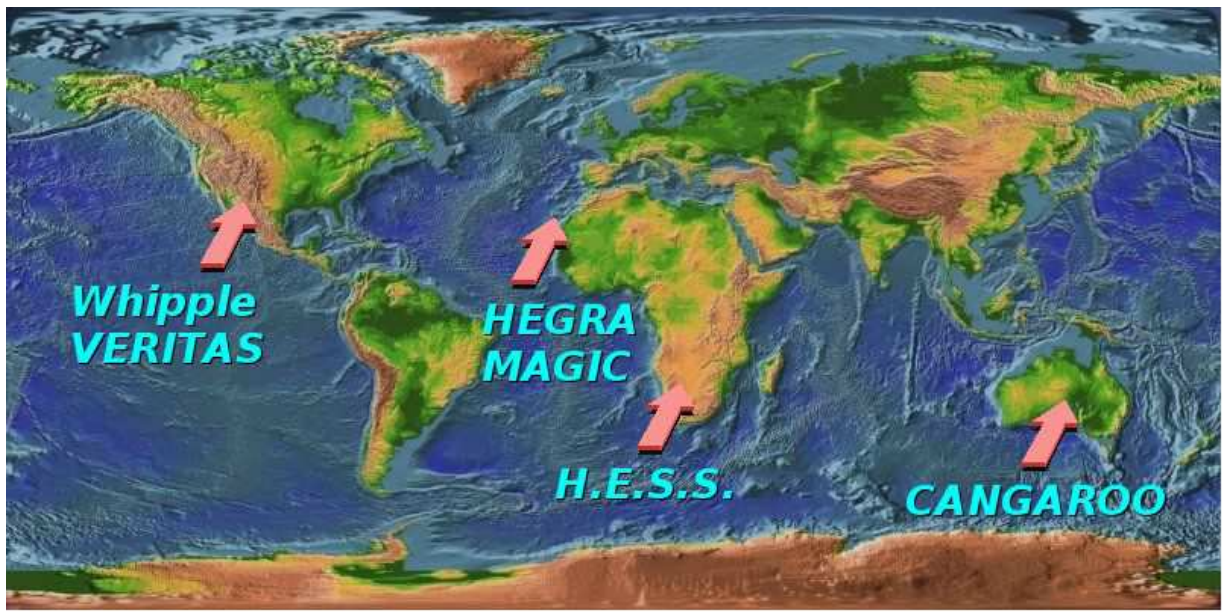}
    \caption[World Map of Imaging Atmospheric Cherenkov Telescopes]{Imaging
    atmospheric Cherenkov telescopes worldwide. The even distribution of
    telescopes between the northern and southern hemisphere provides coverage
    of the full TeV \g-ray sky. Telescopes located in the southern hemisphere
    have a direct view of the galactic center and the galactic plane. (Map
    taken from \cite{Punch:2005}.)}
    \label{fig:IACTelescopes}
\end{figure}

Geographic location is important for IACTs, since it determines the observable
sky regions. Systems located in the southern hemisphere like H.E.S.S. have a
direct view of the galactic center and the galactic plane where the majority of
galactic TeV \g-ray sources are located. The map in
Fig.~\ref{fig:IACTelescopes} shows the second generation IACTs worldwide. The
even distribution of telescopes between the northern and southern hemisphere is
advantageous, since it provides full coverage of the TeV \g-ray sky.

\chapter{The H.E.S.S. Experiment} \label{chp:H.E.S.S.}

\begin{figure}[th!]
  \includegraphics[width=\textwidth]{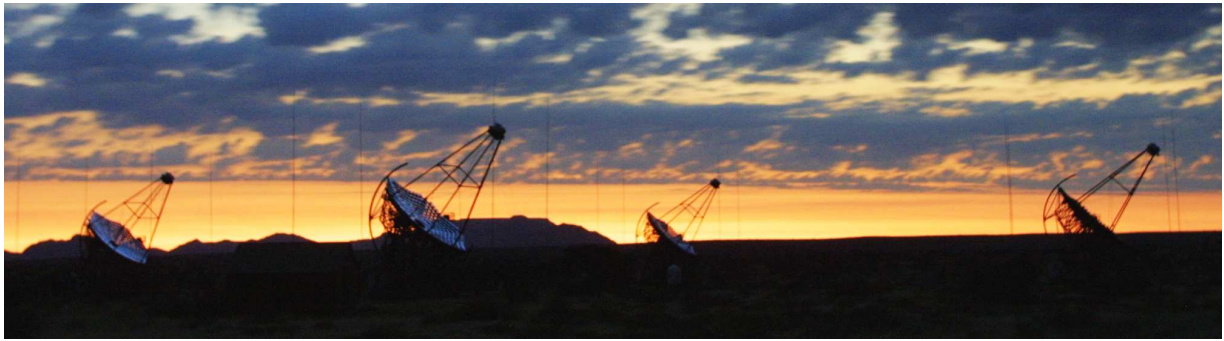}
  \caption[Photo of the H.E.S.S. Array]{The H.E.S.S. array on the Khomas
    Highlands in Namibia. (Photo from \citet{Eifert}.)}
  \label{fig:H.E.S.S.Array}
\end{figure}

H.E.S.S. is the name of the new imaging atmospheric Cherenkov observatory
located in Namibia (Fig.~\ref{fig:H.E.S.S.Array}). It is an acronym for High
Energy Stereoscopic System. H.E.S.S. was founded through the international
collaboration of around 100 scientists from about 20 European and African
institutes under the leadership of the Max-Planck Institute for Nuclear Physics
in Heidelberg, Germany. The name was also chosen in honor of the Austrian
physicist Victor Franz Hess, who laid the foundations of modern astroparticle
physics by his discovery of cosmic rays in 1912 and who, as a result, was
awarded a Nobel Prize in 1936. H.E.S.S. is an IACT of the second generation
which succeeds the IACT experiment HEGRA. It has been developed as well as is
operated partially by the same people. H.E.S.S. is very sensitive in an energy
range from 0.2 to 50\,TeV. It is able to detect a \g-ray point source that has
a flux of $\rm 2.0 \times 10^{-13} cm^{-2}s^{-1}$, corresponding to only 1\% of
the flux from the Crab Nebula \footnote{The Crab Nebula is the standard candle
in VHE \g-ray astronomy.} with a significance of 5 $\sigma$ in about 25 hours
or a source of similar strength within 30 seconds (\citet{CrabPaper}). One of
the basic concepts of H.E.S.S. is the technique of stereoscopy, as is reflected
in its name. Stereoscopy provides improved shower reconstruction and increased
rejection rates for background events. In its first phase (H.E.S.S. I), the
array consists of four identical Cherenkov telescopes (CT\,1, CT\,2, CT\,3,
CT\,4). In the second phase (H.E.S.S. II), the array will be supplemented by an
additional telescope located in the center of the array which will have a
larger mirror surface and increased sensitivity. H.E.S.S. II is currently under
development. H.E.S.S. I started observation in June 2002 using its first
telescope, and in the meantime the full telescope array was gradually
completed. Since early 2004 observations have been made using the array of all
four telescopes in stereoscopic mode. In 2006, H.E.S.S. had already confirmed
most of the approximately 10 TeV \g-ray sources known before and had discovered
about 20 new ones. Within the first two years of its operation H.E.S.S.
exceeded most scientists' expectations (cf. \citet{hofmann_ICRC2005} and
\citet{PlaneScanI}).

\section{The Site}
\begin{figure}[ht]
  \begin{minipage}[c]{0.5\linewidth}
    \includegraphics[width=\textwidth]{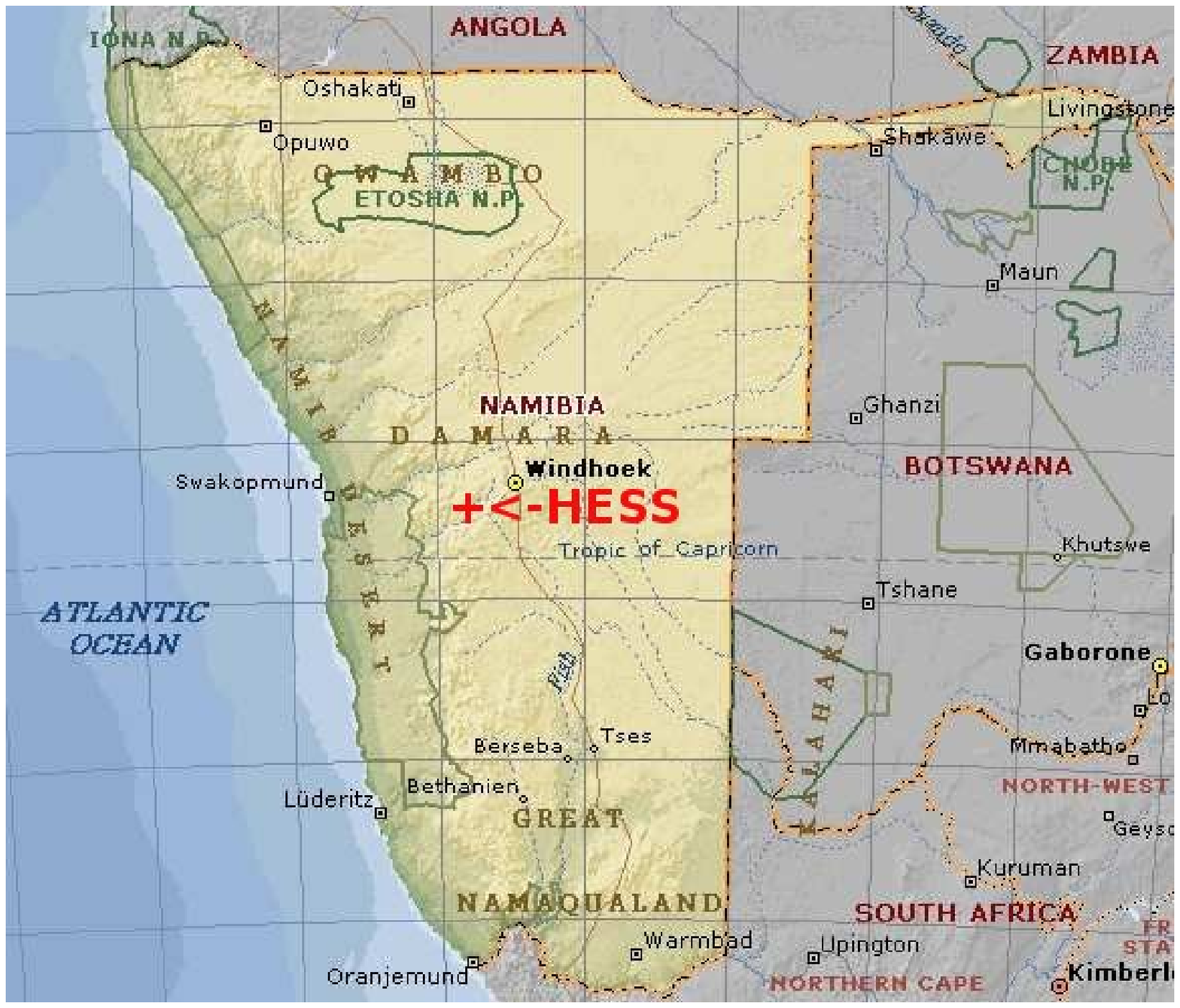}
  \end{minipage}\hfill
  \begin{minipage}[c]{0.45\linewidth}
    \caption[Map of Namibia Showing the H.E.S.S. Site]{Map of Namibia showing
      the location of the H.E.S.S. site. (Map taken from \citet[pg.
      40]{Schlenk}.)}
    \label{fig:MapNamibia}
  \end{minipage}
\end{figure}

The H.E.S.S. site is located in the Khomas Highland of Namibia
(Fig.~\ref{fig:MapNamibia}). The geographic location of the center of the
telescope array is 16$^\circ$30'00.8''\,E, 23$^\circ$16'18.4''\,S at 1800\,m
asl. There were several reasons why this location was selected. First, the dry
climate of the Khomas Highlands allows for observations to be made throughout
the year, with a total observation time of approximately 1600 hours per year at
good and stable atmospheric conditions. The many km$^2$ of sparsely populated
area surrounding the H.E.S.S. site provide a minimum of night sky background
also. Yet the capital of Namibia, Windhoek, lies at a distance of 100\,km
northwest from the H.E.S.S. site and thus provides the necessary infrastructure
to maintain the observatory at reasonable costs. Finally, H.E.S.S.' location
in the southern hemisphere permits the observation of the galactic plane and
the galactic center at high zenith angles, which is when the telescopes are most
sensitive. The galactic plane is particularly important for observation because
it hosts most galactic \g-ray sources and the super massive black hole Sgr A*.
Besides the four Cherenkov telescopes, the site also contains the optical
robotic telescope ROTSE, a control building with a workshop, a generator
house and a residence building at a distance of 1\,km from the observatory
buildings, separated from them by a hill.

\section{The Telescopes}
\begin{figure}[t!]
  \centering
  \includegraphics[width=.7\textwidth]{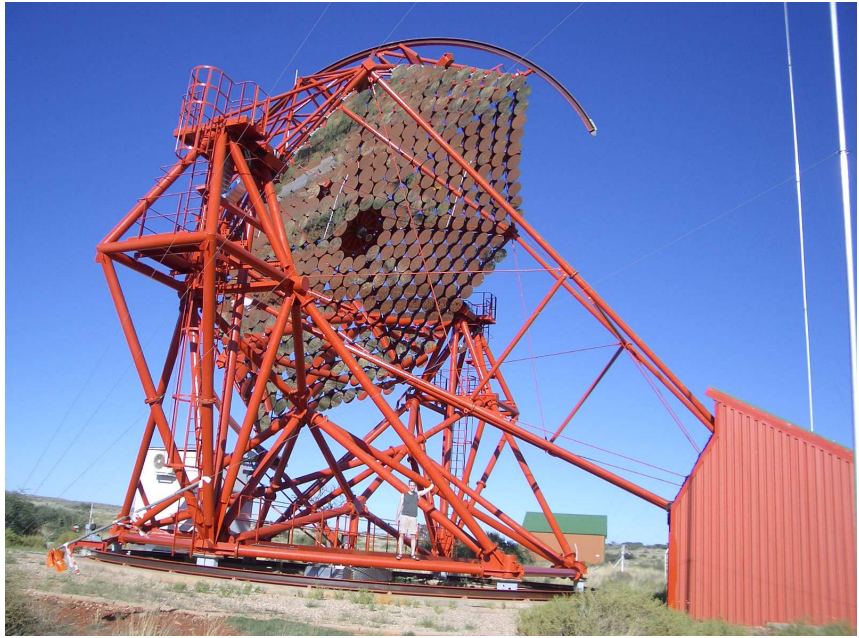}
  \caption[Photo of a H.E.S.S. Telescope (CT\,1)]{H.E.S.S. telescope (CT\,1) in
    the parking position. During the daytime the camera is protected in the
    shelter on the right. The scale is demonstrated by the person in the front.
    (Photo provided by Frank Breitling, 2004.)}
  \label{fig:Telescope}
\end{figure}

The four telescopes are placed in the corners of a square, which has a length
of 120\,m and its diagonals oriented in north-south and east-west directions.
Each telescope is made with a rigid steel structure and has a total weight of
50 tons. Each consists of a reflector dish that is 13\,m in diameter and a
camera mounted in the focal plane of the telescope at a distance of 15\,m from
the dish. The dish is mounted at a height of 13\,m above the ground to the
support structure. The telescope can reach a maximum height of 28\,m for
observations at the zenith. Fig.~\ref{fig:Telescope} shows CT\,1 in the parking
position. During the daytime the telescopes are parked and the cameras are
protected in their shelters against light, heat and dust. The scale of the
telescope is demonstrated by the person standing in front on the structure.

\subsection{Davies-Cotton Design}
Each dish has a total reflector area of 107\,m$^2$. If shadowing by the camera
support structure is taken into account the effective reflector area is reduced
to about 95\,m$^2$. The reflector is composed of 380 circular mirror facets,
each with a diameter of 60\,cm and a reflectivity of 80\%. Each mirror is
mounted on a support unit containing two actuators which allow individual
alignment. The reflector follows a Davies-Cotton design (\citet{DaviesCotton}),
which means that the dish and its mirror facets are spherical and have a focal
length that is identical to the focal length of the dish. An advantage of the
Davies-Cotton design is the cost efficiency of its manufacture. For in
comparison to other designs, all of the mirrors in this one are identical.
Although the Davies-Cotton design suffers from spherical aberration, this is
not critical for H.E.S.S. The reason for this is that the residual point spread
function of the reflector dish after the alignment of the individual mirrors is
well contained in a camera pixel with a size of 0.16$^\circ$. Also, the time
dispersion due to aberration of approximately 5\,ns is not critical for the
readout of the camera image. Detailed information about the telescope mirror,
its alignment and optical characteristics are given in \citet{Bernloehr:2003vd}
and \citet{Cornils:2003ve}.

\subsection{Pointing Accuracy}
Each telescope has an alt-az tracking with a slew speed of 100$^\circ$
min$^{-1}$. The tracking position is measured by shaft encoders with a digital
step size of 10'' and is maintained with an accuracy of 30''. Certain
conditions can negatively affect the actual pointing position of the
telescope's structure, mainly the camera support structure, which can bend due
to its weight. But also wind pressure or dirt on the rail of the tracking
system can eventually contribute to a loss of pointing accuracy of a few arc
seconds. To monitor the deviations, each telescope is equipped with two CCD
cameras: a Sky CCD and a Lid CCD. The Sky CCD is located in the right side of
the dish and can monitor the telescopes field of view (FOV). The Lid CCD is
located in the center of the dish and can monitor stars that are reflected by
the dish onto the closed camera lid. From simultaneous observations by these
two cameras, a pointing model has been developed (\citet{GillessenThesis})
which describes the actual pointing position as a function of the tracking
position. It is used to apply corrections off-line during the data analysis and
is able to limit the systematic pointing error to 20''.

\section{The Camera}
The H.E.S.S. cameras are described in great detail by \citet{Camera}. A camera
consists of 60 drawers which contain a total of 960 pixels. Each pixel consists
of a photo-multiplier tube (PMT) with a quantum efficiency of 20-30\% in the
wavelength range of 300-700\,nm. The front of the PMTs is equipped with a layer
of hexagonal Winston cones in a honeycomb arrangement which reduces the light
insensitive area between neighboring pixels to about 5\%. The FOV of each pixel
is 0.16$^\circ$ and contributes to the total FOV of 5$^\circ$ of the camera.
Each PMT is calibrated to respond with an amplification of $2\times10^5$
electrons for each collected photo-electron (p.e.). The PMT signal is fed into
three different channels: the low gain, the high gain and the trigger channel.
The low and high gain channels provide a linear response from 1 to 1600\,p.e.
Their signal is stored in an Analogue Ring Sampler developed by the ANTARES
experiment. It samples the signal at 1\,GHz over time windows of 16\,ns. If an
event has been triggered, the corresponding buffer is digitized and sent to the
central data acquisition system. It takes about 610\,$\mu$s until the data is
transferred and new data can be recorded. This is the dead-time of the camera.
The resulting upper limit for the camera's acquisition rate is 1.6\,kHz. The
core camera's electronics are located in a crate behind the layer of PMTs. It
contains, among others things, the sockets for the drawers, the readout and
trigger cards, a central processor unit, the bus systems, a 100\,Mbits/s
network interface, four power supplies, 16 temperature sensors and about 80
computer controlled fans. In addition, each camera is equipped with a global
positioning system providing event times with an accuracy of $\mu$s. The
electronics constitute the camera's data acquisition system. It is controlled
by a Linux operating system written in programming language $C$. The camera's
electronics and PMTs are housed in a container that is $\rm
2\,m\times2\,m\times1.6$\,m, with a total weight of about 900\,kg and a lid in
the front and in the back. Its total power consumption is about 5\,kW.

While the individual pixels constitute a first level trigger, the camera
trigger system as a whole constitutes the second level trigger. It consists of
38 overlapping trigger sectors, with each containing 64 pixels. The typical
trigger condition requires four neighboring pixels to exceed a threshold
5\,p.e. within a time window of 2\,ns. It takes about 70\,ns to build a trigger
signal which is fast enough to read out the data from the Analogue Ring
Sampler. The camera trigger significantly reduces the number of background
events. Depending on the trigger configuration and the zenith angle of the
observations, the typical camera trigger rate is about $(200\pm50)$\,Hz.

\begin{figure}[t!]
  \includegraphics[width=.3299\textwidth]{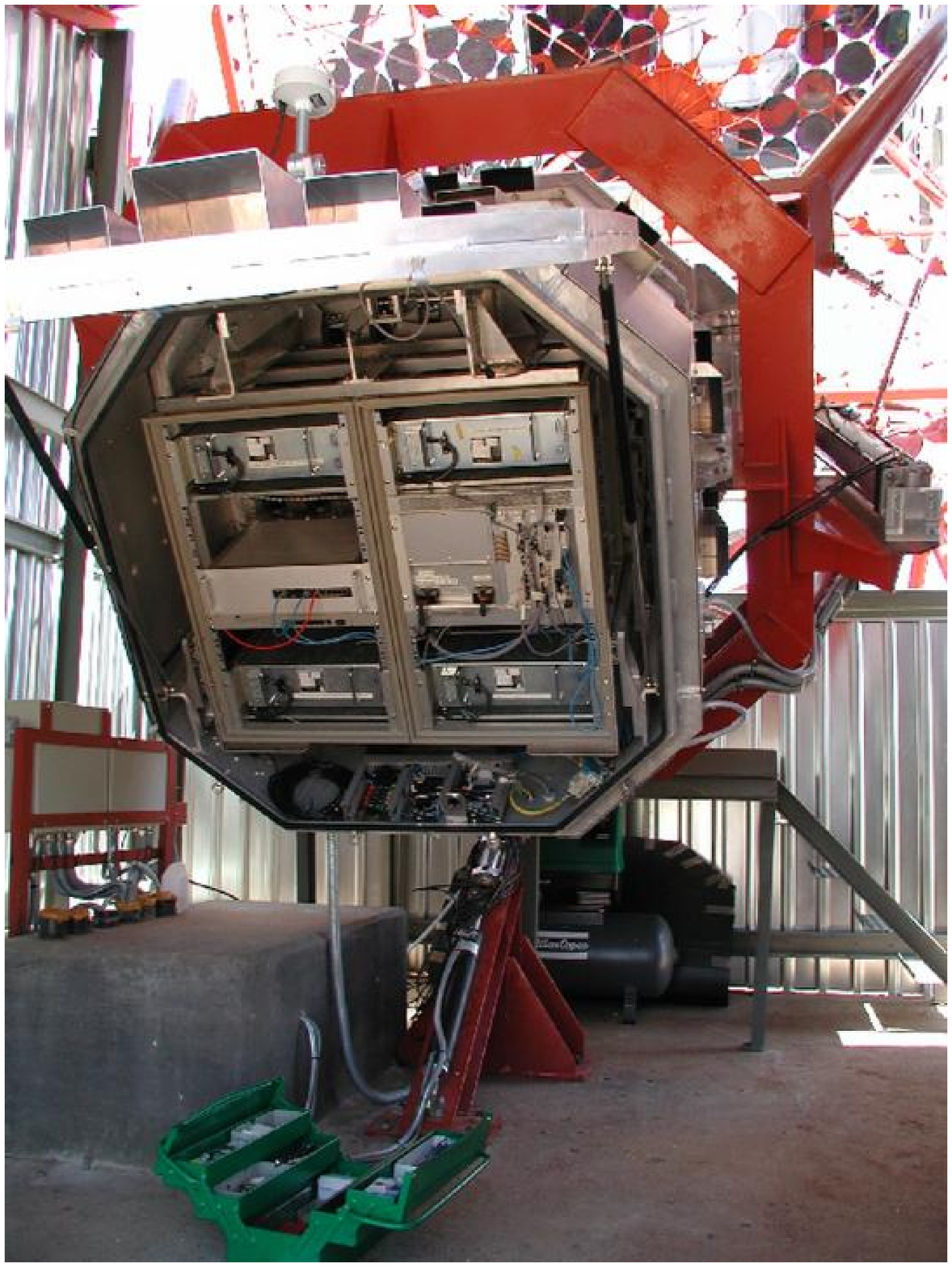}
  \includegraphics[width=.3299\textwidth]{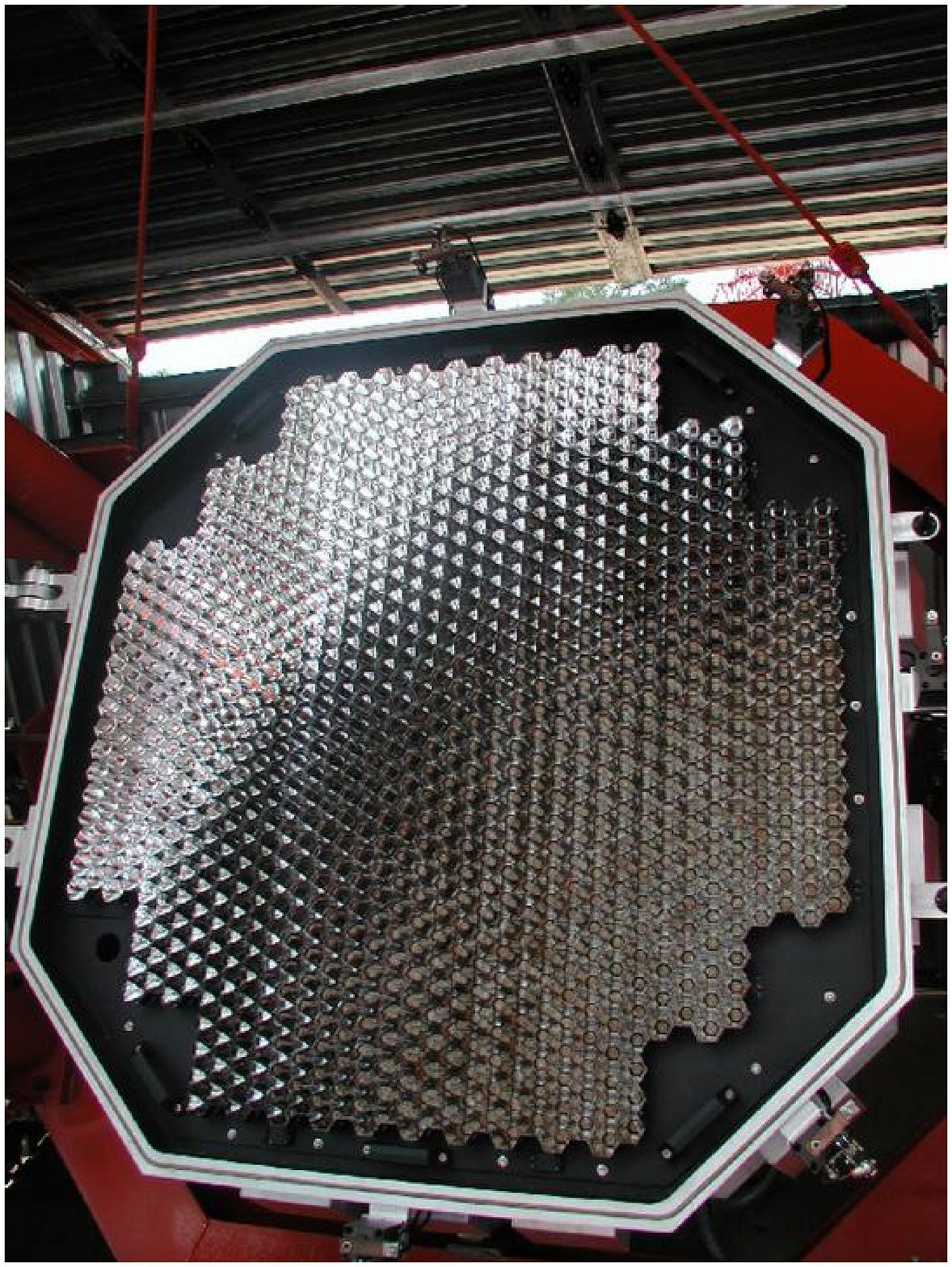}
  \includegraphics[width=.3299\textwidth]{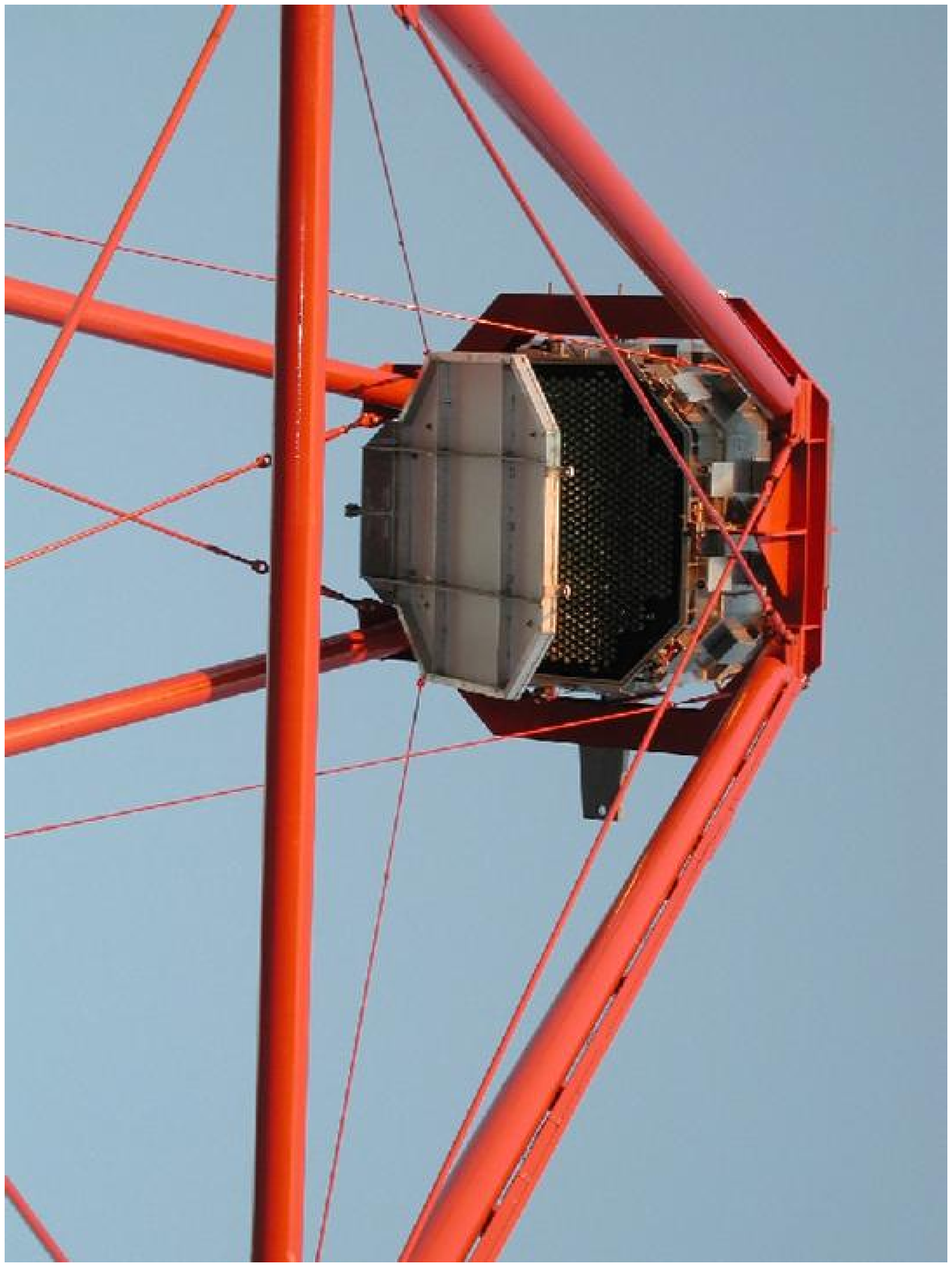}
  \caption[Views of the Second Camera]{Views of the second H.E.S.S. camera. The
    rear view (left) shows the crate with the four power supplies, the bus
    system and the network interface. The front view (middle) shows the 960
    Winston cones. When the camera has reached observation position, the lid
    opens for observations (right). (Figures taken from \citet{Camera}.)}
  \label{fig:Camera}
\end{figure}

\section{The Central Trigger System}
In addition to the individual camera triggers, H.E.S.S. has a central trigger
system (CTS) which constitutes a third level trigger. A detailed description of
this system is given by \citet{CentralTrigger}. The CTS is designed to identify
stereoscopic events by coincidence of individual telescope triggers. The
standard central trigger condition requires a minimum of two telescope
triggers. Higher trigger multiplicities provide events of higher reconstruction
quality but at a cost of a reduced sensitivity for events of lower energy. The
stereoscopic trigger condition effectively reduces the background, e.g. muon
events, which trigger the cameras. In addition, the CTS reduces the dead-time
of the individual cameras, sending reset signals to triggered cameras to stop
the readout process if no coincidence with other telescope triggers is
observed. The CTS also measures the system's dead-time as well as assigns
unique numbers to events which allows for individual camera images to be
combined as a single event. Depending on the zenith angle of the observation,
the typical camera trigger rate is about (300$\pm$50)\,Hz. All of the CTS's
hardware is located in a crate (Fig.~\ref{fig:DAQ}) in the control building.

\section{Atmospheric Monitoring}
H.E.S.S. has instruments for monitoring the atmosphere and for providing
specific information about atmospheric conditions during its observations. A
detailed description of this is given by \citet{Weather}. Atmospheric data is
displayed in the control room and informs the observers about the applicable
atmospheric conditions. It is also sent to the central data acquisition system
and recorded for off-line data analysis. The atmospheric data provides
important information for data quality selection. The different monitoring
devices involved are briefly described in the following subsections.

\subsection{Radiometers}
H.E.S.S. has five radiometers --- one on every telescope and one scanning
radiometer in the center of the array. The telescope radiometers measure 
infrared radiation from the sky in the FOV in a transmission window between 8
to 14\,$\mu$m and calculate the temperature of the atmosphere through
comparison with black body radiation. Since clouds reflect ambient light,
their spectrum differs significantly from the spectrum of the clear sky and thus 
clouds can easily be detected. The scanning radiometer works the same way but
scans the whole sky for the presence of clouds and approaching weather fronts.

\subsection{Ceilometer}
The ceilometer (Fig.~\ref{fig:WeatherStation}) consists of a LIDAR
(Light Detection and Ranging) system which emits short laser pulses
(\citet{Weather2}). It is located next to the scanning radiometer. By measuring
the amount of backscattered light and time of flight, it can provide a detailed
density profile of aerosol in the atmosphere and detect layers of clouds up to
7.5\,km. The correlation between the amount of aerosol in the atmosphere and
the trigger rate has been investigated by \citet{LIDAR}. The aerosol absorption
in the atmosphere affects all events, reducing the measured light intensity and
thus shifting the energy scale.

\subsection{Weather Station}
\begin{figure}[ht]
  \centering
  \includegraphics[width=0.7\textwidth]{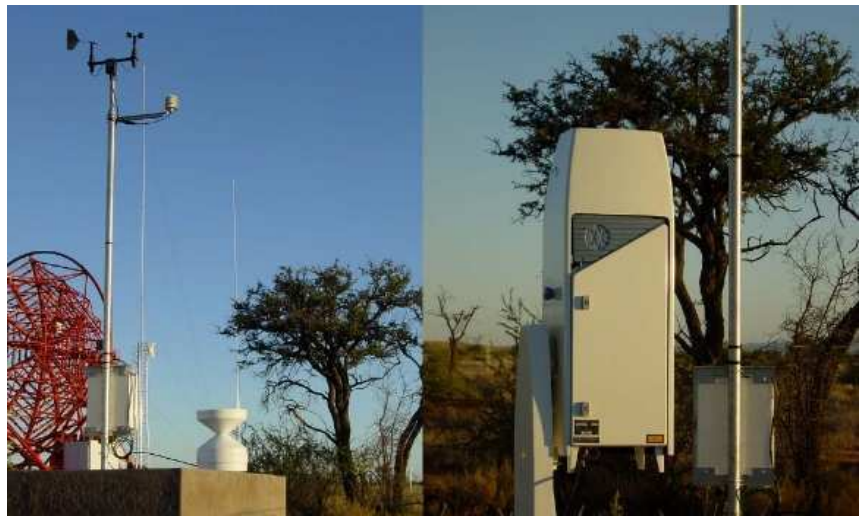}
  \caption[Instruments for Atmospheric Monitoring]{Instruments
    at the H.E.S.S site for atmospheric monitoring: weather station (left) and
    ceilometer (right). (Photos taken from \citet[pg. 46]{Schlenk}.)}
  \label{fig:WeatherStation}
\end{figure}

Fig.~\ref{fig:WeatherStation} shows the weather station which is located
in the center of the array. It consists of a thermometer, hygrometer,
barometer, anemometer and pluviometer. Data from these instruments is
continuously monitored, displayed in the control room and also recorded.

\section{The Central Data Acquisition System}
\begin{figure}[ht]
  \begin{minipage}[c]{0.45\linewidth}
    \includegraphics[width=\textwidth]{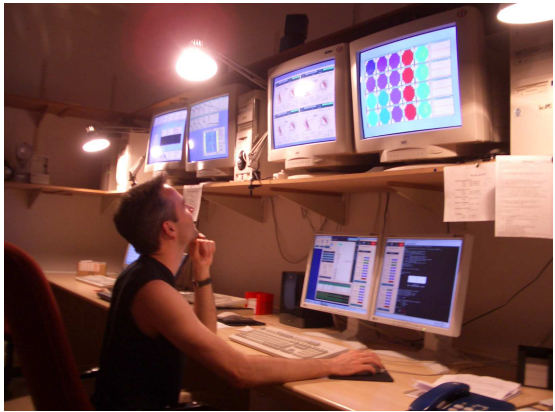}
  \end{minipage}\hfill
  \begin{minipage}[c]{0.52\linewidth}
    \caption[Interior of the Control Room]{Interior of the control room. The
    displays provide the observers with the latest monitoring information.
    Terminals provide access to the central data acquisition system and control
    to the H.E.S.S. array. (Photo provided by Frank Breitling, 2004.)}
    \label{fig:ControlRoom}
  \end{minipage}
\end{figure}

\noindent The central data acquisition system (DAQ) is described in
\citet{DAQ_ICRC27} and \citet{DAQ_ICRC28}. It connects the individual H.E.S.S.
components and manages the storage of data. It also provides the necessary
interface for the control of the system. The DAQ is controlled through the DAQ
front end in the control room (Fig.~\ref{fig:ControlRoom}). Displays in the
control room provide the observers with important monitoring information for
observations such as trigger rates, camera images, PMT currents, significances
of \g-signals, the system load, weather and atmospheric information. Three
computers provide the graphical user interface to schedule, configure and start
observations as well as to stop them. The observations are typically scheduled
in runs of 28 minutes.

The DAQ consists of object-oriented software in the computer language C++,
which is based on the software packages ROOT (\citet{ROOT}) and omniORB
(\citet{omniORB}). During data taking, data is received from the individual
components, then combined to events and stored in the ROOT file format.

\begin{figure}[h]
  \begin{minipage}[c]{0.32\linewidth}
    \caption[Hardware of the Central Data Acquisition System]{The Linux PC farm
      in the control building is the core component of the central data
      acquisition system. The crate which houses the central trigger hardware
      is partially visible on the left. (Photo taken by Frank Breitling,
      2004.)}
    \label{fig:DAQ}
  \end{minipage}\hfill
  \begin{minipage}[c]{0.65\linewidth}
    \includegraphics[width=\textwidth]{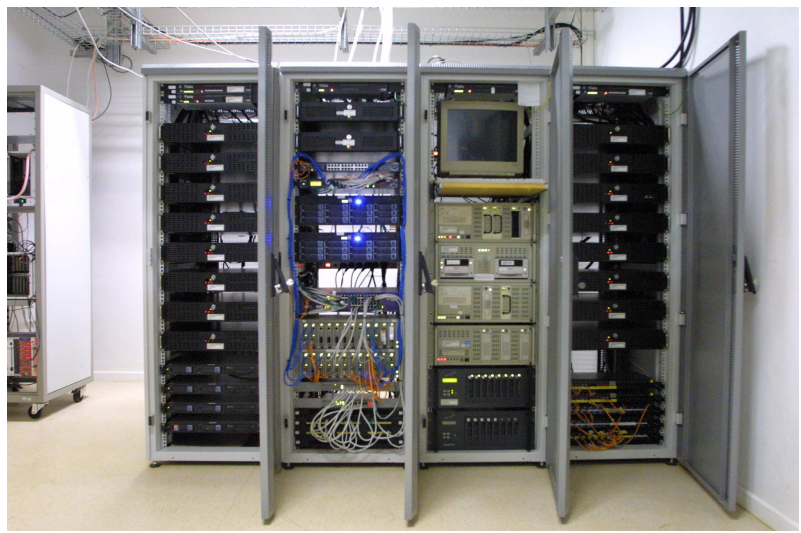}
  \end{minipage}
\end{figure}

The software is run on a Linux computer farm consisting of about 20 Intel 386
compatible PCs, a gigabit network and two RAID arrays with a capacity of
several terabyte each (Fig.~\ref{fig:DAQ}). The hardware is located in the
control building. New data is written to gigabyte tapes and shipped to European
computing centers where it is calibrated and prepared for data analysis.

\chapter{The H.E.S.S. Standard Analysis} \label{chp:StandardAnalysis}

After raw data has been recorded, it can be analyzed and scientific information
can be extracted. This process of data analysis requires a good knowledge of
the hardware, software and methods that can be applied in order to extract the
desired information. The H.E.S.S. standard analysis has been developed to
accomplish this task. It uses the well-established methods of imaging
atmospheric Cherenkov astronomy which can reliably extract results from raw
data in a computationally efficient way. The accuracy of the results has been
analyzed in detailed Monte Carlo studies, through alternative analysis methods
(\citet{Tluczykont:2006}) and through direct comparison with results from other
experiments (\citet{CrabPaper}).

Several computational steps are necessary before physical quantities are
obtained. Many of them require additional data from Monte Carlo simulations.
The H.E.S.S. standard analysis (\citet{H.E.S.S.Software}) performs these steps.
It consists of a set of software packages based on the programming language C++
and the data analysis framework ROOT (\citet{ROOT}). ROOT provides the
statistical tools, which are used in the data analysis, e.g. for the fitting of
functions. In the data analysis, some results rely on others so that all steps
have to be performed in a certain order. This order is illustrated by the
flowchart in App.~\ref{app:AnalysisChain}. Another discussion of the methods
and the accuracy of the H.E.S.S. standard analysis can be found in
\citet{CrabPaper}.

The first part of the analysis consists of reconstructing \g-ray showers. In a
subsequent part statistical methods are applied, sky maps are generated and the
energy spectrum is determined. In this chapter, the individual steps of data
analysis are discussed and they are verified through application to H.E.S.S.
data from the Crab Nebula. Sky coordinates are given in right ascension (RA)
and declination (Dec) of the J2000 equatorial coordinate system unless stated
otherwise.

\section{Monte Carlo Simulations} \label{sec:MonteCarlo}
Since many results of data analysis rely on data from Monte Carlo simulations,
the later is essential for the analysis. For example, Monte Carlo simulations
are required for calibration, background reduction, shower reconstruction and
determinations of effective areas, i.e. determination of the energy spectrum.
Monte Carlo data for the standard analysis is simulated in two steps: first a
shower simulation and then a detector simulation.

\subsection{Shower Simulation}
Shower simulations are produced with CORSIKA, a program for Cosmic Ray
Simulations for Kascade (\citet{CORSIKA}). CORSIKA can simulate atmospheric
shower cascades and the corresponding Cherenkov light emission for various
particles such as photons, protons and leptons. The simulations include many
details, such as the effect of the local magnetic field of the earth on the
charged particles in the shower cascade, in order to provide very realistic
results. Also, the different atmospheric transmission profiles generated with
MODTRAN (\citet{MODTRAN}) can be included in the simulation. For H.E.S.S
simulations, two different atmospheric models are used which both describe the
atmosphere at the H.E.S.S. site sufficiently well enough, as confirmed by
trigger studies (\citet{CentralTrigger}). The desert model reproduces a clear
atmosphere with little haze and a boundary layer starting at 1800 m above sea
level. The maritime model reproduces a more humid atmosphere with a boundary
layer starting at sea level. The desert model is generally preferred in most
analyses.

\subsection{Detector Simulation}
The response of the H.E.S.S. array to simulated Cherenkov light is simulated
with the detector simulation Sim Hessarry which was developed by
\citet{sim_hessarray}. It calculates the PMT response to Cherenkov light for
each of the telescopes. Sim Hessarry is a very accurate detector simulation
which takes into account:

\begin{itemize}
\item the reflector geometry, mirror reflectivity and pointing of the
telescopes
\item shadowing by the camera support structure and point spread function,
\item the transmission of the Winston cones in front of the PMTs,
\item the quantum efficiency of the PMTs,
\item the electronic response of the PMTs and
\item the telescope multiplicity requirement of the central trigger.
\end{itemize}

Sim Hessarry also simulates the camera's response in different observation
modes. The standard modes are the on/off and the wobble mode. In the on/off
mode, the camera is pointed directly towards the source for an on-run and in a
source-free direction for an off-run. The off-run is used to determine the
background. In wobble mode, the camera direction is offset from the source
direction by the wobble offset $(\theta_w)$. $\theta_w$ is chosen such that the
source is still enclosed within the FOV. The advantage of this observation mode
is that it provides regions that are needed for background estimation within a
single run. This has the advantage that systematic errors on a run by run basis
are reduced. For H.E.S.S. observations, the standard wobble offset is
$\theta_w=\pm0.5^\circ$ in Dec and the corresponding value
$\theta_w=\pm0.5^\circ/\cos(\angle \rm Dec)$ in RA. The alternating signs of
the wobble offset provide compensation for linear gradients in acceptance and
hence they reduce systematic errors.

\subsection{Monte Carlo Data} \label{sec:MCfiles}
The Monte Carlo data used in this work is based on a simulation from August
2005. The data has been calculated with the $desert$ atmospheric model and
partially also with the $maritime$ model. These simulations consist of
simulated \g-ray showers from point sources. The point sources were simulated
at zenith angles \footnote{The zenith angle $(\Theta)$ is the angle between the
  pointing direction of the array and the zenith, i.e. $\Theta=90^\circ-\angle
  Alt$, where $\angle Alt$ is the altitude angle.} of 30, 40, 45, 50, 55, 60,
63, 65 and 67$^\circ$ and at azimuth angles of 0 and 180$^\circ$ for sources in
a northern direction, like the Crab Nebula, and in a southern direction, like
\MSH. Wobble offsets $(\theta_w)$ were simulated for 0, 0.5, 1, 1.5, 2 and
2.5$^\circ$. CT\,3 was simulated with a reduced efficiency of 8\%, as measured
in early 2003. At each zenith and offset angle, about $3\times10^5$ showers
reached the sensitive detector area with a radius of 1000\,m from the center of
the H.E.S.S. array. About 10\% of these events passed the cuts. The energy
spectrum of the simulated \g-ray showers reaches from 20\,GeV to 100\,TeV, with
a photon index $\Gamma=2$. Since the shower simulation consumes most of the
computation time required, the efficiency of the simulations can be increased
if the same simulated shower is used multiple times in the detector simulation.
The only parameter that is varied is the impact position with respect to the
telescope. This method can reduce the computational effort by orders of
magnitudes. The Monte Carlo data is converted and stored in the ROOT file
format for an easy integration into data analysis.

Simulations of extended sources have been obtained from the point source Monte
Carlo simulations by scattering events according to a Gaussian distribution.
The width and length of the distribution were chosen according to the standard
deviation of the extension to be simulated. Extended simulations were used for
the determination of the collection areas. Although this method is marginally
less accurate than a full simulation it is preferable due to its higher
computational efficiency.

\section{Shower Reconstruction}
The central part of the data analysis is the shower reconstruction. This
consists of several steps, some of which also require the Monte Carlo data.
This section describes the individual steps that are taken in order to extract
shower information from raw data.

\subsection{Camera Calibration} \label{sec:Calibration}
Calibration is the first step in shower reconstruction, which involves
estimating the number of photo electrons (p.e.) that have produced a camera
event image. A raw image of a camera event is shown in
Fig.~\ref{fig:CameraImages} (upper left). Pixels below a certain threshold were
later removed from the data and therefore do not show up in this image. The
number of p.e. $(A)$ that hit a PMT is calculated from the ADC counts of the
PMT's high-gain $(A^{\rm HG})$ and low-gain $(A^{\rm LG})$ channels. If $A$ is
less than 150\,p.e., $A^{\rm HG}$ is used. If $A$ is greater than 200\,p.e.,
$A^{\rm LG}$ is used. For intermediated values, $A$ is determined from the
weighted average of $A^{\rm HG}$ and $A^{\rm LG}$ as
\begin{equation}
  A = (1-\epsilon) \times \rm ADC^{HG} + \epsilon \times ADC^{LG},
\end{equation}
where $\epsilon=\rm (ADC^{HG}-150)/(200-150)$.

The amplitudes $A^{\rm HG}$ and $A^{\rm LG}$ are given by
\begin{equation}
  A^{\rm HG}=\frac{{\rm ADC^{HG}}-p^{\rm HG}}{\rm \gamma_{e}^{ADC}}\times FF
\end{equation}
for the high-gain channel and by
\begin{equation}
  A^{\rm LG}=\frac{{\rm ADC^{LG}}-p^{\rm LG}}{\gamma_{\rm e}^{\rm ADC}}
  \times (HG/LG) \times FF
\end{equation}
for the low-gain channel, where ADC$^{\rm HG}$ and ADC$^{\rm LG}$ are the
recorded ADC counts for the high- and low-gain channels. $p^{\rm HG}$ and
$p^{\rm LG}$ are the pedestals of these channels measured in ADC counts. $\rm
\gamma_e^{ADC}$ is the conversion factor between ADC counts and p.e. $HG/LG$ is
the amplification ratio of the high-gain to the low-gain channel and $FF$ is
the flat-filed coefficient. These parameters are determined with special
calibration runs as described in detail by \citet{CalibrationPaper} and
\citet{CalibrationLoic}. Since the parameters can vary with time, the
calibration has to be repeated regularly to guarantee a reliable conversion.

\subsubsection{The Pedestals $p^{\rm HG}$ and $p^{\rm LG}$}
The pedestal is defined as the mean ADC value recorded in the absence of
Cherenkov light. It is primarily determined by the night-sky background (NSB)
and secondarily by electronic noise which has a strong temperature dependence.
The pedestals are determined at intervals of 20\,s from the events of an
observation run, where pixels affected by Cherenkov light are excluded.
Typically only 20 pixels of an image contain Cherenkov light.
Fig.~\ref{fig:CameraImages} (top right) shows the pedestals during run 20301.
Since the width of the pedestal distribution changes with the NSB, the pedestal
distribution can also be used to determine the level of NSB and to reject noisy
pixels.

In addition, every two nights the pedestals are monitored in special electronic
pedestal runs, which are performed with a closed camera lid. Deviations from
the nominal value can clearly be measured by these runs.

\subsubsection{The Conversion Factor $\gamma_{\rm e}^{\rm ADC}$}
\begin{figure}[t]
  \begin{minipage}[c]{0.6\linewidth}
    \includegraphics[width=.98\textwidth]{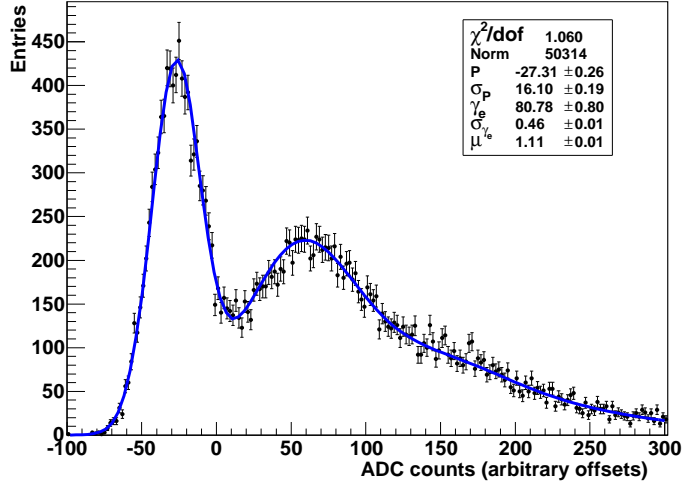}
  \end{minipage}\hfill
  \begin{minipage}[c]{0.4\linewidth}
    \caption[ADC Counts of a Single p.e. Calibration Run]{ADC counts of a
    single p.e. calibration run. The first peak corresponds to the pedestal
    counts and the second to the single p.e. counts. The distance between both
    peaks determines the conversion factor $\rm \gamma_{e}^{ADC}$. The height
    of the single p.e. peak is determined by the intensity of the LED pulser
    and obeys the Poisson statistic. (Figure taken from
    \citet{CalibrationPaper}.)}
  \label{fig:ADCtoPe}
  \end{minipage}
\end{figure}

The offset of a single p.e. peak from the pedestal determines the conversion
factor ($\gamma_{\rm e}^{\rm ADC}$). To obtain a histogram of ADC counts that
can show the pedestal and the single p.e. peak (Fig.~\ref{fig:ADCtoPe}), the
camera is illuminated by a faint LED pulser with an intensity of about 1\,p.e.
per pixel and a frequency of 70\,Hz. The LED pulser is installed in the camera
shelter, where the camera is parked during a single p.e. run, in order to be
shielded from the NSB while the lid is opened to observe the LED pulses. The
single p.e. runs have a duration of two minutes and are taken every two nights.
Since the pedestal ADC counts are Gaussian distributed with a standard
deviation $\sigma_P$, the ADC counts of a signal of $n$\,p.e. ($n \in
\mathbb{N}$) are also Gaussian distributed with a standard deviation
$\sqrt{n}\sigma_{\gamma_{\rm e}}$ and a mean position $P+n\gamma_{\rm e}^{\rm
ADC}$. The number of p.e. follows a Poisson distribution. Therefore, the
conversion factor $\gamma_{\rm e}^{\rm ADC}$ can be found by the fit to the
function $f$ to the ADC histogram of the single p.e. runs. $f$ is give as
\begin{eqnarray}
  f(x) &=& N \times \Bigg( \frac{e^{-\mu}}{\sqrt{2\pi}\sigma_P}
  \exp\left[-\frac12\left(\frac{x-P}{\sigma_P}\right)^2\right] \\
  &&+ \kappa \sum^{m\gg1}_{n=1} \frac{e^{-\mu}}{\sqrt{2\pi}\sigma_{\gamma_{\rm
        e}}}\frac{\mu^n}{n!} \exp\left[-\frac12\left( \frac{x-(P+n\gamma_{\rm
        e}^{\rm ADC})}{\sqrt{n}\sigma_{\gamma_{\rm e}}}\right)^2\right] \Bigg),
\end{eqnarray}
where $x$ is the number of ADC counts and $N$ and $\kappa$ are normalization
constants.

\subsubsection{The Amplification Ratio $HG/LG$}
The amplification ratio ($HG/LG$) can be determined from a comparison of the
ADC counts of the high-gain ($C_{\rm H}$) and low-gain channels ($C_{\rm L}$).
Since the pedestals for the high-gain ($P_{\rm H}$) and low-gain channels
($P_{\rm L}$) are also available as described above,
\begin{equation}
  \frac{\rm HG}{\rm LG} = \frac{C_{\rm H}-P_{\rm H}}{C_{\rm L}-P_{\rm L}}.
\end{equation}
The amplification ratio is calculated from regular observation runs for all
usable pixels with an intensity between 15-200\,p.e.

\subsubsection{The Flat-Filed Coefficient $FF$}
Since the calibration of the PMTs is not absolute, differences in the
efficiency of the photocathodes or the Winston cones can result in different
PMT efficiencies. Deviations with an RMS of about 10\% have been observed. The
$FF$s compensate for this difference and provide a uniform camera response. 
$FF$s are determined in special flat-field runs roughly every two days. In a
flat-field run, the camera is illuminated by an LED flasher which is mounted on
the mirror dish. At a distance of 15\,m from the camera, the flasher provides
a homogeneous illumination from 10-200\,p.e. within a solid angle of 10$^\circ$
at the PMTs' peak efficiency in the range of 390 to 420\,nm. The $FF$ of each
pixel is determined as the inverse ratio of the pixel amplitude to the mean
amplitude of all pixels averaged over a run. By definition the mean of the
$FF$s is equal to 1. The distribution of the $FF$s gives an estimate of the
uniformity of the camera. The relative accuracy after a correction with $FF$s
is $<1$\%. An image after calibration is shown in Fig.~\ref{fig:CameraImages}
(middle left).

\subsubsection{Identification of Unusable Channels and Broken Pixels}
\label{sec:BrokenPixels}
To obtain a correct calibration, it is also important to eliminate channels
that do not provide correct information and therefore falsify the image. There
are several reasons for unusable channels: missing calibration coefficients,
synchronization problems with the analog ring sampler's memory, high voltage
variation in the PMTs due to hardware failures or merely the presence of bright
stars. If the high-gain and the low-gain channels of a pixel are unusable, the
pixel is considered ``broken''. The amount of unusable channels can reach up to
4\% per run. The number of broken pixels is less. Fig.~\ref{fig:CameraImages}
(bottom) shows problematic pixels in a run.

\subsection{Muon Calibration} \label{sec:AbsoluteCalibration}
Camera calibration guarantees images of high quality, but it does not provide
an absolute calibration of intensity since important system components such as
the optical components are not included. An absolute calibration can be
achieved by an analysis of muon rings. This relies on the interdependence of
the opening angle of the Cherenkov light emitted by muons in the atmosphere and
muon energy. Since the mirror dish focuses parallel rays onto the same point in
the focal plane, muons appear as rings in the camera image. The radius of the
rings reflects the muon's energy and hence the number of emitted Cherenkov
photons and the amplitude of the muon ring. Muon rings also provide an
alternative method for the determination of the flat-field coefficients, since
neighboring pixels in a muon ring are expected to have a similar amplitude.
This is explained in \citet{CalibrationPaper}. Details about muon calibration
can be found in \citet{MuonsICRC} and \citet{BolzThesis}. Muon runs are taken
regularly to monitor the H.E.S.S.' photon efficiency, which decreases by about
5\% percent per year (\citet[pg. 89]{BolzThesis}). The degradation is partially
compensated for by regular hardware maintenance and upgrades every few months,
keeping the deviations from the nominal values sufficiently low. Nevertheless,
corrections for this absolute efficiency have been developed by
\citet{MPIK:MuonCorrections} and can be implemented in the future.

\subsection{Image Cleaning}
\begin{figure}
  \begin{minipage}[c]{0.5\linewidth}
    \includegraphics[width=\textwidth]{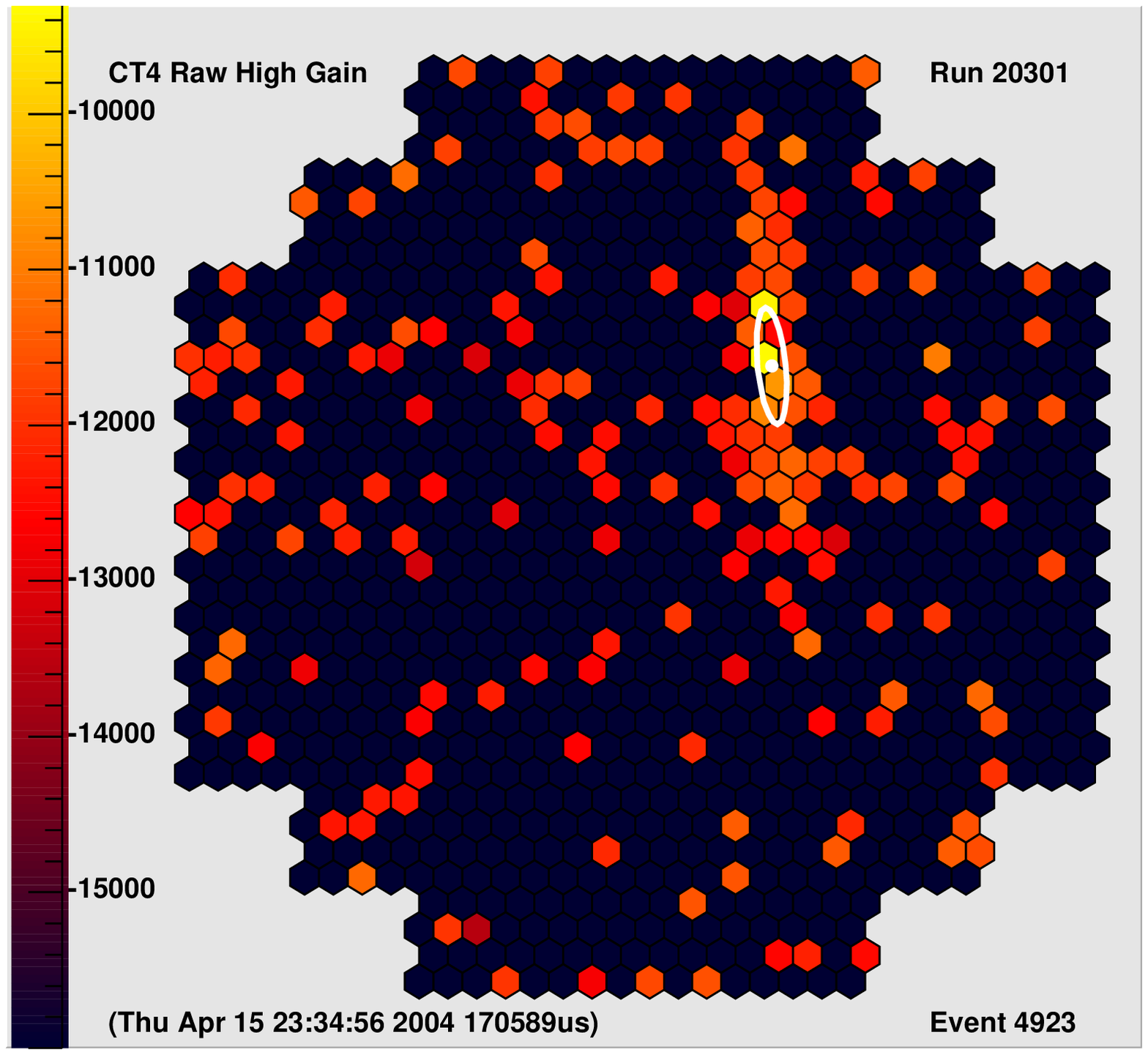}
  \end{minipage} \hfill
  \begin{minipage}[c]{0.5\linewidth}
    \includegraphics[width=\textwidth]{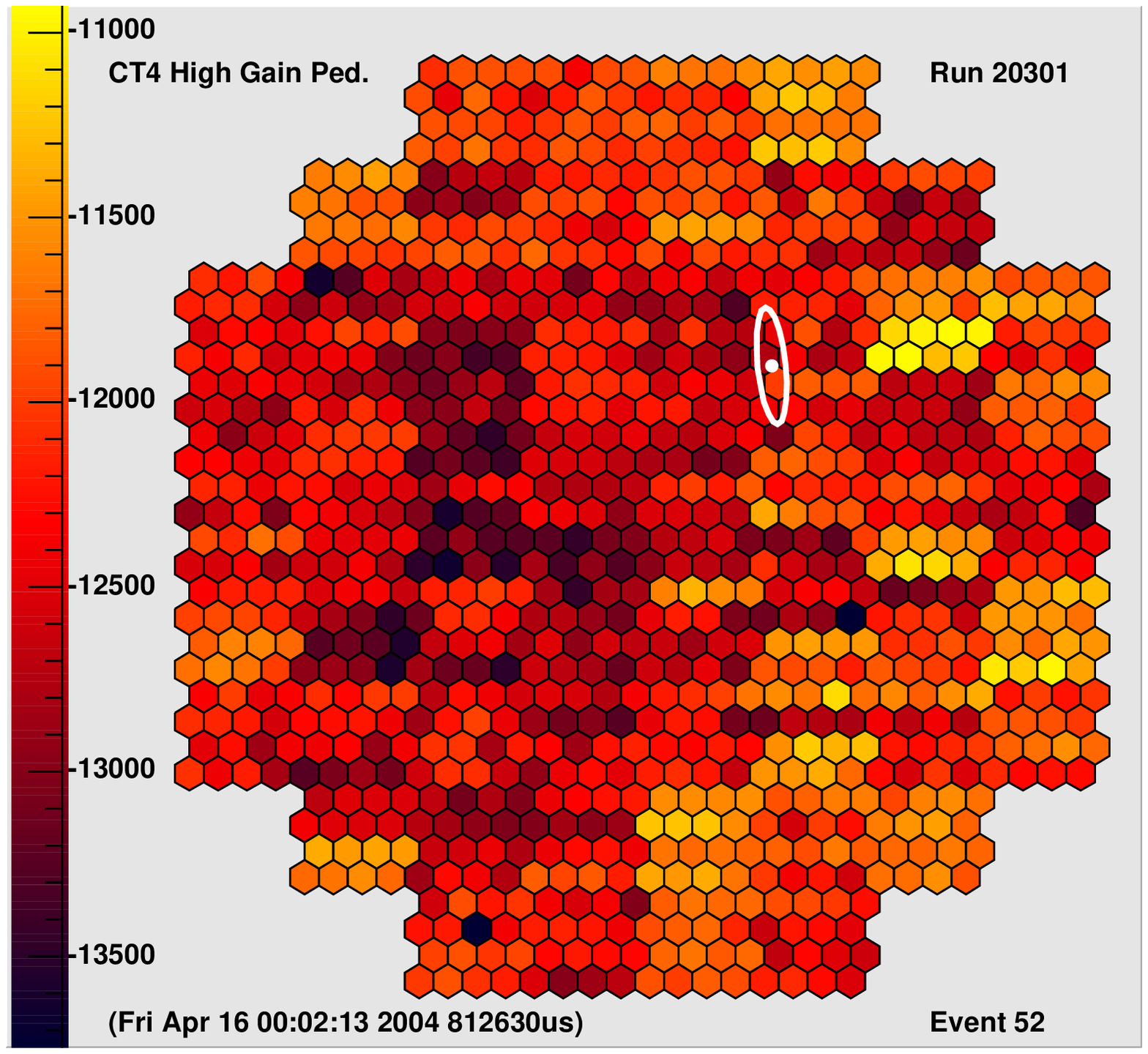}
  \end{minipage} \hfill
  \begin{minipage}[c]{0.5\linewidth}
    \includegraphics[width=\textwidth]{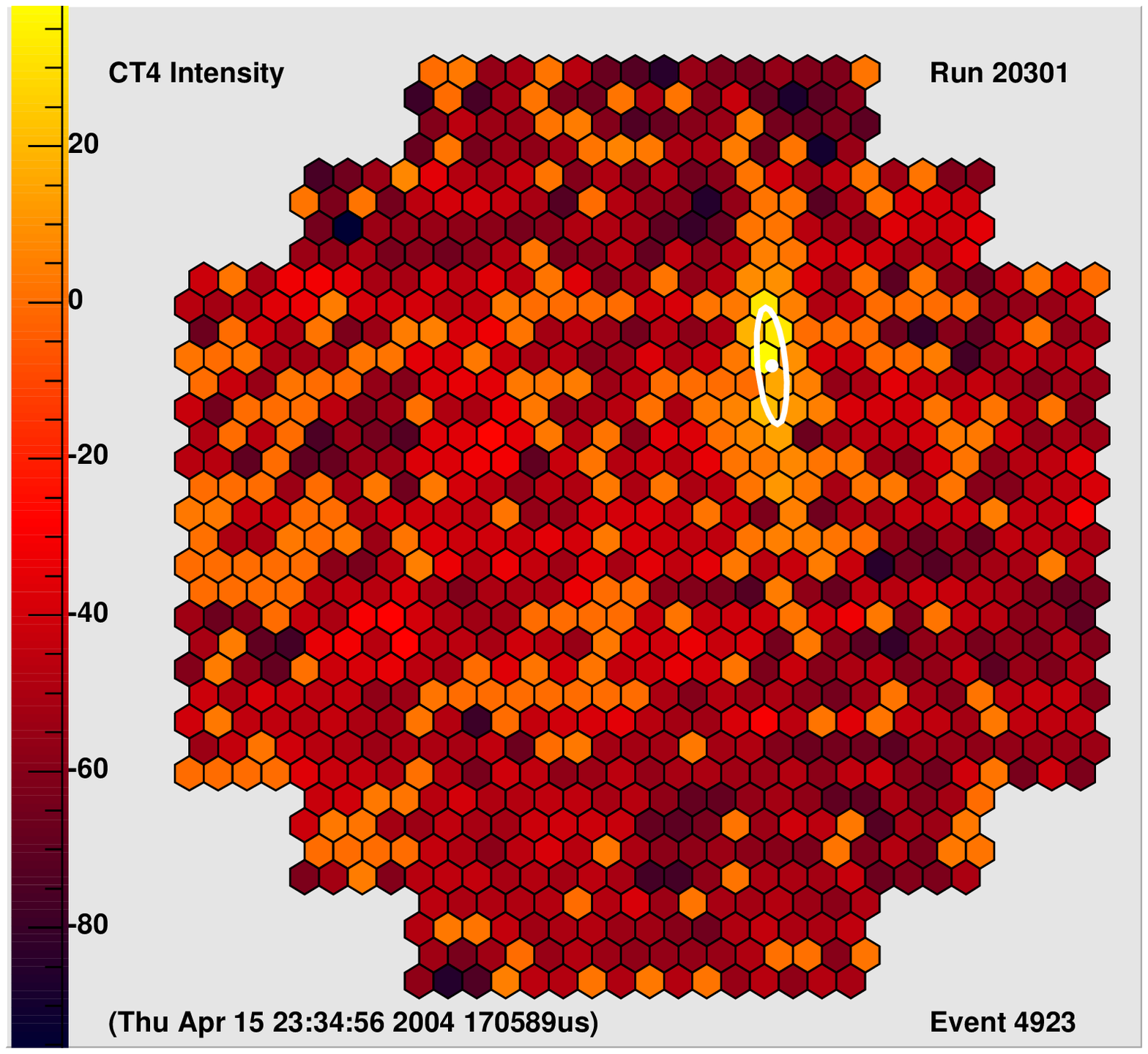}
  \end{minipage} \hfill
  \begin{minipage}[c]{0.5\linewidth}
    \includegraphics[width=\textwidth]{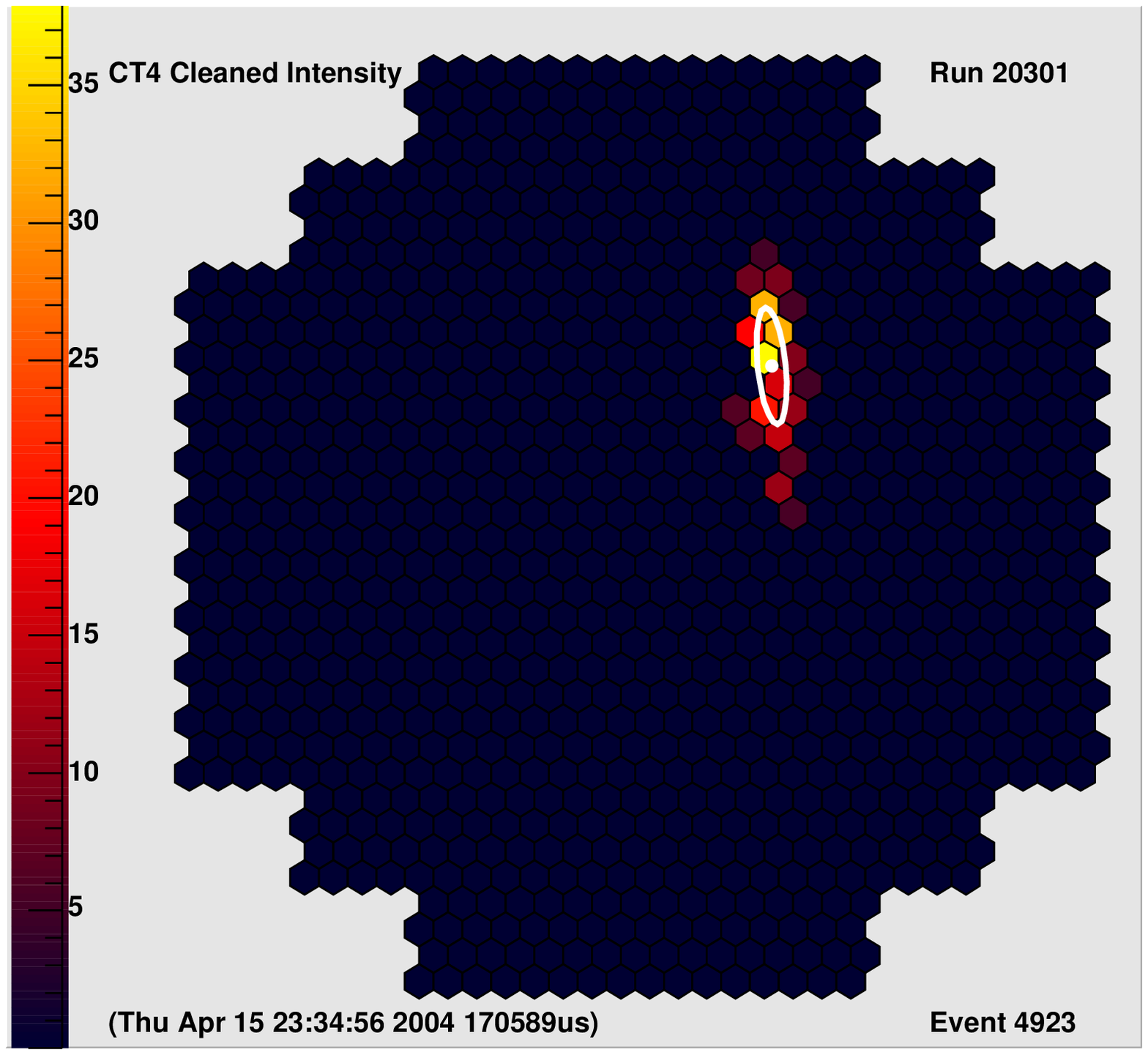}
  \end{minipage} \hfill
  \begin{minipage}[c]{0.5\linewidth}
    \includegraphics[width=\textwidth]{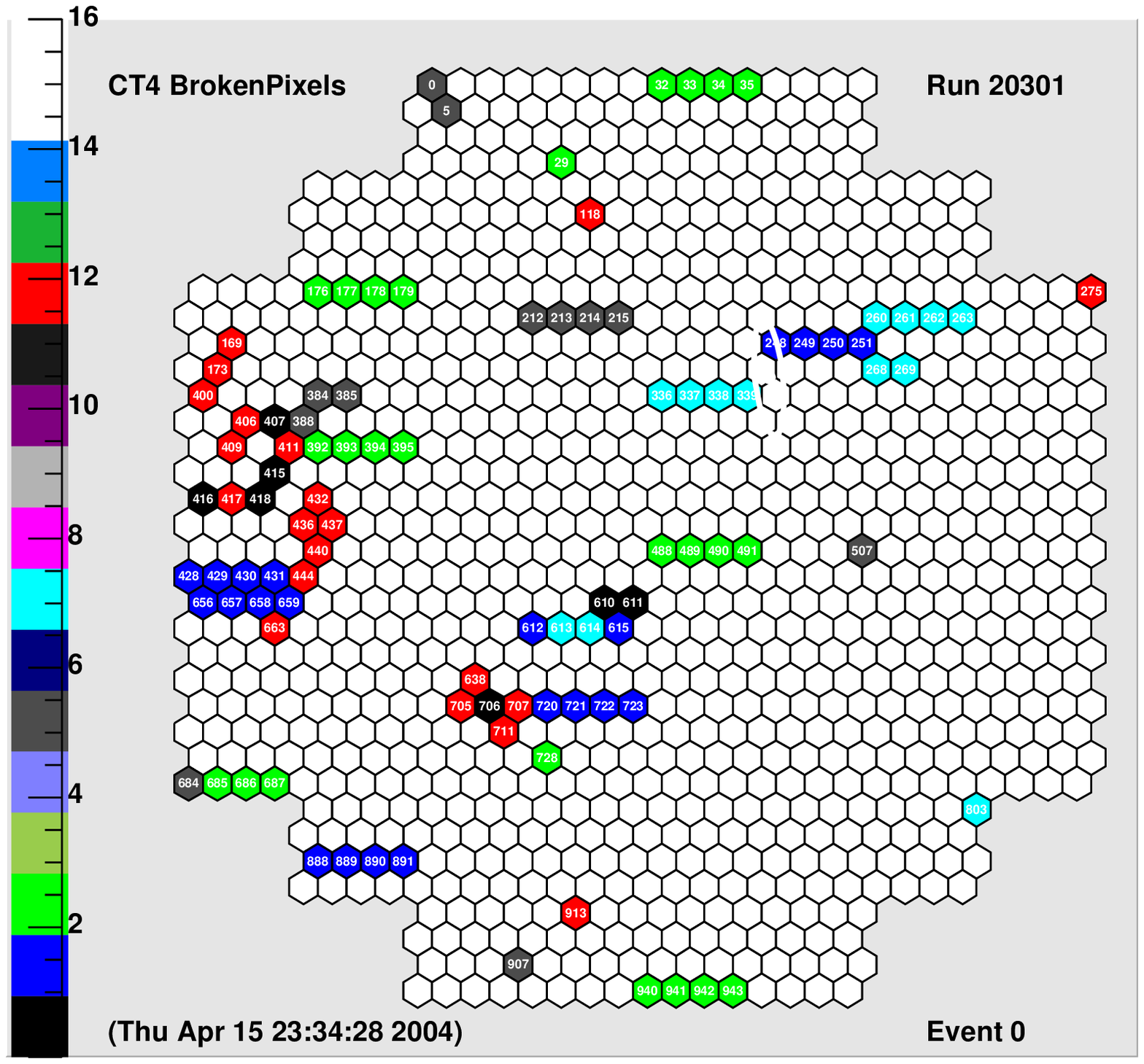}
  \end{minipage} \hfill
  \begin{minipage}[c]{0.47\linewidth}
    \caption[Camera Image at Different Steps in the Analysis]{Camera images of
     CT\,4, run 20301, event 4923 at different steps in the analysis: the zero
     suppressed high-gain ADC counts (top left), the pedestal ADC counts (top
     right), the corresponding pixel intensities in p.e. after calibration
     (middle left), the cleaned image after (5,10) tail cuts (middle right) and
     the unusable channels and broken pixels during the run (bottom). The
     Hillas ellipse with its center of gravity is indicated in each image.}
    \label{fig:CameraImages}
  \end{minipage}
\end{figure}

The purpose of image cleaning is to remove the pixels from an image which do
not represent Cherenkov light from the shower. The cleaning is realized with a
tail cut algorithm that determines which pixels will be kept and which will be
removed from the image. The tail cut algorithm has two cut parameters ($l$,$h$)
--- the low and the high thresholds of the pixel amplitude. The standard values
are $l$=5\,p.e. and $h$=10\,p.e. The tail cuts remove all pixels with an
amplitude below $l$ and keep all pixels with an amplitude above $h$, if they
have a neighboring pixel of at least $l$. They will also keep any pixels within
the range of $l$ and $h$ if they have a neighboring pixel with an amplitude
exceeding $h$. The result is a cleaned shower image, as shown in
Fig.~\ref{fig:CameraImages} (middle right).

\subsection{Image Parameterization according to Hillas}
\begin{figure}[ht]
  \begin{minipage}[c]{0.5\linewidth}
    \includegraphics[width=\textwidth]{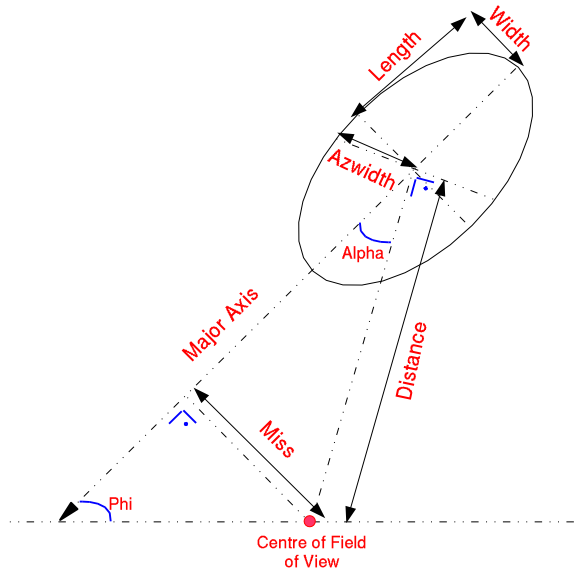}
  \end{minipage}\hfill
  \begin{minipage}[c]{0.45\linewidth}
    \caption[Geometric Representation of Hillas Parameters]{A geometric
      representation of Hillas parameters. The ellipse represents a \g-ray
      shower image. (Sketch taken from \citet[pg. 80]{ErginThesis}.)}
    \label{fig:HillasParameters}
  \end{minipage}
\end{figure}

The standard analysis uses Hillas parameters, which are derived from cleaned
camera images. Hillas parameters were first introduced by \citet{Hillas}. They
are the moments of an image as described in App.~\ref{app:HillasParameters}.
Their geometric representation is shown in Fig.~\ref{fig:HillasParameters}.
They provide a powerful means for shower reconstruction and the rejection of
background events. The relevant parameters for analysis are the image's
amplitude ($IA$, Eqn.~\ref{eqn:size}), local distance ($LD$,
Eqn.~\ref{eqn:dist}), width ($W$, Eqn.~\ref{eqn:width}), length ($L$,
Eqn.~\ref{eqn:width}) and angle ($\phi$, Eqn.~\ref{eqn:phi}).

\subsection{Geometric Reconstruction} \label{sec:Reconstruction}
Stereoscopic IACTs allow geometric shower reconstruction (\citet{Konopelko1997,
Hofmann1999}) because the shower direction and impact point are uniquely
defined by the shower images of two telescopes. With more than two telescope
images, shower geometry is even over-defined and reconstruction accuracy
increases significantly. The increase results from the fact that more than two
telescopes build an array, which guarantees that the majority of shower images
will be taken at a favorable angle $\ll180^\circ$ and distance $<50$\,m. The
actual reconstruction is based on the geometric optics of IACTs. Reconstruction
of the shower direction and the core position, i.e. the shower impact point at
the ground, is done in two different coordinate systems that are associated
with the telescope array (cf. \citet{GillessenThesis, ErginThesis}).

\subsubsection{Reconstruction of the Shower Direction}
\label{sec:ShowerDirection}
\begin{figure}[tbh!]
  \begin{minipage}[c]{0.5\linewidth}
    \includegraphics[width=\textwidth]{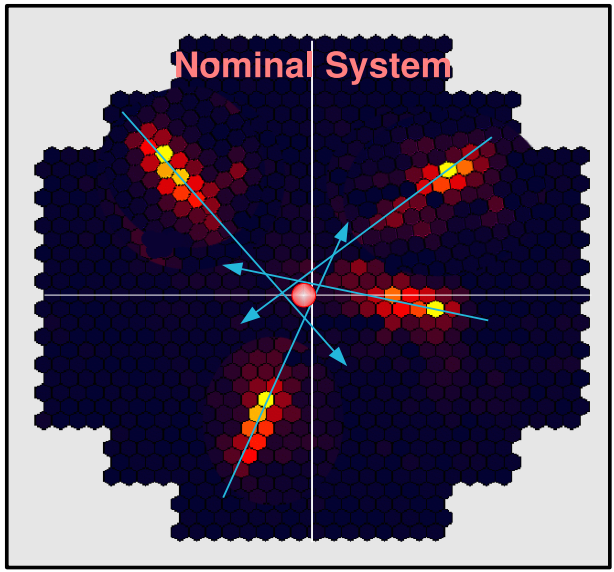}
  \end{minipage}\hfill
  \begin{minipage}[c]{0.47\linewidth}
    \caption[Reconstruction of Shower Direction]{Reconstruction of shower
     direction from four shower images in the nominal system. The intersection
     of the elongated major axis determines the shower direction. (Figure taken
     from \citet[pg. 85]{ErginThesis}.)}
  \label{fig:ShowerDirection}
  \end{minipage}
\end{figure}

Shower direction is reconstructed in the nominal system. The nominal system is
a two\hyp{}dimensional coordinate system perpendicular to the line of sight of
the array. It represents the camera coordinate system in an imaginary nominal
focal plane with a focal length of 1\,m. Distances in this plane are measured
in radians and correspond directly to sky coordinates. Moreover, in the nominal
system the projected shower direction is found along the elongated major axis
of the Hillas ellipse. Hence the shower direction is determined by the
intersection of the elongated major axis of a camera images of an event. In a
perfect reconstruction the lines intersect in one point $(x,y)$. However, due
to limited reconstruction accuracy, the pairwise intersection points
$(x,y)_{i,j}$ of the lines $l_i$ and $l_j$ $(i,j=1,2,3,4)$ differ as shown in
Fig.~\ref{fig:ShowerDirection}. So a unique definition of the shower direction
$(x,y)$ is required. Different definitions are possible. The H.E.S.S. standard
analysis has adopted the following definition which is based on the weighted
sum of each intersection. The intersection point is
\begin{equation}
  (x,y) = \frac{\sum_{i,j}w_{i,j}(x,y)_{i,j}}{\sum_{i,j}w_{i,j}},
  \label{eqn:ShowerDirection}
\end{equation}
with the weight
\begin{equation}
  w_{ij} = \frac{|\sin(\phi_i-\phi_j)|}{(\frac1{IA_i}+\frac1{IA_j})
    (\frac1{L_i/W_i}+\frac1{L_j/W_j})},
\end{equation}
where $w_{i,j}$ is an empirical weight. The weight is defined by three terms
accounting for three different quality criteria. The term in the numerator
favors the intersection of orthogonal lines, which have a smaller error factor
than parallel lines. The $IA$ term in the denominator puts more weight on images
of higher image amplitude which provide a higher accuracy. The $L/W$ term
suppresses images with a small length to width ratio, where the major axis is
less well-defined. Determining the shower direction is necessary for the
generation of sky maps and subsequent statistical analysis.

\subsubsection{Reconstruction of the Shower Core Position}
\begin{figure}[tbh!]
  \begin{minipage}[c]{0.48\linewidth}
    \fbox{\includegraphics[width=\textwidth]{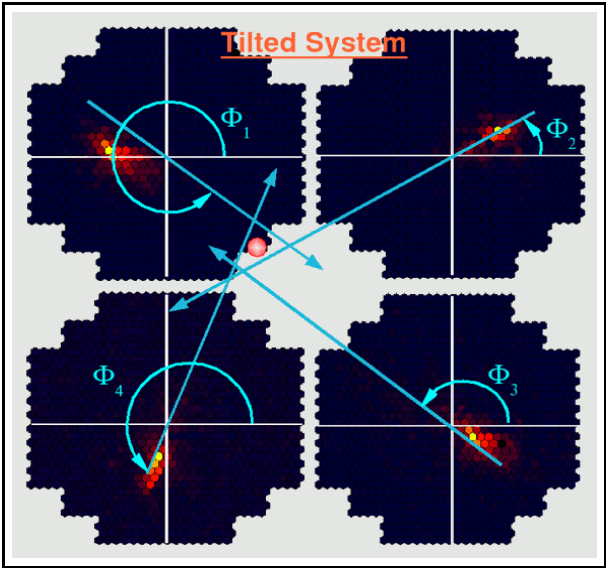}}
  \end{minipage}\hfill
  \begin{minipage}[c]{0.47\linewidth}
    \caption[Reconstruction of the Shower Core Position]{Reconstruction of the
    shower core position from four camera images in the tilted system. The
    intersection of the elongated major axis determines the shower core
    position. The camera images are greatly magnified in comparison to their
    distance. (Figure taken from \citet[pg. 86]{ErginThesis}.)}
    \label{fig:CorePosition}
  \end{minipage}
\end{figure}

The shower core position is determined in a tilted system, similar to the
shower direction. The tilted system is a three-dimensional coordinate system
perpendicular to the line of sight of the array. Each camera image has a
distinct position determined by the center of the telescope. Since the shower
direction is found along the major axis of the Hillas ellipse, the intersection
of the elongated major axis of all images in the tilted system defines the
shower core position (Fig.~\ref{fig:CorePosition}). Transformation to the
ground system provides the shower impact point at the ground. The core position
is also defined by the weighted sum of the pairwise intersection $(x,y)_{i,j}$
given by Eqn.~\ref{eqn:ShowerDirection}, but with different empirical weights.
Either
\begin{equation}
  w_{ij}=|\sin(\phi_i-\phi_j)| \label{eqn:CorePosition}
\end{equation}
or
\begin{equation}
  w_{ij}=\frac{|\sin(\phi_i-\phi_j)|}{(\frac1{IA_i}+\frac1{IA_j})}.
\end{equation}
is used. The difference is minor. For the analyses presented here,
Eqn.~\ref{eqn:CorePosition} has been used. The shower core position is needed
to calculate the impact distance, which is the distance of the core position to
the telescope. This parameter is important for the energy reconstruction of the
shower.

\subsection{Angular Resolution and Point Spread Function (PSF)} \label{sec:PSF}
\begin{figure}[ht!]
  \begin{minipage}[c]{0.55\linewidth}
    \includegraphics[width=\textwidth]{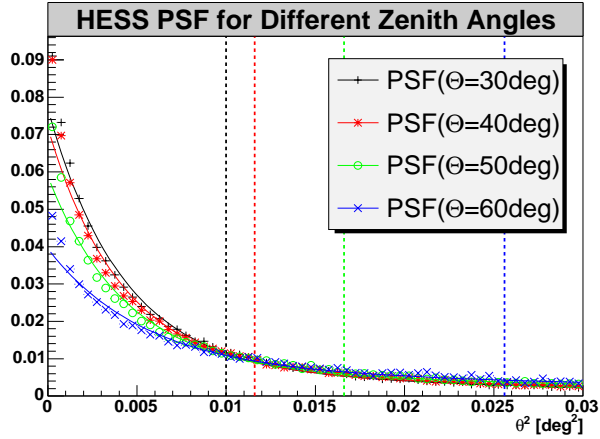}
  \end{minipage}\hfill
  \begin{minipage}[c]{0.47\linewidth}
    \caption[$\theta^2$ Distributions of the H.E.S.S.
      PSF]{$\theta^2$-distributions of the H.E.S.S. PSF as determined from
      point source simulations at different zenith angles $(\Theta)$,
      $\theta_w=0.5^\circ$, south orientation and $IA>80$\,p.e. The
      distributions are fit with the sum of two Gaussian functions (solid
      lines). The containment radii of 68\% are indicated by the dashed
      vertical lines.}
    \label{fig:PSF}
  \end{minipage}
\end{figure}

The angular resolution of H.E.S.S. is determined by the point spread function
(PSF), which is given by the accuracy of the geometric reconstruction of the
shower direction. The PSF can be determined from Monte Carlo simulations. The
H.E.S.S. PSF has been simulated and parameterized for the relevant
configurations. The parameterization is used where information about the PSF is
required, e.g. to indicated the PSF in sky maps. Fig.~\ref{fig:PSF} shows
sample distributions of the reconstructed shower direction versus the square of
the angular distance ($\theta$) from the true value. These
$\theta^2$-distributions are obtained from point source Monte Carlo simulations
for different zenith angles. The distributions in Fig.~\ref{fig:PSF} have been
normalized such that the integral is one. The parameterization ($PSF$) is given
by the sum of two Gaussian functions as
\begin{equation}
  PSF(\theta) = 
  A \left[ \exp\left(\frac{-\theta^2}{2\sigma^2_1}\right)
    +A_{rel}\exp\left(\frac{-\theta^2}{2\sigma^2_2}\right) \right],
  \label{eqn:PSF}
\end{equation}
with the standard deviations $\sigma_1$ and $\sigma_2$, the relative amplitude
$A_{rel}$ and the normalization constant $A$. The fit functions are shown by
the solid lines of the same color. Small deviations can be seen near the
origin. The representation versus $\theta^2$ should not be confused with a
profile of the PSF which is linear in $\theta$. The width of the PSF can be
expressed by a single value: the 68\% containment radius $(r_{\rm 68\%})$. It
is defined as the radius of a circle at the signal regions, which contains 68\%
of events in case of a point source.

The PSF depends on the zenith angle $(\Theta)$, wobble offset $(\theta_w)$,
azimuthal orientation $(Az)$ during observations (due to the magnetic field of
the earth) and image amplitude cut ($IA$). For example, the change of the image
amplitude cut from 80 to 400\,p.e. results in a reduction of the width of the
PSF by $\sim$50\%. The PSF also depends on the \g-ray energy spectrum, but this
dependence is small and negligible for the data discussed in this work. A
summary of different PSF fit parameters and $r_{68\%}$ is given in
Tbl.~\ref{tbl:PSF}.

\begin{table}[h]
  \centering
  \caption[PSF Parameters for Different Configurations]{Fit parameters and 68\%
    containment radius $(r_{68\%})$ of the H.E.S.S. PSF model for different
    zenith angles ($\Theta$), wobble offsets ($\theta_w$), azimuthal
    orientations $(Az)$ and image amplitude cuts ($IA$). The relative error of
    the parameters $\sigma_1$, $\sigma_2$, $A_{rel}$ and $r_{68\%}$ are 0.5\%,
    0.5\%, 2\% and 1\% respectively. The parameterizations for different
    configurations have been numbered to simplify a later reference.}
  \bigskip
  \begin{tabular}{cccccc}
    \hline \hline
    No. & $\Theta [^\circ]$, $\theta_w [^\circ]$, $Az$, $IA$ [p.e.]
    & $A_{rel}$ & $\sigma_1[^\circ]$ & $\sigma_2[^\circ]$ &$r_{68\%}[^\circ]$\\
    \hline
    1 & 30, 0.5, south, 80  & 0.0601 & 0.0477 & 0.111 & 0.112 \\
    2 & 40, 0.0, south, 80  & 0.175  & 0.0481 & 0.117 & 0.121 \\
    3 & 40, 0.5, south, 80  & 0.170  & 0.0477 & 0.117 & 0.120 \\
    4 & 40, 1.0, south, 80  & 0.166  & 0.0472 & 0.116 & 0.118 \\
    5 & 40, 0.5, south, 400 & 0.0655 & 0.0333 & 0.0785& 0.0632\\
    6 & 50, 0.5, south, 80  & 0.206  & 0.0489 & 0.125 & 0.139 \\
    7 & 50, 0.5, north, 80  & 0.174  & 0.0458 & 0.121 & 0.129 \\
    \hline \hline
\end{tabular}
\label{tbl:PSF}
\end{table}

\subsection{Energy Reconstruction}
\begin{figure}[t!]
  \begin{minipage}[t]{0.5\linewidth}
    \includegraphics[width=\textwidth]{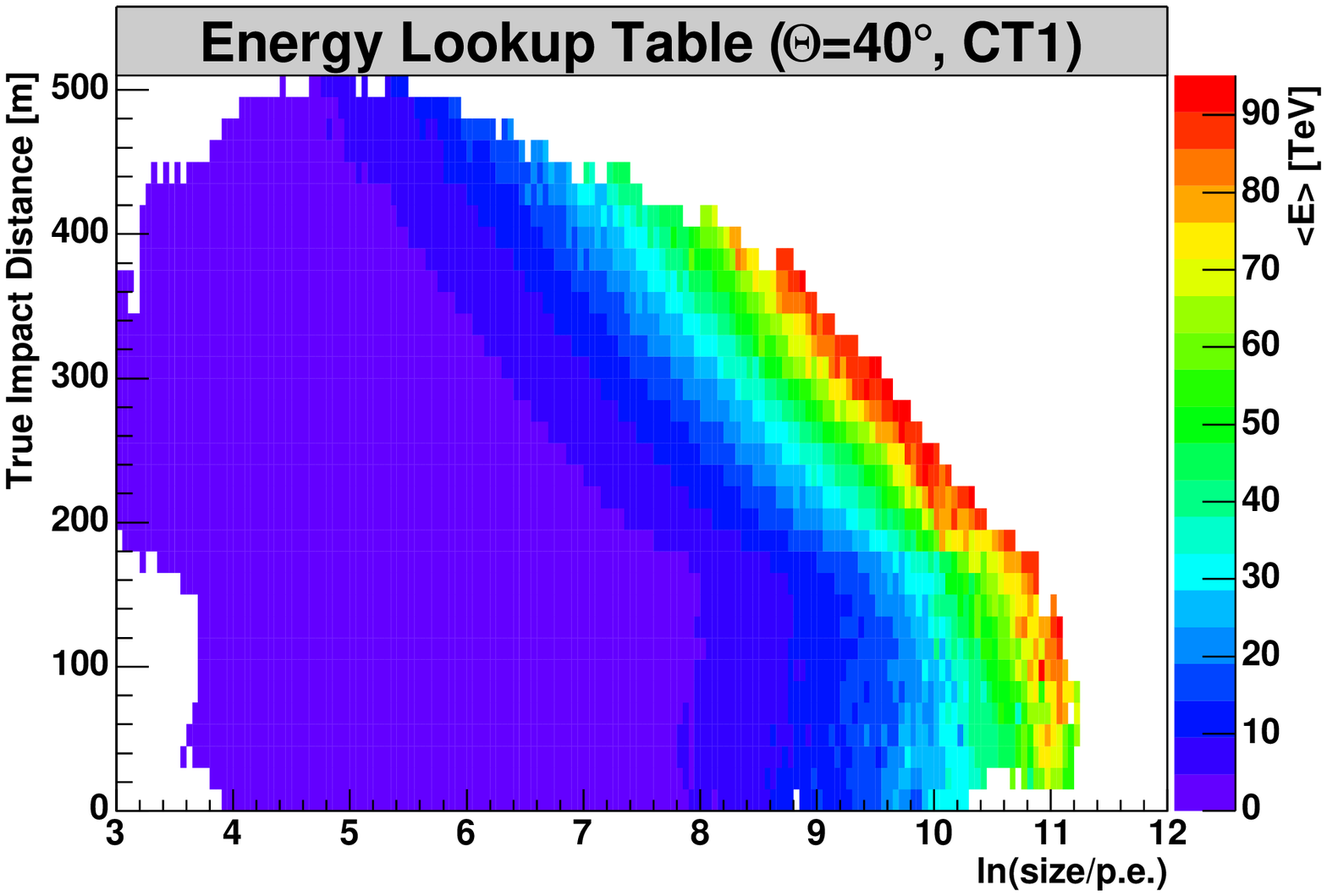}
    \caption[Energy Lookup Table]{Energy lookup table of CT\,1 for the zenith
      angle of $40^\circ$. The mean reconstructed energy is represented as a
      function of the image amplitude and the simulated impact distance.}
    \label{fig:EnergyLookup}
  \end{minipage}\hfill
  \begin{minipage}[t]{0.5\linewidth}
    \includegraphics[width=\textwidth,height=.7\textwidth]{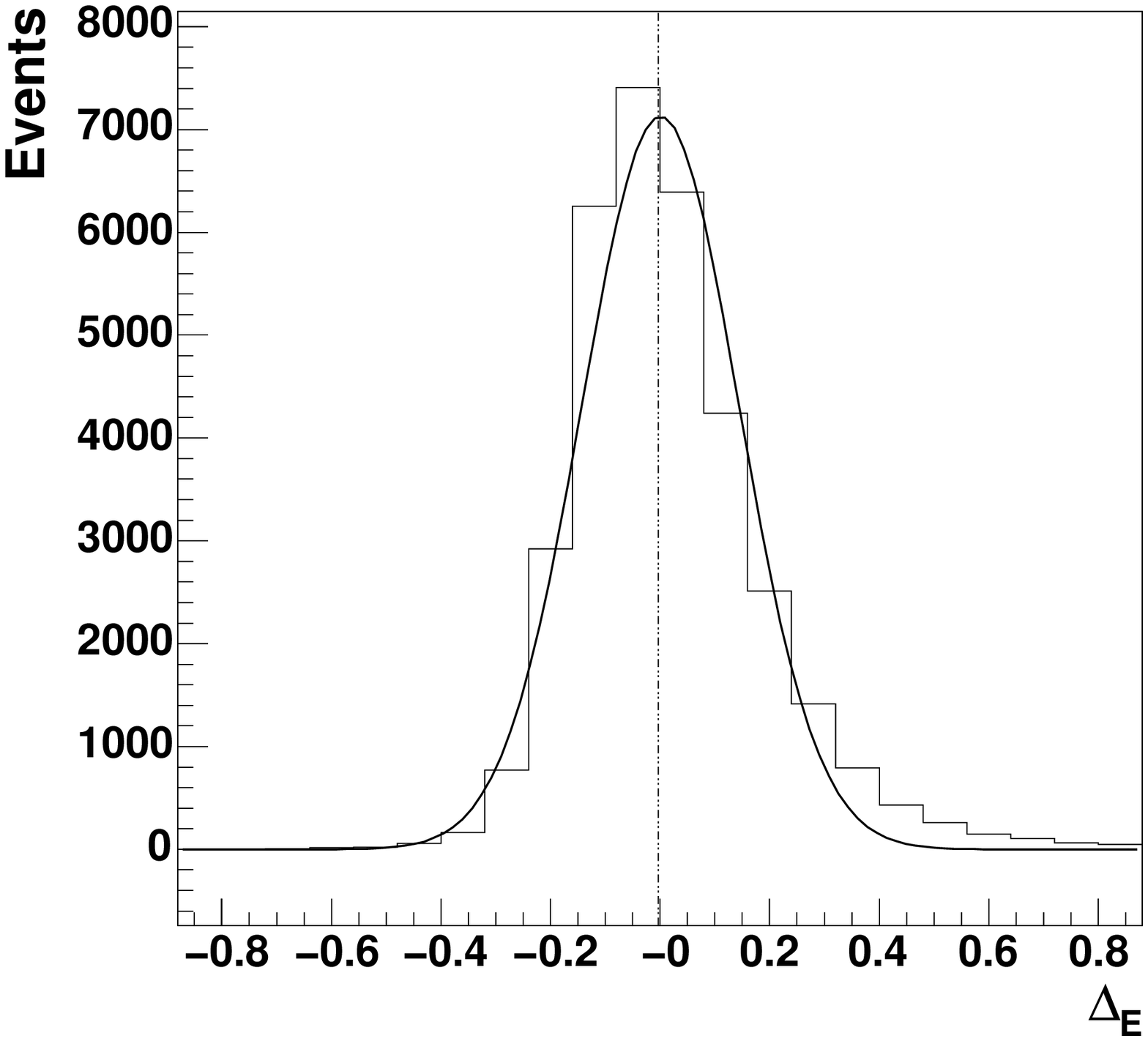}
    \caption[Energy Resolution]{Distribution of the relative error in the
      reconstructed energy per event for simulated \g-rays. (Figure taken from
      \citet{CrabPaper}.)}
    \label{fig:EnergyResolution}
  \end{minipage}
\end{figure}

Energy reconstruction relies on the dependence of image amplitude ($IA$) on
\g-ray energy $(E)$, impact distance $(b)$ and zenith angle $(\Theta)$. This
dependence of $IA(E,b,\Theta)$ can be used to calculate the energy
$E(IA,b,\Theta)$ as a function of image amplitude, impact distance and zenith
angle. This is accomplished with Monte Carlo simulations. For each telescope
and certain zenith angles, lookup tables of the energy as a function of $IA$
and true $b$ are created. For a shower recorded by a telescope with the image
amplitude $IA_{\rm Tel}$, the impact parameter $b_{\rm Tel}$ and the zenith
angle $\Theta$, the energy $E_{\rm Tel}(IA_{\rm Tel},b_{\rm Tel},\Theta)$ can
be obtained from the corresponding lookup table. Values for intermediate zenith
angles are found by linear interpolation in $\cos(\Theta)$. The shower energy
is determined as the arithmetic mean of the energies which have been determined
for each telescope. Fig.~\ref{fig:EnergyLookup} shows the lookup table of CT\,1
for a zenith angle of 40$^\circ$.

The relative error $(\Delta_E)$ for a simulated \g-ray shower with Monte Carlo
true energy $E_{\rm true}$ and reconstructed energy $E_{\rm reco}$ is defined
as $\Delta_E=(E_{\rm reco}-E_{\rm true})/E_{\rm true}$. Above the energy
threshold, $\Delta_E$ has a standard deviation $(\sigma_E)$ of $\sim$15\%
equivalent to the energy resolution of a single event.
Fig.~\ref{fig:EnergyResolution} shows $\sigma_E$ as a function of $E_{\rm
true}$ for simulations with a power law energy spectrum ($\Gamma=2.6$) above
440\,GeV at a zenith angle of 50$^\circ$. The RMS is 16\% and the width of the
Gaussian fit 14\%.

\section{Statistical Methods}
Statistical methods are necessary to find and characterize the \g-ray signals
in the data. The relevant techniques are discussed here and are demonstrated
with data from the Crab Nebula.

\subsection{Run Selection and Quality Criteria} \label{sec:RunSelection}
Before starting a data analysis it is important to verify the data quality. For
example, bad weather or technical problems are a few reasons why data of low
quality could be a part of a data set. Since this data mainly adds background
to the analysis and produces systematic errors, this data should be excluded
from the analysis. The following empirical quality criteria have been developed
in order to decide whether an observation run should be included in an analysis
or not.

\subsubsection{Trigger Rates}
\begin{figure}[ht]
  \begin{minipage}[c]{0.5\linewidth}
    \includegraphics[width=\textwidth]{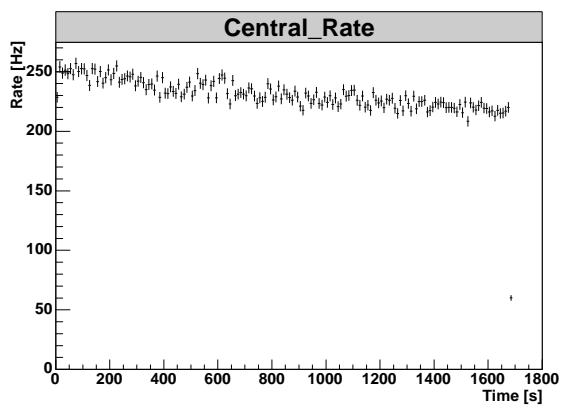}
  \end{minipage}\hfill
  \begin{minipage}[c]{0.5\linewidth}
    \caption[Stability of the Trigger Rate]{Trigger rate of run 20418. The
      stability is typical for a run included in an analysis. The slight slope
      reflects the zenith angle dependence of the trigger rate.}
    \label{fig:TriggerRate}
  \end{minipage}
\end{figure}

The system trigger rate ($R$) is a very sensitive measure for the quality of
atmospheric conditions. Since the hadronic background is very constant,
deviations from the typical trigger rate can indicate variations of the
transmissibility of the atmosphere (Fig.~\ref{fig:TriggerRate}). Since such
data cannot be handled in the standard analysis, these runs have to be
excluded. The individual trigger criteria for including a run are:

\begin{itemize}
\item Root mean square (RMS) of $R<30\%$
\item A relative change of $R<30\%$
\item RMS of $R$ after subtraction of the fit of $R$ by a straight line $<10\%$
\item $R>R_{\rm thres}$. The threshold $R_{\rm thres}(\Theta,t)$ is derived
  from reference observations taking into account the zenith angle
  (\citet{CentralTrigger}) and exponential efficiency loss over time (t)
  (\citet{H.E.S.S.Software}).
\end{itemize}

\subsubsection{Broken Pixels}
Broken pixels and unusable channels (Sec.~\ref{sec:BrokenPixels}) can lead to
systematic errors in reconstruction. The criteria for including a run consist of:

\begin{itemize}
\item Number of broken pixels during a run for any camera $<120$
\item Number of pixels with only the high voltage turned off $<50$.
\end{itemize}

\subsubsection{Dropped Events}
Sometimes not all of the triggered events can be recorded due to technical problems.
If the number of dropped events exceeds 5\%, the problem is considered as
severe and the run is excluded from the analysis.

\subsubsection{Tracking Accuracy}
To maintain the system's pointing accuracy of 20 arc seconds, runs with a
standard deviation of tracking more than 10 arc seconds are excluded
from the analysis.

\subsection{Event Selection} \label{sec:Cuts}
The selection of events with certain characteristics is useful to separate 
\g-ray showers from the background. The event selection is accomplished with
special selection cuts. These cuts reduce only a small fraction of the \g-ray
showers, but a large fraction of the background. An important aspect of the event
selection is \g\ / hadron separation. For example, while the typical system
trigger rate is $\sim$300\,Hz, the actual \g-ray rate is less than
$\sim$0.1\,Hz. However, with the right selection cuts this small signal to
noise ratio can be increased by a few orders of magnitude. A short description
of useful selection cuts is given below.

\subsubsection{Quality Cuts}
The main purpose of quality cuts is to provide a reliable shower reconstruction
and an efficient background reduction. Three such cuts are used: the cuts on
the trigger multiplicity, those on the image amplitude and those on the local distance.

The trigger multiplicity cut selects all events with a minimum number of
triggered telescopes. Its default value is two providing stereoscopic events,
only. The image amplitude cut improves the accuracy of shower reconstruction.
It removes images of small image amplitude that do not have a well-defined
major axis and therefore are difficult to reconstruct. The standard value is
80\,p.e. The local distance ($LD$) cut controls the distance between the center
of gravity of an image and the camera center in the nominal system. A standard
value of 2$^\circ$ guarantees that the majority of the images are not truncated
by the camera edges, which would falsify the reconstruction. The values of the
quality cuts are listed in Tbl.~\ref{tbl:StandardCuts} and \ref{tbl:Cuts}.

\subsubsection{Mean Reduced Scaled Cuts}
\begin{figure}[t]
  \begin{minipage}[t]{0.5\linewidth}
    \includegraphics[width=\textwidth]{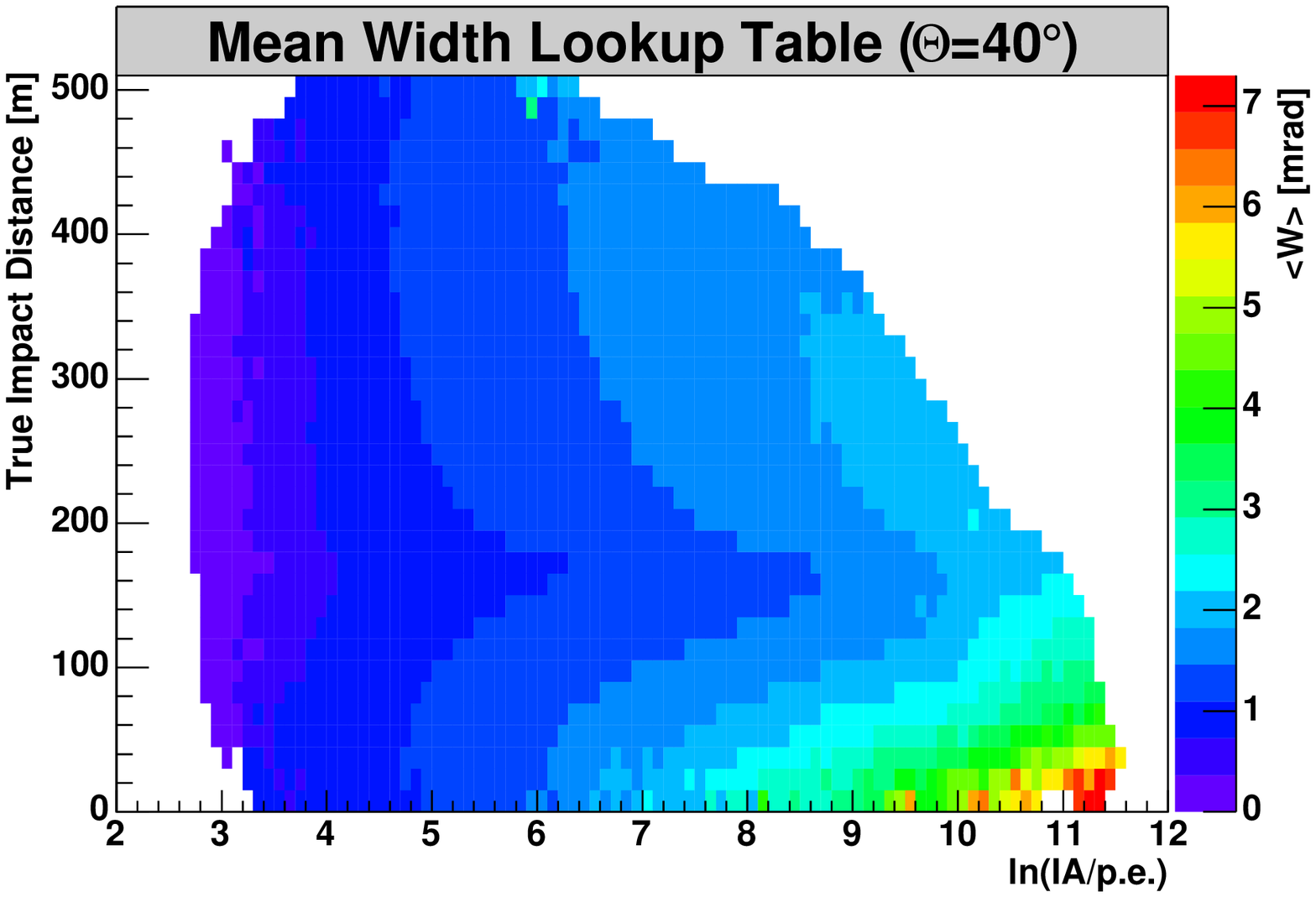}
    \caption[Mean Width Lookup Table]{Lookup table of mean width for a zenith
      angle of $40^\circ$.}
    \label{fig:MSWLookup}
  \end{minipage}\hfill
  \begin{minipage}[t]{0.5\linewidth}
    \includegraphics[width=\textwidth]{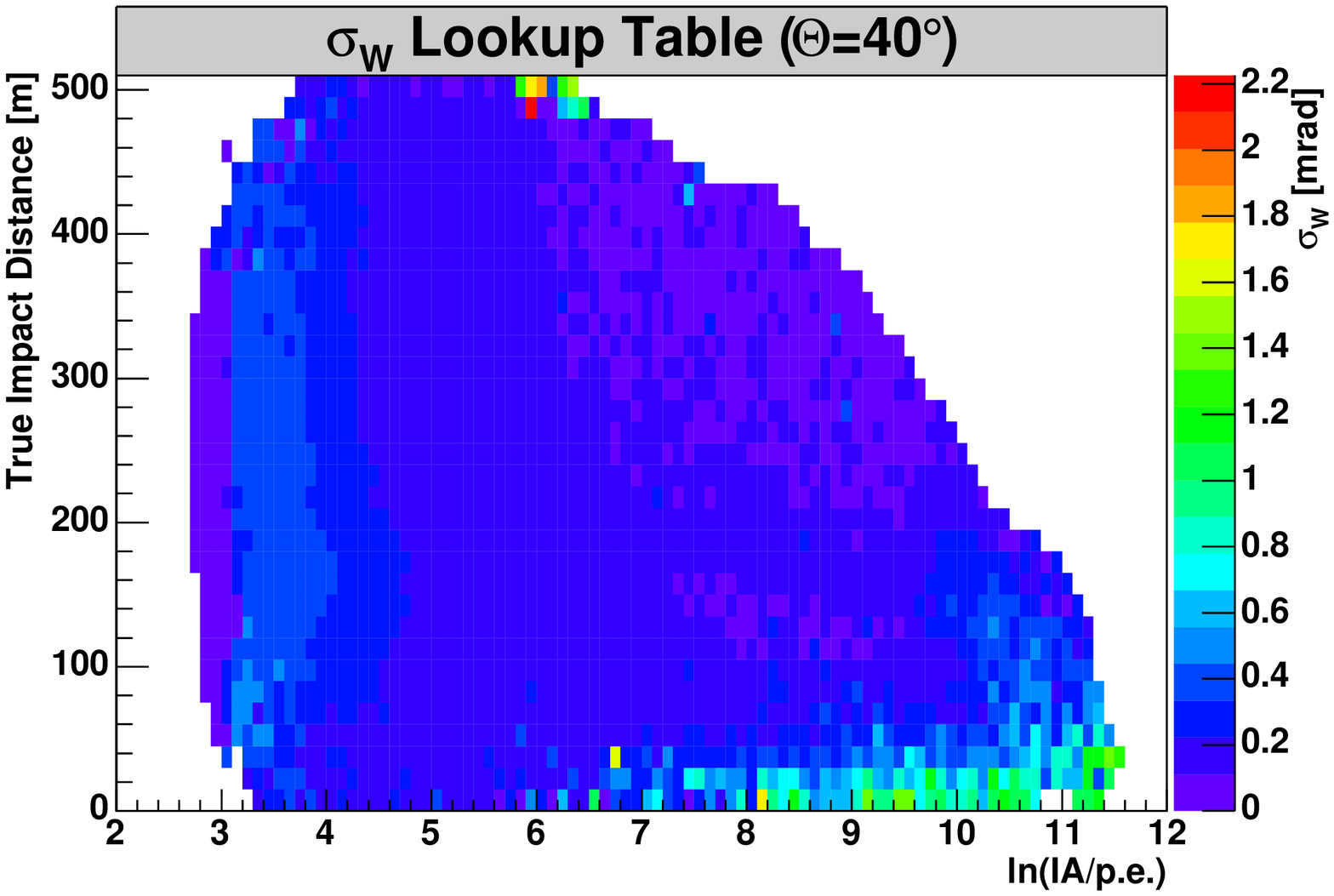}
    \caption[$\sigma_{W}$ Lookup Table]{Lookup table of the standard deviation
      of the width for a zenith angle of $40^\circ$.}
    \label{fig:SigmaMSWLookup}
  \end{minipage}
\end{figure}

\begin{figure}[h]
  \begin{minipage}[c]{0.5\linewidth}
    \includegraphics[width=\textwidth]{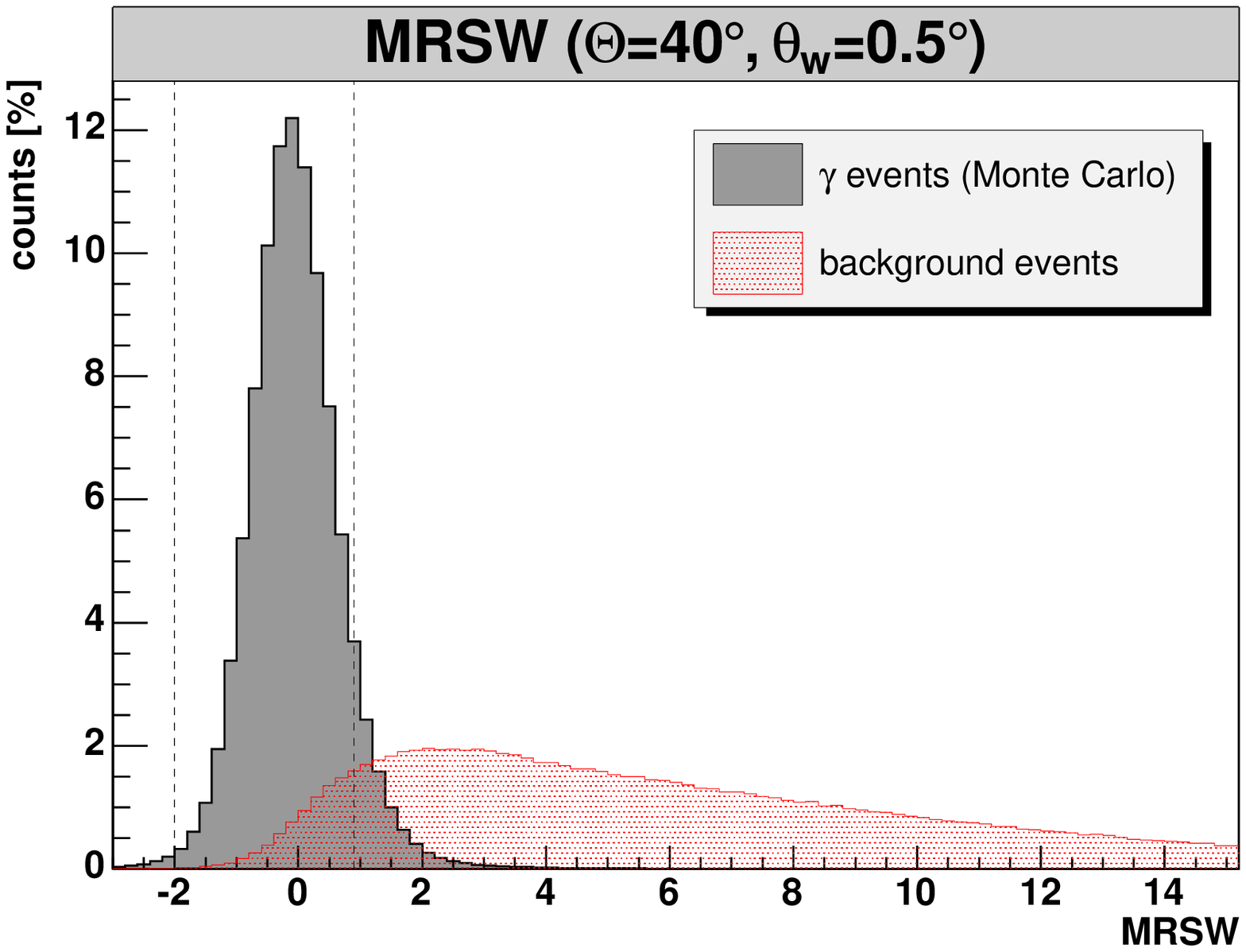}
    \caption[$MRSW$ Distribution]{{\it Mean reduced scaled width}
      distribution for background and Monte Carlo \g-ray data at a zenith angle
      of 40$^\circ$ and a wobble offset of 0.5$^\circ$.}
  \label{fig:MRSWDistributions}
  \end{minipage}\hfill
  \begin{minipage}[c]{0.5\linewidth}
    \includegraphics[width=\textwidth]{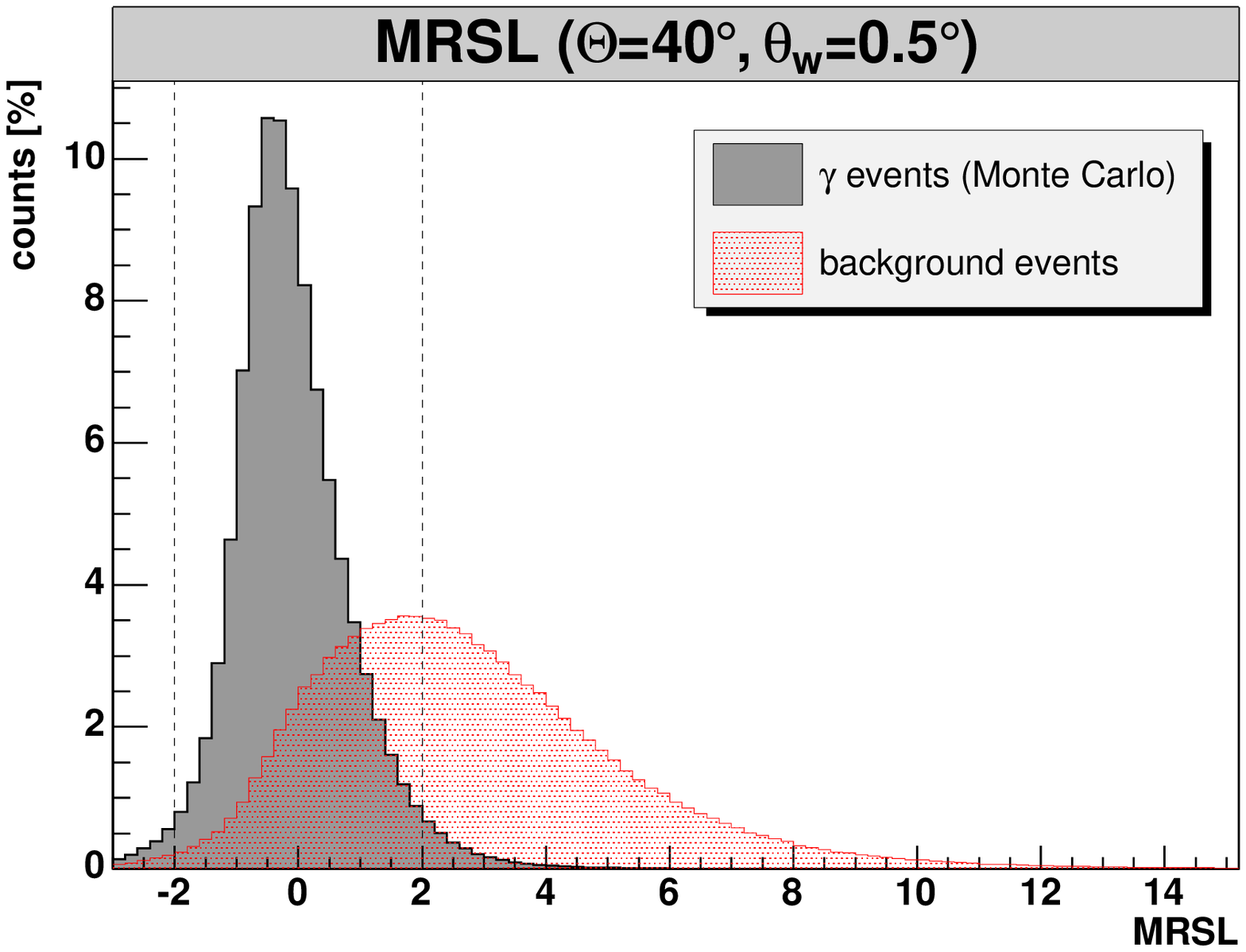}
    \caption[$MRSL$ Distribution]{{\it Mean reduced scaled length}
      distribution for background and Monte Carlo \g-ray data at a zenith angle
      of 40$^\circ$ and a wobble offset of 0.5$^\circ$.}
  \label{fig:MRSLDistributions}
  \end{minipage}
\end{figure}

The Hillas parameters of width $(W)$ and length $(L)$ provide efficient means
for \g\ / hadron separation and background suppression. The separation is
achieved through the different shapes of \g\ and hadronic shower images (cf.
Sec.~\ref{sec:AirShowers}). However, a separation cannot be achieved with a cut
on the width or length alone, since width and length also depend on the image
amplitude ($IA$), the impact distance $(b)$ and the zenith angle $(\Theta)$.
Nevertheless, Monte Carlo simulations can reproduce this complex dependence and
provide this information in lookup tables with the mean width ($<W>$) and
length ($<L>$) and the corresponding standard deviation ($\sigma_W$ and
$\sigma_L$) as a function of the natural logarithm of the image amplitude and
the impact parameter. These lookup tables are calculated for different zenith
angles where intermediate values are obtained by linear interpolation in
$\cos(\Theta)$. Fig.~\ref{fig:MSWLookup} and \ref{fig:SigmaMSWLookup} show the
lookup tables for a zenith angle of 40$^\circ$ with a wobble offset of
0.5$^\circ$. With these lookup tables, one can calculate the \emph{mean reduced
scaled width} ($MRSW$) which is defined as
\begin{equation}
  MRSW=\frac1{N_{\rm Tel}}\sum_{i=0}^{N_{\rm Tel}}\frac{W(IA_i,b_i,\Theta_i)-
    \langle W(IA_i,b_i,\Theta_i) \rangle}
  {\sigma_{W}(IA_i,b_i,\Theta_i)},
  \label{MRSW}
\end{equation}
where $N_{\rm Tel}$ is the total number of telescopes. The \emph{mean reduced
scaled length} ($MRSL$) is defined analogously. Hence the $MRSW$ and $MRSL$
express the deviation of width and length of a shower image from its
expectation value. The deviation is measured in units of standard deviations.
Events with small $MRSW$ and $MRSL$ parameters indicate high similarity to
\g-ray air showers and vice versa. Therefore, cuts on high $MRSW$ and $MRSL$
parameters reduce the number of background events.
Fig.~\ref{fig:MRSWDistributions} and \ref{fig:MRSLDistributions} show the
$MRSW$ and $MRSL$ distributions of simulated \g-ray showers and background
showers from source free observations at a zenith angle of $\sim40^\circ$ and
at a wobble offset of 0.5$^\circ$. The \g-ray showers are approximately
Gaussian-distributed around zero with a standard deviation of one. The
background events are asymmetrically distributed, with a tail extending to high
values. The mean reduced scaled standard cuts are indicated by the vertical
lines. They remove a large fraction of background events while keeping most
\g-events, resulting in an effective \g\ / hadron separation. The mean reduced
scaled cuts for different cut configurations are listed in
Tbl.~\ref{tbl:StandardCuts} and~\ref{tbl:Cuts}.

\subsection{Background Models} \label{sec:BackgroundModels}
Although a large fraction of background events can be reduced with cuts, some
\g-ray-like background events cannot be distinguished and separated from \g-ray
events and contaminate the true \g-ray counts. Nevertheless, the true strength
of a \g-ray signal can still be determined if this contribution from the
background is taken into account. The strength of the background is determined
from background regions, which are defined by a background model. The signal
and background regions are referred to as ON- and OFF-regions. Various
background models exist. The region- and the ring-background models are used in
the standard analysis. Both use a circular ON-region with a radius $\theta$ and
a size proportional to $\theta^2$. The OFF-regions are located at some distance
from the ON-region where virtually no \g-ray showers are observed. The H.E.S.S.
FOV of 5$^\circ$ is large enough to allow for simultaneous ON- and OFF-regions
in the same run. This is an advantage, since systematic errors on a run by run
basis are reduced. Also, bright stars can contribute to systematic errors. As
shown by \citet{Puehlhofer}, stars brighter than a magnitude of about six can
cause dips in acceptance at the star position. The dips are presumably caused
by reduced image amplitudes from switched-off pixels. To reduce these
systematic errors, stars in the ON- and OFF-regions should be avoided. The
advantage of the region-background model is its compensation for the radial
gradient of acceptance in the FOV, which makes it the first choice for spectral
analysis and robust against systematic effects. On the other hand, the
advantage of the ring-background model is its provision of sky maps.

\subsubsection{The Region-Background Model}
\begin{figure}[b]
  \begin{minipage}[c]{0.6\linewidth}
    \includegraphics[width=\textwidth]{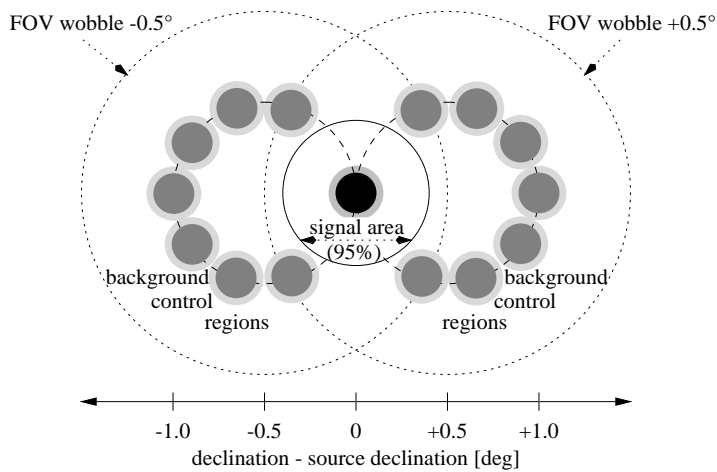}
  \end{minipage}\hfill
  \begin{minipage}[c]{0.4\linewidth}
    \caption[Region-Background Model]{Illustration of the region-background
    model applied to two runs with wobble offsets of $\pm0.5^\circ$ in Dec
    and seven OFF-regions (grey) each. The ON-region (black) is located in the
    center. The containment radius of 95\% defines an area where no OFF-region
    is contained. (Figure taken from \citet[pg. 149]{PuehlhoferThesis}.)}
  \label{fig:RegionBackgroundModel}
  \end{minipage}
\end{figure}

The region-background model can be applied if the data was taken in wobble
mode. The ON-region is chosen at the source region. The distance to the center
of the FOV is given by the wobble offset $(\theta_w)$. It radius $(\theta)$ is
limited by this distance $(\theta \le \theta_w)$. The OFF-regions are placed on
a circle given by the center of the FOV and a radius equal to the wobble
offset. They are preferably arranged symmetrically to the ON-region with
respect to the center of the FOV. This choice provides a very similar
acceptance for the ON- and OFF-regions independent of the radial gradient of
the acceptance. The number of OFF-regions may vary. More OFF-regions provide
higher background statistics. The maximum number of OFF-regions is limited by
their radius and wobble offset. In addition, a certain distance to the
ON-region has to be preserved to avoid a contamination of the OFF-regions with
\g-ray events. Fig.~\ref{fig:RegionBackgroundModel} illustrates the geometric
layout for two runs with a wobble offset of $\pm0.5^\circ$ in Dec and seven
OFF-regions each. The distance to the ON-region is given by the 95\%
containment radius of the source. The number of OFF-regions $(N_{\rm
OFF-regions})$ determines the normalization constant $\alpha$.
\begin{equation}
  \alpha=\frac1{\rm N_{\rm OFF-regions}}.
\end{equation}

\subsubsection{The Ring-Background Model}
\begin{figure}[b]
  \begin{minipage}[c]{0.6\linewidth}
    \includegraphics[width=\textwidth]{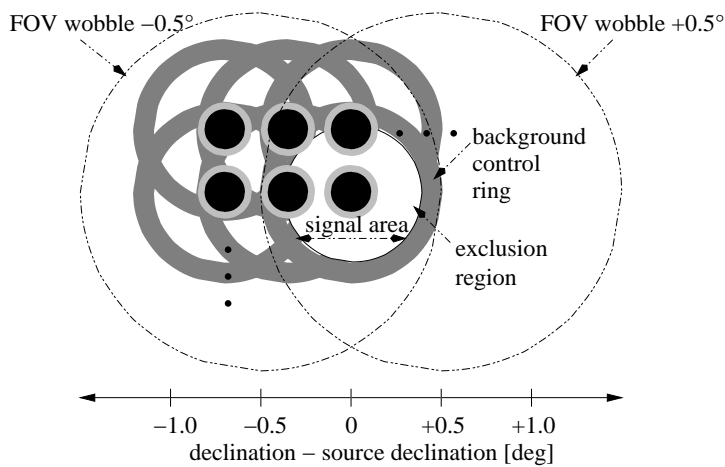}
  \end{minipage}\hfill
  \begin{minipage}[c]{0.4\linewidth}
    \caption[Ring-Background Model]{Illustration of the ring-background model.
    At each sky position a pair of signal- (black) and ring-regions (grey) is
    defined. Here only six pairs of rings are shown. Regions containing
    \g\ radiation, e.g. the signal region, are excluded from the
    ring-background.}
  \label{fig:RingBackgroundModel}
  \end{minipage}
\end{figure}

The ring-background model can be applied to data regardless of its observation
mode. The ON-region is circular and has a radius $(\theta)$. The OFF-region is
a ring which encloses the ON-region. The area of the ring is about seven times
the area of the ON-region. The radius of the ring is chosen close to the signal
region but not so close as to be significantly contaminated by the signal in
the ON-region. A typical value for point sources is $0.5^\circ$. To obtain sky
maps, this background model is applied to each bin in a sky map. This is
illustrated in Fig.~\ref{fig:RingBackgroundModel} for six different sky
positions. The results obtained at the source position provide the statistics
of the source. Sectors which contain \g-ray regions are excluded from the
ring-background. For example, the ON-region is always excluded. However, it is
not only the ratio between the ON- and OFF-region that determines the
normalization constant $\alpha$, but it is also the acceptance. Therefore
$\alpha$ is determined as
\begin{equation}
  \alpha=\frac{\int_{\rm ON-region}\epsilon(x,y)dx\,dy}{\int_{\rm
      OFF-region}\epsilon(x,y)dx\,dy}, \quad \mbox{ with}~x,y \notin R_{\rm ex}
  \label{eqn:RingAlpha}
\end{equation}
where $x$ and $y$ are the sky coordinates and $\epsilon(x,y)$ is the acceptance
for the data set. $\epsilon(x,y)$ is given as
\begin{equation}
  \epsilon(x,y)=\frac1{\sum_{\rm runs}t_{\rm run}}\sum_{\rm runs}\epsilon_{\rm
    run}(x,y)t_{\rm run},
\end{equation}
where $(t_{\rm run})$ is the live-time and $\epsilon_{\rm run}(x,y)$ is the
acceptances for each run. The live-time is the dead-time-corrected observation
time and $\epsilon_{\rm run}(x,y)$ is obtained from a fit to the radial
acceptance profile.

Due to the integration over the ON-region, the resolution of a sky map provided
by the ring-background is limited by the size of the ON-region. The resolution
can be increased if the ON-region is reduced to the size of a bin. The excess
maps discussed in this work are of this type. However, $alpha$ is still
calculated with the original ON-region.

A disadvantage of the ring-background model is the energy dependence of the
acceptance, which makes spectral analysis more difficult. Also, for
technical reasons the calculations are obtained from count maps where some
accuracy is lost through the binning. This is the reason why small differences
from the region-background model can be observed. However, the difference is
insignificant and negligible.

\subsection{Excess and Significance}
With a background model one can calculate signal statistics. The number of
true \g-ray events is given by the \g-ray excess $(N_\gamma)$, which is
determined as
\begin{equation}
  N_\gamma=N_{\rm ON}-\alpha N_{\rm OFF}, \label{eqn:Excess}
\end{equation}
where $N_{\rm ON}$ and $N_{\rm OFF}$ are the number of events in the ON- and
OFF-region and $\alpha$ is the normalization constant given by the background
model. The error of excess $(\Delta N_\gamma)$ is obtained according to
Gaussian error propagation as
\begin{equation}
  \Delta N_\gamma=\sqrt{\Delta N_{\rm ON}^2+(\alpha \Delta N_{\rm OFF})^2},
  \label{eqn:ExcessError}
\end{equation}
where $\Delta N_{\rm ON}=\sqrt{N_{\rm ON}}$ and $\Delta N_{\rm
OFF}=\sqrt{N_{\rm OFF}}$ provide good approximations for statistical
errors. The \g-ray rate is obtained by division with the live-time $(t)$, i.e.
\begin{equation}
  \mathrm{rate}=\frac{N_\gamma}{t}. \label{eqn:Rate}
\end{equation}
The significance of the \g-ray signal in the ON-region according to \citet{LiMa}
is
\begin{equation}
  S=\sqrt2 \Bigg[ N_{\rm ON} \ln \bigg(\frac{1+\alpha}\alpha \frac{N_{\rm
        ON}}{N_{\rm ON}+N_{\rm OFF}} \bigg) + N_{\rm OFF} \ln \bigg( (1+\alpha)
    \frac{N_{\rm OFF}}{N_{\rm ON}+N_{OFF}} \bigg) \Bigg]^{1/2}.
  \label{eqn:LiMa}
\end{equation}
In \g-ray astronomy, the significance is an important value since until recent
years only a few sources had been known, but a lot are now being detected. The
detection of a source is only claimed if it is observed with a significance of
at least five standard deviations $(S\ge5)$. A 5$\sigma$ detection implies the
probability for the signal being caused by a statistical fluctuation of less
than $\sim6\times10^{-5}$.

The significance depends on the cuts used in the analysis, which can be
optimized to give the maximum significance. The optimized cut configurations of
the standard analysis are presented in \citet{CrabPaper} and are listed in
Tbl.~\ref{tbl:Cuts}. They vary with the strength and the spectrum of the
source. The {\it extended} cut configuration is optimized for a flux of 10\% of
the Crab Nebula with a similar spectrum from an extended source. The set of
$hard$ cuts is optimized for sources with a flux of 1\% of the Crab flux and a
photon index $(\Gamma)$ of 2.0. The $loose$ cuts are optimized for strong
sources similar to the Crab Nebula with a $\Gamma$ of 3.0. The cuts used here
are based on these cut configurations and differ only in the $\theta^2$-cut,
which has been adapted for an extended source $(\sigma_w=0.1^\circ)$.

\begin{table}[htb!]
  \centering
  \caption[Common Cut Parameters of the Standard Analysis]{Cuts parameters
  which are used in all cut configurations of the standard analysis.}
  \bigskip
  \begin{tabular}{ccccc}
    \hline \hline
    Trigger Mult.   & $LD$      & $MRSL$ & $MRSL$ & $MRSW$ \\
    $[$No. of tels] & [$\deg.$] & Min.   & Max.   & Min. \\
    \hline
    2               & 2.0       & -2.0   & 2.0    & -2.0 \\
    \hline \hline
\end{tabular}
\label{tbl:StandardCuts}
\end{table}

\begin{table}[htb!]
  \centering
  \caption[Different Cut Configurations of the Standard Analysis]{Different cut
  configurations of the standard analysis. $\theta^2_{\rm std}$ denotes the
  standard cut values and $\theta^2$ the modified values which have been
  adjusted for the analysis of \MSH. Additional cuts parameters which are
  common to all configurations are listed in Tbl.~\ref{tbl:StandardCuts}.}
  \medskip
  \begin{tabular}{lcccc}
    \hline \hline
    Cut-Configuration &$MRSW$ min.&$IA$ min.&$\theta^2_{\rm std}$&$\theta^2$ \\
                      &           & [p.e.]  & [$\deg.^2$   ]     &[$\deg.^2$]\\
    \hline
    {\it extended}    & 0.9       & 80      & 0.16               & 0.09 \\
    {\it hard}        & 0.7       & 200     & 0.01               & 0.09 \\
    {\it loose}       & 1.2       & 40      & 0.04               & 0.09 \\
    \hline \hline
\end{tabular}
\label{tbl:Cuts}
\end{table}

\subsection{Source Position and Size} \label{sec:PositionSize}
The \g-ray excess map obtained from the ring-background model can be fit with
the two\hyp{}dimensional Gaussian function $G_{\sigma_x,\sigma_y, \alpha_0,
\delta_0, \omega,N} (\alpha,\delta)$ to determine the position, size and
orientation of a \g-ray signal. $G$ is obtained through the convolution of a
two-dimensional Gaussian with the PSF as described in
App.~\ref{app:GaussianFit}. This way it is possible to distinguish between the
influence of the PSF and the true i.e. intrinsic width of the \g-ray source.
The parameterization of the H.E.S.S. PSF according to Eqn.~\ref{eqn:PSF} is
determined with Monte Carlo simulations (cf. Sec.~\ref{sec:PSF}) and summarized
in Tbl.~\ref{tbl:PSF}. The parameters $\alpha_0$ and $\delta_0$ represent the
fit position in RA and Dec. The parameters $\sigma_x$ and $\sigma_y$ are the
standard deviations of the intrinsic width and length of the \g-ray signal.
$\omega$ is the angle between the major axis of the fit function and the
RA-axis.

\subsection{Analysis of Data from the Crab Nebula}
To demonstrate and verify the methods described above, they are applied to
H.E.S.S. data from the Crab Nebula taken from January 25, 2004 to March 4,
2005. The data was taken in wobble mode with offsets of $\theta_w=\pm0.5^\circ$
in RA and Dec at (J2000) (${\rm 15^h14^m27^s}$, $-59^\circ16'18''$). The mean
zenith angle of the observations is 50.0$^\circ$. The complete run list of the
30 runs with a live-time of 12.4\,h after run selection is given in
Tbl.~\ref{tbl:CrabRuns} of App.~\ref{app:RunLists}. The analysis was carried
out with {\it extended} cuts (Tbl.~\ref{tbl:Cuts}).

\begin{figure}[ht!]
  \begin{minipage}[c]{0.5\linewidth}
    \includegraphics[width=\textwidth]{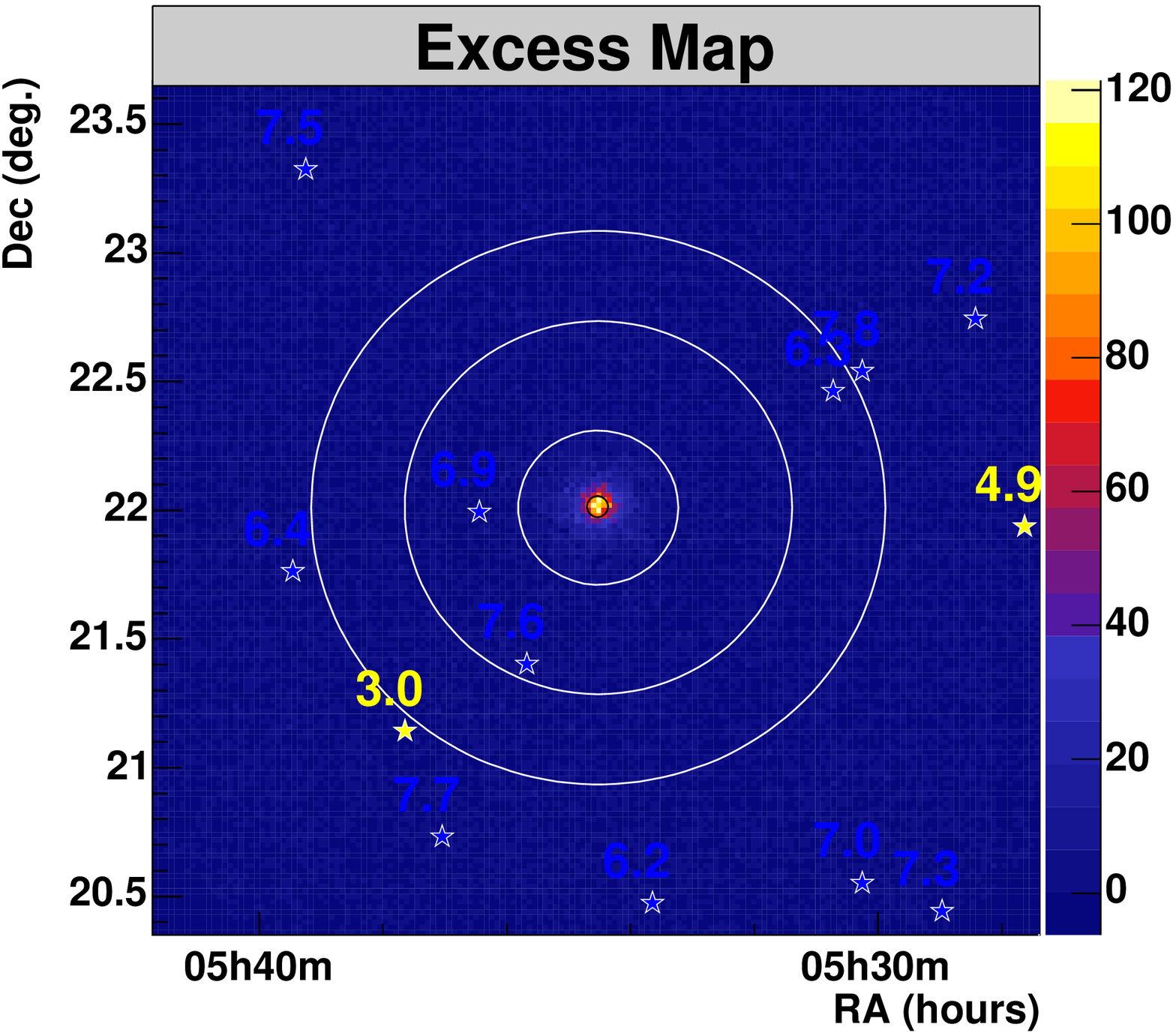}
    \caption[Crab Nebula Excess Map]{Excess map of the Crab Nebula showing the
      configuration of the ring-background model as applied in the analysis.
      Stars are indicated with their magnitudes.}
  \label{fig:CrabExcessMap}
  \end{minipage}\hfill
  \begin{minipage}[c]{0.5\linewidth}
    \includegraphics[width=\textwidth]{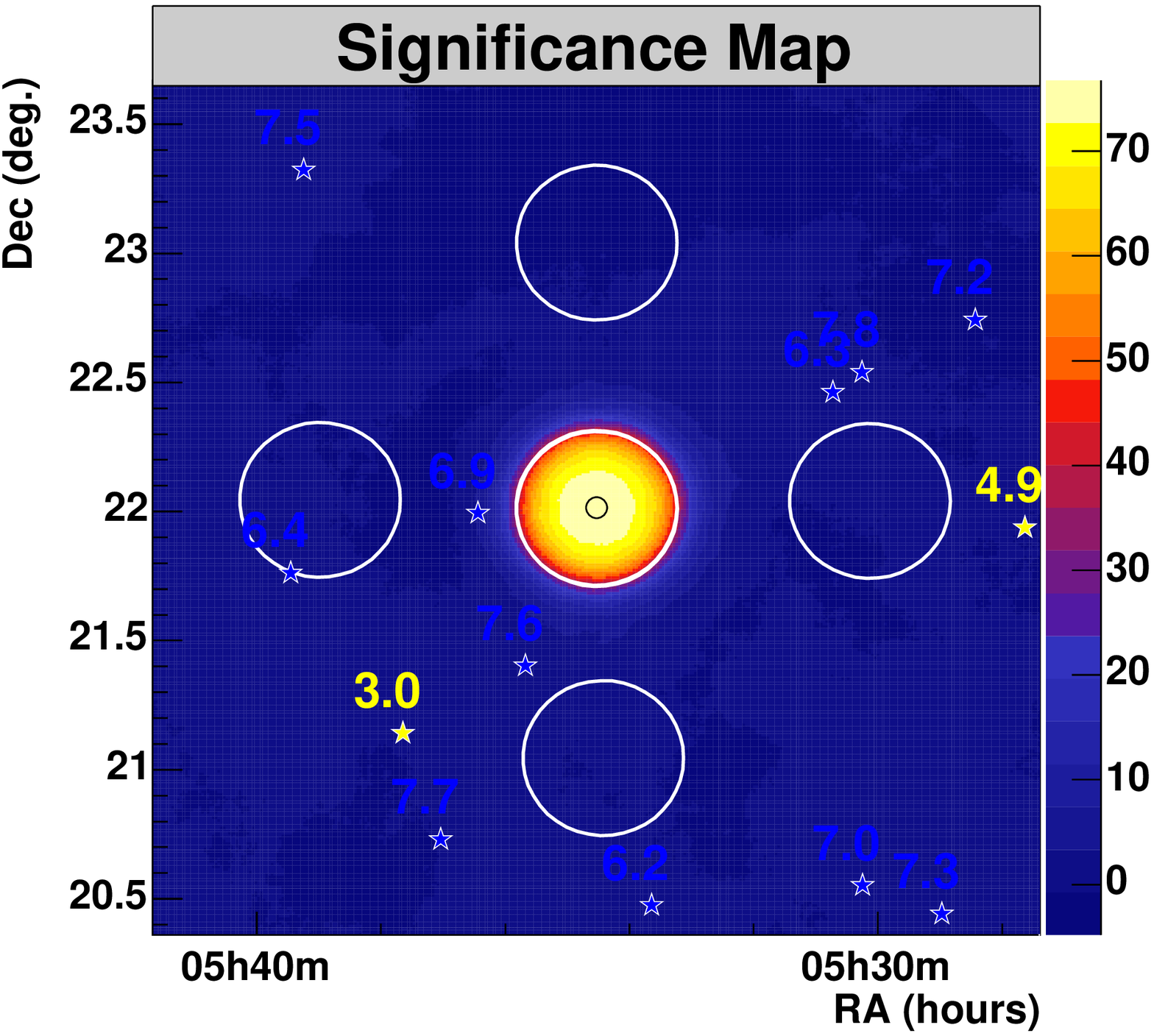}
    \caption[Crab Nebula Significance Map]{Significance map of the Crab Nebula
      showing the configuration of the region-background model as applied in
      the analysis. Stars are indicated with their magnitudes.}
  \label{fig:CrabSignificanceMap}
  \end{minipage}
\end{figure}

Fig.~\ref{fig:CrabExcessMap} and Fig.~\ref{fig:CrabSignificanceMap} show the
excess and the significance map as obtained with the ring-background model. The
configuration of the ring-background and the region-background model are
indicated. The ON- and OFF-regions have been chosen to not contain stars
brighter than a magnitude of six. The ON-region has a radius of 0.3$^\circ$
explaining the broad peak of similar radius in the significance map. The inner
and outer radii of the ring-background are 0.725$^\circ$ and 1.075$^\circ$
respectively. Also the region-background model was applied using one OFF-region
for each of the four wobble offsets. The statistics of both models are
summarized in Tbl.~\ref{tbl:CrabAnalysis}. A very significant \g-ray signal is
detected at the position of the Crab Nebula.

\begin{figure}[ht!]
  \begin{minipage}[c]{0.5\linewidth}
    \includegraphics[width=\textwidth]{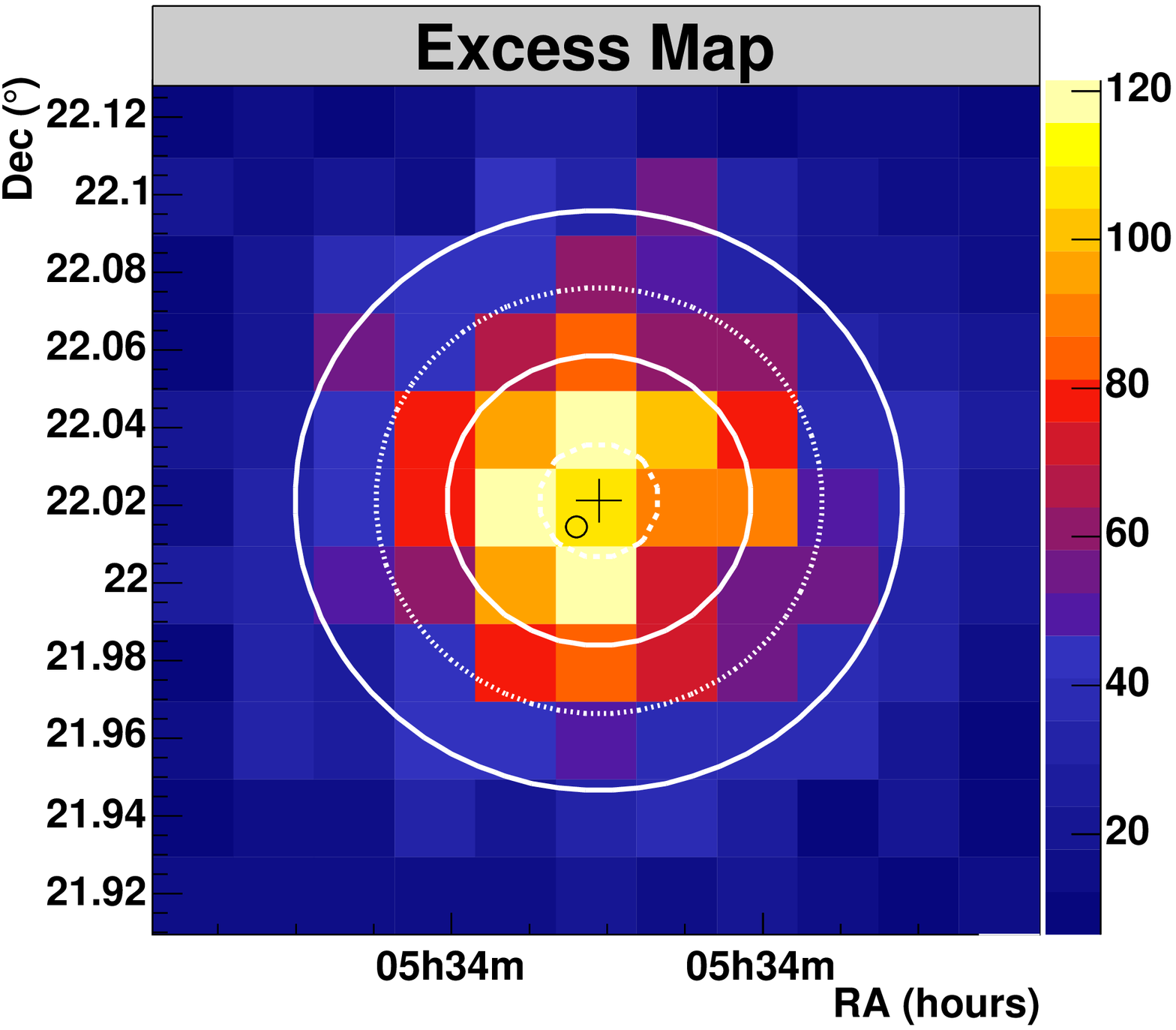}
    \caption[Crab Nebula Contour Lines of the Fit Function]{Excess map of the
      Crab Nebula with the contour lines (white) of the Gaussian fit function.
      The fit position of the centroid and the systematic errors are
      represented by the black cross. The position of the Crab Pulsar
      PSR\,B0531+21 is indicated by the black circle.}
    \label{fig:CrabSourceFit2D}
  \end{minipage}\hfill
  \begin{minipage}[c]{0.5\linewidth}
    \includegraphics[width=\textwidth]{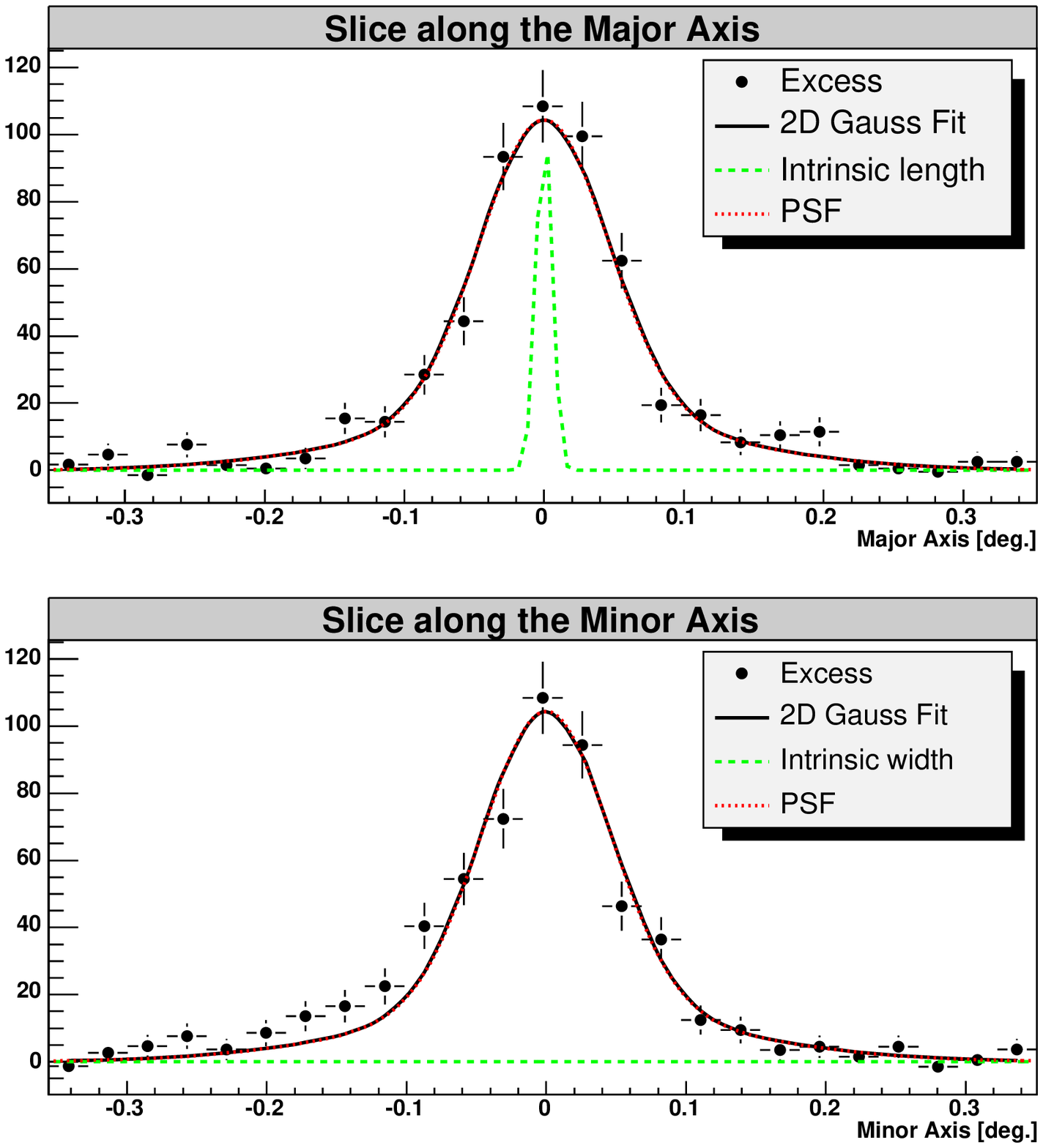}
    \caption[Crab Nebula Slices of the Excess]{Slices along two orthogonal axes
      of the \g-ray excess. The Gaussian fit function, the PSF and the
      intrinsic width and length are indicated. Within statistical errors the
      intrinsic width and length are zero (cf.
      Tbl.~\ref{tbl:CrabPositionSize}).}
    \label{fig:CrabSourceFit1D}
  \end{minipage}
\end{figure}

The results from a fit of the excess map with the Gaussian fit function
$G_{\sigma_x,\sigma_y,\alpha_0,\delta_0,\omega,N}(\alpha,\delta)$ are shown in
Tbl.~\ref{tbl:CrabPositionSize}. The fit position is found close to the pulsar
position as shown in the excess map of Fig.~\ref{fig:CrabSourceFit2D}. The
intrinsic width and length are close to zero. The data and the fit function
along two orthogonal axes are shown in Fig.~\ref{fig:CrabSourceFit1D}. The
extension of the excess can be explained by the component of the PSF alone.
Taking the system's pointing accuracy of $20''$ into account, the best fit
position of the Crab Nebula is (J2000) $(\rm 5^h34^m30\fs6 \pm4\fs5_{stat}
\pm1\fs4_{syst}, 22^\circ1'9\farcs8 \pm3\farcs4_{\rm stat} \pm20''_{\rm
syst})$. For comparison, the position of the Crab Pulsar (PSR\,B0531+21) as
determined from radio observations is (${\rm 5^h34^m31\fs97}$,
$22^\circ0'52\farcs07$).

\begin{table}[ht!]
  \centering
  \caption[Signal Statistics of the Crab Nebula Data]{Signal statistics as
    obtained from the ring-background and the region-background models for 30
    runs from the Crab Nebula with a total live-time of 12.4\,h. The marginally
    different event statistics in the ring-background model are explained by
    binning of the data.}
  \bigskip
  \begin{tabular}{lcc}
    \hline \hline
    & Ring-background & Region-background    \\
    \hline
    $N_{\rm ON}$                   & 11479           & 11512 \\ 
    $N_{\rm OFF}$                  & 25485           & 4380  \\
    $\alpha$                       & 0.180           & 1     \\
    $N_\gamma$                     & 6899 $\pm$ 111  & 7132 $\pm$ 126 \\
    $S [\sigma]$                   & 76.3            & 57.6  \\
    $\frac{S}{\sqrt{t}}
    [\frac{\sigma}{\sqrt{\rm h}}]$ & 21.7            & 16.4 \\
    signal$/$noise                 & 1.50            & 1.63  \\
    rate [min$^{-1}$]              & 9.3 $\pm$ 0.12  & 9.62 $\pm$ 0.17 \\
    \hline \hline
  \end{tabular}
  \label{tbl:CrabAnalysis}
\end{table}

\begin{table}[ht!]
  \centering
  \caption[Position and Size of the Crab Nebula]{Fit parameters of the
    two-dimensional Gaussian function
    $G_{\sigma_x,\sigma_y,\alpha_0,\delta_0,\omega,N}(\alpha,\delta)$
    (App.~\ref{app:GaussianFit}) for the \g-ray excess map of the Crab Nebula.
    The parameterization of the PSF is found in Tbl.~\ref{tbl:PSF}.}
  \bigskip
  \begin{tabular}{l|c}
    \hline \hline
    Parameter                     & Value \\
    \hline
    Position RA ($\alpha_0$)      & $\rm5^h34^m30\fs6\pm4\fs5$,
    $(83.627^\circ \pm 0.001^\circ)$ \\
    Position Dec ($\delta_0$)     & $22^\circ1'9.84''\pm3\farcs4$,
    $(22.019^\circ \pm 0.001^\circ)$ \\
    Length ($\sigma_x$)           & $0.3'\pm0.3'$, $(0.005^\circ\pm0.005^\circ)$\\
    Width ($\sigma_y$)            & $0'\pm0.3'$, $(0^\circ\pm0.005^\circ)$\\
    PSF parameterization          & no. 7 (Tbl.~\ref{tbl:PSF}) \\
    \hline \hline
  \end{tabular}
  \label{tbl:CrabPositionSize}
\end{table}

\clearpage

\section{Spectroscopy} \label{sec:Spectrum}
Spectroscopy is an important aspect in astronomy because it can provide new
information about an astrophysical object and the processes of a source
region. If $\Phi$ is the flux from the source, then its energy spectrum is
given by the differential flux $\frac{d\Phi}{dE}$. Often the energy spectrum
obeys a power law of the form
\begin{equation}
  \frac{d\Phi}{dE}=\phi_{\rm 1TeV}\left(\frac{E}{\rm 1TeV}\right)^{-\Gamma},
  \label{eqn:PowerLaw2}
\end{equation}
where $\phi_{\rm 1TeV}$, $\Gamma$ and $E$ denote the differential flux at
1\,TeV, the photon index and the energy, respectively. The integral flux above
1\,TeV can be obtained by integration of the differential flux as
\begin{equation}
  \Phi(E>{\rm 1T  eV})=\int_{\rm 1TeV}^{\infty}\phi_{\rm 1TeV}
  \left(\frac{E}{\rm 1TeV}\right)^{-\Gamma}dE=\frac{\phi_{\rm
  1TeV}}{1-\Gamma}\rm\,TeV.
  \label{eqn:Flux}
\end{equation}
The integral flux is used for comparison of the flux measured by experiments
with different energy thresholds. The error $\Delta\Phi(E>{\rm 1TeV})$ is
obtained by Gaussian error propagation.\footnote{According to Gaussian error
propagation, the error $\sigma_g$ of the function $g(x,y)$ is given by
\begin{equation}
  \sigma_g=\sqrt{
    \left(\frac{dg}{dx}\right)^2 \sigma_{x}^2 +
    \left(\frac{dg}{dy}\right)^2 \sigma_{y}^2 +
    2\frac{dg}{dx} \frac{dg}{dy} \mathrm{cov}(x,y) },
\end{equation}
where $\sigma$ stands for the standard error and cov$(x,y)$ for the covariance
between $x$ and $y$. Here $g$, $x$ and $y$ have to be substituted by
$\frac{d\Phi}{dE}$, $\phi_{\rm 1TeV}$ and $\Gamma$, respectively.} If
$\sigma_{\rm 1TeV}$, $\sigma_\Gamma$ and cov$(\sigma_{\rm 1TeV},\Gamma)$ denote
the error of the differential flux, the error of the photon index and the
covariance, then
\begin{equation}
  \sigma_{\Phi}(E>{\rm 1TeV})=\sqrt{
    \left(\frac{1}{1-\Gamma}\right)^2\sigma_{\phi_{\rm 1TeV}}^2 +
    \left(\frac{\phi_{\rm 1TeV}}{(1-\Gamma)^2}\right)^2\sigma_{\Gamma}^2+
    2\frac{\phi_{\rm 1TeV}}{(1-\Gamma)^3}\mathrm{cov}(\phi_{\rm 1TeV},\Gamma)},
  \label{eqn:FluxError}
\end{equation}
where the standard errors and the covariance are obtained from a $\chi^2$-fit
to the energy spectrum.

The energy spectrum can be determined from the number of \g-ray excess events
$(N_\gamma)$ provided by the background model. If the detector's effective area
$(A)$ is known, the differential flux is given by
\begin{equation}
  \frac{d\Phi}{dE}=\frac1{tA}\frac{dN_\gamma}{dE}
  =\frac1{t\Delta E}\left[\sum_{i=1}^{N_{\rm ON}}\frac1{A_i}-\alpha\sum_{j=1}^{N_{\rm OFF}}\frac1{A_j}\right].
  \label{eqn:DifferentialFlux}
\end{equation}
Here $t$ is the live-time of the observation, $N_{\rm ON}$ and $N_{\rm OFF}$
are the number of events in the ON- and OFF-regions and $A_i$ is the effective
area for an event $i$. The normalization constant $\alpha$ as well as the ON-
and OFF-regions are determined by the background model. Since the usage of the
ring-background model is difficult in spectral analysis, in this work only the
region-background model is used for spectroscopy. Again, the error of the flux
$\sigma\left(\frac{d\Phi}{dE}\right)$ is found by Gaussian error propagation
for Eqn.~\ref{eqn:DifferentialFlux} as
\begin{equation}
  \sigma\left(\frac{d\Phi}{dE}\right)=
  \frac{\sigma_\gamma}{t\Delta E} =
  \frac{1}{t\Delta E} \sqrt{ \sum_{i=1}^{N_{\rm ON}} 
    \frac1{A_i^2} + \frac{1}{A_i^4}\sigma_{A_i}^2
    -\alpha \sum_{i=1}^{N_{\rm OFF}}\frac1{A_j^2}
    +\frac{1}{A_j^4}\sigma_{A_j}^2}.
  \label{eqn:DifferentialFluxError}
\end{equation}
Here $\sigma(A_i)$ is the error of the effective area for an event $i$. The
covariance is zero, since the error of the event statistics and the effective
area are uncorrelated. Eqn.~\ref{eqn:DifferentialFluxError} takes the
statistical error of the events and the effective area for each event into
account.

In the analysis, a histogram with differential flux and a logarithmic binning
is filled with the excess given by Eqn.~\ref{eqn:DifferentialFlux}. The bin
size is chosen by an empiric rule, according to the significance of the source.
Six bins per decade are typical for strong sources. The spectral parameters
$\phi_{\rm 1TeV}$ and $\Gamma$ are obtained from a least-square fit to the
flux. The fit range covers all bins from the first bin which is above the safe
energy threshold up to the last significant bin with a relative error of less
than 95\%, i.e. two standard deviations. The bin errors represent the 68\%
Feldman-Cousins' confidence interval which is discussed in
App~\ref{app:FeldmanCousins}. These asymmetric errors are also taken into
account in the $\chi^2$-fit. If the fit value for a bin is greater (or smaller)
than the bin content, then the positive (or negative) error is chosen for the
corresponding bin error of the $\chi^2$ fit.

\subsection{Effective Area}
The effective area, also known as collection area, of a detector is the
corresponding area of an imaginary detector with an efficiency of 100\% that
would detect the same event rate. The effective area $(A)$ of the H.E.S.S.
detector with a sensitive area $(H)$ of about $5\times10^5\,$m$^2$ is a
function of the energy $(E)$, the zenith angle $(\Theta)$, the azimuth angle
$(\phi)$ due to the magnetic field of the earth and the wobble offset
$(\theta_w)$. $A(E,\Theta,\phi,\theta_w)$ is determined by Monte Carlo
simulations from the number of detected events past cuts
$N(E,\Theta,\phi,\theta_w)$ and the total number of simulated events $N_{\rm
total}(E,\Theta,\phi,\theta_w)$ as
\begin{equation}
  A(E,\Theta,\phi,\theta_w)=H\frac{N(E,\Theta,\phi,\theta_w)}{N_{\rm
      total}(E,\Theta,\phi,\theta_w)}.
\end{equation}
Since $A$ is dependent on the cuts, $A$ differs for each cut configuration. To
reduce systematic errors from the bias of the energy reconstruction discussed
in the next section, $A$ is determined as a function of the reconstructed
$(E_{\rm reco})$ instead of the true shower energy $(E_{\rm true})$. After $A$
is determined, it is fit by the empirical fit function
\begin{equation}
  f(E) = p_0 \cdot e^{(p_1\cdot E)} + p_2E^4 + p_3E^3 + p_4E^2 + p_5E + p_6.
\end{equation}
$A(E_{\rm reco})$ has been calculated for the Monte Carlo files of
Sec.~\ref{sec:MCfiles}. Intermediate values are determined by linear
interpolation in $f$, $\cos(\Theta)$ and the wobble offset $\theta_w$.

\begin{figure}[tbh!]
  \begin{minipage}[t]{0.5\linewidth}
    \includegraphics[width=\textwidth]{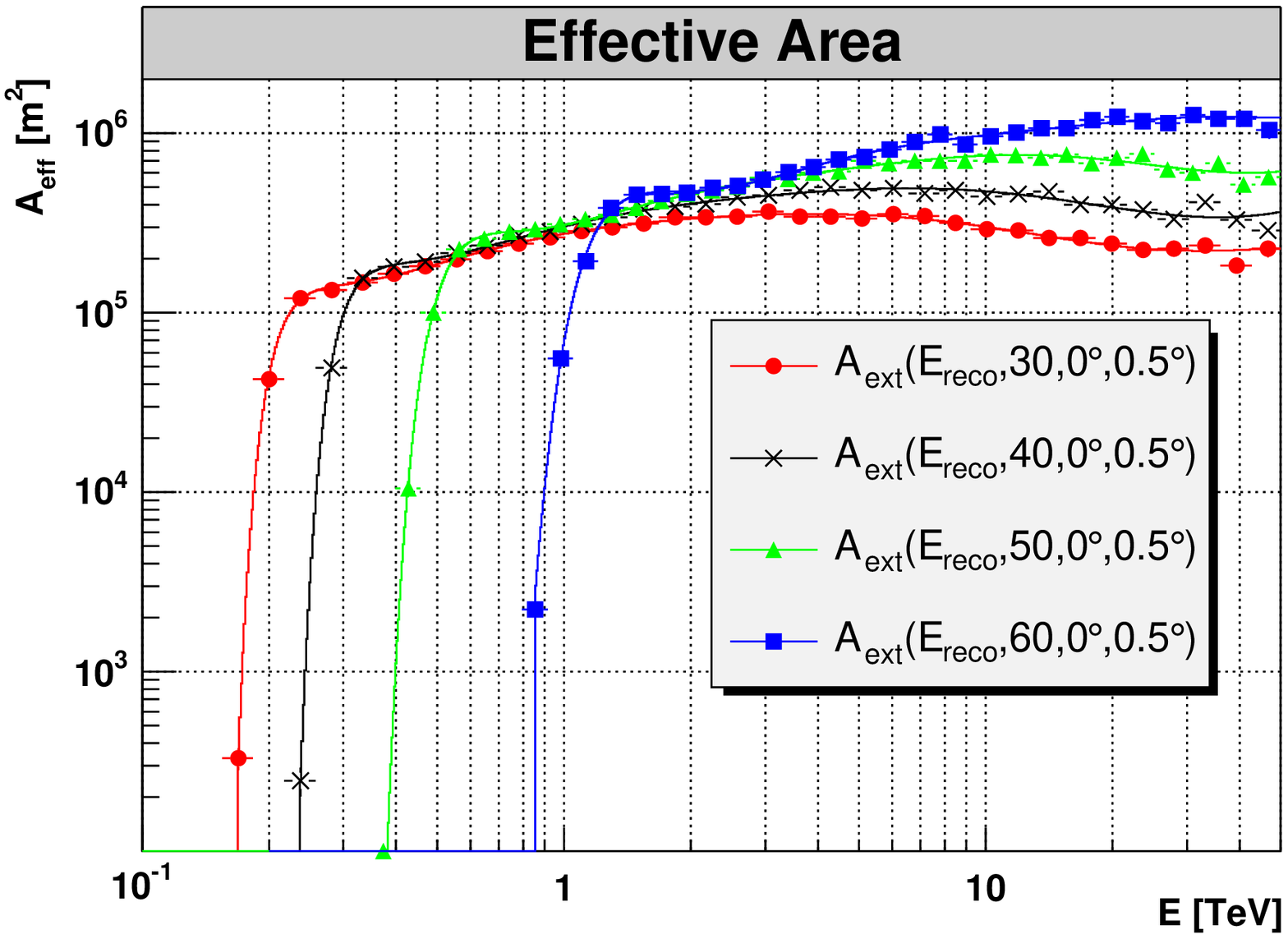}
    \caption[Effective Area for Different Zenith Angles]{Effective area for
      different zenith angles for the extended cuts used in the analysis of
      Crab Nebula data.}
    \label{fig:EffectiveAreasZenith}
  \end{minipage}\hfill
  \begin{minipage}[t]{0.5\linewidth}
    \includegraphics[width=\textwidth]{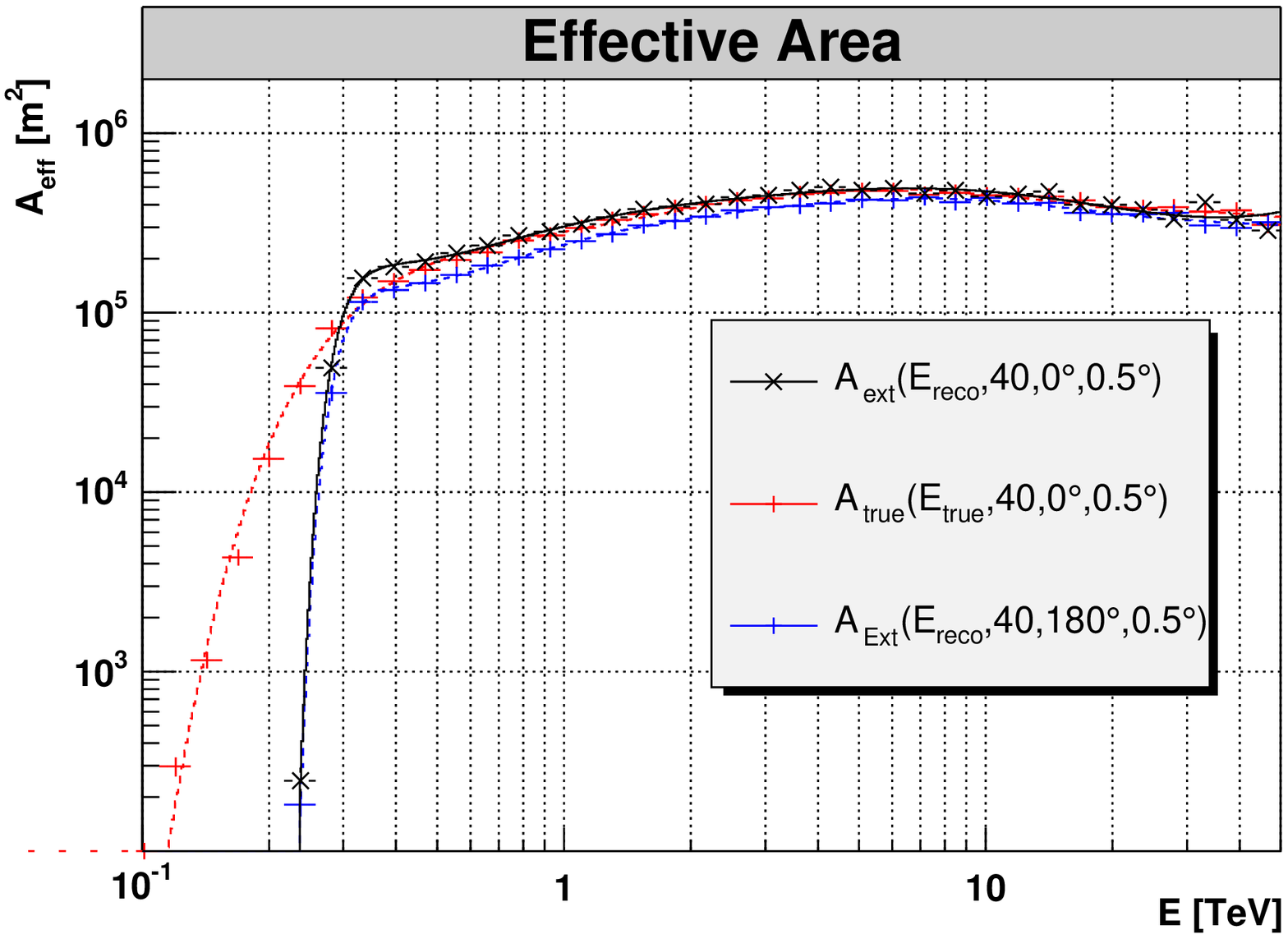}
    \caption[Effective Area for Different Cut Configurations]{Effective area
      for different cut configurations as described in the text. The difference
      between the effective area of true and reconstructed energy is apparent.}
    \label{fig:EffectiveAreas}
  \end{minipage}\hfill
\end{figure}

Fig.~\ref{fig:EffectiveAreasZenith} shows the effective area of the {\it
extended} cuts $(A_{\rm ext})$ for different zenith angles.
Fig.~\ref{fig:EffectiveAreas} shows $A_{\rm ext}$ in comparison with different
effective areas. $A_{\rm true}$ is the effective area as a function of the true
instead of the reconstructed energy. The difference is apparent at low
energies. $A_{\rm Ext}$ is the effective area for the {\it extended} cuts and
an extended source with a standard deviation in width ($\sigma_w$) and length
($\sigma_l$) of 0.04$^\circ$ and 0.11$^\circ$ respectively. Since its
difference from $A_{\rm ext}$ is only minor, it is not apparent in the figure.

\subsection{Energy Bias and Systematic Errors} \label{sec:SpectrumAccuracy}
\begin{figure}[t!]
  \begin{minipage}[t]{0.5\linewidth}
    \includegraphics[width=\textwidth]{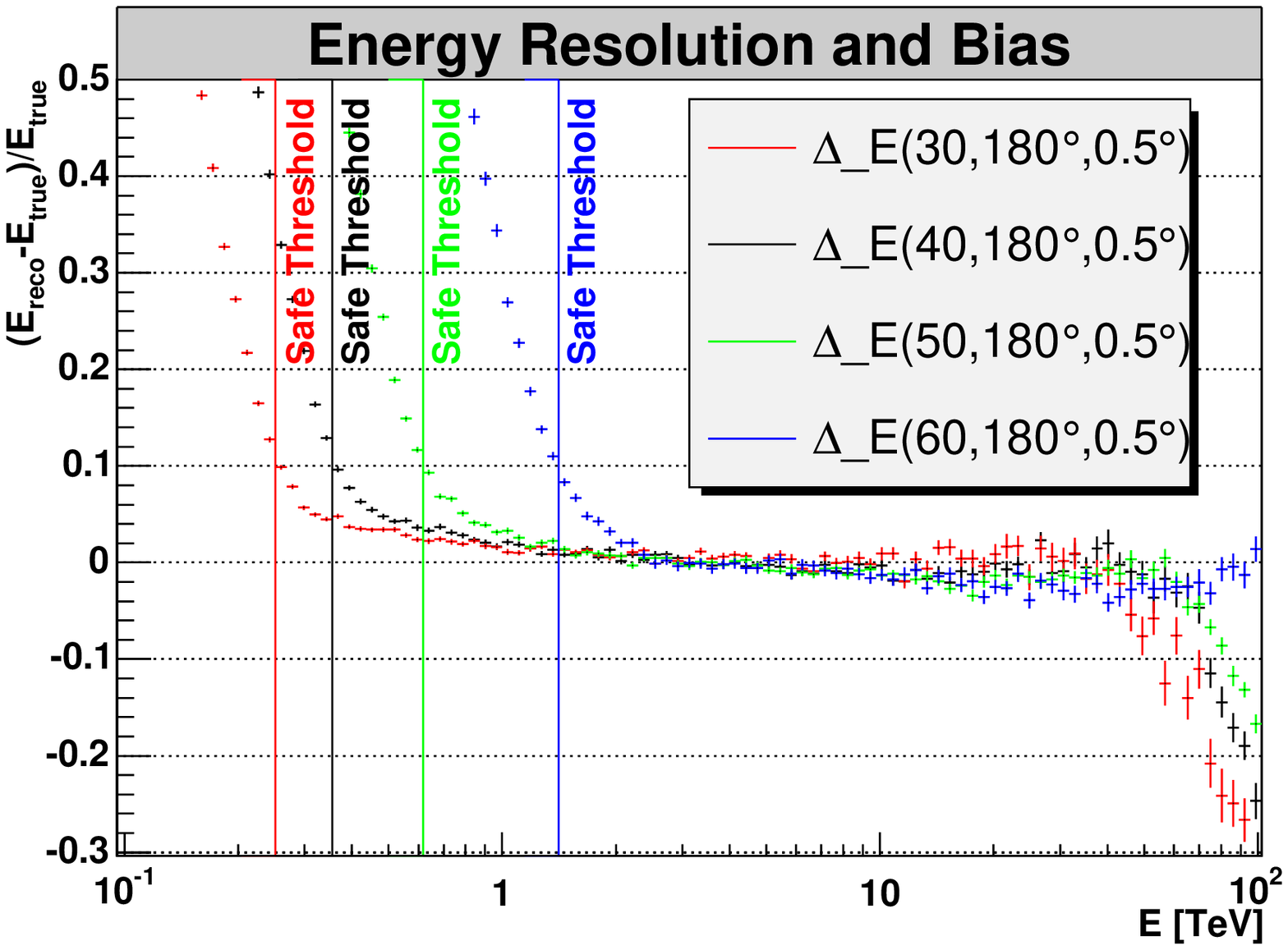}
    \caption[Bias in the Energy Reconstruction]{Bias in the energy
        reconstruction $(E_{\rm reco}-E_{\rm true})/E_{\rm true}$ and safe
        energy threshold for various zenith angles. The bias above the safe
        energy threshold is less than 10\%.}
    \label{fig:EnergyBias}
  \end{minipage}\hfill
  \begin{minipage}[t]{0.5\linewidth}
    \includegraphics[width=\textwidth]{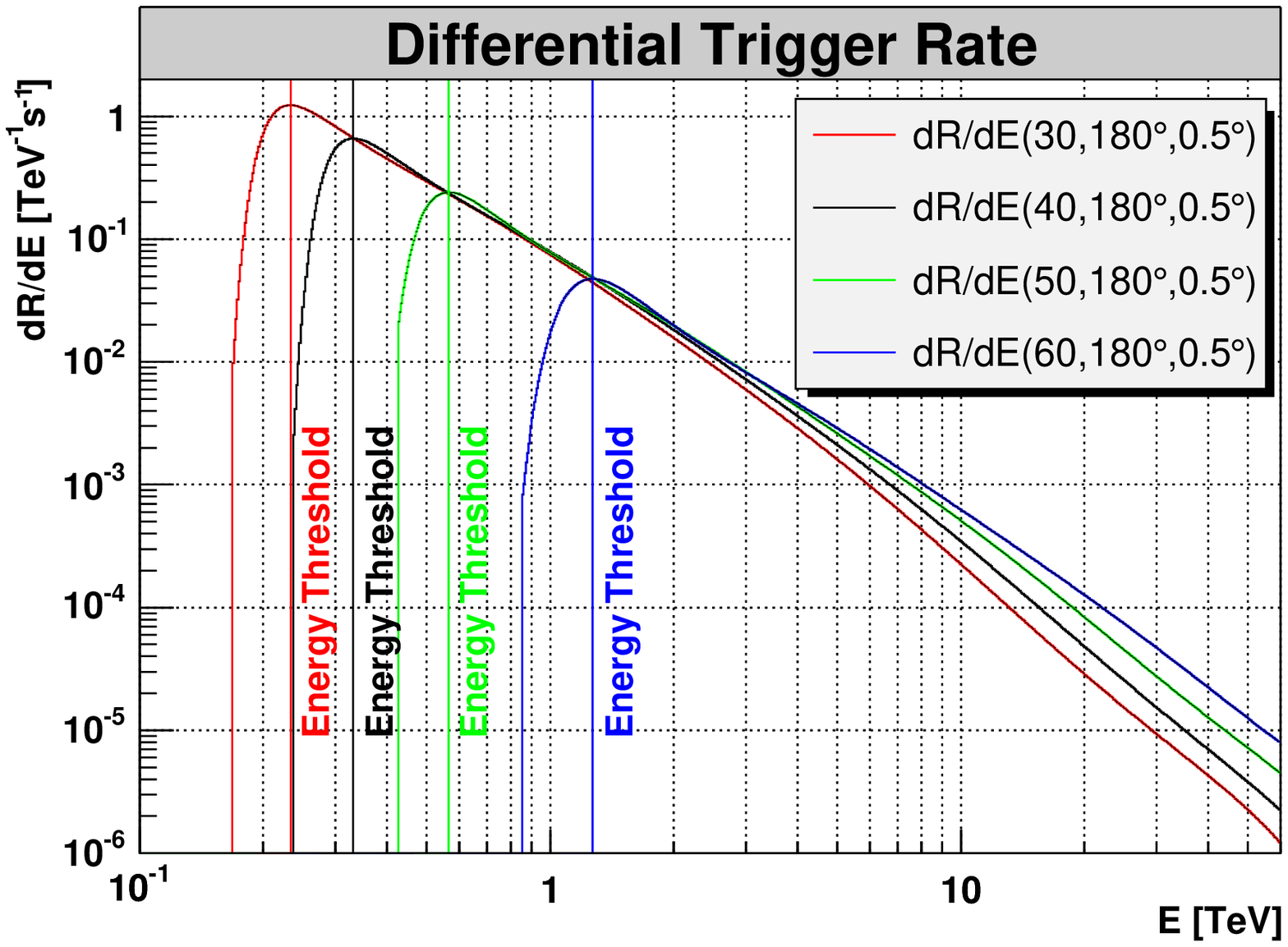}
    \caption[Energy Threshold for Spectroscopy]{An alternative definition of
    the energy threshold as the peak value of the differential event rate
    $dR/dE$ past cuts. For comparision, the energy threshold is indicated for
    the same zenith angles as in Fig.~\ref{fig:EnergyBias}. The safe threshold
    is slightly higher.}
    \label{fig:EnergyThreshold}
  \end{minipage}
\end{figure}

The accuracy of spectral reconstruction is determined by the energy
reconstruction. The mean error of the reconstructed and the true energy
$(\Delta_E=(E_{\rm reco}-E_{\rm true})/E_{\rm true})$ is shown in
Fig.~\ref{fig:EnergyBias} as determined from simulations. A bias of $\Delta_E$
towards both ends of the H.E.S.S. energy range can be seen. The reason for this
is that it is a bias in the event selection which favors events within the
H.E.S.S. energy range and suppresses events at the ends. To avoid the resulting
systematic errors in $A$ and the energy spectrum, $A$ is calculated as a
function of reconstructed instead of the true Monte Carlo energy. A
disadvantage is that $A$ becomes dependent on the energy spectrum of the
simulations. If this dependence was strong, each of the sources would require
their own $A$ which could be determined in an iterative approximation of the
simulated to the true spectrum, as described in \citet{HEGRA:1999}. Fortunately
this is not required in most cases. Another problem is the steepening of $A$
near the energy threshold, which makes this energy range very sensitive to
systematic errors. To avoid this problem, the energy range for spectroscopy is
restricted to the range with an energy bias of less than 10\%, which defines
the safe energy threshold. Fig.~\ref{fig:EnergyBias} shows the safe energy
threshold for different zenith angles. The safe energy threshold is a bit
higher than the maximum of the differential event rate $dR/dE$ past cuts, which
is commonly defined as the energy threshold (\citet{Hofmann1999}). Using
Eqn.~\ref{eqn:PowerLaw2}, the energy threshold is given by
\begin{equation}
  \frac{dR}{dE}=A\frac{d\Phi}{dE}=A\phi_{\rm 1TeV}\left(\frac{E}{\rm
    1TeV}\right)^{-\Gamma}.
\end{equation}
Fig.~\ref{fig:EnergyThreshold} shows this alternative definition of the energy
threshold to the safe energy threshold. It provides similar results but it is
not used in the standard analysis.

\begin{figure}[b!]
  \begin{minipage}[c]{0.5\linewidth}
    \includegraphics[width=.9\textwidth]{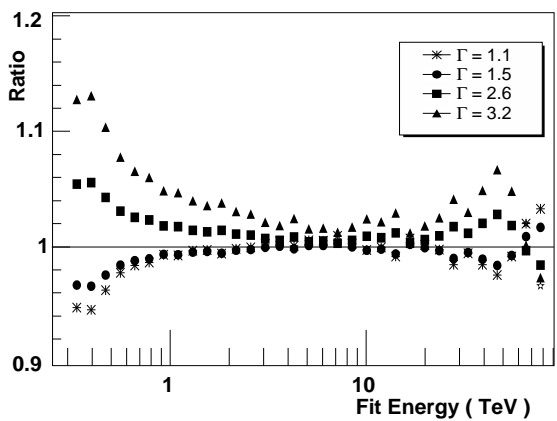}
  \end{minipage}\hfill
  \begin{minipage}[c]{0.5\linewidth}
    \caption[Ratio of Reconstructed to True Effective Area]{The ratio of the
      reconstructed to the true effective area per energy bin, for photon
      indices from 1.1 to 3.2, based on Monte Carlo simulations at a zenith
      angle of 45$^\circ$ and with a photon index of 2.0. (Figure taken from
      \citet{CrabPaper}.)}
  \label{fig:SpectralAccuracy}
  \end{minipage}
\end{figure}

The systematic error due to differences in the spectrum between data and
simulations was studied in detail in \citet{CrabPaper} and \citet{BergeThesis}.
It has been verified that spectra with a photon index ($\Gamma$) ranging from
1.7 to 2.9 for sources of different strengths can precisely be determined with
Monte Carlo simulations with a $\Gamma$ of 2.0. Fig.~\ref{fig:SpectralAccuracy}
shows the ratio of the reconstructed to the true $A$ for energy spectra with
$\Gamma$ ranging from 1.1 to 3.2. Near the safe energy threshold at 440 GeV,
the differential flux for a source with a $\Gamma$ of 2.6 (Crab Nebula) is
overestimated by 5\%, while the differential flux for a $\Gamma$ of 1.5 is
underestimated by 4\%. For energies well above the threshold, the energy bias
is less than 5\% for a wide range of photon indices. Thus the $A$ of the
reconstructed energy are well suited for spectroscopy of most galactic and many
extragalactic sources.

\subsection{Energy Spectrum of the Crab Nebula}
\begin{figure}[ht]
  \begin{minipage}[c]{0.55\linewidth}
  \includegraphics[width=.99\textwidth]{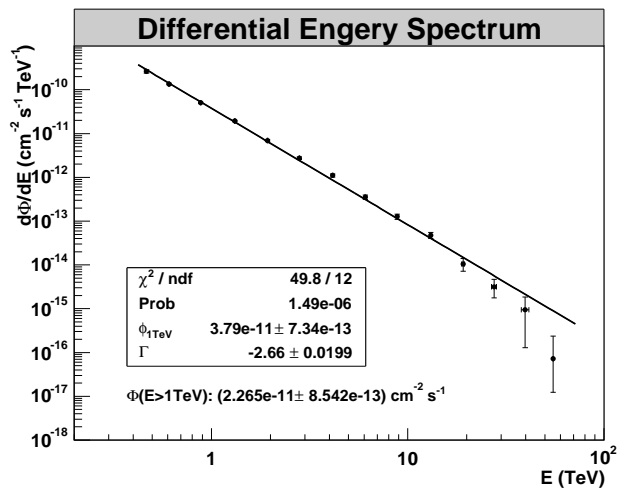}
  \end{minipage}\hfill
  \begin{minipage}[c]{0.45\linewidth}
    \caption[Energy Spectrum of the Crab Nebula]{Energy spectrum of the Crab
    Nebula, fit by a power law. The spectrum is determined using the {\it
    extended} cuts, the region-background model and the effective area $A_{\rm
    ext}$. The exponential cutoff is apparent at higher energies.}
  \label{fig:CrabSpectrum}
  \end{minipage}
\end{figure}

To verify the precision of the spectral reconstruction, the energy spectrum of
the Crab Nebula data was analyzed. The results were obtained using the {\it
extended} cuts, the region-background model and the effective area $A_{\rm
ext}$. The differential flux histogram with six bins per decade and a fit range
from 0.413\,GeV to 73.4\,TeV is shown in Fig~\ref{fig:CrabSpectrum}. The fit to
a simple power law (Eqn.~\ref{eqn:PowerLaw2}) is also shown. A good agreement
with the results from other experiments is found, which are summarized in
Tbl.~\ref{tbl:CrabSpectrum}. Deviations from a power law are apparent at higher
energies, as expected for an exponential cutoff in the Crab Nebula spectrum, as
observed by \citet{CrabPaper}, \citet{HEGRA:2004Crab} and \citet{CrabWhipple}.

\begin{table}[h!]
  \centering
  \caption[Energy Spectra of the Crab Nebula as Determined by Different
    Experiments]{Energy spectra of the Crab Nebula (including statistical
    errors only), determined from a fit to a power law by different
    experiments: $^1$~this work, $^2$~\citet{HEGRA:2004Crab},
    $^3$~\citet{CrabWhipple} and $^4$~\citet{CrabCAT}.}
  \bigskip
  \begin{tabular}{lccc}
    \hline \hline
    Experiment & $\phi_{\rm 1TeV}$ & $\Gamma$ &
    $\Phi(E>$1TeV) \\
    & \scriptsize $[\rm 10^{-11}\rm cm^{-2}s^{-1}TeV^{-1}]$ & & 
    \scriptsize $[10^{-11}\rm cm^{-2}s^{-1}]$ \\
    \hline
    H.E.S.S.$^1$ & 2.84$\pm$0.05 & 2.61$\pm$0.02 & 1.76$\pm$0.04 \\
    HEGRA$^2$    & 2.83$\pm$0.04 & 2.62$\pm$0.02 & 1.74$\pm$0.04 \\
    Whipple$^3$  & 3.20$\pm$0.17 & 2.49$\pm$0.06 & 2.1 $\pm$0.02 \\
    CAT$^4$      & 2.20$\pm$0.05 & 2.80$\pm$0.03 & 1.22$\pm$0.03 \\
    \hline \hline
  \end{tabular}
  \label{tbl:CrabSpectrum}
\end{table}

\chapter{Imaging with H.E.S.S.} \label{chp:Deconvolution}
With the H.E.S.S. data, the first spatially-resolved TeV \g-ray sky maps of
many \g-ray sources became available. These maps provided a new picture of many
TeV \g-ray sources and therefore obtained a lot attention (cf.
\citet{Hess1713Nature, PlaneScanI}). The H.E.S.S. \g-ray maps provide images at
the highest, previously not attainable energy range, and consequently
information for the theoretical understanding of \g-ray sources. An example is
the sky map of RX J1713.7-3946 and its relations to the origin of cosmic
radiation (\citet{Hess1713Nature}). This chapter discusses imaging with
H.E.S.S., its limitations and the application of smoothing and image
deconvolution.

\section{Limitations of H.E.S.S. \g-Ray Maps} \label{sec:Limitations}
H.E.S.S. \g-ray maps were introduced in the previous chapter. While it is
desirable to obtain these maps with high resolution, the later is limited by
the angular resolution of the shower reconstruction and the event statistics.

\subsection{Angular Resolution}
H.E.S.S.' angular resolution is limited by the accuracy of shower
reconstruction, which determines the point spread function (PSF).
Mathematically the count map $(I)$ is the result of the convolution of the true
image $(O)$ with the PSF ($PSF$), i.e.
\begin{eqnarray}
  I(x,y) &=& \int_{x_1=-\infty}^{\infty} \int_{y_1=-\infty}^{\infty}
  PSF(x-x_1,y-y_1)O(x_1,y_1)dx_1,dy_1 \\
  &=& (P * O)(x,y),
  \label{eqn:Convolution}
\end{eqnarray}
where $x$ and $y$ are the coordinates of the two-dimensional image and $*$ is a
short hand notation for the convolution operator (\citet{Starck:2002}).

The convolution of a H.E.S.S. sky map can be studied using simulations.
Fig.~\ref{fig:DecoMCEmission} shows a \g-ray map $(O)$ which represents a true
\g-ray signal and a constant background. The signal is modeled with a
two-dimensional Gaussian distribution given by Eqn.~\ref{eqn:2DGaussian} with
the standard deviations $\sigma_w=0.04^\circ$ and $\sigma_l=0.11^\circ$ in the
longitudinal $(\lambda)$ and latitudinal $(\beta)$ directions. The signal
region $(\theta<0.3^\circ)$ contains 4000 \g\ and 14000 background events. The
peak intensity is 19.3\,counts, where 14.3 counts can be attributed to the
signal and the rest to the uniform background of 5.0 counts/bin. These numbers
where chosen according to Tbl.~\ref{tbl:StatisticsMSH} to provide a realistic
simulation. The map has a bin size of $0.01^\circ\times0.01^\circ$.

Fig.~\ref{fig:DecoMCConvolution} shows the map $I$ as a result of the
convolution of the map $O$ with the PSF. The PSF was modeled with the double
Gaussian parameterization no. 3 of Tbl.~\ref{tbl:PSF}. The excess in $I$ is
significantly broader after convolution. This is also reflected by the peak
value. After background subtraction, the signal peak intensity is only $\sim$6
counts, which is less than half the peak intensity of the true excess in
$O$. A profile of the PSF is indicated in the bottom right corner.

\subsection{Event Statistic} 
Another problem of the H.E.S.S. \g-ray maps is their limited event statistic.
The total number of events is a product of observation time, sensitivity and
strength of the source. Sky maps of the strongest sources consist of at most a
few thousand events. Therefore they are affected by Poisson noise in counting
statistics. The level of noise in each bin is determined by the number of
counts. Fig.~\ref{fig:DecoMC80_0} shows a simulation of a count map, which was
obtained from Fig.~\ref{fig:DecoMCConvolution} by replacing each bin content by
an integer value given by the Poisson statistic. This count map shows the
combined effects of convolution and statistical noise. Its difference from the
true map $O$ is obvious. The intensity profile of the PSF is indicated in the
bottom right corner.

\section{Image Smoothing} \label{sec:Smoothing}

\begin{figure}[bh!]
  \setlength{\abovecaptionskip}{-.05cm} \setlength{\belowcaptionskip}{.5cm}
  \begin{minipage}[t]{0.5\linewidth}
    \includegraphics[width=\textwidth]{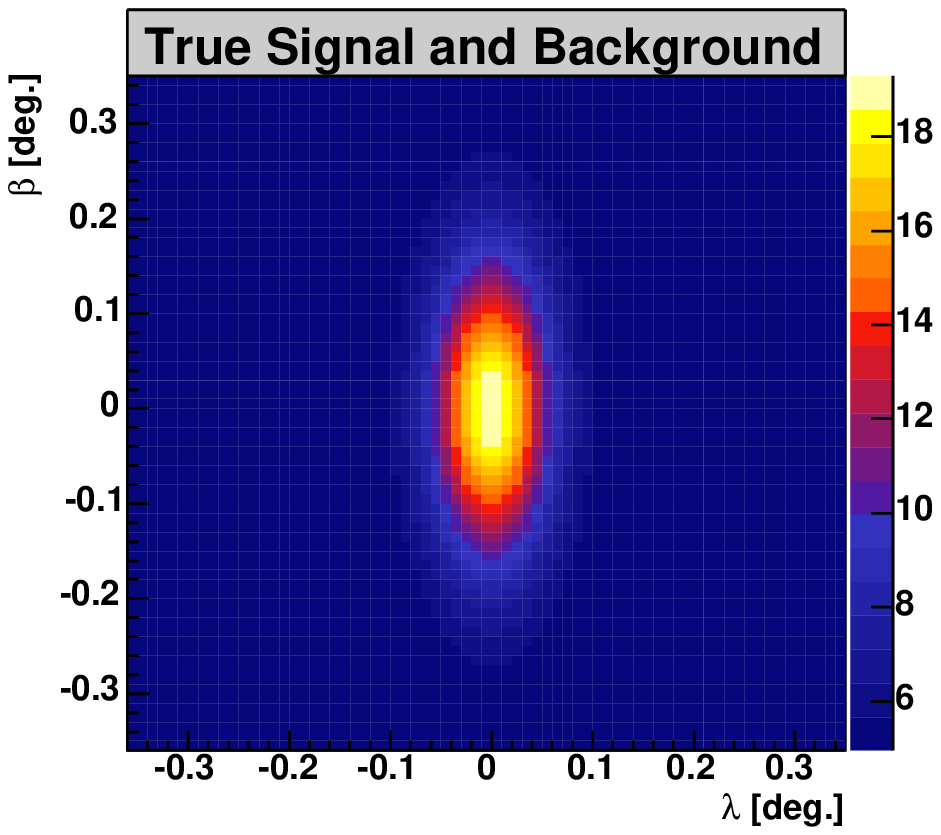}
    \caption[Simulated Map $(O)$ of the True Emission]{Simulated map $(O)$ of
      the true emission with a Gaussian signal and constant background of 5
      counts/bin.}
    \label{fig:DecoMCEmission}
  \end{minipage} \hfill
  \begin{minipage}[t]{0.5\linewidth}
    \includegraphics[width=\textwidth]{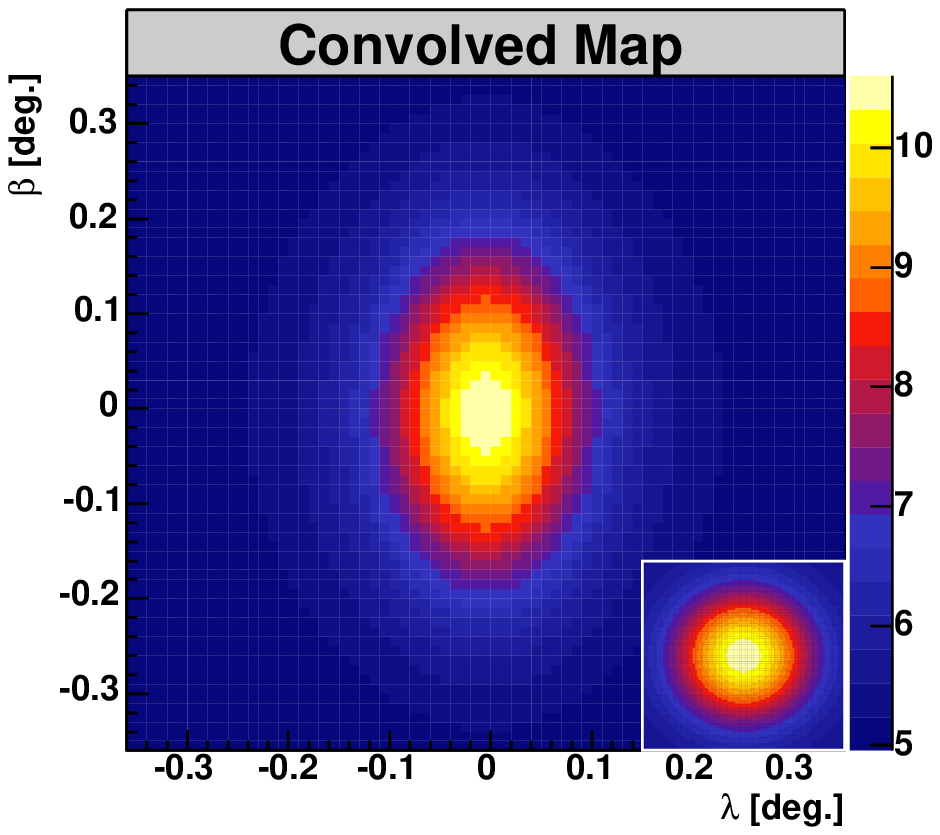}
    \caption[Map of True Emission $(O)$ Convolved with the PSF]{Map of
      Fig.~\ref{fig:DecoMCEmission} convolved with the PSF. A profile of the
      PSF is shown at the bottom right.}
    \label{fig:DecoMCConvolution}
  \end{minipage} \hfill
  \begin{minipage}[t]{0.5\linewidth}
    \includegraphics[width=\textwidth]{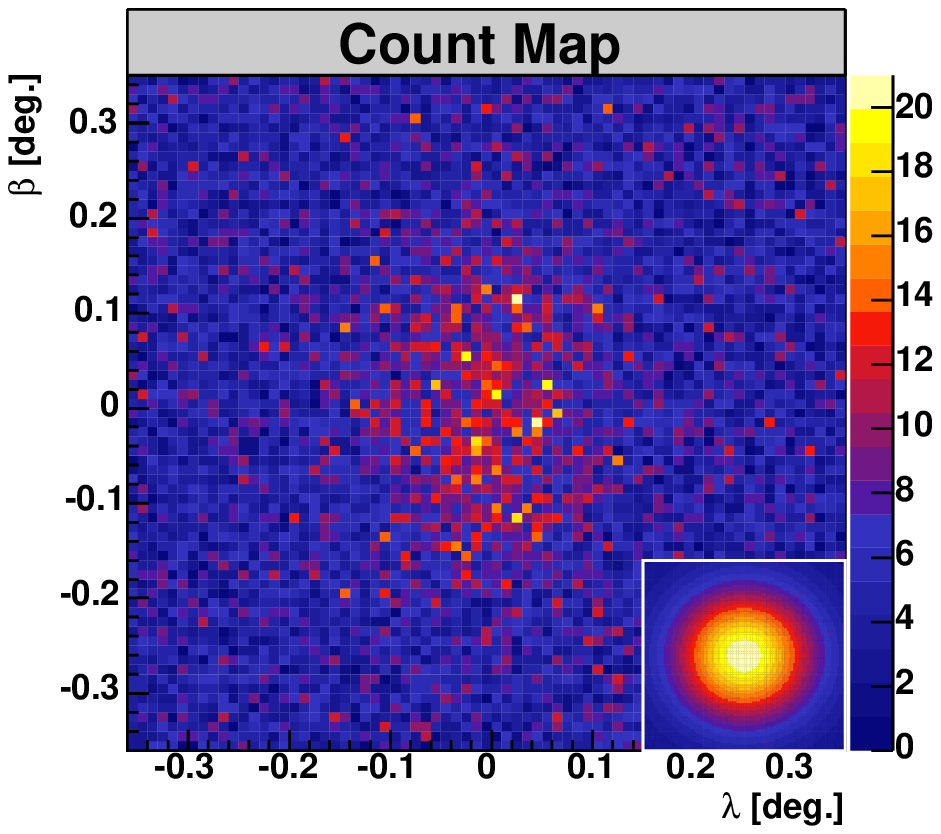}
    \caption[Simulated Count Map $(I)$ Including Poisson Noise]{Simulated count
      map $(I)$ including Poisson noise corresponding to
      Fig.~\ref{fig:DecoMCConvolution}. The PSF is shown at the bottom right.}
    \label{fig:DecoMC80_0}
  \end{minipage} \hfill
  \begin{minipage}[t]{0.5\linewidth}
    \includegraphics[width=\textwidth]{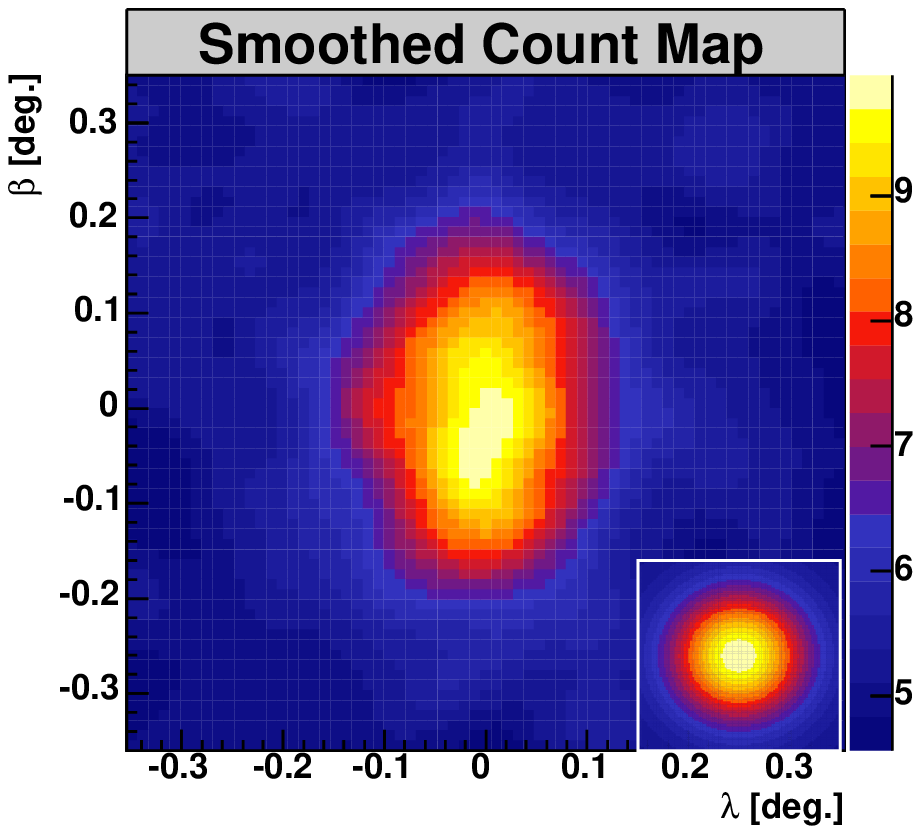}
    \caption[Smoothed Simulated Count Map]{Count map of
      Fig.~\ref{fig:DecoMC80_0} smoothed with a Gaussian function
      $(\sigma_s=0.03^\circ)$. The PSF is shown at the bottom right.}
    \label{fig:DecoMCSmooth}
  \end{minipage} \hfill
\end{figure}

To produce clearer images Poisson noise is often reduced by smoothing.
Smoothing is an effective method for reducing the statistical noise through
weighted averaging between neighboring bins. Here the weights are chosen
according to a Gaussian smoothing function, which is a common choice. If the
smoothed map is $I$, the true map $O$ and the smoothing function $PSF$, then
smoothing is also expressed by Eqn.~\ref{eqn:Convolution}. This equation
illustrates that smoothing is a convolution. In this work, the integral of
$PSF$ is normalized to one so that the total number of counts is preserved. A
smoothed map appears less noisy and more homogeneous, but also has a lower
resolution and fewer details because a convolution is implied.

Fig.~\ref{fig:DecoMCSmooth} shows the simulated count map of
Fig.~\ref{fig:DecoMC80_0} smoothed with a Gaussian function where
$\sigma_s=0.03^\circ$. The excess appears much more clearly and has a more
regular contour. It can therefore be identified more easily. A fit of the
excess by the Gaussian function $G$ (Eqn.~\ref{eqn:ConvolutionGauss2D}) yields
the standard deviations of $\sigma_w=0.09^\circ$ and
$\sigma_l=0.14^\circ$, which is more than twice the initial value of
the true emission in $O$.
\clearpage

\section{Image Deconvolution}
An alternative method for reducing the noise in sky maps is image
deconvolution. It is a technique which does not imply a convolution, and is in
fact defined as the operation which inverses the convolution. Hence,
deconvolution appears to be a promising technique for the restoration of images
which are affected by noise and a limited resolution. Mathematically, image
deconvolution is defined as solving Eqn.~\ref{eqn:Convolution} for $O$. The
relation between the count map $I$, the PSF $(PSF)$ and the noise term $N$ is
\begin{eqnarray}
  I(x,y) &=& \int_{x_1=-\infty}^{\infty} \int_{y_1=-\infty}^{\infty}
  PSF(x-x_1,y-y_1)O(x_1,y_1)dx_1,dy_1 + N(x,y) \nonumber \\
  &=& (P * O)(x,y)+N(x,y)
  \label{eqn:Deconvolution}
\end{eqnarray}
(cf. \citet{Starck:2002}). Unfortunately it is not possible to determine the
exact solution of $O$ in the presence of noise and with a limited knowledge of
the PSF. However, even the approximate solutions can often provide very
valuable results.

\subsection{The Richardson-Lucy Algorithm}
To find an approximate solution for $O$ in Eqn.~\ref{eqn:Deconvolution},
several methods have been developed. For Poisson noise, as in the H.E.S.S.
count maps, the Richardson-Lucy (RL) algorithm is used and has become a
well-established approach. It is a numerical, iterative method for finding an
approximate solution for $O$ and was first proposed by \citet{Richardson:1972}
and \citet{Lucy:1974}. The RL algorithm is defined through the iteration
\begin{equation}
  O^{i+1}(x,y) = \left[\frac{I(x,y)}{(PSF*O^i)(x,y)}*PSF^T(x,y)\right]O^i(x,y),
  \label{eqn:Richardson-Lucy}
\end{equation}
where $i$ denotes the $i$th step of iterations and $PSF^T$ the transpose of
$PSF$ (\citet{Starck:2002}). A detailed discussion of the RL algorithm is found
in the book by \citet{Bertero:InverseProbs}.

The RL algorithm has some important properties which make it very useful. One
of these properties is the conservation of the total image intensity, which
guarantees that the flux at any step in the iterations is conserved. Another
property guarantees a positive intensity in any region of the image and any
step of iteration. Therefore unphysical results with negative intensities are
excluded. As common as it is for iterative deconvolution techniques, the RL
algorithm also has the property of semi-convergence. Semi-convergence means
that there exist an optimal number of iterations $i_{\rm opt}$ for which the
difference between the restored images $O^{i_{\rm opt}}$ and the true image $O$
reaches a minimum. $i_{\rm opt}$ depends on individual imaging situations.
Therefore, no simple stopping rule is known to decide when the optimal number
of iterations is reached. However $i_{\rm opt}$ can often be determined from
simulations.

\subsection{Application to H.E.S.S. Data}
To investigate the use of the RL algorithm for the H.E.S.S. data, the RL
algorithm was applied to H.E.S.S. count maps. The deconvolution of the count
maps was done numerically with GDL, the ``GNU Data Language'' (\citet{GDL}),
which is a free interpreter of IDL, the ``Interactive Data Language''
(\citet{IDL}). The RL algorithm was provided by the {\tt Max\_Likelihood}
procedure from the ``IDL Astronomy Library'' (\citet{IDLAstroLib}) shown in
App.~\ref{app:Richardson-Lucy}. This procedure requires convolutions which
require Fourier transforms at each step of iteration. A Fourier transform is
efficiently realized with the ``Fast Fourier Transform'' algorithm. It can be
applied if the size of the data set is given by powers of two. Therefore, 128
bins have been chosen for both dimensions of the count maps. The bin size is
$0.01^\circ\times0.01^\circ$ and the scale represents \g-ray counts. The
necessary PSF parameters were taken from the double Gaussian parameterization
in Tbl.~\ref{tbl:PSF}.

\subsubsection{Deconvolution of the Sky Map from the Crab Nebula}
\begin{figure}
  \begin{minipage}[c]{0.5\linewidth}
    \includegraphics[width=\textwidth]{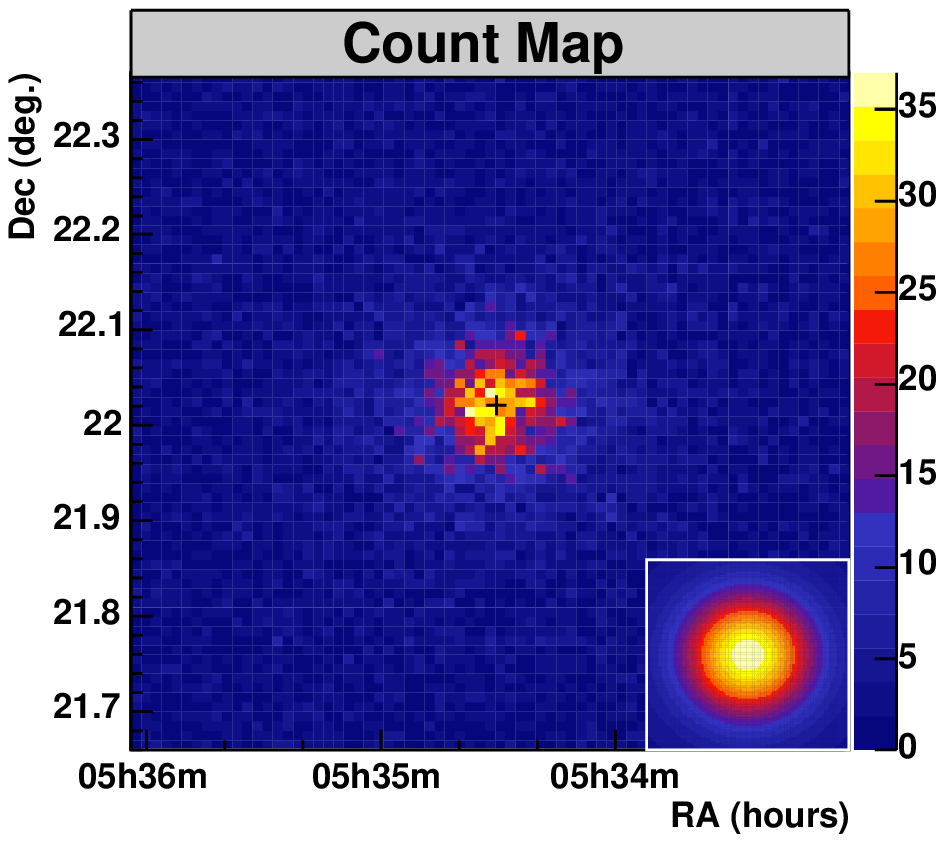}
  \end{minipage} \hfill
  \begin{minipage}[c]{0.5\linewidth}
    \includegraphics[width=\textwidth]{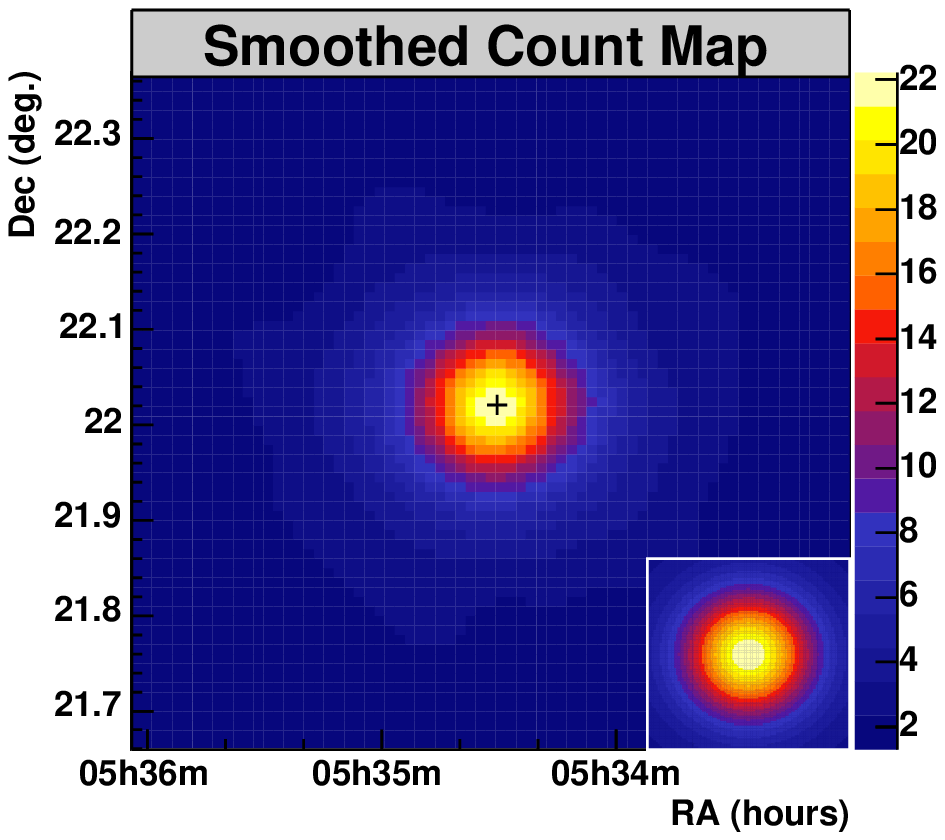}
  \end{minipage} \hfill
  \begin{minipage}[c]{0.5\linewidth}
    \includegraphics[width=\textwidth]{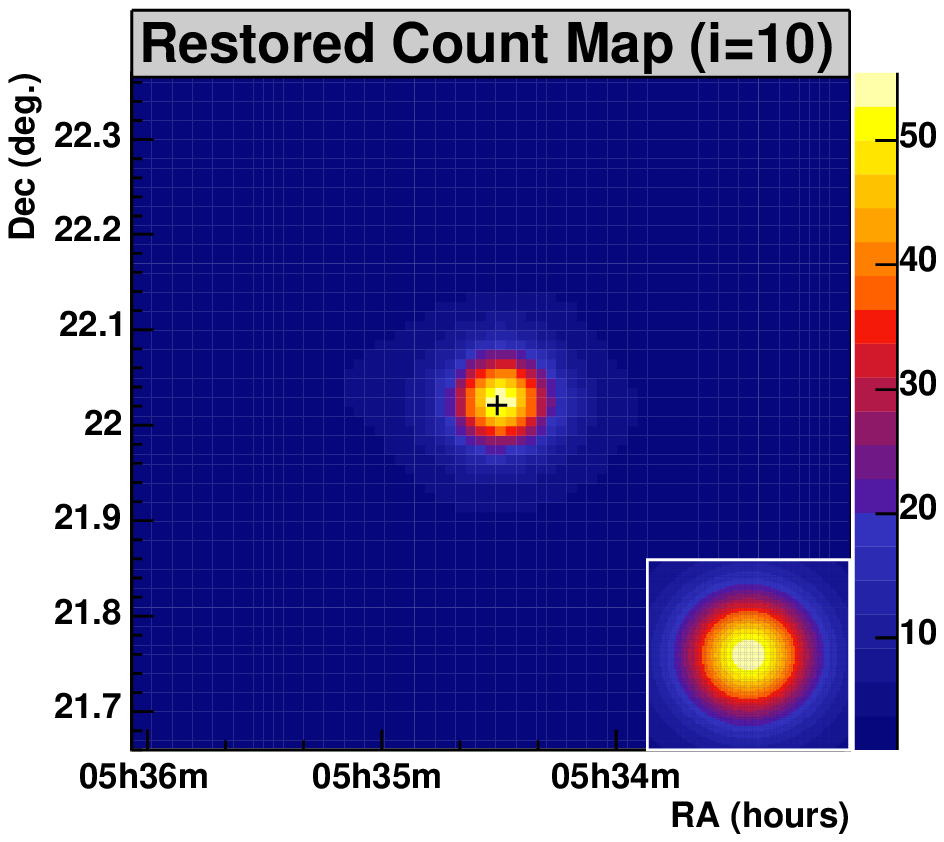}
  \end{minipage} \hfill
  \begin{minipage}[c]{0.5\linewidth}
    \includegraphics[width=\textwidth]{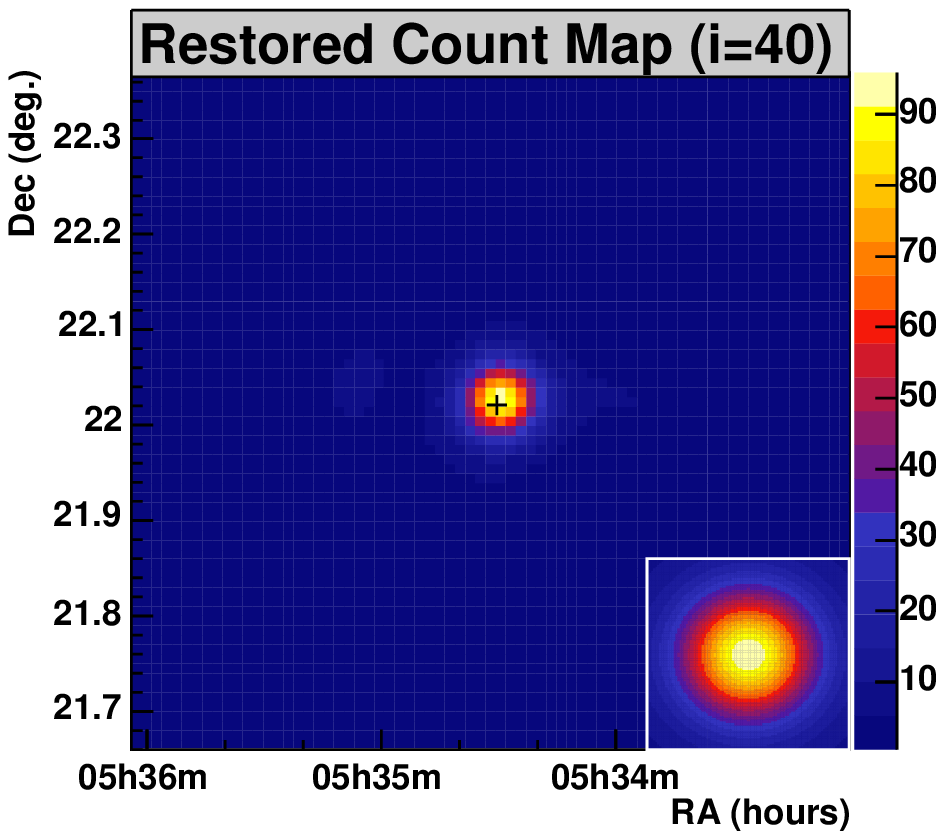}
  \end{minipage} \hfill
  \begin{minipage}[c]{0.5\linewidth}
    \includegraphics[width=\textwidth]{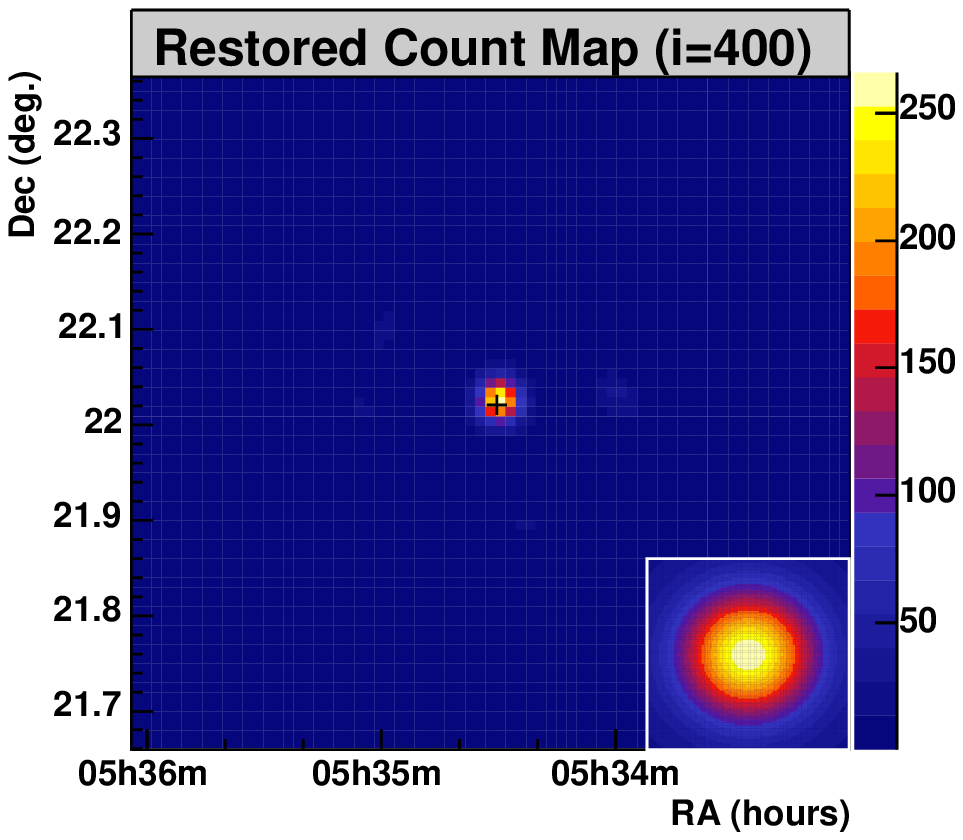}
  \end{minipage} \hfill
  \begin{minipage}[c]{0.45\linewidth}
    \caption[Restored Count Map of the Crab Nebula]{Count map of the Crab
      Nebula (point source) restored with the Richardson-Lucy algorithm for
      different numbers of iterations $(i)$. With increasing $i$, the width of
      the excess reduces to only fraction of its initial value. The original
      count map and smoothed count map are shown for comparison at the upper
      left and upper right. The centroid of the excess of the count map is
      indicated as a reference position by the black cross. A profile of the
      PSF is indicated at the bottom right.}
    \label{fig:DecoCrab}
  \end{minipage} \hfill
\end{figure}

As a first example, the RL algorithm was applied to the count map of the Crab
Nebula data. The analysis of this data was discussed in
Chp.~\ref{chp:StandardAnalysis}. The count map is shown in
Fig.~\ref{fig:DecoCrab} (upper left). It consists of $\sim$7000 excess events
(Tbl.~\ref{tbl:CrabAnalysis}) and represents a strong point source. The three
other count maps show the restorations of the RL algorithm after 10, 40 and 400
iterations. The statistical noise of the restored images is clearly reduced.
The width of the emission region reduces with the number of iterations to a
small fraction of its initial width. The count map smoothed by convolution with
a Gaussian function $(\sigma_s=0.03^\circ)$ is shown in the upper right. It
also shows a clearly reduced noise but with an increased width. The centroid of
the excess from Tbl.~\ref{tbl:CrabPositionSize} is indicated by the black cross
in all maps as a reference position. It coincides with the centroid of the
restored signal. The systematic pointing error of $20''$ corresponds to half
the bin width. The H.E.S.S. PSF is indicated in all plots in the bottom right
corner. The scale represents \g-ray counts. The amplification of noise is
negligible and statistical artefacts are not observed. The example of the Crab
Nebula demonstrates the efficiency of the RL algorithm to smooth statistical
fluctuations and to mitigate the influence of the PSF in H.E.S.S. count maps of
strong point sources.

\subsubsection{Deconvolution of the Sky Map from \MSH}
\begin{figure}
  \setlength{\abovecaptionskip}{-.12cm}
    \includegraphics[width=0.5\textwidth]{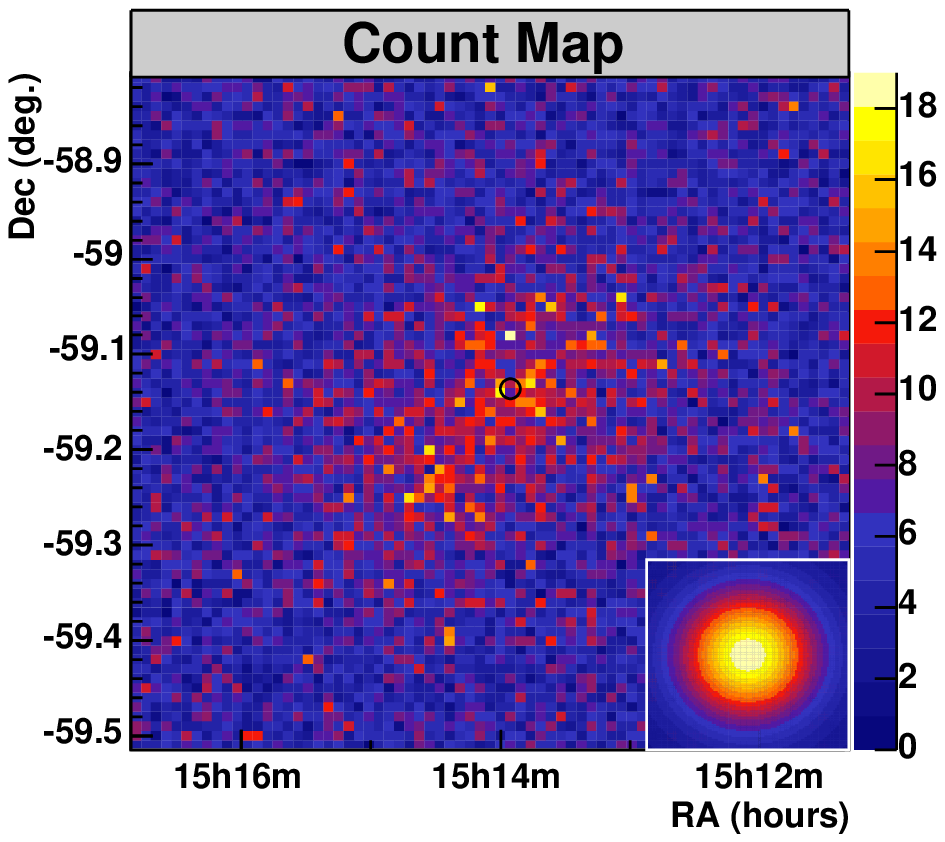}
    \includegraphics[width=0.5\textwidth]{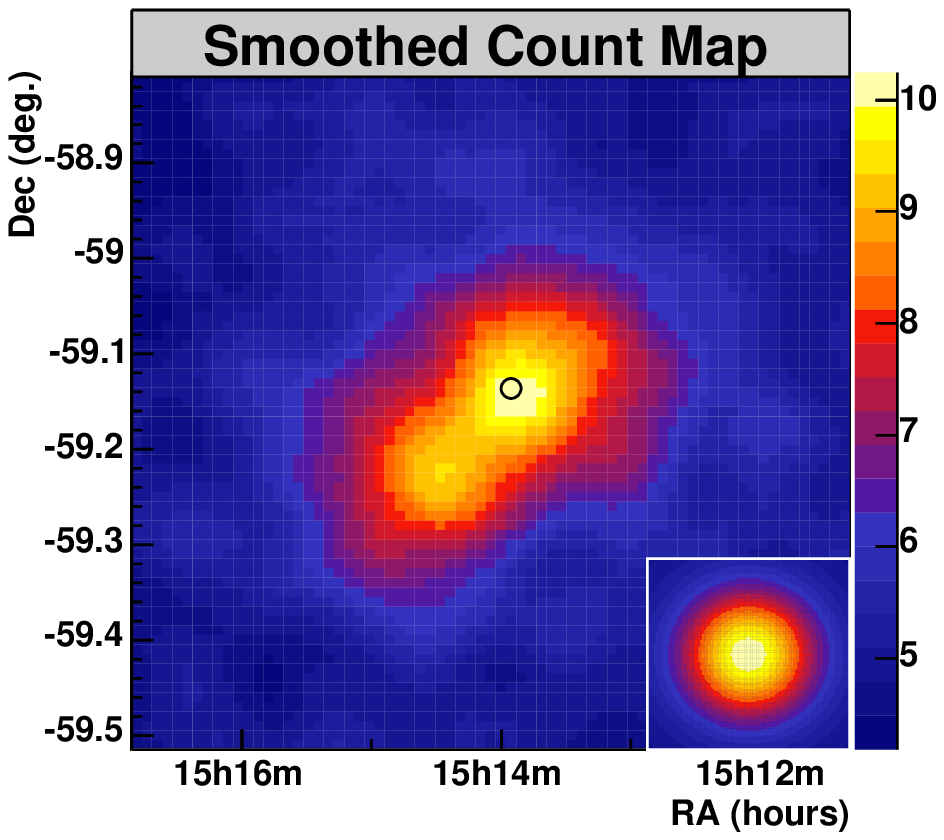}
    \includegraphics[width=0.5\textwidth]{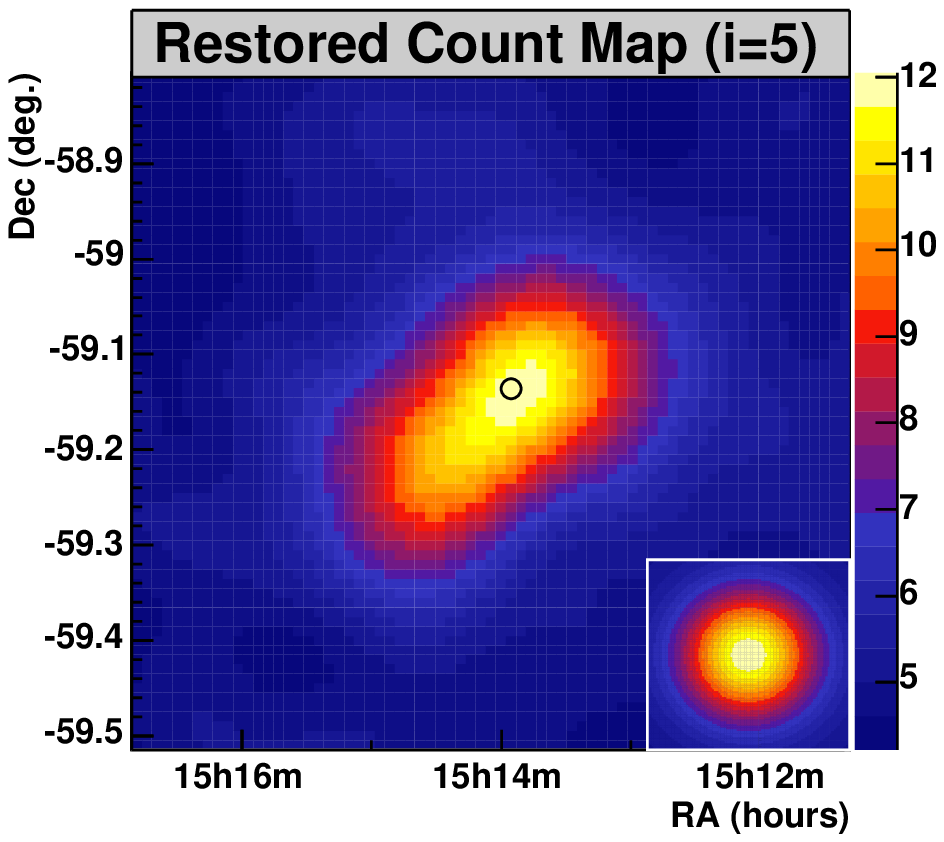}
    \includegraphics[width=0.5\textwidth]{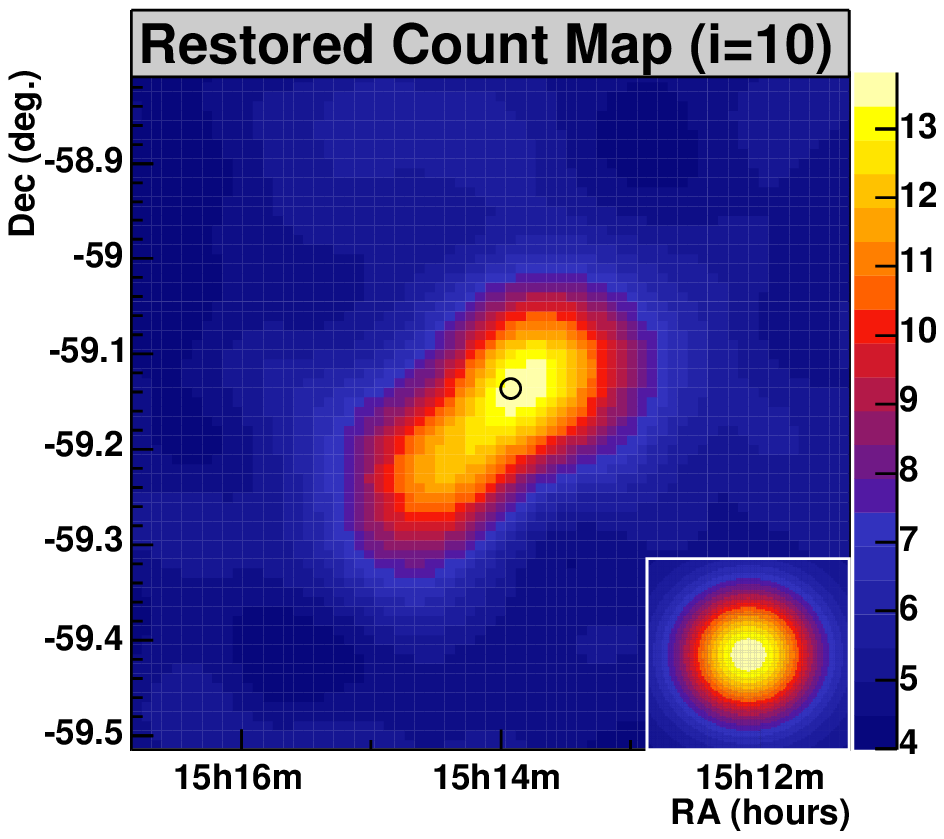}
    \includegraphics[width=0.5\textwidth]{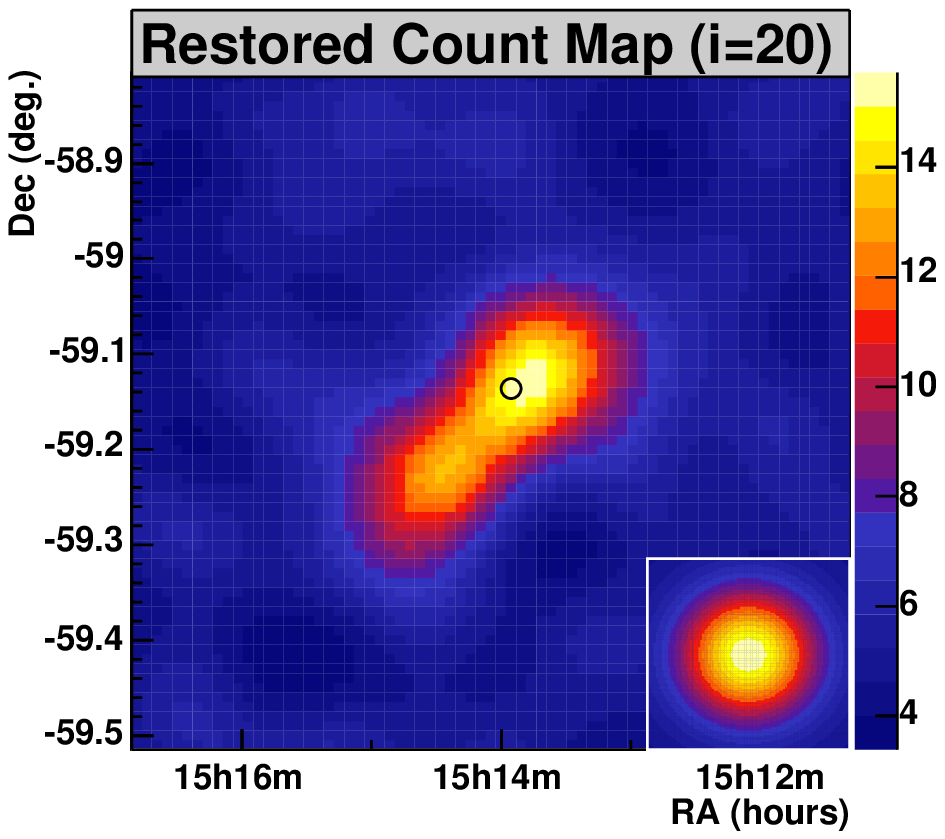}
    \includegraphics[width=0.5\textwidth]{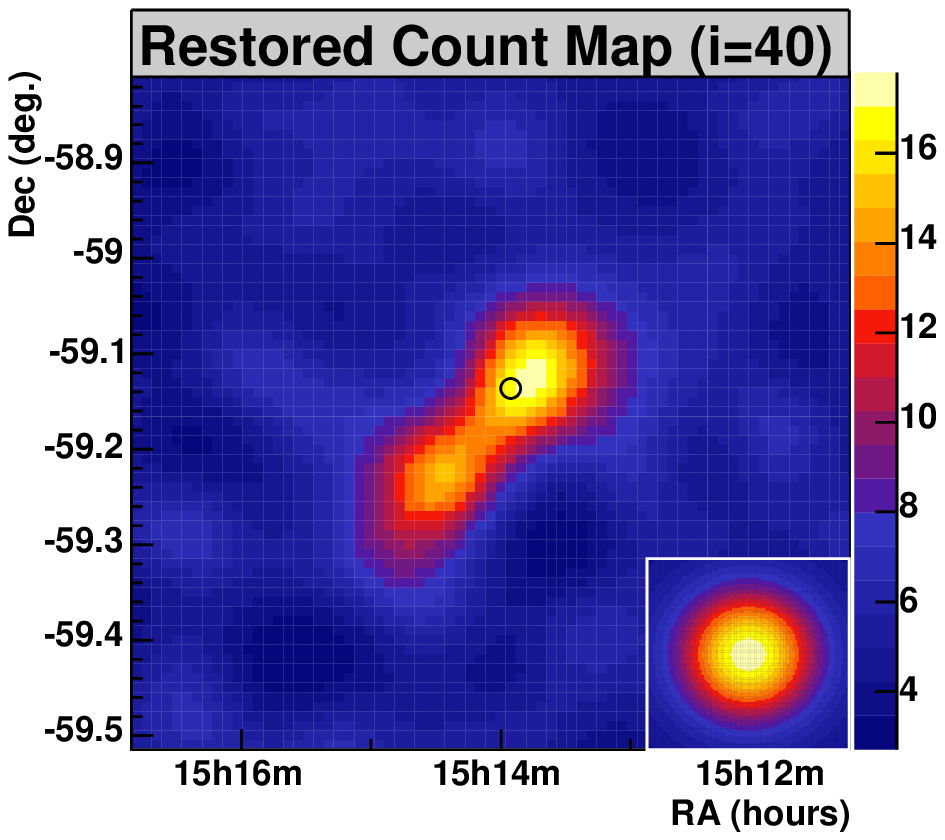}
  \caption[Restored Count Map of \MSH\ ($IA>$80\,p.e.)]{Restored count map of
    \MSH\ ($IA>$80\,p.e.) with the RL algorithm for different numbers of
    iterations $(i)$. The position of \PSR\ (black circle) and the PSF are
    indicated.}
  \label{fig:DecoMSH80}
\end{figure}

\begin{figure}
  \setlength{\abovecaptionskip}{-.12cm}
  \begin{minipage}[c]{0.5\linewidth}
    \includegraphics[width=\textwidth]{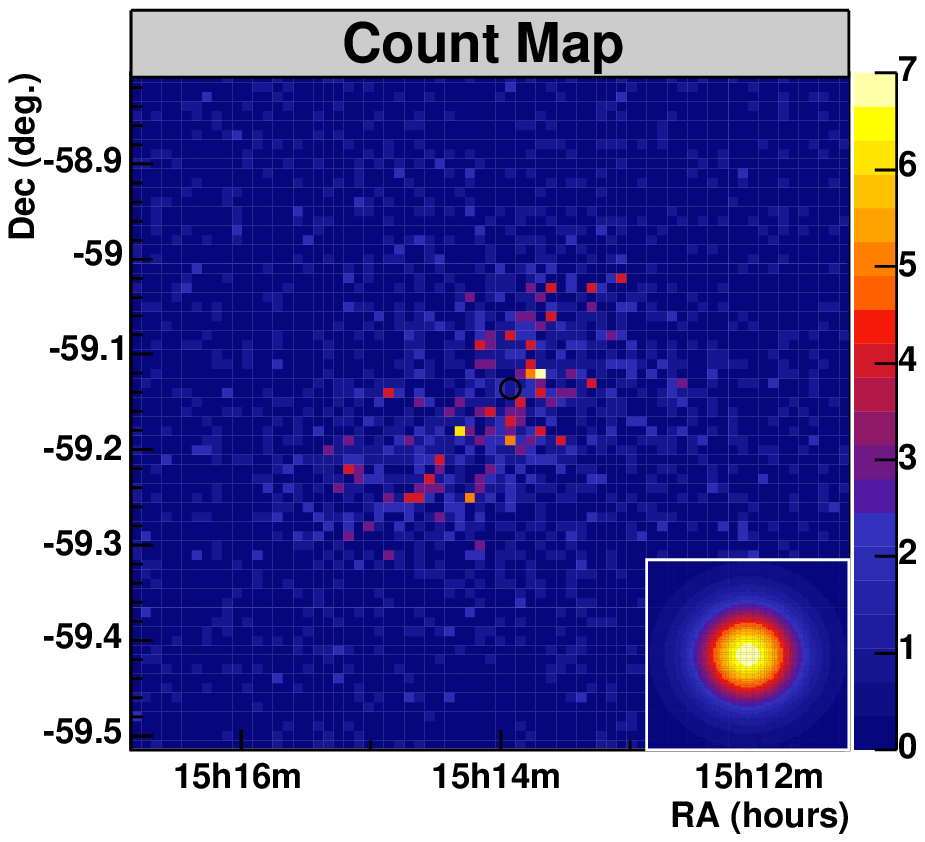}
  \end{minipage}\hfill
  \begin{minipage}[c]{0.5\linewidth}
    \includegraphics[width=\textwidth]{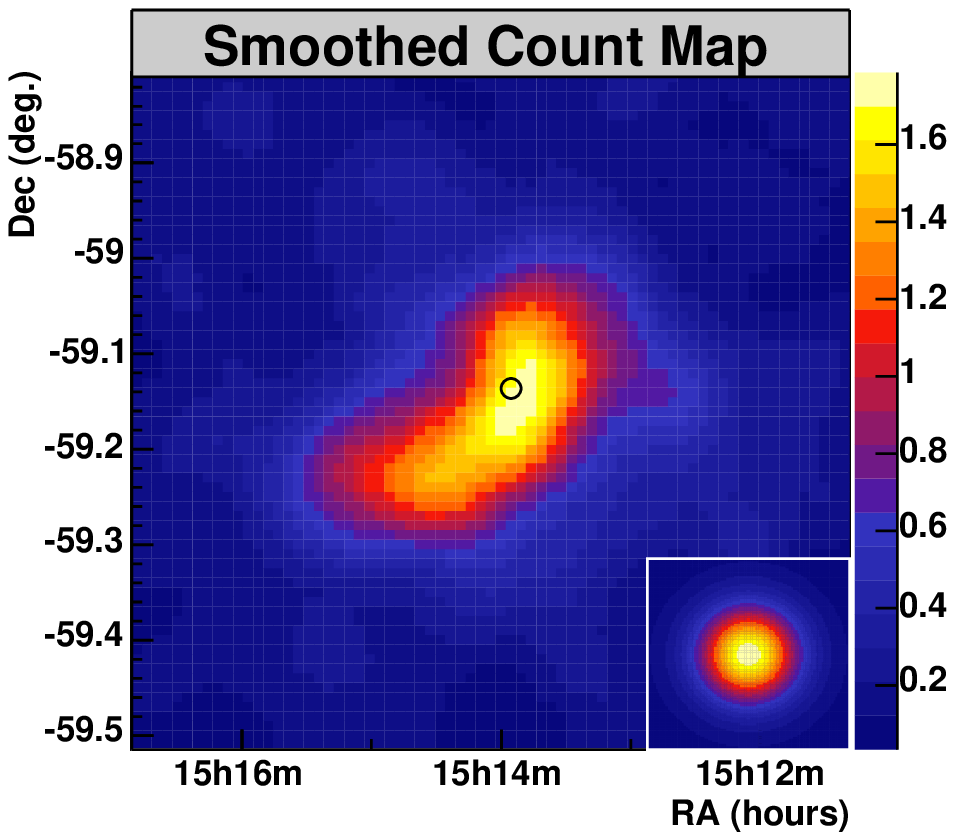}
  \end{minipage}\hfill
  \begin{minipage}[c]{0.5\linewidth}
    \includegraphics[width=\textwidth]{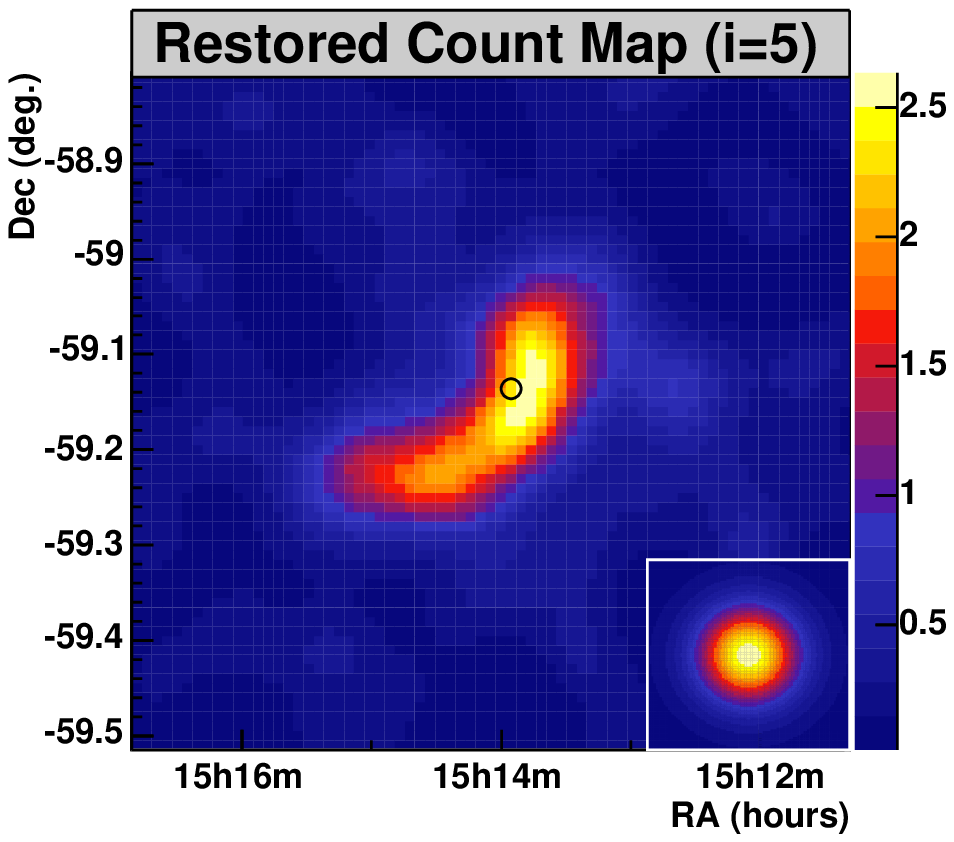}
  \end{minipage}\hfill
  \begin{minipage}[c]{0.5\linewidth}
    \includegraphics[width=\textwidth]{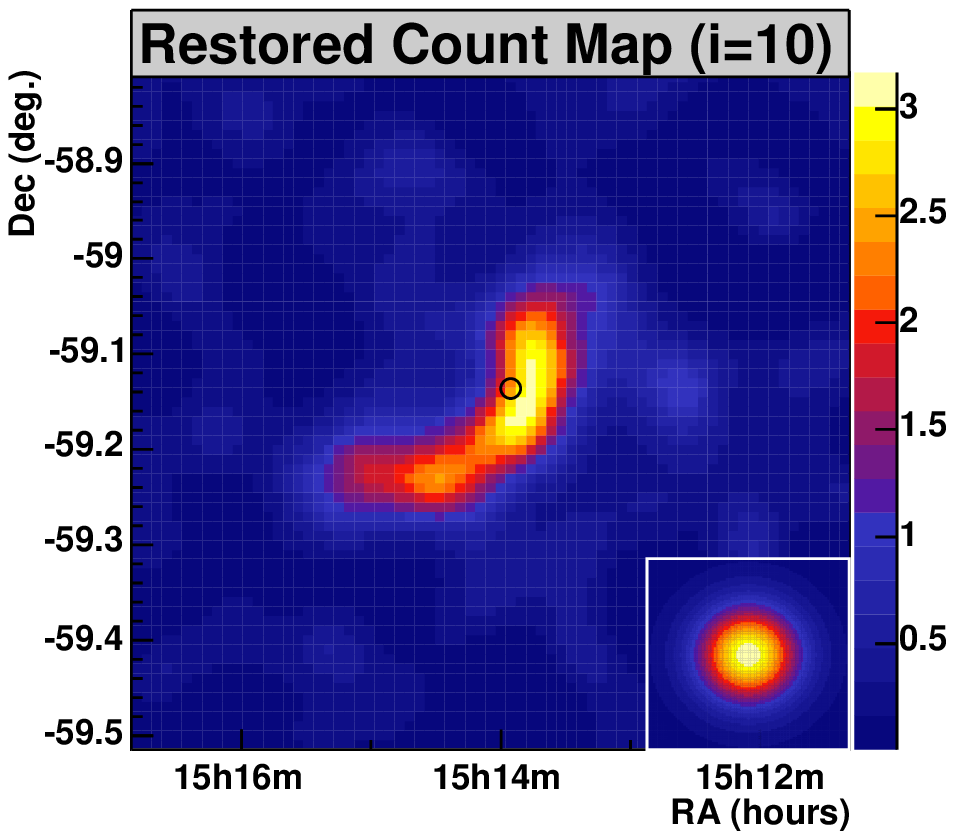}
  \end{minipage}\hfill
  \begin{minipage}[c]{0.5\linewidth}
    \includegraphics[width=\textwidth]{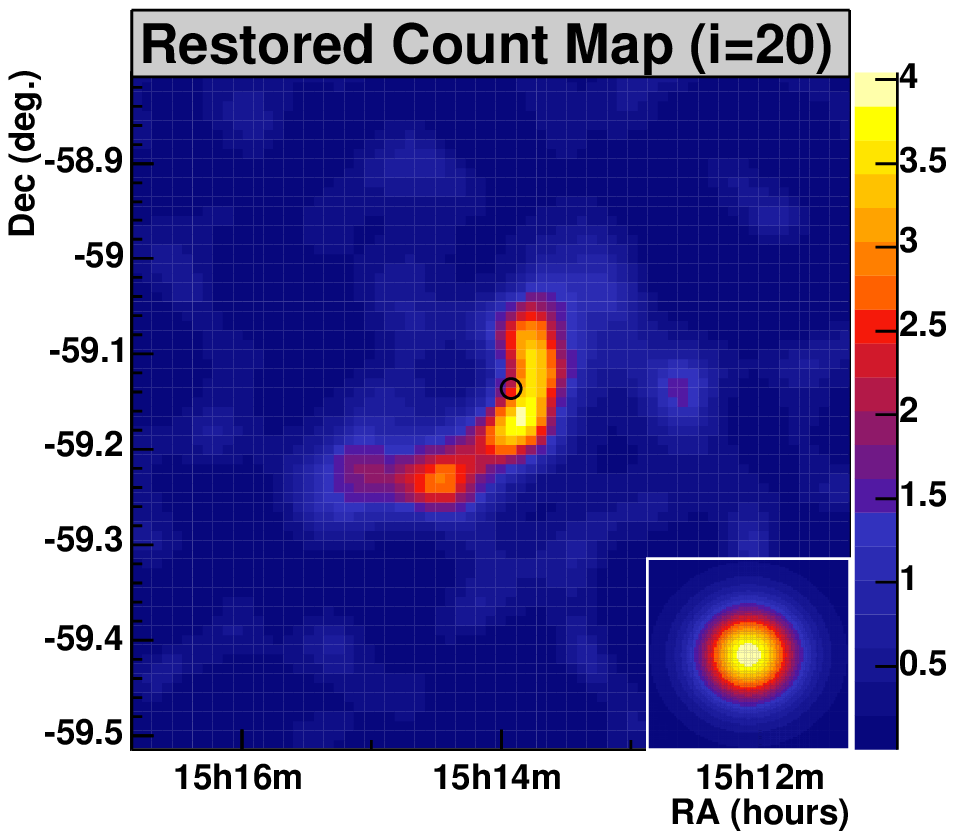}
  \end{minipage}\hfill
  \begin{minipage}[c]{0.5\linewidth}
    \includegraphics[width=\textwidth]{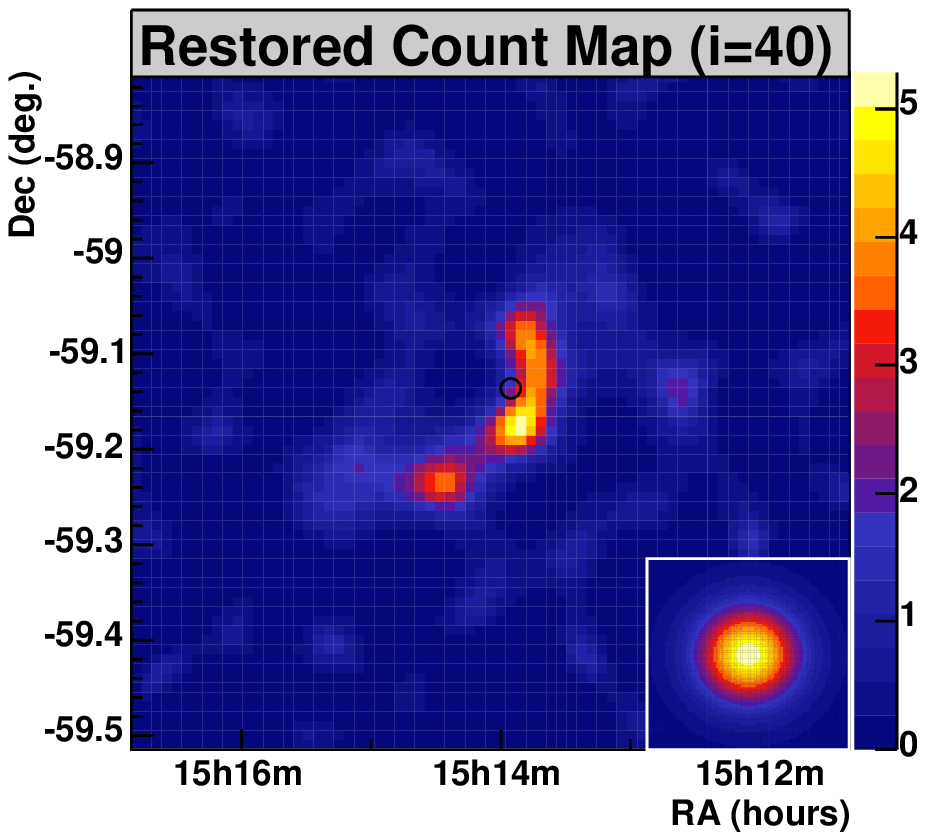}
  \end{minipage}\hfill
  \begin{minipage}[c]{\linewidth}
  \caption[Restored Count Map of \MSH\ ($IA>$400\,p.e.)]{Restored count map of
    \MSH\ ($IA>$400\,p.e.) with the RL algorithm for different numbers of
    iterations $(i)$. The position of \PSR\ (black circle) and the PSF are
    indicated.}
    \label{fig:DecoMSH400}
  \end{minipage}
\end{figure}

As a second example the RL algorithm was applied to two count maps of data from
\MSH. The analysis of this data is described in Chp.~\ref{chp:MSHAnalysis}. The
first count map is obtained with the standard image amplitude ($IA$) cut of
80\,p.e. and the second with a $IA$ cut of 400\,p.e. The corresponding energy
thresholds are 280\,GeV and 900\,GeV, respectively. The statistics of the two
maps are given in Tbl.~\ref{tbl:StatisticsMSH}. The restoration with the RL
algorithm is shown in Fig.~\ref{fig:DecoMSH80} and Fig.~\ref{fig:DecoMSH400}
for 5, 10, 20 and 40 iterations. Again, the statistical noise is significantly
reduced. The width of the emission region reduces with the number iterations,
but the background fluctuations increase. For comparison, the count map and the
smoothed map after convolution with a Gaussian function $(\sigma_s=0.03^\circ)$
are also shown in the upper right. The H.E.S.S. PSF is indicated in all plots
in the bottom right corner. The scale represents \g-ray counts. The position of
\PSR\ is shown by the black circle. The systematic pointing error of
$\pm0.005^\circ$ is comparable to the bin width. Again the examples of the
\MSH\ count map demonstrate the efficiency of the RL algorithm to smooth
statistical fluctuations and to reduce the width of an image when applied to
H.E.S.S. count maps of strong, extended sources. However, the figures with many
iterations, especially Fig.~\ref{fig:DecoMSH400}, show a distinctive appearance
which could be questioned for its physical reality. This question will be
addressed in the error analysis of the next section.

\vspace{-4.5cm}

\clearpage

\subsection{Error Analysis}
The error of the restoration was determined from forward-folding of
simulations. First the true map $O$ was simulated to represent emission and
background in the source region. This map was then convolved with an
appropriate model of the PSF. Then noise was added by scattering each bin
content according to Poisson statistics in order to obtain the simulated count
map $I$. These steps are the same as described in Sec.~\ref{sec:Limitations}.
The simulated count map was then restored using the RL algorithm to generate a
restored map. To distinguish between the true signal and the statistical
artifacts amplified in the restoration, the deconvolution was repeated 20 times
for each simulated true image $O$ with different random numbers for the noise.

The true image $(O)$ was then adjusted to minimize the difference between the
restored simulated count maps and the restored data. The event statistics were
provided by the data analysis and taken from Tbl.~\ref{tbl:StatisticsMSH}. In
the case of the 80\,p.e. map, the shape of $O$ was already defined by the
mathematical model of the fit function. In the case of the 400\,p.e. map, $O$
was iteratively adjusted until it resembled the restored data sufficiently
well. The bin size of the simulated maps and the maps of the data was the same.

The comparison of the 20 different restored maps with each other and with the
true map $O$ provided an estimate of the error of the restoration. The
following quantities are useful for the error analysis.

\subsubsection{The Restoration Error}
The restoration error $E^i(x,y)$ of the pixel $(x,y)$ after $i$ iterations is
given by the difference between the true and the restored image as
$E^i(x,y)=O(x,y)-O^i(x,y)$. Since here $N=20$ different simulations with
different random numbers of noise are compared, $O^i(x,y)$ was replaced by
the arithmetic mean of the simulations
$\overline{O}^i(x,y)=1/N\sum_n^NO^i(x,y)$ to provide the mean restoration error
$\overline{E}^i(x,y)$. $\overline{E}^i(x,y)$ is then given as
\begin{equation}
  \overline{E}^i(x,y)=O(x,y)-\overline{O}^i(x,y). \label{eqn:Ei}
\end{equation}
$\overline{E}^i(x,y)$ expresses the mean difference between the true and the restored
map.

\subsubsection{The Standard Deviation of the Restoration}
Similar to the restoration error, the standard deviation of the restoration for
each pixel $S^i(x,y)$ can be defined. If $N$ is the number of simulations with
different random numbers, then
\begin{equation}
  S^i(x,y)=\left(\sum_n^N\frac{[O_n^i(x,y)-\overline{O}^i(x,y)]^2}{N+1}\right)
  ^{\frac12}.
  \label{eqn:Si}
\end{equation}
$S^i(x,y)$ represents the noise of each pixel of the restored map.

\subsubsection{Distribution of Noise of the Restoration}
The noise in the restoration is determined by the deviation from the mean
restoration value. The deviation of each pixel $D^i(x,y)$ from the mean
restoration value is given as
\begin{equation}
  D^i(x,y)=O^i(x,y)-\overline{O}^i(x,y). \label{eqn:Di}
\end{equation}
The distribution of $D^i$ for all pixels $(x,y)$ is a measure of the overall
noise distribution of the restoration at step $i$. In contrast to $S^i(x,y)$,
the deviation does not refer to a particular pixel or region but to the
deviations within the whole map. The distribution is a measure of the accuracy
of the restoration.

\subsubsection{The Relative Norm of the Restoration Error}
According to \cite[pg. 110]{Bertero:InverseProbs}, the relative norm of the
restoration error $(\epsilon_i)$ is defined as
\begin{equation}
  \epsilon_i=\frac{\|O^i(x,y)-O(x,y)\|}{\|O(x,y)\|},
  \label{eqn:RestorationError}
\end{equation}
where $||O(x,y)||$ is the Euclidean norm of $O(x,y)$ defined as
\begin{equation}
  \|O(x,y)\| = \left( \sum_{k,l} |O(x_k,y_l)|^2 \right)^\frac{1}{2}.
\end{equation}
Hence $\epsilon_i$ is a positive number which expresses the difference between
the true $(O)$ and the restored image $(O^i)$ after $i$ iterations. In the case
of a perfect restoration, $\epsilon_i=0$. Due to the semi-convergence of the RL
algorithm, the sequence of $\epsilon_i$ has a minimum. It defines the optimal
number of iterations $(i_{\rm opt})$ which provides the smallest relative
restoration error $(\epsilon_{\rm opt})$. In principle, $\epsilon_i$ is an
ideal measure for the agreement between the true image and the restoration.
However, for images with compact signal regions and non-negligible background,
$\epsilon_i$ varies with the size of the image. The reason for this is that a
variation of the image size will only change the size of the background area,
and the deconvolution of background will only amplify noise and hence increase
the restoration error. Therefore, other measures such as those discussed above
are also necessary to assess the restoration.

Since $N=20$ simulations with different random numbers of noise are
considered here, $\overline{\epsilon}_i$ is used as the arithmetic mean of the
individual $\epsilon_i$.

\subsubsection{Errors of the 80\,p.e. Model}
The true map $O$ of the 80\,p.e. map is modeled with a Gaussian distribution
($\sigma_w=0.04^\circ$, $\sigma_l=0.11^\circ$) according to the fit of the
corresponding excess map (Tbl.~\ref{tbl:MSHPositionSize}). The statistics in
the signal region with a radius $\theta$ of 0.3$^\circ$ is given by 4000 excess
and 14000 background events (Tbl.~\ref{tbl:StatisticsMSH}), yielding a peak
excess of 14.3. The simulation is the same as the one discussed in reference to
Fig.~\ref{fig:DecoMCEmission} and was repeated 20 times with different random
numbers of statistical noise. The PSF was chosen according to parameterization
no. 3 of Tbl.~\ref{tbl:PSF}. Since the performance of the RL algorithm is
invariant under rotations of an image, the orientation of the excess
distribution was not adjusted in the simulations. The width and length of the
excess are oriented along the $\beta-$ and $\lambda-$axis, respectively.

\begin{figure}
  \begin{minipage}[c]{0.5\linewidth}
    \includegraphics[width=\textwidth]{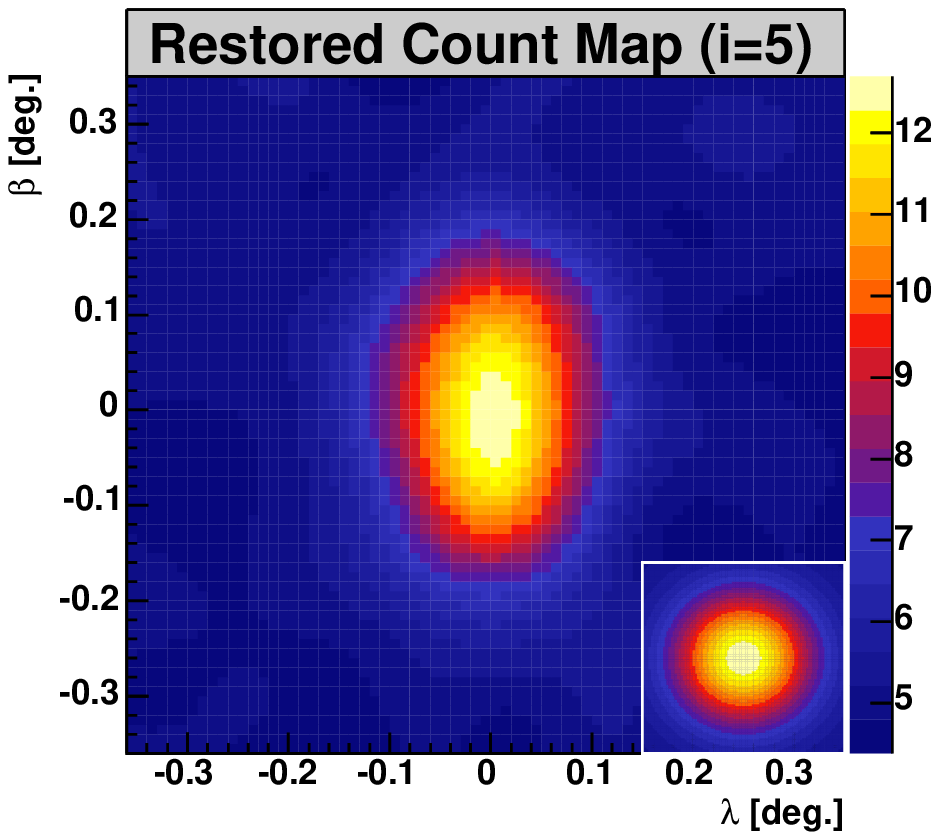}
  \end{minipage} \hfill
  \begin{minipage}[c]{0.5\linewidth}
    \includegraphics[width=\textwidth]{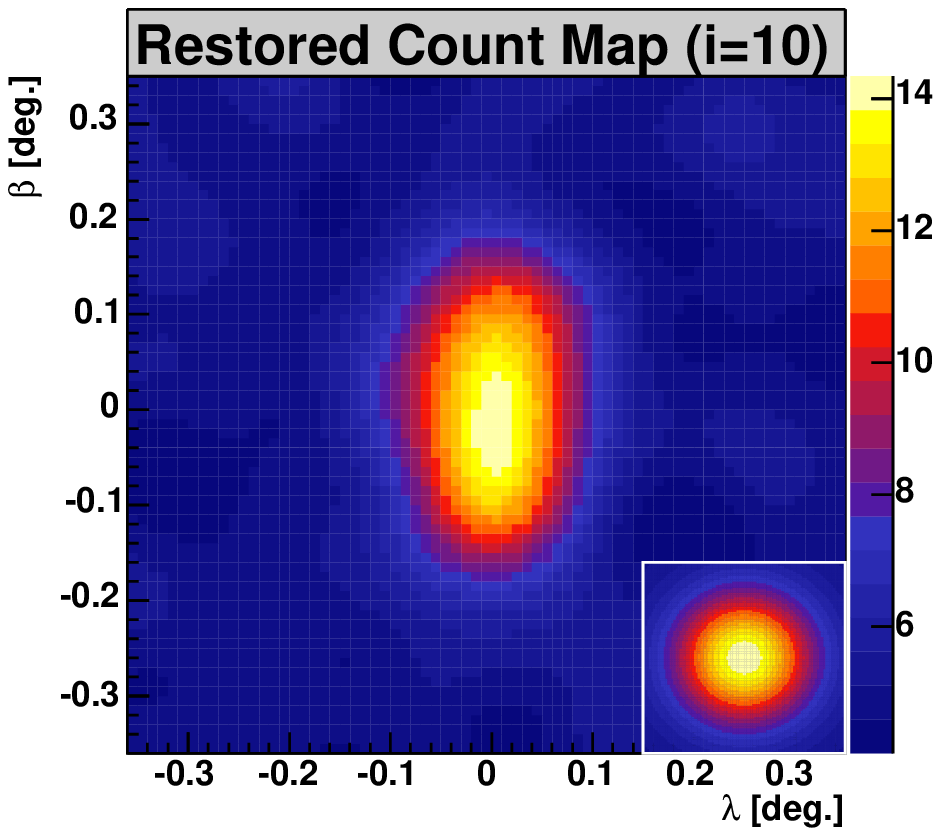}
  \end{minipage} \hfill
  \begin{minipage}[c]{0.5\linewidth}
    \includegraphics[width=\textwidth]{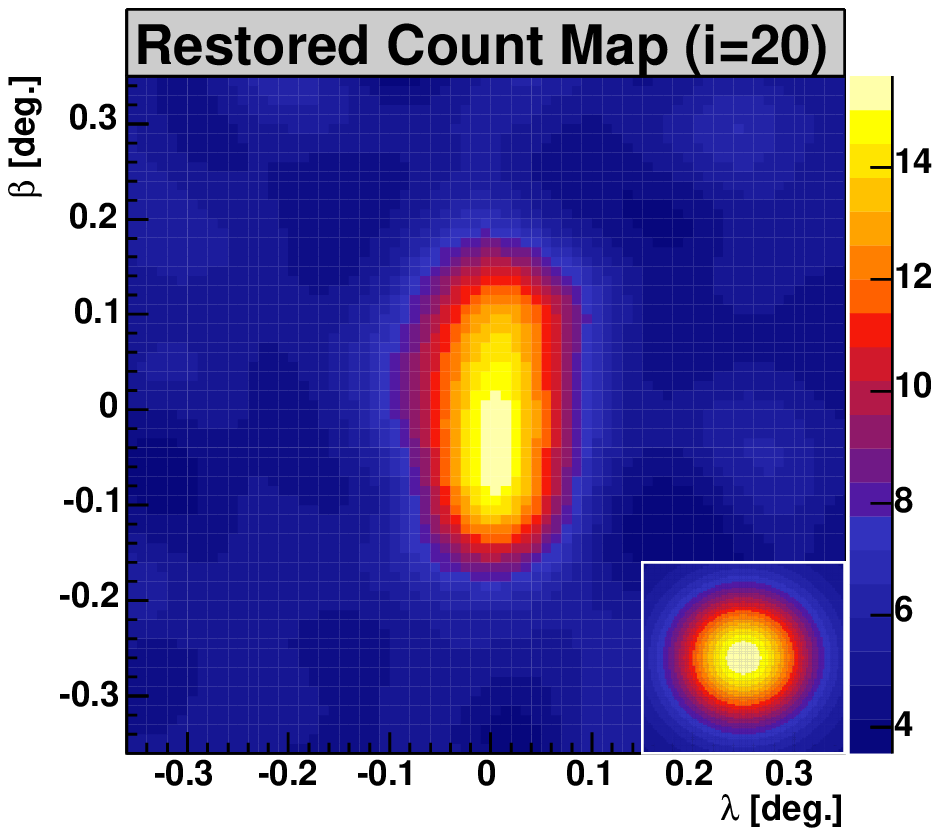}
  \end{minipage} \hfill
  \begin{minipage}[c]{0.5\linewidth}
    \includegraphics[width=\textwidth]{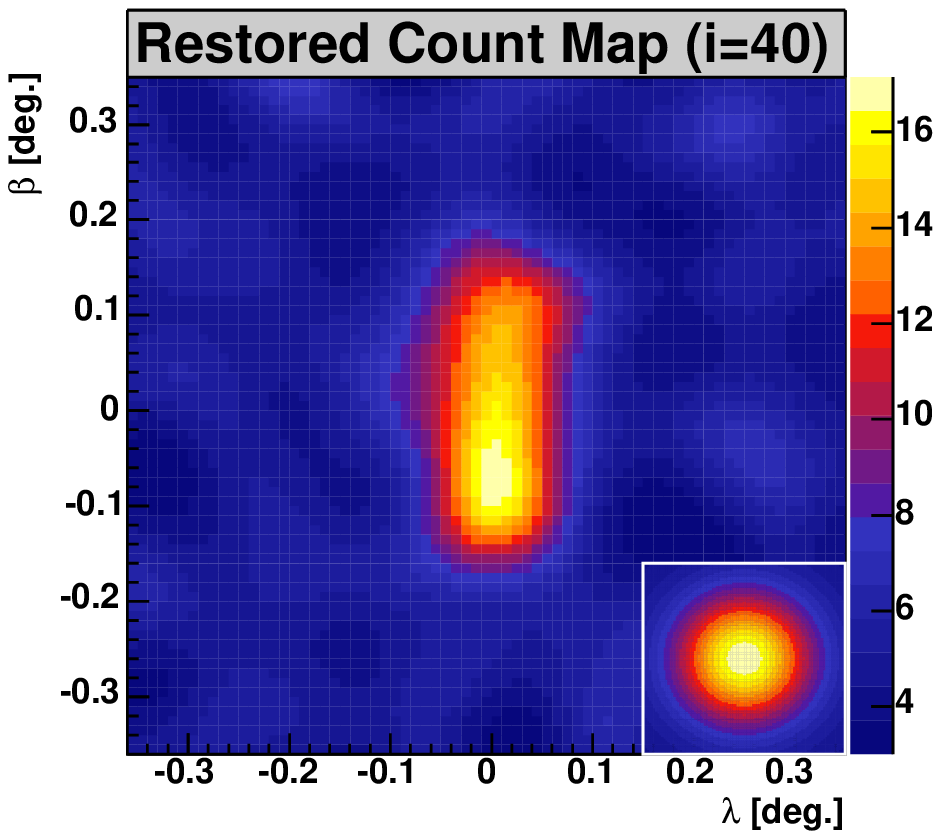}
  \end{minipage} \hfill
  \begin{minipage}[c]{0.5\linewidth}
    \includegraphics[width=\textwidth]{images/Deco80_smooth}
  \end{minipage} \hfill
  \begin{minipage}[c]{0.45\linewidth}
    \caption[Restored Simulated Count Map ($IA>$80\,p.e.)]{Restored simulated
      count map of Fig.~\ref{fig:DecoMCEmission} ($IA>$80\,p.e.) with the
      Richardson-Lucy algorithm for different numbers of iterations $(i)$. A
      profile of the PSF is indicated at the bottom right. With increasing $i$,
      the width of the emission region reduces and a slight asymmetry of the
      excess appears. The smoothed count map is shown for comparison.}
    \label{fig:DecoMC}
  \end{minipage} \hfill
\end{figure}

The restoration of one of the 20 simulated count maps is shown in
Fig.~\ref{fig:DecoMC} for 5, 10, 20 and 40 iterations. It is similar to the
restored count map of the data in Fig.~\ref{fig:DecoMSH80}. The width of the
emission region reduces similarly. For comparison, the last plot shows the
count map that was smoothed by convolution with a Gaussian function
($\sigma_s=0.03^\circ$). It has the largest width.

\begin{figure}
  \begin{minipage}[c]{0.5\linewidth}
    \includegraphics[width=\textwidth]{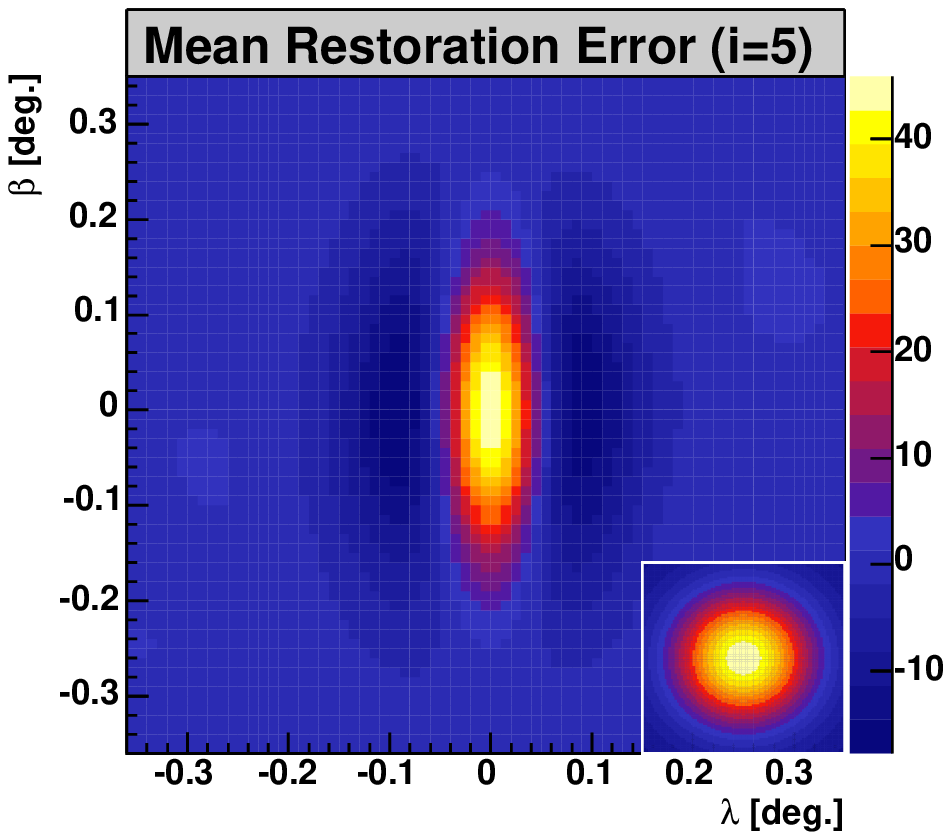}
  \end{minipage} \hfill
  \begin{minipage}[c]{0.5\linewidth}
    \includegraphics[width=\textwidth]{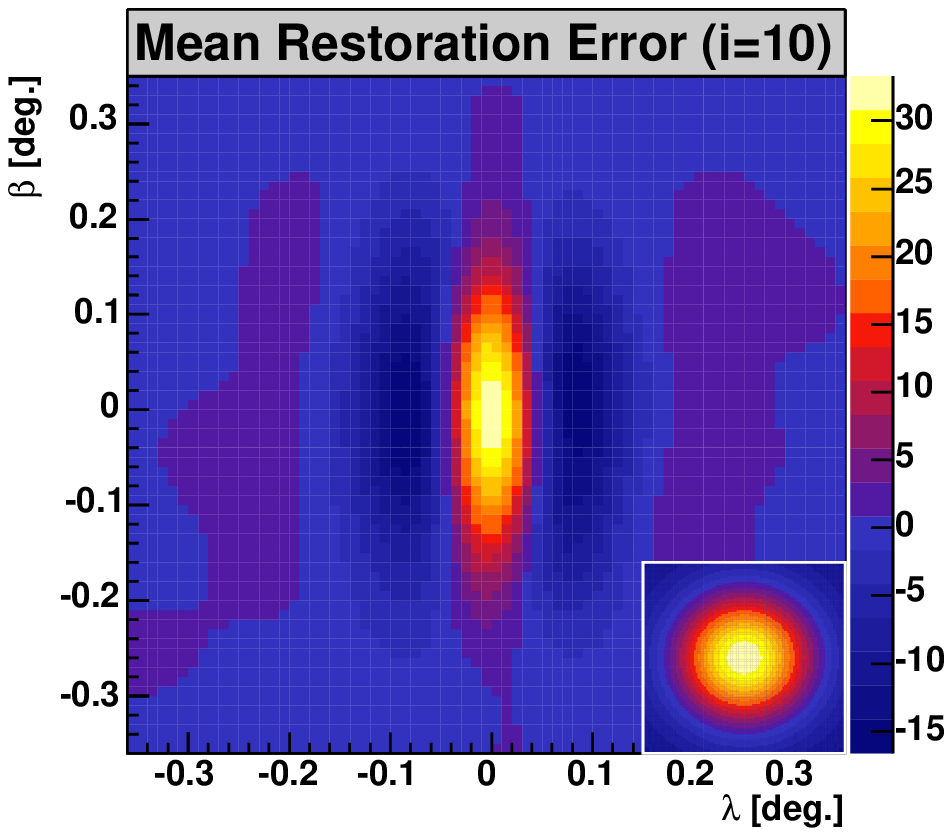}
  \end{minipage} \hfill
  \begin{minipage}[c]{0.5\linewidth}
    \includegraphics[width=\textwidth]{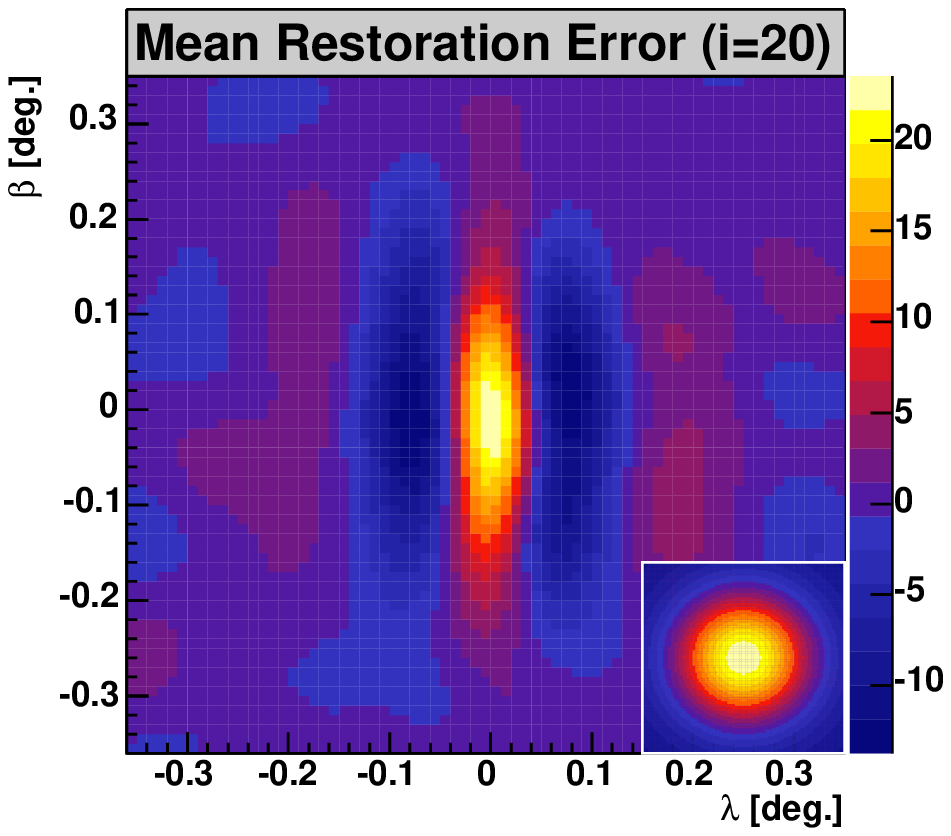}
  \end{minipage} \hfill
  \begin{minipage}[c]{0.5\linewidth}
    \includegraphics[width=\textwidth]{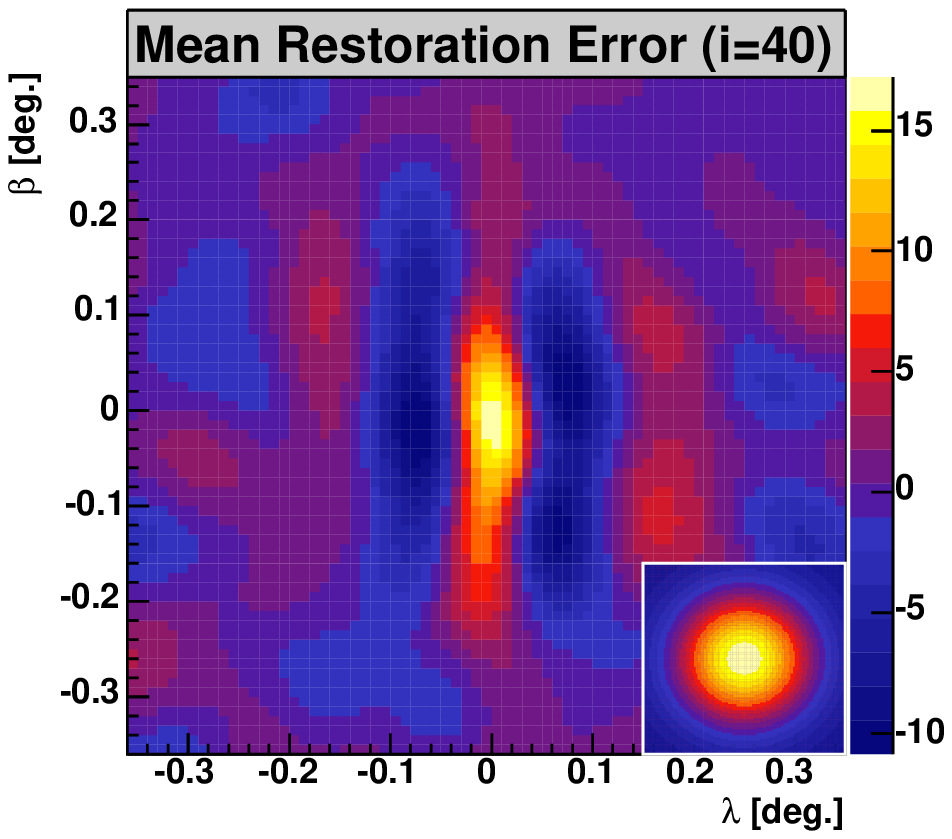}
  \end{minipage} \hfill
  \begin{minipage}[c]{0.5\linewidth}
    \includegraphics[width=\textwidth]{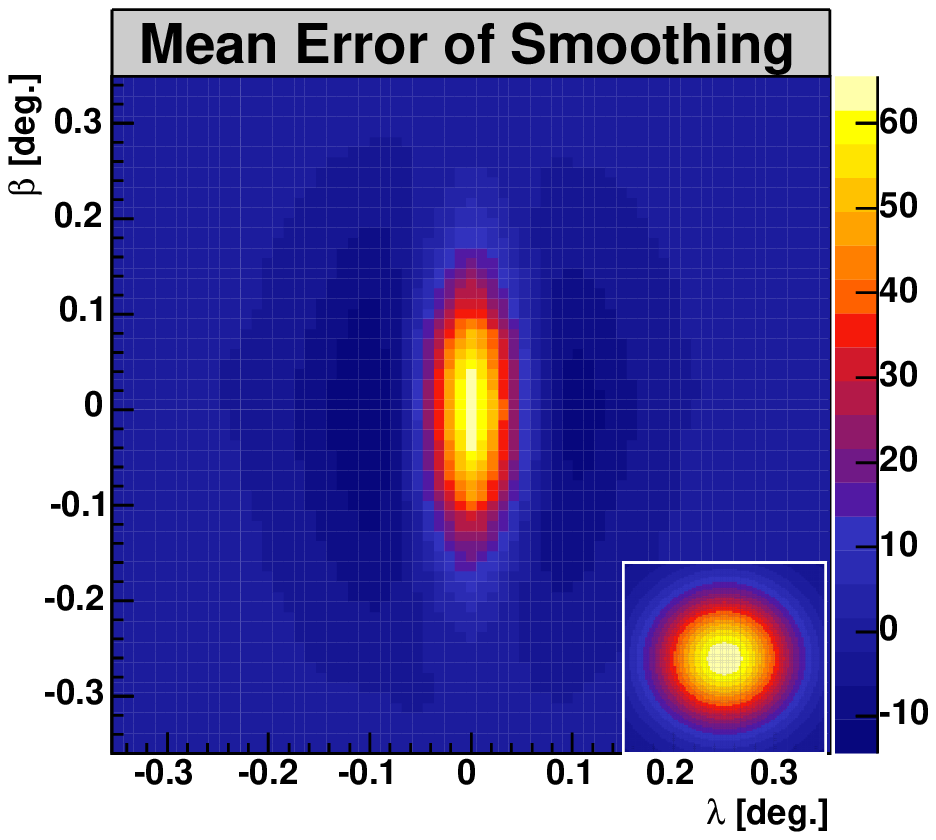}
  \end{minipage} \hfill
  \begin{minipage}[c]{0.45\linewidth}
    \caption[Mean Restoration Error ($IA>$80\,p.e.)]{The mean restoration error
      $(E^i(x,y))$ between the restored and the simulated image $(O)$. The mean
      is obtained from the restoration of 20 different count maps. The scale is
      in percentages of the peak intensity of the simulated excess. $E^i(x,y)$
      decreases with increasing iterations, while background fluctuations
      increase. The largest deviations are found in the center region.
      $E^i(x,y)$ is highest for the smoothed count map, which is shown for
      comparison.}
    \label{fig:DecoError}
  \end{minipage} \hfill
\end{figure}

Fig.~\ref{fig:DecoError} shows the mean error $\overline{E}^i(x,y)$
(Eqn.~\ref{eqn:Ei}) of the 20 different restorations. The scale is chosen in
percentages of the peak intensity of the simulated excess of 14.3 counts. The
largest error $(\overline{E}^i_{max})$ is found at the center of the excess,
where the restored intensity is systematically too small. On both sides of the
center region, at a distance of about 0.1$^\circ$ to 0.2$^\circ$, the excess is
systematically too high. A very similar error distribution is seen in the
smoothed map. $\overline{E}^i(x,y)$ decreases with an increasing number of
iterations and the restoration approaches the true image $O$. For example,
after 40 iterations $\overline{E}^i_{max}$ is smaller than $\sim$15\%. In
comparison, the $\overline{E}^i_{max}$ of the smoothed count map shows the
largest error of $\sim$65\%.

\begin{figure}
  \begin{minipage}[c]{0.5\linewidth}
    \includegraphics[width=\textwidth]{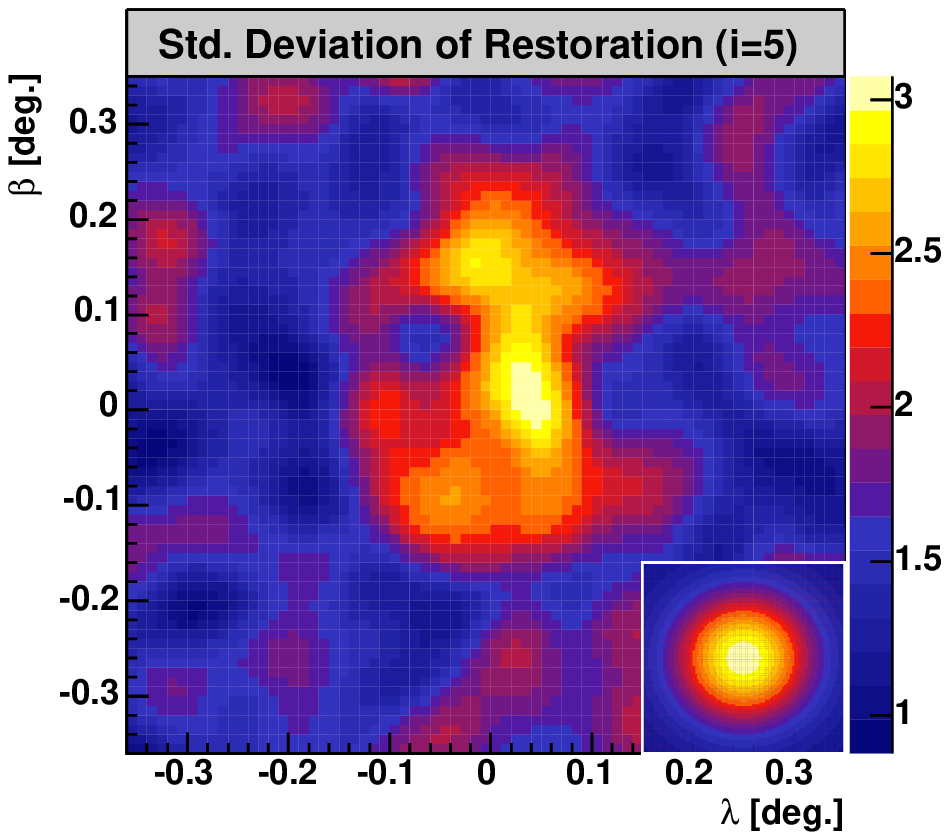}
  \end{minipage} \hfill
  \begin{minipage}[c]{0.5\linewidth}
    \includegraphics[width=\textwidth]{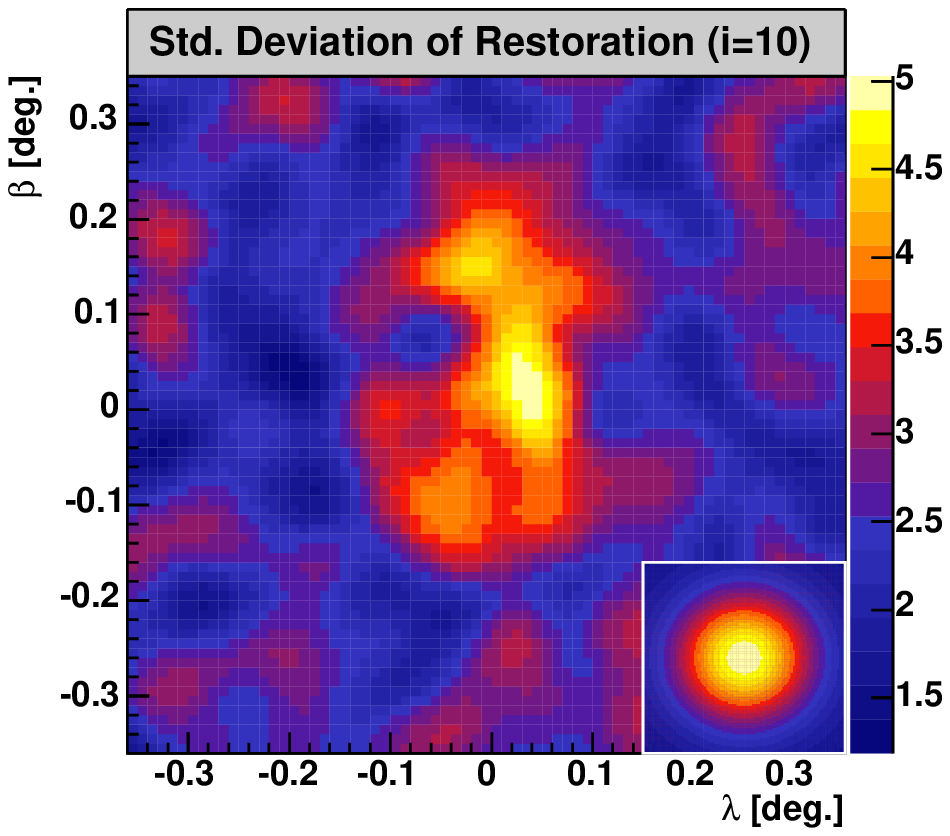}
  \end{minipage} \hfill
  \begin{minipage}[c]{0.5\linewidth}
    \includegraphics[width=\textwidth]{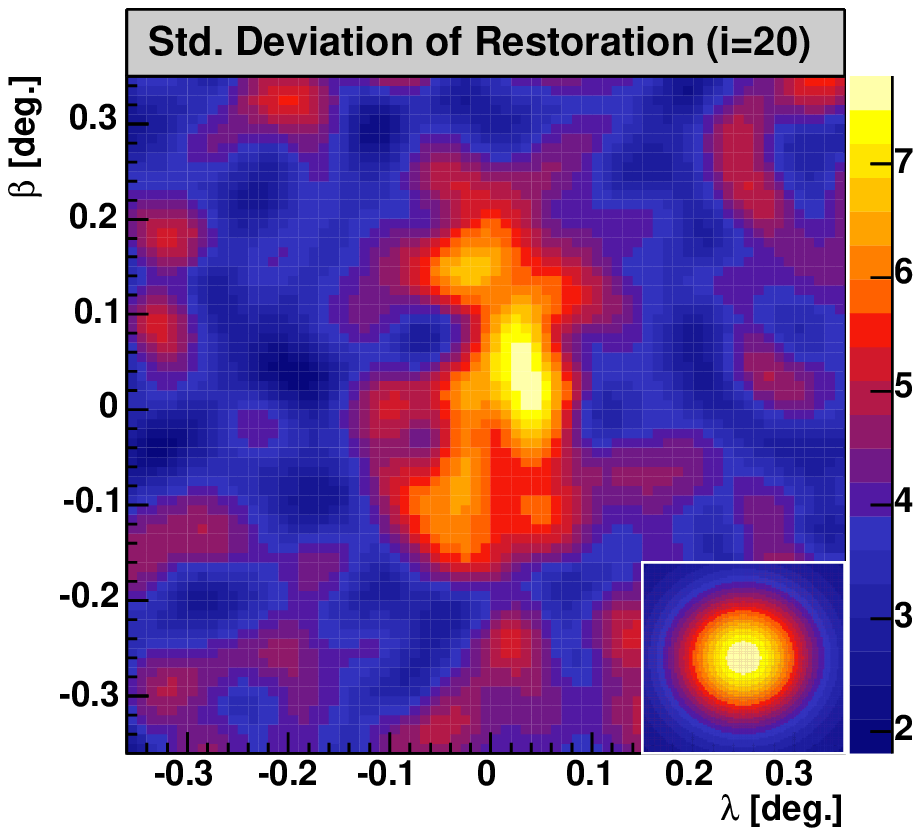}
  \end{minipage} \hfill
  \begin{minipage}[c]{0.5\linewidth}
    \includegraphics[width=\textwidth]{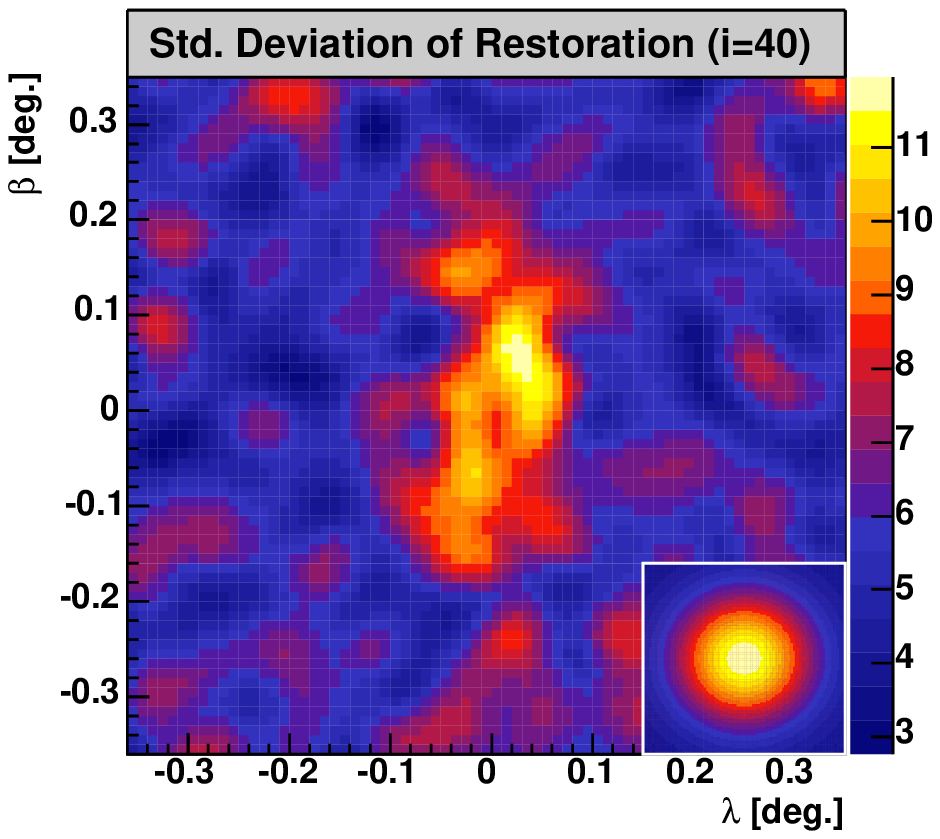}
  \end{minipage} \hfill
  \begin{minipage}[c]{0.5\linewidth}
    \includegraphics[width=\textwidth]{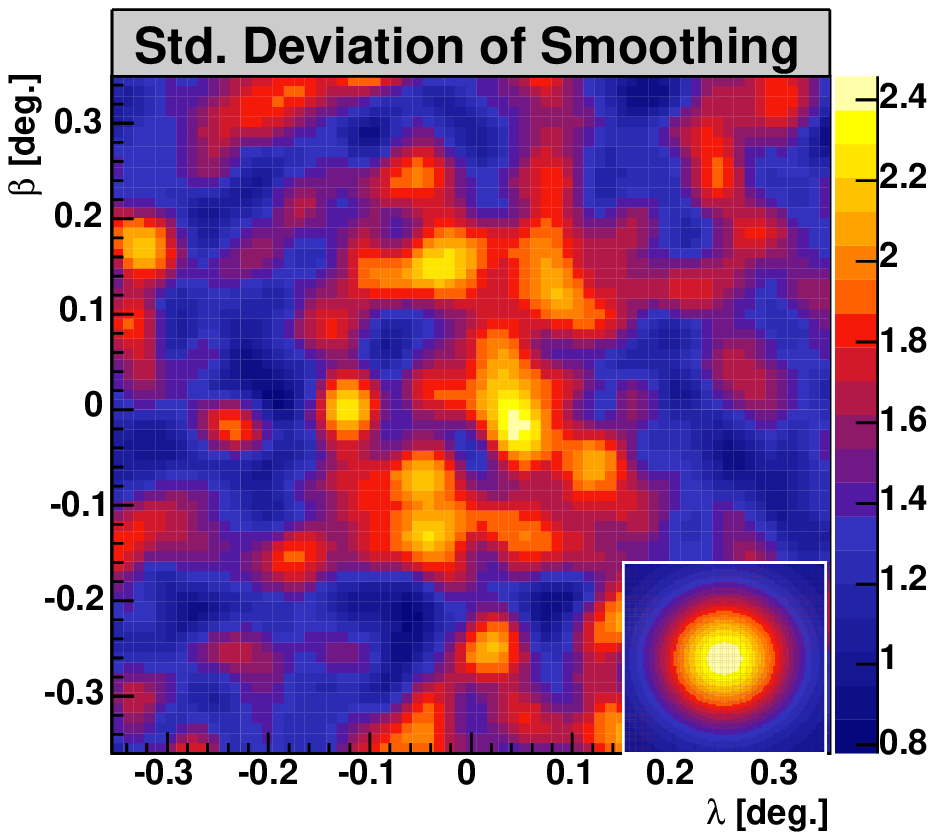}
  \end{minipage} \hfill
  \begin{minipage}[c]{0.45\linewidth}
    \caption[Standard Deviation of the Restoration ($IA>$80\,p.e.)]{Standard
      deviation of the restoration $(S^i(x,y))$ of 20 different restored count
      maps. The scale is given in percentages of the peak intensity of the
      simulated excess. $S^i(x,y)$ increases with the number of iterations. The
      highest values are reached in the center region. After 40 iterations,
      $S^i(x,y)$ reaches a level of $\sim$10\%. $S^i(x,y)$ of the smoothed
      count map is comparatively small.}
    \label{fig:DecoStdDev}
  \end{minipage} \hfill
\end{figure}

Fig.~\ref{fig:DecoStdDev} shows the standard deviation $S^i(x,y)$
(Eqn.~\ref{eqn:Si}) as obtained from the 20 different simulations. Again, the
scale is chosen in percentages of the peak amplitude of the simulated true
emission. An increase of the standard deviation with the number of iterations
can be seen. The largest deviations $S^i_{max}$ are found at the center
region. After 40 iterations they reach a level of $\sim$10\%. For comparison, the standard deviation in the smoothed count map is also shown. It has the
smallest deviations --- only $\sim$3\%.

\begin{figure}
  \begin{minipage}[c]{0.5\linewidth}
    \includegraphics[width=\textwidth]{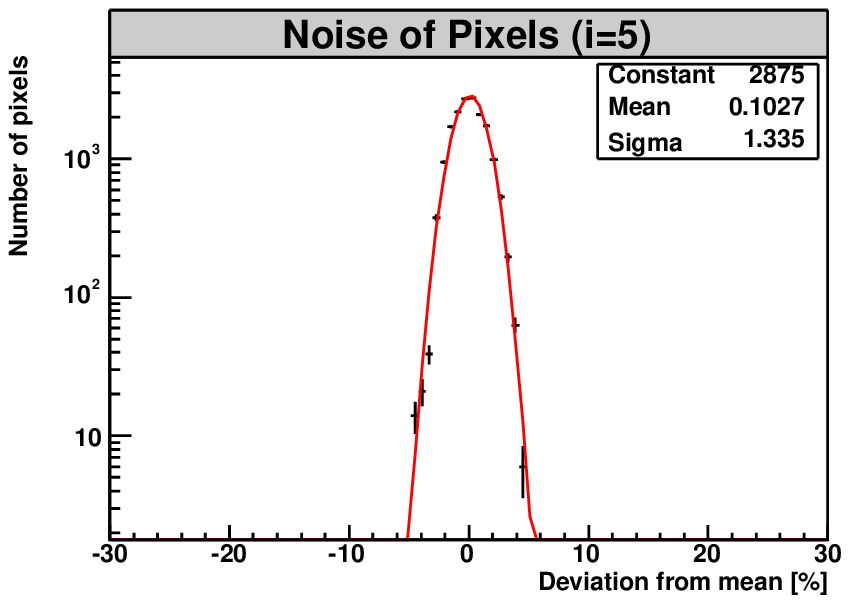}
  \end{minipage} \hfill
  \begin{minipage}[c]{0.5\linewidth}
    \includegraphics[width=\textwidth]{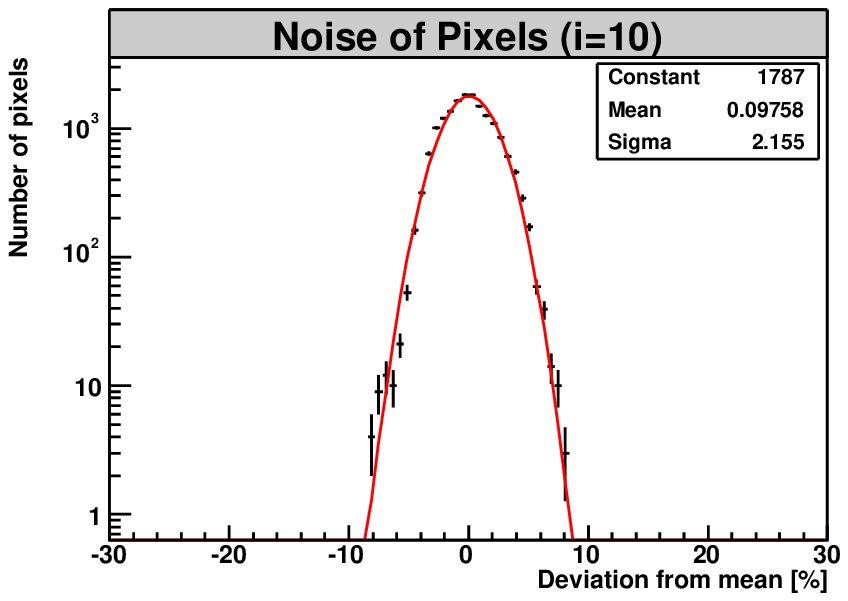}
  \end{minipage} \hfill
  \begin{minipage}[c]{0.5\linewidth}
    \includegraphics[width=\textwidth]{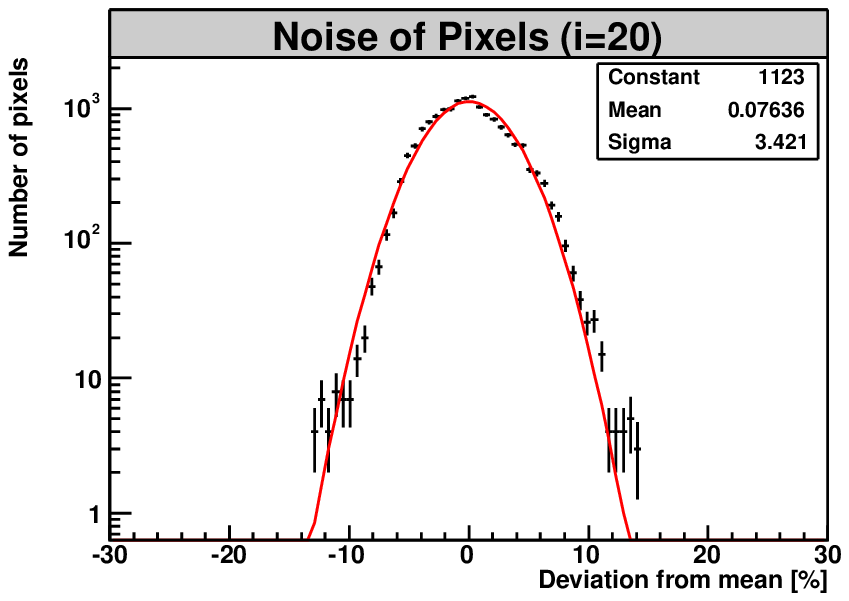}
  \end{minipage} \hfill
  \begin{minipage}[c]{0.5\linewidth}
    \includegraphics[width=\textwidth]{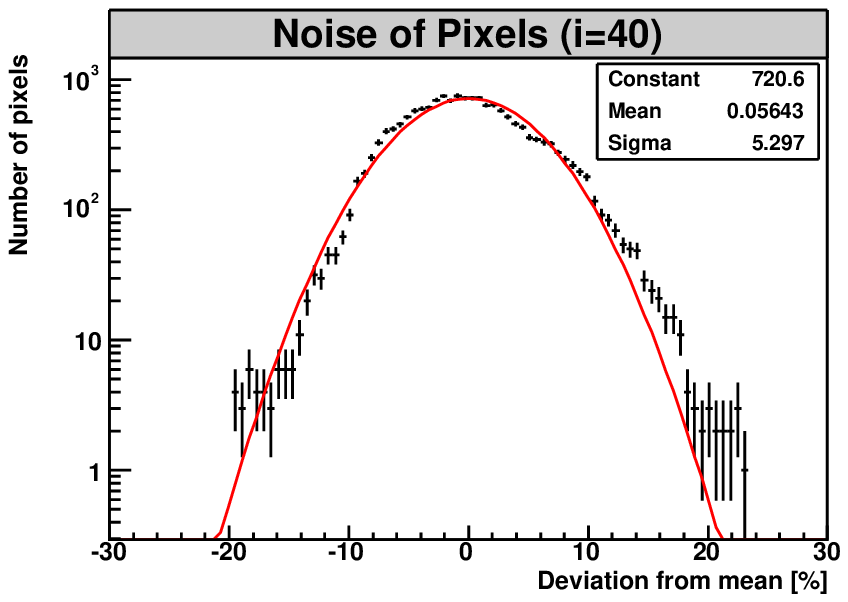}
  \end{minipage} \hfill
  \begin{minipage}[c]{0.5\linewidth}
    \includegraphics[width=\textwidth]{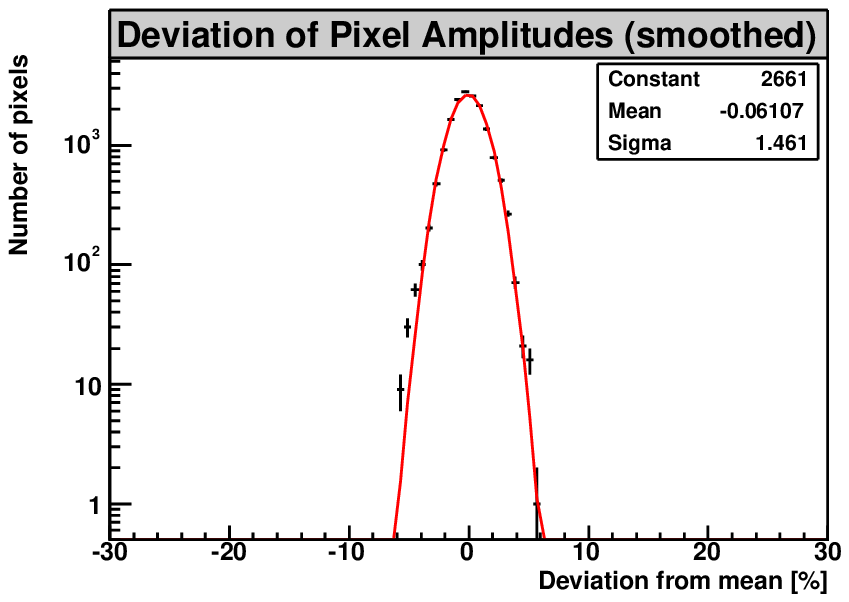}
  \end{minipage} \hfill
  \begin{minipage}[c]{0.48\linewidth}
    \includegraphics[width=\textwidth]{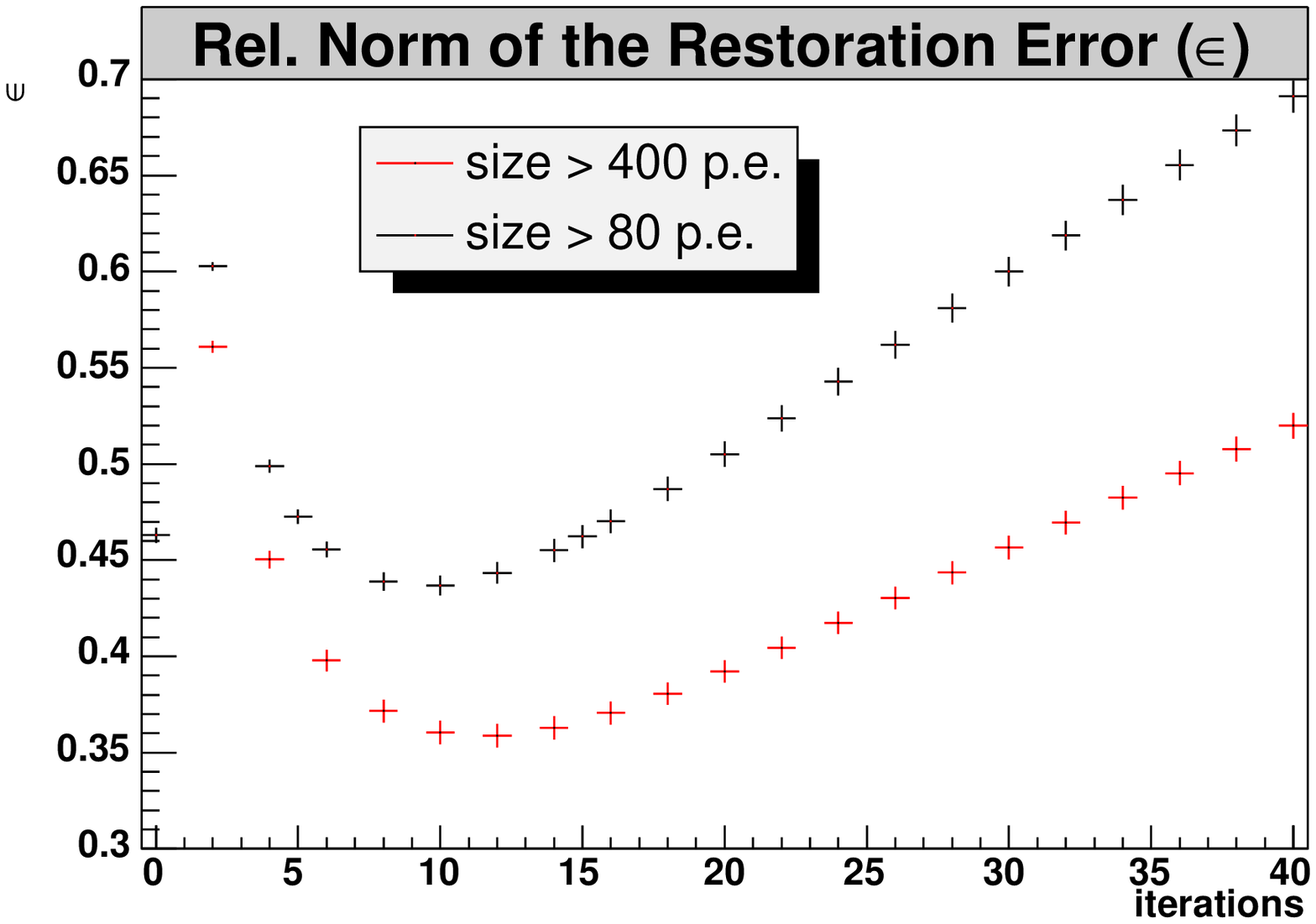}   
  \end{minipage}\hfill
    \caption[Distribution of Noise and the Restoration Error
      ($IA>$80\,p.e.)]{Distribution showing each pixel's deviation from its
      expectation value. The deviation is measured in percentages of the peak
      intensity of the simulated excess and represents the noise. A fit to a
      Gaussian distribution is shown (red). The last plot shows the mean
      relative norm of the restoration error $(\overline{\epsilon}_i)$ versus
      the number of iterations $i$. The minimum is found for $i_{\rm
      opt}\sim10$ iterations.}
    \label{fig:RestorationError}
\end{figure}

Fig.~\ref{fig:RestorationError} shows the distributions of the pixel's noise
$D^i(x,y)$ (Eqn.~\ref{eqn:Di}) for different numbers of iterations $i$. The
distributions of the 20 simulations are added to one plot. The noise is
measured in percentages of the peak intensity of the excess of the true image
$O$. The noise of the restoration is approximately Gaussian distributed with a
zero-mean as shown by a fit to a Gaussian function (red line). Only for higher
numbers of iterations are small deviations from a Gaussian distribution found.
The level of noise can be characterized by the standard deviations of 1.3, 2.2,
3.4 and 5.3\% for 5, 10, 20 and 40 iterations, respectively. The noise of the
map after 5 iterations is comparable to the noise of the smoothed map.

The last plot of Fig.~\ref{fig:RestorationError} shows the mean relative norm
of the restoration error $(\overline{\epsilon}_i)$
(Eqn.~\ref{eqn:RestorationError}). $\overline{\epsilon}_i$ is calculated for
the full region of $128\times128$ bins and as the mean of the 20 different
simulations of the 80\,p.e. map (red marker; the black marker refers to the
400\,p.e. map which is discussed later). The minimum is obtained after about 10
iterations. The statistical error is $\sim$2\%. Within the range of about
$\pm$5 iterations from the minimum, $\epsilon_{\rm opt}$ changes by less than
5\%, showing that the restorations change only marginally in the interval
$5<i<15$.

\begin{figure}
  \begin{minipage}[c]{0.5\linewidth}
    \includegraphics[width=\textwidth]{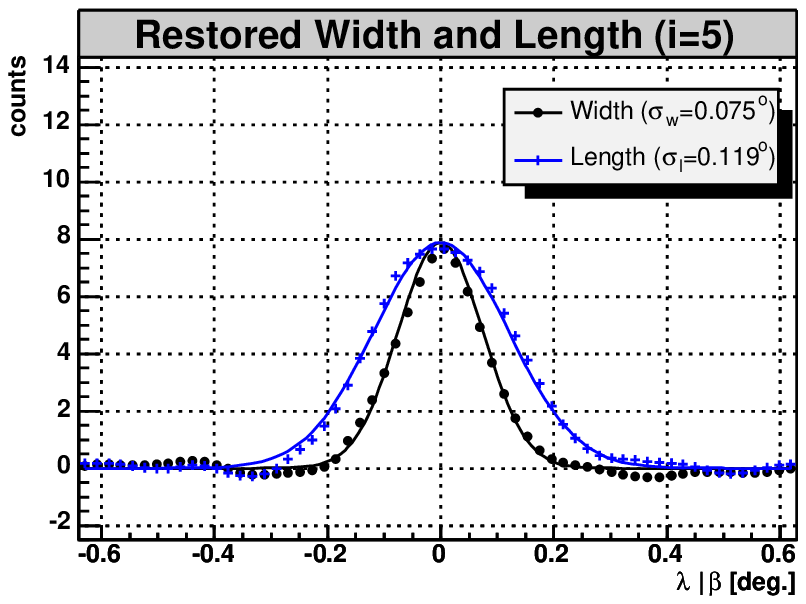}
  \end{minipage} \hfill
  \begin{minipage}[c]{0.5\linewidth}
    \includegraphics[width=\textwidth]{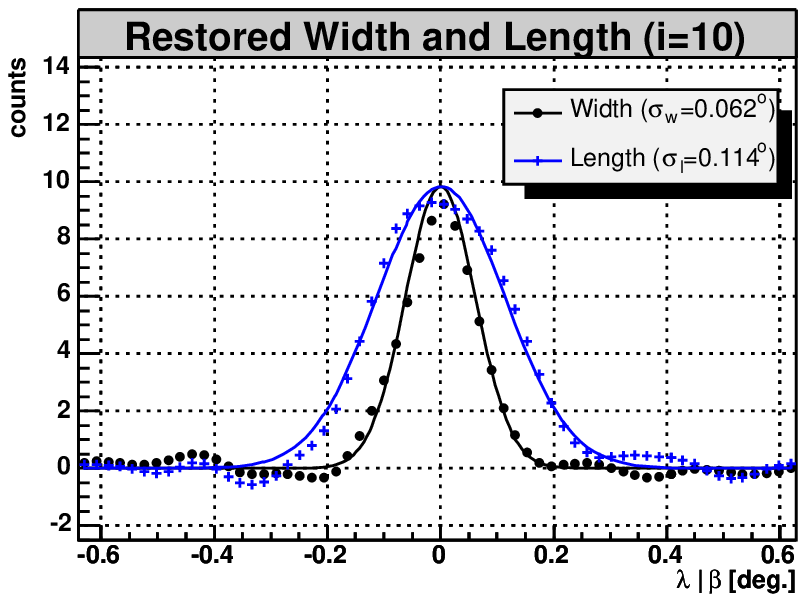}
  \end{minipage} \hfill
  \begin{minipage}[c]{0.5\linewidth}
    \includegraphics[width=\textwidth]{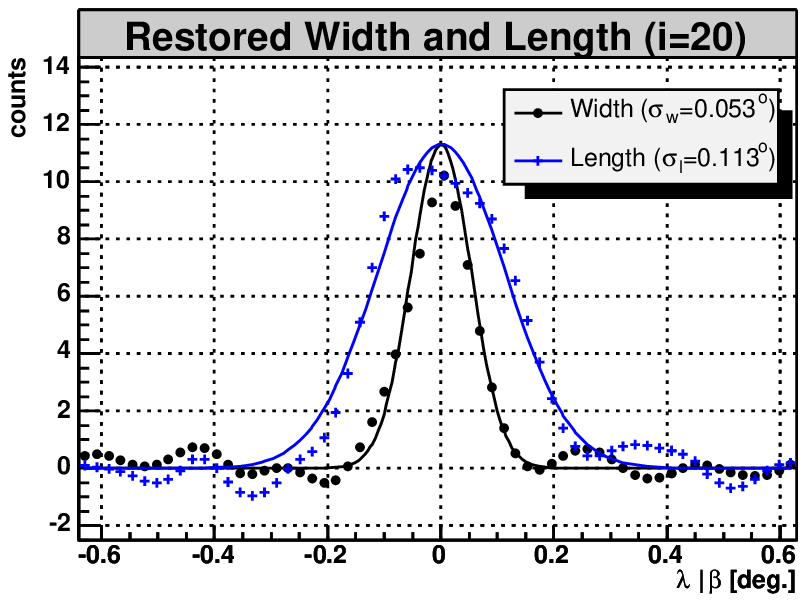}
  \end{minipage} \hfill
  \begin{minipage}[c]{0.5\linewidth}
    \includegraphics[width=\textwidth]{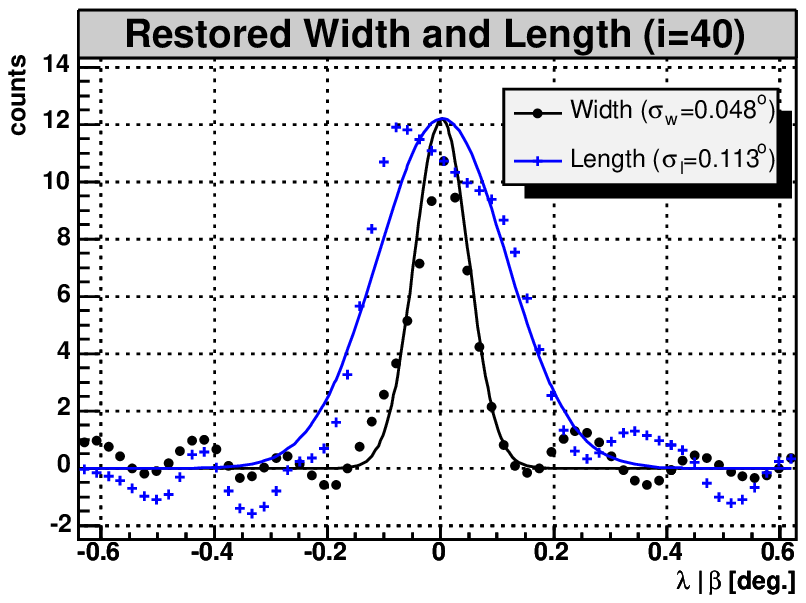}
  \end{minipage} \hfill
  \begin{minipage}[c]{0.5\linewidth}
    \includegraphics[width=\textwidth]{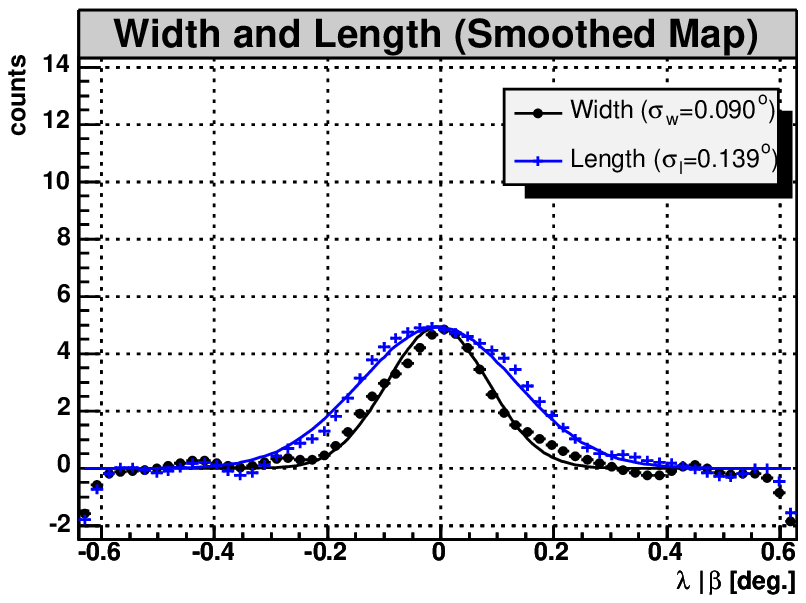}
  \end{minipage} \hfill
  \begin{minipage}[c]{0.48\linewidth}
    \includegraphics[width=\textwidth]{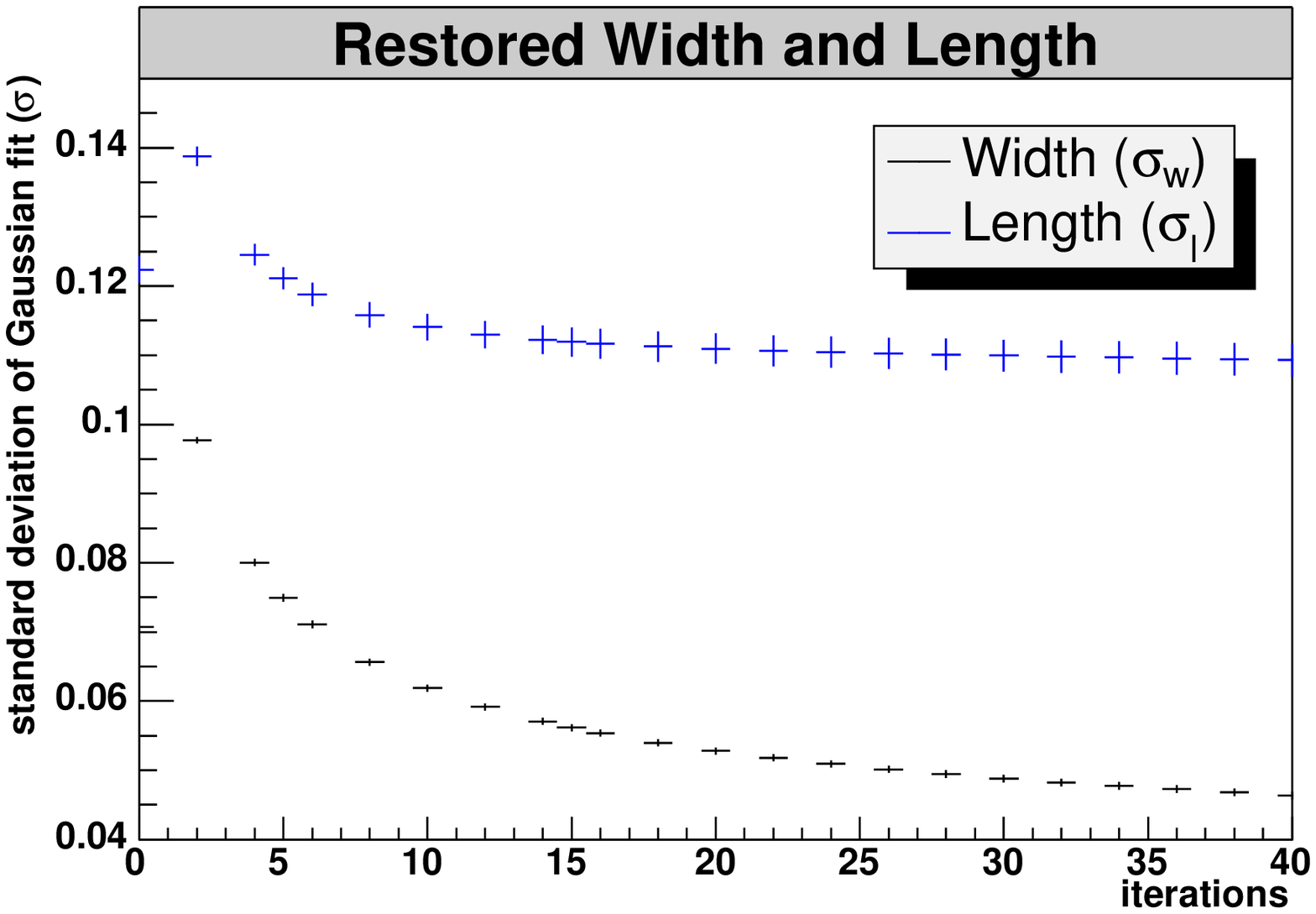}
  \end{minipage} \hfill
    \caption[Profiles of the Restored Count Maps ($IA>$80\,p.e.)]{Profiles of
      the restored count maps of Fig.~\ref{fig:DecoMC} showing the width and
      length of the excess. With increasing number of iterations $(i)$, the
      width and length decreases while the amplification of noise increases.
      The width and length of the smoothed count map are shown for comparison.
      The bottom right plot shows the width and length measured in standard
      deviation vs. the number of iterations.}
    \label{fig:DecoFit}
\end{figure}

Fig.~\ref{fig:DecoFit} shows the corresponding excess profiles of the restored
maps of Fig.~\ref{fig:DecoMC} after background subtraction. The distributions
show the width (black) and length (blue) along the $\lambda$- and $\beta$-axis
of the excess, respectively. The solid line represents the Gaussian fit
function (Eqn.~\ref{eqn:2DGaussian}). The fit value of the standard deviations
$(\sigma_w, \sigma_l)$ of the Gaussian function is shown in each plot. For
example, for 5, 10, 20 and 40 iterations, the width represents 180, 155, 130
and 120\% of the simulated excess distribution in $O$, while the smoothed map
(plot 5) has a width of 220\%. Small waves in the profiles show the
amplification of noise increasing with the number of iterations. The smoothed
map shows only little noise. The last plot shows the successive decrease of the
width and length versus the number of iterations.

The individual errors discussed here are summarized in
Tbl.~\ref{tbl:RestorationError}.

\clearpage

\subsubsection{Restoration of the 400\,p.e. Map}
\begin{figure}[ht!]
  \setlength{\abovecaptionskip}{-.05cm}
  \setlength{\belowcaptionskip}{.5cm}
  \begin{minipage}[t]{0.5\linewidth}
    \includegraphics[width=\textwidth]{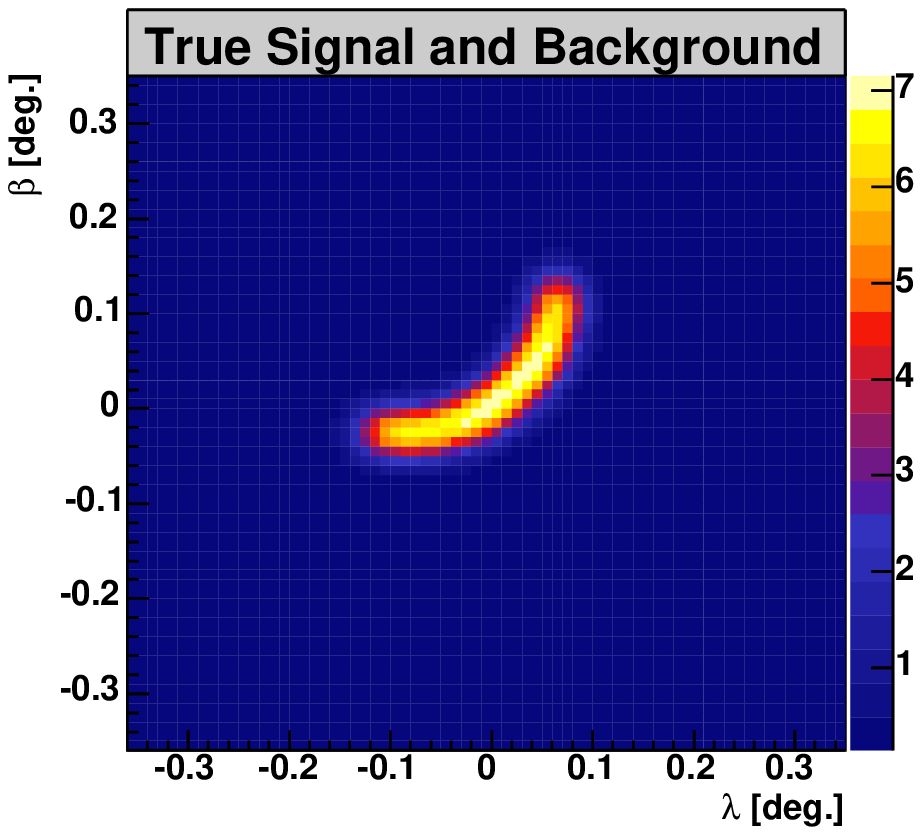}
    \caption[Simulated Map $(O)$ of the True Emission
      ($IA>$400\,p.e.)]{Simulated map $(O)$ of the true emission with arc-like
      signal and constant background of 0.14 counts/bin ($IA>$400\,p.e.).}
    \label{fig:DecoMCEmission400}
  \end{minipage} \hfill
  \begin{minipage}[t]{0.5\linewidth}
    \includegraphics[width=\textwidth]{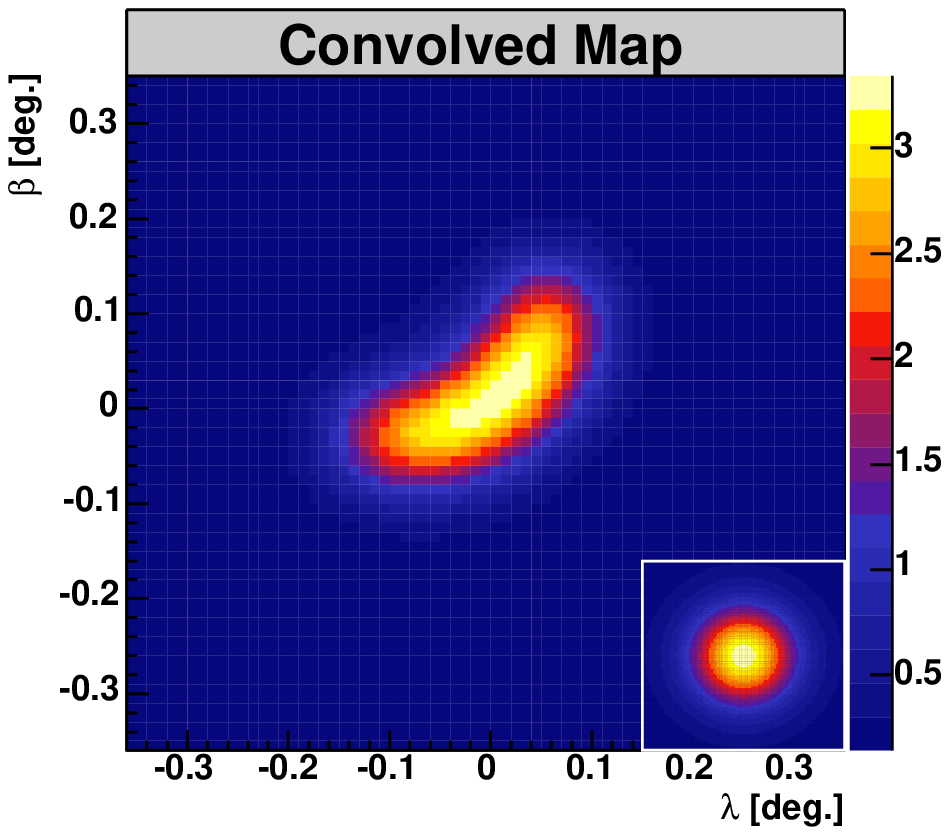}
    \caption[Map of the True Emission $(O)$ Convolved with the PSF
      ($IA>$400\,p.e.)]{Map of Fig.~\ref{fig:DecoMCEmission400} convolved with
      the PSF. A profile of the PSF is shown at the bottom right.}
    \label{fig:DecoMCConvolution400}
  \end{minipage} \hfill
  \begin{minipage}[t]{0.5\linewidth}
    \includegraphics[width=\textwidth]{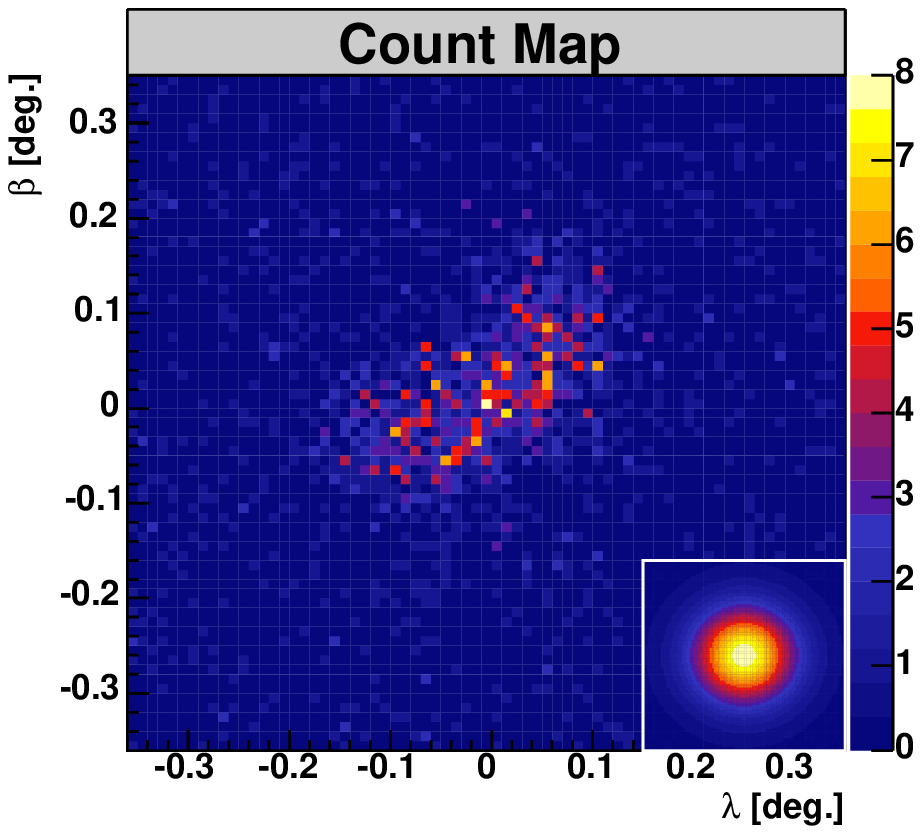}
    \caption[Simulated Count Map $(I)$ Including Poisson Noise
      ($IA>$400\,p.e.)]{Simulated count map $(I)$ including Poisson noise
      corresponding to Fig.~\ref{fig:DecoMCConvolution400}. The PSF is shown at
      the bottom right.}
    \label{fig:DecoMC400_0}
  \end{minipage} \hfill
  \begin{minipage}[t]{0.5\linewidth}
    \includegraphics[width=\textwidth]{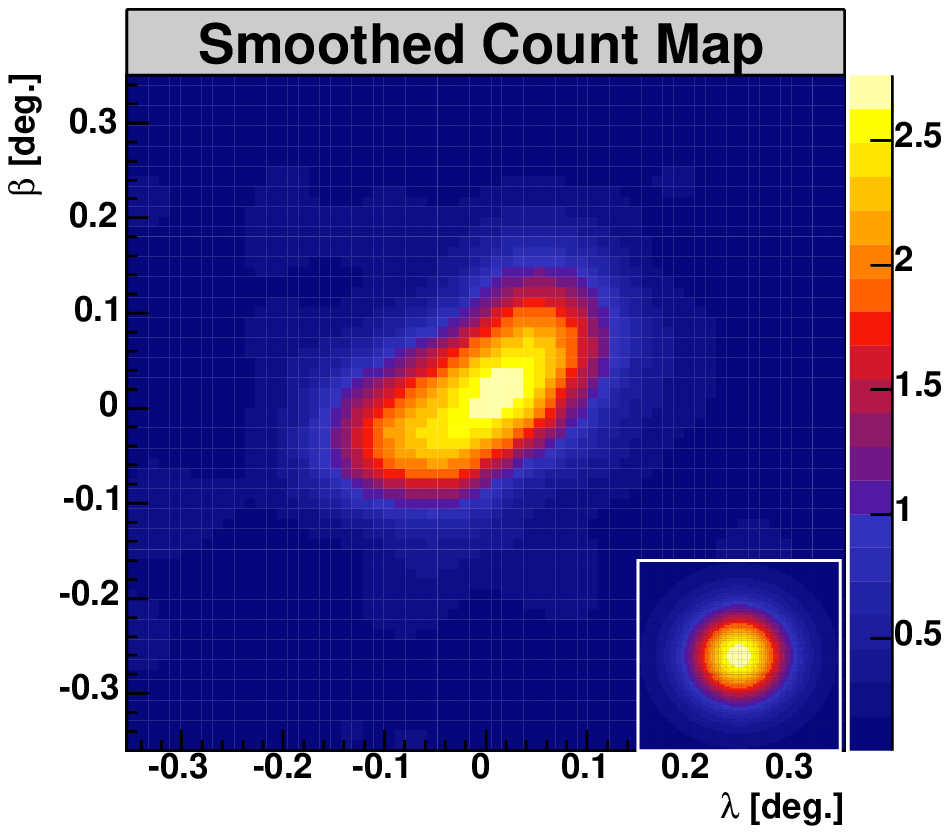}
    \caption[Smoothed Simulated Count Map ($IA>$400\,p.e.)]{Count map of
      Fig.~\ref{fig:DecoMC400_0} smoothed with a Gaussian function
      $(\sigma_s=0.03^\circ)$. The PSF is shown at the bottom right.}
    \label{fig:DecoMC400_smooth}
  \end{minipage} \hfill
\end{figure}

The curved excess of the 400\,p.e. map was modeled by an arc consisting of all
points on a ring ($0.14^\circ<r<0.16^\circ$) around the origin which lie in the
fourth quadrant of the coordinate system. In addition, the arc is convolved
with a Gaussian function with a standard deviation of $0.02^\circ$ and shifted
to the middle of the map. This model was found through a comparison of the
restored count maps of the simulation and those of the data. The statistics in
the signal region with a radius $\theta$ of 0.3$^\circ$ are given by 1000
excess and 400 background events (Tbl.~\ref{tbl:StatisticsMSH}), yielding a
peak excess of 7 counts in $O$ and a background of 0.14 counts/bin. The
simulated maps are shown in Fig.~\ref{fig:DecoMCEmission400} to
\ref{fig:DecoMC400_smooth}. The count map is only one of 20 with different
random numbers of noise which have been restored. The PSF was modeled according
to parameterization no. 5 in Tbl.~\ref{tbl:PSF}.

\begin{figure}[ht!]
  \begin{minipage}[c]{0.5\linewidth}
    \includegraphics[width=\textwidth]{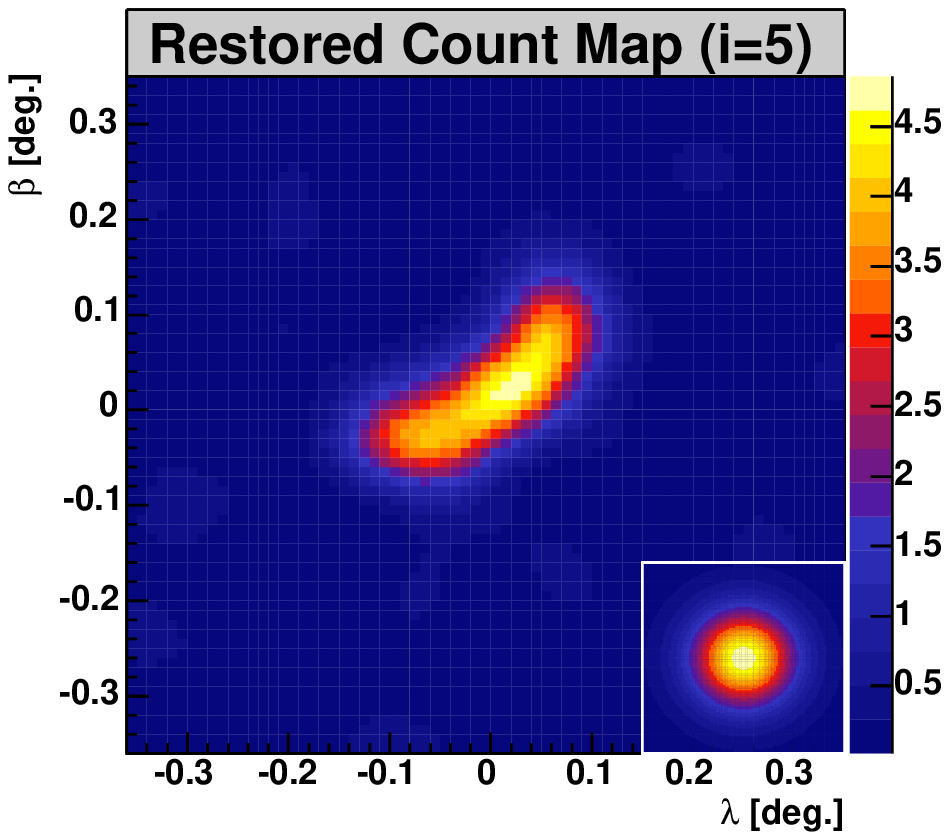}
  \end{minipage} \hfill
  \begin{minipage}[c]{0.5\linewidth}
    \includegraphics[width=\textwidth]{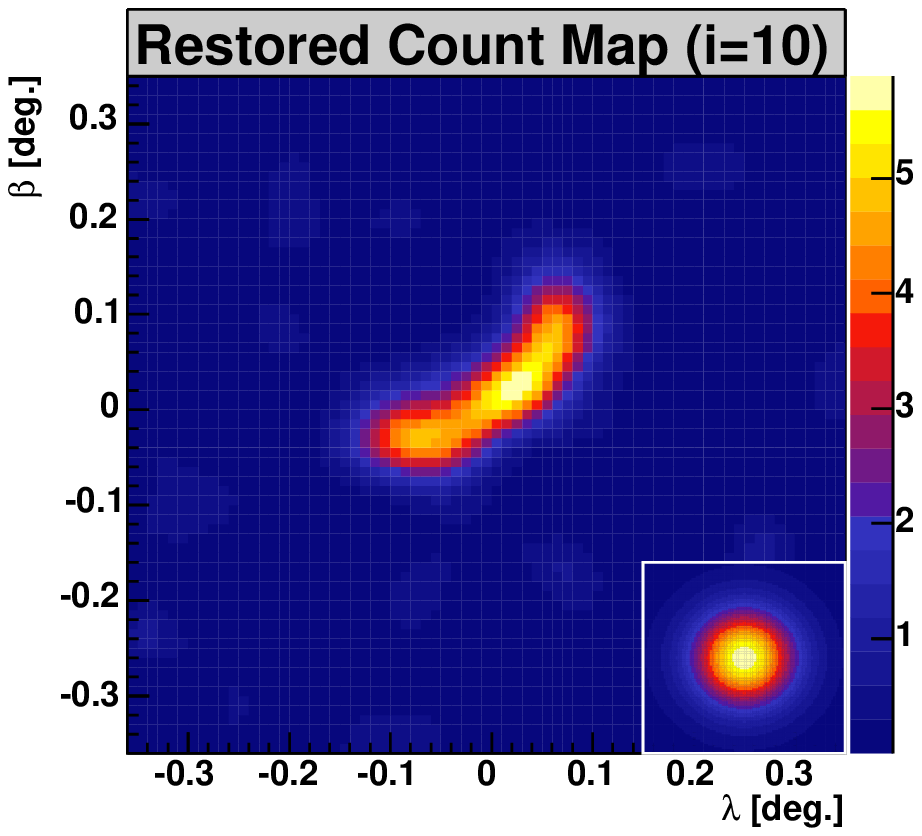}
  \end{minipage} \hfill
  \begin{minipage}[c]{0.5\linewidth}
    \includegraphics[width=\textwidth]{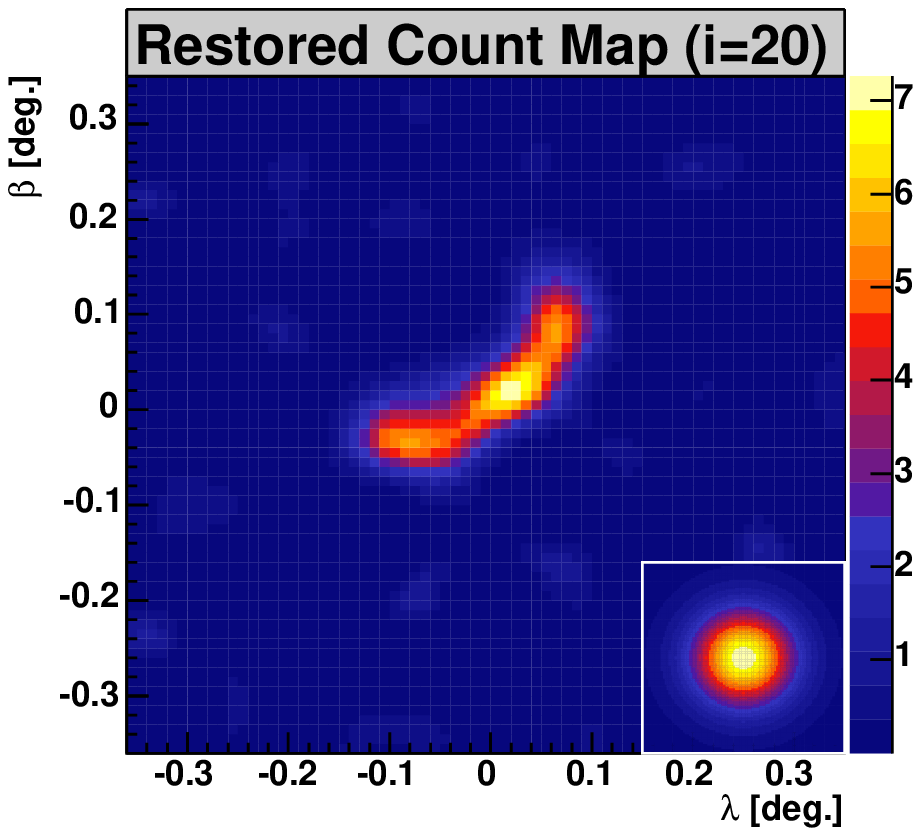}
  \end{minipage} \hfill
  \begin{minipage}[c]{0.5\linewidth}
    \includegraphics[width=\textwidth]{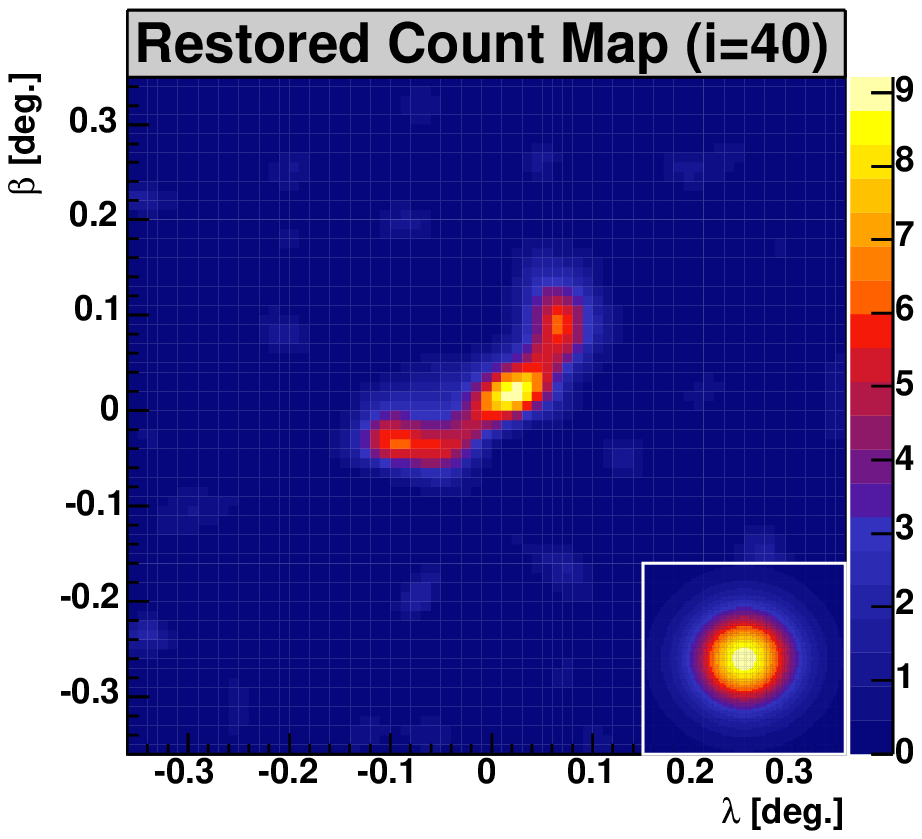}
  \end{minipage} \hfill
  \begin{minipage}[c]{0.5\linewidth}
    \includegraphics[width=\textwidth]{images/Deco400_smooth}
  \end{minipage} \hfill
  \begin{minipage}[c]{0.45\linewidth}
    \caption[Restored Simulated Count Map ($IA>$400\,p.e.)]{Restored simulated
      count map of Fig.~\ref{fig:DecoMCEmission400} ($IA>$400\,p.e.) using the
      Richardson-Lucy algorithm for different numbers of iterations $(i)$. A
      profile of the PSF is indicated at the bottom right. With increasing $i$,
      the width of the emission region reduces and the noise increases. The
      smoothed count map is shown for comparison.}
    \label{fig:DecoMC400}
  \end{minipage} \hfill
\end{figure}

A typical restoration of one of the 20 simulated count maps with the RL
algorithm is shown by the images in Fig.~\ref{fig:DecoMC400} for 5, 10, 20 and
40 iterations. The restorations are similar to the restoration of the count map
of the data in Fig.~\ref{fig:DecoMSH400}. The width of the emission region
reduces accordingly. For comparison, the last plot shows the count map that was
smoothed by convolution with a Gaussian function ($\sigma_s=0.03^\circ$). It
shows the widest extension.

\begin{figure}[ht!]
  \begin{minipage}[c]{0.5\linewidth}
    \includegraphics[width=\textwidth]{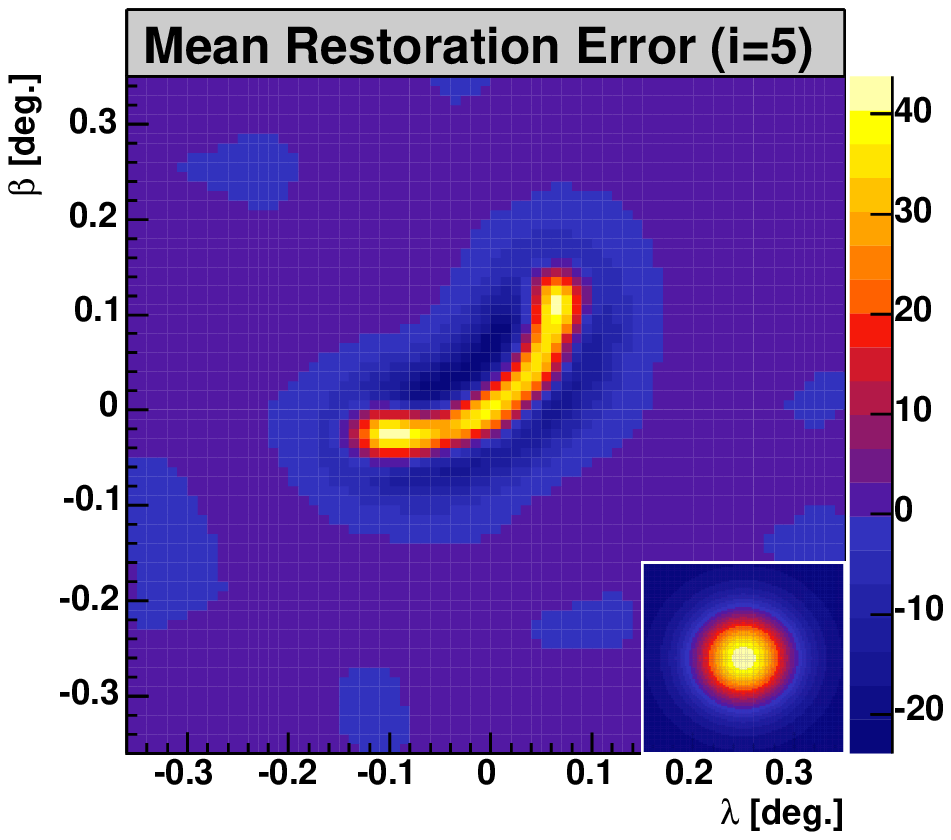}
  \end{minipage} \hfill
  \begin{minipage}[c]{0.5\linewidth}
    \includegraphics[width=\textwidth]{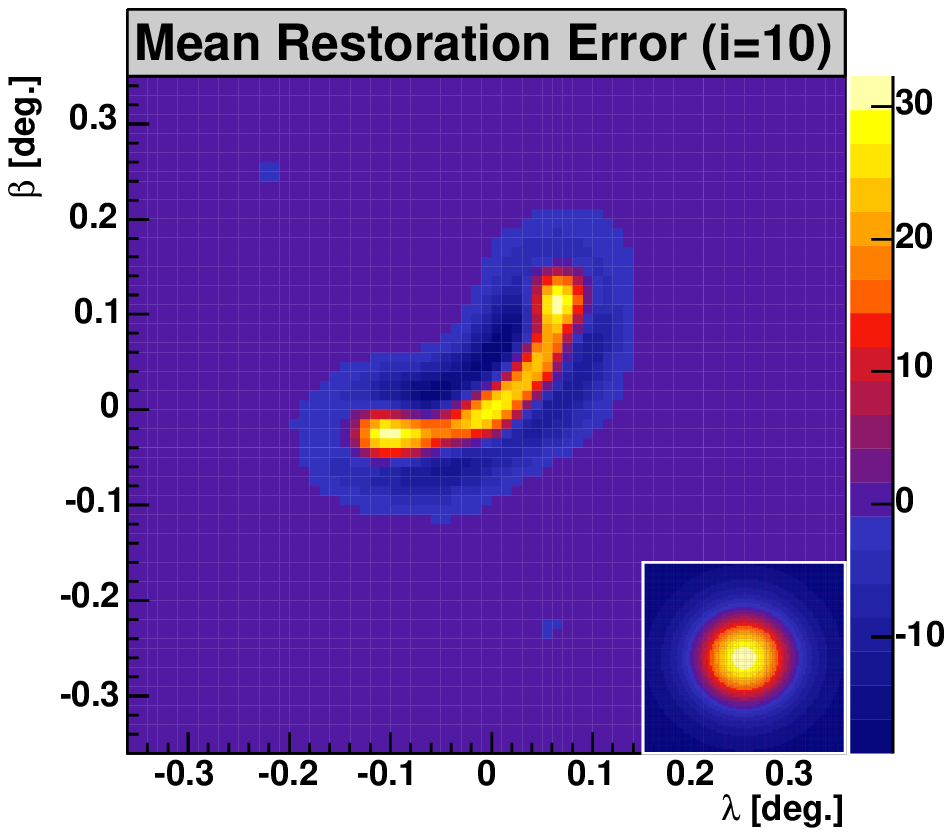}
  \end{minipage} \hfill
  \begin{minipage}[c]{0.5\linewidth}
    \includegraphics[width=\textwidth]{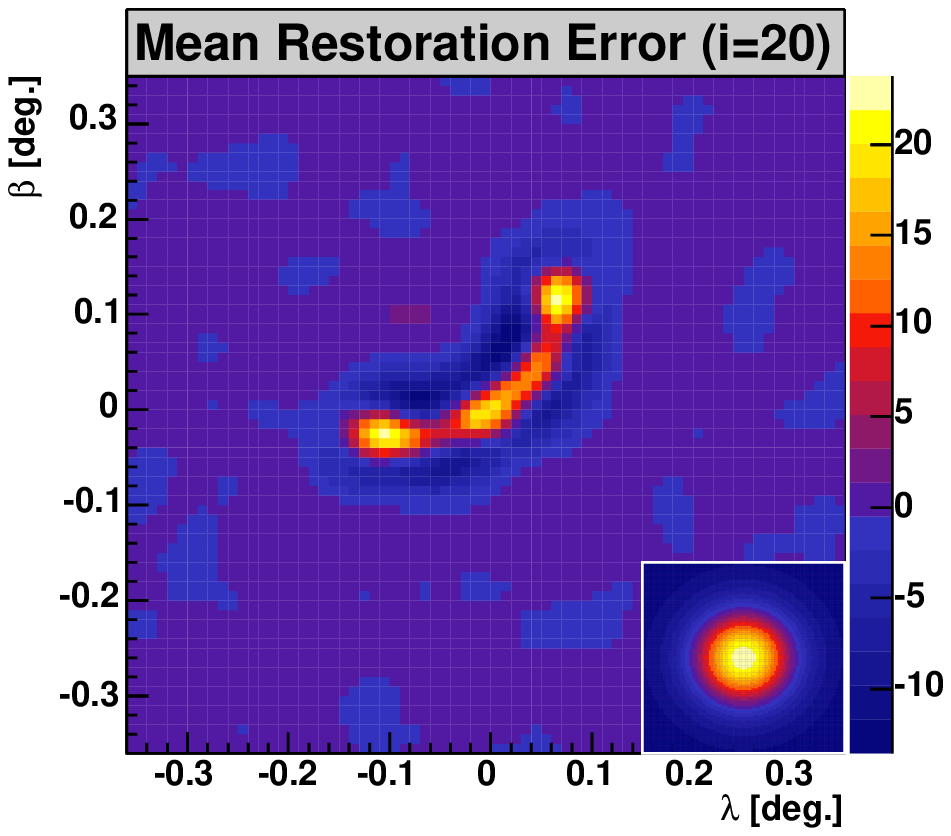}
  \end{minipage} \hfill
  \begin{minipage}[c]{0.5\linewidth}
    \includegraphics[width=\textwidth]{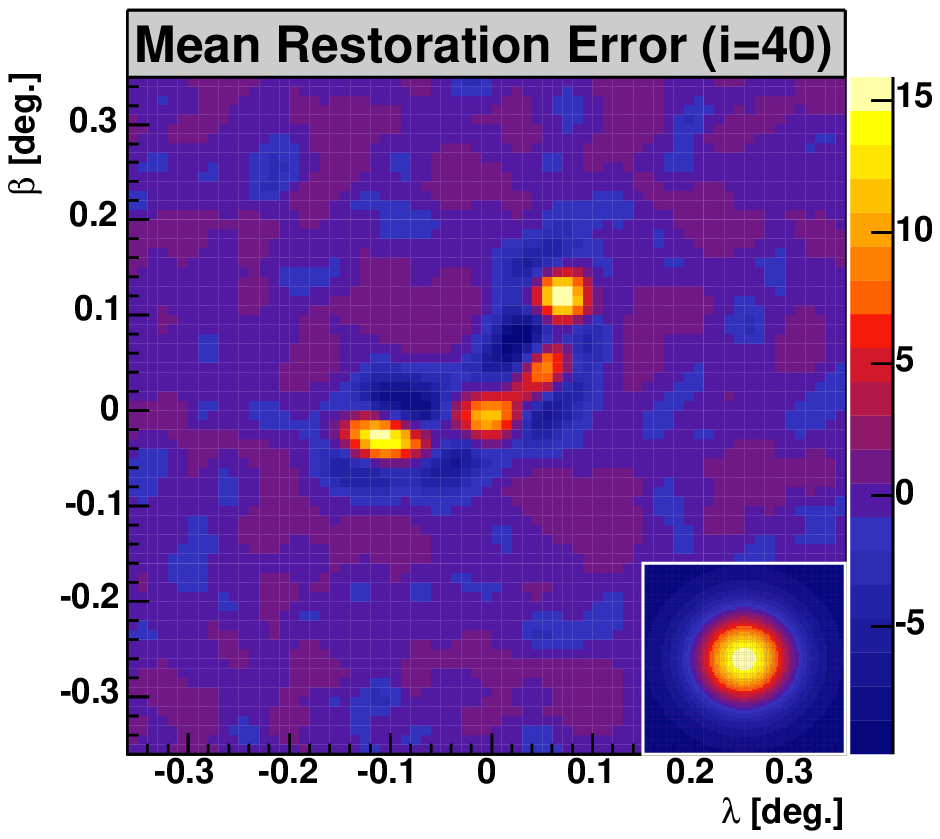}
  \end{minipage} \hfill
  \begin{minipage}[c]{0.5\linewidth}
    \includegraphics[width=\textwidth]{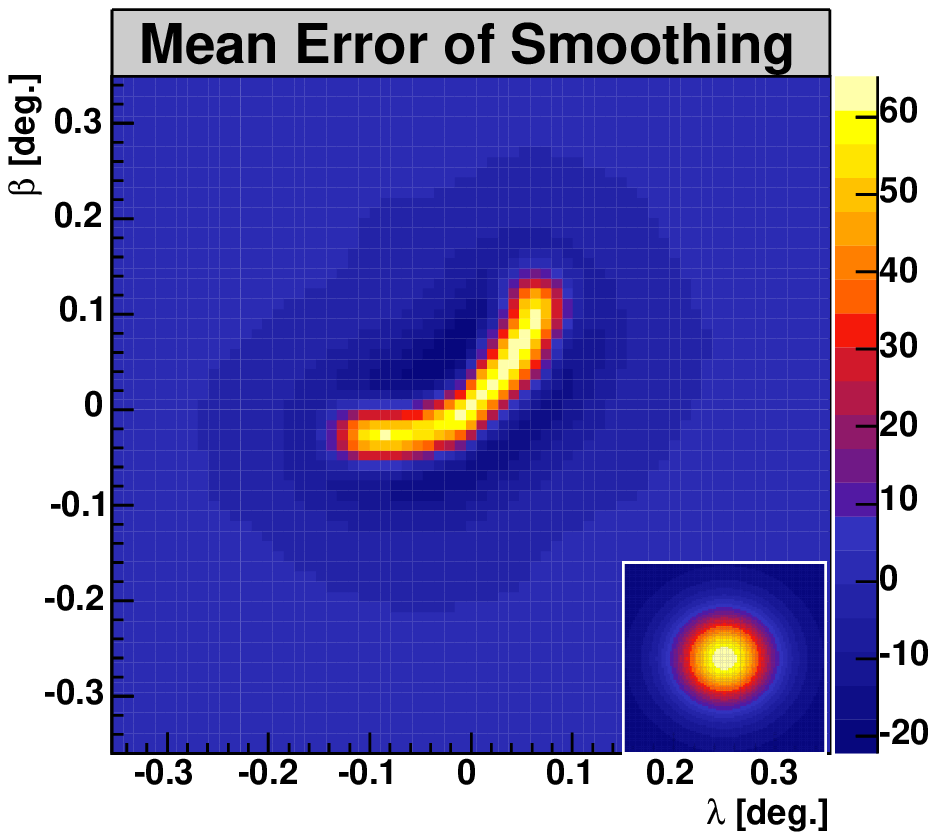}
  \end{minipage} \hfill
  \begin{minipage}[c]{0.45\linewidth}
    \caption[Mean Restoration Error ($IA>$400\,p.e.)]{Mean restoration error
      $(E^i(x,y))$ between the restored and the simulated image $(O)$. The mean
      is obtained from the restoration of 20 different count maps. The scale is
      in percentages of the peak intensity of the simulated excess. $E^i(x,y)$
      decreases with increasing iterations; background fluctuations increase.
      $E^i(x,y)$ is greatest for the smoothed count map, which is shown for
      comparison.}
    \label{fig:DecoError400}
  \end{minipage} \hfill
\end{figure}

Fig.~\ref{fig:DecoError400} shows the mean error $\overline{E}^i(x,y)$
(Eqn.~\ref{eqn:Ei}) of the 20 different restorations. The scale is chosen in
percentages of the peak intensity of the simulated excess (7 counts). The
largest error $(\overline{E}^i_{max})$ is found along the arc, where the
restored intensity is systematically too small. On both sides close to the arc,
the restored excess is systematically too high. A very similar error
distribution is seen for the smoothed map. $\overline{E}^i(x,y)$ decreases with
an increasing number of iterations as the restoration approaches the true image
$O$. For example, after 40 iterations, $\overline{E}^i_{max}$ is smaller than
$\sim$15\%. In comparison, the $\overline{E}^i_{max}$ of the smoothed count map
shows an error of $\sim$65\%.

\begin{figure}[ht!]
  \begin{minipage}[c]{0.5\linewidth}
    \includegraphics[width=\textwidth]{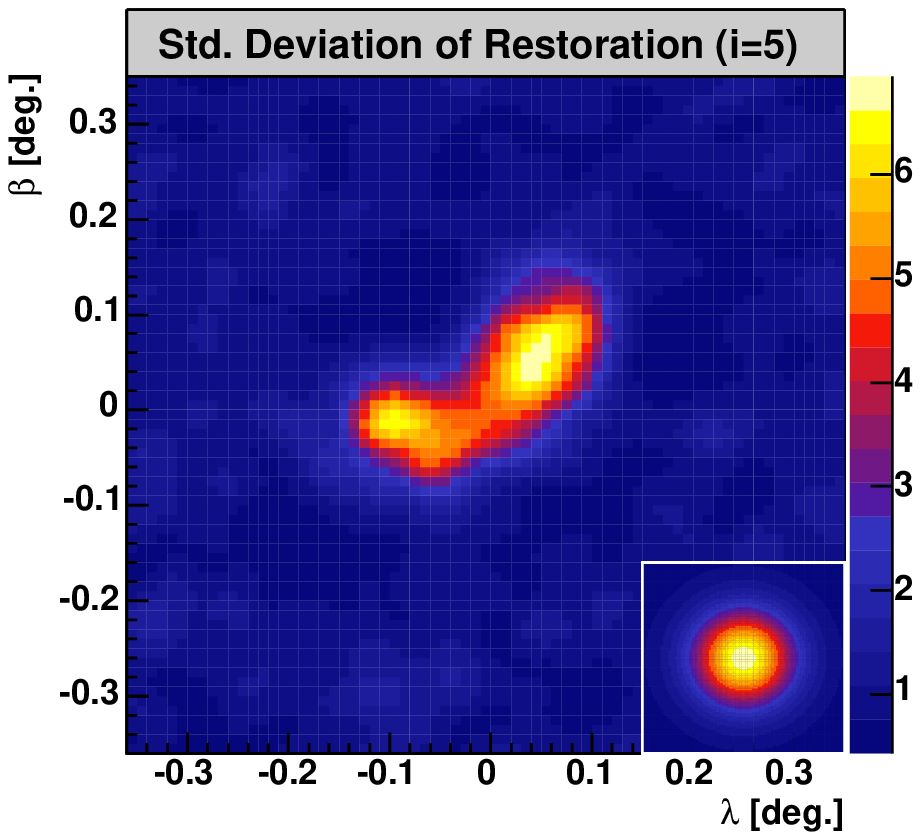}
  \end{minipage} \hfill
  \begin{minipage}[c]{0.5\linewidth}
    \includegraphics[width=\textwidth]{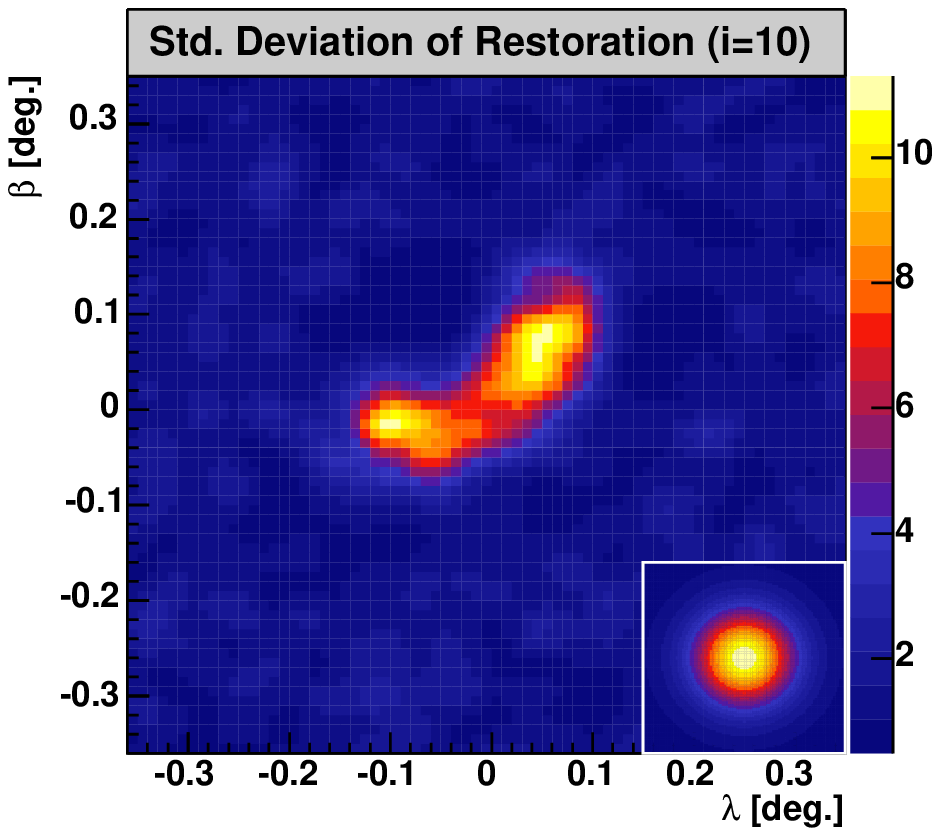}
  \end{minipage} \hfill
  \begin{minipage}[c]{0.5\linewidth}
    \includegraphics[width=\textwidth]{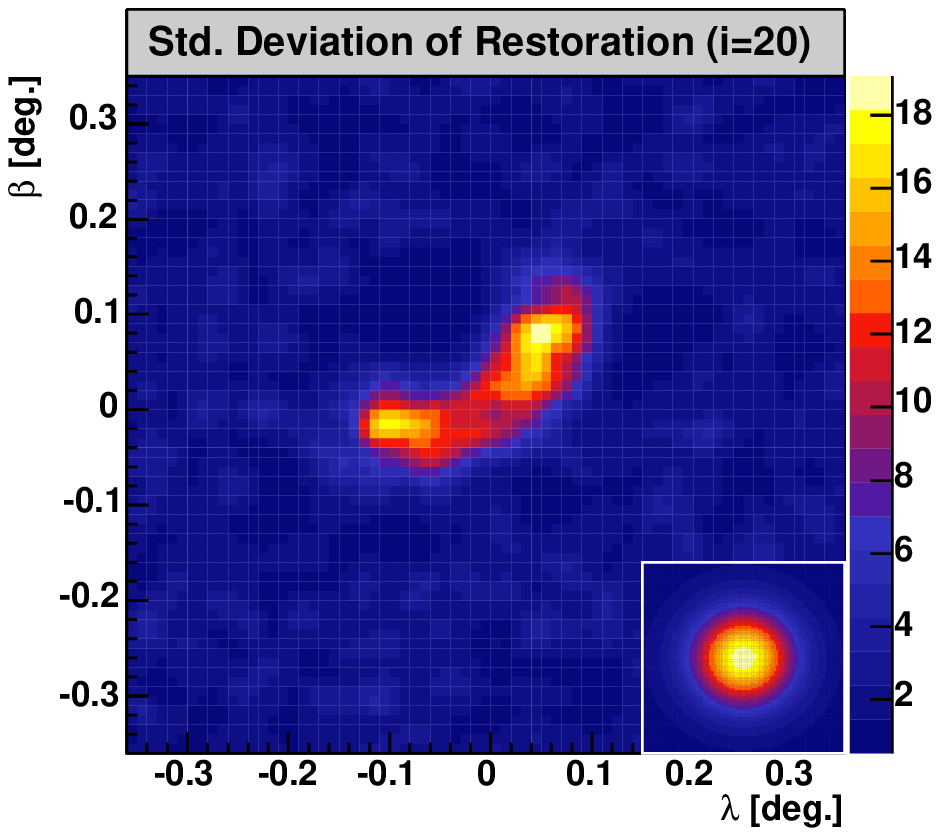}
  \end{minipage} \hfill
  \begin{minipage}[c]{0.5\linewidth}
    \includegraphics[width=\textwidth]{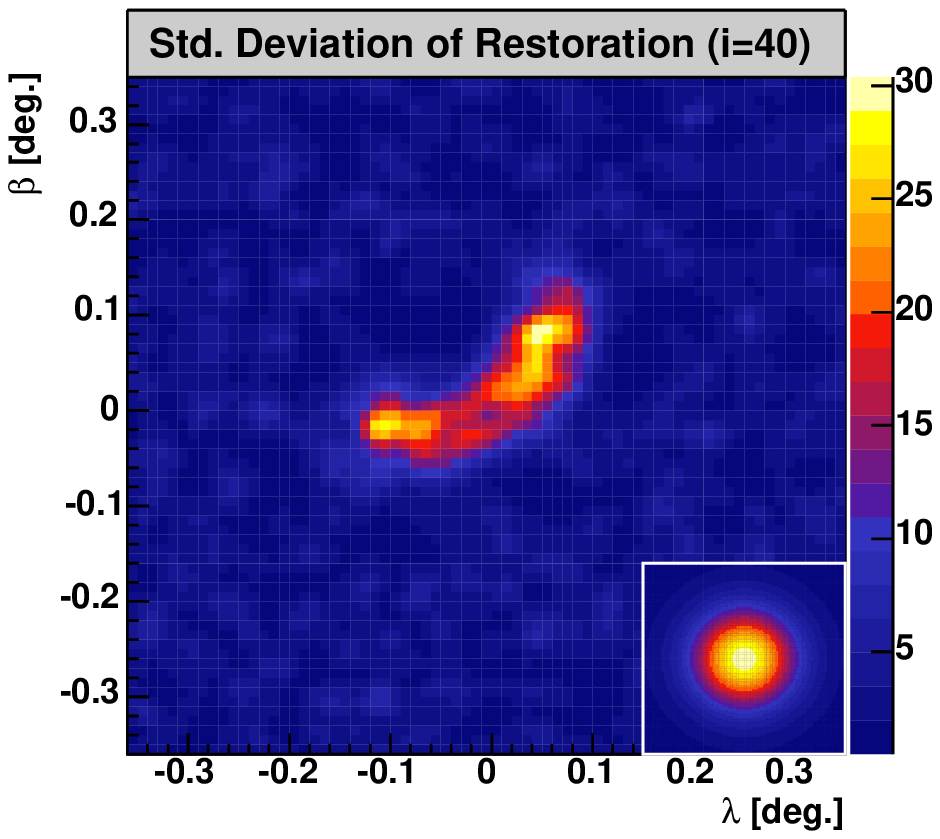}
  \end{minipage} \hfill
  \begin{minipage}[c]{0.5\linewidth}
    \includegraphics[width=\textwidth]{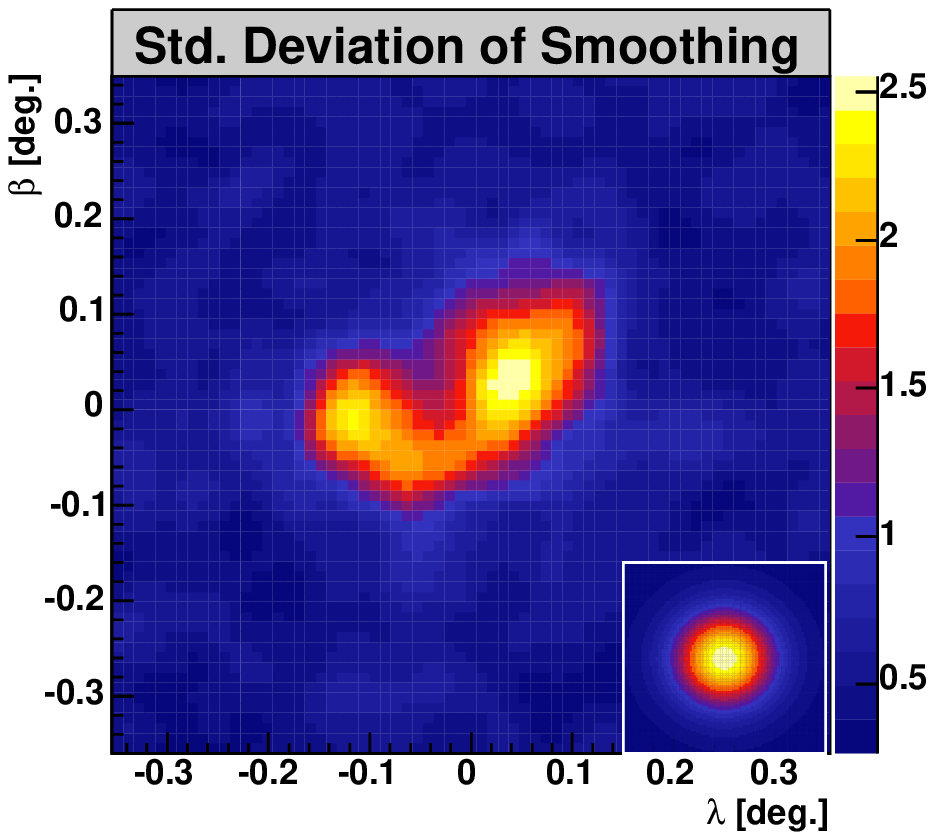}
  \end{minipage} \hfill
  \begin{minipage}[c]{0.45\linewidth}
    \caption[Standard Deviation of the Restoration ($IA>$400\,p.e.)]{Standard
      deviation $(S^i(x,y))$ of 20 different restored count maps. The scale is
      given in percentages of the peak intensity of the simulated excess. With
      an increasing number of iterations, $S^i(x,y)$ increases. The highest
      values are reached in the center region. After 40 iterations, $S^i(x,y)$
      reaches a level of $\sim$30\%. $S^i(x,y)$ of the smoothed count map is
      comparatively small.}
    \label{fig:DecoStdDev400}
  \end{minipage} \hfill
\end{figure}

Fig.~\ref{fig:DecoStdDev400} shows the standard deviation $S^i(x,y)$
(Eqn.~\ref{eqn:Si}) as obtained from the 20 different simulations. Again, the
scale is chosen in percentages of the peak amplitude of the simulated true
emission. An increase of the standard deviation with the number of iterations
can be seen. The largest deviations $S^i_{max}$ are found at the arc. After
40 iterations, they reach a level of $\sim$30\%. For comparison, the
standard deviation in the smoothed count map is also shown. It has the smallest
deviations --- only $\sim$3\%.

\begin{figure}[ht!]
  \begin{minipage}[c]{0.5\linewidth}
    \includegraphics[width=\textwidth]{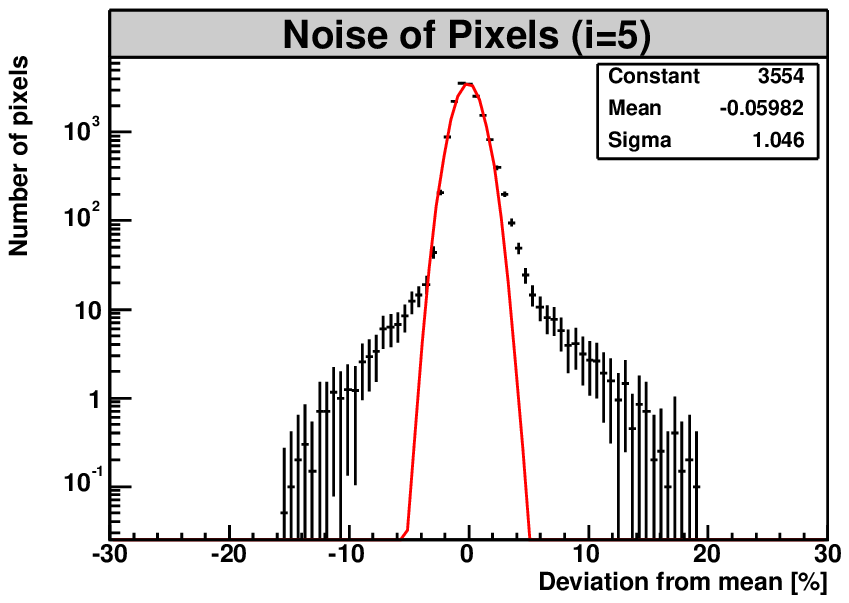}
  \end{minipage} \hfill
  \begin{minipage}[c]{0.5\linewidth}
    \includegraphics[width=\textwidth]{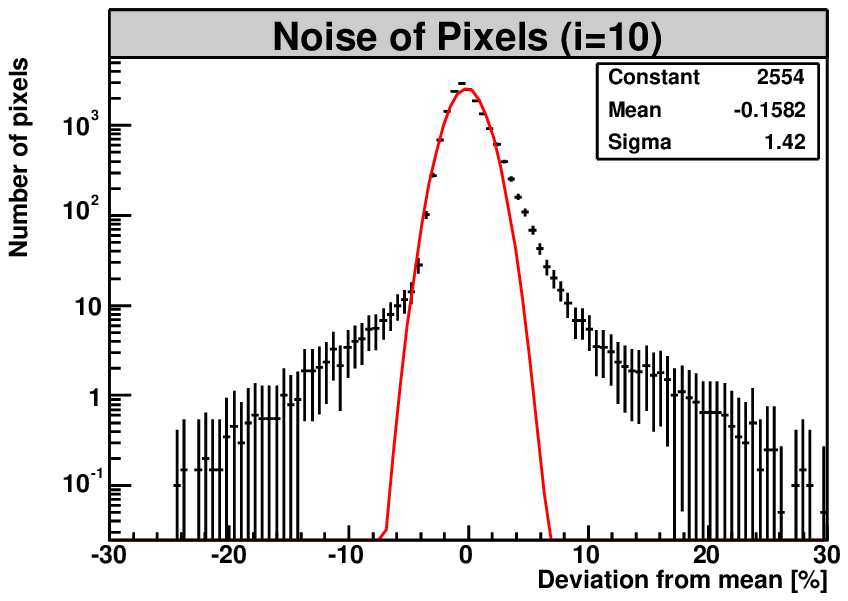}
  \end{minipage} \hfill
  \begin{minipage}[c]{0.5\linewidth}
    \includegraphics[width=\textwidth]{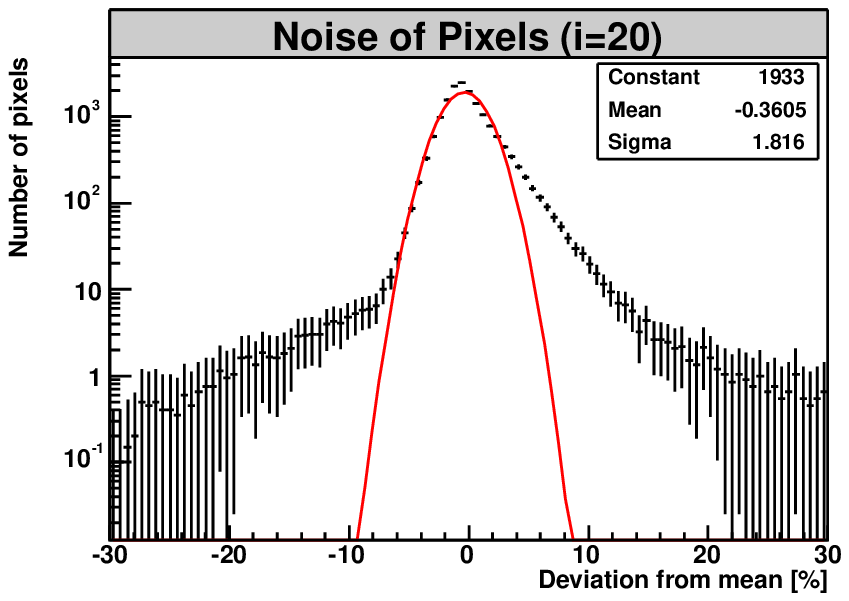}
  \end{minipage} \hfill
  \begin{minipage}[c]{0.5\linewidth}
    \includegraphics[width=\textwidth]{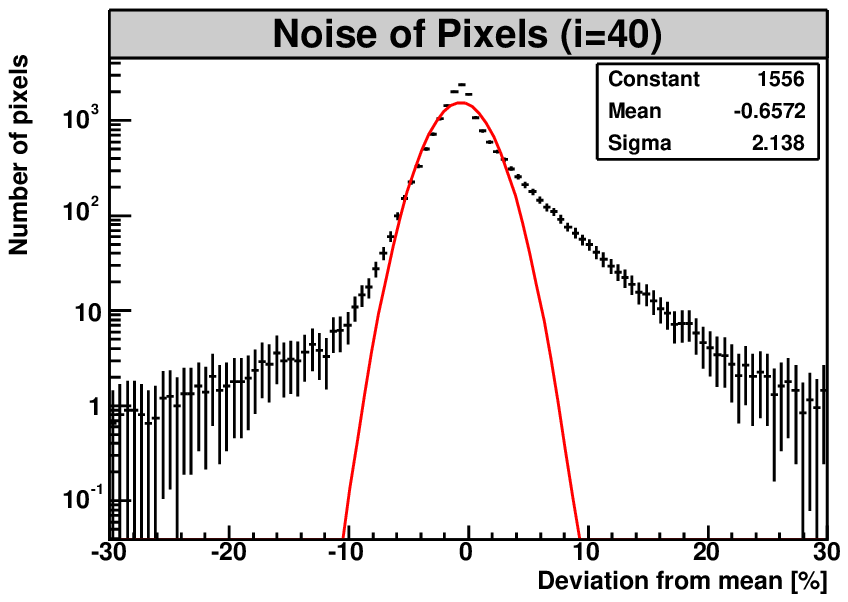}
  \end{minipage} \hfill
  \begin{minipage}[c]{0.5\linewidth}
    \includegraphics[width=\textwidth]{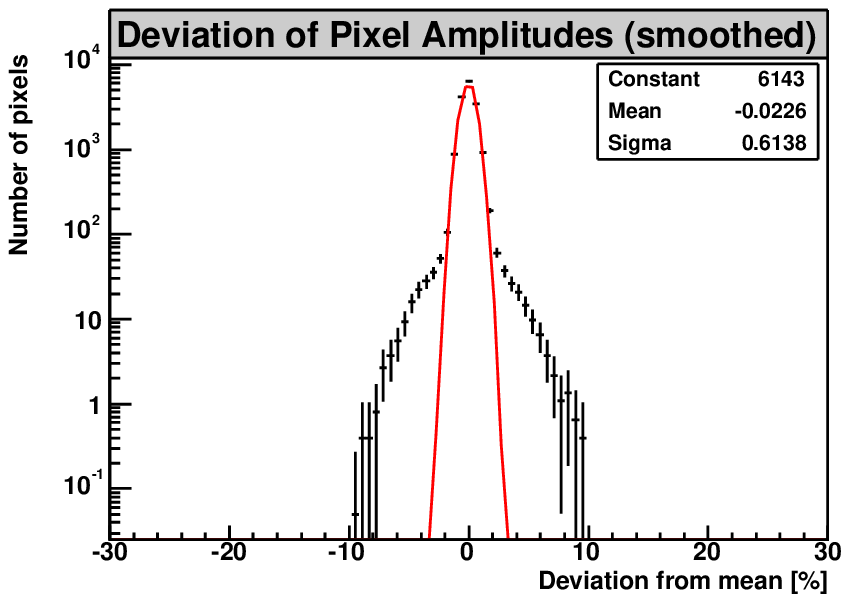}
  \end{minipage} \hfill
  \begin{minipage}[c]{0.5\linewidth}
    \includegraphics[width=\textwidth]{images/DecoError}   
  \end{minipage}\hfill
    \caption[Distribution of Noise and the Restoration Error
      ($IA>$400\,p.e.)]{Distribution showing each pixel's deviation from its
      expectation value after certain numbers of iterations $i$. The deviation
      is measured in percentages of the peak intensity of the simulated excess
      and represents the noise. A fit to a Gaussian distribution is shown
      (red). The last plot shows the mean relative norm of the restoration
      error $(\overline{\epsilon}_i)$ versus the number of iterations $i$. The
      minimum is found at $i_{\rm opt}\sim10$ iterations.}
    \label{fig:RestorationError400}
\end{figure}

Fig.~\ref{fig:RestorationError400} shows the distributions of the pixel noise
$D^i(x,y)$ (Eqn.~\ref{eqn:Di}) for different numbers of iterations $i$. The
distributions of the 20 simulations are added to one plot. The noise is
measured in percentages of the peak intensity of the excess of the true image
$O$. The distribution is fitted with by a Gaussian function with a zero-mean
(red line). The noise is clearly not Gaussian distributed. The standard
deviations of the distributions are 1.1, 1.5, 2.2 and 3.1\% for 5, 10, 20 and
40 iterations, respectively. The smoothed map has the smallest standard
deviation of 0.7\%. The deviations from a Gaussian distribution of noise is an
indication of statistical artefacts which are created in the restoration.

The last plot of Fig.~\ref{fig:RestorationError} shows the mean relative norm
of the restoration error $(\overline{\epsilon}_i)$
(Eqn.~\ref{eqn:RestorationError}). $\overline{\epsilon}_i$ is calculated for
the full region of $128\times128$ bins and as the mean of the 20 different
simulations of the 80\,p.e. map (black marker). The minimum is obtained after
about 10 iterations. The statistical error is $\sim$2\%. Within the range of
about $\pm$5 iterations from the minimum, $\epsilon_{\rm opt}$ changes by less
than 5\%, showing that the restorations for $5<i<15$ will provide restoration
errors similar to $\epsilon_{\rm opt}$.

The individual errors discussed here are summarized in
Tbl.~\ref{tbl:RestorationError}.

\clearpage

\begin{table}[h]
  \centering
  \caption[Errors of the Restoration for the Count Map of \MSH\
    ($IA>$80\,p.e.)]{Errors of the restoration of the count map of \MSH\
    ($IA>$80\,p.e.). The values $E_{max}^i$, $E_{max}^i$ and $\sigma_{D,i}$ are
    measured in percentages of the peak excess counts. $\sigma_{w,i}$ refers to
    the width of the excess of the true map $O$.}
  \bigskip
  \begin{tabular}{lccccc}
    \hline \hline
                               & Smoothed & $i=5$ & $i=10$ & $i=20$ & $i=40$\\
    \hline
    $\overline{E}^i_{max}$ [\%]    & 65   & 45    & 30     & 25     & 15    \\
    $S^i_{max}$ [\%]               & 3    & 3     & 5      & 8      & 12    \\
    $\sigma_{D, i}$ [\%]           & 1.5  & 1.3   & 2.2    & 3.4    & 5.3   \\
    $\overline{\epsilon}_i$        & 0.47 & 0.47  & 0.44   & 0.51   & 0.69  \\
    $\sigma_{w,i}$ [\%]            & 220  & 180   & 155    & 130    & 120   \\
    \hline \hline
  \end{tabular}
  \label{tbl:RestorationError}
\end{table}

\begin{table}[h]
  \centering
  \caption[Errors of the Restoration for the Count Map of \MSH\
    ($IA>$400\,p.e.)]{Errors of the restoration of the count map of \MSH\
    ($IA>$400\,p.e.). The values $E_{\rm max}^i$, $S_{\rm max}^i$ and
    $\sigma_{D,i}$ are measured in percentages of the peak excess counts.}
  \bigskip
  \begin{tabular}{lccccc}
    \hline \hline
                               & Smoothed & $i=5$ & $i=10$ & $i=20$ & $i=40$\\
    \hline
    $\overline{E}^i_{max}$ [\%]    & 65   & 40    & 30     & 20     & 15    \\
    $S^i_{max}$ [\%]               & 3    & 8     & 13     & 20     & 30    \\
    $\sigma_{D, i}$ [\%]           & 0.7  & 1.2   & 1.8    & 2.5    & 3.4   \\
    $\overline{\epsilon}_i$        & 0.47 & 0.42  & 0.36   & 0.39   & 0.49  \\
    \hline \hline
  \end{tabular}
  \label{tbl:RestorationError400}
\end{table}

\subsubsection{Systematic Errors of the Simulations}
While previous discussion refers to the statistical errors produced by Poisson
noise, it is also possible to estimate the systematic errors of the restored
sky maps. The major source for systematic errors has been found in deviations
of the morphology. The influence of the morphology on the restorations has been
investigated with simulations. Before, unphysical images were excluded and the
set of possible maps of $O$ were reduced to a realistic subset of smooth maps
with simple geometric structures. With these restrictions, only small
variations were possible in order to reproduce count maps of similar
appearance. The parameters which were investigated included the size as well as
the aspect ratio of the excess distributions. The results suggest that an
estimate for the systematic error would be less than $\sim$10\% and $\sim$20\%
of the peak intensity of the excess for the 80 and 400\,p.e. maps,
respectively.

Other potential sources of systematic errors have also been investigated,
including the number of excess and background events and the size of the PSF.
No systematic errors related to these parameters have been found.

\section{Conclusion}
Two methods of image restoration for H.E.S.S. count maps have been discussed
here: smoothing by convolution with a Gaussian function and the Richardson-Lucy
algorithm for image deconvolution. Both methods can significantly reduce the
statistical noise in count maps and reveal morphological details which are
hidden by statistical noise.

Convolution is a straightforward approach which provides stable results which
are only little affected by statistical fluctuations. Simulations show that the
smoothed map for the 80 and 400\,p.e. maps of \MSH\ have a restoration error of
about (65$\pm$3)\%.

On the other hand, the Richardson-Lucy algorithm is more sensitive to noise in
count maps and therefore requires simulations for error analysis. With
simulations, it is possible to estimate the quality and error of the
restorations and to limit the statistical fluctuation to an acceptable level.
At the expense of increasing noise, the RL algorithm can provide a high
restoration of morphological details. Depending on the objective of the
analysis, one can choose between a restoration with less details and small
errors or with higher details but also more statistical artefacts. For the
analysis of the morphology of \MSH\ in this work, preference was given to
smaller errors. 10 iterations with the RL algorithm were considered a good
compromise. In the case of the 80\,p.e., this provides a stable restoration
with a mean error of at most (30$\pm$5)\% from the true value. The noise in the
restored map is Gaussian distributed and has a standard deviation of 2.2\%. The
400\,p.e. map after 10 iterations with the RL algorithm provides a relative
error of the excess of at most (30$\pm$13)\%. These maps are in agreement with
the maps obtained by smoothing. The simulations of the \g-ray maps of
\MSH\ have shown that the application of the RL can be useful and provide
\g-ray maps of high resolution.

\chapter{Search for Pulsed Emission from Pulsar} \label{chp:PeriodicAnalysis}

When \g\ radiation from pulsars is detected, the question arises whether a part
of this radiation is pulsed. The detection of pulsed \g\ radiation from pulsars
is of special interest for the development of pulsar models, since it can
provide information about the emission process and constrain parameters.
Unfortunately the detection of pulsed TeV \g\ radiation from a pulsar wind
nebula is more difficult since the nebula often emits a significant amount of
constant \g\ radiation. If at all, the pulsed flux will only be observable as
one component of the \g-ray flux from the source region. This chapter will
introduce methods for the analysis of H.E.S.S. data for pulsed emission,
statistical tests for evaluation and the calculation of flux upper limits.

\section{Pulsar Light Curves and Phasograms}
Methods for finding pulsed emission from pulsars rely on an analysis of the
pulsar light curve. The pulsar light curve can be represented in a phasogram. A
phasogram shows the light curve during one period, usually obtained as an
average over many periods. To produce a phasogram from H.E.S.S. data, the
standard analysis is first carried out to reconstruct the \g-ray showers as
described in Sec.~\ref{chp:StandardAnalysis}. Then the events contained in a
small signal region which encompasses the pulsar are filled in an phasogram.
Since the event statistic is limited, between 16 to 25 bins are chosen. The
phase of each event is calculated from its time stamp and the pulsar ephemeris.
A H.E.S.S. phasogram for the Crab Pulsar is shown in
Fig.~\ref{fig:HESSCrabLightCurve}.

\section{Ephemerides}
Pulsar ephemeris contain a set of parameters which can describe the pulsar
phase as a function of time $(\phi(t))$. Since pulsars behave like a slowly
decelerating sphere, the first terms of the Taylor series
\begin{equation}
  \phi(t) = \phi_0 +f\cdot(t-T_0) +\frac{1}{2}\dot{f}\cdot(t-T_0)^2
  \label{eqn:Phase}
\end{equation}
already provide a description of the pulsar phase with high precision. The
parameters in the Taylor series already constitutes an ephemeris. In this
equation $\phi_0$ is the phase at the reference time $T_0$ and $f$ is the
frequency of rotation of the pulsar. Since the pulsar phase is often very
precisely determined by radio observation, ephemeris are usually derived from
radio measurements. Given the ephemeris of a pulsar, one can calculate its
phase at any time (t). However, pulsars do not exactly obey
Eqn.~\ref{eqn:Phase} --- they are also subject to other processes such as
glitches. Therefore, an ephemeris is only valid within a certain validity
period which can range from a few days, as in the case of the Vela Pulsar, to
many years, as in the case of \PSR. The validity period is therefore also part
of the ephemeris. There are some other parameters necessary for a precise
definition of an ephemeris. An example is shown in Tbl.~\ref{tbl:GROFormat}.

Several formats for ephemeris exist. A common format for pulsar ephemeris is
the GRO format. It was first used by the Compton Gamma-Ray Observatory
community and is now the standard format of the Australian Pulsar Timing Data
Archive (\citet{ATNF}) from the Australia Telescope National Facility (ATNF).
Since several ephemerides have been taken from this archive for the analysis of
H.E.S.S. data, this format was implemented to the H.E.S.S. standard analysis.
The GRO format consists of one line containing 13 parameters in the case of a
single pulsar and of two lines with 10 additional parameters in the case of a
binary system. An illustration of the ephemeris format and parameters is
given in Tbl.~\ref{tbl:GROFormat}.

\begin{table}[tbh!]
  \centering
  \caption[GRO Ephemeris Format]{Structure and contents of the GRO ephemeris
    format as documented at \citet{ATNF}. The basic parameters are required for
    any pulsar. The binary parameters are additional parameters for pulsars in
    binary systems.}
    \label{tbl:GROFormat}
  \bigskip 
  \begin{tabular}{l|c|p{.68\textwidth}}
    \hline \hline
    \bf Character & \bf Symbol & \bf Meaning \\
    \hline
    \multicolumn{2}{l}{Basic Parameters} & \\
    \hline
    1-8 & & Pulsar name (truncated if a J2000 name) \\
    10-21 & $\alpha$ & J2000 right ascension [hh mm ss.sss] \\
    23-34 & $\delta$ & J2000 declination [-dd mm ss.ss] \\
    36-40 & $T_{\rm min}$ & Start of validity range [MJD] \\
    42-46 & $T_{\rm max}$ & End of validity range [MJD] \\
    48-62 & $t_{GEO}$ & TDB epoch of pulse frequencies and infinite frequency UTC pulse TOA at geocenter [MJD] \\
    64-80 & $f$ & Pulse frequency at the Solar system barycenter [Hz] \\
    82-93 & $\dot{f}$ & 1st time derivative of barycentric pulse frequency [s$^{-2}$] \\
    96-104 & $\ddot{f}$& 2nd time derivative of barycentric pulse frequency [s$^{-3}$] \\
    106-109 & & RMS residual of fit in milliperiods \\
    111 & & Letter code indicating origin of data (A = Australia) \\
    115-119 & & Planetary system ephemeris used for barycenter correction \\
    121-130 & & Full J2000 pulsar name \\

    \hline
    \multicolumn{2}{l}{Binary Parameters} & \\
    \hline
    1-8 & & Pulsar name (truncated if a J2000 name) \\
    10-25 & $P_b$ & Orbital period (at the Solar system barycenter) [s] \\
    26-37 & x & Semi-major axis of pulsar orbit [s] \\
    39-48 & e & Orbital eccentricity \\
    50-63 & $T_0$ & TDB epoch of periastron passage [MJD] \\
    65-74 & $\omega$ & Longitude of periastron [deg] \\
    76-82 & $\dot{\omega}$ & Rate of periastron advance [deg/yr] \\
    84-91 & $\gamma$ & Time dilation and gravitational redshift term [s] \\
    93-102 & $\dot{P_b}$ & First time derivative of orbital period \\
    104 & & Letter code indicating origin of data (A = Australia) \\
    \hline \hline
  \end{tabular}
\end{table}

\section{Time of Flight Corrections}
Although the pulsar phase is described by the ephemeris, one cannot directly
apply Eqn.~\ref{eqn:Phase} to the event time provided by the time stamp in
order to calculate the correct pulsar phase $(\phi(t))$. The reasons for this are mainly
the changing distance between the pulsar and the observatory, but also e.g.
relativistic effects. Therefore it is necessary to correct the event time
before Eqn.~\ref{eqn:Phase} is applied. These corrections are described below.

\subsection{Solar System Barycenter Correction}
\begin{figure}[ht]
  \begin{minipage}[c]{0.5\linewidth}
    \includegraphics[width=\textwidth]{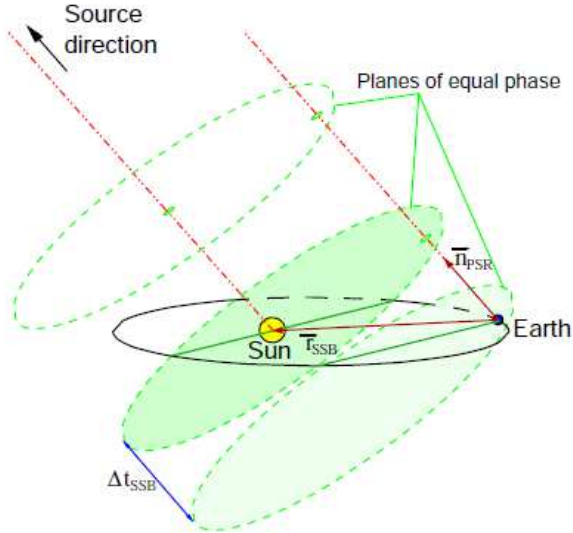}
  \end{minipage}\hfill
  \begin{minipage}[c]{0.45\linewidth}
    \caption[Illustration of the Solar System Barycenter
    Correction]{Illustration of the solar system barycenter correction. Planes
    of equal phases are perpendicular to the line of sight of the pulsar. Thus
    the observed pulse phase changes with the position of the earth. A similar
    phase shift is introduced by the orbit of a pulsar in a binary system.
    (Figure taken from \cite[pg. 44]{Schmidt:Diplomarbeit}.)}
  \label{fig:SSBCorrection}
  \end{minipage}
\end{figure}

The solar system barycenter (SSB) correction compensates for the changing
distance between the pulsar and the observatory caused by the motion of the
earth. The situations is illustrated in Fig.~\ref{fig:SSBCorrection}.
Eqn.~\ref{eqn:Phase} is correct, if applied to the arrival times at the SSB,
since the SSB provides a reference frame of constant velocity in good
approximation. Arrival times of the same pulse recorded at the observatory have
a time difference $(\Delta t_{\rm SSB})$ with respect to the SSB depending on
the earth's position. The relation between the arrival time at the SSB $(t_b)$
and at the observatory $(t)$ is given as
\begin{equation}
  t_b = t - \Delta t_{\rm SSB}
  = t - \frac{\vec{r}_{\rm SSB}(t)\cdot\hat{n}_{\rm PSR}}{c},
\end{equation}
where $r_b$ is a vector to the SSB from the phase center of the observer,
$\hat{n}_{\rm PSR}$ is a unit vector pointing from the SSB to the pulsar and
$c$ is the speed of light. $r_b$ is calculated as
\begin{equation}
  \vec{r}_{\rm SSB}(t)=\vec{e}_b(t)+\vec{e}_r(t),
\end{equation}
where $\vec{e}_b(t)$ is the vector from the geocenter to the SSB and
$\vec{e}_r(t)$ is the vector from the geocenter to the observer. $\vec{e}_r(t)$
is determined by
\begin{equation}
  \vec{e}_r(t) = r_\oplus \cdot
  \begin{pmatrix}
    \cos \lambda(t) \cos \phi_z \\
    \sin \lambda(t) \cos \phi_z \\
    \sin \phi_z \\
  \end{pmatrix},
\end{equation}
where $r_\oplus$ is the earth's radius and $(\lambda,\theta_z)$ are the
geographic coordinates of the observer. The coordinates of the SSB and the
geocenter are obtained from the DE200 solar system ephemeris (\citet{DE200}).
The DE200 ephemerides provide the earth's position for times between the years
1980 and 2020 with an accuracy of within a few meters corresponding to a time
resolution of nanoseconds, which is more than sufficient for time of flight
corrections. Further details regarding the coordinate and time transformations
for H.E.S.S. can be found in \citet{GillessenThesis}.

The magnitude of $(\Delta t_{\rm SSB})$ can be estimated with a simple example as follows.
A geostationary observatory is orbiting the sun with an approximate velocity
$v_{\rm earth} \sim$ 30\,km/s. If a pulsar with a typical frequency of about
$f_{\rm P}$ = 30\,Hz emits pulses, then the distance between two pulses is
$\lambda_{\rm P} = c/f_{\rm P} \approx$ 10000\,km. The earth travels this
distance in $T_{\rm P} = \lambda_{\rm P}/v_{\rm earth}$ = 333\,s. Consequently,
if the earth is moving directly towards or away from the pulsar, the pulsar
phase can shift by one period in about 5 minutes.

\subsection{Binary Correction}
If the pulsar is part of a binary system, then phase shifts also arise due to
the orbital movement of the pulsar around its companion. The phase shift can be
measured and the orbit can be parameterized and summarized in the pulsar
ephemeris. A mathematical model (BT), which describes the time shift as
observed at the SSB, was derived by Blandford and Teukolsky (\citet{BT}). The BT
model defines the transformation from the time of arrival at the SSB $(t_b)$ to
the pulsar proper time $(T)$. The BT formula contains Keplerian elements common
for the description of binary systems and relativistic terms as follows:
\begin{eqnarray}
  t_b - t_0 & = & T + \{ x \sin \omega (\cos E - e)
  + [ x \cos \omega (1-e^2)^{1/2} + \gamma ] \sin E \} \times \nonumber \\
  && \{ 1 - \frac{2 \pi}{P_b}
    [ x \cos \omega (1-e^2)^{1/2} \cos E - x \sin \omega \sin E ] \times
    \nonumber \\
    && (1-e \cos E)^{-1} \}.
  \label{eqn:BT}
\end{eqnarray}
Here $P_b$, $e$ and $\omega$ are the binary orbital period, orbital
eccentricity and longitude of the periastron respectively. The longitude of the
periastron is defined as the angle between the periastron and the ascending
node\footnote{The ascending node is the point in the orbit of an object when it
crosses the ecliptic (i.e. celestial equator) while moving from south to
north.}. $x$ is the projected semi-major axis\footnote{The projected semi-major
axis is the semi-major axis of the apparent ellipse, i.e. of the projection of
the actual elliptical orbit onto a plane perpendicular to the line of sight of
the observer.} of the pulsar orbit in time units. $\gamma$ measures the
combined effect of gravitational redshift and time dilation. $t_0$ is an
arbitrary reference time. The eccentric anomaly $E$ is defined by Kepler's
equation,
\begin{equation} 
  E - e \sin E = \frac{2 \pi}{P_b} (t_b - T_0),
\end{equation} 
in which $T_0$ is a reference time of periastron passage, measured in
the TDB system.

\subsection{TEMPO and CRASH}
The timing corrections for the H.E.S.S. data are done with the Coordinate
Transformation Software for H.E.S.S. (CRASH) (\citet{H.E.S.S.Software}), which
is part of the H.E.S.S. analysis software. CRASH implements the
DE200 solar system ephemeris (\citet{DE200}) and provides the CRASH pulsar
class, which can calculate the pulsar phase at the SSB based on pulsar
ephemerides. CRASH can read pulsar ephemerides of different formats. The
implementations were verified with TEMPO (\citet{TEMPO}), the standard program
for pulsar timing in radio astronomy. TEMPO can deduce pulsar rotation as well as
astrometric and binary parameters by fitting models to pulsar data. It also
contains an implementation of the BT model. Information about the
implementation of the GRO format in CRASH and its verification can be found in
\citet{Breitling}. The calculation of the pulsar phase including the SSB
correction is done in the H.E.S.S. standard analysis chain, which can also
produce phasograms of arbitrary energy ranges.

\begin{figure}[ht]
  \setlength{\abovecaptionskip}{0cm}
  \begin{minipage}[c]{0.5\linewidth}
    \includegraphics[width=\textwidth]{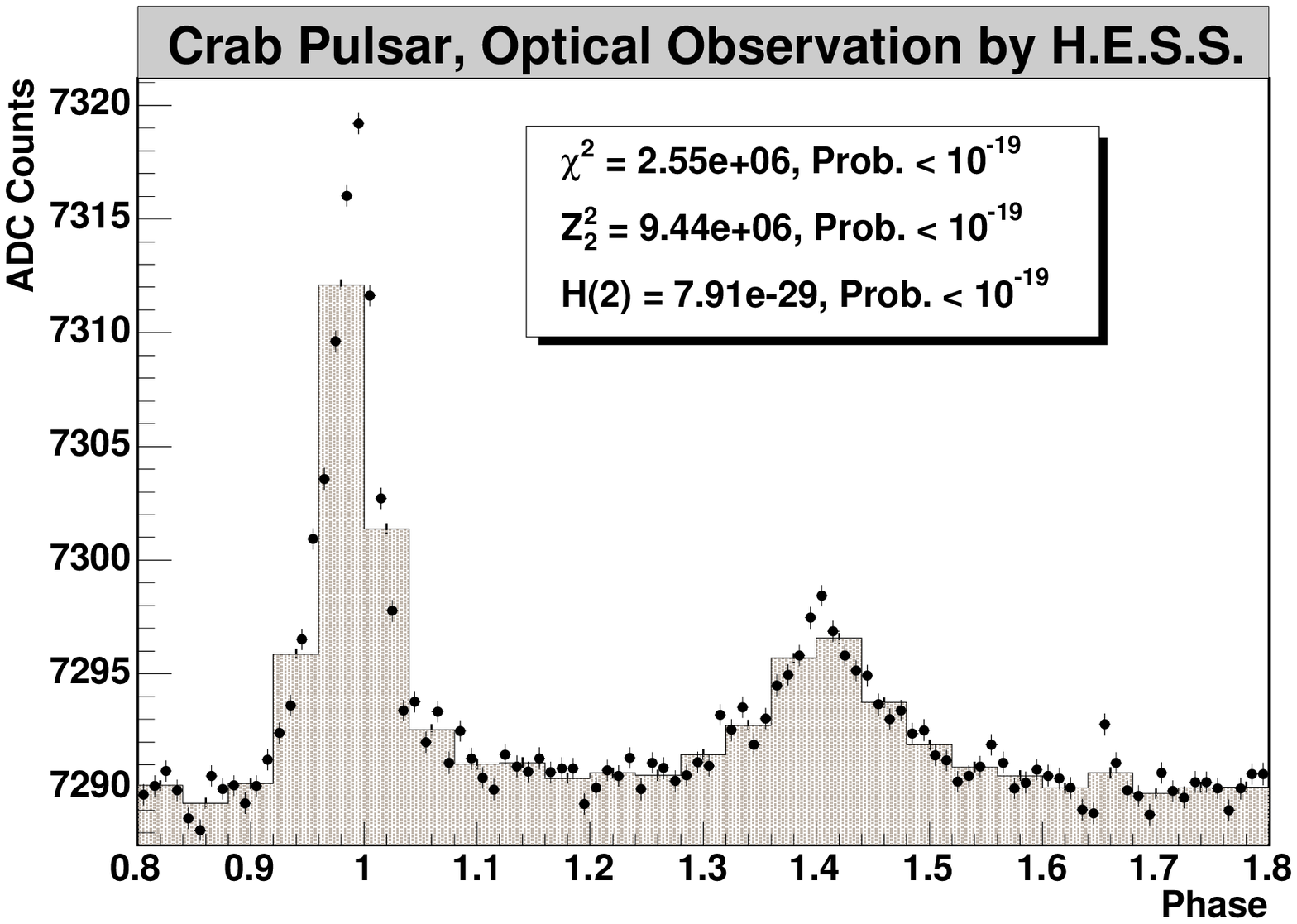}
  \end{minipage}
  \begin{minipage}[c]{0.5\linewidth}
    \includegraphics[width=\textwidth]{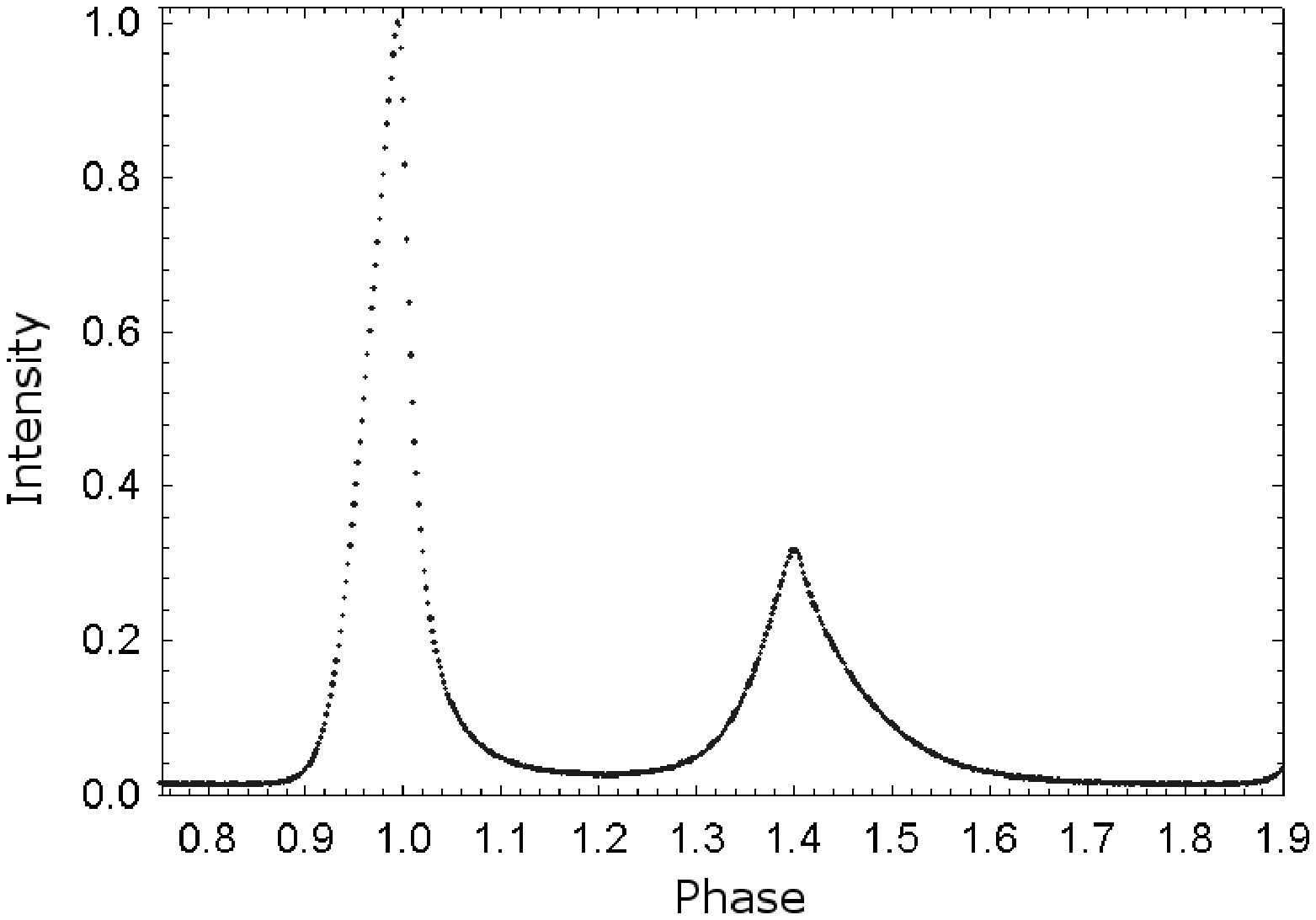}
  \end{minipage}
  \begin{minipage}[t]{0.5\linewidth}
    \caption[H.E.S.S. Optical Light Curve of the Crab Pulsar]{Phasogram for
    100\,s of the optical H.E.S.S. data from the Crab Pulsar obtained with the
    standard analysis. The same data is represented with a binning of 25 and
    100 bins.}
  \label{fig:HESSCrabLightCurve}
  \end{minipage}\hfill
  \begin{minipage}[t]{0.5\linewidth}
    \caption[OTIMA Optical Light Curve of the Crab Pulsar]{Optical light curve
    of the Crab Pulsar as measured with high precision by OPTIMA. (Figure taken
    from \citet{Kanbach:2005kf}.)}
  \label{fig:OptimaCrabLightCurve}
  \end{minipage}
\end{figure}

The implemented timing capabilities of the H.E.S.S. standard analysis were
tested with optical H.E.S.S. data from the Crab Pulsar (PSR\,B0531+21). In this
test, one telescope was equipped with a special PMT camera which was developed
for the detection of optical light curves from the Crab Pulsar. Further details
about this experiment can be found in \citet{Hinton:OpticalCrab} and
\citet{Masterson:2003ICRC}. The ephemeris for this analysis was taken from the
Australian Pulsar Timing Data Archive and is shown in
Tbl.~\ref{tbl:EphemerisCrab}. Fig.~\ref{fig:HESSCrabLightCurve} shows the
phasogram with the mean ADC counts of 100\,s ($2.9\times10^6$ events) of
optical data taken on November 23, 2003 (21:37 UTC). The histogram of 25 bins
(grey) is overlaid with a histogram with 100 bins which can resolve the pulse
shape more clearly. For comparison, the optical light curve as obtained by
OPTIMA in January 2002 is shown in Fig.~\ref{fig:OptimaCrabLightCurve}. OPTIMA
is a high-speed photo-polarimeter for optical pulsar measurements of high
precision (\citet{Kanbach:2005kf}). A good agreement between the two light
curves is found.

\begin{table}[htb!]
  \centering
  \caption[GRO Ephemeris of the Crab Pulsar]{GRO ephemeris of the Crab pulsar
    taken from the ATNF archive (\citet{ATNF}). The parameters are documented
    in Tbl.~\ref{tbl:GROFormat}. The validity range is given in Modified Julian
    Dates and corresponds to a range from January 31, 2004 to July 3, 2005.}
    \label{tbl:EphemerisCrab}
  \bigskip 
  \begin{tabular}{lc}
    \hline \hline
    Character & Value \\
    \hline
    $\alpha$      & 05 34 31.972 \\
    $\delta$      & 22 00 52.07 \\
    $T_{\rm min}$ & 52944\,MJD \\
    $T_{\rm max}$ & 52975\,MJD \\
    $t_{GEO}$     & 52960.000000296\,MJD \\
    $f$           & 29.8003951530036\,s$^{-1}$ \\
    $\dot f$      & -3.73414D-10\,s$^{-2}$ \\
    $\ddot f$     & 1.18D-20\,s$^{-3}$\\
    \hline \hline
  \end{tabular}
\end{table}


\newpage

\section{Statistical Test} \label{sec:Tests}
For many pulsar observations with H.E.S.S., in particular for observation of
PWNs, the detection of pulsed emission requires the identification of a small
pulsed component from a strong constant background signal. The situation is
complicated by the fact that each pulsar has its own characteristic light curve
which can differ with wavelength (cf. Fig.~\ref{fig:PulsarLightCurves}). Thus
the position and shape of the pulse are not known and its identification is
more difficult. For detecting an unknown light curve above a strong background,
statistical tests can be applied. Common tests are the $\chi^2$, the $Z^2_m$
and the H test. They are tests of uniformity of a phasogram. For weak pulse
shapes the tests differ in their power. The power of a test is its efficiency
to reject the hypothesis of a uniform phase distribution, if this is not true.
A detailed discussion of these tests and their application can be found in
\citet{DeJager1989} and \citet{DeJager1994}. Here they are briefly introduced.

\subsection{The $\chi^2$ Test}
Pearson's $\chi^2$ test is applied by fitting a constant to the phasogram. The
number of degrees of freedom is $n-1$, where $n$ is the number of bins in the
phasogram. The typical choice for H.E.S.S. data is $16<n<25$. The $\chi^2$
value or rather the corresponding $\chi^2$ probability then determine whether
the hypothesis of a uniform distribution can be rejected or not. The
$\chi^2$ test is most efficient for single and narrow pulse shapes but less for
other, especially sinusoidal, shapes. Another disadvantage is the dependence on
the binning.

\subsection{The $Z^2_m$ Test}
The $Z^2_m$ test can be considered as a complementary test to the
$\chi^2$ test, in the sense that it is more sensitive to sinusoidal signals
with the periodicity $m$. $Z^2_m$ is defined through the sum of trigonometric
moments $\alpha_j$ and $\beta_j$ as
\begin{equation}
  Z^2_m = 2N\sum^m_{j=1}{(\alpha^2_j+\beta^2_j)},
\end{equation}
with
\begin{equation}
  \alpha_j = \frac1N\sum^N_{i=1}{\cos(j\,\phi_i)}
\end{equation}
and
\begin{equation}
  \beta_j = \frac1N\sum^N_{i=1}{\sin(j\,\phi_i)}.
\end{equation}
Here $N$ is the number of events in the phasogram and $\phi_i$ is the phase of
event $i$. $Z^2_m$ is $\chi^2$ distributed for $2m$ degrees of freedom. The
probability is calculated accordingly. $m=2$ is a good choice for the detection
of wider pulse shapes, which is often used for the analysis of H.E.S.S. data.
Another advantage of the
$Z^2_m$ test, besides sensitivity to sinusoidal signals, is its independence from the binning of the phasogram.

\subsection{The H test}
The H test is an improved version of the $Z^2_m$ test which has an increased
sensitivity for arbitrary pulse shapes. It is very useful when no a priori
information about the pulse shape is available. The H test is as powerful as
the $\chi^2$ test and more powerful than the $Z^2_m$ test in the case of more
than two pulse peaks (cf. \citet{DeJager1989} and \citet{DeJager1994}). $H$ is
defined through the $Z^2_m$ test as follows:
\begin{equation}
  H = \max_{1<m<20} (Z^2_m-4m+4)
\end{equation}
The selection of the maximum value of $H$ accounts for its increased power in
comparison to the $Z^2_m$ test. In the case of the absence of a signal, $H$ is
distributed by an exponential function and the probability $P$ for obtaining a
lager value for $H$ is determined as
\begin{equation}
  P(>H) = \exp(-0.4 H).
\end{equation}
This test is a favorable choice for the analysis of most H.E.S.S. pulsar data,
since often no priori information about the pulse shape is available.

\subsection{Application of Tests to the Optical Crab Pulsar Data}
These three tests have been applied to the optical H.E.S.S. data from the Crab
Pulsar shown in Fig.~\ref{fig:HESSCrabLightCurve}. The results are listed in
the panel on the figure. For all three tests, the probabilities for a uniform
light curve are close to zero as is expected for data with such a clear
pulse shape as in the Crab Pulsar. The numerical values of the probabilities
are less than the numerical accuracy of $10^{-19}$.

\section{Calculation of Upper Limits} \label{sec:UpperLimits}
If pulsed emission cannot be detected, one can still determine an upper limit
of the pulsed flux in a phase region. The phase region is referred to as an
On-region and the remainder as an Off-region. The On-region is usually chosen
according to the light curve at other wavelengths where a pulsed signal has
been observed.

Eqn.~\ref{eqn:DifferentialFlux} and Eqn.~\ref{eqn:DifferentialFluxError} can
provide the flux $\Phi$ and the corresponding error $\sigma_\Phi$ within a
given energy range. A common choice for H.E.S.S. data is the energy range above
the threshold energy or above 1\,TeV. $\Phi$ and $\sigma_\Phi$ are sufficient
to calculate the upper limit of the flux $(UL_\Phi)$ for a confidence level
(CL). There have been different methods proposed for the calculation of upper
limits. Here the unified approach by \citet{FeldmanCousins} was chosen, which is
described in more detail in App.~\ref{app:FeldmanCousins}. $UL_\Phi$ is
determined with an upper limit function $F_{UL}(\Phi,\sigma_\Phi,\rm CL)$ as
described in App.~\ref{app:FeldmanCousins}. Examples for the application of the
H.E.S.S. standard analysis and the calculation of upper limits according to
Feldman and Cousins are found in \citet{HDGSPulsarULs}, where H.E.S.S. data of
young pulsars are analyzed.

\chapter{Detection of \MSH} \label{chp:MSHAnalysis}
After the introduction and verification of the H.E.S.S. data analysis
techniques, they are now applied to the H.E.S.S. data from \MSH. The
observation, the data, the analysis and the results are subject of this
chapter. The detection, position, energy spectrum and morphology of \MSH\ as
well as the analysis for pulsed emission from \PSR\ are discussed.

\section{Observation}
\begin{figure}[tbh!]
  \begin{minipage}[c]{0.5\linewidth}
    \includegraphics[width=\textwidth]{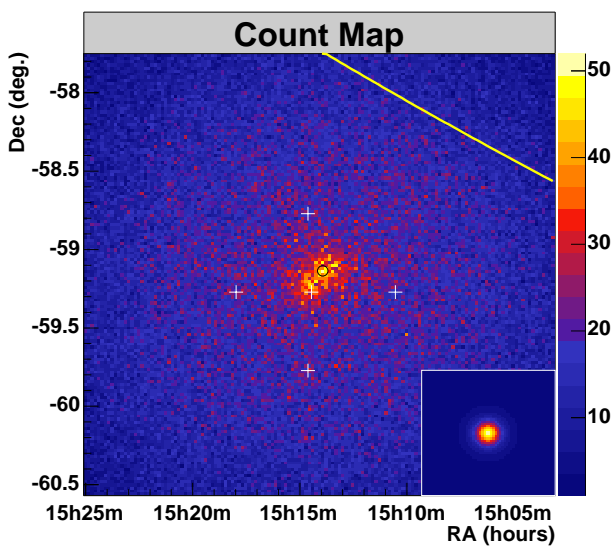}
  \end{minipage}\hfill
  \begin{minipage}[c]{0.45\linewidth}
    \caption[Count Map of the H.E.S.S. Data for \MSH]{Count map of the H.E.S.S.
      data from \MSH\ passing the quality criteria. White crosses represent the
      target position (center) and the four observation positions for the
      different wobble offsets. The position of \PSR\ is marked (black circle).
      The galactic plane is indicated by the yellow line in the upper right.
      The PSF is shown in the lower right corner.}
    \label{fig:ObsPos}
  \end{minipage}
\end{figure}

\MSH\ was repeatedly observed in 2004 from March 26 to July 20. A total of 78
stereoscopic observation runs with a duration of mostly 28 minutes each were
taken during this period. The data was taken in wobble mode with all four
telescopes fully operational. The system trigger required a telescope
multiplicity of at least two. The observation period was favored by good
weather providing data of high quality. Fig.~\ref{fig:ObsPos} shows the
corresponding count map of \g-ray candidates after run selection and cuts. The
galactic plane is indicated by the yellow line in the upper right region. It
passes the field of view at the approximate distance of 1$^\circ$ from source
position. The position of \PSR\ as determined from radio observations is
represented by the black circle near the center of the map.

\subsection{Observation Position}
The data was taken at the target position of (J2000) (${\rm 15^h14^m27^s}$,
$-59^\circ16'18''$) and wobble offsets of $\pm0.5^\circ$ in RA and
$\pm0.5^\circ$ in Dec, resulting in four different observation positions. The
wobble positions and the target position are indicated by white crosses in
Fig.~\ref{fig:ObsPos}. Tbl.~\ref{tbl:ObservationSummary} shows the number of
runs after run selection for each wobble position. The number of positive and
negative wobble offsets is approximately equal for each wobble direction
reducing systematic errors. The full run list is found in
App.~\ref{app:RunLists}.

\begin{table}[tbh!]
  \centering
  \caption[Summary of Observation Runs of \MSH]{Summary of observation runs
    taken at the four different wobble offsets shown in Fig.~\ref{fig:ObsPos}.}
  \bigskip
  \begin{tabular}{lccc}
    \hline \hline
    Obs. Pos. & No. of runs &  <Zenith angle> [$^\circ$] & live-time [h] \\
    \hline
    Dec $+0.500^\circ$ & 23 & 36.7  &  9.93 \\
    Dec $-0.500^\circ$ & 24 & 37.7  &  9.83 \\
    RA  $+0.978^\circ$ &  7 & 36.7  &  2.98 \\
    RA  $-0.978^\circ$ &  8 & 36.7  &  3.40 \\
    \hline
                       & 62 & 37.1  & 26.14 \\
    \hline \hline
  \end{tabular}
  \label{tbl:ObservationSummary}
\end{table}

\subsection{Run Selection}
The runs were selected according to the quality criteria described in
Sec.~\ref{sec:RunSelection}. The result of the run selection is summarized in
Tbl.~\ref{tbl:RunSelection}. Run by run statistics are given in
App.~\ref{app:RunLists}. The low rejection ratio of 16 out of 78 runs
($\sim$20\%) is owed to good weather conditions during the observation period.
The total observation time was 35 hours of which 29 hours had sufficient data
quality. After dead-time correction, data with a live-time of 26.14 hours
remained.

\begin{table}[tbh!]
  \centering
  \caption[Results of the Run Selection]{Results of the run selection according
  to the quality criteria of Sec.~\ref{sec:RunSelection}.}
  \bigskip
  \begin{tabular}{lcc}
    \hline \hline
    & No. of runs &  Observation time [h]\\
    \hline
    selected & 62 & 29.1 \\
    rejected & 16 &  6.2 \\
    \hline
    total    & 78 & 35.3 \\
    \hline \hline
  \end{tabular}
  \label{tbl:RunSelection}
\end{table}

\subsection{Zenith Angle Distribution}
\begin{figure}[ht!]
  \begin{minipage}[c]{0.55\linewidth}
    \includegraphics[width=\textwidth]{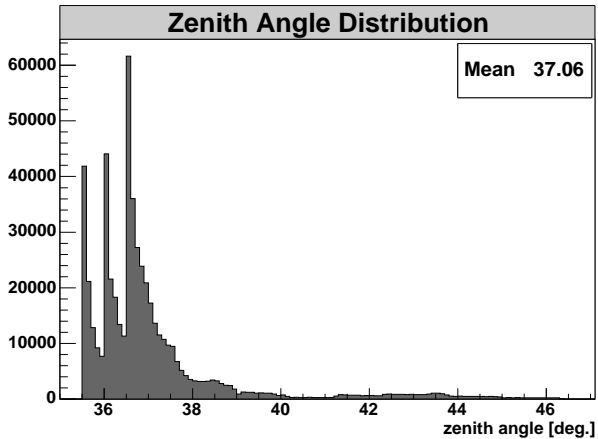}
  \end{minipage}\hfill
  \begin{minipage}[c]{0.45\linewidth}
    \caption[Zenith Angle Distribution of \g-Ray Candidates]{Zenith angle
      distribution of \g-ray candidates from the H.E.S.S. data for \MSH. The
      peaks indicate the different zenith angles of culmination of \MSH\ for
      each month of the observation period.}
    \label{ZADistribution}
  \end{minipage}
\end{figure}

The mean zenith angle of the selected data was $37.1^\circ$.
Fig.~\ref{ZADistribution} shows the zenith angle distribution of the \g-ray
candidate events. The three peaks result from different zenith angles of
culmination for each month of the observation period.

\section{Detection of the \g-Ray Signal} \label{sec:MSHstatistics}
The selected data has been processed with the standard analysis chain described
in Chp.~\ref{chp:StandardAnalysis}. The {\it extended} cuts
(Tbl.~\ref{tbl:Cuts}) have been applied with both the ring-background and
region-background model (Sec.~\ref{sec:BackgroundModels}). The ON-region was
chosen near the centroid of the excess $({\rm 15^h14^m4\fs8}, -59^\circ9'36'')$
as obtained from a fit of a Gaussian function (Tbl.~\ref{tbl:MSHPositionSize}).
With a radius of $0.3^\circ$, the ON-region includes most of the \g-ray events.
To avoid systematic errors from bright stars, the OFF-regions have been chosen
to exclude stars with a magnitude brighter than seven. Also, the ON-region is
free of bright stars and therefore this configuration of the background models
provides good conditions for a reliable analysis.

\subsection{Ring-Background Model} \label{sec:Ring-Background}
Fig.~\ref{fig:Ring} shows the excess map of \MSH\ which was obtained with the
ring-background model. ON- and OFF-regions as applied here are shown by the
white circles. The ring radii are 0.725 and $1.075^\circ$. Stars with a
magnitude brighter than 7.8 are shown with the labels of their magnitudes. The
ring-background yields a clear detection of a \g-ray excess with a significance
of 28 standard deviations according to \citet{LiMa}. The corresponding
significance map is shown in Fig.~\ref{fig:Region}.

The analysis was repeated with the same configuration but an increased minimum
image amplitude cut of 400\,p.e. This cut implies a higher energy threshold of
about 900\,GeV and therefore a higher reconstruction accuracy and smaller PSF.
Although this cut reduces the number of excess events to $\sim$25\%, the signal
to noise ratio and the significance is increased. The 400\,p.e. map reveals a
slightly different picture of \MSH\ (Fig.~\ref{fig:SmoothMSH400}), which is
discussed in Sec.~\ref{sec:Morphology}. The statistics of both analyses are
summarized in Tbl.~\ref{tbl:StatisticsMSH}.

\subsection{Region-Background Model}
\begin{figure}[t!]
  \begin{minipage}[t]{0.5\linewidth}
    \includegraphics[width=\textwidth]{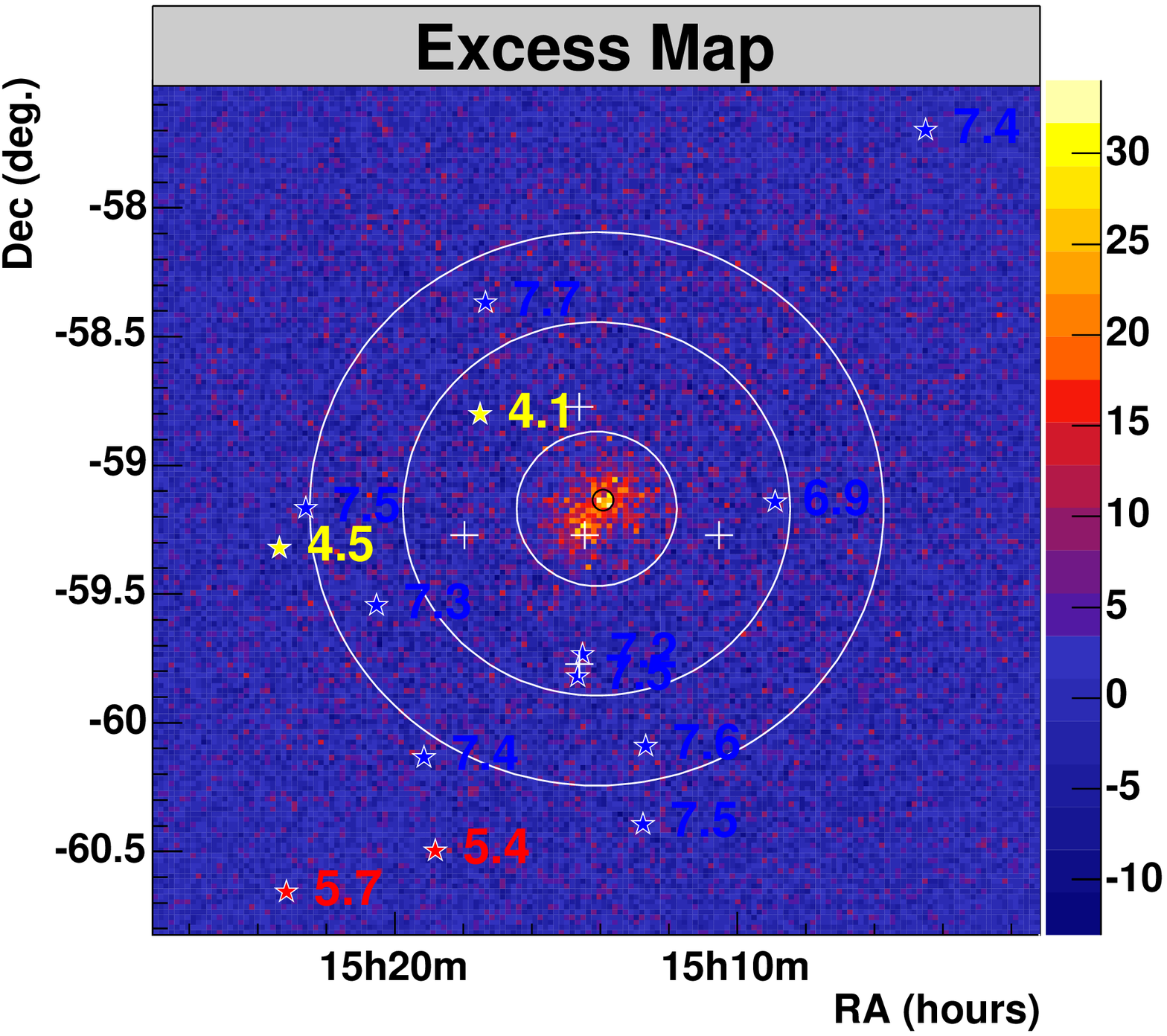}
    \caption[Excess Map Showing the Configuration of the Ring-Background]
      {Excess map obtained with the ring-background model showing the ON- and
      OFF-region by white circles. Bright stars are indicated. They lie outside
      the ON- and OFF-regions. \PSR\ is represented by the black circle.}
    \label{fig:Ring}
  \end{minipage}\hfill
  \begin{minipage}[t]{0.5\linewidth}
    \includegraphics[width=\textwidth]{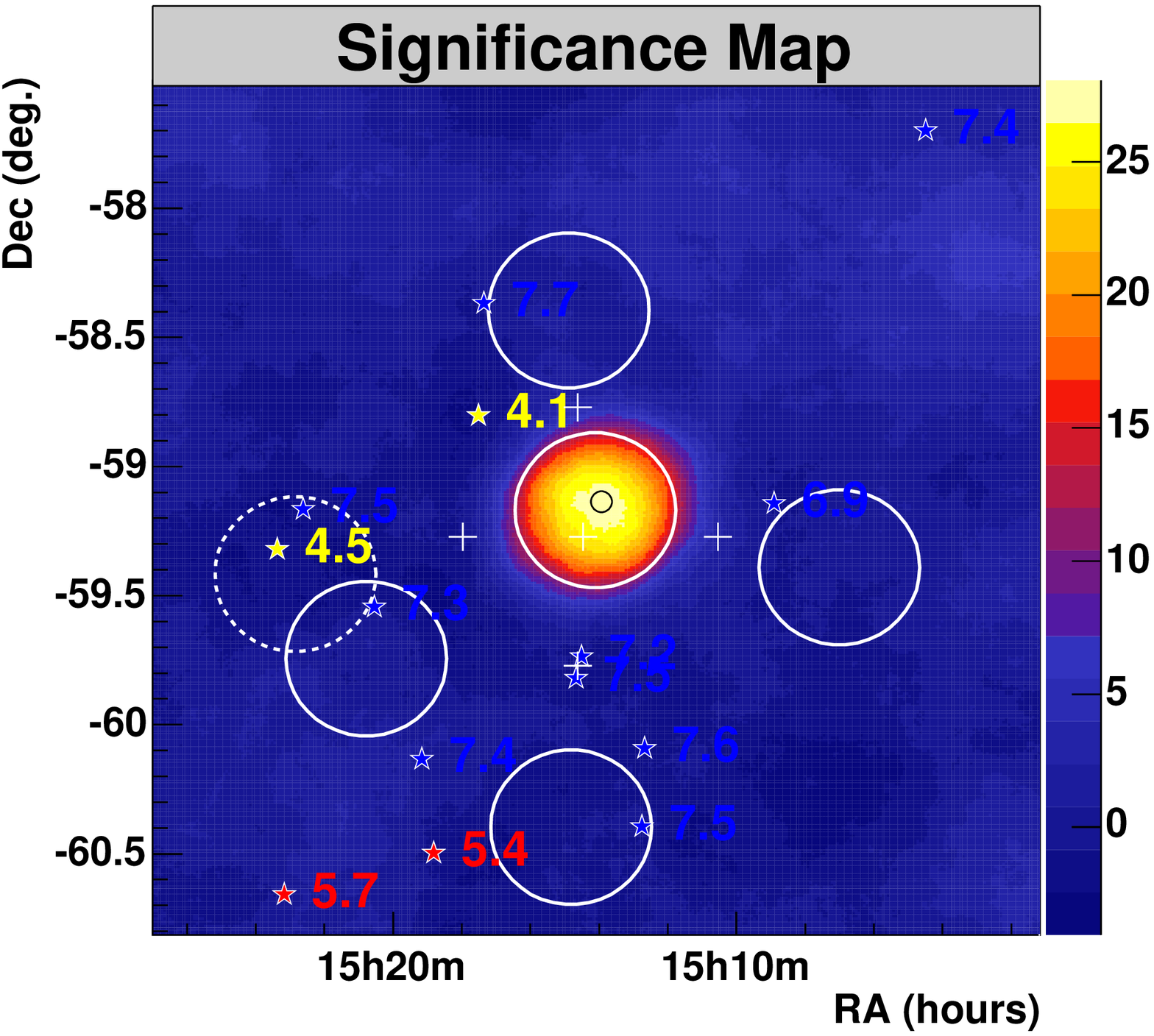}
    \caption[Significance Map Showing the Configuration of the
      Region-Background]{Significance map obtained with the ring-background
      model. It shows the ON- and OFF-regions of the region-background model
      (white circles). Bright stars are indicated. The dashed circle shows the
      position of the left OFF-region before its shift to a new position to
      exclude a star of magnitude 4.5. \PSR\ is represented by the black
      circle.}
  \label{fig:Region}
  \end{minipage}
\end{figure}

The analysis was also repeated with the region-background model. The
configuration of the ON- and OFF-regions is shown in the significance map of
Fig.~\ref{fig:Region}, which was produced by the ring-background described in
the previous section. One OFF-region was chosen for each observation position
opposite to the ON-region, resulting in a total of four different OFF-regions.
The restriction to one OFF-region was consequence of the partially small
distances from the ON-region to the four wobble positions. To exclude a star of
magnitude 4.5, the OFF-region to the left was shifted in a clockwise direction
along the arc of constant acceptance. Its previous position is shown by the
dashed circle. In the final configuration no stars brighter than a magnitude of
7.3 are included in the ON- and OFF-regions. Again, a clear detection of a
\g-ray signal is found with a significance of 23 standard deviations. The
statistics are given in Tbl.~\ref{tbl:StatisticsMSH}. These results are in good
agreement with the results of the ring-background model. The same configuration
was used for the determination of the energy spectrum.

%

\begin{table}[h!]
  \centering
  \caption[Signal Statistics of the \MSH\ Data]{Signal statistics as obtained
    with the standard analysis and the {\it extended} cuts
    (Tbl.~\ref{tbl:Cuts}) for different background models and image amplitude
    ($IA$) cuts.}
  \bigskip
  \begin{tabular}{lccc}
    \hline \hline
    & Ring-background      & Region-background    & Ring-background \\
    & ($IA>80$\,p.e.)      & ($IA>80$\,p.e.)      & ($IA>400$\,p.e.)\\
    \hline
    $N_{\rm ON}$               & 17651         & 17760         & 1333  \\ 
    $N_{\rm OFF}$              & 74895         & 13711         & 2078  \\
    $\alpha$                   & 0.186         & 1             & 0.196 \\
    $N_\gamma$                 & $3752\pm133$  & $4049\pm177$  & $925\pm38$ \\
    $S [\sigma]$               & 27.9          & 22.9          & 31.6 \\
    $S/\sqrt{t}$
    $[\sigma/\sqrt{\rm h}]$    & 5.46          & 4.47          & 6.18 \\
    signal / noise             & 0.27          & 0.30          & 2.27 \\
    $\gamma$-rate [min$^{-1}$] & $2.40\pm0.09$ & $2.58\pm0.11$ &$0.59\pm0.02$\\
    \hline \hline
  \end{tabular}
  \label{tbl:StatisticsMSH}
\end{table}

\clearpage
\section{Position and Size of the \g-ray Excess} \label{sec:MSHPositionSize}


\begin{figure}[b!]
  \begin{minipage}[c]{0.5\linewidth}
    \includegraphics[width=\textwidth]{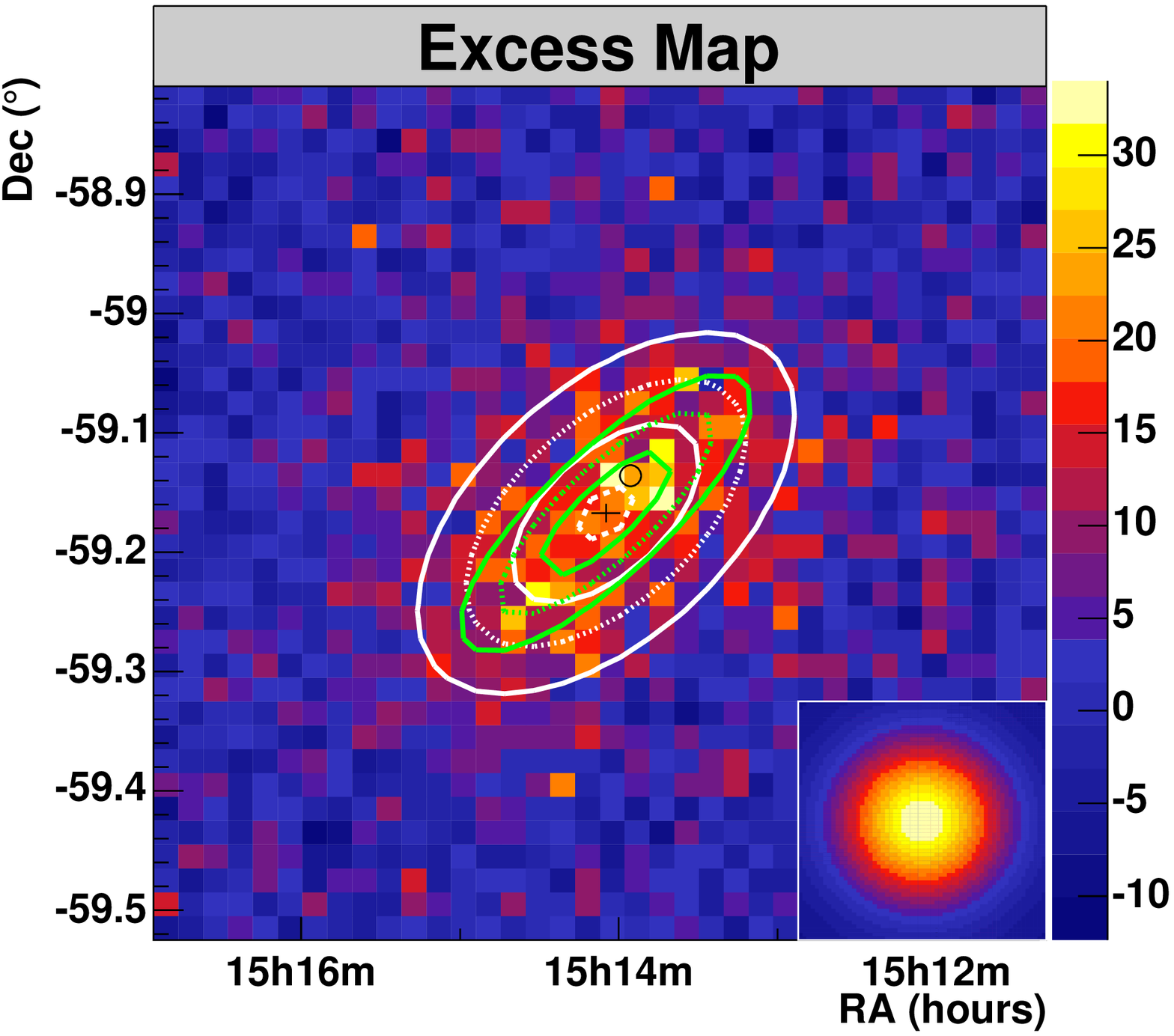}
    \caption[Contour Lines of the Fit Function for \MSH]{\g-ray excess map of
      the \MSH\ region. Contour lines at levels of 25, 50 and 75\% indicate the
      fit function (white, solid) and the function of the intrinsic width
      (green, dashed). The black cross represents the best fit position and the
      statistical error. The systematic error has about the same size. The
      position of \PSR\ is indicated by the black circle, a profile of the PSF
      is shown in the bottom right.}
    \label{fig:SourceFit2D}
  \end{minipage}
  \begin{minipage}[c]{0.5\linewidth}
    \includegraphics[width=\textwidth]{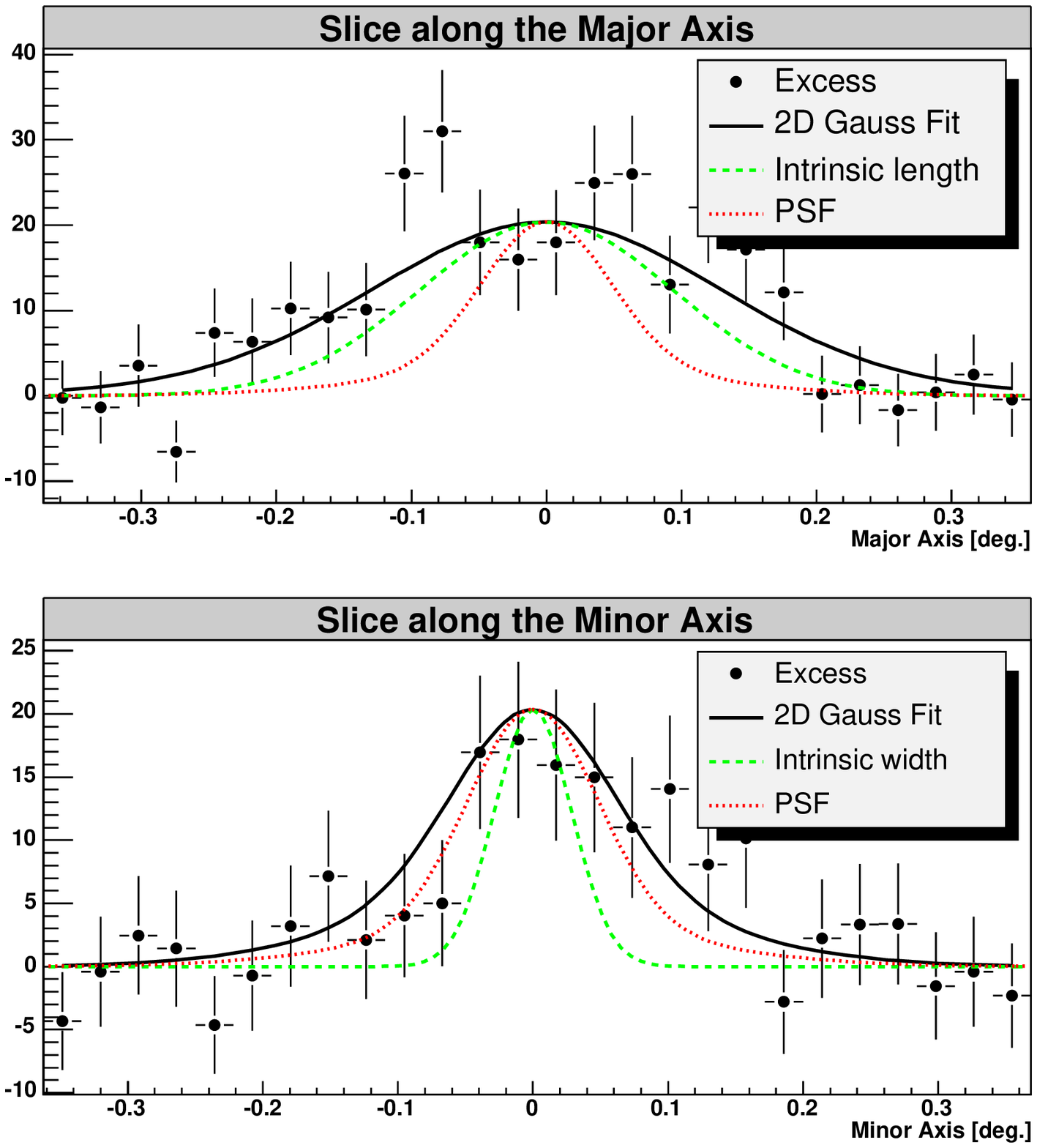}
    \caption[Slices along the Excess of \MSH]{\g-ray excess counts and fit
      functions along the major axis (upper plot) and minor axis (lower plot)
      of the \g-ray excess. The fit function as well as the functions of the
      intrinsic width and of the PSF are shown. \\ \\ \\ \medskip}
    \label{fig:SourceFit1D}
  \end{minipage}
\end{figure}

\MSH\ is located close to the galactic plane at a distance of 1$^\circ$ as
shown in Fig.~\ref{fig:ObsPos}. The position and the intrinsic extension of the
\g-ray excess were determined from a fit of the Gaussian function
$G_{\sigma_x,\sigma_y,\alpha_0,\delta_0,\omega,N}(\alpha, \delta)$ to the
excess map of Fig.~\ref{fig:Ring} as described in Sec.~\ref{sec:PositionSize}.
The explicit formula for $G$ is given in Eqn.~\ref{eqn:ConvolutionGauss2D} of
App.~\ref{app:GaussianFit}. The PSF fit parameters where chosen according to
parameterization number 3 of Tbl.~\ref{tbl:PSF}. The $\chi^2$/dof. of the fit
was 391.2/394, equivalent to a probability of 0.53 for the approximation of the
\g-ray excess by a Gaussian model. The best fit position of the emission was
found at (J2000) (${\rm 15^h14^m6\fs5}\pm2\fs4, -59^{\circ}10'1.2''\pm21''$)
and the H.E.S.S. catalog name \HESS\ was assigned to it. The distance to the
radio position of \PSR\ (${\rm 15^h13^m55\fs620}$, $-59^\circ08'09\farcs0$) is
$2.3'$. The galactic coordinates of the fit position are
($320.324^\circ\pm0.005^\circ$, $-1.200^\circ\pm0.005^\circ$). The emission
region is asymmetric and elongated in the north-west direction, defining the
major and minor axis with a length and width of $6.5'\pm0.5'$ and $2.3'\pm0.4'$
respectively. The angle between the major axis and the RA-axis is
$43^\circ\pm4^\circ$. Tbl.~\ref{tbl:MSHPositionSize} summarizes the fit
results. Fig.~\ref{fig:SourceFit2D} shows the excess map with the contour lines
of the fit function (white, solid) and of the intrinsic Gaussian (green,
dashed) at levels of 25, 50 and 75\%. Fig.~\ref{fig:SourceFit1D} shows slices
along the major and minor axes of the excess map. The fit function and its
components, i.e. intrinsic width or length and the PSF, are shown in addition
to the data.

\begin{table}[th!]
  \centering
  \caption[Position and Size of \MSH]{Parameters obtained from the fit of a
    Gaussian function (Eqn.~\ref{eqn:ConvolutionGauss2D}) to the \g-ray excess
    map of \MSH. The parameterization of the PSF is given in
    Tbl.~\ref{tbl:PSF}.}
  \bigskip
  \begin{tabular}{l|c}
    \hline \hline
    Parameter                       & Value \\
    \hline
    RA ($\alpha_0$) 
    & ${\rm 15^h14^m6\fs5\pm2\fs4}$, $(228.527^\circ\pm0.010^\circ)$ \\
    Dec ($\delta_0$)
    & $-59^\circ10'1\farcs2\pm21''$, $(-59.169^\circ\pm0.005^\circ)$ \\
    Length ($\sigma_x$)         &$6.5'\pm0.5'$, $(0.109^\circ\pm0.008^\circ)$\\
    Width ($\sigma_y$)          &$2.3'\pm0.4'$, $(0.039^\circ\pm0.006^\circ)$\\
    Angle with RA-axis ($\omega$)& $43^\circ\pm4^\circ$ \\
    PSF parameterization no.        & 3 \\
    \hline \hline
  \end{tabular}
  \label{tbl:MSHPositionSize}
\end{table}

\subsection{Systematic Errors}
\subsubsection{Position}
The systematic error of the position is determined by the pointing accuracy of
the telescope system, which is $20''$ in each direction. In RA the error scales
as $\rm 1/\cos(Dec)$ and the relation between a second of arc to second of RA
is $15''\hat=1$\,s. So at the declination of \MSH\ the error in RA is $\rm
20''/\cos(-59.16^\circ)\times1\,s/15''=2.6$\,s. Including systematic errors,
the fit position of the excess is (J2000)
$\mathrm{(15h14^m6\fs5\pm2\fs4_{stat}\pm2\fs6_{syst},
-59^{\circ}10'1\farcs2\pm21''_{stat}\pm20''_{syst})}$.

\subsubsection{Size}
A systematic error of the width and length of the \g-ray excess can result from
an imperfect model of the PSF, which is what was used in the fit function ($G$,
App.~\ref{app:GaussianFit}). Fig.~\ref{fig:PSF} shows that the width of the PSF
is slightly overestimated by the Gaussian parameterization, which would result
in an underestimation of the intrinsic width and length of a \g-ray excess. One
can reduce the PSF parameter $\sigma_2$ by 20\% to obtain a model of the PSF
with a similar accuracy but slightly underestimated width. In this case, the
intrinsic width and length of the \g-ray excess of \MSH\ would be overestimated
by $\sim$5\% and $\sim$1\% respectively. These numbers provide an estimate for
the systematic uncertainty in width and length of the \g-ray excess. With
respect to the statistical errors they are negligible.

\section{Energy Spectrum} \label{sec:SpectrumMSH}

\begin{figure}[bth!]
  \setlength{\abovecaptionskip}{-0.0cm}
  \centering
  \includegraphics[width=.65\linewidth]{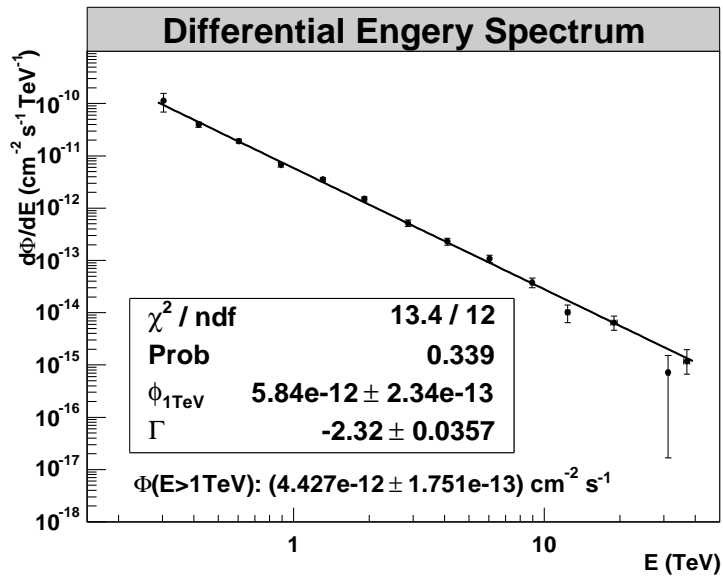}
  \caption[Energy Spectrum and Power Law Fit of \MSH]{Energy spectrum and
    power law fit of the \g-ray emission from the region of \MSH\ with radius
    $\theta=0.3^\circ$ in the energy rang from 0.28 to 40\,TeV.}
  \label{fig:Spectrum}
\end{figure}

\begin{figure}[b!]
  \setlength{\abovecaptionskip}{-0.0cm}
  \begin{minipage}[t]{0.5\linewidth}
    \includegraphics[width=\textwidth,
    height=.88\textwidth]{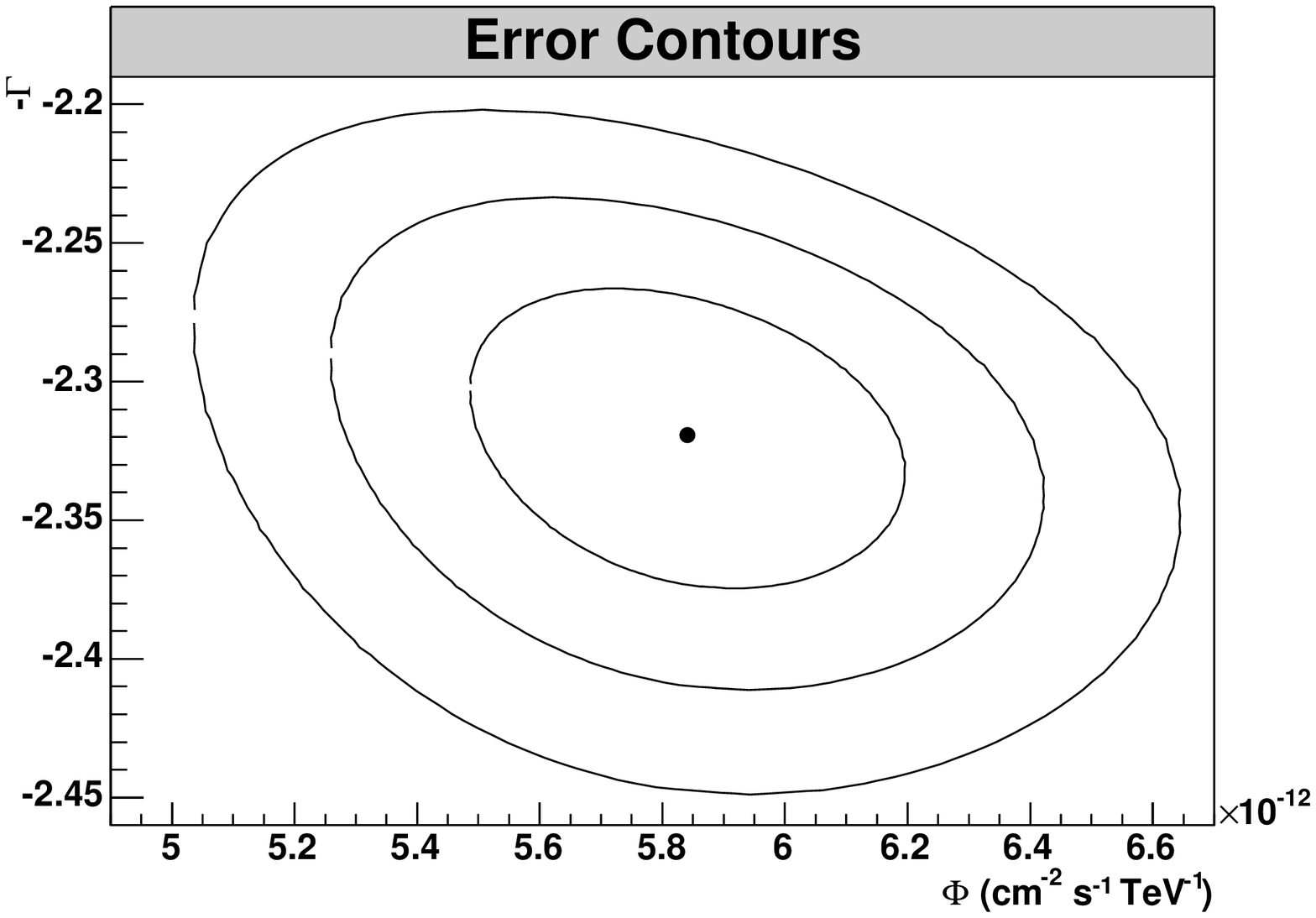}
    \caption[Error Contours of the Power Law Fit of \MSH]{One, two and three
      $\sigma$ error contours for the spectral parameters $\Gamma$ and
      $\phi_{\rm 1TeV}$ as obtained from the $\chi^2$-fit to a power law in
      Fig.~\ref{fig:Spectrum}.}
  \label{fig:ErrorContours}
  \end{minipage}\hfill
  \begin{minipage}[t]{0.5\linewidth}
    \includegraphics[width=\textwidth]{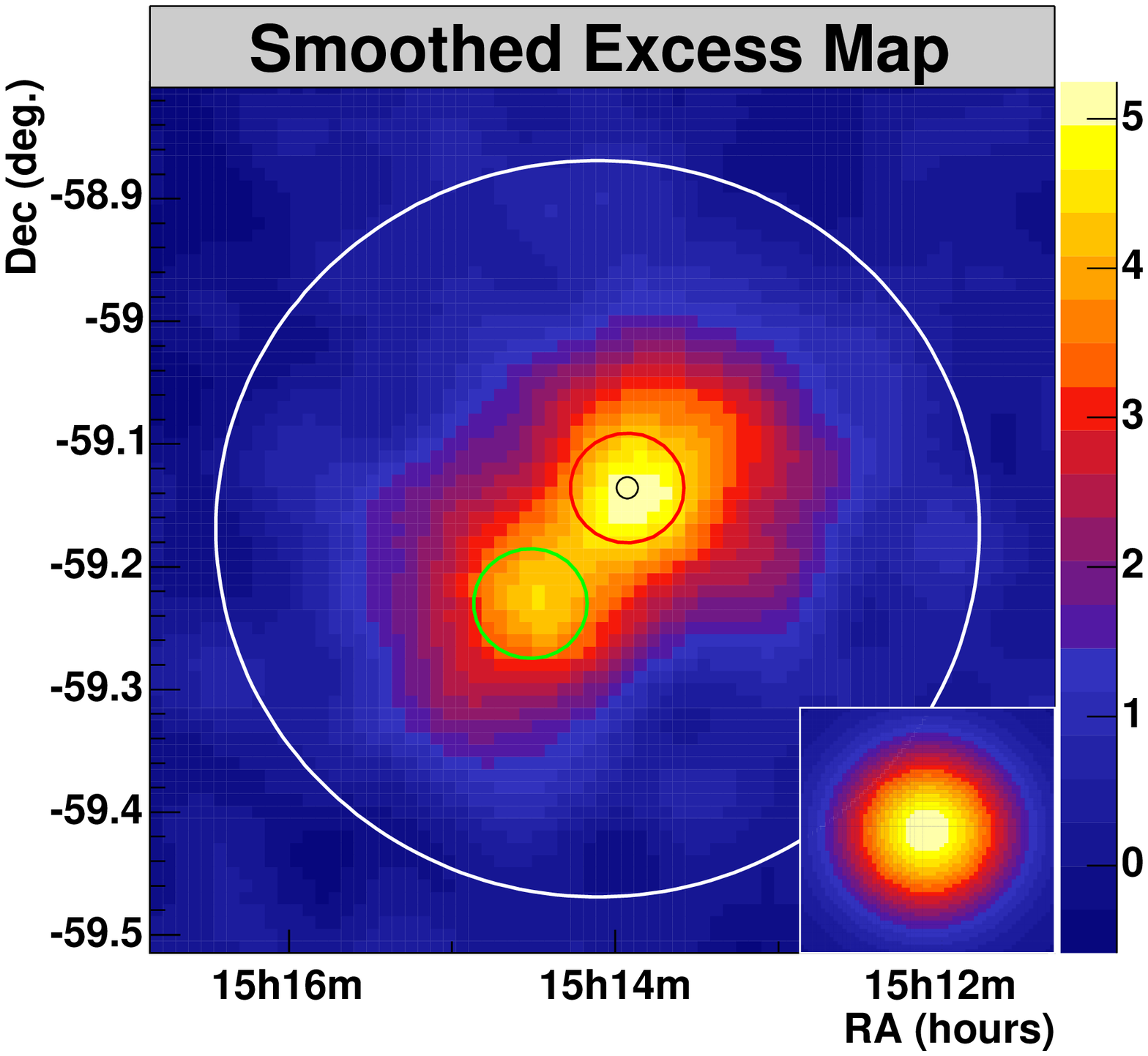}
    \caption[Subregions of the Spectral Analysis]{Excess map showing the \g-ray
      excess region ($\theta=0.3^\circ$, white) and two smaller ON-regions (red
      and green) where the energy spectrum was determined. The position of
      \PSR\ is indicated by the black circle.}
  \label{fig:SpectrumRegions}
  \end{minipage}
\end{figure}

The energy spectrum was determined using the {\it extended} cuts and the
region-background model with the configuration described in
Sec.~\ref{sec:MSHstatistics}. The effective area $(A_{\rm Ext})$ was obtained
from simulations of an extended source with a standard deviation of
0.04$^\circ$ and 0.11$^\circ$ in width and length respectively (cf.
Sec.~\ref{sec:Spectrum}). The spectral parameters were determined by a
$\chi^2$-fit to a power law according to Eqn.~\ref{eqn:PowerLaw2}. It provides
a good description for the energy spectrum with a $\chi^2/dof$ of 13.4/12
corresponding to a probability of 0.34. A photon index ($\Gamma$) of about
$2.32\pm0.04$ and a differential flux at 1\,TeV ($\phi_{\rm 1TeV}$) of
$(5.8\pm0.2) \times \rm 10^{-12}cm^{-2}s^{-1}TeV^{-1}$ were obtained. The
integrated flux above 1\,TeV $(\Phi(E>1$TeV)) is $(4.4\pm0.2) \times \rm
10^{-12}cm^{-2}s^{-1}$, corresponding to about 20\% of the flux from the Crab
Nebula above the same threshold. The safe energy threshold was found at
$\sim$280\,GeV, allowing for a fit range from 280\,GeV to 40\,TeV. The results
are summarized in Tbl.~\ref{tbl:SpectrumSystematics}. The data and the fit
function are shown in Fig.~\ref{fig:Spectrum}.

Fig.~\ref{fig:ErrorContours} shows the one, two and three $\sigma$ error
contours of $\phi_{\rm 1TeV}$ and $\Gamma$ as obtained from the fit with the
{\tt TMinuit} class in ROOT. The contours correspond to a $\chi^2$/dof of
2.3/2, 6.18/2 and 11.83/2 with the $\chi^2$ probabilities of 0.68, 0.95 and
0.99 respectively.

To investigate the emission region for spatial variations of the photon index,
the analysis was repeated for two different smaller signal regions with radius
$\theta=0.045^\circ$ --- one near the center of the emission at the position of
\PSR\ and the other one in some distance at the \X-ray jet axis to the
southeast. The regions are shown in Fig.~\ref{fig:SpectrumRegions}. The area
and position of the background regions, and the effective area, were adjusted
accordingly. The measured energy spectrum was fit to a power law in the same
fit range, but no significant variation of the photon index was found. In the
pulsar region, a $\Gamma$ of $2.22\pm0.08$ with a $\chi^2$/dof of 5.0/6 and in
the jet region, a $\Gamma$ of $2.16\pm0.10$ with a $\chi^2$/dof of 7.0/5 was
obtained. For the interpretation of the results one has to consider that the
regions are slightly correlated due to the PSF and that the ON-regions are
slightly offset from the center of the excess. While the first effect reduces
spectral differences between the regions, the later leads to a spectral
steepening. This is explained by the energy dependence of the PSF, which
spreads low energy events further from the center. However, the effect on the
spectral index is very small.

\subsection{Systematic Errors}
Since the energy spectrum is determined at a late point in the analysis chain,
it accumulates errors from previous analysis steps. Therefore it is important
to consider some systematic uncertainties.

\subsubsection{Atmospheric Model of the Simulations}
\label{sec:AtmosphericModels}
The calculation of the energy spectrum is sensitive to atmospheric variations.
Unfortunately it is very difficult to model the atmospheric variations in all
detail. Therefore, Monte Carlo simulations only contain a model of the mean
atmospheric density profile. The $desert$ and the $maritime$ atmospheric models
(Sec.~\ref{sec:MonteCarlo}) are both appropriate models of the atmosphere at
the H.E.S.S. site, as confirmed with trigger studies by \citet{CentralTrigger}.
In the analysis the effective area for the $desert$ model was used. The
effective area for the $maritime$ model provides slightly different results and
these are summarized in Tbl.~\ref{tbl:SpectrumSystematics}. The differences
provide an estimate for the systematic error as resulting from uncertainties in
the atmospheric model. They are 15\% and 3.5\% for $\phi_{\rm 1TeV}$ and
$\Gamma$ respectively.

\subsubsection{Absolute Calibration}
\begin{figure}[ht!]
  \begin{minipage}[c]{0.44\linewidth}
  \includegraphics[width=\textwidth]{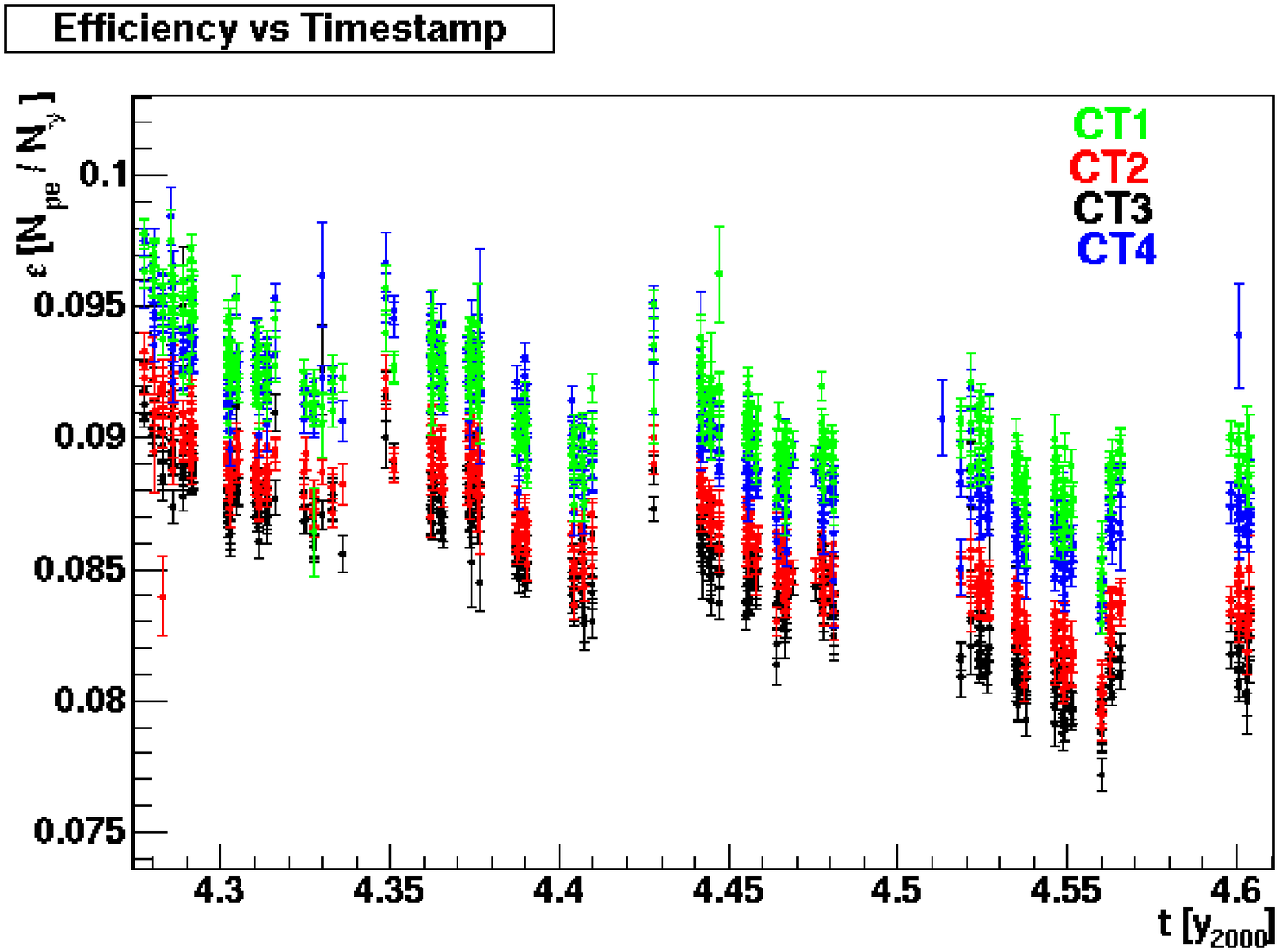}
  \end{minipage}\hfill
  \begin{minipage}[c]{0.5\linewidth}
    \caption[Muon Efficiency During the Observation Period]{Muon efficiency
      during the observation period of \MSH. The time is measured in units of
      years since 2000. The average efficiency is only 87\% of the nominal
      value 0.106. The decrease in the efficiency is about 5\% during the
      observation period. (Figure taken from \citet{MPIK:MuonEfficiency}.)}
    \label{fig:MuonEfficiency}
  \end{minipage}
\end{figure}

The degradation of the absolute photon efficiency of H.E.S.S. and its
measurement with muons was previously discussed in Sec.~\ref{sec:Calibration}.
Fig.~\ref{fig:MuonEfficiency} shows the degradation of the muon efficiency
during the observation of \MSH\ as determined by \citet{MPIK:MuonEfficiency}.
It can be seen that the mean muon efficiency during this period is only 87\% of
the nominal value of 0.106 and also a decrease by 5\% from the beginning to the
end of this period is observed. Further detailed studies by
\citet{MPIK:MuonCorrections} for this observation period have shown that in the
standard analysis the differential flux $d\Phi/dE$ is underestimated by about
13\%, while the photon index is not affected significantly
(\citet{MPIK:MuonCorrections}). These results provide an estimate of the
systematic error of the spectrum related to the absolute calibration.

\subsubsection{Cut Configuration}
The cut configuration also contributes a small systematic uncertainty to the
energy spectrum. To estimate this uncertainty, the analysis was repeated with
$loose$ and $hard$ cuts (cf. Tbl.~\ref{tbl:Cuts}). The results in
Tbl.~\ref{tbl:SpectrumSystematics} show that $\phi_{\rm 1TeV}$ is either
slightly over or underestimated. The systematic error of $\phi_{\rm 1TeV}$ and
$\Gamma$ is estimated as 7\% and 1\%.

\subsubsection{Broken Pixels}
The broken pixels discussed in Sec.~\ref{sec:BrokenPixels} introduce a
systematic uncertainty of about 5\% to each image amplitude, as detailed
studies by \citet{CrabPaper} and \citet{Schwanke:BrokenPixels} have shown. This
value can be directly adopted as the systematic error of the flux normalization
$\phi_{\rm 1TeV}$.

\subsubsection{Uncertainty of Sources Extension} \label{sec:ExtendedSources}
The effective areas were derived from Monte Carlo simulations for extended
sources as described in Sec.~\ref{sec:MCfiles}. The standard error of the
extension is about $0.07^\circ$ (cf. Tbl.~\ref{tbl:MSHPositionSize}) in width
and length. To determine the resulting uncertainty for the energy spectrum, the
spectrum was reproduced with effective areas for an increased
($\sigma_w=0.046^\circ$, $\sigma_l=0.12^\circ$) and reduced
($\sigma_w=0.034^\circ$, $\sigma_l=0.10^\circ$) source extension by
approximately one standard error. Tbl.~\ref{tbl:SpectrumSystematics} shows the
results of a power law fit to the two new energy spectra. The relative
systematic uncertainty for $\phi_{\rm 1TeV}$ is about 2\%, while $\Gamma$ is
unaffected.

\subsubsection{Simulation of Extended Sources}
\begin{figure}[h!]
  \begin{minipage}[c]{0.48\linewidth}
  \includegraphics[width=\textwidth]{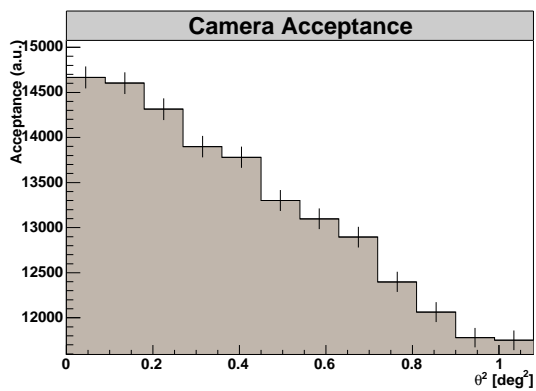}
  \end{minipage}\hfill
  \begin{minipage}[c]{0.5\linewidth}
    \caption[Camera Acceptance]{Camera acceptance versus squared difference
      between the pointing and shower direction ($\theta^2$). A decrease of
      10\% per degree$^2$ is found.}
    \label{fig:Acceptance}
  \end{minipage}
\end{figure}

A marginal systematic error results from the simulation of extended sources
which do not take the radial acceptance gradient in the FOV into account.
Fig.~\ref{fig:Acceptance} shows the radial profile of the system acceptance.
The acceptance decreases by about 1\% per 0.1$^\circ$. The estimated total
event loss for a source with a size of \MSH\ is roughly 1\%, providing also an
estimate for the error of the $\phi_{\rm 1TeV}$.

\subsubsection{Live-time}
There is an uncertainty of 1\% in the determination of live-time. This was
determined by \citet{CentralTrigger}. The error of the live-time translates
directly into an error of the flux normalization.

\subsubsection{Total Systematic Error}
Tbl.~\ref{tbl:SpectrumSystematics} shows the spectra as obtained for the
different analysis configurations discussed above.
Tbl.~\ref{tbl:SystematicErrors} shows the relative systematic errors of $\Phi$
and $\Gamma$. Under the reasonable assumption that the individual systematic
errors are uncorrelated, the total relative systematic error is obtained from
the quadratic mean of individual systematic errors. One finds $\Delta\phi_{\rm
1TeV}/\phi_{\rm 1TeV}=22\%$ and $\Delta\Gamma/\Gamma=4\%$. The total relative
systematic error of the integral flux $(\Delta\Phi/\Phi)(E>1{\rm TeV})=23\%$ is
obtained by error propagation according to Eqn.~\ref{eqn:FluxError}. The
covariance is zero in this case, since the errors in $\Phi$ and $\Gamma$ are
assumed to be independent.

\begin{table}[bth]
  \centering
  \caption[Energy Spectra for Different Configurations of \MSH]{Energy spectra
    as determined from a fit to a power law for the different configurations
    discussed in this section. The meanings of superscript numbers are as
    follows: $^0$ configuration providing the quoted results, $^1$with
    $maritime$ atmospheric model, $^2$ with $hard$ cuts, $^3$ with $loose$
    cuts, $^4$ with decreased $(\sigma_w=0.034^\circ,\sigma_w=0.10^\circ)$
    source extension and $^5$ with increased
    $(\sigma_w=0.046^\circ,\sigma_w=0.12^\circ)$ source extension.}
  \bigskip
  \begin{tabular}{lccccc}
    \hline \hline
    Configuration & $\phi_{\rm 1TeV}$ & $\Gamma$ &
                    $\Phi(E>$1TeV) & $\chi^2$/dof & P($\chi^2$) \\
                 & \scriptsize $[\rm 10^{-12}\rm cm^{-2}s^{-1}TeV^{-1}]$ & & 
    \scriptsize $[10^{-12}\rm cm^{-2}s^{-1}]$ & & \\
    \hline
    {\it extended} $^0$&5.84$\pm$0.23&2.32$\pm$0.04&4.43$\pm$0.18&13.4/12&0.34\\
    \hline
    $maritime$ atm $^1$&6.74$\pm$0.26&2.24$\pm$0.04&5.46$\pm$0.22&10.4/8&0.24\\
    \hline
    $hard$ cuts $^2$ &6.20$\pm$0.27&2.34$\pm$0.04 &4.62$\pm$0.17&22.7/20&0.30\\
    $loose$ cuts $^3$&5.36$\pm$0.28& 2.29$\pm$0.04&4.14$\pm$0.20&10.6/11&0.48\\
    \hline
    size- $^4$       &5.76$\pm$0.23&2.32$\pm$0.04 &4.37$\pm$0.17&13.4/12&0.34\\
    size+ $^5$       &5.97$\pm$0.24&2.32$\pm$0.04 &4.52$\pm$0.18&13.4/12&0.34\\
    \hline \hline
  \end{tabular}
  \label{tbl:SpectrumSystematics}
\end{table}

\begin{table}[ht!]
  \centering
  \caption[Systematic Errors of the Energy Spectrum]{Individual and total
  relative systematic errors for the energy spectrum.}
  \bigskip
  \begin{tabular}{lcc}
    \hline \hline
    Systematic Error & $\Delta\phi_{\rm 1TeV}/\phi_{\rm 1TeV} [\%]$ &
    $\Delta\Gamma/\Gamma [\%]$ \\
    \hline
    Atmospheric model               & 15        & 3.5 \\ 
    Absolute calibration            & 13        & - \\
    Cut configuration               &  7        & 1 \\
    Broken pixels                   &  5        & - \\
    Uncertainty of source extension &  2        & - \\
    Simulation for extended sources &  1        & - \\
    Live-time                       &  1        & - \\
    \hline
    Total                           & 22        & 4 \\
    \hline \hline
  \end{tabular}
  \label{tbl:SystematicErrors}
\end{table}

\section{Light Curve}
\begin{figure}[h!]
  \begin{minipage}[c]{0.45\linewidth}
    \caption[Light Curve of \MSH]{Light curve showing the daily integrated flux
      average $\Phi(E>1$\,TeV) from the region of \MSH. The fit to a constant
      (red line) is in good agreement with a constant \g-ray emission.}
    \label{LightCurve}
  \end{minipage}\hfill
  \begin{minipage}[c]{0.52\linewidth}
    \hspace{0.25cm}\includegraphics[width=\linewidth]{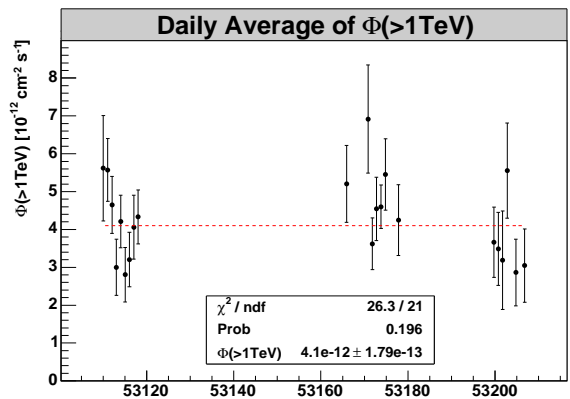}
  \end{minipage}
  \vspace{-1.1cm}
\end{figure}

A light curve of the daily integrated flux average $\Phi(E>1\rm TeV)$ for the
emission from the ON-region ($\theta=0.3^\circ$) is shown in
Fig.~\ref{LightCurve}. It is obtained from a power law fit to the daily data
assuming a constant photon index of $\Gamma=2.3$. A fit of the light curve to a
constant (red line) yields a $\Phi(E>1\rm TeV)$ of
$(4.1\pm0.2)\times10^{-12}\rm cm^{-2}s^{-1}$ and a $\chi^2$/dof of 26.2/21.
This result is in good agreement with a constant \g-ray flux from \MSH.

\section{\g-Ray Morphology} \label{sec:Morphology}
The morphology of \MSH\ is difficult to resolve with H.E.S.S., since the
extension of the \g-ray excess is small and of the same order as the size of
the PSF. Nevertheless, a better knowledge of the \g-ray excess distribution is
desirable for a better picture of the processes of the PWN and for a comparison
with structures resolved in \X-rays. Here an attempt is made to investigate the
\g-ray morphology of \MSH\ in greater detail using different representations,
energy bands and methods.

\begin{figure}[tbh!]
  \setlength{\abovecaptionskip}{-0.03cm}
  \setlength{\belowcaptionskip}{0.2cm}
  \begin{minipage}[t]{0.5\linewidth}
    \includegraphics[width=1.03\textwidth]{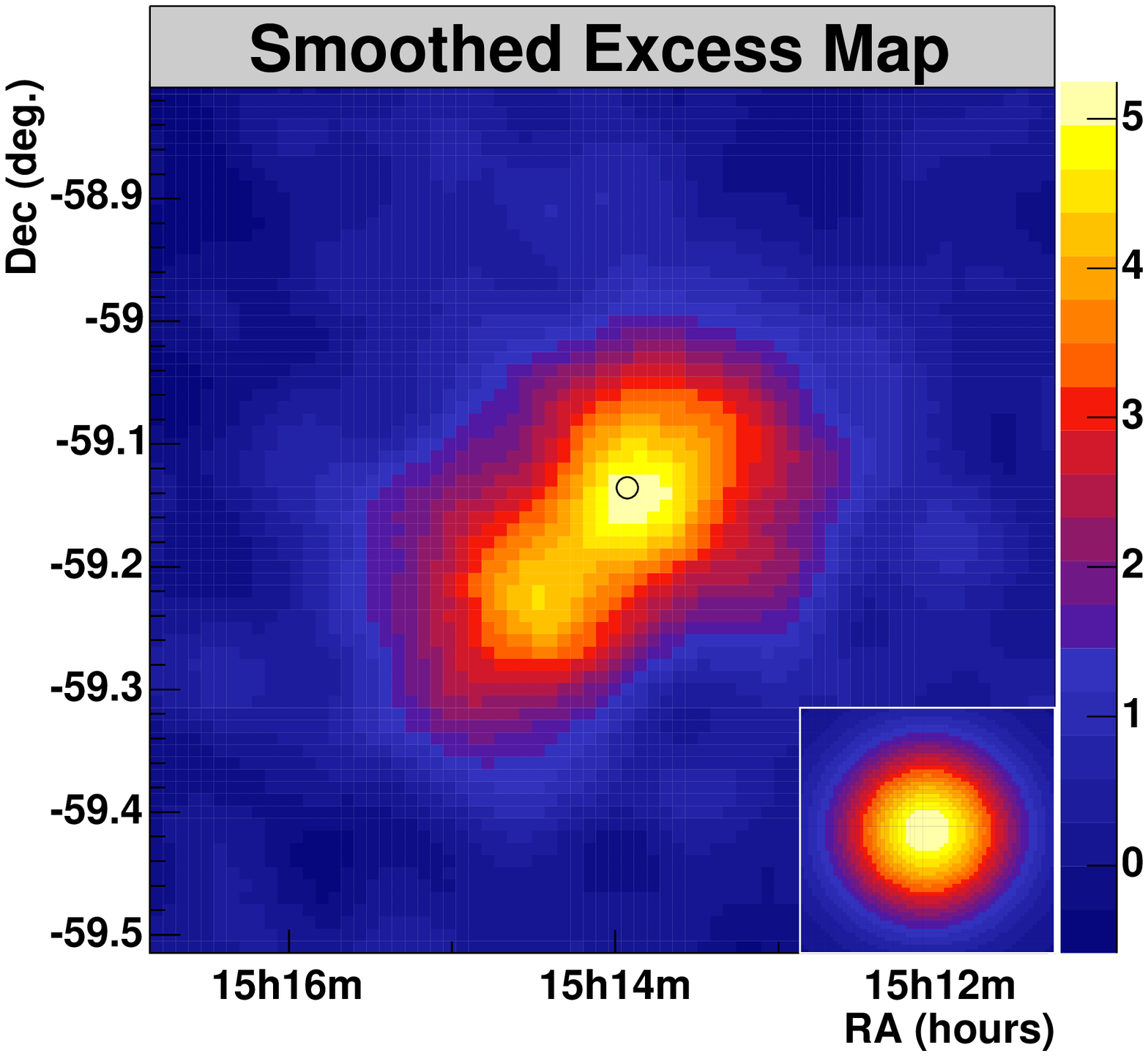}
    \caption[Smoothed Excess Map of \MSH\ ($IA>$80\,p.e.)]{\g-ray excess map
      convolved with a Gaussian ($\sigma=0.03^\circ$). The position of \PSR\ is
      marked by the black circle. The PSF is indicated in the bottom right
      corner.}
    \label{fig:SmoothMSH80}
  \end{minipage}\hfill
  \begin{minipage}[t]{0.5\linewidth}
    \includegraphics[width=\textwidth]{images/DecoAnalysis80_2}
    \caption[Restored Count Map of \MSH\ ($IA>$80\,p.e.)]{\g-ray count map
      after 10 iterations with the Richardson-Lucy algorithm. The position of
      \PSR\ is marked by the black circle. The PSF is indicated in the bottom
      right.}
    \label{fig:DecoMSH80chp08}
  \end{minipage}\hfill
  \setlength{\belowcaptionskip}{-0.3cm}
  \begin{minipage}[t]{0.5\linewidth}
    \includegraphics[width=1.03\textwidth]{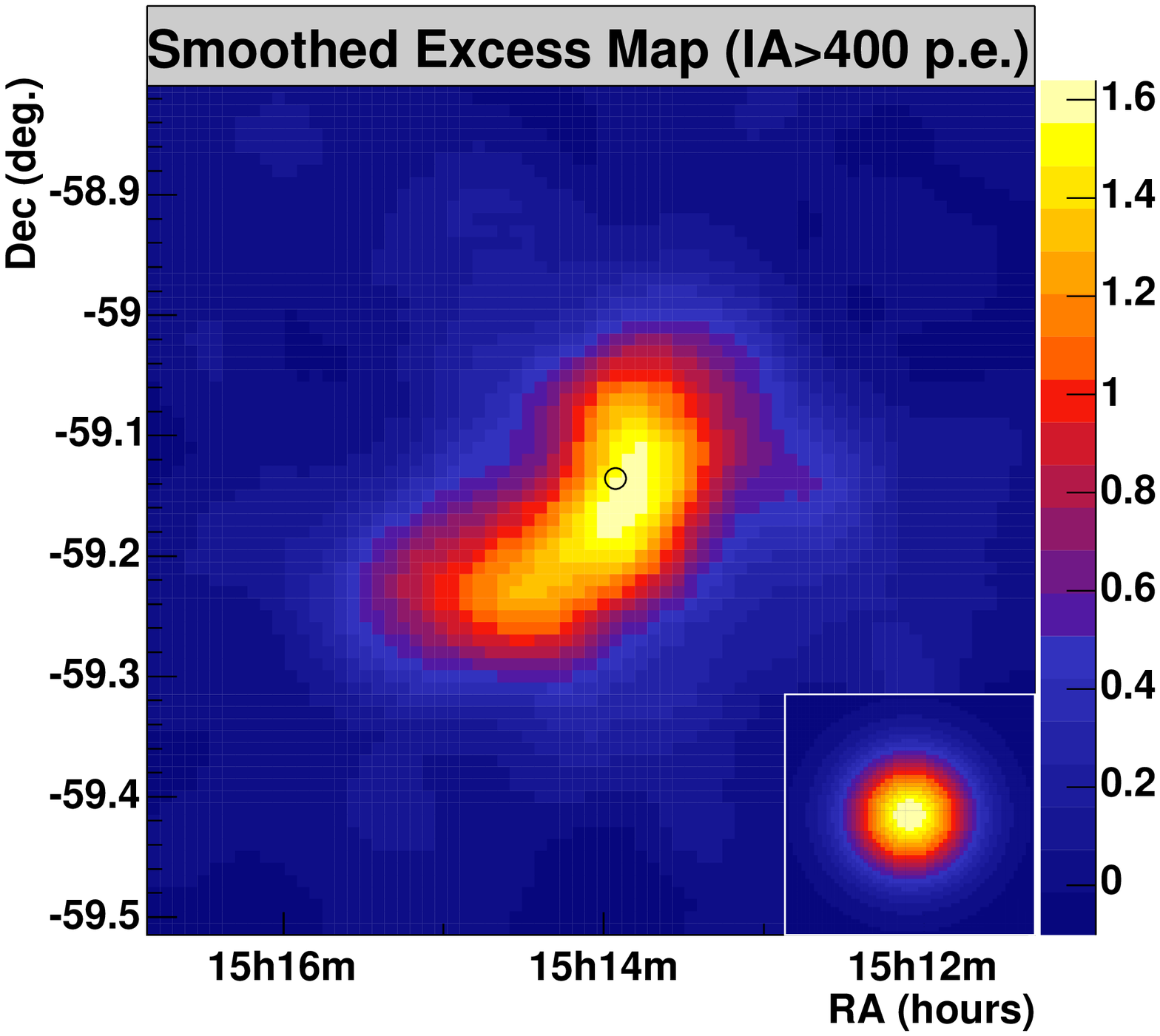}
    \caption[Smoothed Excess Map of \MSH\ ($IA>$400\,p.e.)]{The same as
      Fig.~\ref{fig:SmoothMSH80}, but with a minimum image amplitude of at
      least 400 p.e., reducing the width of the PSF by about 50\%. The PSF is
      indicated in the bottom right corner.}
    \label{fig:SmoothMSH400}
  \end{minipage}\hfill
  \begin{minipage}[t]{0.5\linewidth}
    \includegraphics[width=1\textwidth]{images/DecoAnalysis400_2}
    \caption[Restored Count Map of \MSH\ ($IA>$400\,p.e.)]{The same as
    Fig.~\ref{fig:DecoMSH80chp08}, but with a minimum image amplitude of at
    least 400 p.e. reducing the width of the PSF by about 50\%. The PSF is
    indicated in the bottom right corner.}
    \label{fig:DecoMSH400chp08}
  \end{minipage}\hfill
\end{figure}

\subsection{Sky Maps}
Fig.~\ref{fig:SmoothMSH80} to \ref{fig:DecoMSH400chp08} shows four different
\g-ray maps which differ in their image amplitude cuts and methods of
restoration. An increased image amplitude cut provides an increased resolution
at the expense of an increased energy threshold. Deconvolution (cf.
Chp.~\ref{chp:Deconvolution}) on the other hand can also reduce the apparent
PSF while preserving the original energy range. However, statistical artefacts
are amplified. The method of image smoothing can always provide a conservative
representation. The sky maps are based on the count map described before. A
profile of the PSF is shown in the bottom right of each map.

Fig.~\ref{fig:SmoothMSH80} shows the smoothed excess map with a minimum image
amplitude cut of 80\,p.e. with an energy threshold of about 280\,GeV. The
smoothing Gaussian has a width $\sigma_s=0.03^\circ$. An elliptical \g-ray
excess region with small deviations from a Gaussian distribution can be seen.
\PSR\ (black circle) lies close to the peak intensity. The containment radius
of 68\% is $0.12^\circ$ (Sec.~\ref{sec:PSF}).

Fig.~\ref{fig:DecoMSH80chp08} shows the deconvolved count map. It was obtained
using the same cuts and 10 iterations with the Richardson-Lucy algorithm. The
results are similar to the smoothed excess map of Fig.~\ref{fig:SmoothMSH80},
but the overall width is smaller and the peak intensity is higher since the
spread by the PSF is reduced. To make Fig.~\ref{fig:DecoMSH80chp08} comparable
to Fig.~\ref{fig:SmoothMSH80}, the background of 5.0 counts/bin has to be
subtracted. Then the peak intensities of the smoothed and the restored map
compare as 5 to 9 counts.

A slightly different image is obtained with an image amplitude cut of
400\,p.e., which raises the energy threshold to about 900\,GeV.
Fig.~\ref{fig:SmoothMSH400} shows this excess map smoothed with a Gaussian
function $(\sigma=0.03^\circ)$. Due to the increased energy threshold, the size
of the PSF is reduced to about half its size at 80\,p.e. increasing the
resolution. The 68\% containment radius is $0.063^\circ$ (Sec.~\ref{sec:PSF}).
Although the event statistic is reduced, the signal to noise ratio is higher.
The main emission region in this map appears slightly curved.

Fig.~\ref{fig:DecoMSH400chp08} shows the 400\,p.e. count map after
deconvolution with 10 iterations of the Richardson\hyp{}Lucy algorithm. The
emission region has a significantly reduced width but about the same length as
the corresponding smoothed map. The highest intensity is seen in a compact
region oriented along a northwest-southeast direction which covers the \X-ray
jet axis. The emission region is also slightly curved and extends to the
southwest of the pulsar.

The maps are interesting and complementary to each other: the 80\,p.e. maps for
aspects to a low energy threshold, the 400\,p.e. maps for their increased
resolution. The also suggest a slightly different morphology at low and high
energies motivating an investigation for an energy dependent morphology in the
next section.

\subsection{Energy Bands} \label{sec:EnergyBands}
\begin{figure}[t!]
  \begin{minipage}[c]{0.327\linewidth}
    \includegraphics[width=\textwidth]{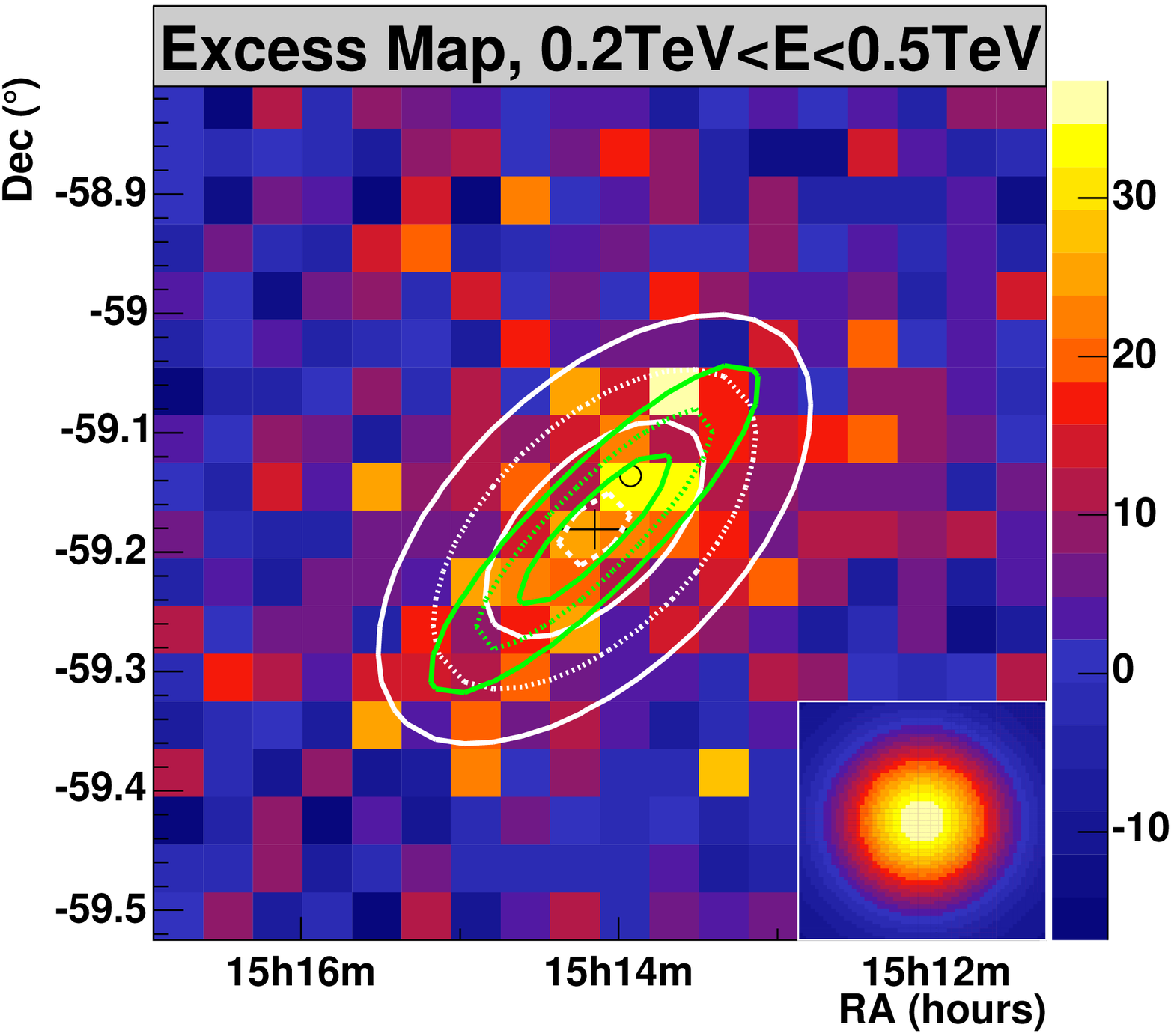}
  \end{minipage} \hfill
  \begin{minipage}[c]{0.327\linewidth}
    \includegraphics[width=\textwidth]{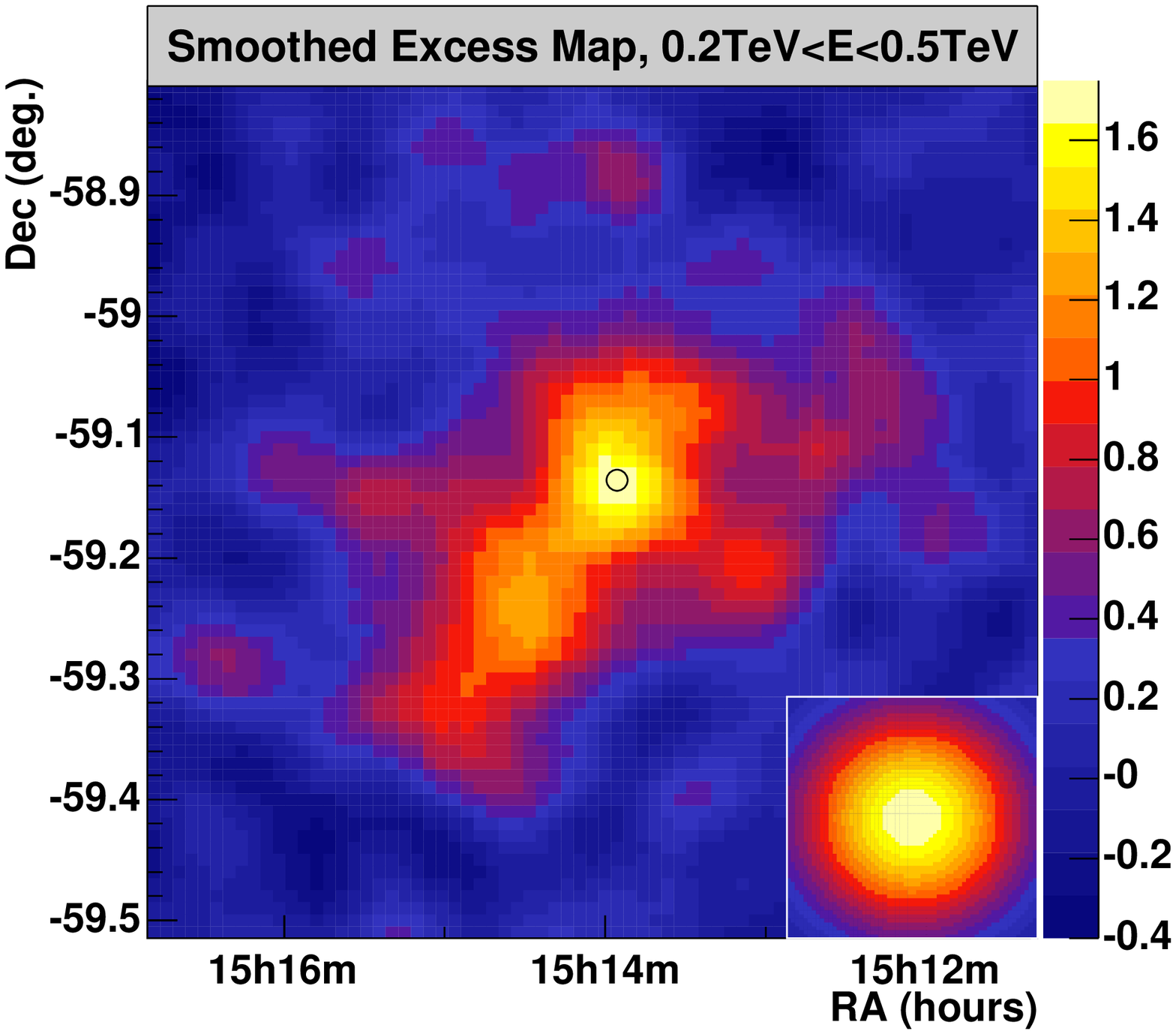}
  \end{minipage} \hfill
  \begin{minipage}[c]{0.327\linewidth}
    \caption[Morphology at Different Energy Bands]{Morphology at different
      energy bands.\\ First column: excess map overlaid with Gaussian fit
      function (white) and the corresponding component of the intrinsic width
      (green).\\ Second column: smoothed excess map ($\sigma_s$=0.03$^\circ$).}
    \label{fig:EnergyBands}
  \end{minipage} \hfill

  \begin{minipage}[c]{0.327\linewidth}
    \includegraphics[width=\textwidth]{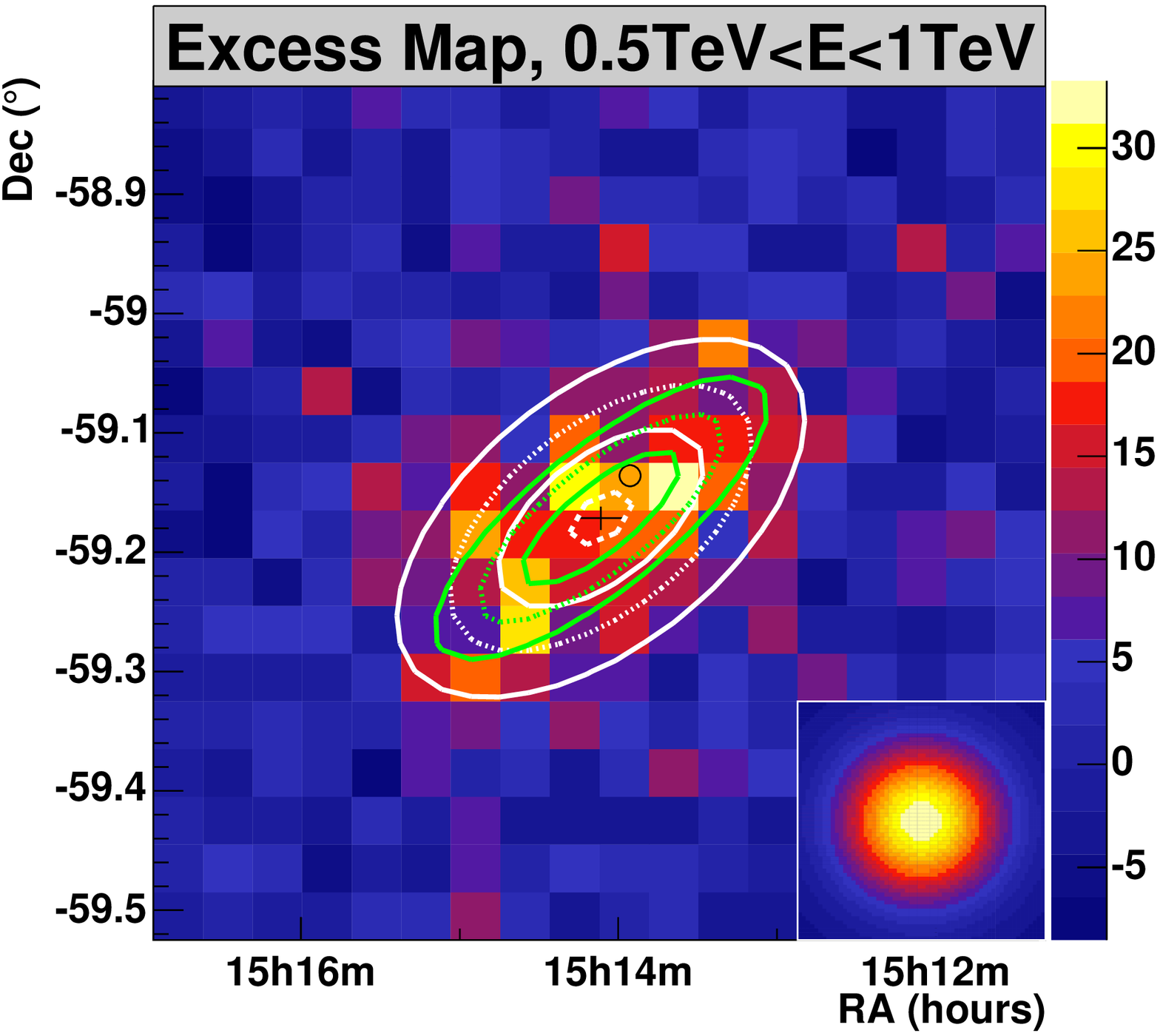}
  \end{minipage} \hfill
  \begin{minipage}[c]{0.327\linewidth}
    \includegraphics[width=\textwidth]{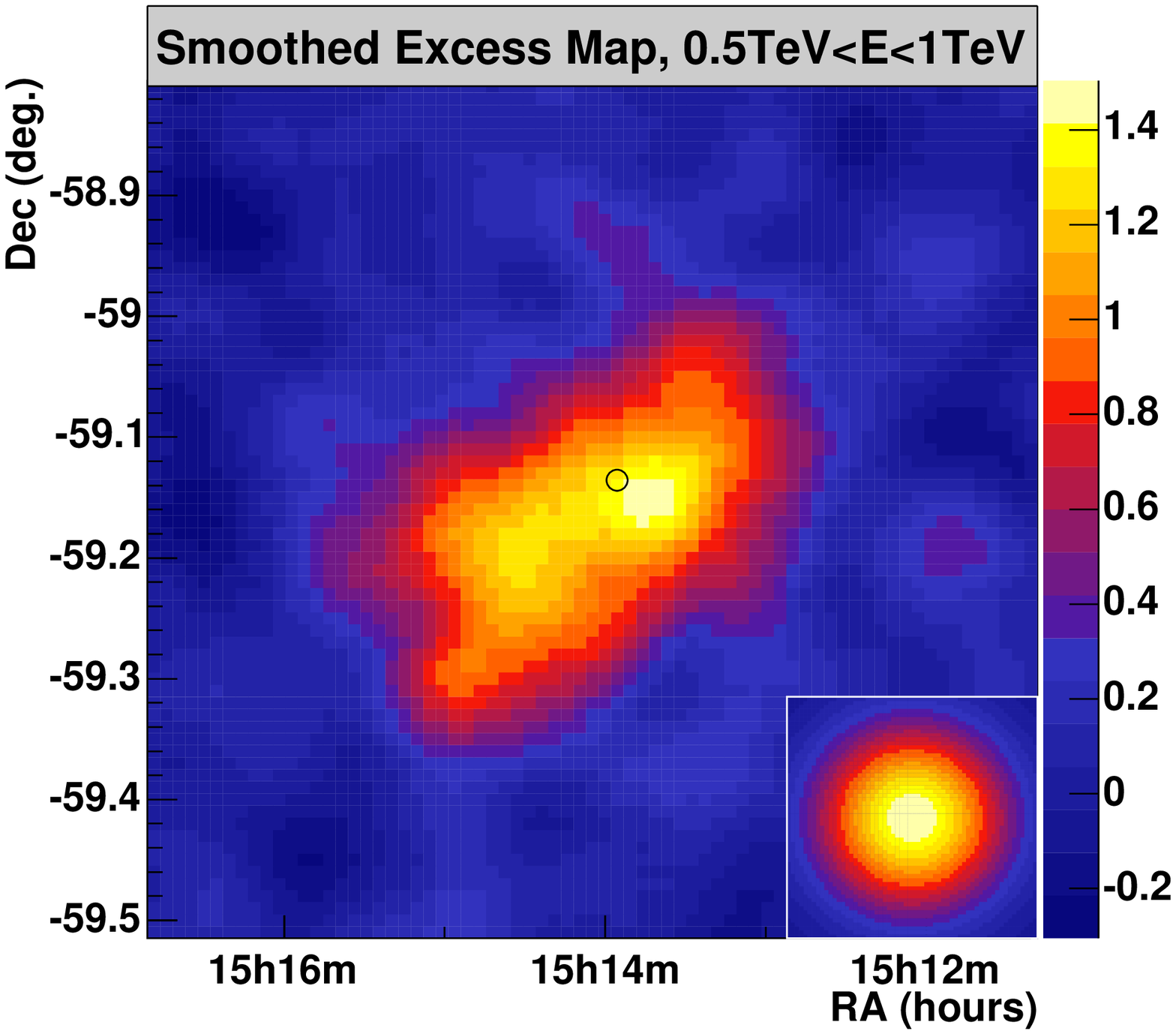}
  \end{minipage} \hfill
  \begin{minipage}[c]{0.327\linewidth}
    \caption*{Third column: excess maps additionally smoothed to the same 68\%
      containment radius of 0.125$^\circ$. The sky maps for $E<5$\,TeV have
      been combined. The PSF is indicated in the bottom right of each plot. The
      position of \PSR\ is marked by the black circle. 
    }
  \end{minipage} \hfill

  \begin{minipage}[c]{0.327\linewidth}
    \includegraphics[width=\textwidth]{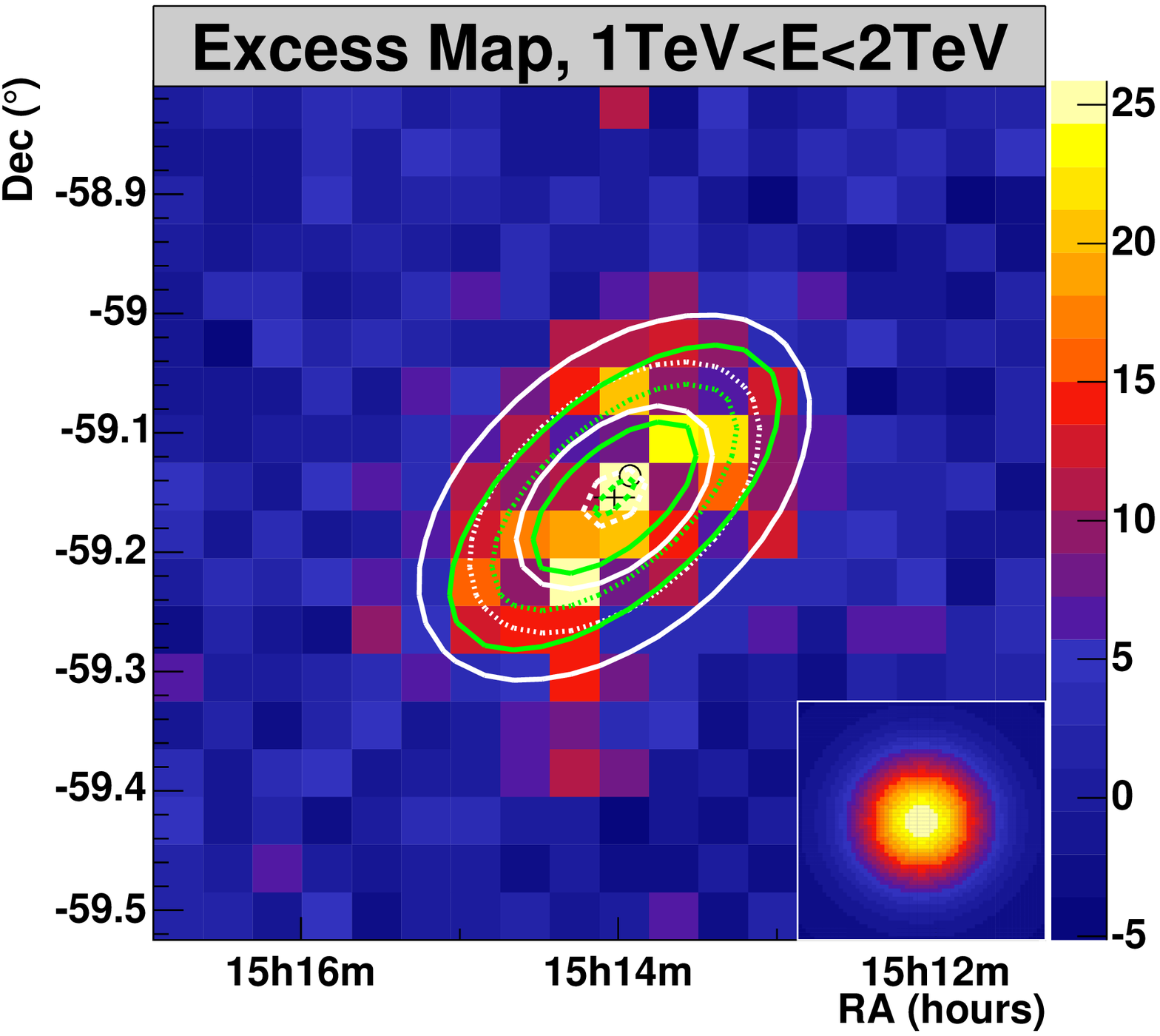}
  \end{minipage} \hfill
  \begin{minipage}[c]{0.327\linewidth}
    \includegraphics[width=\textwidth]{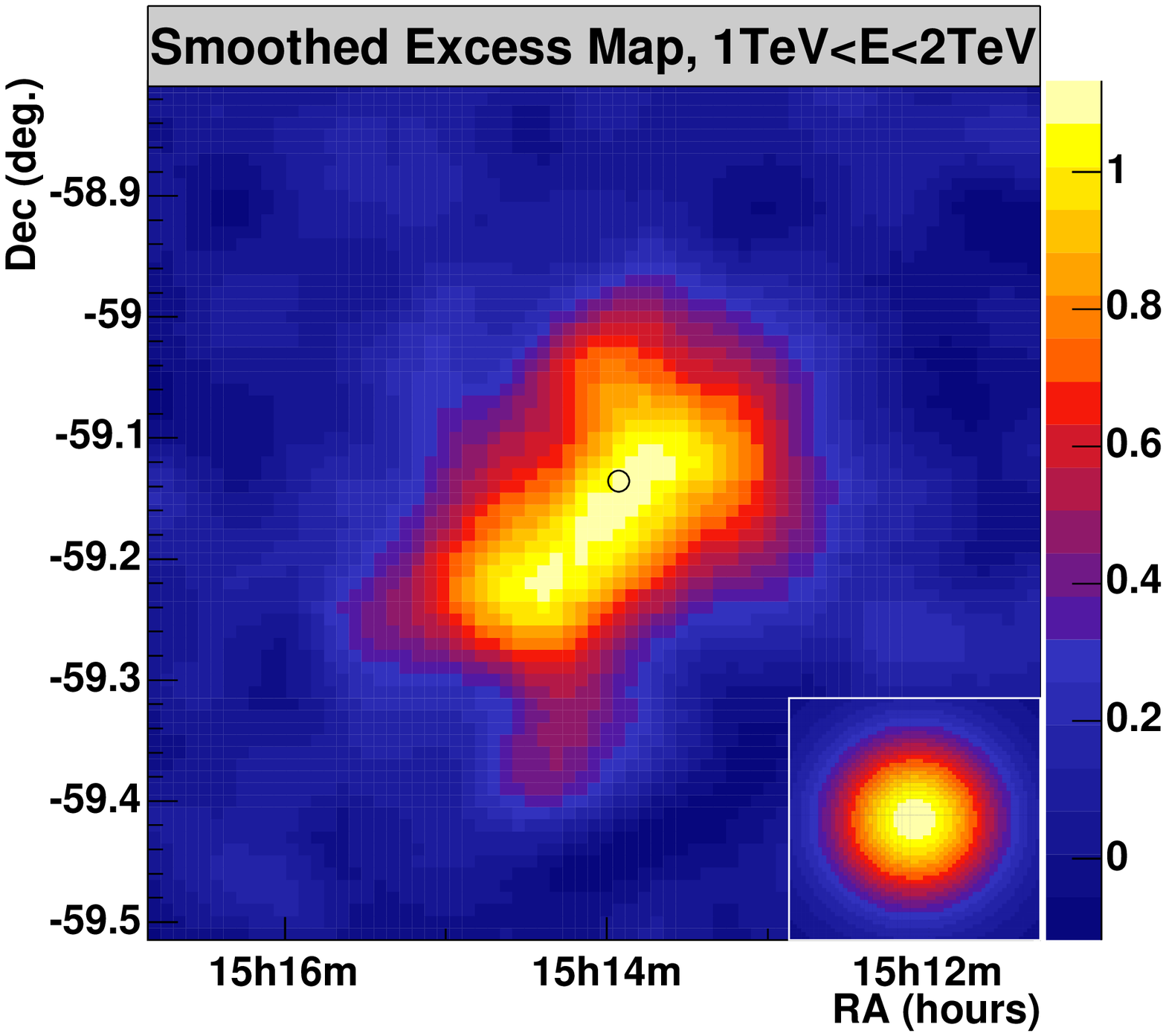}
  \end{minipage} \hfill
  \begin{minipage}[c]{0.327\linewidth}
    \hfill
  \end{minipage} \hfill

  \begin{minipage}[c]{0.327\linewidth}
    \includegraphics[width=\textwidth]{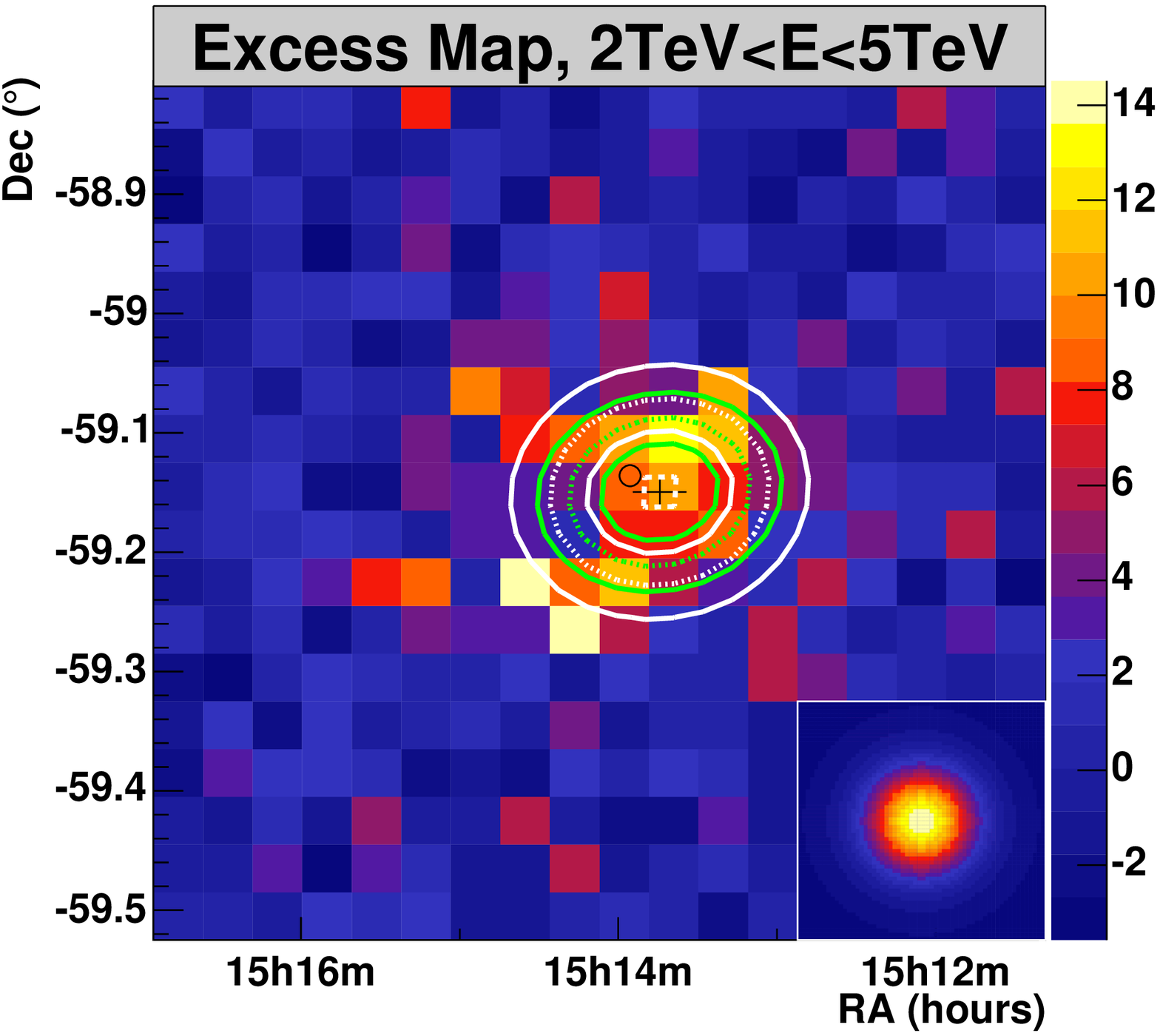}
  \end{minipage} \hfill
  \begin{minipage}[c]{0.327\linewidth}
    \includegraphics[width=\textwidth]{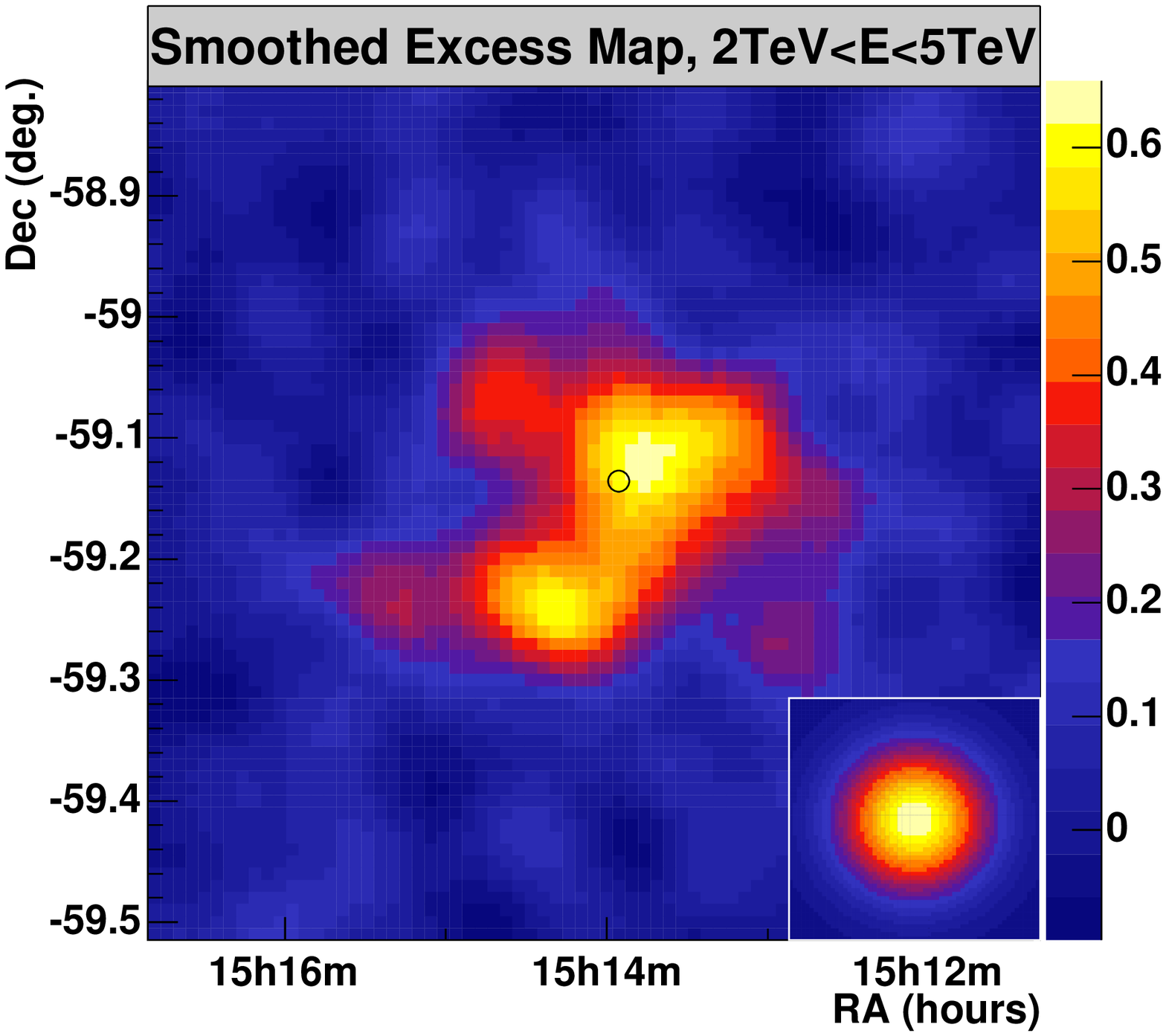}
  \end{minipage} \hfill
  \begin{minipage}[c]{0.327\linewidth}
    \includegraphics[width=\textwidth]{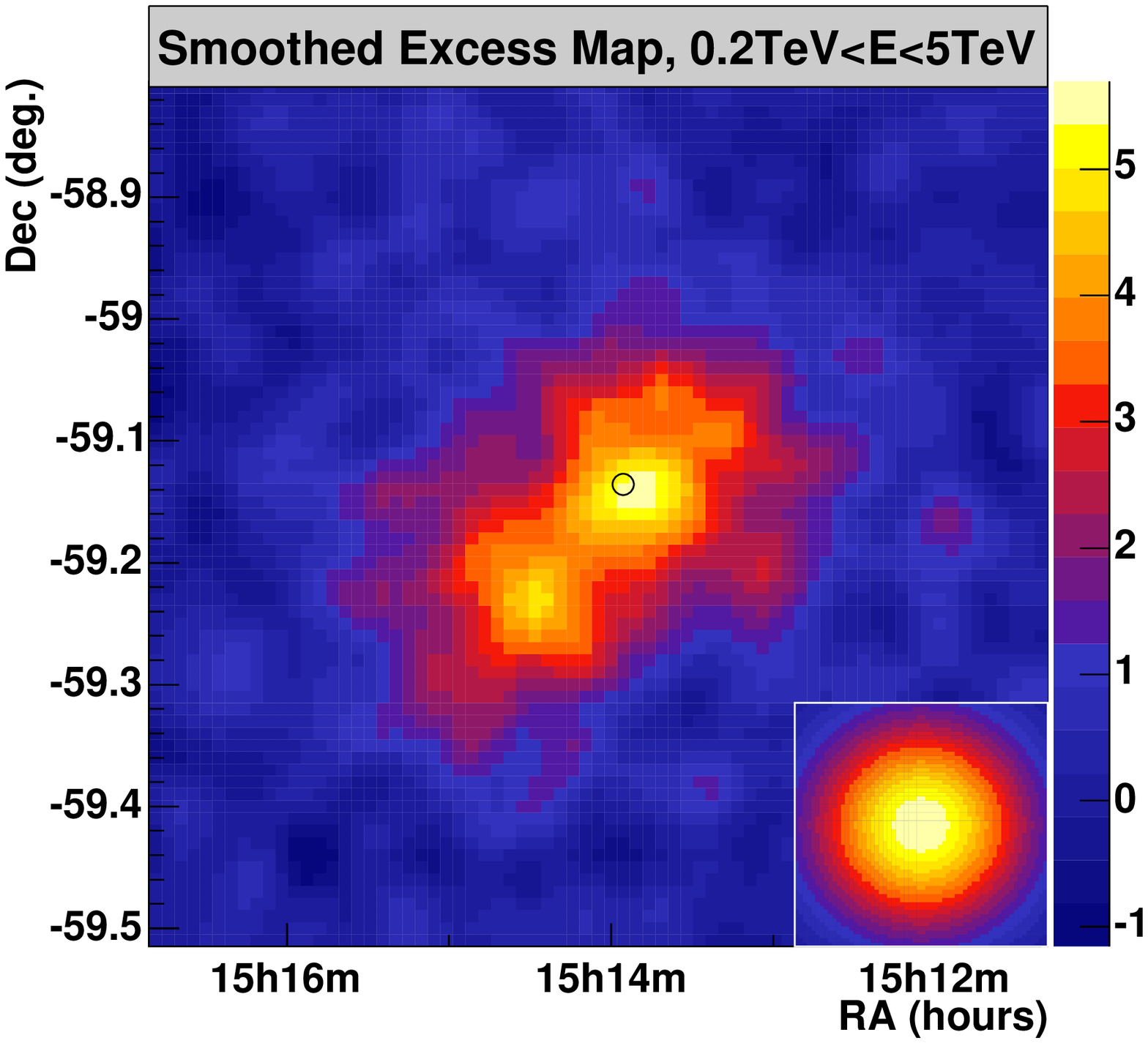}
  \end{minipage} \hfill

  \begin{minipage}[c]{0.327\linewidth}
    \includegraphics[width=\textwidth]{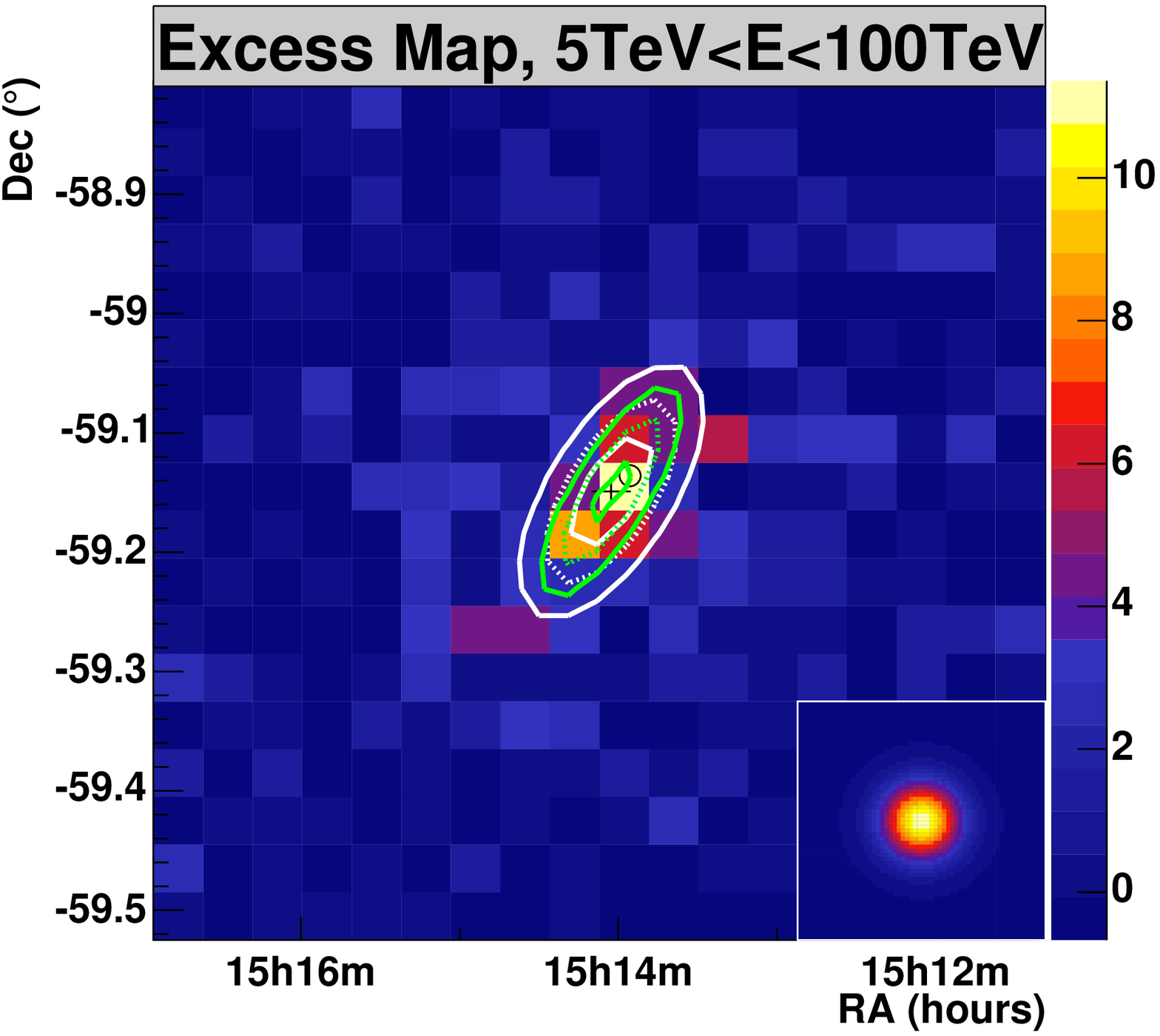}
  \end{minipage} \hfill
  \begin{minipage}[c]{0.327\linewidth}
    \includegraphics[width=\textwidth]{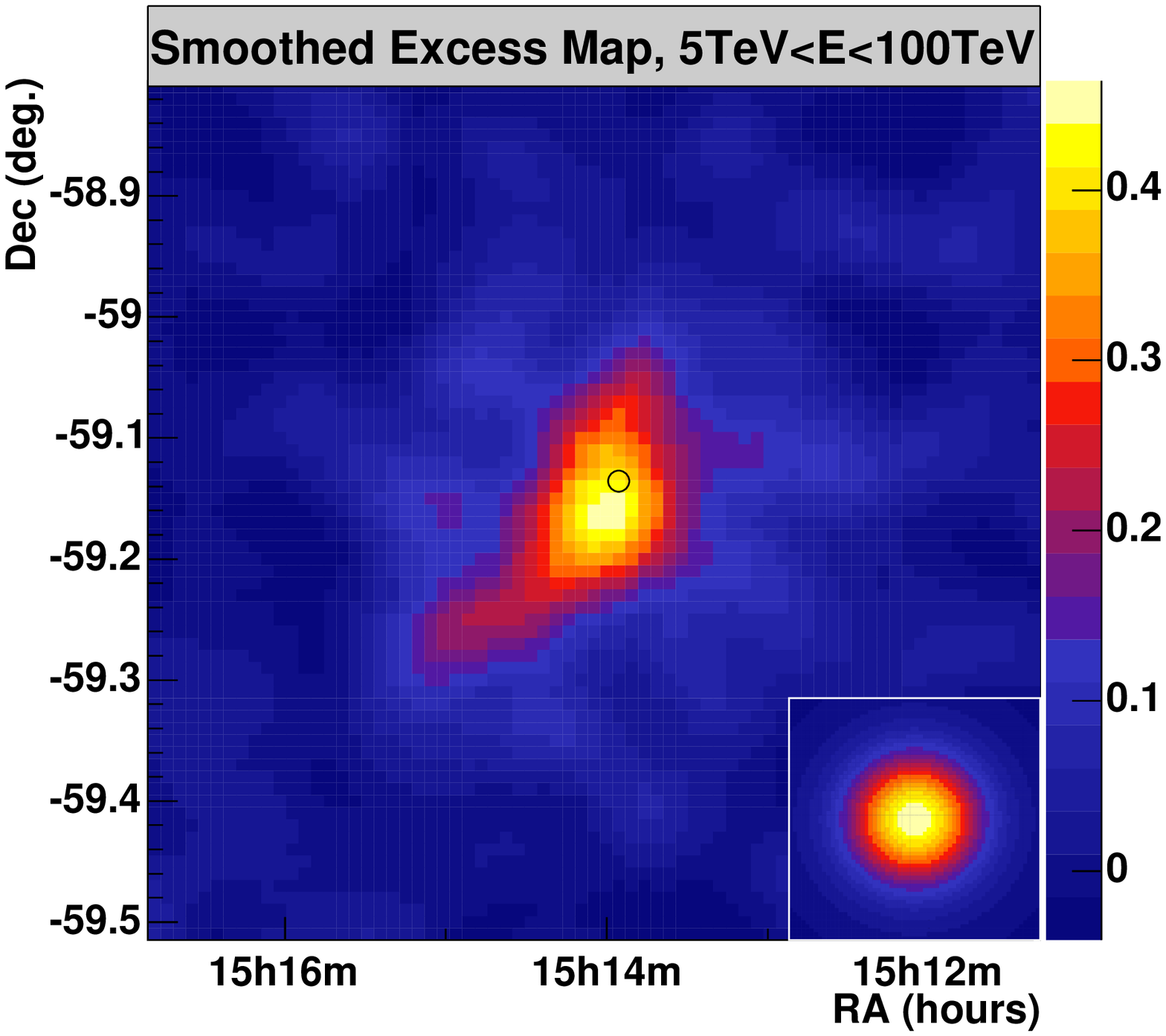}
  \end{minipage} \hfill
  \begin{minipage}[c]{0.327\linewidth}
    \includegraphics[width=\textwidth]{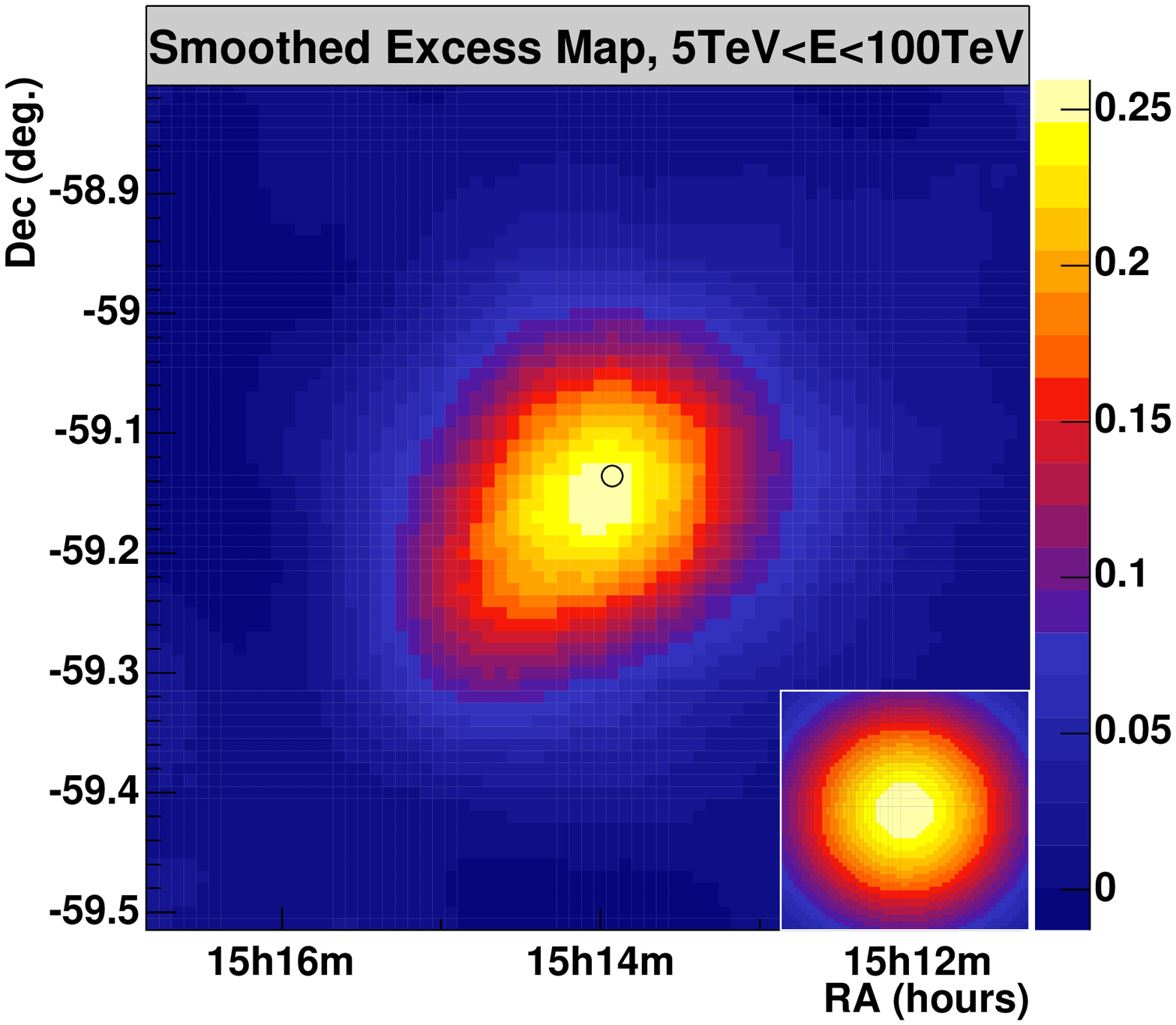}
  \end{minipage} \hfill
\end{figure}

To investigate the energy dependence of the morphology, the data was divided in
five different energy bands with the energy ranges 0.2--0.5, 0.5--1, 1--2, 2--5
and 5--100\,TeV. The event statistics and PSF of each energy band is summarized
in Tbl.~\ref{tbl:EBandsSignificance} and Tbl.~\ref{tbl:EnergyBandsPSF}
respectively. The corresponding excess maps are shown in
Fig.~\ref{fig:EnergyBands} (left column) each with a profile of the PSF in the
bottom right. The intrinsic extension of the \g-ray excess of each energy band
was determined from a fit of the Gaussian function $(G)$ to the map as
described in Sec.~\ref{sec:MSHPositionSize}. The contour lines of $G$ (white)
and the component representing the intrinsic source size (green) are overlaid.
One-dimensional slices of the fit functions are shown in
App.~\ref{app:EnergyBands}.

\begin{figure}[tbh!]
  \begin{minipage}[c]{0.327\linewidth}
    \centering\includegraphics[width=\textwidth]{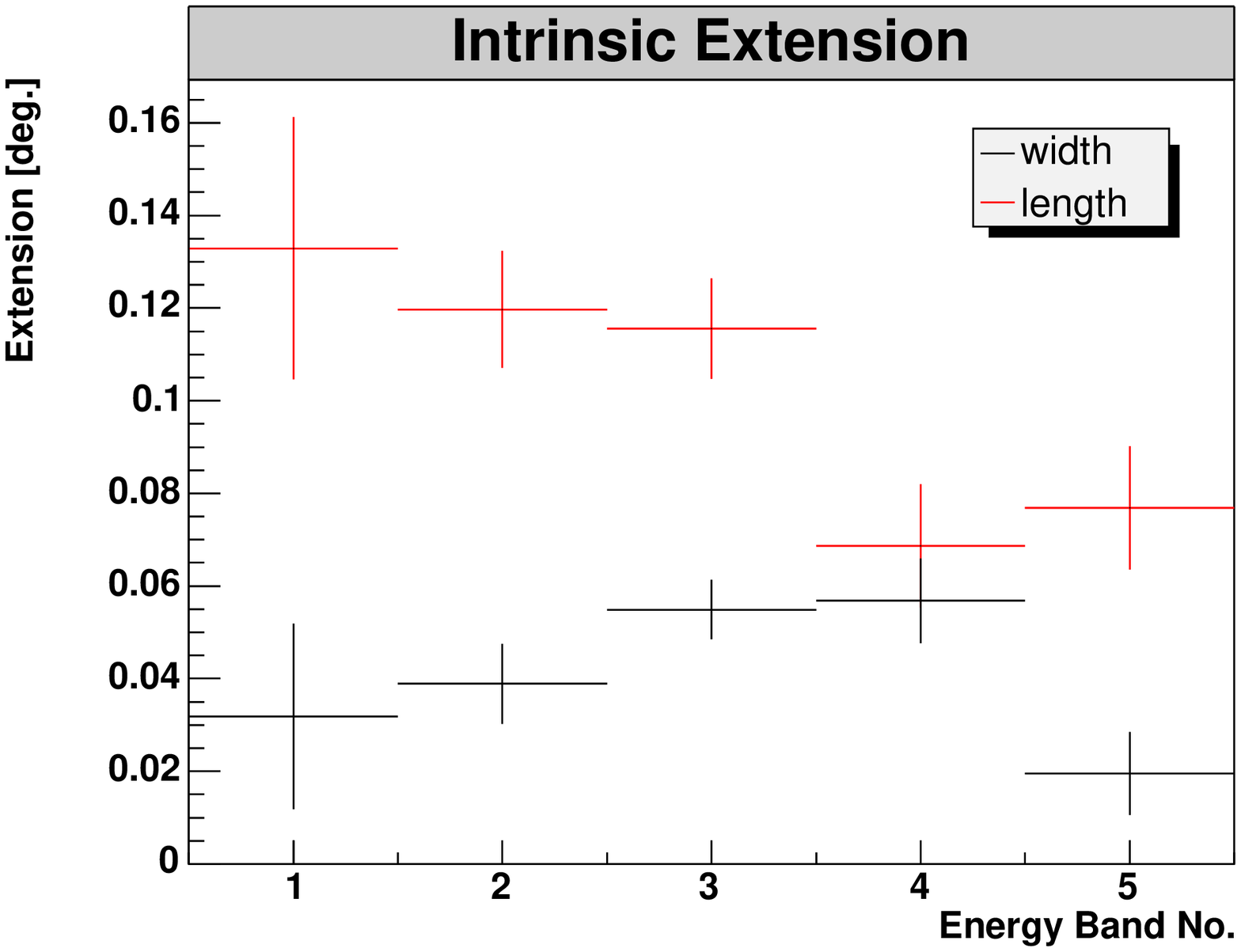}
    \caption[Intrinsic Extension of the Excess for Different Energy
      Bands]{Intrinsic width and length of the Gaussian function $G$
      (App.~\ref{app:GaussianFit}) for the fit to the excess maps of the energy
      bands in Fig.~\ref{fig:EnergyBands}.}
    \label{fig:Width}
  \end{minipage} \hfill
  \begin{minipage}[c]{0.327\linewidth}
    \centering\includegraphics[width=\textwidth]{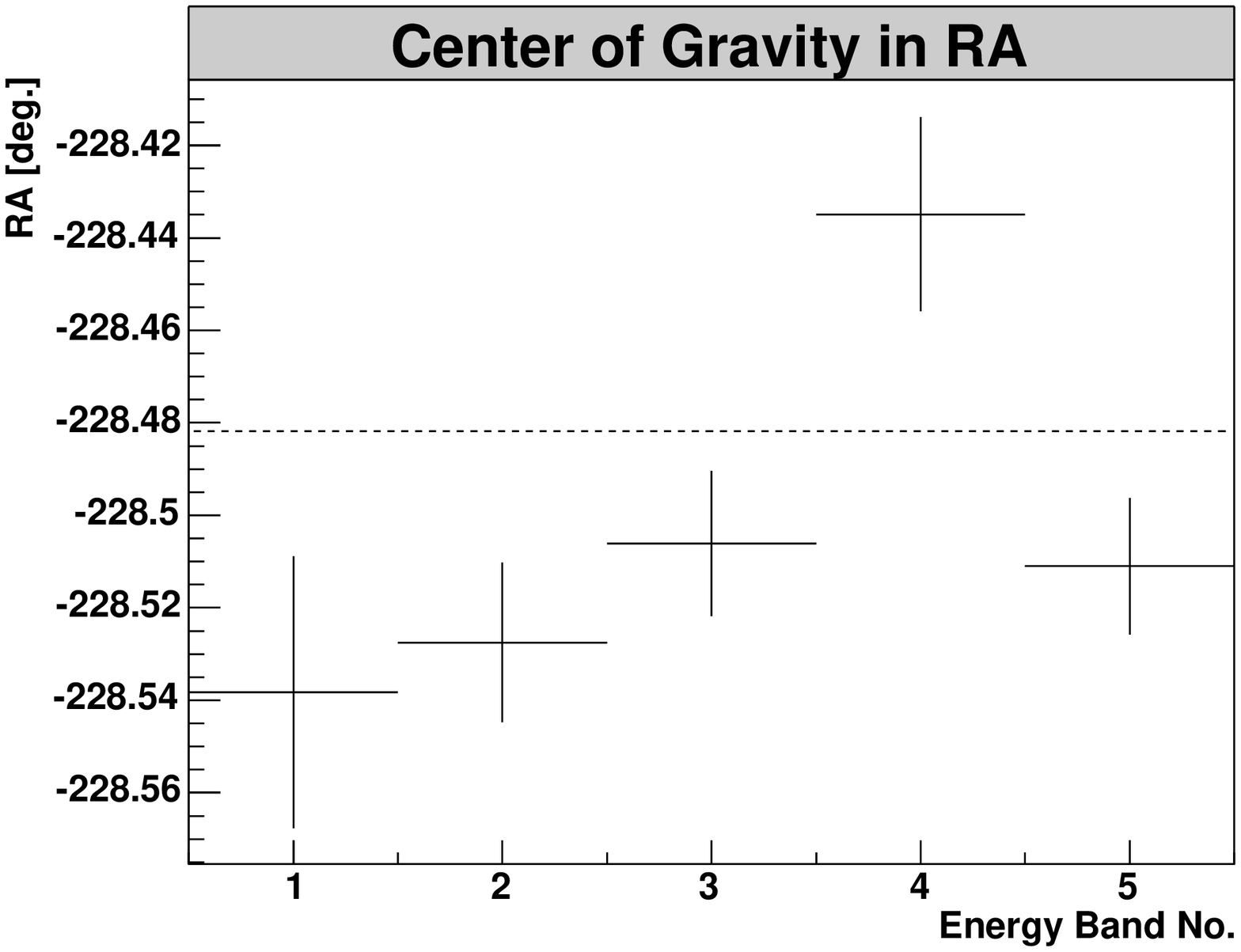}
    \caption[Centroid in RA for Different Energy Bands]{Centroid in RA of the
      Gaussian function $G$ (App.~\ref{app:GaussianFit}) for the fit to the
      excess maps of the energy bands in Fig.~\ref{fig:EnergyBands}.}
    \label{fig:CentroidRA}
  \end{minipage} \hfill
  \begin{minipage}[c]{0.327\linewidth}
    \centering\includegraphics[width=\textwidth]{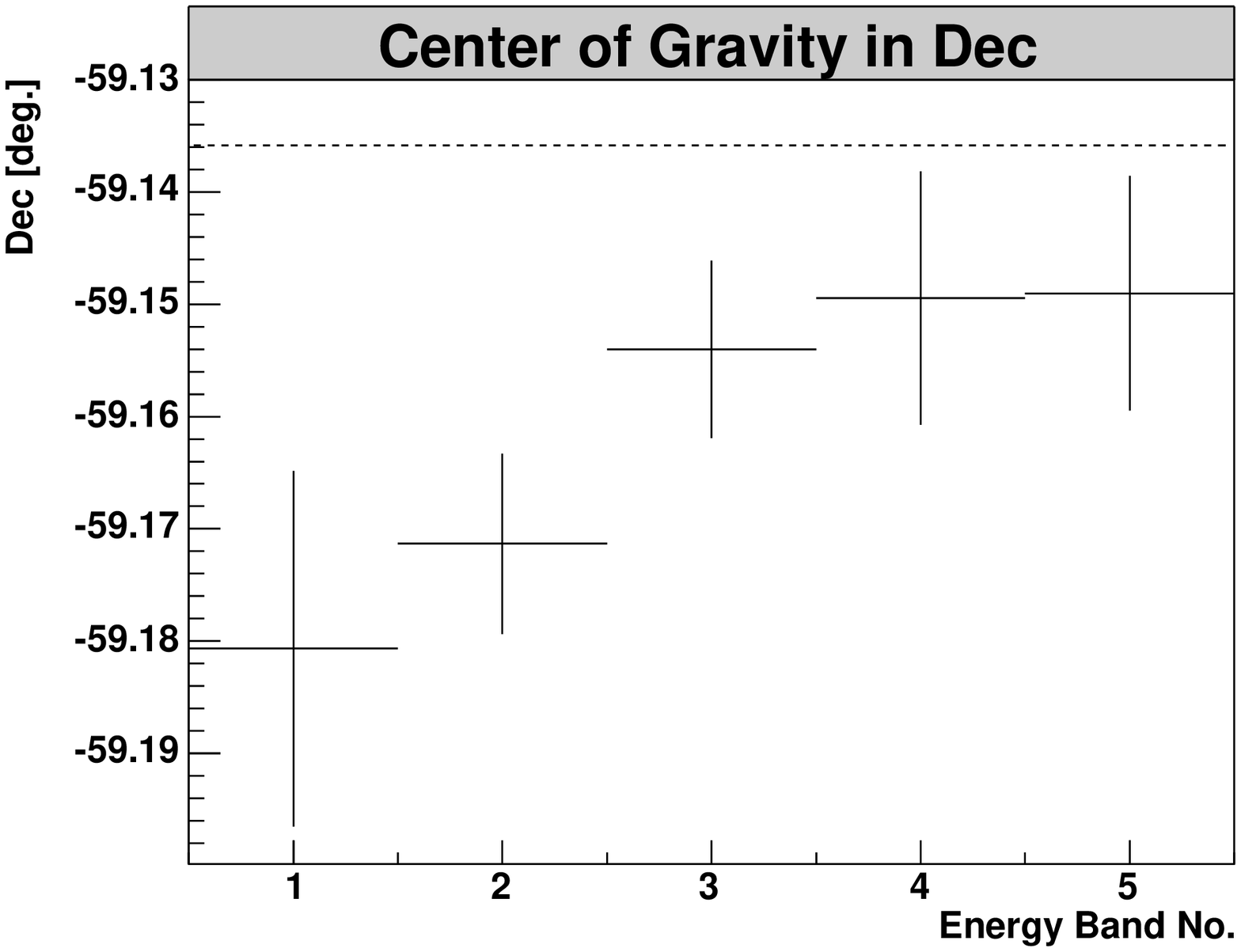}
    \caption[Centroid in Dec for Different Energy Bands]{Centroid in Dec of the
      Gaussian function $G$ (App.~\ref{app:GaussianFit}) for the fit to the
      excess maps of the energy bands in Fig.~\ref{fig:EnergyBands}.}
    \label{fig:CentroidDec}
  \end{minipage}
\end{figure}

Fig.~\ref{fig:Width} show the intrinsic standard deviations of the major and
minor axis versus the different energy bands in order of increasing energy (cf.
Tbl.~\ref{tbl:EBandsSignificance}). A decreasing trend of the longitudinal
source extension is found. Fig.~\ref{fig:CentroidRA} and \ref{fig:CentroidDec}
show the same representation for the centroid of the fit function in RA and
Dec. In both cases the centroid converges towards the pulsar position with
increasing energy. The pulsar position is indicated by the horizontal dashed
lines. This is what would be expected if the pulsar is the source of a \g-ray
producing pulsar wind.

The second column in Fig.~\ref{fig:EnergyBands} shows the excess maps smoothed
with a Gaussian function of standard deviation $\sigma_s=0.03^\circ$. Again the
size of the smoothed emission region decreases with energy, but a immediate
conclusion about the intrinsic source extension is complicated by the energy
dependence of the PSFs (cf. Sec.~\ref{sec:PSF}).

However, a conclusion becomes possible, if the PSF is similar for each energy
band. Therefore count maps with a smaller PSF, have been artificially smoothed
to increase the PSF to the same 68\% containment radius of 0.125$^\circ$. The
difference of the source extensions between high and low energies is more
apparent if the sky maps below 5\,TeV are combined. The required smoothing
values $(\sigma_s)$ are 0.023$^\circ$ and 0.064$^\circ$ respectively (cf.
Tbl.~\ref{tbl:EnergyBandsPSF}). The resulting two sky maps are shown in the
third column of Fig.~\ref{fig:EnergyBands}.

\begin{figure}[h!]
  \begin{minipage}[c]{0.55\linewidth} \hspace{0.25cm}
    \includegraphics[width=\linewidth]{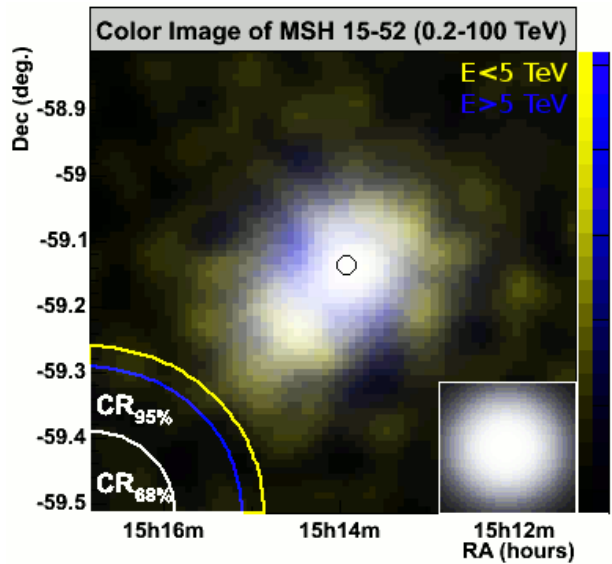}
  \end{minipage} \hfill
  \begin{minipage}[c]{0.44\linewidth}
    \caption[Two Color Image of \MSH]{Two color image of \MSH\ combining the
      maps from Fig.~\ref{fig:EnergyBands} for $E<5$\,TeV (yellow) and
      $E>5$\,TeV (blue). The color palettes have been chosen complementary to
      each other to provide a color neutral (white) appearance for similar
      brightness. However, the color brightness between both energy bands is
      not normalized. Therefore it does not represent the absolute number of
      events but the relative intensity profile of each band. The position of
      \PSR\ is indicated by the circle. The 68\% and 95\% containment radii are
      shown at the bottom left, the profile of the PSF at the bottom right.}
    \label{fig:TwoColorImage}
  \end{minipage}\hfill
\end{figure}

Fig.~\ref{fig:TwoColorImage} shows the combination of these two smoothed maps.
Two complementary color scales, i.e. yellow for $E<5$\,TeV and blue for
$E>5$\,TeV, have been chosen to provide a color neutral, i.e. white, appearance
for similar bright regions. However, the color brightness between both energy
bands is not normalized. Therefore, it does not represent absolute numbers of
events but the relative intensity profile of each band. The 68\% and 95\%
containment radii are shown at the bottom left. The later do not match exactly
since the PSF parameterization changes with energy and therefore prevents this.
Since the central region appears white and the outer region yellow, the \g-ray
emission region is more compact at higher energies.

\clearpage

\begin{table}[h!]
  \centering
  \caption[Statistics and Significance at Different Energy Bands]{Statistics
  and significance $(S)$ for different energy bands as obtained from the
  ring-background model with the same configuration discussed in
  Sec.~\ref{sec:Ring-Background} before. The energy bands have been numbered in
  oder of increasing energy.}
  \bigskip
  \begin{tabular}{ccccccc}
    \hline \hline
    No. & $E$ [TeV]     & $N_{\rm ON}$ & $N_{\rm OFF}$ & $N_\gamma$
    & $\alpha$ & $S[\sigma]$ \\
    \hline
    1 & $0.28$\,TeV$<E<0.5$\,TeV & 10820 & 51594  & 1282 & 0.185  & 11.8 \\
    2 & $0.5$\,TeV$<E<1$\,TeV    &  3615 & 13165  & 1051 & 0.195  & 17.7 \\
    3 & $1$\,TeV$<E<2$\,TeV      &  1875 &  5532  &  814 & 0.192  & 20.3 \\
    4 & $2$\,TeV$<E<5$\,TeV      &  1045 &  3932  &  422 & 0.158  & 14.1 \\
    5 & $5$\,TeV$<E<100$\,TeV    &   296 &   672  &  185 & 0.165  & 13.0 \\
    \hline
      & $0.28$\,TeV$<E<100$\,TeV & 17651 & 74895  & 3752 & 0.186  & 27.9 \\
    \hline \hline
  \end{tabular}
  \label{tbl:EBandsSignificance}
\end{table}

In summary, the energy dependent analysis supports a picture of a decreasing
longitudinal extension of the \g-ray emission along the pulsar jet axis with
increasing energy. This is similar to what is observed at X-rays
(\cite{Goret:2006}) as pointed out in Sec.~\ref{sec:X-rays}.

\newpage

\subsection{Correlation with X-ray Emission} \label{sec:X-rayCorrelation}
The similarity between the \g-ray and \X-ray morphology discussed in
Chp.~\ref{chp:MSHandPSR} motivates further comparisons. According to
Chp.~\ref{sec:PWNSpectra} a correlation between X- and \g-ray emission is
expected for a synchrotron and IC radiation producing wind of VHE electrons.
Here this correlation is investigated using H.E.S.S. \g- and \X-ray data from
the ROSAT and Chandra satellites.

\subsubsection{ROSAT Data}
\begin{figure}[h!]
  \centering
  \begin{minipage}[c]{.35\linewidth}
    \caption[ROSAT X-Ray Contours Overlaid to the H.E.S.S. \g-Ray Map]{ROSAT
      \X-ray contours overlaid on the reconstructed H.E.S.S. \g-ray map of
      Fig.~\ref{fig:SmoothMSH400}. A good correlation between both wavelengths
      can be seen. A profile of the H.E.S.S PSF is indicated in the bottom
      right corner. The position of \PSR\ is marked by the black circle.}
    \label{fig:ROSATCorrelations}
  \end{minipage}\hfill
  \begin{minipage}[c]{.6\linewidth}
    \includegraphics[width=1\textwidth]{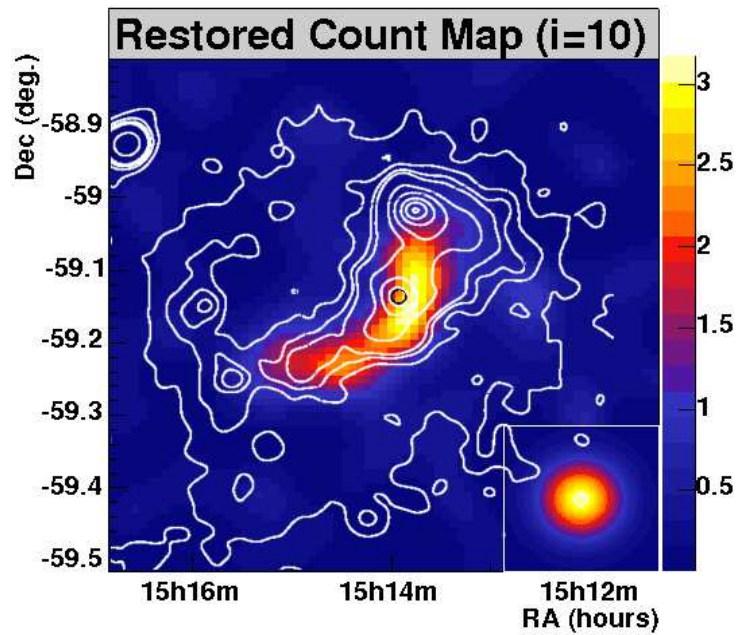}
  \end{minipage}
\end{figure}

The ROSAT \X-ray data was recorded in 1991 and 1992 covering the energy range
from 0.1 to 2.4\,keV. Fig.~\ref{fig:ROSATCorrelations} shows the deconvolved
H.E.S.S. \g-ray map of Fig.~\ref{fig:DecoMSH400chp08} overlaid with the contour
lines of the ROSAT data at the levels of 0.5, 1, 1.5, 2, 5, 10, 20, 30 and 40
in arbitrary units as determined by \citet{Trussoni:1996}. A good correlation
between the \g- and \X-ray data is found. However, it is interesting to note
that the region of \RCW\ shows strong \X-ray emission but no strong \g-ray
emission. Such a difference can indicate a difference of the X-ray production
mechanisms in this region.

\subsubsection{Chandra Data}

\begin{figure}[h!]
  \begin{minipage}[t]{0.5\linewidth}
    \includegraphics[width=\textwidth]{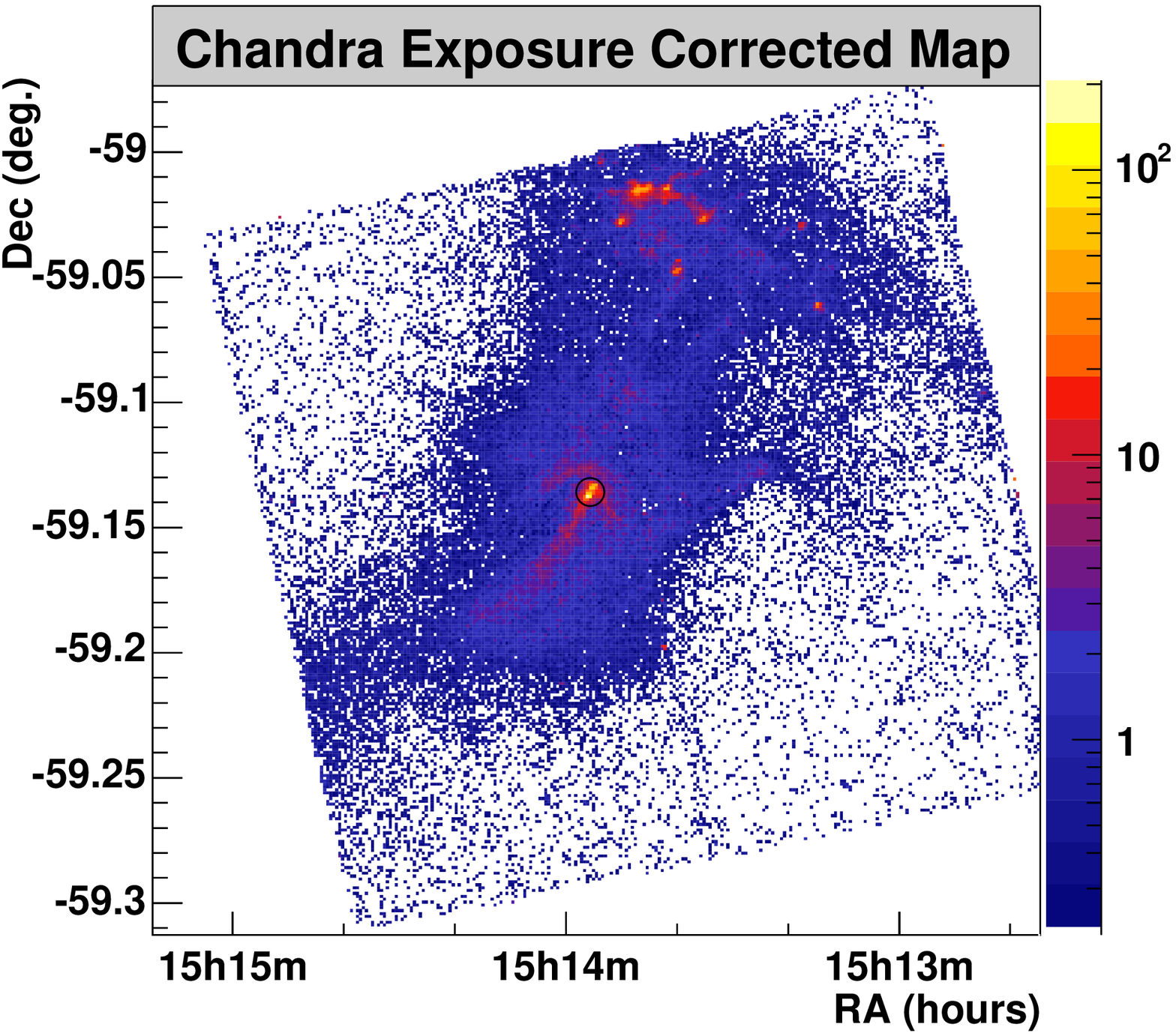}
    \caption[Chandra X-Ray Map of \MSH]{Chandra map of the \X-ray emission in
      the energy range from 0.3 to 10\,keV. Strong \X-ray emission is seen from
      the region of \PSR\ (black circle) and the northwest region of \RCW.}
    \label{fig:ChandraRate}
  \end{minipage}
  \begin{minipage}[t]{0.5\linewidth}
    \includegraphics[width=\textwidth]{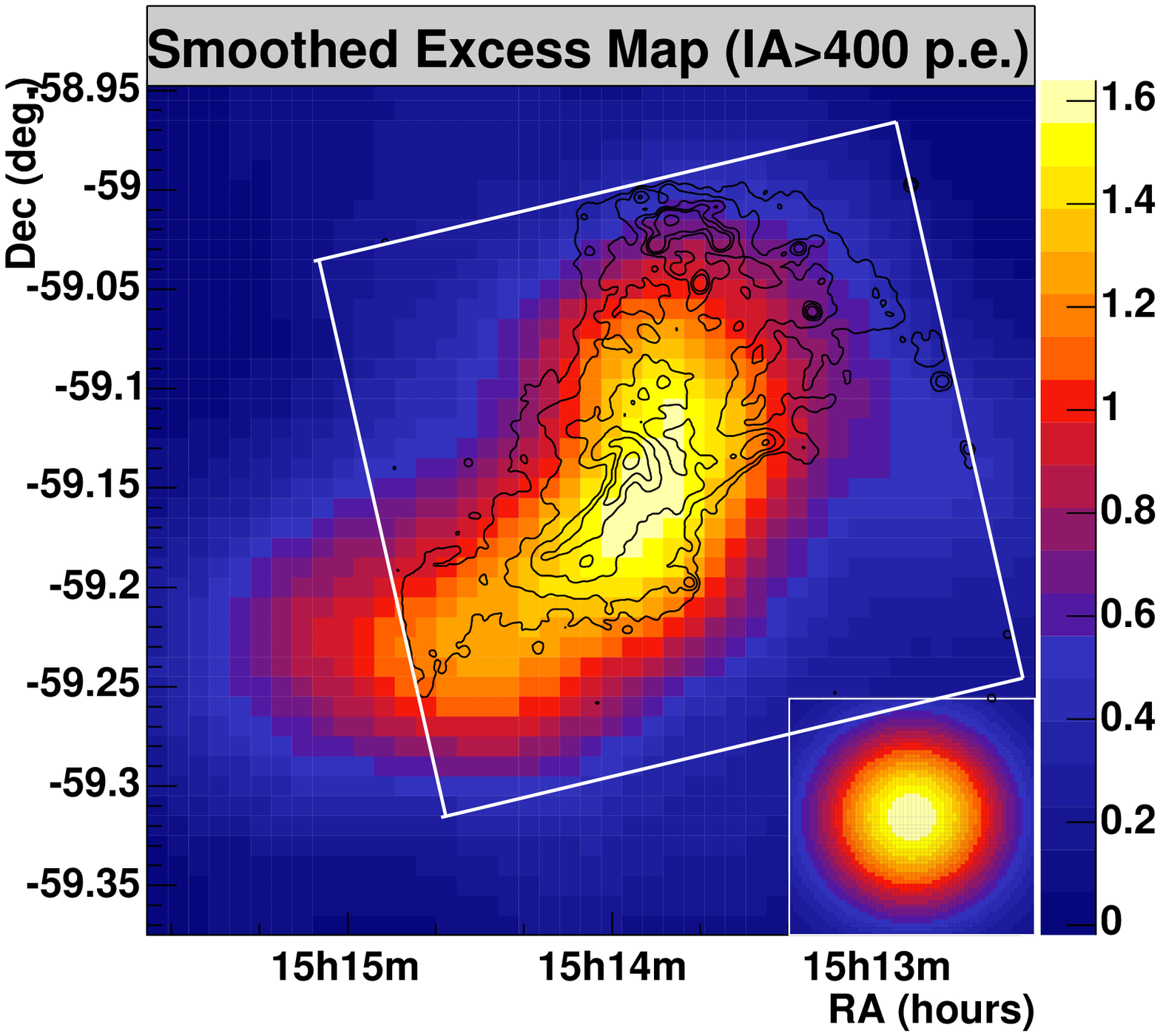}
    \caption[Chandra \X-ray Contours Overlaid to the H.E.S.S. \g-Ray
      Map]{H.E.S.S. \g-ray excess map overlaid with the Chandra \X-ray contours
      of Fig.~\ref{fig:ChandraRate}. The contour lines are chosen at the levels
      of 3, 6, 12, 18, and 30\% of the peak intensity. The PSF is shown (bottom
      right).}
    \label{fig:ChandraContourLines}
  \end{minipage}
\end{figure}

In 2000, the Chandra satellite provided new \X-ray data of higher resolution
and statistics. The data discussed here was recorded on August 14, 2000 with
observation ID 754. It is the same data which was discussed in Chp.
\ref{chp:MSHandPSR} based on the analysis carried out by \citet{Gaensler:2002}.
It has a live-time of 19039\,s, a total number of 222914 \X-ray events and
covers the energy range from 0.3 to 10\,keV. Fig.~\ref{fig:ChandraRate} shows
the exposure-corrected \X-ray map of this data on a logarithmic scale. Details
about the Chandra data analysis and exposure correction are given in
App.~\ref{app:ChandraDataAnalysis}. The Chandra and the ROAST \X-ray maps are
in good agreement. Both show significant \X-ray emission from the region of
\PSR\ and \RCW\ and an elongated structure extending to the southeast.
Moreover, the Chandra map can clearly resolve the jet-like character of this
structure and the clumps of \X-ray emission in the region of \RCW.
Fig.~\ref{fig:ChandraContourLines} shows the contour lines of the Chandra data
at levels of 3, 6, 12, 18 and 30\% overlaid on the H.E.S.S. image of
Fig.~\ref{fig:SmoothMSH80}. The contour lines were obtained from the count map
after smoothing with a Gaussian function of a width of 0.005$^\circ$. The
Chandra field of view is indicated by the white square. Again, the \g-ray
emission seems to be correlated with the \X-ray emission at the \PSR\ region,
however not at the \RCW\ region.

To quantify the correlation between the \g- and \X-ray emission, the
correlation coefficient was calculated for the H.E.S.S. and Chandra data in two
rectangular regions. One of them $(R_{\rm pulsar})$ covers the region of \PSR\
and the other $(R_{\rm \RCW})$ covers the region of \RCW. The correlation
coefficient ($\varrho$) is defined as
\begin{equation}
  \varrho = \frac{\sum^N_{t=1}{(\gamma_i-\overline\gamma)(X_i-\overline X)}}
          {\sqrt{\sum^N_{t=1}(\gamma_i-\overline\gamma)^2\sum^N_{t=1}
              (X_i-\overline X)^2}},
\end{equation}
where $\gamma_i$ and $X_i$ represent the content of bin $i$ of the \g- and
\X-ray maps respectively. $\overline\gamma$ and $\overline X$ denote the
corresponding mean values. $\varrho$ is bound to the interval $(-1<\varrho<1)$,
where the values 1, 0, and -1 express a perfect correlation, no correlation and
perfect anti-correlation respectively.

\begin{figure}[ht!]
  \begin{minipage}[c]{\linewidth}
    \includegraphics[width=0.5\textwidth]{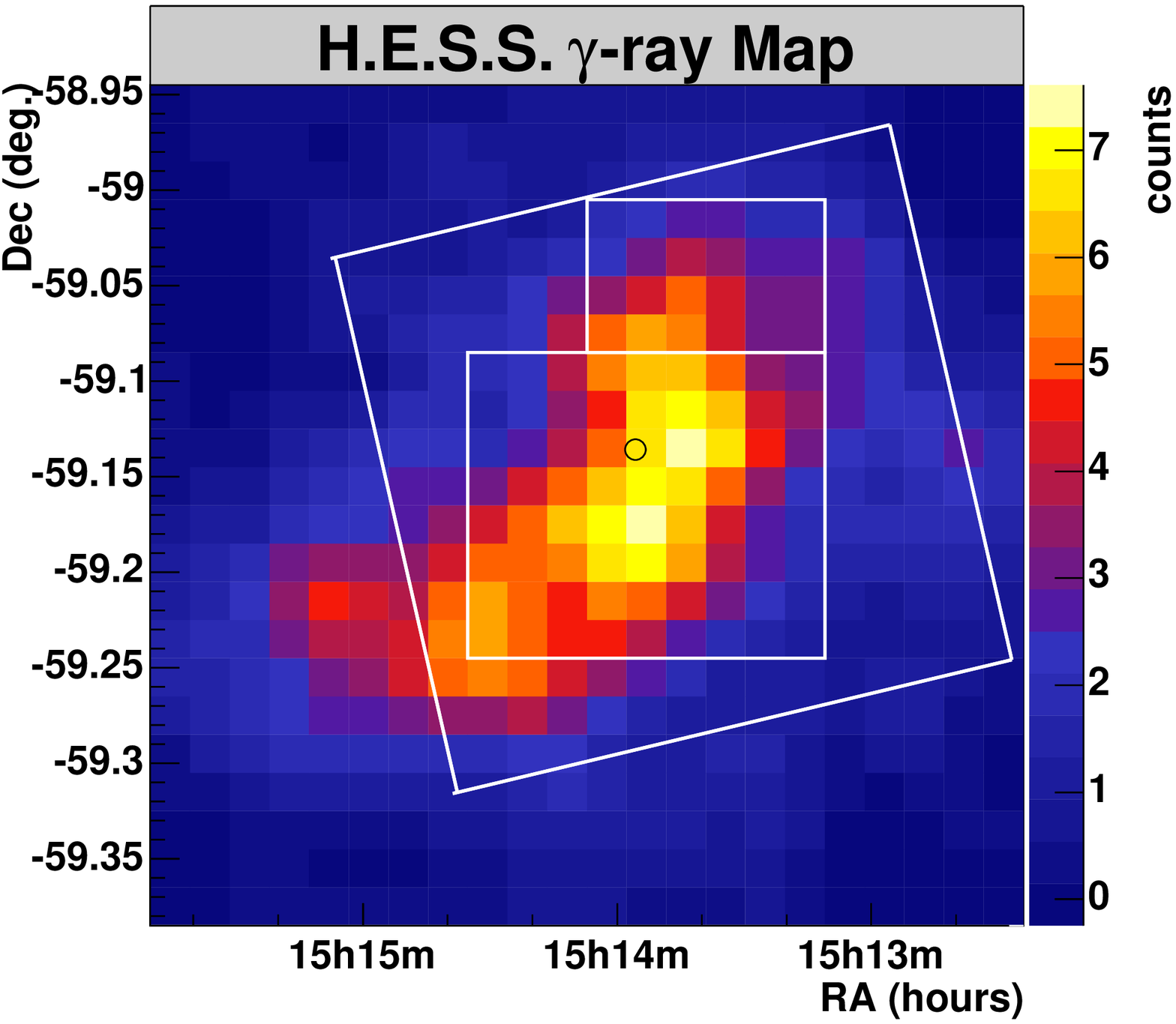}
    \includegraphics[width=0.5\textwidth]{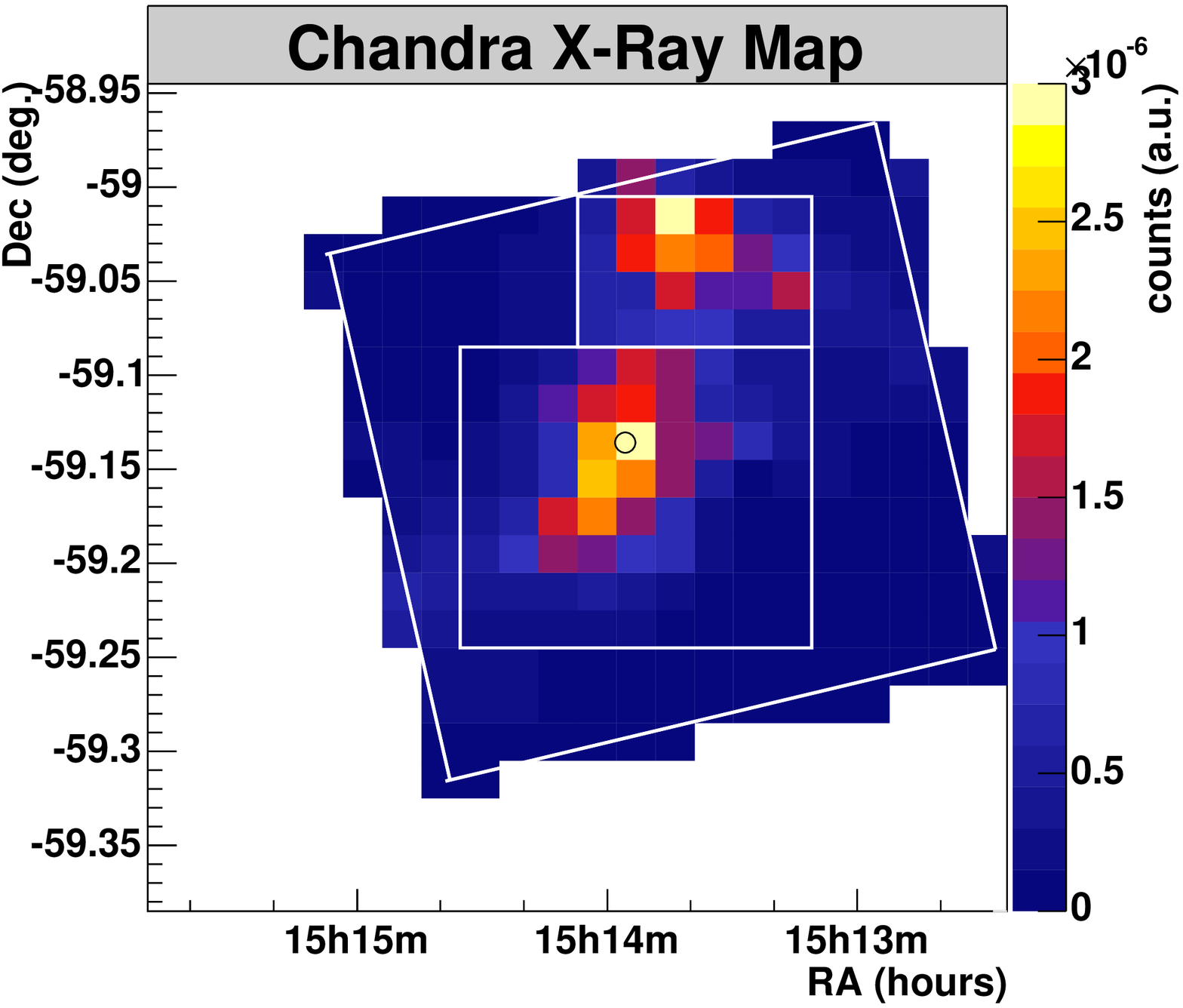}
    \caption[H.E.S.S. \g-Ray and Chandra X-Ray Map with Identical
    Binning]{H.E.S.S. \g-ray excess map (left) and Chandra \X-ray map (right)
    of \MSH\ with identical binning and a bin width of $0.02^\circ \times
    0.02^\circ$. The two adjacent (white) squares define the regions $R_{\rm
    pulsar}$ (9\,bins$\times$8\,bins) and $R_{\RCW}$ (6\,bins$\times4$\,bins)
    where correlation coefficients are determined. The position of \PSR\ is
    indicated by the black circle.}
    \label{fig:ChandraCorrelation}
  \end{minipage}\hfill
\bigskip

  \begin{minipage}[c]{\linewidth}
    \includegraphics[width=0.5\textwidth]{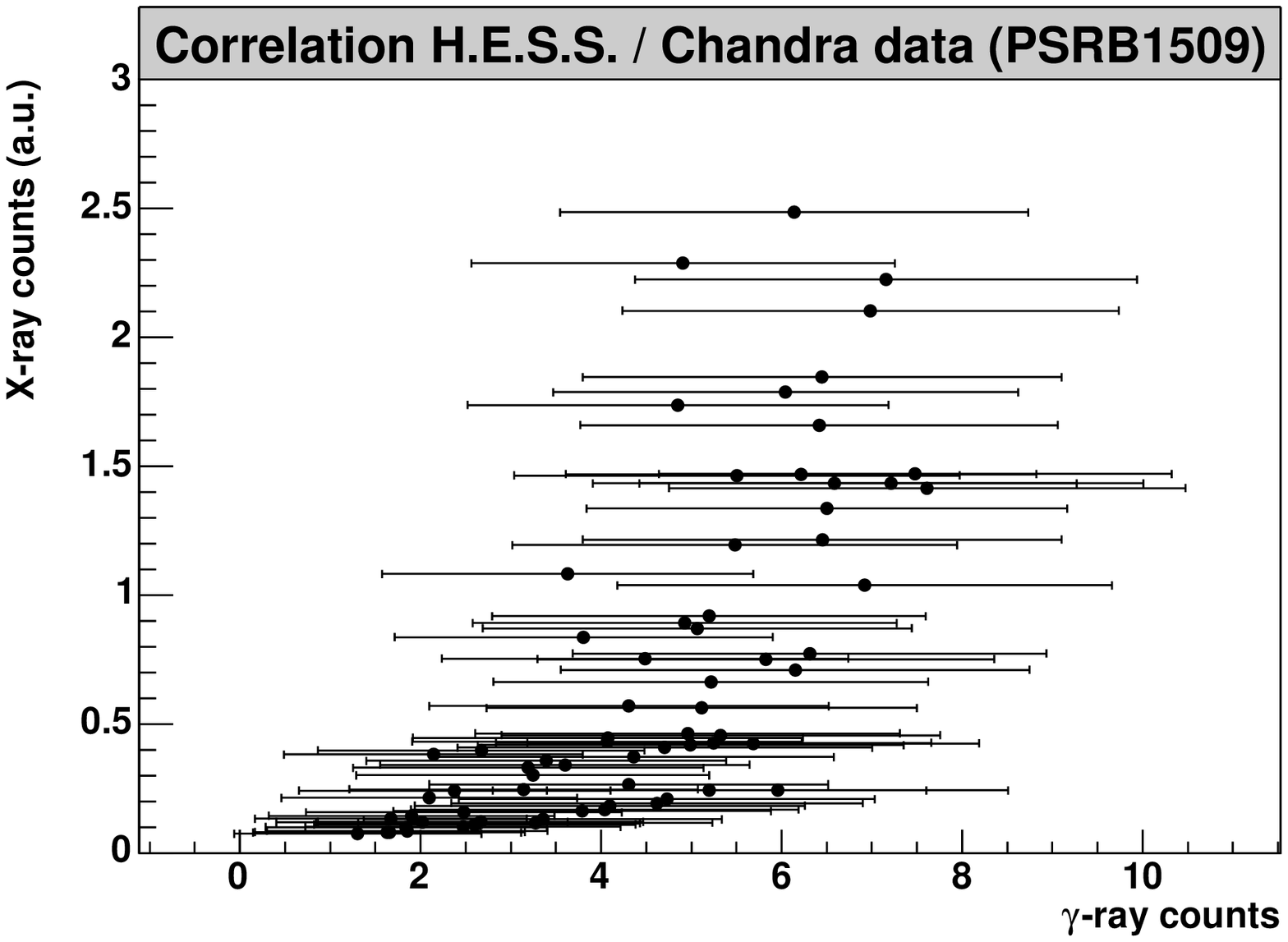}
    \includegraphics[width=0.5\textwidth]{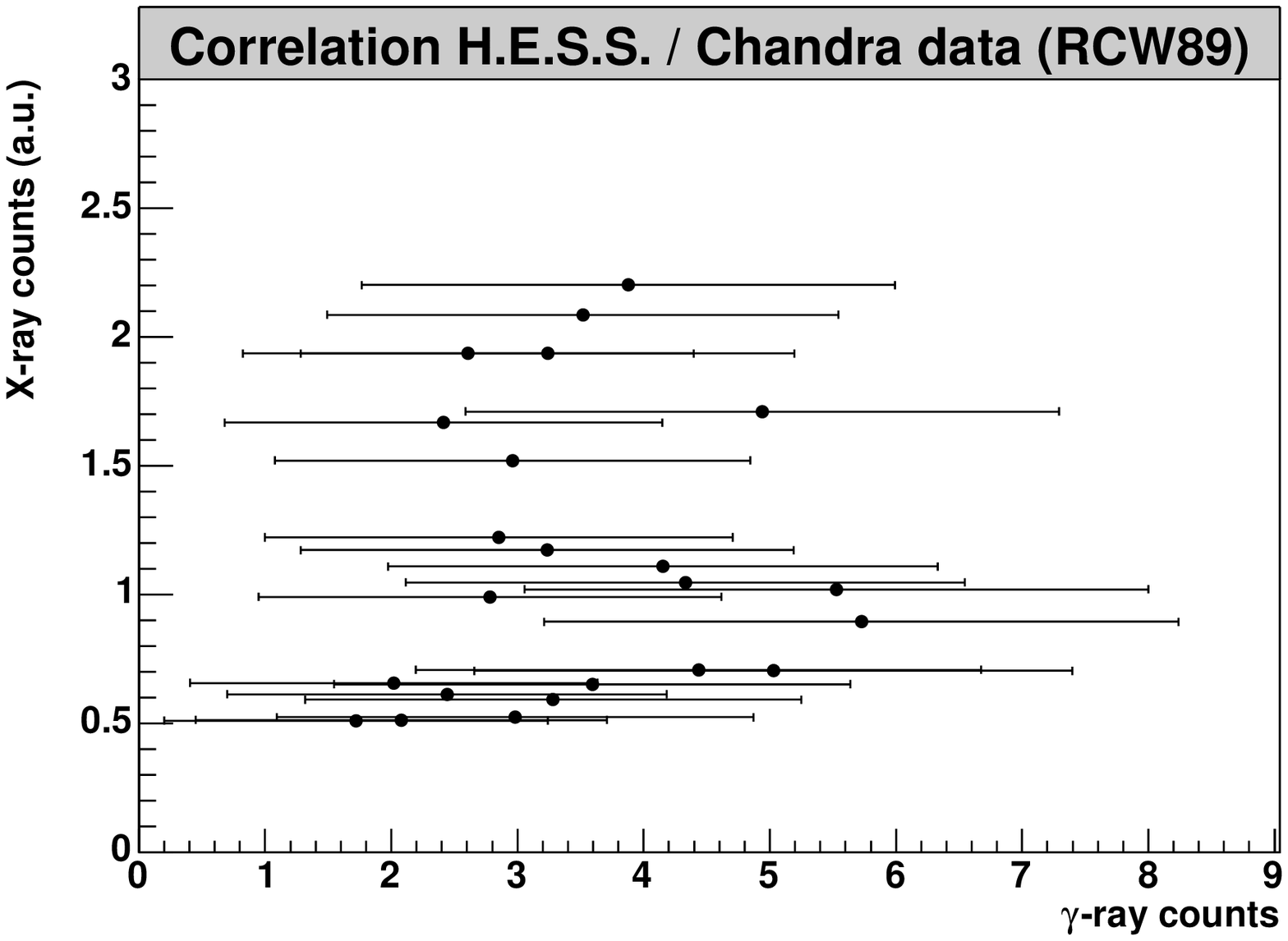}
    \caption[Scatter Plots of the H.E.S.S. \g-Ray and Chandra X-Ray
      Data]{Scatter plots for determining the correlation coefficient between
      the H.E.S.S. \g-ray and the Chandra \X-ray data in the regions of \PSR\ (
      $R_{\rm pulsar}$) and \RCW\ ($R_{\RCW}$) which are indicated in
      Fig.~\ref{fig:ChandraCorrelation}. The peak value is outside the visible
      range of the \X-ray axis in each of the plots.}
    \label{fig:ChandraScatter}
  \end{minipage}
\end{figure}

The errors of the correlation coefficient ($\sigma_{\varrho}^-$,
$\sigma_{\varrho}^+$) due to errors in $\gamma_i$ $(\sigma_{\gamma,i})$ and
$X_i$ $(\sigma_{X,i})$ are asymmetric, as implied by the asymmetric probability
density distribution of $\varrho$ on the finite interval. Therefore they cannot be immediately calculated by Gaussian error propagation. However, Fisher's
Z-Transform (\citet{FishersZ}) provides the appropriate transformation of
$\varrho$ to the Gaussian probability density distribution $\zeta(\varrho)$ on
the infinite interval ($-\infty<\zeta(\varrho)<+\infty$), where Gaussian error
propagation can be applied and Gaussian confidence intervals are valid.
Fisher's Z-Transform ($\zeta$) is defined as
\begin{equation}
  \zeta(\varrho) = \rm atanh(\varrho)
\end{equation}
and has the following properties
\begin{equation}
  \zeta(\varrho-\sigma_{\varrho}^-) = \zeta(\varrho)-\sigma_{\zeta}(\varrho),
\end{equation}
\begin{equation}
    \zeta(\varrho+\sigma_{\varrho}^+) = \zeta(\varrho)+\sigma_{\zeta}(\varrho),
\end{equation}
where $\sigma_{\zeta}$ is the symmetric error of $\zeta$. Hence with the
inverse transformation ($\zeta^{-1}=\tanh$) the asymmetric errors are
\begin{equation}
  \sigma_{\varrho}^-=-\left [\zeta^{-1}\left(\zeta(\varrho)
  -\sigma_{\zeta}(\varrho)\right)-\varrho\right] = -\left[\tanh\left(\rm
  atanh(\varrho)-\sigma_{\zeta}(\varrho)\right)-\varrho\right],
\end{equation}
\begin{equation}
  \sigma_{\varrho}^+=+\left[\zeta^{-1}\left(\zeta(\varrho)+
  \sigma_{\zeta}(\varrho)\right)-\varrho\right] = +\left[\tanh\left(\rm
  atanh(\varrho)+\sigma_{\zeta}(\varrho)\right)-\varrho\right].
\end{equation}
$\sigma_{\zeta}(\varrho)$ can be found by Gaussian error propagation as
\begin{equation}
  \sigma_{\zeta}(\varrho) = \frac{\partial \zeta(\varrho)}
        {\partial \varrho}\sigma_{\varrho} =
        \frac{\partial \rm atanh(\varrho)}{\partial \varrho}\sigma_{\varrho} =
        \frac{\sigma_{\varrho}}{1-\varrho^2}
\end{equation}
and
\begin{equation}
  \sigma_{\varrho}=\sqrt{\sum^N_{t=1} \left|
    \left(\frac{\partial\varrho}{\partial\gamma_i}\sigma_{\gamma,i}\right)^2+
    \left(\frac{\partial\varrho}{\partial X_i}\sigma_{X,i}\right)^2 \right|}.
  \label{eqn:CorrelationError}
\end{equation}
The calculation of the partial derivatives $\partial\varrho/\partial\gamma_i$
and $\partial\varrho/\partial X_i$ is described in
\citet{Schwanke:CorrelationCoefficient}. For the H.E.S.S. data
$\sigma_{\gamma,i} = \sqrt{N_{\rm ON,i}+\alpha^2 N_{\rm OFF,i}}$. For the
Chandra data $\sigma_{X,i}=\sqrt{\tilde{X}_i}$, where $\tilde{X}$ is the number
of \X-ray counts before the exposure correction.
Fig.~\ref{fig:ChandraCorrelation} shows the \g-ray excess and \X-ray count map
with equal binning as well as with the two regions where the correlation
coefficient has been determined. The bin size of the maps has been increased to
$0.02^\circ\times0.02^\circ$. The \g-ray count map has been convolved with a
Gaussian of 0.02$^\circ$ before the equally convolved background map was
subtracted. The regions $R_{\rm pulsar}$ and $R_{\rm \RCW}$ consist of
$9\times8$ and $6\times4$ bins in RA and Dec respectively. The coordinates
$\alpha_{min}$ and $\delta_{min}$, which define the lower left corner of each
region, are given in Tbl.~\ref{tbl:CorrealtionCoefficients}. The corresponding
scatter plots of the \g- and \X-ray maps are shown in
Fig.~\ref{fig:ChandraScatter}. The counts in each bin are sufficiently high
enough to justify the assumption of Gaussian errors.

At the regions of $R_{\rm pulsar}$ and $R_{\rm \RCW}$ correlation coefficients
of $0.58_{-0.17}^{+0.13}$ and $-0.06_{-0.33}^{+0.34}$ are obtained
respectively. They support the speculations about a difference of the X-ray
production mechanisms in the region of \RCW\ in comparison to the region of
\PSR.

\begin{table}[ht!]
  \centering
  \caption[Correlation of H.E.S.S and Chandra Data of \MSH]{Correlation between
    the H.E.S.S. \g-ray and Chandra \X-ray data in the regions of \PSR\
    $(R_{\rm pulsar})$ and \RCW\ $(R_{\RCW})$. The coordinates mark the lower
    left corners of each of the two square regions shown in
    Fig.~\ref{fig:ChandraCorrelation}.}
  \bigskip
  \begin{tabular}{lcc}
    \hline \hline
                   & $R_{\rm pulsar}$       & $R_{\rm \RCW}$          \\
    \hline
    $\alpha_{min}$ & $\rm 15^h14^m35\fs23$  & $\rm 15^h14^m7\fs14$    \\
    $\delta_{min}$ & $-59.245^\circ$        & $-59.085^\circ$         \\
    No. of bins (RA$\times$Dec)&$9\times8$& $6\times4$                \\
    $\varrho$      & $0.58_{-0.17}^{+0.13}$ & $-0.06_{-0.33}^{+0.34}$ \\
    \hline \hline
  \end{tabular}
  \label{tbl:CorrealtionCoefficients}
\end{table}

\clearpage
\renewcommand{\sectionmark}[1]
             {\markright{\thesection\ \ Search for Pulsed Emission}}

\section{Search for Pulsed Emission from \PSR}
\begin{figure}[tbh]
  \begin{minipage}[t]{0.5\linewidth}
    \includegraphics[width=\textwidth]{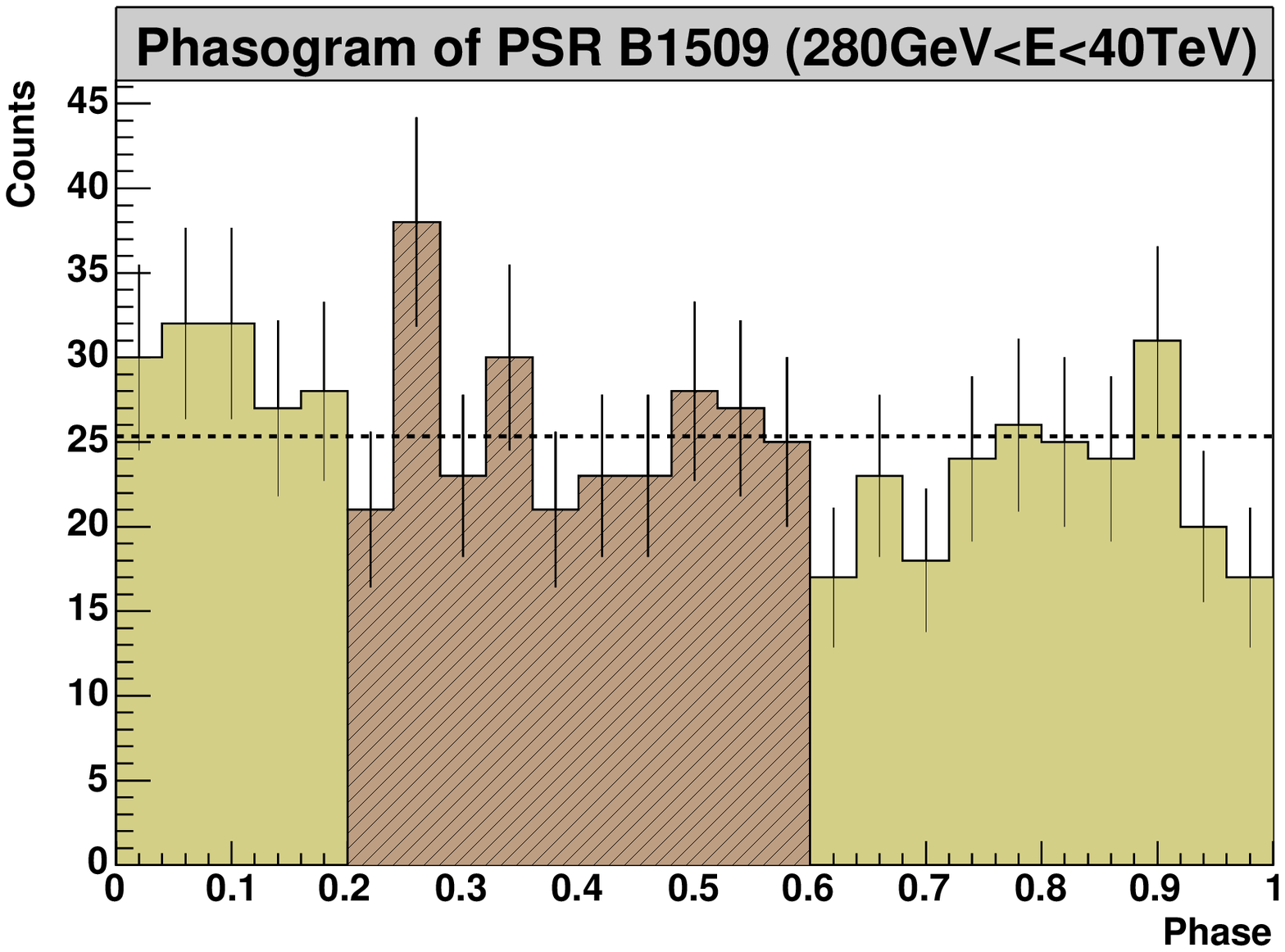}
    \caption[H.E.S.S. Phasogram of \PSR]{Phasogram of the \PSR\ in the energy
    range from 0.28 to 40\,TeV. The hatched phase region [0.2; 0.6] was used
    for the calculation of an upper limit of the pulsed flux. The mean is
    indicated by the horizontal line.}
    \label{fig:Phasogram}
  \end{minipage}\hfill
  \begin{minipage}[t]{0.5\linewidth}
    \includegraphics[width=\textwidth]{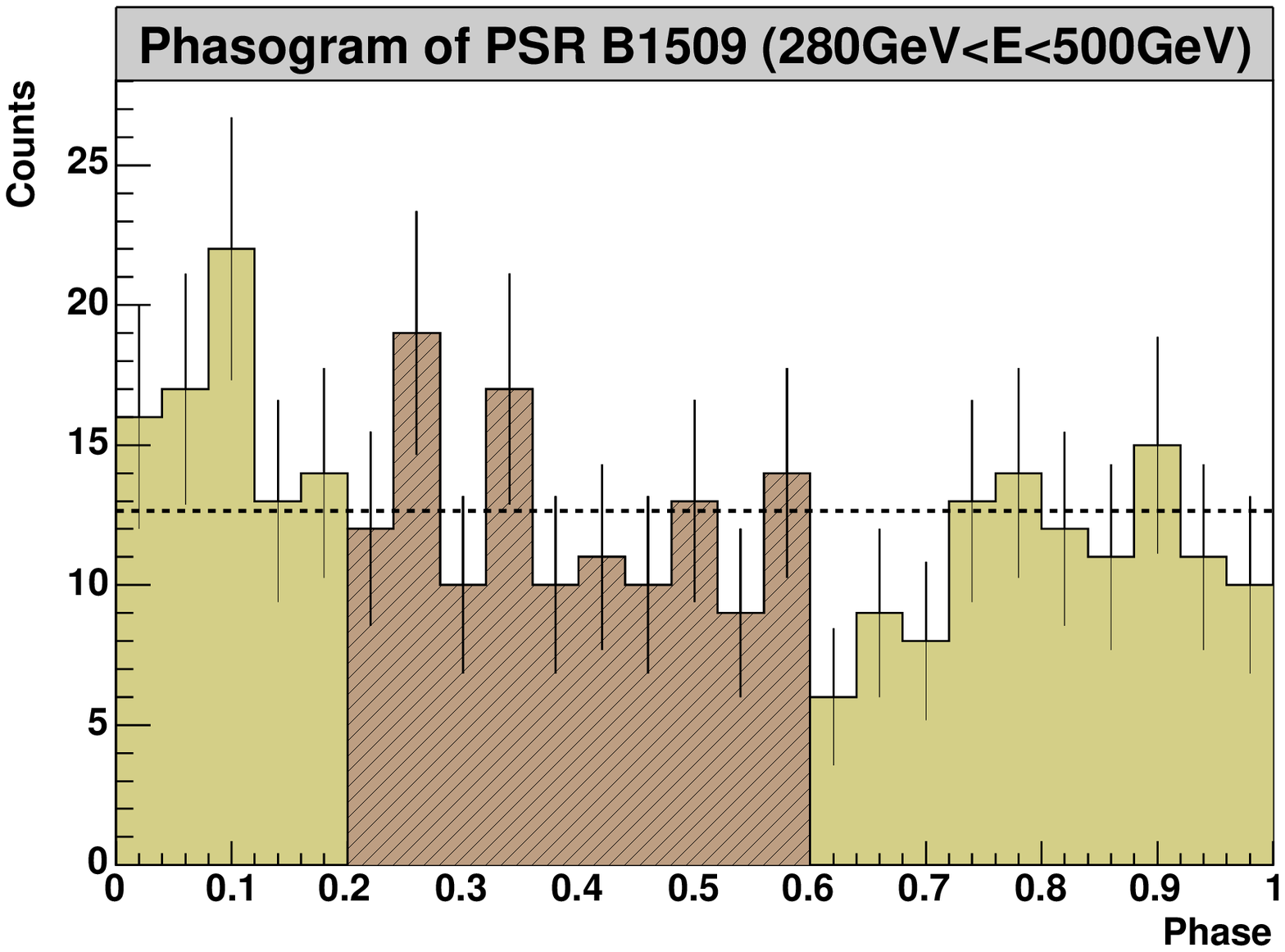}
  \caption[H.E.S.S. Phasogram of \PSR\ (0.28\,TeV$<E<$0.5\,TeV)]{Phasogram as
    described in Fig.~\ref{fig:Phasogram} but for the reduced energy range of
    0.28\,TeV$<E<$0.50\,TeV.}
    \label{fig:Phasogram500}
  \end{minipage}
\end{figure}

Since a strong \g-ray signal from the region of \PSR\ was detected it is
possible and interesting to analyze this signal for pulsed emission with a
period of the pulsar. In contrast to the constant emission seen from the
extended PWN of \MSH, the detection of pulsed emission would indicate emission
processes near the light cylinder of \PSR. Since this has never been observed
for any pulsar at the VHE range accessible to H.E.S.S. it would have been a
very remarkable finding. In order to test for pulsed emission, the data was
analyzed for periodicity according to the radio ephemeris of
Tbl.~\ref{tbl:PSRB1509Ephemeris}, which was taken from the ATNF Data Archive
(\citet{ATNF}). The {\it extended} cut configuration was chosen for this
analysis and applied to a reduced ON-region shown in
Fig.~\ref{fig:SpectrumRegions} (red circle). The small radius
($\theta=0.045^\circ$) reduces the number of \g-rays events which do not
originate near the pulsar. The events in the ON-region have been filled into a
phasogram as described in Chp.~\ref{chp:PeriodicAnalysis}. The resulting
phasogram in Fig.~\ref{fig:Phasogram} shows a uniform distribution. It covers
the full H.E.S.S. energy range from 280\,GeV to 40\,TeV.

Since most pulsar models predict a decrease of the pulsed VHE flux with
increasing energy, the detection of pulsed emission is more likely to occur at
the lower end of the energy spectrum. Consequently, removing events of higher
energy reduces events which are less likely pulsed and therefore increases the
ratio of pulsed \g-ray flux --- at a cost of event statistics. So the analysis
was repeated for a reduced energy range of 280 to 500\,GeV.
Fig.~\ref{fig:Phasogram500} shows the corresponding phasogram. Also this light
curve is consistent with a uniform distribution.

\begin{table}[htb!]
  \centering
  \caption[Ephemeris of \PSR]{Parameters of the ephemeris of \PSR\ from the
    ATNF archive (\citet{ATNF}) which have been used in the periodic analysis.}
  \label{tbl:PSRB1509Ephemeris}
  \bigskip 
  \begin{tabular}{lc}
    \hline \hline
    Parameter              & Value \\
    \hline
    RA (J2000)             & $\rm 15^h13^m55\fs620$ \\
    Dec (J2000)            & $-59^\circ08'9\farcs00$ \\
    validity range [MJD]   & [53035; 53554] \\
    $t_0^{GEO}$ [MJD]      & 53295.000000712 \\
    $f_0$ [s$^{-1}$]       & 6.6088537688620 \\
    $\dot{f}_0$ [s$^{-2}$] & $\rm -6.68672 \times 10^{-11}$ \\
    \hline \hline
  \end{tabular}
\end{table}

\subsection{Tests for Periodicity}
The uniformity of the distribution and the non-detection of pulsed emission was
confirmed using the Pearson $\chi^2$ test, the $Z$ test and the $H$ test as
described in Chp. \ref{chp:PeriodicAnalysis}. The results for the reduced and
full energy ranges are summarized in Tbl.~\ref{tbl:PeriodicTests}. Within
statistical errors, no indication for a pulsed \g-ray emission is found.

\begin{table}[htb!]
  \centering
  \caption[Results of Tests for Pulsed Emission from \PSR]{Results for the
    $\chi^2$, $Z^2_2$ and $H$ test (cf. Chp.~\ref{chp:PeriodicAnalysis}) as
    applied to the phasograms of Fig.~\ref{fig:Phasogram} and
    \ref{fig:Phasogram500}. The corresponding probabilities $(P)$ are also
    listed. They are in agreement with a uniform distribution, i.e. no pulsed
    signal is identified.}
  \bigskip
  \begin{tabular}{lcccccc}
    \hline \hline
    $E$ [TeV] & $\chi^2$ &  $P(\chi^2)$ & $Z^2_2$ & $P(Z^2_2)$ & $H$ & $P(H)$\\
    \hline
    $0.28<E<40$  & 24.7 / 24 & 0.42  & 4.18 & 0.38  & 3.94 & 0.21 \\
    $0.28<E<0.5$ & 24.4 / 24 & 0.44  & 7.68 & 0.10  & 7.51 & 0.05 \\
    \hline \hline
  \end{tabular}
  \label{tbl:PeriodicTests}
\end{table}

\newpage
\subsection{Flux Upper Limit}
An upper limit for the pulsed integrated flux was calculated in the phase
region selected by \citet{Kuiper:1999}, who reported the detection of pulsed
emission from \PSR\ at the energy range of 0.75-10\,MeV in the phase region
[0.2; 0.6] (cf. Sec.~\ref{sec:PSR1509}, Fig.~\ref{fig:Kuiper1999}). The
corresponding On- / Off-statistics are given in Tbl.~\ref{tbl:UpperLimit}. The
upper limit was determined following the unified approach by
\citet{FeldmanCousins} as described in Chp.~\ref{chp:PeriodicAnalysis}. The
upper limit for the integrated flux above the H.E.S.S. energy threshold of
280\,GeV and above 1\,TeV, is shown in the same table for a confidence level
(CL) of 99\%. The upper limits are given with and without the systematic flux
error of 22\%.

\begin{table}[ht!]
  \centering
  \caption[Event Statistics and Flux Upper Limits for \PSR]{Event statistics
    and the upper limit in the phase region [0.2; 0.6] for the phasogram of
    Fig.~\ref{fig:Phasogram}. The upper limits are calculated with
    $(UL_\Phi^{\rm syst})$ and without $(UL_\Phi)$ the systematic error for a
    confidence level of 99\%.}
  \bigskip
  \begin{tabular}{lcccccc}
    \hline \hline
    $E_0$ [TeV] & On  & Off & $\alpha$ & Excess & $UL_\Phi(E>E_0)$ &
    $UL_\Phi^{\rm syst}(E>E_0)$ \\
    & & & & & $[\rm10^{-12}cm^{-2}s^{-1}]$ & $[\rm10^{-12} cm^{-2}s^{-1}]$ \\
    \hline
    0.28     & 259 & 374 & 2/3 & 9.7  &  $<8.3$ (99\% CL) & $<11.0$ (99\% CL)\\
    1.00     &  69 & 106 & 2/3 & -1.6 & $<0.82$ (99\% CL) & $<0.97$ (99\% CL)\\
    \hline \hline
  \end{tabular}
  \label{tbl:UpperLimit}
\end{table}

\renewcommand{\sectionmark}[1]{\markright{\thesection\ \ #1}}

\chapter{Interpretation} \label{chp:Interpretation}

Analysis of the H.E.S.S. TeV \g-ray data of \MSH\ has provided new information
about this exotic object. With this information, further conclusions can be
drawn when the results are interpreted in the context of astrophysical models
and together with data from observations at other wavelengths. Three such
examples are discussed here. They include conclusions about the \g-ray
production in \MSH, about the dominating transport mechanism in the pulsar wind
and the interaction of the pulsar wind with the optical nebula \RCW.

\section{\g-Ray Production in \MSH}
\begin{figure}[ht!]
  \centering \includegraphics[width=.75\textwidth]{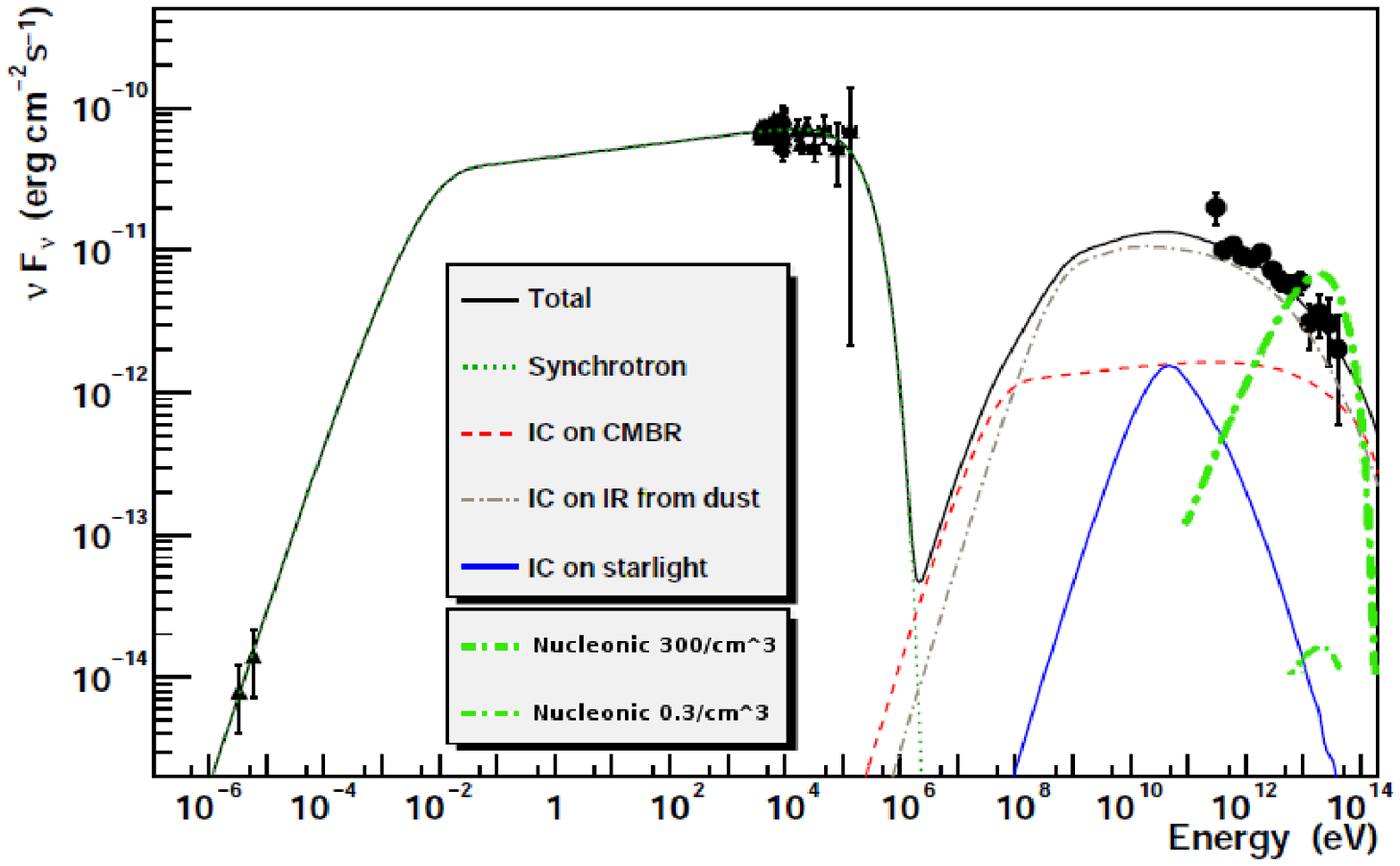}
  \caption[Spectral Energy Distribution of Radiation from \MSH]{Spectral energy
    distribution of radiation from \MSH. The radio, \X-ray and TeV \g-ray
    measurements by ATCA (\citet{Gaensler:2002}), BeppoSAX (\citet{Mineo:2001})
    and H.E.S.S. (\citet{Khelifi:2005ICRC}) are shown along with the simulated
    curves for leptonic and hadronic radiation. The inverse Compton curves are
    obtained for different photon fields. The green thick and thin curves
    represent the \g\ radiation from $\pi^0$ decay of nucleonic interactions in
    high (300\,cm$^{-3}$) and low (0.3\,cm$^{-3}$) density mediums. (Figure
    taken from \citet{Khelifi:2005ICRC}. Green curves added from
    \citet{Bednarek:2003}.)}
  \label{fig:Khelifi2005}
\end{figure}

An important discovery from the H.E.S.S. data is that it agrees well with the
leptonic model of \g-ray production in PWN. For example, \citet{Khelifi:2005}
showed with simulations of GALPROP (\citet{Strong:2000}) that the TeV \g-ray
spectrum can be explained by inverse Compton (IC) scattering of cosmic
microwave background (CMB) radiation and infrared (IR) photons from dust and
starlight by highly accelerated electrons. In these simulations, the parent
energy spectrum of the electrons obeys a power law and is adjusted to reproduce
the IC \g\ radiation measured by H.E.S.S as well as the synchrotron emission
determined by \X-ray and radio measurements by BeppoSAX (\citet{Mineo:2001})
and ATCA (\citet{Gaensler:2002}). The energy density of the IR photons and the
magnetic field inside the nebula were kept free to match the experimental data.
The simulations determine the magnetic field to $\sim$17\,$\mu$G, the IR energy
density to 2.5\,eV\,cm$^{-3}$ for the IR seed photons with wavelengths of
$\sim$100\,$\mu$m and the spectral index of the electrons to $\Gamma_e\sim2.9$.
Fig.~\ref{fig:Khelifi2005} shows the corresponding spectral energy distribution
of the synchrotron and IC radiation, together with the experimental data. The
IC radiation from the individual components and the sum of the seed photon
field of these components is shown. Apparently this model provides a good
description of the data.

An alternative interpretation with a model of TeV \g\ radiation from hadronic
primary particles was introduced in Sec.~\ref{sec:PWNSpectra}.
\citet{Bednarek:2003} have calculated the expected \g-ray spectrum from \MSH\
for nucleonic collisions with a high and low density medium surrounding the
PWN. The green thick and thin dot-dashed curve in Fig.~\ref{fig:Khelifi2005}
represents the spectral energy distribution in high and low density mediums
with average densities of 300\,cm$^{-3}$ and 0.3\,cm$^{-3}$, respectively. Both
curves are taken from Fig.~\ref{fig:Bednarek2003}. It can be seen that in a
high density medium the predicted TeV \g\ radiation exceeds the observed flux.
Therefore the H.E.S.S. data constrains the density of the ambient medium below
300\,cm$^{-3}$. On the other hand, for the low density medium the expected flux
is too low to be visible in the measured TeV spectrum. H\,{\sc i} measurements
by \citet{Dubner:2002} have suggested a density of $\sim$0.4\,cm$^{-3}$ for the
interstellar medium in the southeast and a density of $\sim$12\,cm$^{-3}$ for
the northern region, in particular a density of $\sim$15\,cm$^{-3}$ for the
region of \RCW. Therefore, in the denser regions, in particular the region of
\RCW, one would expect an increased \g-ray flux from hadronic interactions as
pointed out by \citet{Bednarek:2003}. The analysis of the H.E.S.S. data
reveals, however, a decreased \g-ray flux in these regions (cf.
Sec~\ref{sec:Morphology}). This could imply that the nucleon density in the
pulsar wind is too low to contribute significantly to the \g\ emission.

It can be concluded, therefore, that the leptonic radiation model can well
explain the observed TeV spectrum from \MSH. The existence of a significant
electron component is evident by the synchrotron flux observed in the \X-ray
band. However, evidence for a hadronic \g-ray production has not been found and
therefore does not play a major role in the TeV \g-ray production in \MSH.

\section{Transport Mechanism in \MSH} \label{sec:Transport}
A further, interesting conclusion can be drawn concerning the transport
mechanism in \MSH\ if the model by \citet{Kennel:1984} is considered. This
model assumes that particles are highly accelerated near a pulsar and that they
drive the expansion of a PWN into the ambient interstellar medium
(Sec.~\ref{sec:KennelCoroniti}). Moreover, it assumes that pulsar wind consists
of a plasma which is frozen into the magnetic field lines of a pulsar. The
field lines are wrapped in tight spirals around the pulsar, due to the pulsar's
spin. Using this picture of pulsar wind, one can explain the expansion of this
magnetized flow either by diffusion or by convection. Diffusion is the random
propagation of scattering particles with a net transport determined by the
diffusion gradient. Convection, on the other hand, is a collective flow of
particles in a common direction. According to \citet{Okkie:PWNTransport},
\citet{Okkie:2006} and \citet{Okkie:2006IAU}, one can compare the efficiency of
these two transport mechanisms in a PWN and determine the dominant transport
mechanism of the radial flow.

In the model of \citet{Kennel:1984}, the pulsar wind is shocked at the shock
radius $r_s$ inside the PWN of radius $R_N$. This situation is illustrated in
Fig.~\ref{fig:Kennel1984}. The flow velocities $v(R_N)$ and $v(r_s)$ of the
wind have to meet the boundary conditions at the radii $r_s$ and $R_N$, i.e.
$v(R_N)$ is equal to the outer expansion velocity of the nebula and
$v(r_s)=c/\beta_s=c/\sqrt{3}$, where $\beta_s$ is the expected Alphen velocity
of the relativistic plasma (\citet{Kennel:1984}).

Therefore de Jager argued, that a radial flow velocity between $r_s$ and $R_N$
given by
\begin{equation}
  v(r)=\frac{c}{\beta_s}\frac{r_s}{r}
  \label{eqn:Velocity}
\end{equation}
will meet both boundary conditions, and that the $1/r$ dependence is a
reasonable assumption in the case of constant B-field strength (which is
generally fulfilled in good approximation) and due to conservation of magnetic
flux.

Following the argumentation, one can estimate the diffusive and convective flow
in a PWN. If the diffusion time $T_d$ is smaller than the convection time
$T_c$, most energetic particles overtake the bulk convective flow due to
scattering and the transport mechanism is dominated by diffusion (and vice
versa). Hence the criterion for a flow being dominated by diffusion can be
written as
\begin{equation}
  T_c>T_d. \label{eqn:tctd}
\end{equation}
The time for diffusion of an electron to a radius r is given by
\begin{equation}
  T_d(r)=\frac{r^2}{2\kappa_\bot}, \label{eqn:td}
\end{equation}
where $\kappa_\bot$ is the diffusion coefficient for the electron transport
perpendicular to the magnetic field, i.e. in the radial direction. The
diffusion coefficient is given by
\begin{equation}
  \kappa_\bot=\frac13\lambda_\bot v,
\end{equation}
where $\lambda_\bot$ is the mean free path between scatterings in the radial
direction and $v$ is the velocity of the electron. Moreover, in the case of
strong and weak scattering the fundamental limit by \citet{Steenberg:1998}
states that
\begin{equation}
  \lambda_\bot\le\rho_L, \label{eqn:SteenbergLimit}
\end{equation}
where $\rho_L$ is the particle's gyroradius.

The time for convection from $r_s$ to the radius r can be found by integration
of Eqn.~\ref{eqn:Velocity} and is given by
\begin{equation}
  T_c(r) = \int_{r_s}^{r} \frac{dr}{v(r)} =
  \frac{1}{2 \beta_s c} \frac{r^2-r_s^2}{r_s} \sim 
  \frac{1}{2 \beta_s c} \frac{r^2}{r_s}, \quad {\rm if} \quad r_s \ll r.
  \label{eqn:tc}
\end{equation}
Again, $\beta_s$ is the expected Alphen velocity of a relativistic plasma.

Using this equations in Eqn.~\ref{eqn:tctd}, the criterion for a flow to be
dominated by convection is
\begin{equation}
  \lambda_\bot >3\beta_s r_s
\end{equation}
and applying the Steenberg limit of Eqn.~\ref{eqn:SteenbergLimit} yields
\begin{equation}
  \rho_L > 3\beta_s r_s.
  \label{eqn:unequality}
\end{equation}
Sine $r_s$ is determined through Eqn.~\ref{eqn:tc} one obtains by integrating
to the radius of the PWN $(r=R_N)$
\begin{equation}
  r_s=\frac{1}{2\beta_s c}\frac{R_N^2}{T_c(R_N)}.
\end{equation}
Here $T_c(R_N)=1700$\,y is given by the age of the PWN and the radius of the
PWN can be estimated from its angular size of $\theta_N=0.1^\circ$ and its
distance $d=5.2$\,kpc as
\begin{equation}
  R_N=\frac{2\pi \theta_N}{360^\circ}\times d = \rm \frac{2\pi\times
  0.1^\circ}{360^\circ}\times 5.2\,kpc= 9\,pc.
\end{equation}
With these numbers for $R_N$ and $T_c(R_N)$ one obtains
\begin{equation}
  r_s=\rm \frac{1}{2\beta_sc} \frac{(9\,pc)^2}{1700\,y}= \frac{(9\,pc)^2}
  {2\sqrt3 (1700\,ly/3.26\,ly)\,pc}= 0.04\,pc \widehat= 11''. 
\end{equation}
This is a remarkable result, since it predicts an angular size of
$\theta_s=11''$ for the pulsar wind shock radius which coincides with feature 5
of the Chandra \X-ray image in Fig.~\ref{fig:Gaensler2002f5}. Therefore this
theoretical prediction supports an interpretation of the observed ring-like
feature in a distance of $17''$ from the pulsar as shock front --- similar to
the one observed for the Crab Nebula in Fig.~\ref{fig:Weisskopf2000}.

Now the right hand side of Eqn.~\ref{eqn:unequality} provides
\begin{equation}
  3\beta_s r_s=\rm 0.2\,pc= 7 \times 10^{17}\,cm.
\end{equation}
On the other hand, $\rho_L=E_e/eBc$ is determined by the energy of the
electrons $(E_e)$ in the pulsar wind where $E_e$ can be estimated from the IC
\g-ray spectrum measured by H.E.S.S. As discussed in the previous section, most
of the \g\ radiation can be explained by IC scattering of infrared photons with
$E_{ph}=1.24\times10^{-2}$\,eV $(\sim100\,\mu{\rm m})$ and CMB photons with
$E_{ph}=2.35\times10^{-4}$\,eV, as shown in Fig.~\ref{fig:Khelifi2005}. For
these photons the energy of the electrons can be estimated in the Thomson
limit. According to Eqn.~\ref{eqn:ThomsonE}, one finds
\begin{equation}
  E_e= m_ec^2 (E_\gamma/E_{ph})^{1/2}
\end{equation}
and obtains $E_e \sim 2\times(E_{\gamma}/{\rm TeV})^{1/2}$\,TeV for infrared
photons and $E_e \sim 20\times(E_{\gamma}/{\rm TeV})^{1/2}$\,TeV for the CMB
photons (cf. Eqn.~\ref{eqn:OkkieIC}). Therefore IC \g\ radiation in the energy
range from 0.2 and 20\,TeV is mainly produced by electrons of energies between
1 to 10\,TeV and 10 to 100\,TeV (cf. \citet{HESSJ1825II}). So in both cases,
$E_e<100$\,TeV and the gyroradius of the electrons can be estimated as
\begin{equation}
  \rho_L= \frac{E_e}{eBc} < \frac{100\,{\rm TeV}}{e 3\times10^{8}\frac{\rm
  m}{\rm s} \times 17\times10^{-10}{\rm T}} \left(\frac{E_{\gamma}}{\rm
  TeV}\right)^{1/2} \left(\frac{17\,\mu G}{B}\right) = 2\times10^{16}\rm cm.
\end{equation}
Since the gyroradius of the electrons in the pulsar wind is more than an order
of magnitude smaller than required by Eqn.~\ref{eqn:unequality} and
Eqn.~\ref{eqn:tctd} for a diffusion dominated transport, the transport cannot
be dominated by diffusion.

This picture of \MSH\ is consistent with the general model by \cite{Okkie:2006}
according to which PWNs are described as ``filled bags'' which confine their
VHE electron wind by the magnetic field which is wrapped in spirals around the
pulsar. Due to the perpendicular orientation of the magnetic field lines to the
radial direction, diffusion of the VHE electrons wind is suppressed and a PWN
can only release it slowly via convective flows.

\section{Interaction of the Pulsar Wind Nebula with \RCW}
The sky maps of \MSH\ in Sec.~\ref{sec:Morphology} have provided first insights
into the TeV \g-ray morphology of \MSH\ which also allow for some further
conclusions.

First, the compact structure extending southeast of the pulsar coincides in
size and orientation with the jet of the pulsar, which has been observed in
\X-rays. It can therefore can be identified as such, which is the first time
that a jet has been resolved at TeV energies.

\begin{figure}[h!]
    \begin{minipage}[t]{0.368\textwidth}
      \setlength{\abovecaptionskip}{.01cm}
      \vfill\includegraphics[width=\textwidth]{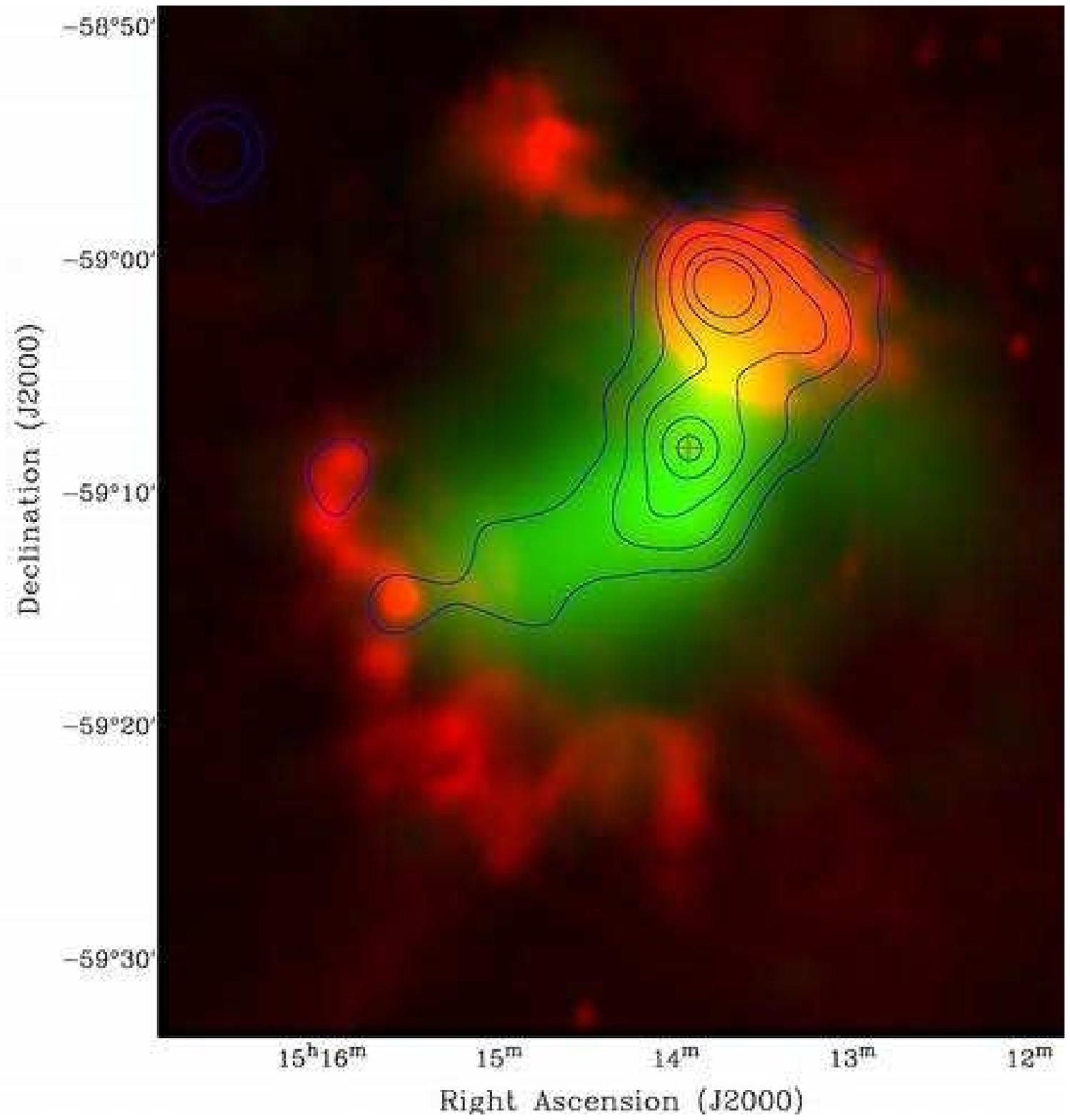}
      \caption[Multiwavelength Data of \MSH]{843\,MHz Molonglo data (red),
        ROSAT \X-ray contours and smoothed H.E.S.S. data (green). (Figure taken
        from \citet{Gaensler:2006}.)}
      \label{fig:Gaensler2006_2}
    \end{minipage}
    \begin{minipage}[t]{0.296\textwidth}
      \setlength{\abovecaptionskip}{.4cm}
      \captionmargin0cm
      \vfill\includegraphics[width=\textwidth]{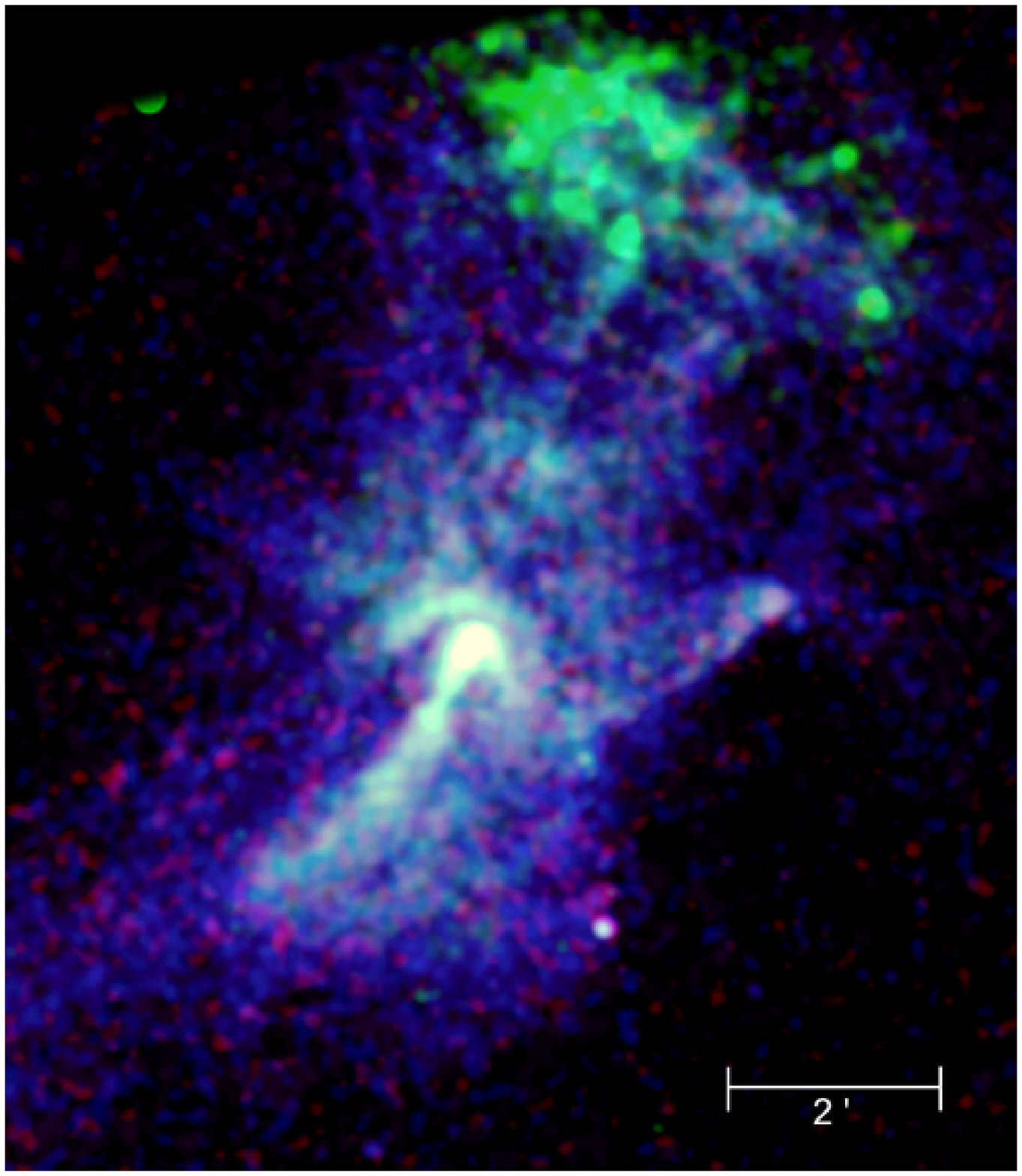}
      \caption[Chandra \X-Ray Data for Different Energy Bands of \MSH]{Chandra
        \X-ray data at different energy bands. (Figure taken from
        \citet{B1509_scale}.)}
      \label{fig:B1509_scale}
  \end{minipage}
    \begin{minipage}[t]{0.331\textwidth}
      \setlength{\abovecaptionskip}{.4cm}
      \vfill\includegraphics[width=\textwidth]{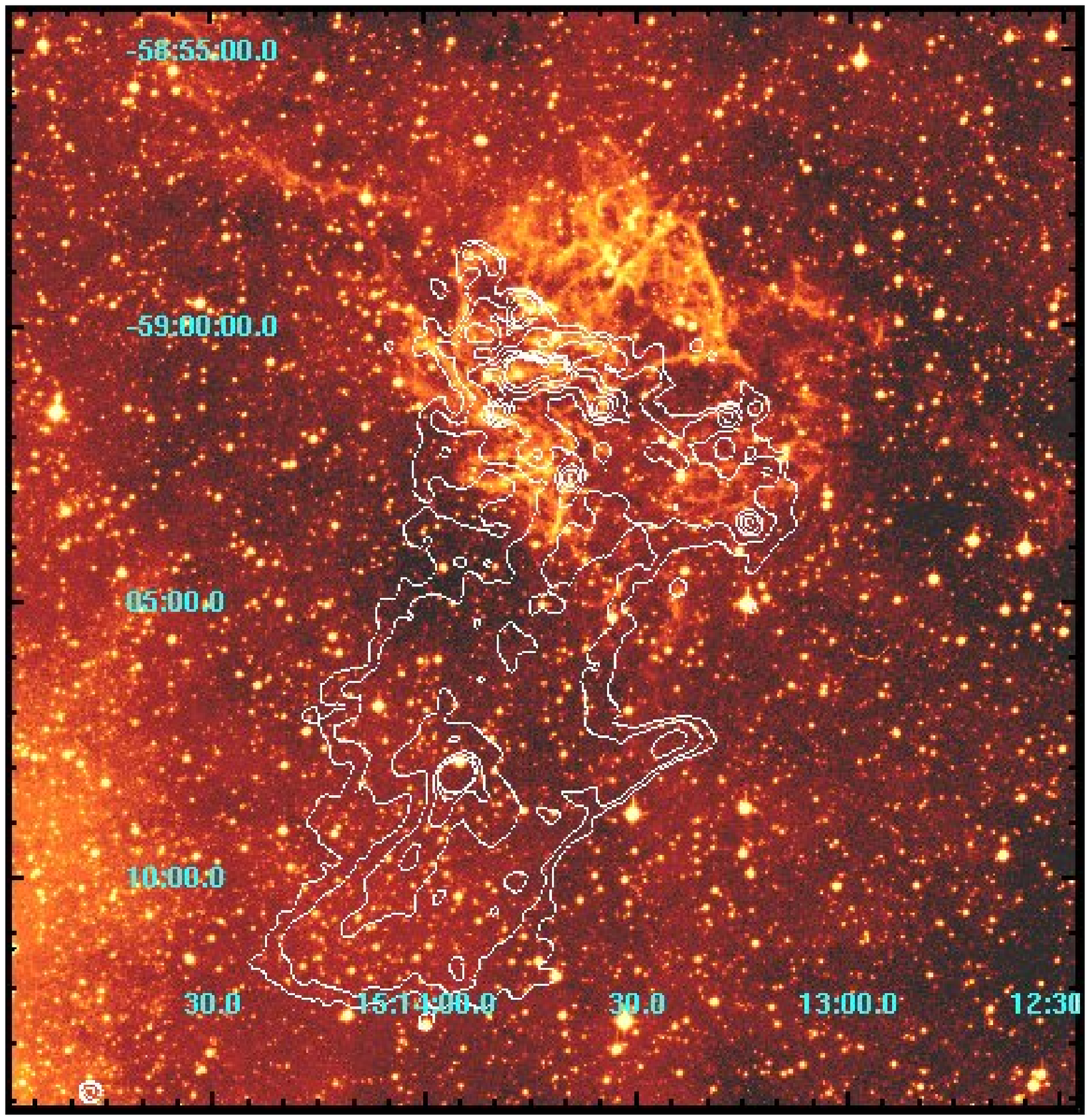}
      \caption[Data of \MSH\ from the COSMOS H-alpha Survey]{Image from the
        COSMOS H-alpha Survey overlaid with Chandra \X-ray contours. (Figure
        taken from \citet{Yatsu:2006}.)}
      \label{fig:Yatsu2006}
    \end{minipage}\hfill
\end{figure}

Second, the region of \RCW\ shows a less intense VHE \g-ray emission than the
region of \PSR, while both regions show a similar intensity in \X-rays. A
possible explanation for this could be that the TeV \g\ radiation is being
absorbed in the dense nebula of \RCW\ or that the PWN has not expanded far into
this region. In the case that the latter is true, one could speculate that
\RCW\ might constitute a kind of barrier for the PWN that also influences its
future evolution. The difference between the two regions is illustrated by the
multiwavelength data of Fig.~\ref{fig:Gaensler2006_2} to \ref{fig:Yatsu2006}.
Fig.~\ref{fig:Gaensler2006_2} shows a superposition of the 843\,MHz Molonglo
data (red) of the surrounding supernova remnant (\citet{Whiteoak:1996}), the
ROSAT \X-ray contours (\citet{Trussoni:1996}) and the smoothed H.E.S.S. data
(green, \citet{Khelifi:2005}). Fig.~\ref{fig:B1509_scale} shows the Chandra
\X-ray data (\citet{Gaensler:2002}) separated into four different energy bands:
6-8 (red), 4-6 (blue), 2-4 (purple) and 0.3-2\,keV (green).
Fig.~\ref{fig:Yatsu2006} shows the data from the COSMOS H-alpha Survey
(\citet{COSMOS}) overlaid with the Chandra \X-ray data.

Third, it is noteworthy that the \g-ray measurements are able to indicate a
difference between the PWN and the optical nebula \RCW\ which is not as obvious
from the \X-ray measurements alone. On the other hand, the resolution of the
\g-ray data is not alone sufficient enough to identify the pulsar jet without
the high-resolution images of the \X-ray data. In this respect the \g-ray and
\X-ray data are complementary to each other. Here multiwavelength data has
provided new insights, which would have been more difficult to achieve with
data from a singe energy band only.

This progress supports hopes that future experiments will make new valuable
contributions to the understanding of more aspects of \MSH. For example
H.E.S.S. II will be able to resolve the VHE \g-ray morphology at lower energies
and with higher resolution than any other experiment before. Hopefully these
new insights will also increase the knowledge and understanding of pulsars and
PWNs in general.

\chapter{Summary}
This work has reported on the discovery of VHE \g\ radiation from the pulsar
wind nebula \MSH\ with H.E.S.S. \MSH\ is also known as supernova remnant \SNR.
It hosts and is powered by \PSR, one of the most energetic pulsars known, which
is embedded in the supernova remnant. As a prime target, \MSH\ was one of the
first objects to be observed with H.E.S.S., which is currently one of the most
sensitive instruments for VHE \g-ray astronomy.

A review of the available data of \MSH\ and \PSR, since its discovery in 1961
to the most recent observations with \X-ray satellites like Chandra and
INTEGRAL shows, that \MSH\ is a complex and only partially understood system,
which poses many open questions to astrophysicists.

Imaging atmospheric Cherenkov technique provides means for conducting \g-ray
astronomy by allowing for the reconstruction of direction and energy of VHE
\g-ray photons using Cherenkov light which is produced when these photons enter
the atmosphere. The detection of Cherenkov light is achieved with telescopes
with large reflecting dishes. Combining multiple telescopes for stereoscopic
observations provides further advantages --- most importantly increased
accuracy and sensitivity.

H.E.S.S. is a stereoscopic system of imaging atmospheric Cherenkov telescopes
which has been developed to have an unprecedented sensitivity and precision. It
is located in Namibia, where it has an ideal view of the galactic center. It
consists of four telescopes, each with a mirror surface of 107\,m$^2$, a height
of 28\,m and a camera with 960 photo-multiplier tubes. H.E.S.S. is sensitive in
the energy range from 0.2 to 50\,TeV and able to detect a \g-ray point source
with a flux of 1\% of the flux from the Crab Nebula in about 25 hours or a
source of similar strength as the Crab Nebula within 30 seconds. H.E.S.S. has
been in operation since June 2002.

Several standard methods of the imaging atmospheric Cherenkov technique are
used in H.E.S.S. data analysis and have been introduced here. They include the
production of Monte Carlo simulations, the reconstruction of \g-ray photons,
statistical methods for the calculation of significance, position and size of a
source and spectroscopy. The methods have been tested with data from the Crab
Nebula and provide reliable results. Image deconvolution using the
Richardson-Lucy algorithm has been developed with Monte Carlo simulations and
applied here to H.E.S.S. data for the first time. In particular for strong
sources, interesting results with an seemingly higher resolution are obtained.
The reconstruction of light curves from pulsars with ephemeris from the archive
of the Australia Telescope National Facility has been discussed and was
verified with optical data from the Crab Nebula.

The methods of data analysis have been applied to the first H.E.S.S. data of
\MSH. The observations were conducted in 2004 from March 26 to July 20. After
run selections, data with a live-time of 26.14\,h was available. \MSH\ was
detected as \HESS\ with a significance up to 32 standard deviation,
corresponding to 6\,$\sigma$/h and a \g-ray rate of $(2.6\pm0.1)$\,min$^{-1}$.
The center of the emission region was found at (J2000) ($\rm 15h14^m6\fs5\pm
2\fs4_{stat}\pm 2\fs6_{syst}$, $-59^{\circ}10'$ $1\farcs2$ $\pm21''_{\rm stat}$
$\pm20''_{\rm syst}$) at a distance of 2.3' from \PSR. The \g-ray excess can be
fit by a Gaussian distribution with a width and length of $\rm
2.3'\pm0.4'_{stat}$ and $\rm 6.5'\pm0.5'_{stat}$ respectively. The major axis
of the emission region has an angle of $43^\circ\pm4^\circ$ to the RA-axis and
is aligned with the jet axis of \PSR, which has been observed in \X-rays.

The energy spectrum obeys a power law of the form $\frac{d\Phi}{dE}=\phi_{\rm
1TeV}\left(\frac{E}{\rm 1TeV}\right)^{-\Gamma}$, with $\rm \phi_{\rm 1TeV} =
(5.8\pm0.2_{stat}\pm1.3_{syst}) \times 10^{-12}cm^{-2}s^{-1}TeV^{-1}$, $\rm
\Gamma = 2.32 \pm0.04_{stat} \pm0.10_{syst}$ and $\rm \Phi(E>1TeV)$ $\rm = (4.4
\pm0.2_{stat} \pm1.0_{syst}) \times 10^{-12}cm^{-2}s^{-1}$. If the integrated
flux above 1\,TeV is compared with it, \MSH\ emits 20\% of the flux of the Crab
Nebula with a harder photon index. No spatial variation of the photon index nor
temporal variability of the flux on the timescales of days within the
observation period was found.

The Richardson-Lucy algorithm for image deconvolution has been applied to the
H.E.S.S. \g-ray maps and images have been obtained which are complementary to
and in agreement with those obtained by the standard smoothing technique. They
show the VHE \g-ray morphology of \MSH. A study of the energy dependence of the
morphology shows a decreasing \g-ray emission along the pulsar jet axis with
increasing energy. The correlation between the \g- and \X-ray emission was
studied based on the \X-ray data from the ROSAT and the Chandra satellite.
While a correlation between \g- and \X-ray data was found in the region of
\PSR, a correlation was not found in the northwest region of the optical nebula
\RCW. Between the H.E.S.S. and Chandra data, the correlation coefficients of
$0.58_{-0.17}^{+0.13}$ and $-0.06_{-0.33}^{+0.34}$ have been obtained for the
regions of \PSR\ and \RCW, respectively.

The \g-ray excess was tested for pulsed emission from \PSR. Since no indication
for a pulsed emission were found, upper limits for the pulsed \g-ray flux were
determined for a confidence level of 99\%. They are 8.3$\rm \times 10^{-12}
cm^{-2}s^{-1}$ and 0.82$\rm \times 10^{-12} cm^{-2}s^{-1}$ for the energy range
above 280\,GeV and 1\,TeV respectively.

The results obtained from the data analysis allow various conclusions to be
drawn about the PWN. For example, it is possible to reproduce the
multiwavelength spectrum from \MSH\ with a population of VHE electrons, where
the \g-ray energies are produced by IC scattering of photons from the
interstellar radiation field. A hadronic production scenario for the \g\
radiation by nuclei of the PWN cannot be excluded, but no evidence for this
assumption has been found.

Moreover, following the argumentation by \citet{Okkie:PWNTransport} which is
based on the model by \citet{Kennel:1984}, it is possible to show that the
transport in the PWN must be dominated by convection rather than diffusion.

Finally, using the correlation between the \g-ray and \X-ray data, it is
possible for the first time to detect a jet in TeV \g\ radiation. In addition,
the missing correlation in the region of the dense optical nebula \RCW\ allows
for speculation as to whether or not IC radiation is produced in this region
and absorbed by the dense nebula of \RCW, or if instead the PWN has not
expanded far into this region and \RCW\ constitutes a kind of barrier for the
PWN. Further experiments such as H.E.S.S. II are expected to provide new data
and new information about the \g-ray morphology of \MSH\ as well as about \PSR\
and PWNs in general.



\renewcommand{\chaptermark}[1]{
  \markboth{\MakeUppercase{\appendixname}~\thechapter\ \ #1}{}}
\renewcommand{\sectionmark}[1]{\markright{\thesection\ \ #1}}

\appendix


\chapter{Flowchart of the H.E.S.S. Standard Analysis Chain}
\label{app:AnalysisChain}

\begin{figure}[ht]
  \centering
  \includegraphics[height=.57\textheight]{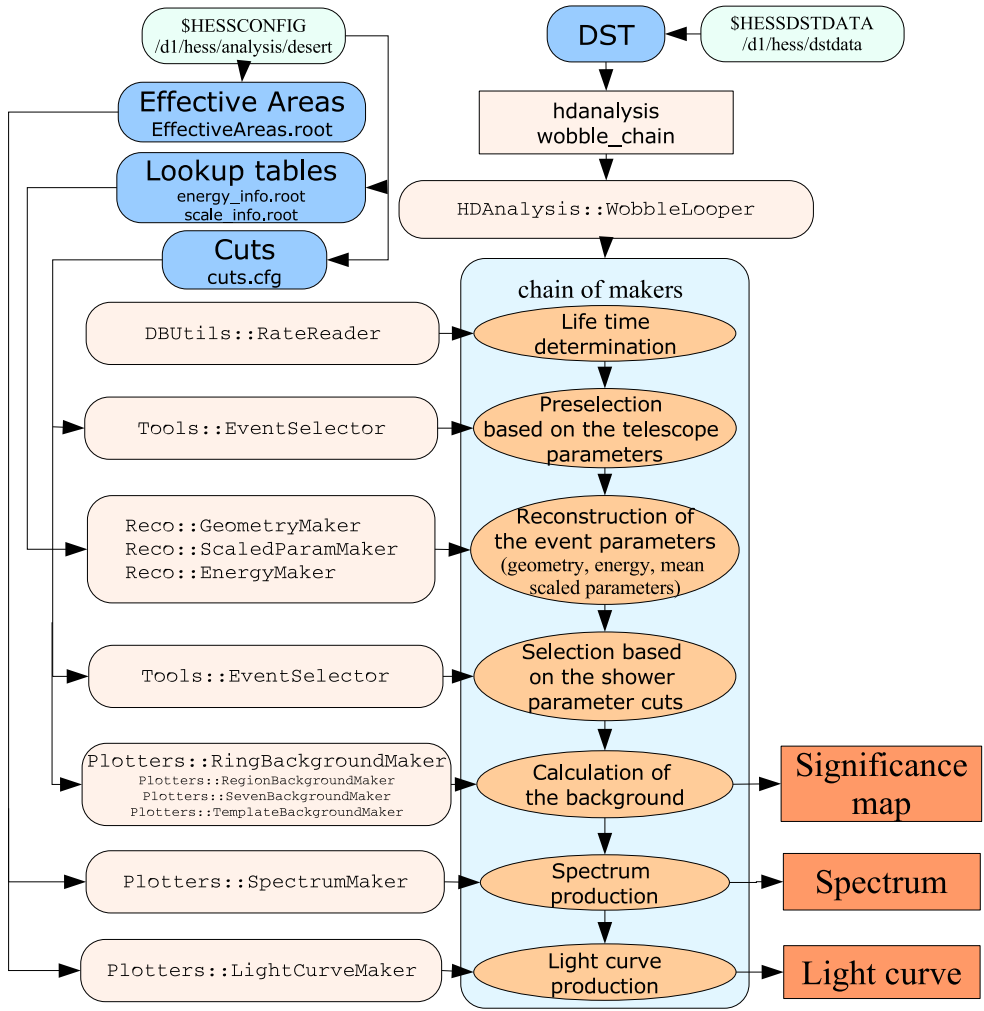}
  \caption[Flowchart of the H.E.S.S. Standard Analysis Chain]{Flowchart of the
    H.E.S.S. standard analysis chain showing the dependencies and the order of
    execution of the individual steps. Further details can be found in
    \cite{hd_manual}. (Flowchart taken from \cite{hd_manual}.)}
  \label{fig:Flowchart}
\end{figure}

\chapter{Hillas Parameters} \label{app:HillasParameters}

The Hillas parameters were first introduced by \cite{Hillas}. They are derived
from the moments of the image amplitude distribution. If $I_i$ is the amplitude
of the pixel $i$ ($i \in \mathbb{N}$) with the coordinates $x_i$ and $y_i$,
then the nth order moment $\langle x^p y^q \rangle$ is defined as
\begin{equation}
  \langle x^p y^q \rangle=\frac{\sum_i I_i x_i^p y_i^q}{\sum_i I_i},
\end{equation}
with $n=p+q$ and $n,p,q \in \mathbb{N}$. Typical Hillas parameters are shown
in Fig.~\ref{fig:HillasParameters}.

\section{0th Order Moment}
The 0th order moment is the image amplitude ($IA$) given as
\begin{equation}
  IA=\sum_i I_i. \label{eqn:size}
\end{equation}
The image amplitude or size represents the total intensity of the image.

\section{1st Order Moments}
The 1st order moments are
\begin{eqnarray}
\langle x \rangle&=&\frac{\sum_i I_i x_i}IA, \\
\langle y \rangle&=&\frac{\sum_i I_i y_i}IA.
\end{eqnarray}
These are the coordinates of the center of gravity. The local distance ($LD$)
is the distance from the center of gravity to the camera's origin. It can be
expressed using the 1st moments as
\begin{equation}
  LD=\sqrt{\langle x \rangle^2 + \langle y \rangle^2}. \label{eqn:dist}
\end{equation}

\section{2nd Order Moments}
The 2nd order moments are
\begin{eqnarray}
  \langle x \rangle^2&=&\frac{\sum_i I_i x_i^2}IA,\\
  \langle y \rangle^2&=&\frac{\sum_i I_i y_i^2}IA, \\
  \langle xy \rangle&=&\frac{\sum_i I_i xy_i}IA,
\end{eqnarray}
which provide the image distributions
\begin{eqnarray}
  \sigma_{xx}&=&\langle x^2 \rangle-\langle x \rangle^2, \\
  \sigma_{yy}&=&\langle y^2 \rangle-\langle y \rangle^2, \\
  \sigma_{xy}&=&\langle xy \rangle-\langle x \rangle \langle y \rangle^2
\end{eqnarray}
and the auxiliary parameters
\begin{eqnarray}
  k&=&\sigma_{y^2}-\sigma_{x^2}, \\
  l&=&\sqrt{k^2+4\sigma_{xy}^2}, \\
  m&=&\langle y^2 \rangle - \langle x^2 \rangle , \\
  n&=&\sqrt{m^2+4\langle xy \rangle^2}, \\
  u_\pm&=&1\pm k/l.
\end{eqnarray}
Then the width $(W)$ and length $(L)$ of the image with respect to the major
and minor axes of the Hillas ellipse of the image can be written as
\begin{eqnarray}
  W&=&\sqrt{(\sigma_{x^2}+\sigma_{y^2}-l)/2}, \label{eqn:width} \\
  L&=&\sqrt{(\sigma_{x^2}+\sigma_{y^2}+l)/2}. \label{eqn:length}
\end{eqnarray}
Finally, the $miss$, the angles $\alpha$ and $\phi$ and the $azwidth$ are given
by
\begin{eqnarray}
  miss&=&\sqrt{(u_+\langle x \rangle^2+u_-\langle y \rangle^2)/2-2\langle xy
    \rangle\sigma_{xy}/l_0}, \\
  \phi&=&tan^{-1}\left(\frac{(k+l)\langle y \rangle+2\sigma_{xy}\langle x
    \rangle}{2\sigma_{xy}\langle y \rangle - (k-l) \langle x \rangle} \right), 
  \label{eqn:phi} \\
  \alpha&=&\left|\sin^{-1}\left(\frac{miss}LD\right)\right|, 
  \label{eqn:alpha} \\
  azwidth&=&\sqrt{(\langle x \rangle^2+\langle y \rangle^2 - n_0)/2}.
\end{eqnarray}

\chapter{Gaussian Fit Function} \label{app:GaussianFit}

The position and size of a \g-ray signal is determined from a fit of a
two-dimensional Gaussian function $G$ to the excess map of the data. $G$ is
obtained by a convolution of the two-dimensional Gaussian function
$g_{\sigma_x,\sigma_y}$ with the H.E.S.S. point spread function (PSF) to
allow a separation between the influence of the PSF and true or intrinsic width
and length of the source. The derivation of $G$ is given here.

If $g$ is a function describing a Gaussian distribution with a zero-mean, then
a one\hyp{}dimensional representation with standard deviation $\sigma_x$ is
\begin{equation}
  g_{\sigma_x}(x) = \frac{1}{\sqrt{2\pi}\sigma_x}
  \exp\left({-\frac{1}{2}\frac{x^2}{\sigma_x^2}}\right).
\end{equation}
The two-dimensional extension is
\begin{equation}
  g_{{\sigma_x},{\sigma_y}}(x,y) = g_{\sigma_x}(x)\cdot g_{\sigma_y}(y) =
  \frac{1}{2\pi\sigma_x \sigma_y} 
  \exp\left({-\frac{1}{2}\left(\frac{x^2}{\sigma_x^2}
    +\frac{y^2}{\sigma_y^2}\right)}\right).
  \label{eqn:2DGaussian}
\end{equation}
Furthermore, with $\ast$ denoting the convolution operator, the convolution of
two Gaussian functions is
\begin{equation}
  g_{\sigma_x}(x) \ast g_{\sigma_1}(x) = 
  \frac{1}{\sqrt{2\pi(\sigma_x^2+\sigma_1^2)}}
  \exp\left({-\frac{1}{2}\frac{x^2}{\sigma_x^2+\sigma_1^2}}\right) =
  g_{\sqrt{\sigma_x^2+\sigma_1^2}}(x)
\end{equation}
and consequently
\begin{eqnarray}
  g_{\sigma_x,\sigma_y}(x,y) \ast g_{\sigma_1,\sigma_1}(x,y)
  &=& [ g_{\sigma_x}(x)\cdot g_{\sigma_y}(y)] \ast [g_{\sigma_1}(x)\cdot
  g_{\sigma_1}(y)] \nonumber \\
  &=& \left[g_{\sigma_x}(x) \ast g_{\sigma_1}(x)\right] \cdot
  [g_{\sigma_y}(y) \ast \cdot g_{\sigma_1}(y)] \nonumber \\
  &=& g_{\sqrt{\sigma_x^2+\sigma_1^2}}(x) \cdot 
  g_{\sqrt{\sigma_y^2+\sigma_1^2}}(y)
  = g_{\sqrt{\sigma_x^2+\sigma_1^2},
    \sqrt{\sigma_y^2+\sigma_1^2}}(x,y)\nonumber\\
  &=& \frac{1}{2\pi\sqrt{\sigma_x^2+\sigma_1^2}\sqrt{\sigma_y^2+\sigma_1^2}}
  \exp\left({-\frac{1}{2}\left(\frac{x^2}{\sigma_x^2+\sigma_1^2}
    +\frac{y^2}{\sigma_y^2+\sigma_1^2}\right)}\right). \nonumber
\end{eqnarray}
According to Sec.~\ref{sec:PSF}, the H.E.S.S. PSF ($PSF$) can be described by
the sum of two radial symmetric Gaussian functions with normalization constants
$A$ and $A_{rel}$ as
\begin{eqnarray}
  PSF(x,y) &=& A [g_{{\sigma_1},{\sigma_1}}(x,y) + A_{rel}
    g_{{\sigma_2},{\sigma_2}}(x,y)] \\
  &=&N_1 g_{{\sigma_1},{\sigma_1}}(x,y)+N_2 g_{{\sigma_2},{\sigma_2}}(x,y)
  \nonumber \\
  &=&\frac{N_1}{2\pi\sigma_1^2}
  \exp\left({-\frac{1}{2}
    \left(\frac{x^2}{\sigma_1^2}+\frac{y^2}{\sigma_1^2}\right)}\right)
  +\frac{N_2}{2\pi\sigma_2^2}
  \exp\left({-\frac{1}{2}
    \left(\frac{x^2}{\sigma_2^2}+\frac{y^2}{\sigma_2^2}\right)}\right).
  \nonumber
  \label{eqn:appPSF}
\end{eqnarray}
The convolution of $g_{\sigma_x,\sigma_y}(x,y)$ with $PSF$ is
\begin{eqnarray}
  G_{\sigma_{x},\sigma_{y}}(x,y)
  &=&g_{{\sigma_x},{\sigma_y}}(x,y)\ast PSF(x,y) \\
  &=&g_{{\sigma_x},{\sigma_y}}(x,y) \ast
  [N_1 g_{{\sigma_1},{\sigma_1}}(x,y) 
    +  N_2 g_{{\sigma_2},{\sigma_2}}(x,y)] \nonumber\\
  &=&N_1 g_{\sqrt{\sigma_x^2+\sigma_1^2},\sqrt{\sigma_y^2+\sigma_1^2}}(x,y) 
    +N_2 g_{\sqrt{\sigma_x^2+\sigma_2^2},\sqrt{\sigma_y^2+\sigma_2^2}}(x,y)
    \nonumber \\
  &=&\frac{N_1}{2\pi\sqrt{\sigma_x^2 +\sigma_1^2} \sqrt{\sigma_y^2+\sigma_1^2}}
  \exp\left(-\frac{1}{2} \left(\frac{x^2}{(\sigma_x^2+\sigma_1^2)}
  +\frac{y^2}{(\sigma_y^2+\sigma_1^2)}\right)\right)
  \nonumber \\
  && +\frac{N_2}{2\pi\sqrt{\sigma_x^2 +\sigma_2^2} 
    \sqrt{\sigma_y^2+\sigma_2^2}}
  \exp\left(-\frac{1}{2} \left(\frac{x^2}{(\sigma_x^2+\sigma_2^2)}
  +\frac{y^2}{(\sigma_y^2+\sigma_2^2)}\right)\right) 
  \nonumber \\
  &=&\frac{N_1}{2\pi\sqrt{\sigma_x^2 +\sigma_1^2} \sqrt{\sigma_y^2+\sigma_1^2}}
  \Bigg[\exp\left(-\frac{1}{2} \left(\frac{x^2}{(\sigma_x^2+\sigma_1^2)}
    +\frac{y^2}{(\sigma_y^2+\sigma_1^2)}\right)\right)
    \nonumber \\
    && +\frac{N_2}{N_1} 
    \frac{\sqrt{\sigma_x^2+\sigma_1^2}\sqrt{\sigma_y^2+\sigma_1^2}}
         {\sqrt{\sigma_x^2 +\sigma_2^2}\sqrt{\sigma_y^2+\sigma_2^2}}
         \exp\left(-\frac{1}{2} \left(\frac{x^2}{(\sigma_x^2+\sigma_2^2)}
         +\frac{y^2}{(\sigma_y^2+\sigma_2^2)}\right)\right)\Bigg]. \nonumber
  \label{eqn:ConvolutionGauss2D}
\end{eqnarray}
According to Eqn.~\ref{eqn:appPSF} the relative amplitude $A_{rel}$ before the
convolution is given as
\begin{equation}
  A_{rel}=\frac{N_2}{N_1}\frac{\sigma_1^2}{\sigma_2^2} \Leftrightarrow 
  \frac{N_2}{N_1} = A_{rel}\frac{\sigma_2^2}{\sigma_1^2}.
  \label{eqn:arel}
\end{equation}
Consequently, by substituting $N_1/N_2$ with Eqn.~\ref{eqn:arel}
$G_{\sigma_{x},\sigma_{y}}(x,y)$ can be written as
\begin{eqnarray}
  G_{\sigma_{x},\sigma_{y}}(x,y)
  &=& \bar A \Bigg[
    \exp\left(-\frac{1}{2} \left(\frac{x^2}{(\sigma_x^2+\sigma_1^2)}
    +\frac{y^2}{(\sigma_y^2+\sigma_1^2)}\right)\right) \\
    && +A_{rel}\frac{\sigma_2^2}{\sigma_1^2}
    \frac{\sqrt{\sigma_x^2+\sigma_1^2}\sqrt{\sigma_y^2+\sigma_1^2}}
         {\sqrt{\sigma_x^2 +\sigma_2^2}\sqrt{\sigma_y^2+\sigma_2^2}}
         \exp\left({-\frac{1}{2}\left(\frac{x^2}{(\sigma_x^2+\sigma_2^2)}
           +\frac{y^2}{(\sigma_y^2+\sigma_2^2)}\right)}\right)
         \Bigg]. \nonumber
\end{eqnarray}
$G_{\sigma_{x},\sigma_{y}}(x,y)$ is centered at the origin, and the major and
minor axes are parallel to the coordinate axes. To fit excess distributions of
arbitrary orientation, a coordinate transformation by the rotation matrix
$M_{\omega}$ provides the necessary rotation by the angle $\omega$.
Additionally, two parameters ($\alpha_0$ and $\delta_0$) are required for
fitting an excess with an arbitrary position. The final fit function
$G_{\sigma_x,\sigma_y,\alpha_0,\delta_0,\omega,N}(\alpha,\delta)$ in equatorial
coordinates $\alpha$ and $\delta$ is obtained from substitution of $x$ and $y$
as
\begin{eqnarray}
  \left( \begin{matrix} x \\ y \end{matrix} \right) = M_{\omega} \left(
  \begin{matrix} \alpha-\alpha_0 \\ \delta-\delta_0 \end{matrix} \right) =
  \left( \begin{matrix} 
    \cos(\omega)\cos(\delta)(\alpha-\alpha_0)-\sin(\omega)(\delta-\delta_0) \\
    \sin(\omega)\cos(\delta)(\alpha-\alpha_0)+\cos(\omega)(\delta-\delta_0)
  \end{matrix} \right).
\end{eqnarray}

\chapter{Upper Limits According to Feldman and Cousins}
\label{app:FeldmanCousins}
The unified approach by \cite{FeldmanCousins} is a method for determining the
confidence intervals (CI) of a measurement for a given confidence level (CL). A
discussion of this approach is also found in the Particle Data Booklet
(\cite{ParticleDataBooklet:2006}, Chp.~31). The Feldman Cousins approach has
two important properties: it provides the correct ``coverage'' and it avoids
``flip-flopping''. Correct coverage means that a measured quantity $x$ lies
within the CI with the same probability that is given by the corresponding CL,
if the measurement is repeated numerous times. Flip-flopping refers to a
problem that arises if different methods for the calculation of the CIs of
upper limits and measurements are applied, which results in an incorrect
coverage. The avoidance of flip-flopping is of particular importance for
measurements of small signals where it is not clear beforehand whether one will
find a signal or an upper limit. Flip-flopping is avoided in the unified
approach by Feldman and Cousins, which provides a unified construction of CI
regardless of whether a signal or an upper limit is obtained. Here the problem
is illustrated and the approach is discussed in the case of a ``bounded
Gaussian''. The same discussion is found in greater detail in e.g.
\cite{FeldmanCousins} and \cite{Schwanke:UpperLimit}.

The classic approach constructs the CI for a CL of $\alpha$ by determining the
parameter $L$ such that
\begin{equation}
  P(x\in [\mu_1,\mu+2]|\mu)=
  \begin{cases}
    \int_{\mu+L}^{\mu+L}P(x|\mu)dx=\alpha, \quad\mbox{in the case of a signal} \\
    \int_{-\infty}^{L}P(x|\mu)dx=\alpha, \quad\mbox{in the case of an upper limit.}
  \end{cases}
\end{equation}
However, the transition from one case to the other implies flip-flopping and
does not guarantee correct coverage.

\begin{figure}[tb!]
  \begin{minipage}[t]{0.5\linewidth}
    \includegraphics[width=\textwidth]{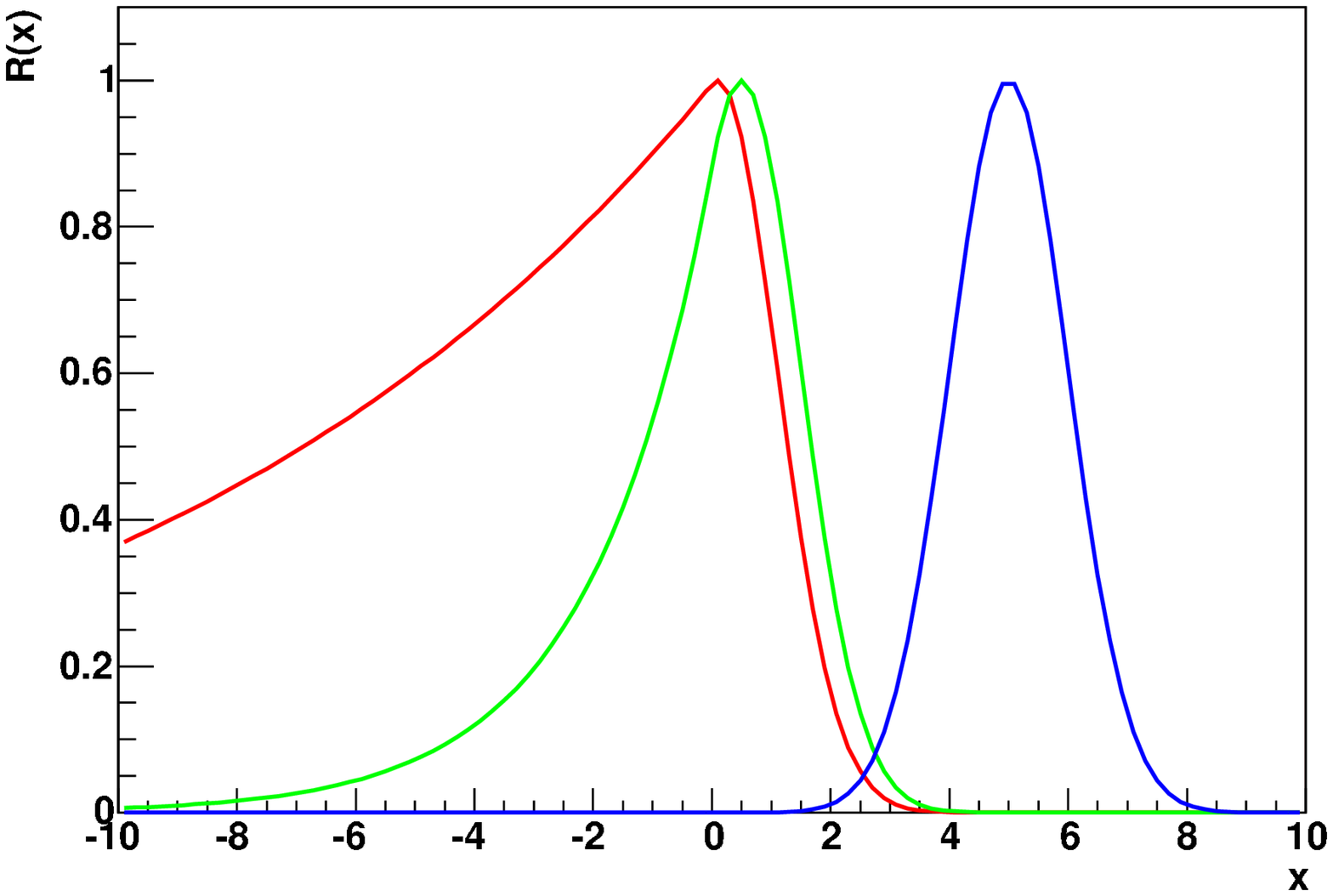}
    \caption[Ordering Parameter $R(x)$ Used in the Feldman Cousins
      Approach]{Ordering parameter $R(x)$ for a bounded Gaussian PDF
      (Eqn.~\ref{eqn:BoundedR}). The red, green and blue curves correspond to
      $\mu=$0.1, 0.5 and 5.0, respectively. (Figure taken from
      \cite{Schwanke:UpperLimit}.)}
  \label{fig:R}
  \end{minipage}\hfill
  \begin{minipage}[t]{0.5\linewidth}
    \includegraphics[width=\textwidth]{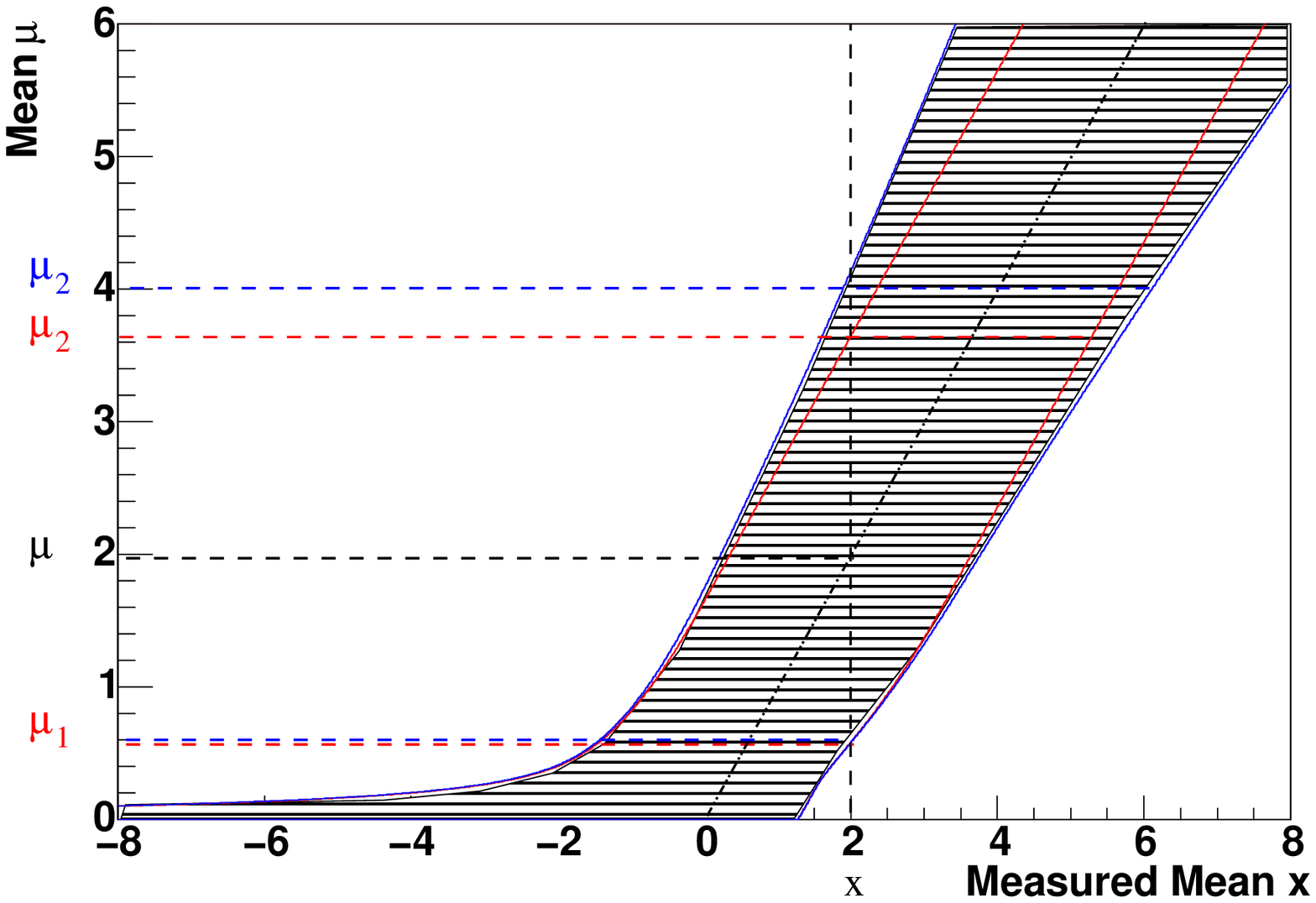}
    \caption[Confidence Belt for a Bounded Gaussian]{Confidence belts for a
      bounded Gaussian (Eqn.~\ref{eqn:BoundedR}) with a CL of 90\% constituted
      by a set of acceptance intervals. The blue confidence belt includes
      systematic errors; the red does not. The corresponding CI $[\mu_1,\mu_2]$
      is determined through the intersections with the vertical line
      representing a measured value $x$. (Figure taken from
      \cite{Schwanke:UpperLimit}.)}
  \label{fig:ConfidenceBelt}
  \end{minipage}
\end{figure}

Therefore, the approach by Cousins and Feldman introduces an ordering parameter
$R$, which specifies in what order the CI is constructed. $R$ is defined
through the Likelihood ratio of the experiment's probability density function
(PDF) $P(x|\mu)$ and $P(x|\hat{\mu})$ as
\begin{equation}
  R=\frac{P(x|\mu)}{P(x|\hat{\mu})},
\end{equation}
where $x$ and $\mu$ are the measured and true values and $\hat{\mu}$ denotes the
value of $\mu$ which maximizes $P(x|\mu)$ and which is physically allowed.
Acceptance intervals $[x_1, x_2]$ are then constructed for all possible
measurements of $x$ to fulfill
\begin{equation}
  P(x\in [x_1,x_2]|\mu)=\int_{x_1}^{x_2}P(x|\mu)dx=\alpha,
\end{equation}
and including the additional conditions
\begin{equation}
  R(x_1)=R(x_2).
\end{equation}
The union of all acceptance intervals of the possible values of $x$ will
provide a so-called confidence belt as shown in Fig.~\ref{fig:ConfidenceBelt}
from which the CI $[\mu_1,\mu_2]$ is determined.


In the case of a Gaussian PDF where the physical values of $\mu$ are restricted to
positive values
\begin{equation}
  P(x|\mu)=\frac{1}{\sqrt{2\pi}}\exp\left(-\frac12(x-\mu)^2\right).
  \label{eqn:GaussianPDF}
\end{equation}
For a measurement $x$, $P(x|\hat{\mu})$ is maximized if $\hat{\mu}=x$. This is
true for positive values of $x$, including zero. If $x$ is negative, $\mu=0$ is
the physically allowed value which maximizes $P(x|\hat{\mu})$, i.e.
\begin{equation}
  R(x) =
  \begin{cases}
    \exp(-\frac12(x-\mu)^2), \quad x \ge 0 \\
    \exp(-\frac12(x-\mu)^2-x^2), \quad x < 0. \\
  \end{cases}
  \label{eqn:BoundedR}
\end{equation}
The corresponding values of $R$ for $\mu=$0.1, 0.5 and 5.0 are shown by the
red, green and blue curves in Fig.~\ref{fig:R}. For values of $\mu$ that are
small in comparison to the standard deviation $\sigma=1$, the asymmetry is
apparent, which is reflected in the acceptance interval. The resulting
confidence belt for a CL of 90\% is shown in Fig.~\ref{fig:ConfidenceBelt} by
the red curves.
The CI $[\mu_1,\mu_2]$ is determined by the intersection of the vertical line
at the measured value $x$ with the confidence belt, as shown in
Fig.~\ref{fig:ConfidenceBelt}. If the vertical line at $x$ does not intersect
the lower boundary at $\mu_1>0$, one does not obtain a lower boundary but only
an upper limit. The Feldman Cousins approach therefore guarantees a smooth
transition between a measurement and an upper limit and it avoids any
discontinuities and flip-flopping.

It is also possible to incorporate systematic errors into this approach.
Therefore it is assumed that the systematic error $(\sigma_s)$ has a Gaussian
PDF
\begin{equation}
  L(s|\sigma_s)=\frac{1}{\sqrt{2\pi}\sigma_s}
  \exp\left(-\frac{(1-s)^2}{2\sigma_s^2}\right),
\end{equation}
where $s=(1\pm e)$ represents a factor by which an actual measurement could
appear multiplied and where $e$ is of the order of $\sigma_s$. With this ansatz
one can formulate a new PDF
\begin{equation}
  F(s|\sigma_s)=\int P(x|s\mu)L(s|\sigma_s)ds
\end{equation}
which replaces the PDF in Eqn.~\ref{eqn:GaussianPDF}. For the correct treatment
one also has to replace $R$ accordingly, i.e.
\begin{equation}
  R=\frac{P(x|\mu)}{P(x|\hat{\mu})}.
\end{equation}
With these modifications, one proceeds as described above. The CI with the same
parameters but with an additional systematic error of 20\% is shown by the blue
curves in Fig.~\ref{fig:ConfidenceBelt}. The new confidence interval is larger.

In the H.E.S.S. standard analysis, the calculation of the CI for a bounded
Gaussian according to the unified approach by Feldman and Cousins is realized
by the class {\it TBoundedGaussian}. This class is initialized with a CL
$\alpha$ and a systematic error $\sigma_s$. The CI is obtained through the
method {\it GetConfidenceInterval}($x$, $\sigma_x$, \&$\mu_1$, \&$\mu_2$),
which returns the boundaries of the CI $\mu_1$ and $\mu_2$ for the measured
value $x$ and the error $\sigma_x$.

\chapter{Lists of Observation Runs} \label{app:RunLists}

\vspace{-1.25cm}
\begin{table}[h!]
  \centering
  \caption[Run List for the Crab Nebula]{Run list of H.E.S.S. wobble
     observation at (J2000) (${\rm 15^h14^m27^s}$, $-59^\circ16'18''$) of the
     Crab Nebula. The observation period lasted from January 25, 2004 to
     March 4, 2005.}
  \medskip
  \begin{tabular}{ccccccc}
    \hline \hline
    & Run no. & Wobble offset & <Altitude>[$^\circ$] & <Rate>[Hz]
    & Events & Duration[s] \\
    \hline
    1  &  18526  & Dec $-0.5^\circ$  & 59.3  & 231 & 437765 &  1682 \\
    2  &  18870  & Dec $+0.5^\circ$  & 46.2  & 294 & 556196 &  1683 \\
    3  &  18874  & Dec $+0.5^\circ$  & 60.1  & 208 & 393848 &  1681 \\
    4  &  18875  & Dec $-0.5^\circ$  & 64.7  & 176 & 341233 &  1683 \\
    5  &  18889  & Dec $+0.5^\circ$  & 58.0  & 213 & 406677 &  1683 \\
    6  &  18890  & Dec $-0.5^\circ$  & 62.5  & 185 & 351518 &  1683 \\
    7  &  18927  & Dec $+0.5^\circ$  & 61.1  & 195 & 372168 &  1684 \\
    8  &  23037  & Dec $-0.5^\circ$  & 58.6  & 113 & 198793 &  1567 \\
    9  &  23062  & RA  $-0.5^\circ$  & 61.9  & 83.3& 147520 &  1687 \\
    10 &  23063  & RA  $+0.5^\circ$  & 57.7  & 101 & 180349 &  1686 \\
    11 &  23304  & RA  $-0.5^\circ$  & 52.4  & 137 & 243054 &  1613 \\
    12 &  23309  & Dec $+0.5^\circ$  & 48.3  & 182 & 348151 &  1686 \\
    13 &  23310  & Dec $-0.5^\circ$  & 50.2  & 171 & 323219 &  1686 \\
    14 &  23526  & Dec $+0.5^\circ$  & 46.3  & 213 & 425821 &  1684 \\
    15 &  23544  & RA  $+0.5^\circ$  & 48.8  & 210 & 435015 &  1686 \\
    16 &  23545  & RA  $-0.5^\circ$  & 46.3  & 220 & 460561 &  1686 \\
    17 &  23546  & Dec $+0.5^\circ$  & 46.0  & 221 & 264014 &   948 \\
    18 &  23547  & Dec $-0.5^\circ$  & 44.8  & 227 & 270707 &   942 \\
    19 &  23577  & RA  $-0.5^\circ$  & 45.3  & 214 & 451231 &  1686 \\
    20 &  23579  & RA  $-0.5^\circ$  & 47.3  & 208 & 246834 &   972 \\
    21 &  23593  & Dec $-0.5^\circ$  & 46.2  & 204 & 402139 &  1687 \\
    22 &  23642  & Dec $-0.5^\circ$  & 51.1  & 193 & 379326 &  1686 \\
    23 &  23662  & Dec $+0.5^\circ$  & 46.0  & 218 & 342204 &  1679 \\
    24 &  23739  & RA  $+0.5^\circ$  & 46.2  & 189 & 373756 &  1686 \\
    25 &  23740  & Dec $-0.5^\circ$  & 44.9  & 192 & 381556 &  1686 \\
    26 &  23741  & Dec $+0.5^\circ$  & 46.2  & 186 & 372221 &  1686 \\
    27 &  23753  & RA  $+0.5^\circ$  & 52.1  & 169 & 316510 &  1686 \\
    28 &  23756  & Dec $+0.5^\circ$  & 46.8  & 199 & 408124 &  1687 \\
    29 &  24138  & Dec $+0.5^\circ$  & 48.1  & 192 & 187203 &   907 \\
    30 &  24412  & RA  $+0.5^\circ$  & 51.8  & 166 & 308661 &  1687 \\
    \hline
    30 &    -    &  -                & 51.5  & 190 &10326374& 47385 \\ 
    \hline \hline
  \end{tabular}
  \label{tbl:CrabRuns}
\end{table}

\begin{table}
  \centering
  \caption[Run List for \MSH\ with Wobble Offset $\theta_w=+0.5^\circ$ in Dec]
    {Run list of all H.E.S.S. observation runs of \MSH\ from the observation
    period of March 26 to July 20, 2004 taken at the wobble position (J2000)
    (${\rm 15^h14^m27^s}$, $-59^\circ16'18''$) $+0.5^\circ$ in Dec. Runs in
    parentheses did not pass the run selection criteria and were not used in
    the analysis.}
  \bigskip
  \begin{tabular}{cccccc}
    \hline \hline
    & Run no. & <Altitude> [$^\circ$] & <Rate> [Hz] & Events & Duration [s] \\
    \hline
    -  &\it (20136) &\it (51.9) &\it (287) &\it (520570) &\it (1682) \\
    1  &  20282  & 53.4  & 259 & 480314 &  1687 \\
    2  &  20301  & 53.7  & 269 & 502145 &  1683 \\ 
    3  &  20303  & 54.3  & 272 & 521903 &  1682 \\ 
    4  &  20323  & 54.4  & 295 & 586611 &  1682 \\ 
    5  &  20325  & 53.5  & 294 & 580954 &  1682 \\ 
    6  &  20343  & 53.2  & 286 & 540909 &  1682 \\ 
    7  &  20345  & 54.4  & 289 & 550983 &  1683 \\ 
    8  &  20366  & 54.4  & 261 & 487784 &  1683 \\ 
    9  &  20368  & 53.3  & 258 & 485038 &  1683 \\ 
    10 &  21085  & 53.1  & 249 & 547372 &  1683 \\
    -  &\it (21181) &\it (54.0) &\it (246) &\it (457715)&\it (1684) \\ 
    11 &  21207  & 53.3  & 233 & 475160 &  1683 \\ 
    12 &  21210  & 54.4  & 237 & 490840 &  1683 \\ 
    13 &  21228  & 45.2  & 199 & 375716 &  1682 \\ 
    -  &\it (21230)& -     & -   & \it (0) &\it (1681) \\ 
    -  &\it (21232)& -     & -   & \it (0) &\it (1681) \\
    14 &  21261  & 51.2  & 227 & 425355 &  1683 \\ 
    15 &  21263  & 54.1  & 239 & 446111 &  1683 \\ 
    16 &  21266  & 53.8  & 238 & 443065 &  1683 \\ 
    17 &  21290  & 54.4  & 227 & 419622 &  1683 \\ 
    18 &  21368  & 54.2  & 223 & 412307 &  1683 \\ 
    19 &  21580  & 54.4  & 222 & 436539 &  1686 \\ 
    20 &  21598  & 53.9  & 227 & 476823 &  1683 \\ 
    21 &  21618  & 54.2  & 207 & 397984 &  1683 \\ 
    -  &\it  (21641)  &\it (54.3)  &\it  (96) &\it (167231) &\it (1684) \\ 
    22 &  21643  & 52.0  & 200 & 390105 &  1686 \\ 
    -  &\it  (21688)  &\it (54.5)  &\it (215) &\it (391629) &\it (1681) \\ 
    -  &\it  (21690)  &  -    &\it (213) &\it  (54357) &\it (1685) \\
    -  &\it  (21711)  & -     & -   & \it (0) &\it  (424) \\ 
    23 &  21739  & 49.8 & 191 & 350838  &  1685 \\
    \hline
    23 &   -     & 53.3 & 245 &10824478 & 40407 \\ 
    \hline \hline
  \end{tabular}
  \label{tbl:ObsPos1}
\end{table}

\begin{table}
  \centering
  \caption[Run List for \MSH\ with Wobble Offset $\theta_w=-0.5^\circ$ in Dec]
    {Run list of all H.E.S.S. observation runs of \MSH\ from the observation
    period of March 26 to July 20, 2004 taken at the wobble position (J2000)
    (${\rm 15^h14^m27^s}$, $-59^\circ16'18''$) $-0.5^\circ$ in Dec. Runs in
    parentheses did not pass the run selection criteria and were not used in
    the analysis.}
  \bigskip
  \begin{tabular}{cccccc}
    \hline \hline
    & Run no. &  <Altitude> [$^\circ$] & <Rate> [Hz] & Events & Duration [s] \\
    \hline
    -  &\it (20137) &\it (49.4) &\it (276) &\it (267475) &\it (902) \\
    -  &\it (20283) &\it (53.3) &\it (260) &\it (457461) &\it (1683) \\ 
    1  &  20302  & 53.4  & 272 & 508825 &  1682 \\
    -  &\it  (20304)  &  -    &\it (267) &\it (503183) &\it (1683) \\
    2  &  20322  & 52.6  & 291 & 568503 &  1682 \\
    3  &  20324  & 53.3  & 293 & 577842 &  1683 \\
    4  &  20344  & 53.2  & 286 & 537738 &  1682 \\
    5  &  20346  & 52.9  & 284 & 532043 &  1683 \\
    6  &  20365  & 52.8  & 258 & 480312 &  1683 \\
    7  &  20367  & 53.3  & 254 & 475231 &  1682 \\
    8  &  21083  & 53.4  & 252 & 220554 &   680 \\
    9  &  21084  & 53.1  & 250 & 334531 &  1022 \\
    10 &  21182  & 53.5  & 246 & 461764 &  1683 \\
    11 &  21206  & 50.8  & 225 & 451413 &  1683 \\
    12 &  21209  & 53.3  & 234 & 478839 &  1683 \\
    13 &  21229  & 47.5  & 207 & 379886 &  1687 \\
    14 &  21231  & 51.8  & 224 & 415204 &  1683 \\
    -  &\it  (21233)  & -     &  -  & \it (0) &\it (1681) \\
    15 &  21258  & 47.5  & 214 & 388676 &  1683 \\
    16 &  21262  & 52.1  & 232 & 431181 &  1683 \\
    17 &  21264  & 53.4  & 236 & 438404 &  1682 \\
    18 &  21291  & 53.2  & 222 & 409893 &  1683 \\
    -  &\it  (21316)  &\it (53.4)  &\it (222) &\it (383159) &\it (1682) \\
    19 &  21366  & 53.4  & 222 & 412774 &  1686 \\
    20 &  21579  & 53.3  & 219 & 414979 &  1682 \\
    21 &  21599  & 53.4  & 225 & 480188 &  1682 \\
    22 &  21689  & 53.0  & 212 & 402123 &  1685 \\
    23 &  21691  & 51.5  & 205 & 609570 &  1206 \\
    -  &\it  (21710)  & -     &  -  & \it (0) &\it  (378) \\
    24 &  21740  & 46.4  & 179 & 336346 &  1682 \\
    \hline
    24 &   -     & 52.3  & 241 &10746819& 39164 \\
    \hline \hline
  \end{tabular}
  \label{tbl:ObsPos2}
\end{table}

\begin{table}
  \centering
  \caption[Run List for \MSH\ with Wobble Offset $\theta_w=+0.5^\circ$ in RA]
    {Run list of all H.E.S.S. observation runs of \MSH\ from the observation
    period of March 26 to July 20, 2004 taken at the wobble position (J2000)
    (${\rm 15^h14^m27^s}$, $-59^\circ16'18''$) $+0.5^\circ$ in RA.}
  \bigskip
  \begin{tabular}{cccccc}
    \hline \hline
    & Run no. & <Altitude> [$^\circ$] & <Rate> [Hz] & Events & Duration [s] \\
    \hline
    1  &  20391  & 54.0  & 283 & 542630 &  1682 \\       
    2  &  20394  & 51.9  & 275 & 528343 &  1682 \\
    3  &  20415  & 52.8  & 280 & 513576 &  1683 \\
    4  &  20417  & 53.9  & 283 & 522969 &  1682 \\
    5  &  20451  & 53.8  & 282 & 543365 &  1684 \\
    6  &  20486  & 52.8  & 272 & 496787 &  1683 \\
    7  &  20488  & 53.9  & 273 & 512023 &  1683 \\
    \hline
    7  &    -    & 53.3  & 277 &3659693 & 11779 \\
    \hline \hline
  \end{tabular}
  \label{tbl:ObsPos3}
\end{table}

\begin{table}
  \centering
  \caption[Run List for \MSH\ with Wobble Offset $\theta_w=-0.5^\circ$ in RA]
    {Run list of all H.E.S.S. observation runs of \MSH\ from the observation
    period of March 26 to July 20, 2004 taken at the wobble position (J2000)
    (${\rm 15^h14^m27^s}$, $-59^\circ16'18''$) $-0.5^\circ$ in RA. Runs in
    parentheses did not pass the run selection criteria and were not used in
    the analysis.}
  \bigskip
  \begin{tabular}{cccccc}
    \hline \hline
    & Run no. & <Altitude> [$^\circ$] & <Rate> [Hz] & Events & Duration [s] \\
    \hline
    1  &  20390  & 53.7  & 282 & 538022 &  1686 \\
    -  &\it  (20392)  &\it (53.7)  &\it (285) &\it (130710) &\it  (407) \\
    2  &  20393  & 53.0  & 280 & 550658 &  1682 \\
    3  &  20416  & 53.9  & 283 & 517250 &  1683 \\
    4  &  20418  & 53.0  & 278 & 523280 &  1683 \\
    5  &  20447  & 53.5  & 283 & 510339 &  1682 \\
    6  &  20452  & 52.5  & 276 & 529727 &  1683 \\
    7  &  20487  & 53.9  & 272 & 503246 &  1683 \\
    8  &  20489  & 53.0  & 269 & 513066 &  1682 \\
    -  &\it (21086) &\it (50.8) &\it (242) &\it (533691) &\it (1682) \\
    \hline
    8  &    -    & 53.3  & 277 &4185588 & 13464 \\
    \hline \hline
  \end{tabular}
  \label{tbl:ObsPos4}
\end{table}

\newpage{\thispagestyle{empty}\cleardoublepage}

\chapter{The Richardson-Lucy Algorithm} \label{app:Richardson-Lucy}

\lstset{basicstyle=\scriptsize\ttfamily, language=IDL, breaklines=true,
  showstringspaces=false}

\begin{lstlisting}[caption={The ``max\_likelihood.pro'' procedure as provided
  by the ``IDL Astronomy Library'' (\cite{IDLAstroLib}) for the Richardson-Lucy
  deconvolution.}, label={_likelihood.pro}]
;+
; NAME:
;	MAX_LIKELIHOOD
;
; PURPOSE:
;	Maximum likelihood deconvolution of an image or a spectrum.
; EXPLANATION:
; 	Deconvolution of an observed image (or spectrum) given the 
;	instrument point spread response function (spatially invariant psf).
;	Performs iteration based on the Maximum Likelihood solution for
;	the restoration of a blurred image (or spectrum) with additive noise.
;	Maximum Likelihood formulation can assume Poisson noise statistics
;	or Gaussian additive noise, yielding two types of iteration.
;
; CALLING SEQUENCE:
;	for i=1,Niter do Max_Likelihood, data, psf, deconv, FT_PSF=psf_ft
;
; INPUTS PARAMETERS:
;	data = observed image or spectrum, should be mostly positive,
;				with mean sky (background) near zero.
;	psf = Point Spread Function of the observing instrument,
;				(response to a point source, must sum to unity).
; INPUT/OUTPUT PARAMETERS:
;	deconv = as input: the result of previous call to Max_Likelihood,
;		(initial guess on first call, default = average of data),
;		as output: result of one more iteration by Max_Likelihood.
;	Re_conv = (optional) the current deconv image reconvolved with PSF
;		for use in next iteration and to check convergence.
;
; OPTIONAL INPUT KEYWORDS:
;      /GAUSSIAN causes max-likelihood iteration for Gaussian additive noise
;		to be used,  otherwise the default is Poisson statistics.
;	FT_PSF = passes (out/in) the Fourier transform of the PSF,
;		so that it can be reused for the next time procedure is called,
;      /NO_FT overrides the use of FFT, using the IDL function convol() instead.
;	POSITIVITY_EPS = value of epsilon passed to function positivity,
;			default = -1 which means no action (identity).
;	UNDERFLOW_ZERO = cutoff to consider as zero, if numbers less than this.
;
; EXTERNAL CALLS:
;	function convolve( image, psf ) for convolutions using FFT or otherwise.
;	function positivity( image, EPS= ) to make image positive.
;
; METHOD:
;	Maximum Likelihood solution is a fixed point of an iterative eq.
;	(derived by setting partial derivatives of Log(Likelihood) to zero).
;	Poisson noise case was derived by Richardson(1972) & Lucy(1974).
;	Gaussian noise case is similar with subtraction instead of division.
; HISTORY:
;	written: Frank Varosi at NASA/GSFC, 1992.
;	F.V. 1993, added optional arg. Re_conv (to avoid doing it twice).
;	Converted to IDL V5.0   W. Landsman   September 1997
;-

pro Max_Likelihood, data, psf, deconv, Re_conv, FT_PSF=psf_ft, NO_FT=noft, $
						GAUSSIAN=gaussian, $
						POSITIVITY_EPS=epsilon, $
						UNDERFLOW_ZERO=under

	if N_elements( deconv ) NE N_elements( data ) then begin
		deconv = data
		deconv[*] = total( data )/N_elements( data )
		Re_conv = 0
	   endif

	if N_elements( under ) NE 1 then under = 1.e-22
	if N_elements( epsilon ) NE 1 then epsilon = -1

	if N_elements( Re_conv ) NE N_elements( deconv ) then $
		Re_conv = convolve( positivity( deconv, EPS=epsilon ), psf, $
						  FT_PSF=psf_ft, NO_FT=noft )

	if keyword_set( gaussian ) then begin

		deconv = deconv + convolve( data - Re_conv, psf, /CORREL, $
						   FT_PSF=psf_ft, NO_FT=noft )
	  endif else begin

		wp = where( Re_conv GT under, npos)
		wz = where( Re_conv LE under, nneg)
		if (npos GT 0) then Re_conv[wp] = ( data[wp]/Re_conv[wp] ) > 0
		if (nneg GT 0) then Re_conv[wz] = 1

		deconv = deconv * convolve( Re_conv, psf, FT_PSF=psf_ft, $
							/CORREL, NO_FT=noft )
	   endelse

	if N_params() GE 4 then $
		Re_conv = convolve( positivity( deconv, EPS=epsilon ), psf, $
						FT_PSF = psf_ft, NO_FT = noft )
 end
\end{lstlisting}

\chapter{PSF and Source Extension at Different Energy Bands}
\label{app:EnergyBands}
Tbl.~\ref{tbl:EnergyBandsPSF} shows the individual best fit parameters (cf.
Eqn.~\ref{eqn:PSF}) and the containment radius $(r_{68\%})$ of the H.E.S.S. PSF
for different energy bands. They have been determined from point source Monte
Carlo data with zenith angle of 40$^\circ$, wobble offset of 0.5$^\circ$ and a
restriction to the corresponding energy range. The variation of the PSF is
explained by the reconstruction accuracy for \g-ray showers which increases
with energy. The parameter $\sigma_s$ denotes the standard deviation of the
Gaussian which has been applied to smooth two maps to a similar PSF with same
$r_{68\%}$.

\begin{figure}[h!]
  \begin{minipage}[c]{0.5\linewidth}
    \includegraphics[width=\textwidth]{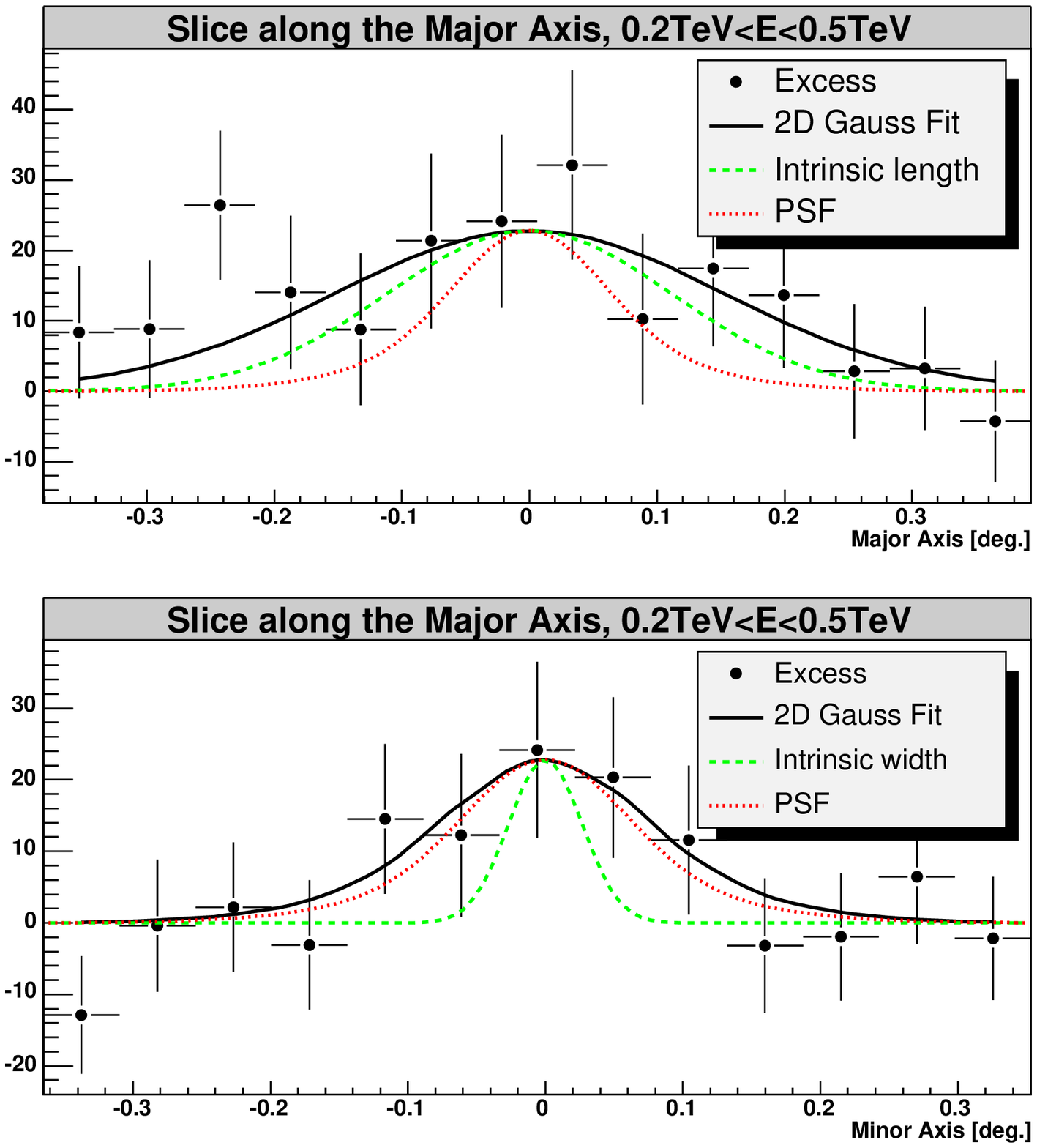}
  \end{minipage} \hfill
  \begin{minipage}[c]{0.5\linewidth}
    \includegraphics[width=\textwidth]{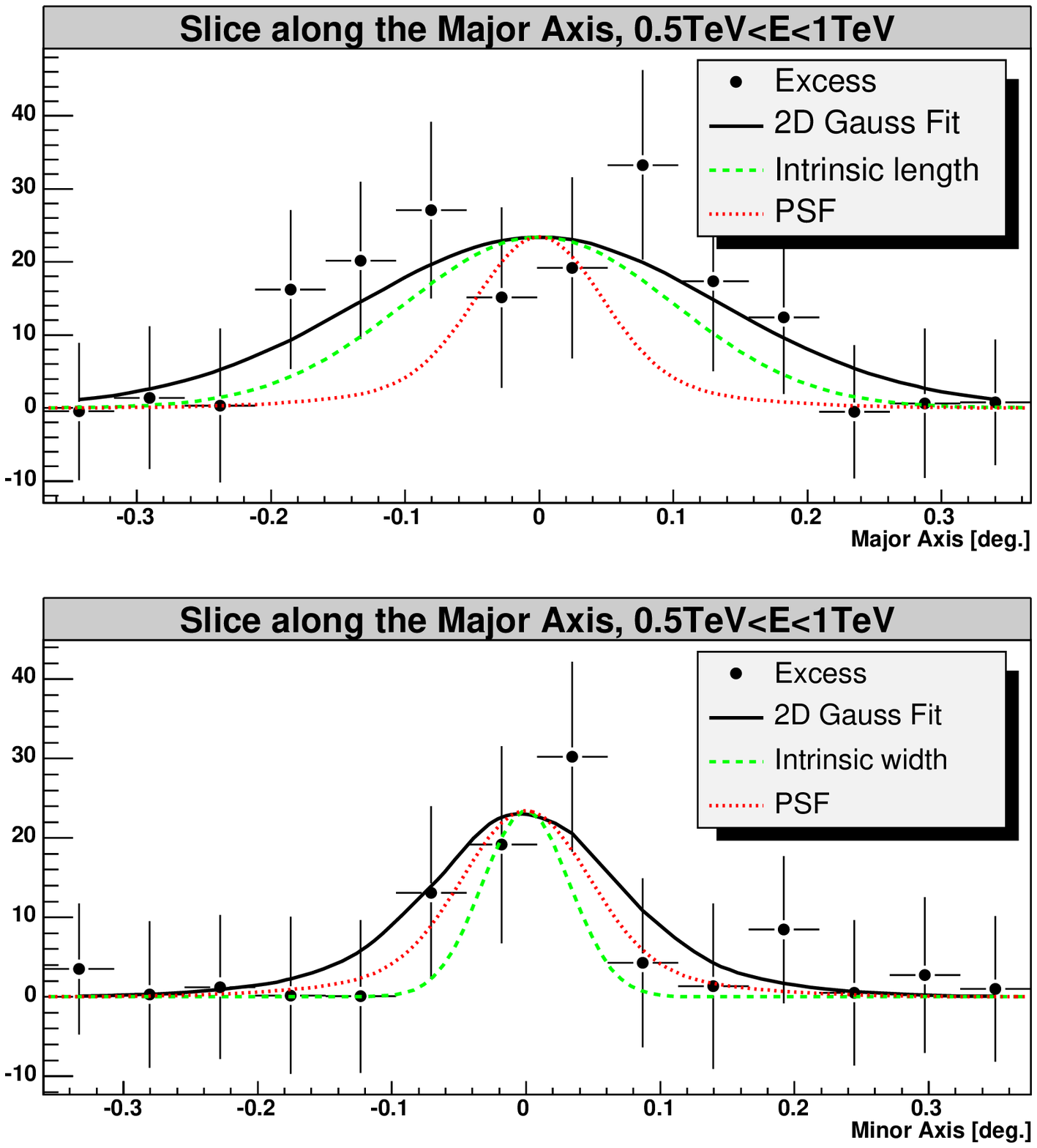}
  \end{minipage} \hfill \bigskip \\
  \begin{minipage}[c]{0.5\linewidth}
    \includegraphics[width=\textwidth]{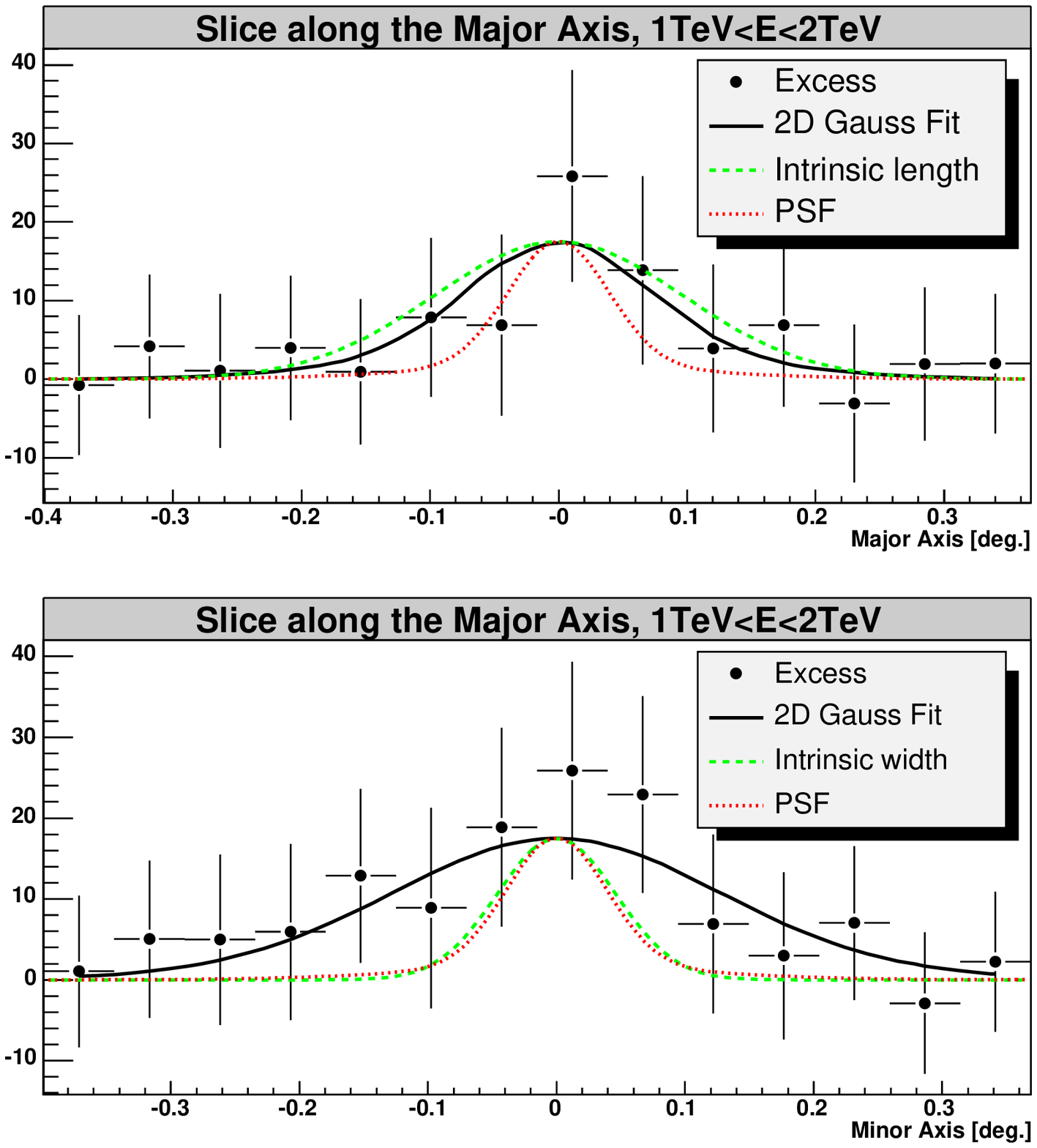}
  \end{minipage} \hfill
  \begin{minipage}[c]{0.5\linewidth}
    \includegraphics[width=\textwidth]{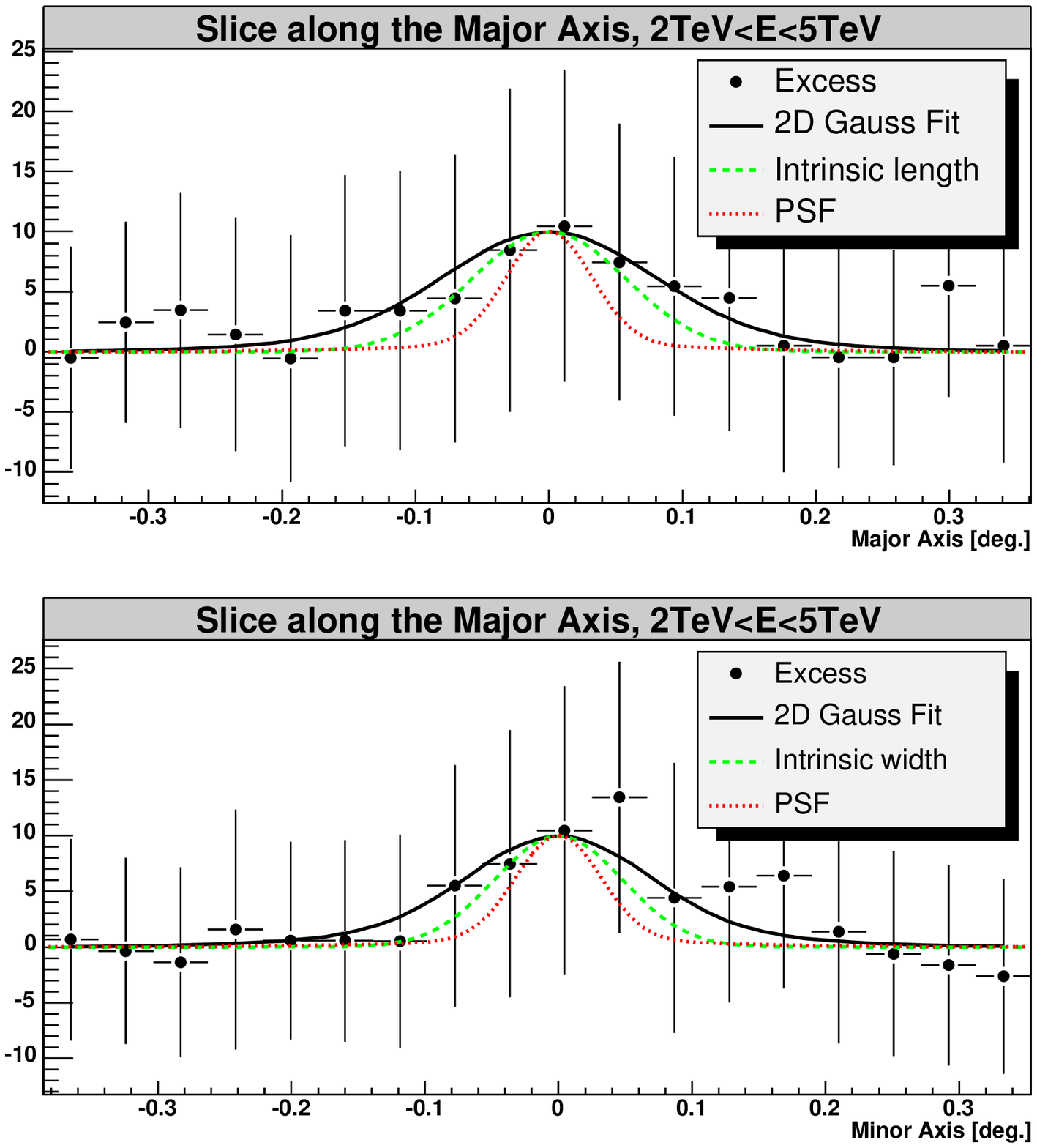}
  \end{minipage} \hfill \bigskip \\
  \begin{minipage}[c]{0.5\linewidth}
    \includegraphics[width=\textwidth]{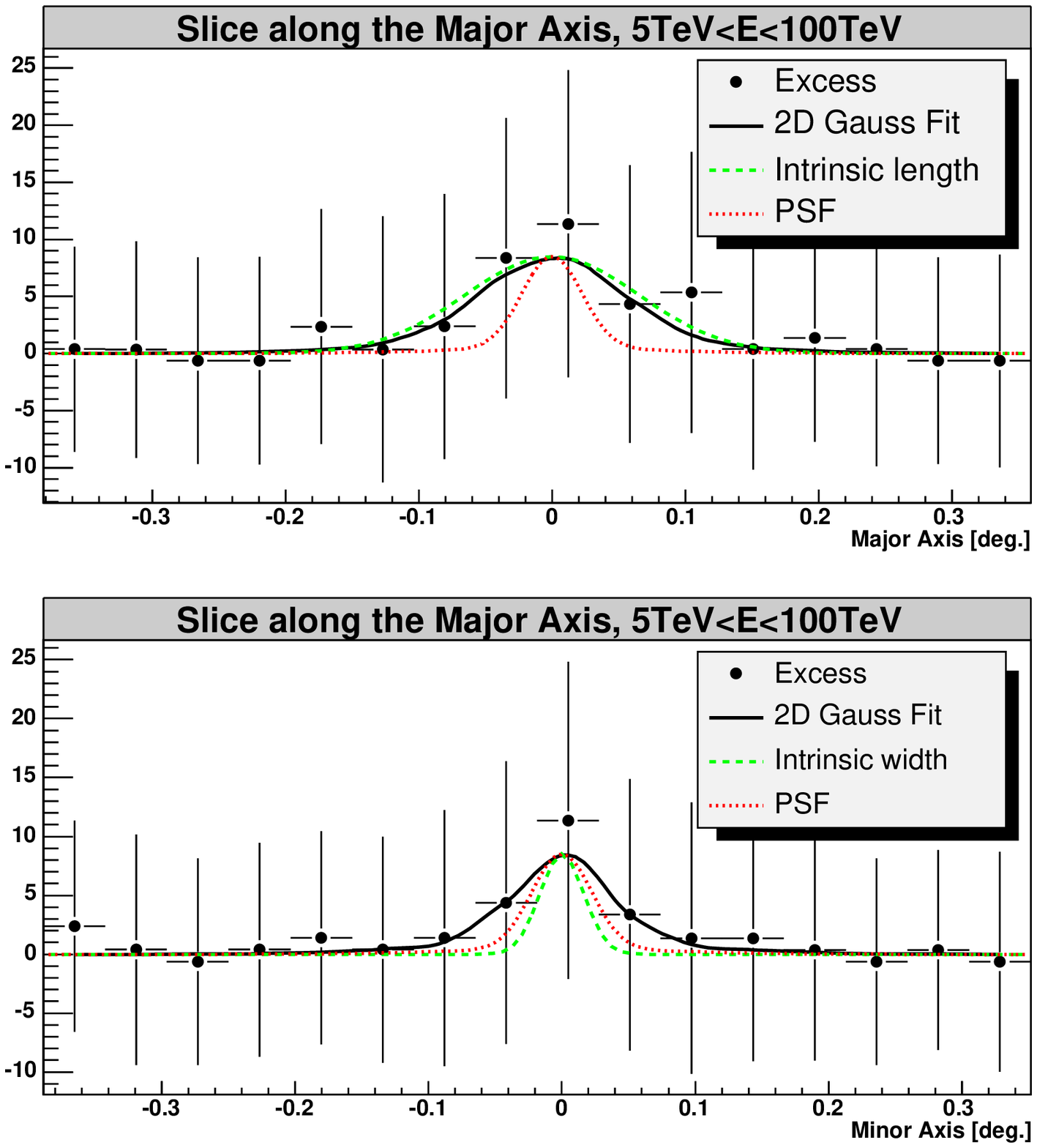}
  \end{minipage} \hfill
  \begin{minipage}[c]{0.43\linewidth}
    \caption[Fit Functions of the Excess of Fig.~\ref{fig:EnergyBands}]{Fit
       functions along the major axis (upper plot) and minor axis (lower plot)
       of the \g-ray excess maps of Fig.~\ref{fig:EnergyBands} (left column).
       Each fit function as well as their components of the intrinsic width and
       the PSF are shown.}
    \label{fig:EnergyBandsFits}
  \end{minipage} \hfill
\end{figure}

Fig.~\ref{fig:EnergyBandsFits} shows slices along the major and minor axis of
the excess maps and the fit function $G$ (cf. App.~\ref{app:GaussianFit}) of
Fig.~\ref{fig:EnergyBands} (left column). In addition to $G$ (black) its two
components of the PSF (red) and the intrinsic width (green) are shown.

\begin{table}[h!]
  \centering
  \caption[PSF at Different Energy Bands]{PSFs at different energy bands shown
    in Fig.~\ref{fig:EnergyBands}. The parameters $A_{rel}$, $\sigma_1$ and
    $\sigma_2$ are determined by Eqn.~\ref{eqn:PSF}. $r_{68\%}$ denotes the
    68\% containment radius and $\sigma_s$ the standard deviation of the
    smoothing Gaussian applied before the parameters were determined. The
    values of the last line in brackets show the PSF standard parameters of the
    unrestricted energy range for comparison.}
  \bigskip
  \begin{tabular}{cccccc}
    \hline \hline
    $E$ [TeV] & $\sigma_s$ & $A_{rel}$ & $\sigma_1[^\circ]$ &$\sigma_2[^\circ]$
    & $r_{68\%}[^\circ]$\\
    \hline
    $0.2$\,TeV$<E<0.5$\,TeV  & 0     & 0.35  & 0.0574 & 0.108 & 0.127 \\
    $0.5$\,TeV$<E<1$\,TeV    & 0     & 0.17  & 0.0479 & 0.110 & 0.110 \\
    $1$\,TeV$<E<2$\,TeV      & 0     & 0.096 & 0.0409 & 0.116 & 0.102 \\
    $2$\,TeV$<E<5$\,TeV      & 0     & 0.057 & 0.0361 & 0.123 & 0.095 \\
    $5$\,TeV$<E<100$\,TeV    & 0     & 0.052 & 0.0275 & 0.0929& 0.067 \\
    $0.2$\,TeV$<E<5$\,TeV    & 0     & 0.181 & 0.0461 & 0.113 & 0.118 \\
    $0.2$\,TeV$<E<5$\,TeV    & 0.023 & 0.219 & 0.0518 & 0.116 & 0.125 \\
    $5$\,TeV$<E<100$\,TeV    & 0.064 & 0.298 & 0.0673 & 0.106 & 0.125 \\
    ($0.2$\,TeV$<E<100$\,TeV & 0     & 0.170 & 0.0477 & 0.117 & 0.120)\\
    \hline \hline
  \end{tabular}
  \label{tbl:EnergyBandsPSF}
\end{table}


\chapter{Chandra Data Analysis} \label{app:ChandraDataAnalysis}
\begin{figure}[b!]
  \begin{minipage}[t]{0.32\linewidth}
    \includegraphics[width=\textwidth]{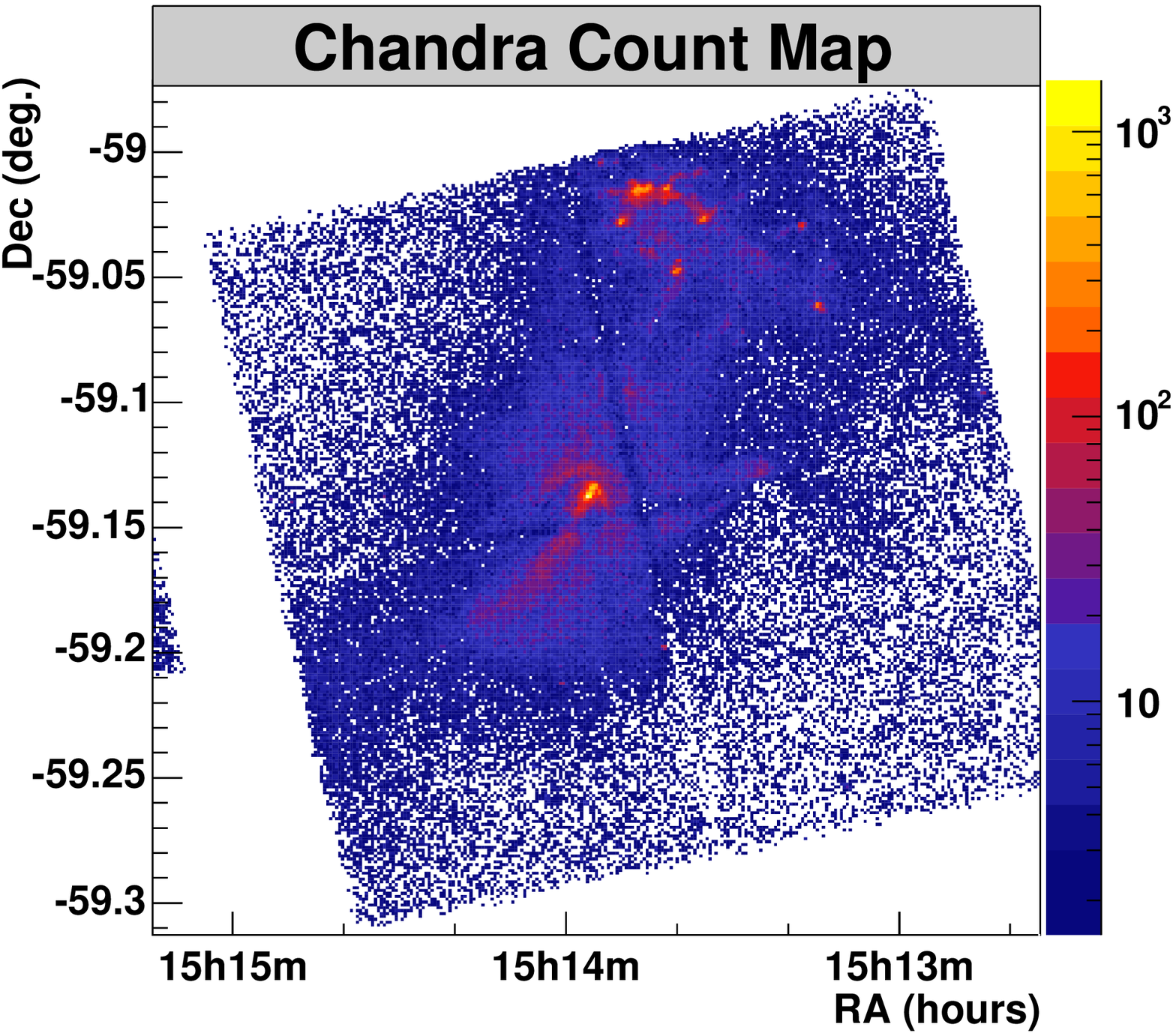}
    \caption[Chandra \X-Ray Count Map of \MSH]{Chandra X-ray count map of the
    ACIS CCD detectors of \MSH\ after calibration and dead-time correction.}
    \label{fig:ChandraCountMap}
  \end{minipage}\hfill
  \begin{minipage}[t]{0.32\linewidth}
    \includegraphics[width=\textwidth]{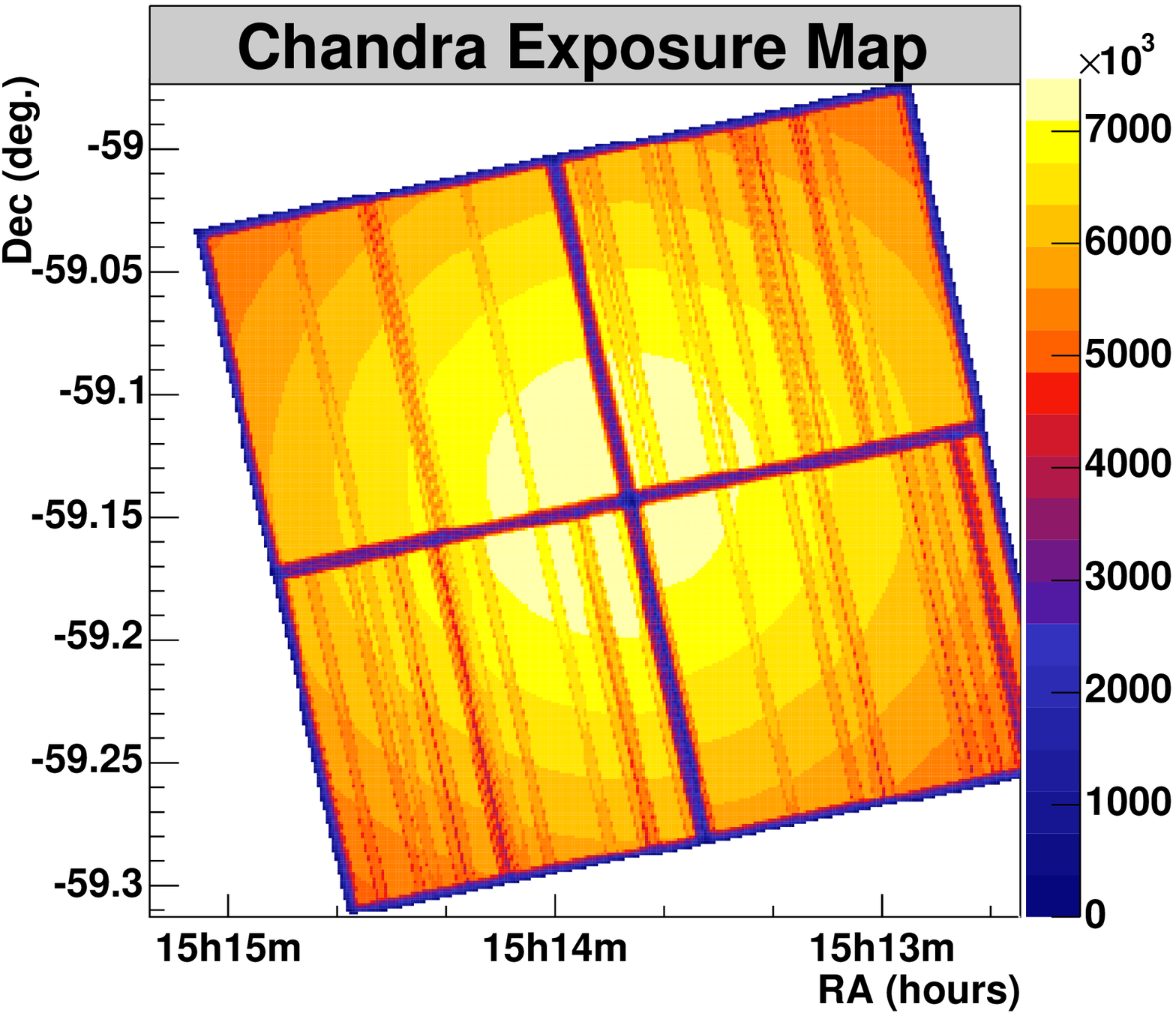}
    \caption[Chandra Exposure Map]{Chandra exposure map for the ACIS CCD
    corresponding to the observation of Fig.~\ref{fig:ChandraCountMap}.}
    \label{fig:ChandraExposureMap}
  \end{minipage}\hfill
 \begin{minipage}[t]{0.32\linewidth}
  \includegraphics[width=\textwidth]{images/ChandraExpcorrLog}
  \caption[Chandra Exposure Corrected \X-ray Map]{The corresponding Chandra
    exposure corrected X-ray map of Fig.~\ref{fig:ChandraCountMap}. The peak
    intensity is found at the position of \PSR\, which is indicated by the
    black circle.}
  \label{fig:ChandraExposureCorrected}
  \end{minipage}\hfill
\end{figure}

The Chandra X-ray data discussed in Sec.~\ref{sec:X-rayCorrelation} was
obtained from the Chandra Data Archive \cite{ChandraDataArchive}. The data was
taken by the four ACIS CCD detectors on 2000 August 14 (observation ID 754)
during a single exposure of 20\,ks. After standard processing at the Chandra
Science Center, the further analysis was carried out with the ``Chandra
Interactive Analysis of Observations'' (CIAO 3.3, \cite{Fruscione:2006}). The
list of bad pixels was provided with the data and taken into account in the
analysis. After dead-time corrections a live-time of 19039\,s with a total of
259770 events remained. A further restriction of the energy range to
0.3-10\,keV, where the detector response is reliable, reduced the data set to a
total of 222914 events. The corresponding count map is shown in
Fig.~\ref{fig:ChandraCountMap}. The exposure corrected count map of
Fig.~\ref{fig:ChandraExposureCorrected} was obtained by a bin-wise devision of
the count map by the exposure map which is shown in
Fig.~\ref{fig:ChandraExposureMap}. The exposure map for this observation period
was created following the guidelines of the Chandra analysis software. The
exposure map clearly shows the four individual ACIS CCD detectors separated by
gaps of reduced exposure. While these gaps are clearly seen in the count map,
they have been compensated in the exposure corrected map.

\fancyhead[LO]{}
\bibliography{bibliography}

\begin{thebibliography}{143}
\providecommand{\natexlab}[1]{#1}
\providecommand{\url}[1]{\texttt{#1}}
\expandafter\ifx\csname urlstyle\endcsname\relax
  \providecommand{\doi}[1]{doi: #1}\else
  \providecommand{\doi}{doi: \begingroup \urlstyle{rm}\Url}\fi

\bibitem[{Aharonian}(2004)]{Aharonian:2004}
F.~A. {Aharonian}.
\newblock \emph{{Very high energy cosmic gamma radiation: a crucial window on
  the extreme Universe}}.
\newblock River Edge, NJ: World Scientific Publishing, 2004.

\bibitem[{Aharonian} and {Atoyan}(1981)]{Aharonian:1981}
F.~A. {Aharonian} and A.~M. {Atoyan}.
\newblock {Compton scattering of relativistic electrons in compact X-ray
  sources}.
\newblock \emph{\apss}, 79:\penalty0 321--336, October 1981.

\bibitem[{Aharonian} and {Bogovalov}(2003)]{Bogovalov:2003}
F.~A. {Aharonian} and S.~V. {Bogovalov}.
\newblock {Exploring physics of rotation powered pulsars with sub-10 GeV
  imaging atmospheric Cherenkov telescopes}.
\newblock \emph{New Astronomy}, 8:\penalty0 85--103, February 2003.
\newblock \doi{10.1016/S1384-1076(02)00200-2}.

\bibitem[{Aharonian} et~al.(1997){Aharonian}, {Hofmann}, {Konopelko}, and {V{\"
  o}lk}]{Konopelko1997}
F.~A. {Aharonian}, W.~{Hofmann}, A.~K. {Konopelko}, and H.~J. {V{\" o}lk}.
\newblock {The potential of ground based arrays of imaging atmospheric
  Cherenkov telescopes. I. Determination of shower parameters}.
\newblock \emph{Astroparticle Physics}, 6:\penalty0 343--368, March 1997.

\bibitem[{Aharonian} et~al.(2004{\natexlab{a}})]{CalibrationPaper}
{H.E.S.S. collaboration,} F.~A. {Aharonian} et~al.
\newblock {Calibration of cameras of the H.E.S.S. detector}.
\newblock \emph{Astroparticle Physics}, 22:\penalty0 109--125, November
  2004{\natexlab{a}}.
\newblock \doi{10.1016/j.astropartphys.2004.06.006}.

\bibitem[{Aharonian} et~al.(2006{\natexlab{a}})]{CrabPaper}
{H.E.S.S. collaboration,} F.~A. {Aharonian} et~al.
\newblock {Observations of the Crab nebula with HESS}.
\newblock \emph{\aap}, 457:\penalty0 899--915, October 2006{\natexlab{a}}.
\newblock \doi{10.1051/0004-6361:20065351}.

\bibitem[{Aharonian} et~al.(2005{\natexlab{a}})]{HESSJ1825I}
{H.E.S.S. collaboration,} F.~A. {Aharonian} et~al.
\newblock {A possible association of the new VHE {$\gamma$}-ray source HESS
  J1825 137 with the pulsar wind nebula G 18.0 0.7}.
\newblock \emph{\aap}, 442:\penalty0 L25--L29, November 2005{\natexlab{a}}.
\newblock \doi{10.1051/0004-6361:200500180}.

\bibitem[{Aharonian} et~al.(2006{\natexlab{b}})]{HESSJ1825II}
{H.E.S.S. collaboration,} F.~A. {Aharonian} et~al.
\newblock {Energy dependent {$\gamma$}-ray morphology in the pulsar wind nebula
  HESS J1825-137}.
\newblock \emph{\aap}, 460:\penalty0 365--374, December 2006{\natexlab{b}}.
\newblock \doi{10.1051/0004-6361:20065546}.

\bibitem[{Aharonian} et~al.(2004{\natexlab{b}})]{Hess1713Nature}
{H.E.S.S. collaboration,} F.~A. {Aharonian} et~al.
\newblock {High-energy particle acceleration in the shell of a supernova
  remnant}.
\newblock \emph{Nature}, 432:\penalty0 75--77, November 2004{\natexlab{b}}.

\bibitem[{Aharonian} et~al.(2005{\natexlab{b}})]{Khelifi:2005}
{H.E.S.S. collaboration,} F.~A. {Aharonian} et~al.
\newblock {Discovery of extended VHE gamma-ray emission from the asymmetric
  pulsar wind nebula in MSH 15-52 with HESS}.
\newblock \emph{Astronomy and Astrophysics}, 435:\penalty0 L17--L20, May
  2005{\natexlab{b}}.

\bibitem[{Aharonian} et~al.(2005{\natexlab{c}})]{PlaneScanI}
{H.E.S.S. collaboration,} F.~A. {Aharonian} et~al.
\newblock {A New Population of Very High Energy Gamma-Ray Sources in the Milky
  Way}.
\newblock \emph{Science}, 307:\penalty0 1938--1942, March 2005{\natexlab{c}}.
\newblock \doi{10.1126/science.1108643}.

\bibitem[{Aharonian} et~al.(2006{\natexlab{c}})]{Tluczykont:2006}
{H.E.S.S. collaboration,} F.~A. {Aharonian} et~al.
\newblock {Discovery of very high energy {$\gamma$}-ray emission from the BL
  Lacertae object H 2356-309 with the HESS Cherenkov telescopes}.
\newblock \emph{\aap}, 455:\penalty0 461--466, August 2006{\natexlab{c}}.
\newblock \doi{10.1051/0004-6361:20054732}.

\bibitem[{Anderson} et~al.(1996){Anderson}, {Kneizys}, {Chetwynd}, {Rothman},
  {Hoke}, {Berk}, {Bernstein}, {Acharya}, {Snell}, {Mlawer}, {Clough}, {Wang},
  {Lee}, {Revercomb}, {Yokota}, {Kimball}, {Shettle}, {Abreu}, and
  {Selby}]{MODTRAN}
G.~P. {Anderson}, F.~X. {Kneizys}, J.~H. {Chetwynd}, L.~S. {Rothman}, M.~L.
  {Hoke}, A.~{Berk}, L.~S. {Bernstein}, P.~K. {Acharya}, H.~E. {Snell},
  E.~{Mlawer}, S.~A. {Clough}, J.~{Wang}, S.~{Lee}, H.~E. {Revercomb},
  T.~{Yokota}, L.~M. {Kimball}, E.~P. {Shettle}, L.~W. {Abreu}, and J.~E.
  {Selby}.
\newblock {Reviewing atmospheric radiative transfer modeling: new developments
  in high- and moderate-resolution FASCODE/FASE and MODTRAN}.
\newblock In Paul~B. Hays and Jinxue Wang, editors, \emph{{Optical
  Spectroscopic Techniques and Instrumentation for Atmospheric and Space
  Research II}}, volume 2830 of \emph{Proc. SPIE}, pages 82--93, October 1996.

\bibitem[Aye et~al.(2003)]{Weather}
M.~Aye, K. et~al.
\newblock {Atmospheric Monitoring for the H.E.S.S. Project}.
\newblock In \emph{Proceedings of the 28th International Cosmic Ray Conference,
  Tsukuba}, page 2879. Univ. Academy Press, Tokyo, 2003.

\bibitem[{Bednarek} and {Bartosik}(2003)]{Bednarek:2003}
W.~{Bednarek} and M.~{Bartosik}.
\newblock {Gamma-rays from the pulsar wind nebulae}.
\newblock \emph{Astronomy and Astrophysics}, 405:\penalty0 689--702, July 2003.
\newblock \doi{10.1051/0004-6361:20030593}.

\bibitem[{Berge}(2006)]{BergeThesis}
D.~{Berge}.
\newblock \emph{{A detailed study of the gamma-ray supernova remnant RX
  J1713.7-3946 with H.E.S.S.}}
\newblock PhD thesis, {Ruprecht-Karls-Universit\"at Heidelberg}, 2006.

\bibitem[Bernloehr et~al.(2003)]{Bernloehr:2003vd}
K.~Bernloehr et~al.
\newblock The optical system of the hess imaging atmospheric cherenkov
  telescopes. i: Layout and components of the system.
\newblock \emph{Astroparticle Physics}, 20:\penalty0 111--128, 2003.

\bibitem[{Bernl\"ohr}(2006)]{Bernloehr}
K.~{Bernl\"ohr}.
\newblock {Monte Carlo Images of Air Showers}.
\newblock \url{http://www.mpi-hd.mpg.de/hfm/~bernlohr/HESS/MC_images}, 2006.

\bibitem[{Bernlohr}(1996)]{Bernloehr:1996}
K.~{Bernlohr}.
\newblock {Low Threshold Particle Arrays}.
\newblock \emph{Space Science Reviews}, 75:\penalty0 185--197, January 1996.

\bibitem[Bernl{\"o}hr(2002)]{sim_hessarray}
K.~Bernl{\"o}hr.
\newblock {CORSIKA and sim hessarray - Simulation of the imaging atmospheric
  Cherenkov technique for the H.E.S.S.~experiment}.
\newblock \emph{unpublished, H.E.S.S.~internal note}, 2002.

\bibitem[{Bertero} and {Boccacci}(1998)]{Bertero:InverseProbs}
M.~{Bertero} and P.~{Boccacci}.
\newblock \emph{{An Introduction to Inverse Problems in Imaging}}.
\newblock Institute of Physics Publishing, Bristol, 1998.

\bibitem[{Bethe}(1953)]{Bethe:1953}
H.~A. {Bethe}.
\newblock Moli\`ere's theory of multiple scattering.
\newblock \emph{Physical Review}, pages 1256--1266, 1953.

\bibitem[Blandford and Teukolsky(1976)]{BT}
R.~Blandford and S.~A. Teukolsky.
\newblock {Arrival-time analysis for a pulsar in a binary system.}
\newblock \emph{Astrophysical Journal}, 205:\penalty0 580--591, April 1976.

\bibitem[{Blondin} et~al.(2001){Blondin}, {Chevalier}, and
  {Frierson}]{Blondin:2001}
J.~M. {Blondin}, R.~A. {Chevalier}, and D.~M. {Frierson}.
\newblock {Pulsar Wind Nebulae in Evolved Supernova Remnants}.
\newblock \emph{\apj}, 563:\penalty0 806--815, December 2001.
\newblock \doi{10.1086/324042}.

\bibitem[{Blumenthal} and {Gould}(1970)]{Blumenthal:1970}
G.~R. {Blumenthal} and R.~J. {Gould}.
\newblock {Bremsstrahlung, Synchrotron Radiation, and Compton Scattering of
  High-Energy Electrons Traversing Dilute Gases}.
\newblock \emph{Reviews of Modern Physics}, 42:\penalty0 237--271, 1970.

\bibitem[{Bogovalov} and {Aharonian}(2000)]{Bogovalov:2000}
S.~V. {Bogovalov} and F.~A. {Aharonian}.
\newblock {Very-high-energy gamma radiation associated with the unshocked wind
  of the Crab pulsar}.
\newblock \emph{\mnras}, 313:\penalty0 504--514, April 2000.

\bibitem[{Bolz}(2004{\natexlab{a}})]{BolzThesis}
O.~{Bolz}.
\newblock \emph{{Absolute Energiekalibration der abbildenden
  Cherenkov-Teleskope des H.E.S.S. Experiments und Ergebnisse erster
  Beobachtungen des Supernova-\"Uberrests RX J1713.7 $-$3946}}.
\newblock PhD thesis, {Ruprecht-Karls-Universit\"at Heidelberg},
  2004{\natexlab{a}}.

\bibitem[{Bolz}(2004{\natexlab{b}})]{MPIK:MuonEfficiency}
O.~{Bolz}.
\newblock Muon analysis update.
\newblock Weekly H.E.S.S. technical meetings at MPIK, November
  2004{\natexlab{b}}.

\bibitem[{Borg\-mei\-er} et~al.(2003){Borg\-mei\-er}, {Komin}, {de Naurois},
  {Schlenker}, {Schwanke}, and {Stegmann}]{DAQ_ICRC28}
C.~{Borg\-mei\-er}, Nu. {Komin}, M.~{de Naurois}, S.~{Schlenker},
  U.~{Schwanke}, and C.~for the H.E.S.S.~collaboration {Stegmann}.
\newblock {The Central Data Acquisition System of the H.E.S.S. Telescope
  System}.
\newblock In \emph{Proceedings of the 28th International Cosmic Ray Conference,
  Tsukuba}, page 2891. Univ. Academy Press, Tokyo, 2003.

\bibitem[Borgmeier et~al.(2001)Borgmeier, Mauritz, and Stegmann]{DAQ_ICRC27}
C.~Borgmeier, K.~Mauritz, and C.~Stegmann.
\newblock {The central data acquisition system for the H.E.S.S. telescope
  system}.
\newblock In \emph{Proceedings of the 28th International Cosmic Ray Conference,
  Hamburg}, page 2896. Copernicus Gesellschaft, 2001.

\bibitem[Breitling et~al.(2004)Breitling, Gillessen, and Konopelko]{Breitling}
F.~Breitling, S.~Gillessen, and A.~Konopelko.
\newblock Search for pulsed gamma-ray emission from binary systems with
  h.e.s.s.
\newblock H.E.S.S. internal note, June 2004.

\bibitem[Brown et~al.(2005)]{Weather2}
A.~M. Brown et~al.
\newblock {Atmospheric monitoring for the H.E.S.S. Cherenkov telescope array by
  transmissometer and LIDAR}.
\newblock In \emph{Proceedings of the 29th International Cosmic Ray Conference,
  Pune}, 2005.

\bibitem[{Brun} et~al.(2006){Brun}, {Rademakers}, {Buncic}, {Fine}, {Canal},
  and {Panacek}]{ROOT}
R.~{Brun}, F.~{Rademakers}, N.~{Buncic}, V.~{Fine}, P.~{Canal}, and
  S.~{Panacek}.
\newblock {ROOT, An Object-Oriented Data Analysis Framework}.
\newblock \url{http://root.cern.ch}, 2006.

\bibitem[{Bruno Kh\'elifi} and {Conor Masterson}(2005)]{MPIK:MuonCorrections}
{Bruno Kh\'elifi} and {Conor Masterson}.
\newblock Efficiency corrections using muon rings.
\newblock Weekly H.E.S.S. technical meetings at MPIK, September 2005.

\bibitem[{Caraveo} et~al.(1994){Caraveo}, {Mereghetti}, and
  {Bignami}]{Caraveo:1994}
P.~A. {Caraveo}, S.~{Mereghetti}, and G.~F. {Bignami}.
\newblock {An Optical Counterpart for PSR 1509-58}.
\newblock \emph{\apjl}, 423:\penalty0 L125+, March 1994.
\newblock \doi{10.1086/187252}.

\bibitem[{Caswell} et~al.(1981){Caswell}, {Milne}, and
  {Wellington}]{Caswell:1981}
J.~L. {Caswell}, D.~K. {Milne}, and K.~J. {Wellington}.
\newblock {High-resolution radio observations of five supernova remnants}.
\newblock \emph{\mnras}, 195:\penalty0 89--99, April 1981.

\bibitem[Cornils et~al.(2003)]{Cornils:2003ve}
R.~Cornils et~al.
\newblock The optical system of the hess imaging atmospheric cherenkov
  telescopes. ii: Mirror alignment and point spread function.
\newblock \emph{Astroparticle Physics}, 20:\penalty0 129--143, 2003.

\bibitem[{Davies} and {Cotton}(1957)]{DaviesCotton}
J.M. {Davies} and E.S. {Cotton}.
\newblock {Design of the quartermaster solar furnace}.
\newblock \emph{Solar Energy}, 1\penalty0 (2-3):\penalty0 16, 1957.

\bibitem[{de Jager}(1994)]{DeJager1994}
O.~C. {de Jager}.
\newblock {On periodicity tests and flux limit calculations for gamma-ray
  pulsars}.
\newblock \emph{Astrophysical Journal}, 436:\penalty0 239--248, November 1994.
\newblock \doi{10.1086/174896}.

\bibitem[{de Jager}(2006{\natexlab{a}})]{Okkie:2006}
O.~C. {de Jager}.
\newblock {Pulsar wind nebulae and beyond: Multi-wavelength observations and
  theory}.
\newblock In W.~{Becker}, editor, \emph{Proceedings of the 363. WE-Heraeus
  Seminar on: Neutron Stars and Pulsars (Posters and contributed talks), 14-19
  May, 2006, Physikzentrum Bad Honnef, Germany}, 2006{\natexlab{a}}.
\newblock to appear.

\bibitem[{de Jager}(2006{\natexlab{b}})]{Okkie:2006IAU}
O.~C. {de Jager}.
\newblock {Gamma-ray and TeV Emission Properties of Pulsars and Pulsar Wind
  Nebulae}.
\newblock \emph{On the Present and Future of Pulsar Astronomy, 26th meeting of
  the IAU, Joint Discussion 2, 16-17 August, 2006, Prague, Czech Republic,
  JD02, \#53}, 2, August 2006{\natexlab{b}}.

\bibitem[{de Jager}(2006{\natexlab{c}})]{Okkie:PWNTransport}
O.~C. {de Jager}.
\newblock Diffusion vs advection in magnetized flow of a pulsar wind nebula.
\newblock H.E.S.S. internal note, February 2006{\natexlab{c}}.

\bibitem[{de Jager} et~al.(1989){de Jager}, {Raubenheimer}, and
  {Swanepoel}]{DeJager1989}
O.~C. {de Jager}, B.~C. {Raubenheimer}, and J.~W.~H. {Swanepoel}.
\newblock {A poweful test for weak periodic signals with unknown light curve
  shape in sparse data}.
\newblock \emph{Astronomy and Astrophysics}, 221:\penalty0 180--190, August
  1989.

\bibitem[{DeLaney} et~al.(2006){DeLaney}, {Gaensler}, {Arons}, and
  {Pivovaroff}]{DeLaney:2006}
T.~{DeLaney}, B.~M. {Gaensler}, J.~{Arons}, and M.~J. {Pivovaroff}.
\newblock {Time Variability in the X-Ray Nebula Powered by Pulsar B1509-58}.
\newblock \emph{\apj}, 640:\penalty0 929--940, April 2006.
\newblock \doi{10.1086/500189}.

\bibitem[{Dodson} et~al.(2003){Dodson}, {Legge}, {Reynolds}, and
  {McCulloch}]{Dodson:2003}
R.~{Dodson}, D.~{Legge}, J.~E. {Reynolds}, and P.~M. {McCulloch}.
\newblock {The Vela Pulsar's Proper Motion and Parallax Derived from VLBI
  Observations}.
\newblock \emph{\apj}, 596:\penalty0 1137--1141, October 2003.
\newblock \doi{10.1086/378089}.

\bibitem[{Du Plessis} et~al.(1995){Du Plessis}, {de Jager}, {Buchner}, {Nel},
  {North}, {Raubenheimer}, and {van der Walt}]{DuPlessis:1995}
I.~{Du Plessis}, O.~C. {de Jager}, S.~{Buchner}, H.~I. {Nel}, A.~R. {North},
  B.~C. {Raubenheimer}, and D.~J. {van der Walt}.
\newblock {The Nonthermal Radio, X-Ray, and TeV Gamma-Ray Spectra of MSH
  15-52}.
\newblock \emph{\apj}, 453:\penalty0 746--+, November 1995.
\newblock \doi{10.1086/176436}.

\bibitem[{Dubner} et~al.(2002){Dubner}, {Gaensler}, {Giacani}, {Goss}, and
  {Green}]{Dubner:2002}
G.~M. {Dubner}, B.~M. {Gaensler}, E.~B. {Giacani}, W.~M. {Goss}, and A.~J.
  {Green}.
\newblock {The Interstellar Medium around the Supernova Remnant G320.4-1.2}.
\newblock \emph{\aj}, 123:\penalty0 337--345, January 2002.
\newblock \doi{10.1086/324736}.

\bibitem[{Duncan} et~al.(1996){Duncan}, {Stewart}, {Haynes}, and
  {Jones}]{Duncan:1996}
A.~R. {Duncan}, R.~T. {Stewart}, R.~F. {Haynes}, and K.~L. {Jones}.
\newblock {The VELA supernova remnant and the GUM nebula: new perspectives at
  2.4 GHz}.
\newblock \emph{\mnras}, 280:\penalty0 252--266, May 1996.

\bibitem[{Eifert}(2005)]{Eifert}
T.~{Eifert}.
\newblock {Search for Pulsed Very High Energy Gamma-Ray Emission from the
  Millisecond Pulsar PSR J0437-4715 with H.E.S.S.}
\newblock Diplomarbeit, Humboldt-Universit\"at zu Berlin, 2005.

\bibitem[{Ergin}(2005)]{ErginThesis}
T.~{Ergin}.
\newblock \emph{{The Energy Spectrum of Very High Energy Gamma Rays from the
  Crab Nebula as measured by the H.E.S.S. Array}}.
\newblock PhD thesis, Humboldt-Universit\"at zu Berlin, 2005.

\bibitem[{Feldman} and {Cousins}(1998)]{FeldmanCousins}
G.~J. {Feldman} and R.~D. {Cousins}.
\newblock {Unified approach to the classical statistical analysis of small
  signals}.
\newblock \emph{Physical Review~D}, 57:\penalty0 3873--3889, April 1998.

\bibitem[Fisher(1925)]{FishersZ}
R.~A. Fisher.
\newblock \emph{Statistical Methods for Research Workers.}
\newblock OliverAndBoyd, - 1925.

\bibitem[{Forot} et~al.(2006){Forot}, {Hermsen}, {Renaud}, {Laurent},
  {Grenier}, {Goret}, {Khelifi}, and {Kuiper}]{Goret:2006}
M.~{Forot}, W.~{Hermsen}, M.~{Renaud}, P.~{Laurent}, I.~{Grenier}, P.~{Goret},
  B.~{Khelifi}, and L.~{Kuiper}.
\newblock {High-Energy Particles in the Wind Nebula of Pulsar B1509-58 as Seen
  by INTEGRAL}.
\newblock \emph{\apjl}, 651:\penalty0 L45--L48, November 2006.
\newblock \doi{10.1086/509077}.

\bibitem[{Fruscione} et~al.(2006){Fruscione}, {McDowell}, {Allen},
  {Brickhouse}, {Burke}, {Davis}, {Durham}, {Elvis}, {Galle}, {Harris},
  {Huenemoerder}, {Houck}, {Ishibashi}, {Karovska}, {Nicastro}, {Noble},
  {Nowak}, {Primini}, {Siemiginowska}, {Smith}, and {Wise}]{Fruscione:2006}
A.~{Fruscione}, J.~C. {McDowell}, G.~E. {Allen}, N.~S. {Brickhouse}, D.~J.
  {Burke}, J.~E. {Davis}, N.~{Durham}, M.~{Elvis}, E.~C. {Galle}, D.~E.
  {Harris}, D.~P. {Huenemoerder}, J.~C. {Houck}, B.~{Ishibashi}, M.~{Karovska},
  F.~{Nicastro}, M.~S. {Noble}, M.~A. {Nowak}, F.~A. {Primini},
  A.~{Siemiginowska}, R.~K. {Smith}, and M.~{Wise}.
\newblock {CIAO: Chandra's data analysis system}.
\newblock In \emph{Observatory Operations: Strategies, Processes, and Systems.
  Edited by Silva, David R.; Doxsey, Rodger E.. Proceedings of the SPIE, Volume
  6270, pp. 62701V (2006).}, July 2006.
\newblock \doi{10.1117/12.671760}.
\newblock URL \url{http://cxc.harvard.edu/ciao/}.

\bibitem[{Funk} et~al.(2004){Funk}, {Hermann}, {Hinton}, {Berge},
  {Bernl{\"o}hr}, {Hofmann}, {Nayman}, {Toussenel}, and
  {Vincent}]{CentralTrigger}
S.~{Funk}, G.~{Hermann}, J.~{Hinton}, D.~{Berge}, K.~{Bernl{\"o}hr},
  W.~{Hofmann}, P.~{Nayman}, F.~{Toussenel}, and P.~{Vincent}.
\newblock {The trigger system of the H.E.S.S. telescope array}.
\newblock \emph{Astroparticle Physics}, 22:\penalty0 285--296, November 2004.
\newblock \doi{10.1016/j.astropartphys.2004.08.001}.

\bibitem[{Gaensler} and {Slane}(2006)]{Gaensler:2006}
B.~M. {Gaensler} and P.~O. {Slane}.
\newblock {The Evolution and Structure of Pulsar Wind Nebulae}.
\newblock \emph{\araa}, 44:\penalty0 17--47, September 2006.
\newblock \doi{10.1146/annurev.astro.44.051905.092528PDF:
  http://arjournals.annualreviews.org/doi/pdf/10.1146/annurev.astro.44.051905.%
092528}.

\bibitem[{Gaensler} et~al.(1999){Gaensler}, {Brazier}, {Manchester},
  {Johnston}, and {Green}]{Gaensler:1999}
B.~M. {Gaensler}, K.~T.~S. {Brazier}, R.~N. {Manchester}, S.~{Johnston}, and
  A.~J. {Green}.
\newblock {SNR G320.4-01.2 and PSR B1509-58: new radio observations of a
  complex interacting system}.
\newblock \emph{\mnras}, 305:\penalty0 724--736, May 1999.

\bibitem[{Gaensler} et~al.(2002){Gaensler}, {Arons}, {Pivovaroff}, and
  {Kaspi}]{Gaensler:2002}
B.~M. {Gaensler}, J.~{Arons}, M.~J. {Pivovaroff}, and V.~M. {Kaspi}.
\newblock {Chandra Observations of Pulsar B1509-58 and Supernova Remnant
  G320.4-1.2}.
\newblock In P.~O. {Slane} and B.~M. {Gaensler}, editors, \emph{ASP Conf. Ser.
  271: Neutron Stars in Supernova Remnants}, pages 175--+, 2002.

\bibitem[{Gillessen}(2004)]{GillessenThesis}
S.~{Gillessen}.
\newblock \emph{{Sub-Bogenminuten-genaue Positionen von TeV-Quellen mit
  H.E.S.S.}}
\newblock PhD thesis, {Ruprecht-Karls-Universit\"at Heidelberg}, 2004.

\bibitem[{Ginzburg} and {Syrovatskii}(1965)]{Ginzburg:1965}
V.~L. {Ginzburg} and S.~I. {Syrovatskii}.
\newblock {Cosmic Magnetobremsstrahlung (synchrotron Radiation)}.
\newblock \emph{\araa}, 3:\penalty0 297--+, 1965.
\newblock \doi{10.1146/annurev.aa.03.090165.001501}.

\bibitem[{Giuliani}(2006)]{Magnetosphere}
A.~{Giuliani}.
\newblock {Il cielo visto da EGRET}.
\newblock
  \url{http://www.iasf-milano.inaf.it/~giuliani/public/tesi/node9.html}, 2006.

\bibitem[Greisen(1965)]{NKG}
K.~Greisen.
\newblock \emph{Prog. Cosmic Ray Physics}, volume vol. 3.
\newblock 1North Holland Publishing Co., 1965.

\bibitem[{Harding} et~al.(1997){Harding}, {Baring}, and
  {Gonthier}]{Harding:1997}
A.~K. {Harding}, M.~G. {Baring}, and P.~L. {Gonthier}.
\newblock {Photon-splitting Cascades in Gamma-Ray Pulsars and the Spectrum of
  PSR 1509-58}.
\newblock \emph{\apj}, 476:\penalty0 246--+, February 1997.
\newblock \doi{10.1086/303605}.

\bibitem[{Heck} et~al.(1998){Heck}, {Knapp}, {Capdevielle}, {Schatz}, and
  {Thouw}]{CORSIKA}
D.~{Heck}, J.~{Knapp}, J.N. {Capdevielle}, G.~{Schatz}, and T.~{Thouw}.
\newblock {CORSIKA: A Monte Carlo Code to Simulate Extensive Air Showers}.
\newblock Technical Report FZKA 6019, Forschungszentrum Karlsruhe Report, 1998.

\bibitem[{HEGRA collaboration, F. A. Aharonian et al.}(1999)]{HEGRA:1999}
{HEGRA collaboration, F. A. Aharonian et al.}
\newblock {The time averaged TeV energy spectrum of MKN 501 of the
  extraordinary 1997 outburst as measured with the stereoscopic Cherenkov
  telescope system of HEGRA}.
\newblock \emph{Astronomy and Astrophysics}, 349:\penalty0 11--28, September
  1999.

\bibitem[{HEGRA collaboration, F. A. Aharonian et al.}(2004)]{HEGRA:2004Crab}
{HEGRA collaboration, F. A. Aharonian et al.}
\newblock {The Crab Nebula and Pulsar between 500 GeV and 80 TeV: Observations
  with the HEGRA Stereoscopic Air Cerenkov Telescopes}.
\newblock \emph{Astrophysical Journal}, 614:\penalty0 897--913, October 2004.

\bibitem[{Heitler}(1954)]{Heitler}
W.~{Heitler}.
\newblock \emph{Quantum Theory of Radiation}.
\newblock Dover Press, 3rd edition, 1954.

\bibitem[{H.E.S.S. collaboration}(2001)]{H.E.S.S.Software}
{H.E.S.S. collaboration}.
\newblock {H.E.S.S. Software}.
\newblock \url{http://www-eep.physik.hu-berlin.de/hess/}, 2001.

\bibitem[{Hester} et~al.(2002){Hester}, {Mori}, {Burrows}, {Gallagher},
  {Graham}, {Halverson}, {Kader}, {Michel}, and {Scowen}]{Hester:2002}
J.~J. {Hester}, K.~{Mori}, D.~{Burrows}, J.~S. {Gallagher}, J.~R. {Graham},
  M.~{Halverson}, A.~{Kader}, F.~C. {Michel}, and P.~{Scowen}.
\newblock {Hubble Space Telescope and Chandra Monitoring of the Crab
  Synchrotron Nebula}.
\newblock \emph{\apjl}, 577:\penalty0 L49--L52, September 2002.
\newblock \doi{10.1086/344132}.

\bibitem[{Hillas}(1985)]{Hillas}
A.~M. {Hillas}.
\newblock {Cerenkov light images of EAS produced by primary gamma}.
\newblock In \emph{Proceedings of the 19th International Cosmic Ray
  Conference}, pages 445--448, August 1985.

\bibitem[{Hillas} et~al.(1998){Hillas}, {Akerlof}, {Biller}, {Buckley},
  {Carter-Lewis}, {Catanese}, {Cawley}, {Fegan}, {Finley}, {Gaidos},
  {Krennrich}, {Lamb}, {Lang}, {Mohanty}, {Punch}, {Reynolds}, {Rodgers},
  {Rose}, {Rovero}, {Schubnell}, {Sembroski}, {Vacanti}, {Weekes}, {West}, and
  {Zweerink}]{CrabWhipple}
A.~M. {Hillas}, C.~W. {Akerlof}, S.~D. {Biller}, J.~H. {Buckley}, D.~A.
  {Carter-Lewis}, M.~{Catanese}, M.~F. {Cawley}, D.~J. {Fegan}, J.~P. {Finley},
  J.~A. {Gaidos}, F.~{Krennrich}, R.~C. {Lamb}, M.~J. {Lang}, G.~{Mohanty},
  M.~{Punch}, P.~T. {Reynolds}, A.~J. {Rodgers}, H.~J. {Rose}, A.~C. {Rovero},
  M.~S. {Schubnell}, G.~H. {Sembroski}, G.~{Vacanti}, T.~C. {Weekes},
  M.~{West}, and J.~{Zweerink}.
\newblock {The Spectrum of TeV Gamma Rays from the Crab Nebula}.
\newblock \emph{Astrophysical Journal}, 503:\penalty0 744--+, August 1998.

\bibitem[{Hinton} et~al.(2006){Hinton}, {Hermann}, {Kr{\"o}tz}, and
  {Funk}]{Hinton:OpticalCrab}
J.~{Hinton}, G.~{Hermann}, P.~{Kr{\"o}tz}, and S.~{Funk}.
\newblock {Precision measurement of optical pulsation using a Cherenkov
  telescope}.
\newblock \emph{Astroparticle Physics}, 26:\penalty0 22--27, August 2006.
\newblock \doi{10.1016/j.astropartphys.2006.04.008}.

\bibitem[{Hofmann}(2005)]{hofmann_ICRC2005}
W.~{Hofmann}.
\newblock {H.E.S.S. Highlights}.
\newblock In \emph{Proceedings of the 29th International Comsic Ray Conference,
  Pune, India}, volume~10, pages 97--114, 2005.

\bibitem[{Hofmann} et~al.(1999){Hofmann}, {Jung}, {Konopelko}, {Krawczynski},
  {Lampeitl}, and {P{\" u}hlhofer}]{Hofmann1999}
W.~{Hofmann}, I.~{Jung}, A.~{Konopelko}, H.~{Krawczynski}, H.~{Lampeitl}, and
  G.~{P{\" u}hlhofer}.
\newblock {Comparison of techniques to reconstruct VHE gamma-ray showers from
  multiple stereoscopic Cherenkov images}.
\newblock \emph{Astroparticle Physics}, 12:\penalty0 135--143, November 1999.

\bibitem[{Horns} et~al.(2006){Horns}, {Aharonian}, {Santangelo}, {Hoffmann},
  and {Masterson}]{Horns:2006}
D.~{Horns}, F.~{Aharonian}, A.~{Santangelo}, A.~I.~D. {Hoffmann}, and
  C.~{Masterson}.
\newblock {Nucleonic gamma-ray production in <ASTROBJ>Vela X</ASTROBJ>}.
\newblock \emph{Astronomy and Astrophysics}, 451:\penalty0 L51--L54, June 2006.
\newblock \doi{10.1051/0004-6361:20065116}.

\bibitem[Kanbach et~al.(2005)Kanbach, Slowikowska, Kellner, and
  Steinle]{Kanbach:2005kf}
G.~Kanbach, Agnieszka Slowikowska, S.~Kellner, and H.~Steinle.
\newblock New optical polarization measurements of the crab pulsar.
\newblock \emph{AIP Conf. Proc.}, 801:\penalty0 306--311, 2005.

\bibitem[{Kaplan} and {Moon}(2006)]{Kaplan:2006}
D.~L. {Kaplan} and D.-S. {Moon}.
\newblock {A Near-Infrared Search for Counterparts to Three Pulsars in Young
  Supernova Remnants}.
\newblock \emph{\apj}, 644:\penalty0 1056--1062, June 2006.
\newblock \doi{10.1086/503794}.

\bibitem[{Kaspi} et~al.(1994){Kaspi}, {Manchester}, {Siegman}, {Johnston}, and
  {Lyne}]{Kaspi:1994}
V.~M. {Kaspi}, R.~N. {Manchester}, B.~{Siegman}, S.~{Johnston}, and A.~G.
  {Lyne}.
\newblock {On the spin-down of PSR B1509-58}.
\newblock \emph{\apjl}, 422:\penalty0 L83--L86, February 1994.
\newblock \doi{10.1086/187218}.

\bibitem[{Kelner} et~al.(2006){Kelner}, {Aharonian}, and
  {Bugayov}]{Kelner:2006}
S.~R. {Kelner}, F.~A. {Aharonian}, and V.~V. {Bugayov}.
\newblock {Energy spectra of gamma rays, electrons, and neutrinos produced at
  proton-proton interactions in the very high energy regime}.
\newblock \emph{Physical Review~D}, 74\penalty0 (3):\penalty0 034018--+, August
  2006.
\newblock \doi{10.1103/PhysRevD.74.034018}.

\bibitem[{Kennel} and {Coroniti}(1984)]{Kennel:1984}
C.~F. {Kennel} and F.~V. {Coroniti}.
\newblock {Confinement of the Crab pulsar's wind by its supernova remnant}.
\newblock \emph{\apj}, 283:\penalty0 694--709, August 1984.
\newblock \doi{10.1086/162356}.

\bibitem[{{Kh\'elifi}, B. et al, (H.E.S.S.
  collaboration)}(2005)]{Khelifi:2005ICRC}
{{Kh\'elifi}, B. et al, (H.E.S.S. collaboration)}.
\newblock {VHE observations of pulsar wind nebulae with H.E.S.S.}
\newblock In \emph{Proceedings of the 29th International Cosmic Ray
  Conference}, volume~4, pages 127--130, 2005.

\bibitem[Klages et~al.(2005)]{hd_manual}
S.~Klages et~al.
\newblock {Brief overview of the Heidelberg H.E.S.S. analysis}.
\newblock \url{http://www.mpi-hd.mpg.de/hfm/HESS/intern/}, 2005.

\bibitem[{Kuiper} et~al.(1999){Kuiper}, {Hermsen}, {Krijger}, {Bennett},
  {Carrami{\~n}ana}, {Sch{\"o}nfelder}, {Bailes}, and
  {Manchester}]{Kuiper:1999}
L.~{Kuiper}, W.~{Hermsen}, J.~M. {Krijger}, K.~{Bennett}, A.~{Carrami{\~n}ana},
  V.~{Sch{\"o}nfelder}, M.~{Bailes}, and R.~N. {Manchester}.
\newblock {COMPTEL detection of pulsed gamma -ray emission from PSR B1509-58 up
  to at least 10 MeV}.
\newblock \emph{\aap}, 351:\penalty0 119--132, November 1999.

\bibitem[Landsman(2004)]{IDLAstroLib}
W.~Landsman.
\newblock Idl astronomy library.
\newblock \url{http://idlastro.gsfc.nasa.gov}, 2004.

\bibitem[{Le Gallou} et~al.(2005){Le Gallou}, {Nolan}, {Masterson}, and
  {Spangler}]{LIDAR}
R.~{Le Gallou}, S.J. {Nolan}, C.~{Masterson}, and D.~{Spangler}.
\newblock {Atmospheric quality and H.E.S.S.\ array performance: study using the
  ceilometer and correction method for $\gamma$-ray data}.
\newblock H.E.S.S. internal note, May 2005.

\bibitem[{Leroy} et~al.(2003){Leroy}, {Bolz}, {Guy}, {Jung}, {Redondo},
  {Rolland}, {Tavernet}, {Aye}, {Berghaus}, {Bernl\"ohr}, {Chadwick},
  {Chitnis}, {de Naurois}, {Djannati-Ata\"i}, {Espigat}, {Hermann}, {Hinton},
  {Kh{\' e}lifi}, {Kohnle}, {LeGallou}, {Masterson}, {Pita}, {Saitoh}, {Th{\'
  e}oret}, and {Vincent}]{MuonsICRC}
N.~{Leroy}, O.~{Bolz}, J.~{Guy}, I.~{Jung}, I.~{Redondo}, L.~{Rolland}, J.-P.
  {Tavernet}, K.-M. {Aye}, P.~{Berghaus}, K.~{Bernl\"ohr}, P.M. {Chadwick},
  V.~{Chitnis}, M.~{de Naurois}, A.~{Djannati-Ata\"i}, P.~{Espigat},
  G.~{Hermann}, J.~{Hinton}, B.~{Kh{\' e}lifi}, A.~{Kohnle}, R.~{LeGallou},
  C.~{Masterson}, S.~{Pita}, T.~{Saitoh}, C.~{Th{\' e}oret}, and P.~for the
  H.E.S.S.\~collaboration {Vincent}.
\newblock {Calibration results for the first two H.E.S.S.\ array telescopes.}
\newblock In \emph{Proceedings of the 28th International Cosmic Ray Conference,
  Tsukuba}, page 2895. Univ. Academy Press, Tokyo, 2003.

\bibitem[{Li} and {Ma}(1983)]{LiMa}
T.-P. {Li} and Y.-Q. {Ma}.
\newblock {Analysis methods for results in gamma-ray astronomy}.
\newblock \emph{\apj}, 272:\penalty0 317--324, September 1983.
\newblock \doi{10.1086/161295}.

\bibitem[{Lo} and {Pope}(1998)]{omniORB}
S.~{Lo} and S.~{Pope}.
\newblock {The Implementation of a High Performance ORB over Multiple Network
  Transports}.
\newblock In \emph{Proc. Middleware '98}, pages 157--172. The Lake District,
  England, ISBN 1-885233-088-0, 1998.
\newblock URL \url{http://omniorb.sourceforge.net}.

\bibitem[{Longair}(1992)]{Longair1}
M.~S. {Longair}.
\newblock \emph{{High energy astrophysics. Vol.1: Particles, photons and their
  detection}}.
\newblock Cambridge, UK: Cambridge University Press, 2nd edition, 1992.

\bibitem[{Longair}(1994)]{Longair2}
M.~S. {Longair}.
\newblock \emph{{High energy astrophysics. Vol.2: Stars, the Galaxy and the
  interstellar medium}}.
\newblock Cambridge, UK: Cambridge University Press, 2nd edition, 1994.

\bibitem[{Lucy}(1974)]{Lucy:1974}
L.~{Lucy}.
\newblock {An iteration technique for the rectification of observed
  distributions}.
\newblock \emph{Astronomical Journal}, 79:\penalty0 745, 1974.

\bibitem[{Lyne} and {Graham-Smith}(1998)]{Lyne:PulsarAstronomy}
A.~G. {Lyne} and F.~{Graham-Smith}.
\newblock \emph{{Pulsar astronomy}}.
\newblock Pulsar astronomy / Andrew G.~Lyne and Francis Graham-Smith.~
  Cambridge, U.K.; New York : Cambridge University Press, 1998.~(Cambridge
  astrophysics series ; 31) ISBN 0521594138, 1998.

\bibitem[{Manchester} et~al.(1982){Manchester}, {Tuohy}, and
  {Damico}]{Manchester:1982}
R.~N. {Manchester}, I.~R. {Tuohy}, and N.~{Damico}.
\newblock {Discovery of radio pulsations from the X-ray pulsar in the supernova
  remnant G320.4-1.2}.
\newblock \emph{\apjl}, 262:\penalty0 L31--L33, November 1982.
\newblock \doi{10.1086/183906}.

\bibitem[{Manchester} et~al.(2006)]{ATNF}
R.~N. {Manchester} et~al.
\newblock Australian pulsar timing data archive.
\newblock \url{http://www.atnf.csiro.au/people/pulsar/archive/}, 2006.

\bibitem[{Martini} et~al.(2004){Martini}, {Persson}, {Murphy}, {Birk},
  {Shectman}, {Gunnels}, and {Koch}]{Martini:2004}
P.~{Martini}, S.~E. {Persson}, D.~C. {Murphy}, C.~{Birk}, S.~A. {Shectman},
  S.~M. {Gunnels}, and E.~{Koch}.
\newblock {PANIC: a near-infrared camera for the Magellan telescopes}.
\newblock In A.~F.~M. {Moorwood} and M.~{Iye}, editors, \emph{Ground-based
  Instrumentation for Astronomy. Edited by Alan F. M. Moorwood and Iye
  Masanori. Proceedings of the SPIE, Volume 5492, pp. 1653-1660 (2004).}, pages
  1653--1660, September 2004.
\newblock \doi{10.1117/12.551828}.

\bibitem[{Masterson} and {The CAT Collaboration}(2001)]{CrabCAT}
C.~{Masterson} and {The CAT Collaboration}.
\newblock {Observations of the Crab Nebula with the CAT Imaging Atmospheric {\v
  C}erenkov Telescope}.
\newblock In F.~A. {Aharonian} and H.~J. {V{\"o}lk}, editors, \emph{Proc. of
  High Energy Gamma-Ray AStronomy 2000}, pages 753--756. American Institute of
  Physics Conference Series, 2001.

\bibitem[{Masterson} and {H.~E.~S.~S.~Collaboration}(2003)]{Masterson:2003ICRC}
C.~P. {Masterson} and {H.~E.~S.~S.~Collaboration}.
\newblock {Optical Observations of the Crab Pulsar Using the First H.E.S.S.
  Cherenkov Telescope}.
\newblock In \emph{Proceedings of the 28th International Cosmic Ray
  Conference}, pages 2987--+, July 2003.

\bibitem[{Mills}(1981)]{Mills:1981}
B.~Y. {Mills}.
\newblock {The Molonglo Observatory synthesis telescope}.
\newblock \emph{Proceedings of the Astronomical Society of Australia},
  4:\penalty0 156--159, 1981.

\bibitem[{Mills} et~al.(1961){Mills}, {Slee}, and {Hill}]{Mills:1961}
B.~Y. {Mills}, O.~B. {Slee}, and E.~R. {Hill}.
\newblock {A Catalogue of Radio Sources between Declinations -50$\deg$ and
  -80$\deg$}.
\newblock \emph{Australian Journal of Physics}, 14:\penalty0 497--+, 1961.

\bibitem[{Mineo} et~al.(2001){Mineo}, {Cusumano}, {Maccarone}, {Massaglia},
  {Massaro}, and {Trussoni}]{Mineo:2001}
T.~{Mineo}, G.~{Cusumano}, M.~C. {Maccarone}, S.~{Massaglia}, E.~{Massaro}, and
  E.~{Trussoni}.
\newblock {The hard X-ray emission from the complex SNR MSH 15-52 observed by
  BeppoSAX}.
\newblock \emph{\aap}, 380:\penalty0 695--703, December 2001.
\newblock \doi{10.1051/0004-6361:20011576}.

\bibitem[{NASA}(2006{\natexlab{a}})]{EGRET:3rdCatalog}
{NASA}.
\newblock {CGRO SSC >> Surveying the Universe with EGRET 1991-1996}.
\newblock
  \url{http://cossc.gsfc.nasa.gov/docs/cgro/cossc/egret/3rd_EGRET_Cat.html},
  2006{\natexlab{a}}.

\bibitem[{NASA}(2006{\natexlab{b}})]{NASA:PulsarLightCurves}
{NASA}.
\newblock {Light Curves of the Crab Pular, PSR B1509 and the Vela Pulsar}.
\newblock \url{http://imagine.gsfc.nasa.gov/docs/science/know_l2/pulsars.html},
  2006{\natexlab{b}}.

\bibitem[{NASA/CXC}(2006)]{ChandraDataArchive}
{NASA/CXC}.
\newblock The chandra data archive.
\newblock \url{http://cxc.harvard.edu/cda/}, 2006.

\bibitem[{NASA/CXC/MIT/B.Gaensler et al.}(2006)]{B1509_scale}
{NASA/CXC/MIT/B.Gaensler et al.}
\newblock More images of b1509-58 in snr g320.4-1.2.
\newblock \url{http://chandra.harvard.edu/photo/2001/1175/more.html}, May 2006.

\bibitem[{NASA/HEASARC}(2006)]{NeutronStar}
{NASA/HEASARC}.
\newblock Internal structure of a neutron star.
\newblock
  \url{http://heasarc.gsfc.nasa.gov/docs/objects/binaries/neutron_star_structu%
re.html}, 2006.

\bibitem[{Parker} et~al.(2005){Parker}, {Phillipps}, {Pierce}, {Hartley},
  {Hambly}, {Read}, {MacGillivray}, {Tritton}, {Cass}, {Cannon}, {Cohen},
  {Drew}, {Frew}, {Hopewell}, {Mader}, {Malin}, {Masheder}, {Morgan}, {Morris},
  {Russeil}, {Russell}, and {Walker}]{COSMOS}
Q.~A. {Parker}, S.~{Phillipps}, M.~J. {Pierce}, M.~{Hartley}, N.~C. {Hambly},
  M.~A. {Read}, H.~T. {MacGillivray}, S.~B. {Tritton}, C.~P. {Cass}, R.~D.
  {Cannon}, M.~{Cohen}, J.~E. {Drew}, D.~J. {Frew}, E.~{Hopewell}, S.~{Mader},
  D.~F. {Malin}, M.~R.~W. {Masheder}, D.~H. {Morgan}, R.~A.~H. {Morris},
  D.~{Russeil}, K.~S. {Russell}, and R.~N.~F. {Walker}.
\newblock {The AAO/UKST SuperCOSMOS H{$\alpha$} survey}.
\newblock \emph{\mnras}, 362:\penalty0 689--710, September 2005.
\newblock \doi{10.1111/j.1365-2966.2005.09350.x}.

\bibitem[Puehlhofer(2004)]{Puehlhofer}
G.~Puehlhofer.
\newblock Background estimation, excess counts and significance determination
  when using a relative acceptance correction.
\newblock H.E.S.S. internal note, March 2004.

\bibitem[{P\"uhlhofer}(2001)]{PuehlhoferThesis}
G.~{P\"uhlhofer}.
\newblock \emph{{TeV-$\gamma$-Emission des Supernova-\"Uberrestes Cassiopia A:
  Erster Nachweis mit dem HEGRA-Cherenkov-Teleskop-System}}.
\newblock PhD thesis, {Ruprecht-Karls-Universit\"at Heidelberg}, 2001.

\bibitem[{Punch}(2005)]{Punch:2005}
M.~{Punch}.
\newblock {Review of Ground-Based Cherenkov Telescopes}.
\newblock In \emph{Proceedings of Very High Energy Phenomena in the Universe,
  Moriond, Italy}, March 2005.
\newblock URL \url{http://moriond.in2p3.fr/J05/schedule.html}.

\bibitem[{Research Systems, Inc. (RSI)}(2006)]{IDL}
{Research Systems, Inc. (RSI)}.
\newblock {IDL The Data Visualization \& Analysis Platform}, 2006.
\newblock URL \url{http://www.ittvis.com/idl/}.

\bibitem[{Richardson}(1972)]{Richardson:1972}
W.~H. {Richardson}.
\newblock {Bayesian-based iterative method of image restoration}.
\newblock \emph{Optical Society of America Journal A}, 62:\penalty0 55--59,
  1972.

\bibitem[Rolland(2003)]{CalibrationLoic}
L.~Rolland.
\newblock Callibration in paris, method and results.
\newblock H.E.S.S. internal note, December 2003.

\bibitem[{Rots} et~al.(1998){Rots}, {Jahoda}, {Macomb}, {Kawai}, {Saito},
  {Kaspi}, {Lyne}, {Manchester}, {Backer}, {Somer}, {Marsden}, and
  {Rothschild}]{Rots:1998}
A.~H. {Rots}, K.~{Jahoda}, D.~J. {Macomb}, N.~{Kawai}, Y.~{Saito}, V.~M.
  {Kaspi}, A.~G. {Lyne}, R.~N. {Manchester}, D.~C. {Backer}, A.~L. {Somer},
  D.~{Marsden}, and R.~E. {Rothschild}.
\newblock {Rossi X-Ray Timing Explorer Absolute Timing Results for the Pulsars
  B1821-24 and B1509-58}.
\newblock \emph{\apj}, 501:\penalty0 749--+, July 1998.
\newblock \doi{10.1086/305836}.

\bibitem[{Ruderman}(1974)]{Ruderman:1974}
M.~A. {Ruderman}.
\newblock {Possible Consequences of Nearby Supernova Explosions for Atmospheric
  Ozone and Terrestrial Life}.
\newblock \emph{Science}, 184:\penalty0 1079--1081, June 1974.

\bibitem[{Saito} et~al.(1997){Saito}, {Kawai}, {Kamae}, and
  {Shibata}]{Saito:1997}
Y.~{Saito}, N.~{Kawai}, T.~{Kamae}, and S.~{Shibata}.
\newblock {Search for X-ray Pulsation from Rotation-Powered Pulsars with ASCA}.
\newblock In C.~D. {Dermer}, M.~S. {Strickman}, and J.~D. {Kurfess}, editors,
  \emph{AIP Conf. Proc. 410: Proceedings of the Fourth Compton Symposium},
  pages 628--+, 1997.

\bibitem[{Sako} et~al.(2000){Sako}, {Matsubara}, {Muraki}, {Ramanamurthy},
  {Dazeley}, {Edwards}, {Gunji}, {Hara}, {Hara}, {Holder}, {Kamei}, {Kawachi},
  {Kifune}, {Kita}, {Masaike}, {Mizumoto}, {Mori}, {Moriya}, {Muraishi},
  {Naito}, {Nishijima}, {Ogio}, {Patterson}, {Rowell}, {Sakurazawa}, {Sato},
  {Susukita}, {Suzuki}, {Tamura}, {Tanimori}, {Thornton}, {Yanagita},
  {Yoshida}, and {Yoshikoshi}]{Sako:2000}
T.~{Sako}, Y.~{Matsubara}, Y.~{Muraki}, P.~V. {Ramanamurthy}, S.~A. {Dazeley},
  P.~G. {Edwards}, S.~{Gunji}, T.~{Hara}, S.~{Hara}, J.~{Holder}, S.~{Kamei},
  A.~{Kawachi}, T.~{Kifune}, R.~{Kita}, A.~{Masaike}, Y.~{Mizumoto}, M.~{Mori},
  M.~{Moriya}, H.~{Muraishi}, T.~{Naito}, K.~{Nishijima}, S.~{Ogio}, J.~R.
  {Patterson}, G.~P. {Rowell}, K.~{Sakurazawa}, Y.~{Sato}, R.~{Susukita},
  R.~{Suzuki}, T.~{Tamura}, T.~{Tanimori}, G.~J. {Thornton}, S.~{Yanagita},
  T.~{Yoshida}, and T.~{Yoshikoshi}.
\newblock {Very High Energy Gamma-Ray Observations of PSR B1509-58 with the
  CANGAROO 3.8 Meter Telescope}.
\newblock \emph{\apj}, 537:\penalty0 422--428, July 2000.
\newblock \doi{10.1086/308998}.

\bibitem[{Schaefer}(1995)]{Schaefer:1995}
B.~E. {Schaefer}.
\newblock {`Supernova' 185 is Really a Nova Plus Comet P/Swift-Tuttle}.
\newblock \emph{\aj}, 110:\penalty0 1793--+, October 1995.
\newblock \doi{10.1086/117650}.

\bibitem[{Schellens}(2006)]{GDL}
M.~et~al. {Schellens}.
\newblock {GDL --- GNU Data Language}.
\newblock \url{http://gnudatalanguage.sourceforge.net/}, 2006.

\bibitem[{Schlenker}(2005)]{Schlenk}
S.~{Schlenker}.
\newblock \emph{Very High Energy Gamma Rays from the Binary Pulsar
  PSR\,B1259-63}.
\newblock PhD thesis, Humboldt-Universit\"at zu Berlin, 2005.

\bibitem[{Schmidt}(2005)]{Schmidt:Diplomarbeit}
F.~{Schmidt}.
\newblock {Search for Pulsed TeV Gamma-Ray Emission from Pulsars with H.E.S.S.}
\newblock Diplomarbeit, Humboldt-Universit\"at zu Berlin, 2005.

\bibitem[{Schmidt} et~al.(2005){Schmidt}, {Breitling}, {Gillessen},
  {Konopelko}, {Lohse}, {Schlenker}, {Schwanke}, {Stegmann}, and {Hess
  Collaboration}]{HDGSPulsarULs}
F.~{Schmidt}, F.~{Breitling}, S.~{Gillessen}, A.~{Konopelko}, T.~{Lohse},
  S.~{Schlenker}, U.~{Schwanke}, C.~{Stegmann}, and {Hess Collaboration}.
\newblock {Search for Pulsed TeV Gamma-Ray Emission from Young Pulsars with
  H.E.S.S.}
\newblock In \emph{AIP Conf. Proc. 745: High Energy Gamma-Ray Astronomy}, pages
  377--381, February 2005.

\bibitem[Schwanke and Lohse(2005)]{Schwanke:CorrelationCoefficient}
U.~Schwanke and T.~Lohse.
\newblock Calculating the error of correlation coefficients.
\newblock H.E.S.S. internal note, February 2005.

\bibitem[Schwanke and Lohse(2004)]{Schwanke:UpperLimit}
U.~Schwanke and T.~Lohse.
\newblock Calculation of upper limits and measurement errors for small signals.
\newblock H.E.S.S. internal note, Sep 2004.

\bibitem[Schwanke et~al.(2004)]{Schwanke:BrokenPixels}
U.~Schwanke et~al.
\newblock Recovering the intensities of broken pixels using interpolation and
  fitting techniques.
\newblock H.E.S.S. internal note, June 2004.

\bibitem[{Seward} and {Harnden}(1982)]{Seward:1982}
F.~D. {Seward} and F.~R. {Harnden}, Jr.
\newblock {A new, fast X-ray pulsar in the supernova remnant MSH 15-52}.
\newblock \emph{\apjl}, 256:\penalty0 L45--L47, May 1982.
\newblock \doi{10.1086/183793}.

\bibitem[{Seward} et~al.(1983){Seward}, {Harnden}, {Murdin}, and
  {Clark}]{Seward1983}
F.~D. {Seward}, F.~R. {Harnden}, P.~{Murdin}, and D.~H. {Clark}.
\newblock {MSH 15-52 - A supernova remnant containing two compact X-ray
  sources}.
\newblock \emph{Astrophysical Journal}, 267:\penalty0 698--710, April 1983.
\newblock \doi{10.1086/160907}.

\bibitem[{Standish}(1982)]{DE200}
E.~M. {Standish}.
\newblock {Orientation of the JPL Ephemerides, DE 200/LE 200, to the dynamical
  equinox of J 2000}.
\newblock \emph{\aap}, 114:\penalty0 297--302, October 1982.

\bibitem[{Starck} et~al.(2002){Starck}, {Pantin}, and {Murtagh}]{Starck:2002}
J.~L. {Starck}, E.~{Pantin}, and F.~{Murtagh}.
\newblock {Deconvolution in Astronomy: A Review}.
\newblock \emph{Publications of the Astronomical Society of the Pacific},
  114:\penalty0 1051--1069, October 2002.

\bibitem[{Steenberg}(1998)]{Steenberg:1998}
C.~D. {Steenberg}.
\newblock PhD thesis, University of Potchefstroom, South Africa, June 1998.

\bibitem[{Strom}(1994)]{Strom:1994}
R.~G. {Strom}.
\newblock {Supernova 185 its Associated Remnant and PSR:1509-58}.
\newblock \emph{\mnras}, 268:\penalty0 L5+, May 1994.

\bibitem[{Strong} et~al.(2000){Strong}, {Moskalenko}, and
  {Reimer}]{Strong:2000}
A.~W. {Strong}, I.~V. {Moskalenko}, and O.~{Reimer}.
\newblock {Diffuse Continuum Gamma Rays from the Galaxy}.
\newblock \emph{\apj}, 537:\penalty0 763--784, July 2000.
\newblock \doi{10.1086/309038}.

\bibitem[{Sturrock}(1971)]{Sturrock:1971}
P.~A. {Sturrock}.
\newblock {A Model of Pulsars}.
\newblock \emph{\apj}, 164:\penalty0 529--+, March 1971.

\bibitem[{Taylor} et~al.(2004){Taylor}, {Manchester}, {Nice}, et~al.]{TEMPO}
J.~H. {Taylor}, R.~N. {Manchester}, D.~{Nice}, et~al.
\newblock Tempo.
\newblock \url{http://www.atnf.csiro.au/research/pulsar/tempo/}, 2004.

\bibitem[{Thompson} et~al.(1999){Thompson}, {Bailes}, {Bertsch}, {Cordes},
  {D'Amico}, {Esposito}, {Finley}, {Hartman}, {Hermsen}, {Kanbach}, {Kaspi},
  {Kniffen}, {Kuiper}, {Lin}, {Lyne}, {Manchester}, {Matz},
  {Mayer-Hasselwander}, {Michelson}, {Nolan}, {{\"O}gelman}, {Pohl},
  {Ramanamurthy}, {Sreekumar}, {Reimer}, {Taylor}, and {Ulmer}]{Thompson:1999}
D.~J. {Thompson}, M.~{Bailes}, D.~L. {Bertsch}, J.~{Cordes}, N.~{D'Amico},
  J.~A. {Esposito}, J.~{Finley}, R.~C. {Hartman}, W.~{Hermsen}, G.~{Kanbach},
  V.~M. {Kaspi}, D.~A. {Kniffen}, L.~{Kuiper}, Y.~C. {Lin}, A.~{Lyne},
  R.~{Manchester}, S.~M. {Matz}, H.~A. {Mayer-Hasselwander}, P.~F. {Michelson},
  P.~L. {Nolan}, H.~{{\"O}gelman}, M.~{Pohl}, P.~V. {Ramanamurthy},
  P.~{Sreekumar}, O.~{Reimer}, J.~H. {Taylor}, and M.~{Ulmer}.
\newblock {Gamma Radiation from PSR B1055-52}.
\newblock \emph{Astrophysical Journal}, 516:\penalty0 297--306, May 1999.
\newblock \doi{10.1086/307083}.

\bibitem[{Trussoni} et~al.(1996){Trussoni}, {Massaglia}, {Caucino},
  {Brinkmann}, and {Aschenbach}]{Trussoni:1996}
E.~{Trussoni}, S.~{Massaglia}, S.~{Caucino}, W.~{Brinkmann}, and
  B.~{Aschenbach}.
\newblock {ROSAT PSPC observations of the supernova remnant MSH 15-52.}
\newblock \emph{\aap}, 306:\penalty0 581--+, February 1996.

\bibitem[{van der Swaluw}(2001)]{Swaluw:2001}
E.~{van der Swaluw}.
\newblock \emph{Supernova Remnants, Pulsar Wind Nebulae and Their Interaction}.
\newblock PhD thesis, Universiteit Utrecht, 2001.

\bibitem[{Vincent} et~al.(2003)]{Camera}
P.~{Vincent} et~al.
\newblock {Performance of the H.E.S.S. Cameras}.
\newblock In \emph{Proceedings of the 28th International Cosmic Ray
  Conference}, pages 2887--+, July 2003.

\bibitem[{Weekes} et~al.(1989){Weekes}, {Cawley}, {Fegan}, {Gibbs}, {Hillas},
  {Kowk}, {Lamb}, {Lewis}, {Macomb}, {Porter}, {Reynolds}, and
  {Vacanti}]{Weekes:1989}
T.~C. {Weekes}, M.~F. {Cawley}, D.~J. {Fegan}, K.~G. {Gibbs}, A.~M. {Hillas},
  P.~W. {Kowk}, R.~C. {Lamb}, D.~A. {Lewis}, D.~{Macomb}, N.~A. {Porter}, P.~T.
  {Reynolds}, and G.~{Vacanti}.
\newblock {Observation of TeV gamma rays from the Crab nebula using the
  atmospheric Cerenkov imaging technique}.
\newblock \emph{\apj}, 342:\penalty0 379--395, July 1989.
\newblock \doi{10.1086/167599}.

\bibitem[{Weisskopf} et~al.(2000){Weisskopf}, {Hester}, {Tennant}, {Elsner},
  {Schulz}, {Marshall}, {Karovska}, {Nichols}, {Swartz}, {Kolodziejczak}, and
  {O'Dell}]{Weisskopf:2000}
M.~C. {Weisskopf}, J.~J. {Hester}, A.~F. {Tennant}, R.~F. {Elsner}, N.~S.
  {Schulz}, H.~L. {Marshall}, M.~{Karovska}, J.~S. {Nichols}, D.~A. {Swartz},
  J.~J. {Kolodziejczak}, and S.~L. {O'Dell}.
\newblock {Discovery of Spatial and Spectral Structure in the X-Ray Emission
  from the Crab Nebula}.
\newblock \emph{\apjl}, 536:\penalty0 L81--L84, June 2000.
\newblock \doi{10.1086/312733}.

\bibitem[{Whiteoak} and {Green}(1996)]{Whiteoak:1996}
J.~B.~Z. {Whiteoak} and A.~J. {Green}.
\newblock {The MOST supernova remnant catalogue (MSC).}
\newblock \emph{Astronomy and Astrophysics Supplement}, 118:\penalty0 329--380,
  August 1996.

\bibitem[{Wigmans}(2000)]{Wigmans:2000}
R.~{Wigmans}.
\newblock \emph{{Calorimetry: Energy Measurement in Particle Physics}}, volume
  107 of \emph{International Series of Monographs on Physics}.
\newblock Oxford Univerisity Press, September 2000.

\bibitem[{Yao} et~al.(2006)]{ParticleDataBooklet:2006}
W.-M. {Yao} et~al.
\newblock {Review of Particle Physics}.
\newblock \emph{{Journal of Physics G}}, 33:\penalty0 1+, 2006.
\newblock URL \url{http://pdg.lbl.gov}.

\bibitem[{Yatsu} et~al.(2006){Yatsu}, {Kawai}, {Kataoka}, {Tamura}, and
  {Brinkmann}]{Yatsu:2006}
Y.~{Yatsu}, N.~{Kawai}, J.~{Kataoka}, T.~{Tamura}, and W.~{Brinkmann}.
\newblock {Chandra observation of RCW 89 at two epochs}.
\newblock In A.~{Wilson}, editor, \emph{ESA SP-604: The X-ray Universe 2005},
  pages 379--380, January 2006.

\end{thebibliography}
\bibliographystyle{plainnat}  


\chapter*{Abbreviations}

\begin{tabular}{ll}
  \vspace{.15cm}
  {\bf Abbreviation} & {\bf Meaning}    \\
  ADC & analog-to-digital conversion \\
  CCD & charge-coupled device \\
  CL & confidence level \\
  CMB & cosmic microwave background \\
  CT & Cherenkov telescope \\
  Dec & declination \\
  dof & degrees of freedom \\
  FOV & field of view \\
  H.E.S.S. & High Energy Stereoscopic System \\
  $IA$ & image amplitude \\
  IACT & imaging atmospheric Cherenkov telescop \\
  IC & inverse Compoton \\
  LED & light emitting diode \\ 
  MHD & magneto-hydrodynamic \\
  MJD & Modified Julian Date \\
  $MRSL$ & mean reduced scaled length \\
  $MRSW$ & mean reduced scaled width \\
  PMT & photo multiplier \\
  PWN & pulsar wind nebula \\
  PDF & probability density function \\
  PSF & point spread function \\
  RA & right ascension \\
  RL & Richardson-Lucy \\
  RMS & root mean square \\
  SNR & supernova remnant \\
  VHE & very high energy (10\,GeV$<E<$100\,TeV) \\
\end{tabular}



\chapter*{Acknowledgement}
I would like to thank everybody who has supported this work, in particular:
\begin{itemize}
\item Prof. Okkie de Jager for helpful discussions and new ideas for the
understanding of PWNs

\item Alexander Konopelko for his support with the topic of \MSH\ and his
  interest in image deconvolution applied to VHE \g-ray astronomy

\item my colleagues T\"ul\"un Ergin, Nukri Komin, Stefan Schlenker and Fabian
  Schmidt for interesting discussions, questions and advice

\item Conor Masterson, Bruno Kh\'elifi and Konrad Bernl\"ohr for good
  communications regarding the experiment, software development and data
  analysis

\item the Namibian locals, Toni Hanke, Eben Tjingaete and Maveipi Kandjii, Rosi
  and Winston for great times during observations in Namibia

\item Prof. Thomas Lohse for financial support

\item Dick Manchester from the Australia Telescope National Facility for recent
  pulsar ephemeris for the timing analysis of H.E.S.S. pulsar data

\item my family and friends for moral support.
\end{itemize}


\include{cv}


\selectlanguage{german}
\chapter*{Selbst"andigkeitserkl"arung}

Hiermit erkl"are ich, die vorliegende Arbeit selbst"andig ohne fremde Hilfe.
verfasst und nur die angegebene Literatur und erlaubte Hilfsmittel verwendet
zu haben.

\vspace{3cm}

Frank Breitling


\end{document}